\documentclass[a4paper,11pt,twoside]{book}

\usepackage{ae,lmodern}
\usepackage[english, french]{babel}
\usepackage[utf8]{inputenc}
\usepackage[T1]{fontenc}
\usepackage{xspace}
\usepackage{xcolor}
\usepackage{sidecap}
\usepackage{rotating}
\usepackage{numprint}

\usepackage{graphicx}
\usepackage{bm}
\usepackage[final]{pdfpages}
\usepackage[toc]{glossaries}
\usepackage{acronym}
\usepackage{xspace}
\usepackage{listings}
\usepackage{threeparttable}
\usepackage{pdflscape}
\usepackage{booktabs}
\usepackage{tabularx}
\usepackage{listings}
\usepackage[]{algorithm2e}
\usepackage{lipsum}
\usepackage{physics}
\usepackage{amsthm}
\usepackage{Config/aas_macro}
\usepackage{bigints}
\usepackage[titletoc,page]{appendix}

\theoremstyle{definition}
\newtheorem*{definition}{Définition}

\usepackage{mathtools}

\usepackage{setspace}
\usepackage[T1]{fontenc}
\usepackage[utf8]{inputenc}
\usepackage{lmodern}
\usepackage[english,french]{babel}
\usepackage{textgreek}

\usepackage{fixltx2e}

\usepackage{geometry} 
\reversemarginpar

\usepackage{amsmath,amssymb,amsfonts,amsthm}
\usepackage{mathrsfs}

\usepackage[locale = FR]{siunitx} 
\DeclareSIUnit\vitesse{\meter\per\second}
\usepackage{eurosym}
\DeclareSIUnit{\octet}{o}

\usepackage{graphicx,array,tikz,multirow}
\usepackage{caption,subcaption} 
\usepackage{svg,float}
\usepackage{booktabs,paralist}
\newcolumntype{x}[1]{>{\centering\arraybackslash\hspace{0pt}}p{#1}}
\usepackage[section]{placeins}
\usepackage{hanging}

\usepackage{fancyhdr,emptypage} 
\let\cleardoublepage\clearpage
 
\fancypagestyle{plain}{ 
    \fancyhead{}\fancyfoot[C]{\thepage}

}

\usepackage[Conny]{fncychap}
\ChTitleAsIs
\renewcommand{\thesection}{\arabic{section}}

\usepackage[francais,nohints,tight]{minitoc}		
\dominitoc

\usepackage[nottoc]{tocbibind} 
\usepackage[titles]{tocloft}

\usepackage{textcomp} 
\usepackage{xcolor} 
\usepackage{epigraph} 
\usepackage{titling}
\usepackage{lipsum} 
\usepackage{csquotes} 

\usepackage{xspace}
\usepackage{afterpage}

\usepackage[textsize=footnotesize]{todonotes}

\usepackage{bookmark}
\usepackage{acronym}
\usepackage[nameinlink,french]{cleveref} 
\Crefname{figure}{Fig.}{Figs.} 
\crefname{figure}{fig.}{figs.}
\Crefname{equation}{Eq.}{Eqs.}
\crefname{equation}{eq.}{eqs.}
\Crefname{table}{Table.}{Tables.}
\crefname{table}{table.}{tables.}

\definecolor{color_ref}{rgb}{0.1, 0.1, 0.45} 
\definecolor{color_link}{RGB}{0,0,255}
\definecolor{curcolor}{RGB}{0,0,255} 

\hypersetup{
	colorlinks=true, 
	pdfstartview=FitV, 
	urlcolor=color_link, 
	linkcolor= color_link, 
	citecolor=color_ref 
}

\usepackage[hyperref=true,natbib=true,
			backref=true,date=year,
			backend=biber,
			url=false,doi=false,isbn=false,%
			minbibnames=6, 
			maxbibnames=6, 
			maxcitenames=1,mincitenames=1, 
			maxalphanames=1, 
			style=alphabetic,
			sorting=nyt]{biblatex}

\renewbibmacro{in:}{} 

\DeclareLabelalphaTemplate{
  \labelelement{
    \field[final]{shorthand}
    \field{labelname}
    \field{label}
  }
  \labelelement{\literal{,\addhighpenspace}}
  \labelelement{\field{year}}
}

\AtEveryBibitem{%
    \clearfield{note} 
    \clearlist{language} 
}
\newcommand{\sujet}[1]{\renewcommand{\sujet}{#1}}
\newcommand{\auteur}[1]{\renewcommand{\auteur}{#1}}
\newcommand{\encadrant}[1]{\renewcommand{\encadrant}{#1}}

\newcommand{\degree}{^{\circ}}

\usepackage[nomaketitle]{Config/psl-cover/psl-cover}

\sujet{Astrophysique Extrême avec GRAVITY : \\sursauts énergétiques aux abords de l'horizon des événements du trou noir central de la Galaxie}
\author{Nicolas Aimar}
\encadrant{Directeur \textsc{De Recherche}}

\institute{l'Observatoire de Paris}
\doctoralschool{Astronomie et Astrophysique d'île-de-France}{127}
\specialty{Astronomie et Astrophysique}
\laboratory{LESIA}
\date{03 Octobre 2023}

\pslassetspath{Config/psl-cover}

\jurymember{1}{Eric \textsc{Gourgoulhon}}{Directeur de recherche, LUTh}{} 
\jurymember{2}{Didier \textsc{Barret}}{Directeur de recherche, IRAP}{Rapporteur}
\jurymember{3}{Katia \textsc{Ferrière}}{Directrice de recherche, IRAP}{Rapportrice}
\jurymember{4}{Benoît \textsc{Cerutti}}{Chargé de recherche, IPAG}{Examinateur}
\jurymember{5}{Pierre-Olivier \textsc{Petrucci}}{Directeur de recherche, IPAG}{Examinateur}
\jurymember{6}{Peggy \textsc{Varnière}}{Chargée de recherche, APC}{Examinatrice}
\jurymember{7}{Frédéric \textsc{Vincent}}{Chargé de Recherche, LESIA}{Directeur de Thèse}
\jurymember{8}{Thibaut \textsc{Paumard}}{Chargé de Recherche, LESIA}{Directeur de Thèse}
\jurymember{9}{Guy \textsc{Perrin}}{Astronome, LESIA}{Invité}

\frabstract{
Le centre de la Voie Lactée abrite un trou noir supermassif de 4 millions de masses solaires nommé Sagittarius~A*. Cet objet est observé depuis plus de 20 ans en infrarouge proche. Cela a permis de confirmer des effets de la Relativité Générale. De plus, les observations récurrentes ont permis de détecter des sursauts, c'est-à-dire un flux beaucoup plus important que la moyenne, avec une variabilité de l'ordre de 30 min à 1h. En 2018, GRAVITY, utilisant les 4 grands télescopes du VLTI/ESO, a observé un mouvement orbital de la source de ces sursauts. Dans cette thèse, on s'intéresse à la modélisation de ces sursauts, à travers des modèles plus ou moins complexe, dont un basé sur le phénomène de reconnexion magnétique. Cette dernière correspond à un changement brutal de la configuration magnétique autour du trou noir, libérant une grande quantité d'énergie. On s'intéresse aussi au problème de la polarisation en Relativité Générale.
}

\enabstract{
The centre of the Milky Way hosts a supermassive black hole of 4 million solar masses called Sagittarius~A*. This object has been observed for more than 20 years in the near infrared. This has confirmed some effects of General Relativity. In addition, recurrent observations have made it possible to detect flares, i.e. a much larger flux than the average, with a variability of the order of 30 minutes to 1 hour. In 2018, GRAVITY, using the 4 large telescopes of the VLTI/ESO, observed an orbital motion of the source of these flares. This thesis focuses on the modelling of these flares, using models of varying complexity, including one based on the phenomenon of magnetic reconnection. The latter corresponds to an abrupt change in the magnetic configuration around the black hole, releasing a large amount of energy. We are also looking at the problem of polarisation in General Relativity.
}

\frkeywords{Transfert radiatif, Phénomènes d'accrétion, Interférométrie, Trous noirs.}
\enkeywords{Radiative transfer, Accretion flow, Interferometry, Black holes.}

\title{\sujet}
\hypersetup{
pdftitle={Thesis Year},
pdfsubject={\sujet},
pdfauthor={\theauthor},
pdfkeywords={\thefrkeywords},
}
\RequirePackage[normalem]{ulem} 
\RequirePackage{color}\definecolor{RED}{rgb}{1,0,0}\definecolor{BLUE}{rgb}{0,0,1} 
\providecommand{\DIFaddbegin}{} 
\providecommand{\DIFaddend}{} 
\providecommand{\DIFdelbegin}{} 
\providecommand{\DIFdelend}{} 
\providecommand{\DIFaddbeginFL}{} 
\providecommand{\DIFaddendFL}{} 
\providecommand{\DIFdelbeginFL}{} 
\providecommand{\DIFdelendFL}{} 
\newcommand{\DIFscaledelfig}{0.5}
\RequirePackage{settobox} 
\RequirePackage{letltxmacro} 
\newsavebox{\DIFdelgraphicsbox} 
\newlength{\DIFdelgraphicswidth} 
\newlength{\DIFdelgraphicsheight} 
\LetLtxMacro{\DIFOincludegraphics}{\includegraphics} 
\newcommand{\DIFaddincludegraphics}[2][]{{\color{blue}\fbox{\DIFOincludegraphics[#1]{#2}}}} 
\newcommand{\DIFdelincludegraphics}[2][]{
\sbox{\DIFdelgraphicsbox}{\DIFOincludegraphics[#1]{#2}}
\settoboxwidth{\DIFdelgraphicswidth}{\DIFdelgraphicsbox} 
\settoboxtotalheight{\DIFdelgraphicsheight}{\DIFdelgraphicsbox} 
\scalebox{\DIFscaledelfig}{
\parbox[b]{\DIFdelgraphicswidth}{\usebox{\DIFdelgraphicsbox}\\[-\baselineskip] \rule{\DIFdelgraphicswidth}{0em}}\llap{\resizebox{\DIFdelgraphicswidth}{\DIFdelgraphicsheight}{
\setlength{\unitlength}{\DIFdelgraphicswidth}
\begin{picture}(1,1)
\thicklines\linethickness{2pt} 
{\color[rgb]{1,0,0}\put(0,0){\framebox(1,1){}}}
{\color[rgb]{1,0,0}\put(0,0){\line( 1,1){1}}}
{\color[rgb]{1,0,0}\put(0,1){\line(1,-1){1}}}
\end{picture}
}\hspace*{3pt}}} 
} 
\LetLtxMacro{\DIFOaddbegin}{\DIFaddbegin} 
\LetLtxMacro{\DIFOaddend}{\DIFaddend} 
\LetLtxMacro{\DIFOdelbegin}{\DIFdelbegin} 
\LetLtxMacro{\DIFOdelend}{\DIFdelend} 
\DeclareRobustCommand{\DIFaddbegin}{\DIFOaddbegin \let\includegraphics\DIFaddincludegraphics} 
\DeclareRobustCommand{\DIFaddend}{\DIFOaddend \let\includegraphics\DIFOincludegraphics} 
\DeclareRobustCommand{\DIFdelbegin}{\DIFOdelbegin \let\includegraphics\DIFdelincludegraphics} 
\DeclareRobustCommand{\DIFdelend}{\DIFOaddend \let\includegraphics\DIFOincludegraphics} 
\LetLtxMacro{\DIFOaddbeginFL}{\DIFaddbeginFL} 
\LetLtxMacro{\DIFOaddendFL}{\DIFaddendFL} 
\LetLtxMacro{\DIFOdelbeginFL}{\DIFdelbeginFL} 
\LetLtxMacro{\DIFOdelendFL}{\DIFdelendFL} 
\DeclareRobustCommand{\DIFaddbeginFL}{\DIFOaddbeginFL \let\includegraphics\DIFaddincludegraphics} 
\DeclareRobustCommand{\DIFaddendFL}{\DIFOaddendFL \let\includegraphics\DIFOincludegraphics} 
\DeclareRobustCommand{\DIFdelbeginFL}{\DIFOdelbeginFL \let\includegraphics\DIFdelincludegraphics} 
\DeclareRobustCommand{\DIFdelendFL}{\DIFOaddendFL \let\includegraphics\DIFOincludegraphics} 
\RequirePackage{listings} 
\RequirePackage{color} 
\lstdefinelanguage{DIFcode}{ 
  moredelim=[il][\color{red}\sout]{\%DIF\ <\ }, 
  moredelim=[il][\color{blue}\uwave]{\%DIF\ >\ } 
} 
\lstdefinestyle{DIFverbatimstyle}{ 
	language=DIFcode, 
	basicstyle=\ttfamily, 
	columns=fullflexible, 
	keepspaces=true 
} 
\lstnewenvironment{DIFverbatim}{\lstset{style=DIFverbatimstyle}}{} 
\lstnewenvironment{DIFverbatim*}{\lstset{style=DIFverbatimstyle,showspaces=true}}{} 

\begin{document}
\pslcover{} 

\frontmatter

\chapter*{Remerciements}
\markboth{Remerciements}{Remerciements}
\vspace*{1em}

Je souhaite tout d'abord remercier les membres du jury d'avoir accepté de prendre le temps de lire ma thèse et de participer à la soutenance. Je sais qu'il est difficile de trouver du temps dans des agendas déjà très denses, j'apprécie donc beaucoup cet effort.\\

Je souhaite évidemment remercier mes Directeurs de Thèse, Frédéric Vincent et Thibaut Paumard, qui m’ont conseillé, guidé et soutenu pendant toutes ces années de recherche. Leur disponibilité et leurs observations avisées m’ont été précieuses dans la réalisation du présent travail. Merci pour leur confiance.\\

Je voudrais aussi remercier chaleureusement, l'ensemble de mes collègues du Bâtiment~5 et de l'équipe Centres GalactiqueS, chercheurs et enseignants-chercheurs (en particulier Guy Perrin et Daniel Rouan), postdocs (Miguel Montargès, Gernot Heißel), doctorants (Karim Abd~El~Dayem, Garance Bras), ingénieurs (Vincent Lapeyrere, Eitan Perchevis) ainsi que les stagiaires, pour tous les conseils qu'ils m'ont apportés et leur bonne humeur. Cela a été un vrai plaisir de partager ces trois ans avec vous (malgré les restrictions Covid). Je remercie aussi, particulièrement, l'équipe de doctorants du Bâtiment 16 (exoplanètes). Je remercie aussi particulièrement, mon collègue de bureau Miguel Montargès pour son aide et son support quotidien.\\

Les collaborations étant le cœur de la recherche scientifique, je remercie l'ensemble de mes collaborateurs, dans l'ordre chronologique, Anton Dmytriiev (LUTh puis North-West University South Africa), Andreas Zech (LUTh) avec qui j'ai construit le modèle de plasmoïde, Raphaël Mignon-Risse (APC), Peggy Varnière (APC) avec qui j'ai collaboré pour une interface simulations-Gyoto, Ileyk El~Mellah (IPAG puis Universidad de Santiago de Chile), Benoit Cerutti (IPAG) qui m'ont aidé concernant les simulations (GR)PIC, Bart Ripperda (Princeton University) pour les discussions très intéressantes concernant les simulations GRMHD, Maciek Wielgus (MPIfR) pour son aide concernant la polarisation et les sursauts en radio, Sebastiano von~Fellenberg (MPIfR), Michi Baudöck (MPIfR) pour notre collaboration sur une analyse statistique des sursauts. Enfin, je remercie, l'ensemble des membres de la collaboration GRAVITY et le personnel de l'ESO sans qui ce travail n'aurait pas eu lieu.\\

Je remercie infiniment l'ensemble de mes relecteurs, à savoir Frédéric Vincent, Thibaut Paumard, Guy Perrin, ma mère, Isabelle Filatriau, et Mathéo Collin pour leur aide dans la correction de cette thèse. Dédicace aussi à mes compagnons de rédaction, Mathilde Mâlin et Achrène Dyrek.\\

Je n'aurais pas réussi mes études et cette thèse sans le soutien de mes proches, en particulier ma famille, dont mon père Patrick Aimar et ma mère Isabelle Filatriau et de mes amis. Ne pouvant tous les citer, je remercie particulièrement Maxime Defrances, Mathéo Collin, Maïder Lay, Valentine Maillard, Mathilde Mâlin, Quentin Jacopin, Anna Luashvilli, Thomas Gouze, Benjamin Robert pour leur soutien et leur confiance au quotidien.\\

Cette thèse n'aurait pas pu être réalisée sans le financement de l'École Doctorale 127 Astronomie et Astrophysique d'Île-de-France et les moyens mis à ma disposition par l'Observatoire de Paris et PSL.\\

Enfin, je voudrais remercier toutes les personnes qui m'ont aidé pour mon orientation professionnelle et mon parcours académique. Parmi elles, je remercie particulièrement, Rémi Cabanac (OMP) et Jihane Moultaka (IRAP) car leur rencontre a conforté mon souhait de faire de l'astrophysique mon métier. Je remercie aussi les différentes associations d'astronomie auxquelles j'ai adhéré et qui ont entretenu ma passion pour l'astronomie et l'astrophysique.\\

Je n'en serai pas là aujourd'hui sans l'ensemble de ces personnes et plein d'autres que je n'ai pas citées, alors du fond du cœur, merci !

\clearpage

	\thispagestyle{plain}
	\section*{\Huge Résumé en Français}
La Voie Lactée, comme a priori toutes les galaxies structurées, abrite en son cœur un trou noir supermassif, d'environ 4,3 millions de masses solaires, nommé Sagittarius A* (Sgr~A*). Sa taille, liée à sa masse, et sa proximité d'environ 8,3 kpc, en font le trou noir avec la plus grande taille angulaire dans le ciel ($\sim 20\, \mu$as), ce qui fait de Sgr~A* la cible d'études idéales de ce genre d'objets. Les trous noirs sont les objets les plus compacts de l'Univers, avec un champ gravitationnel extrême proche de leur horizon. La description de ces objets et de leur environnement proche nécessite la prise en compte de la Relativité Générale, introduite en 1915 par Albert Einstein.

Depuis plus de 20 ans, Sgr~A* et son environnement sont la cible de nombreuses campagnes d'observations à différentes longueurs d'ondes (radio, infrarouge, rayons X). Le suivi en infrarouge proche (NIR) des orbites des étoiles-S contenues dans la seconde d'angle autour de Sgr~A* a permis de prouver certains effets prédits par la Relativité Générale, tels que la précession de Schwarzschild. Les observations en rayons X et en NIR ont montré que Sgr~A* présente une importante variabilité du flux émis par le flot d'accrétion, avec des sursauts dont le flux peut atteindre jusqu'à $\sim$100 fois le flux médian. L'avènement de l'optique adaptative et de l'interférométrie optique, notamment avec les quatre grands télescopes du VLTI et l'instrument GRAVITY, ont permis de mettre en évidence un mouvement orbital de l'origine de trois sursauts observés en 2018.

De nombreux modèles ont été envisagés pour expliquer les sursauts de Sgr~A*, mais l'observation d'un mouvement orbital a fortement contraint ces modèles. Parmi eux, le modèle analytique de point chaud est largement utilisé avec différents degrés de complexité et d'hypothèses. En parallèle du développement des modèles analytiques, de nombreuses simulations d'accrétion autour de trous noirs ont été réalisées avec un intérêt particulier pour le phénomène de reconnexion magnétique qui apparaît comme un scénario plausible pour expliquer l'origine des sursauts de Sgr~A*.

Dans cette thèse, nous étudions différents modèles pour les sursauts de Sgr~A* à l'aide du code de tracé de rayons \textsc{Gyoto}, allant d'un modèle analytique de point chaud avec une variabilité intrinsèque à un modèle semi-analytique basé sur la reconnexion magnétique.
Le premier modèle est très utile pour comprendre les effets de la Relativité (Restreinte et Générale) sur les observables (astrométries et courbes de lumière), ainsi que l'influence de la variabilité intrinsèque sur celles-ci.
Le second modèle est motivé par un phénomène physique particulier, la reconnexion magnétique, et est construit à partir des résultats des simulations numériques. Dans ce modèle, la vitesse azimutale du point chaud est libre d'être super-Képlérienne, en raison de l'entraînement du site de reconnexion par les lignes de champ magnétique. Cette propriété permet de répondre à une contrainte observationnelle des sursauts de 2018 observés par GRAVITY que les modèles précédents ne parvenaient pas à expliquer.
De plus, nous étudions également l'impact de la modélisation de l'état quiescent combiné aux sursauts sur les observables. La contribution de celui-ci dans les calculs d'astrométrie se traduit par un décalage entre la position du trou noir et le centre de l'orbite apparente du sursaut, ce qui constitue une autre conclusion des observations des sursauts de 2018.

En plus des astrométries et des courbes de lumière, GRAVITY a mesuré la polarisation des sursauts de 2018. Le code de tracé de rayons \textsc{Gyoto} est maintenant capable de calculer la polarisation des images. La nouvelle version du code a été validée en comparant les résultats avec un autre code de tracé de rayons, \textsc{ipole}.

Le modèle basé sur la reconnexion magnétique montre des résultats très encourageants et peut être encore amélioré pour prendre en compte la polarisation, ainsi que les propriétés multi-longueurs d'onde des sursauts de Sgr~A*.

	\vspace*{0.5cm}
	{\noindent\large\textbf{Mots clés :} Transfert radiatif ; Phénomènes d'accrétion ; Interférométrie ; Trous noirs.}
\clearpage

\begin{otherlanguage}{english} 
	\thispagestyle{plain}
	\section*{\Huge Abstract}
The Milky Way, like presumably all structured galaxies, harbors a supermassive black hole at its core, approximately 4.3 million times the mass of the Sun, named Sagittarius A* (Sgr~A*). Its size, determined by its mass, and its proximity of about 8.3 kpc make it the black hole with the largest angular size in the sky ($\sim 20 \mu$as), making it the ideal target for studying this type of object. Black holes are the most compact objects in the Universe, with an extreme gravitational field near their event horizon. Describing these objects and their immediate environment requires taking into account General Relativity, introduced in 1915 by Albert Einstein.

For over 20 years, Sgr~A* and its environment have been the subject of numerous observation campaigns at various wavelengths (radio, IR, X-rays). Tracking the orbits of S-stars in NIR within one arcsecond around Sgr~A* has provided evidence for certain effects predicted by General Relativity, such as Schwarzschild precession. X-ray and NIR observations have shown that Sgr~A* exhibits significant variability in the emitted flux from the accretion flow, with flares that can reach up to $\sim$100 times the median flux. The advent of adaptive optics and optical interferometry, particularly with the four large telescopes of the VLTI and the GRAVITY instrument, have revealed an orbital motion of the origin of three flares observed in 2018.

Numerous models have been proposed to explain the flares of Sgr~A*, but the observation of orbital motion has strongly constrained these models. Among them, the analytical hot spot model is widely used with varying degrees of complexity and assumptions. In parallel with the development of analytical models, numerous simulations of accretion around black holes have been realized, with a particular focus on the phenomenon of magnetic reconnection, which appears as a plausible scenario to explain the origin of the flares of Sgr~A*.

In this thesis, we study different models for the flares of Sgr~A* using the ray-tracing code \textsc{Gyoto}, ranging from an analytical hot spot model with intrinsic variability to a semi-analytical model based on magnetic reconnection. The first model is very useful for understanding the effects of Relativity (Special and General) on observables (astrometry and light curves), as well as the influence of intrinsic variability on them. The second model is motivated by a specific physical phenomenon, magnetic reconnection, and is constructed based on the results of numerical simulations. In this model, the azimuthal velocity of the hot spot is free to be super-Keplerian, due to the dragging of the reconnection site by the magnetic field lines. This property aswer to an observational constraint of the 2018 flares observed by GRAVITY that previous models failed to explain. Additionally, we also study the impact of modeling the quiescent state combined with the flares on the observables. Its contribution in astrometry calculations results in a shift between the position of the black hole and the center of the hot spot's apparent orbit, which is another conclusion from the observations of the 2018 flares.

In addition to astrometry and light curves, GRAVITY has measured the polarization of the 2018 flares. The \textsc{Gyoto} ray-tracing code is now capable of calculating the polarization of images. The new version of the code has been validated by comparing the results with another ray-tracing code, \textsc{ipole}.

The model based on magnetic reconnection shows very promising results and can be further improved to account for polarization, as well as the multi-wavelength properties of the flares of Sgr~A*.

	\vspace*{0.5cm}
	{\noindent\large\textbf{Keywords:} Radiative transfer; Accretion flow; Interferometry, Black holes.}
\end{otherlanguage}

\newpage

\begin{singlespace} 
\phantomsection 
{
\hypersetup{linkcolor=black}
\setcounter{tocdepth}{2}
\tableofcontents\newpage
}
\thispagestyle{plain}
\mbox{}
\newpage
\end{singlespace}

\mainmatter
\setcounter{page}{1}

\chapter*{Introduction}\label{chap:intro}
\addcontentsline{toc}{part}{Introduction}
\markboth{Introduction}{Introduction}

Les phénomènes physiques peuvent être décrits par le biais de quatre interactions fondamentales, les forces nucléaires forte et faible, l'électromagnétisme et la gravité. Si le modèle standard arrive à bien décrire les trois premières, la meilleure théorie pour décrire la gravitation à l'heure actuelle est la relativité générale (RG) introduite par Albert Einstein en 1915. Dans cette théorie, la gravitation n'est plus une force, mais la conséquence de la courbure de l'espace-temps par un objet massif. Elle a rapidement été confrontée avec succès à des observations telles que l'anomalie de l'avance du périhélie de Mercure autour du Soleil, c'est-à-dire le changement d'orientation de l'ellipse dans son plan orbital, ainsi que la déviation par le Soleil de la lumière provenant d'étoiles d'arrière-plan observées lors d'éclipses. Cette théorie a été développée et testée depuis plus d'un siècle par des expériences de plus en plus poussées, de l'échelle du système solaire à celle de la cosmologie, sans jamais échouer.

Alors que la relativité générale est bien vérifiée dans le régime des champs gravitationnels faibles, elle n'est pas aussi bien contrainte dans les champs gravitationnels forts, comme aux abords des trous noirs (black hole en anglais, BH). Même si la métrique des trous noirs de Schwarzschild (sans rotation) était la première solution exacte des équations d'Einstein trouvée par Karl Schwarzschild en 1915, quelques mois après la publication de la RG, l'observation et la preuve de l'existence de ces objets sont assez récentes. En effet, en raison de leur nature "sombre", leur détection/observation n'est possible que par l'émission de la matière environnante accrétée par le trou noir ou par leur influence gravitationnelle. Cette dernière offre un bon moyen de contraindre la RG sans incertitudes astrophysiques à travers les effets de lentille gravitationnelle ou de précession orbitale, mais le premier est très rare, car il demande un alignement très précis, le second, plus commun, notamment au centre de la Voie Lactée (voir Chapitre~\ref{chap:GC}), contraint les effets à grandes distances. Au contraire, l'étude de la matière environnant les trous noirs permet de sonder la RG aux abords de l'horizon des événements, mais dépend de nombreux phénomènes physiques de cette même matière en champ fort.

L'observation de flux d'accrétion autour des trous noirs est plus facile que d'observer l'influence gravitationnelle de ces derniers grâce aux binaires-X pour des trous noirs de masse stellaire et aux noyaux actifs de galaxie (AGN) pour des trous noirs supermassifs (SMBH). L'échelle de temps dynamique du flux d'accrétion dépend de la masse du trou noir, allant de quelques millisecondes ou moins pour les trous noirs de masse stellaire à quelques jours ou semaines pour les SMBH comme M87* ou les AGN rendant l'étude de certains phénomènes plus ou moins complexe. L'échelle de temps dynamique de Sagittarius A* (Sgr A*), le SMBH au centre de notre galaxie qui est de l'ordre de quelques dizaines de secondes, et sa proximité (voir Chapitre \ref{chap:GC}) en font un candidat idéal pour l'étude de la physique des trous noirs.

Cette thèse se concentre sur la physique d'accrétion de Sagittarius A* et plus particulièrement sur ses éruptions tant du point de vue observationnel que théorique.

\part{Contexte}
\chapter{Le Centre de notre Galaxie}\label{chap:GC}
\markboth{Le Centre de notre Galaxie}{Le Centre de notre Galaxie}
{
\hypersetup{linkcolor=black}
\minitoc 
}

\section{Une vue globale du Centre Galactique}
\begin{figure}
    \centering
    \resizebox{.8\hsize}{!}{\includegraphics{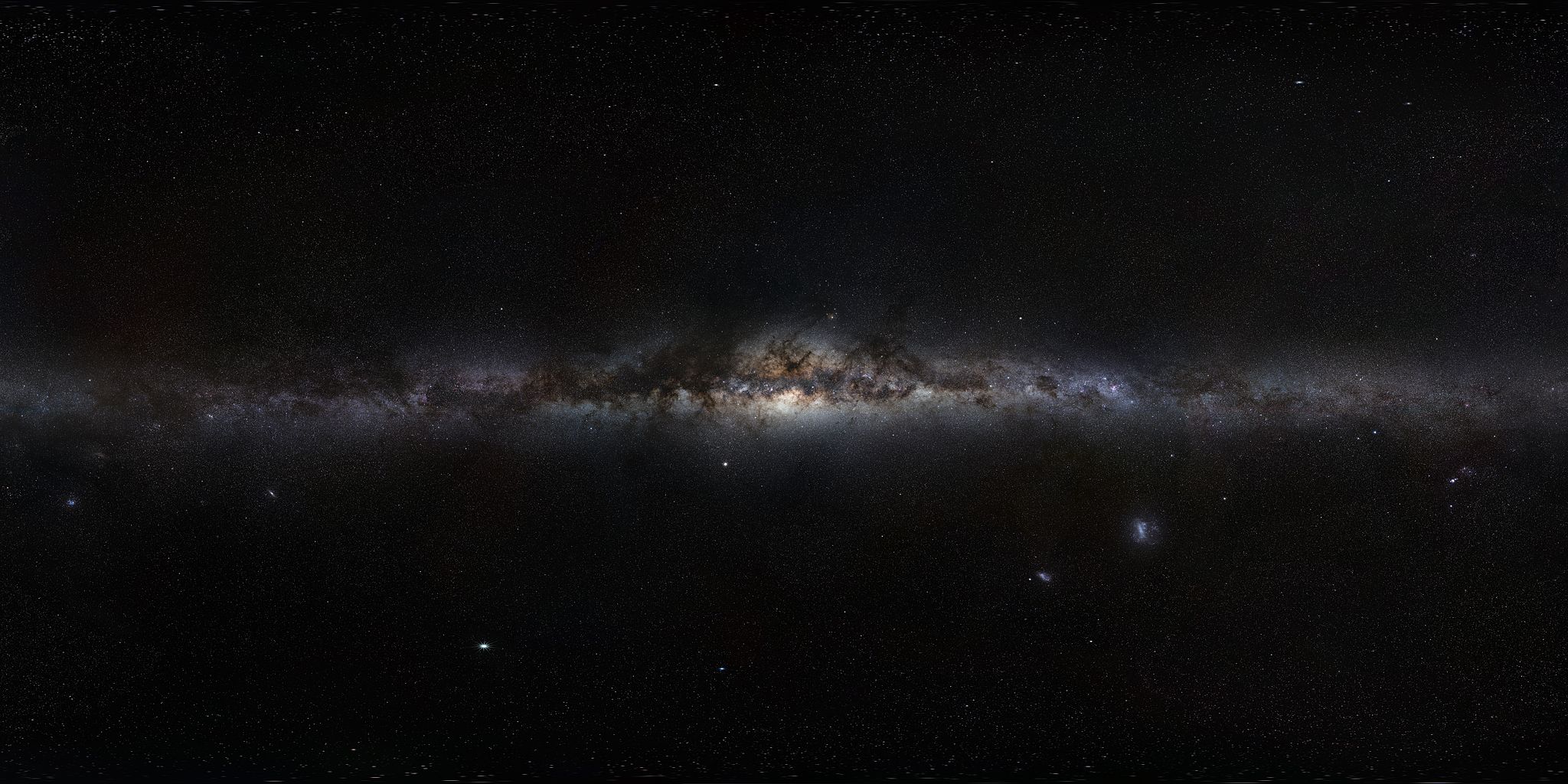}}
    \caption{Panorama à 360° de la sphère céleste dans le domaine visible, centrée sur le centre de la Voie Lactée. La présence de nuages moléculaires et de poussière dans le plan de la galaxie atténue fortement la lumière émise en arrière-plan et notamment celle du centre galactique. Crédit : ESO/S. Brunier}
    \label{fig:Voie_Lactée_visible}
\end{figure}

\begin{figure}
    \centering
    \resizebox{0.8\hsize}{!}{\includegraphics{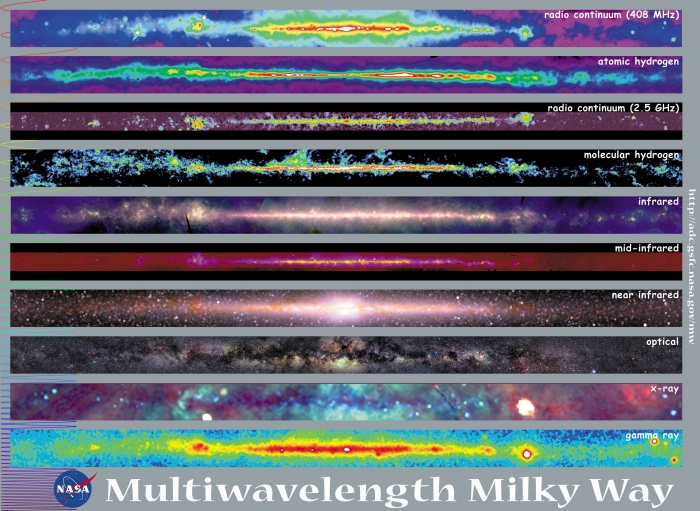}}
    \caption{Images à 360° du plan de la Voie Lactée à différentes longueurs d'ondes. De \textit{haut} en \textit{bas} : radio (480 MHz), raie de l'hydrogène atomique (21 cm), radio (2,4-2,7 GHz), hydrogène moléculaire (115 GHz), composite infrarouge (12 $\mu$m en bleu, 60 $\mu$m en vert et 100 $\mu$m en rouge), infrarouge moyen (7-11 $\mu$m), composite infrarouge proche (1,25 $\mu$m en bleu, 2,2 $\mu$m en vert et 3,5 $\mu$m en rouge), visible (0,4-0,6 $\mu$m), rayon-X doux (0,25 keV en bleu, 0,75 keV $\mu$m en vert et 1,5 keV en rouge) et rayon $\gamma$ (>300 MeV). Ces images proviennent de différents relevés faits avec des instruments terrestres et spatiaux. Crédit : \href{https://www.nasa.gov/goddard}{NASA Goddard Space Flight Center}.}
    \label{fig:Voie_Lactée_multilambda}
\end{figure}

Comme dans toutes les galaxies structurées, la densité de matière et d'étoiles de notre galaxie croît au fur et à mesure que l'on se rapproche de son centre. Alors que la densité d'étoiles au voisinage du Soleil est de l'ordre de 0,2 étoile par parsec cube, celle du centre galactique est estimée à 10 millions d'étoiles par parsec cube, dans le parsec central. De plus, le système solaire se situant dans le plan de la galaxie à environ 8,3 kpc du centre galactique~\cite{Gravity2020a}, la lumière issue de ce dernier doit traverser une grande quantité de matière interstellaire composée de gaz et de poussières et est donc fortement atténuée dans le domaine visible, $A_V \approx 30$~\cite{Scoville2003}, comme le montre la~Figure~\ref{fig:Voie_Lactée_visible}. Cependant, les longueurs d'ondes plus grandes (infrarouges, radios) et plus courtes (rayons~X durs, rayons $\gamma$) sont moins impactées par l'absorption interstellaire et tracent différentes composantes~(Fig.~\ref{fig:Voie_Lactée_multilambda}). La radio permet d'observer le gaz, notamment l'hydrogène sous forme moléculaire ($115$ GHz) ou atomique ($\lambda=21$ cm). Parmi les nombreuses fenêtres d'observation en radio, on retient celle à 230 GHz utilisée par la collaboration \textit{Event Horizon Telescope} (EHT, voir ci-dessous). L'infrarouge peut être divisé en trois catégories, IR lointain entre 15 $\mu$m et 1 mm, IR moyen de 2,5 à 15 $\mu$m et IR proche entre 800 nm et 2,5 $\mu$m. L'infrarouge lointain et moyen sont majoritairement dominés par la poussière interstellaire, les molécules complexes et les étoiles géantes rouges. L'infrarouge proche est dominé quant à lui par les étoiles géantes froides. Alors que la poussière absorbe une grande partie du rayonnement dans le domaine visible, ce n'est pas le cas dans l'IR proche avec $A_{Ks} \approx 2,5$~\cite{Fritz2011}, ce qui en fait une fenêtre idéale pour l'observation du centre galactique. En rayon X, l'émission étendue correspond à du gaz chaud auquel s'ajoutent des sources ponctuelles~\cite{Muno2003} correspondant à des objets compacts (pulsars, magnétars, trous noirs, etc) ou à des restes d'explosion d'étoiles, entre autres. Comme pour le domaine visible, les rayons~X doux sont beaucoup absorbés par le milieu interstellaire, néanmoins, c'est moins le cas pour les rayons~X durs. Enfin, les rayons $\gamma$ tracent les sources les plus énergétiques comme le pulsar du Crabe à droite de l'image en rayons gamma au bas de la Fig.~\ref{fig:Voie_Lactée_multilambda}. L'émission diffuse est due, entre autres, aux collisions entre des rayons cosmiques avec les molécules d'hydrogène du gaz interstellaire. En plus des sources principales dans chaque gamme de longueur d'onde citée précédemment, certaines sources émettent dans un grand nombre de domaines. En effet, les particules énergétiques plongées dans un champ magnétique émettent du rayonnement synchrotron (plus de détails dans le Chap.~\ref{chap:modele_hotspot+jet}) visible en radio, IR, rayons X et, dans certaines conditions, jusqu'au rayon $\gamma$. Ce type de rayonnement est d'un intérêt majeur dans le cas de Sagittarius~A*.

Ainsi, le centre galactique a été la cible de nombreux télescopes terrestres et spatiaux en radio, infrarouge et rayons~X principalement. La région centrale de la galaxie tire son nom du domaine radio. En effet, en 1955, une source très brillante a été détectée dans la constellation du Sagittaire \cite{Pawsey1955} et a naturellement été nommée Sagittarius A (même si la première détection imprécise, qui est aussi la première détection d'une source astrophysique, date de 1933 par Karl Jansky). Le panneau en haut de la Figure~\ref{fig:GC_scales} montre un champ large (quelques degrés carrés) du centre galactique observé par \href{https://www.sarao.ac.za/gallery/meerkat/}{MeerKAT} composé de différentes sources dont Sagittarius A ainsi que des restes de supernovae (SNR) et des filaments traçant le champ magnétique. Des observations ultérieures ont révélé que la région de Sagittarius A, de quelques parsecs de rayon, soit quelques minutes d'arc, était composée d'un amas dense d'étoiles, mais aussi de gaz neutre et ionisé avec des températures très élevées~($\sim 10^6$ K) émettant des rayons~X \citep{Baganoff2001,Baganoff2003,Muno2004}. Le VLA (Very Large Array) a permis de mettre en évidence que cette source a une structure complexe avec de nombreuses composantes. Toute la région centrale est entourée, en projection, par un reste de supernovae nommé Sagittarius~A East d'un rayon d'environ huit parsecs, dont on estime l'âge à environ 10.000 ans \citep{Maeda2002}. À l'intérieur, deux structures se dégagent : la première est un tore composé de nuages de gaz moléculaire dense ["circum-nuclear disk" (CND)] en orbite à 1,5-4 pc \citep{Lo1983}, la seconde est une structure apparaissant sous la forme d'une spirale nommée Sagittarius A West surnommée "minispiral" (voir panneau du milieu à droite de la Figure~\ref{fig:GC_scales}), composée de gaz et de poussières qui orbitent\footnote{L'aspect en spirale de ces structures est un effet de projection, il s'agit bien de gaz/poussière qui orbite le centre de la Galaxie.} la source compacte non résolue nommée Sagittarius A* avec une vitesse de 1.000 km/s~\cite{Paumard2004}.

En plus du gaz, cette région présente une forte concentration d'étoiles que l'on peut séparer en plusieurs groupes selon leurs propriétés.

\section{Population stellaire du parsec central de notre Galaxie}
\subsection{L'amas d'étoiles vieilles et froides}
\begin{figure}
    \centering
    \resizebox{0.8\hsize}{!}{\includegraphics{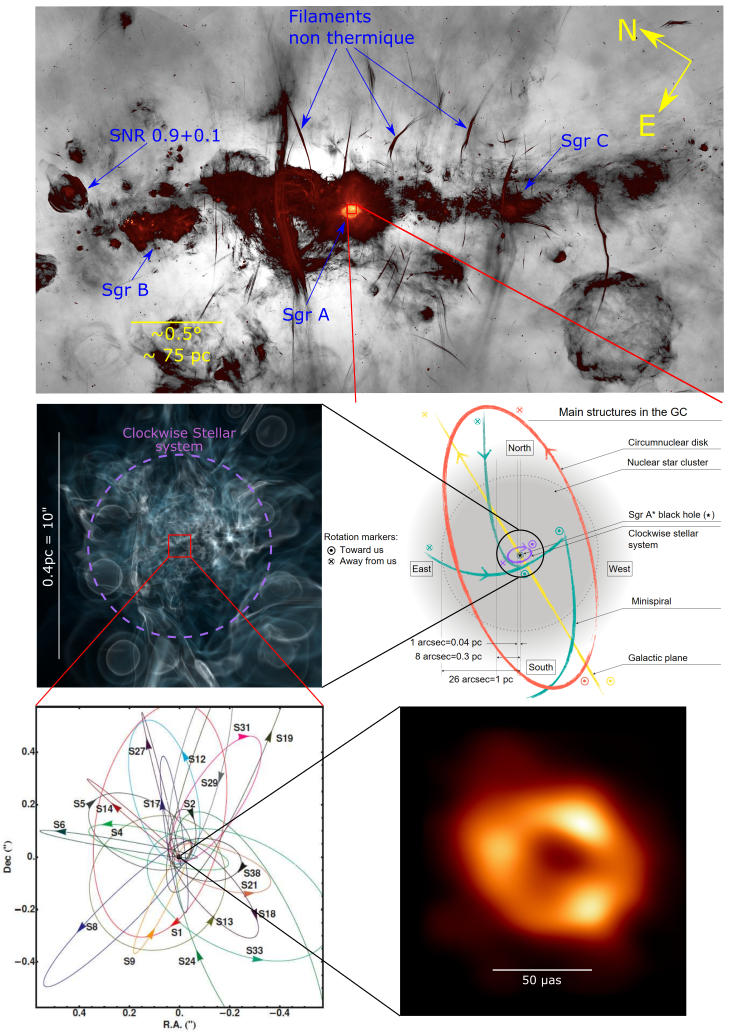}}
    \caption{\textbf{(haut)} : Image radio créé à partir d'observation \href{https://www.sarao.ac.za/gallery/meerkat/}{MeerKAT} avec un champ large du centre galactique, avec Sagittarius A la position exacte du centre de la Galaxie. Crédit : \href{https://www.sarao.ac.za/media-releases/new-meerkat-radio-image-reveals-complex-heart-of-the-milky-way/}{\cite{Heywood2022} South African Radio Astronomy Observatory (SARAO)}. \textbf{(milieu à droite)} : Schéma détaillant les différentes composantes des quelques parsecs centraux de notre Galaxie. Crédit : adapté de \cite{Murchikova2019}. \textbf{(milieu à gauche)} : Simulation hydrodynamique de l'environnement de Sgr~A* avec un apport de matière par le vent de 30 étoiles Wolf-Rayet. Crédit : \cite{Ressler2018, Ressler2023}. \textbf{(bas à gauche)} : Orbite des étoiles S dans la seconde d'angle autour de Sagittarius A*. Crédit : \cite{Gillessen2009}. \textbf{(bas à droite)} : Image reconstruite de Sgr~A* crée par la collaboration EHT (plus de détails dans la Fig.~\ref{fig:EHT_img}). Crédit : \href{https://www.eso.org/public/france/news/eso2208-eht-mw/}{EHT Collaboration}. Les axes sont en coordonnées équatorial.}
    \label{fig:GC_scales}
\end{figure}

Une première population d'étoiles qui se distinguent des autres est constituée d'étoiles froides ($\sim 3500 K$) et vieilles ($>10^9$ ans) de masses intermédiaires (entre $0,5$ et $4\, M_\odot$). Ce sont majoritairement des géantes rouges brûlant de l'hélium accompagnées de quelques étoiles AGB (\textit{Asymptotic Giant Branch}) présentant une importante perte de masse. Elles représentent 96\% des étoiles observées dans le parsec central de la galaxie~\cite{Genzel2010}. À celles-ci s'ajoutent quelques supergéantes rouges plus massives~\citep{Blum2003}. L'ensemble des quelque 6000 étoiles observées forme un amas sphérique en rotation solide dans le même sens de rotation que la galaxie \citep{Trippe2008, Schodel2009}. Cette population correspond au "\textit{Nuclear star cluster}" dans le panneau du milieu à droite de la Figure~\ref{fig:GC_scales}.

\subsection{Les étoiles jeunes et massives}
La deuxième classe, composée de jeunes étoiles ($\approx 6 \times 10^6$ ans) de type O-B et Wolf-rayet (WR; dont le vent stellaire alimente en matière l'objet compact Sgr~A* comme le montre le panneau du milieu à gauche de la Fig.~\ref{fig:GC_scales}) située dans les 25 secondes d'arc ($\sim$1 pc) autour de Sagittarius A*, est regroupée via un paramètre commun, leur vitesse radiale qui suggère un champ de vitesse cohérent à grande échelle. La plupart des étoiles au Nord se déplacent vers nous alors que celles au Sud s'éloignent. Ce mouvement est en opposition avec le sens de rotation de la Galaxie et de la population précédente. L'analyse 3D du mouvement de ces étoiles révèle des mouvements d'ensemble cohérent divisés en deux groupes. Ces deux groupes d'étoiles ont un mouvement assimilable à une rotation dans deux disques distincts avec des inclinaisons différentes et des sens de rotation opposés~\citep{Genzel2000, Paumard2006}. Le disque le plus important et le mieux défini, comprenant le groupe IRS~16, présente un mouvement orbital circulaire dans le sens horaire (voir panneau du milieu à droite de la Figure~\ref{fig:GC_scales}) avec une inclinaison de $\sim 120 \degree$ par rapport au plan du ciel, une orientation de $\sim 120 \degree$ par rapport au Nord (positif vers l'Est) et une épaisseur relativement importante $h/R\sim 0.1$~\citep{Paumard2006, Lu2009}. Ce disque est communément appelé "\textit{Clockwise Disk}" ou "\textit{Clockwise Stellar system}" dans le panneau du millieu à droite de la Fig.~\ref{fig:GC_scales}. Les étoiles du second disque, présentant une rotation dans le sens trigonométrique dans le plan du ciel, communément appelé "\textit{Counter Clockwise Disk}" (CCW), ont majoritairement des orbites excentriques, notamment l'amas dense IRS~13E~\citep{Paumard2006}. Néanmoins, le mouvement des étoiles du disque CCW est moins structuré que celui du disque CW~\citep{Kocsis2011}. Le regroupement des étoiles du disque CCW, et donc son existence, est débattue et n'apparait donc pas dans les panneaux de la Fig.~\ref{fig:GC_scales}. Malgré ces différences orbitales, les propriétés physiques de ces étoiles, dont leur âge, sont assez similaires, suggérant une formation commune et locale à partir d'un nuage de gaz dense, mais probablement pas issu d'un disque simple~\citep{Nayakshin2006}. Le mécanisme de formation de ces étoiles n'est pas encore totalement résolu et est encore aujourd'hui une question ouverte qui va au-delà du cadre de cette thèse. Le lecteur peut se référer à \cite{Genzel2010} pour une discussion plus approfondie de cette question qui est fondamentale pour la physique du centre galactique.

\subsection{L'amas des étoiles S}
La seconde d'arc qui entoure Sagittarius A*, correspondant à 0,04 pc, est peuplée d'étoiles massives ($\sim 3,5-20 M_\odot$) majoritairement de type B, des étoiles froides ainsi que quelques étoiles de type~O~\citep{Gillessen2009} regroupées sous le nom d'étoiles S (S-Stars). Ces étoiles sont les plus proches du centre galactique et de sa source compacte connue à ce jour. Les propriétés spectroscopiques de ces étoiles (profil de raies, magnitude absolue, vitesse de rotation) sont comparables à celles des étoiles de type B dans le voisinage du Soleil~\citep{Eisenhauer2005}. Ainsi, l'âge estimé des étoiles S à partir de la durée de vie des étoiles de ce type dans la séquence principale est compris entre 6 et 400 millions d'années. Il est intéressant de noter que la limite inférieure correspond à l'âge approximatif des étoiles de type O/WR des disques. Cependant, contrairement aux étoiles citées précédemment qui sont réparties dans des plans orbitaux plus ou moins bien définis, la distribution du moment angulaire des étoiles S est isotrope~\citep{Ghez2005}. L'analyse du mouvement des étoiles S a permis de déterminer les paramètres orbitaux de ces étoiles~\citep{Schodel2003, Eisenhauer2005, Ghez2005, Gillessen2009} montrant une grande proportion d'orbites très excentriques avec Sagittarius~A* comme foyer commun, comme le montre le panneau en bas à gauche de la Figure~\ref{fig:GC_scales}. La masse et la distance de Sagittarius A* ont ainsi pu être estimées à partir de ces orbites à $M_{BH}=(4,297 \pm 0.016) \times 10^6 M_\odot$ pour une distance de $R_0 = (8277 \pm 9)$ pc~\citep{GRAVITY2022a}. Il est à noter que la détermination de la masse, ainsi que les incertitudes, dépendent de la valeur (et incertitudes) de la distance $R_0$. En effet, la masse est déterminée à partir des lois de Kepler et de la détermination du demi-grand axe physique. Or, la détermination de ce dernier dépend de la séparation angulaire mesurée et de la distance. La masse et la distance sont donc dégénérées. L'usage conjoint d'astrométries de précision avec GRAVITY et de spectroscopie permet de lever cette dégénérescence. Cette masse étant contenue à l'intérieur du périastre\footnote{Distance la plus courte entre l'étoile et Sgr~A*.} de l'orbite de l'étoile S2~\citep{Gravity2020a} soit dans une sphère de moins de 1 mpc de rayon, on en déduit que Sagittarius A* est un candidat trou noir supermassif (ou équivalent), ce qui fait de ce dernier la cible idéale d'étude de ces objets.

Parmi les étoiles S, l'une d'entre elle, l'étoile S2, est d'une importance particulière du fait de sa proximité à Sgr~A* lors de son passage au périastre, de sa période orbitale relativement courte ($\sim$ 16 ans) et de son importante luminosité ($K \approx 14$). Cette étoile est observée et suivie (position par rapport au centre de la galaxie) depuis 1992~\cite{Hofmann1993}, on a donc observé deux orbites de cette étoile autour de Sgr~A* avec une grande précision (avec les mesures de GRAVITY) ce qui permet de mettre en évidence la précession de Schwarzschild~\cite{Gravity2020a}, c'est-à-dire le changement d'orientation de l'ellipse dans son plan orbital, une prédiction de la Relativité Générale (RG). De plus, \citet{Gravity2020a} a pu estimer la masse de Sgr~A* et la proportion maximale de masse étendue dans un rayon de 1 arcsec autour de l'objet central à moins de 3.000 $M_\odot$.

\section{La source compacte Sagittarius A*}
\subsection{La nature de Sagittarius A*}
La découverte des quasars (\textit{quasi-stellar objects} que l'on peut traduire par "objets quasi-stellaires") dans les années 1950 a tout d'abord rendu les astronomes perplexes du fait de l'énergie colossale de ces objets. En effet, ces sources brillantes dans le domaine radio se situent à des distances cosmologiques mesurées grâce à leur décalage vers le rouge. Ainsi, leur luminosité intrinsèque peut atteindre plusieurs milliers de fois la luminosité de la Voie Lactée~\citep{Wu2015}. Ces phénomènes ont été attribués à l'accrétion de matière par un trou noir supermassif au cœur des galaxies~\cite{Frank2002} et sont désormais une sous-catégorie des noyaux actifs de galaxies (AGN \textit{Active Galactic Nuclei}). La possibilité que chaque galaxie (hormis les galaxies irrégulières) abrite un trou noir supermassif en son cœur a naturellement été émise dans les années 1990~\cite{Merritt2013} et est largement admise de nos jours.

Ainsi, lors de la découverte de la source compacte Sagittarius~A* par~\citet{Balick1974} au cœur de la Voie Lactée, l'hypothèse d'un trou noir supermassif comme nature de Sgr~A* a été évoquée, mais n'a pas fait l'unanimité sans mesure de sa masse. La première estimation de cette dernière a été réalisée via des mesures de vitesse radiale du gaz ionisé dans le parsec central de la galaxie (minispirale) pour une valeur de $\sim$ 4 millions de masses solaires~\citep{Wollman1977}. Néanmoins, cette masse étant contenue dans un rayon d'un parsec, d'autres explications étaient encore possibles. La découverte des étoiles S, en particulier l'étoile S2 (ou S02), a permis une nouvelle estimation de la masse de Sgr~A* utilisant cette fois la troisième loi de Kepler à partir des mesures de période et demi-grand axe de ces étoiles. La valeur la plus récente utilisant cette méthode est de $M_{BH}=(4,297 \pm 0.016) \times 10^6 M_\odot$ pour une distance de $R_0 = (8277 \pm 9)$~pc~\citep{GRAVITY2022a}. Les deux valeurs sont comparables, même si la seconde est plus précise grâce aux meilleurs moyens d'observation, mais elle implique que la masse soit contenue dans une sphère dont le rayon est inférieur au péri-astre de S29, à savoir 100 Unités Astronomiques (UA)~\citep{GRAVITY2022a} (correspondant à deux fois la distance entre le Soleil et la Ceinture de Kuiper). Alors que via le théorème du Viriel la possibilité d'un amas stellaire très dense était encore possible, elle est totalement incompatible avec la densité obtenue via la 3$^\mathrm{ème}$ loi de Kepler. De plus, les observations en radio à 3,5 mm ne permettent pas de résoudre Sgr~A*, donnant ainsi une limite supérieure à sa taille de $\sim$ 1 UA~\citep{Shen2005}. Combinée à la masse mesurée, cette dernière contrainte permet de définir la densité minimale de Sgr~A* à $6,5 \times 10^{21} M_{\odot}.pc^{-3}$ (avec $M_{\odot}$ la masse du Soleil), ce qui fournit une preuve solide que Sagittarius A* est un objet compact, vraisemblablement un trou noir supermassif. Il faut toutefois noter que Sgr~A* n'est pas forcément composé de toute la masse contenue dans le péri-astre de S29. En effet, il peut y avoir des étoiles de faible luminosité ou plus généralement de la masse "sombre" (dans le sens qui n'émet pas ou peu de lumière) étendue comme des résidus stellaires ou de la matière noire. La limite supérieure de la distribution de la masse étendue (autre que le trou noir supermassif) a été évaluée par~\cite{GRAVITY2022a} à $\leq 3000 M_{\odot}$ soit $\leq 0,1 \%$ de $M_{BH}$.

La nature exacte de l'objet compact est cependant encore débattue. En effet, l'hypothèse la plus simple pour expliquer la nature de Sgr~A* est un trou noir supermassif au sens de la RG, mais d'autres possibilités, plus exotiques, comme un trou de ver, un trou noir avec cheveux scalaires, etc., sont compatibles avec les contraintes actuelles. Les effets différentiels des différents modèles par rapport au modèle standard de trou noir sont très faibles et demandent une précision extrême, au-delà des capacités des instruments actuels. Pour le reste de cette thèse, nous considérons l'hypothèse la plus simple, à savoir que Sgr~A* est un trou noir supermassif au sens de la RG.

\subsection{Le flot d'accrétion de Sgr~A*}
Par définition, les trous noirs (ou leurs équivalents) n'émettent pas (ou peu) de rayonnement. La lumière reçue de Sgr~A* provient donc exclusivement (ou en grande majorité) de son environnement direct, à savoir son flot d'accrétion. Les flots d'accrétion autour d'objets compacts (étoiles à neutrons ou trous noirs) peuvent être séparés en deux grandes catégories. La première est définie par des disques géométriquement minces et optiquement épais avec une émission très intense~\cite{Shakura1973}, évacuant la chaleur du fluide accumulée par frottement visqueux. Ce genre de configuration se retrouve usuellement autour d'étoiles à neutrons ou des trous noirs stellaires aspirant la matière d'un compagnon comme dans les binaires~X. La seconde catégorie regroupe les disques géométriquement épais et optiquement mince, on parle de flots d'accrétion radiativement inefficaces (\textit{Radiatively Inefficent Accretion Flow} RIAF en anglais)~\cite{Yuan2014}. En effet, dans cette configuration, la chaleur n'est pas évacuée par le rayonnement comme précédemment et s'accumule dans le fluide, le faisant gonfler comme un gâteau dans un four. La température de ce type de flot d'accrétion peut donc être très élevée et atteindre $10^{9-10}$ K. Ce type de flot d'accrétion est plus adapté pour expliquer les flots de faible luminosité comme ceux de Sgr~A* et M87* ou d'AGN à fortes luminosités.

\subsection{Observations multi-longueurs d'onde de Sgr~A*}
\subsubsection{Observations radio}
La découverte de Sagittarius A* s'est faite dans le domaine radio comme dit précédemment. Il a fait l'objet de vastes campagnes d'observation dans ce domaine, à différentes fréquences et utilisant différentes techniques. Bien que l'on puisse l'observer avec des télescopes uniques, le signal comprend d'autres sources parasites dues à la taille importante du faisceau, qui est l'équivalent de la \textit{Fonction d'Étalement du Point} (PSF en anglais pour \textit{Point Spred function}) en optique. Pour réduire la taille du faisceau, il faut soit augmenter la fréquence, soit augmenter la taille des télescopes. Cependant, cette dernière ne peut pas être augmentée indéfiniment. Pour contourner cette limitation, il faut changer de technique et faire de l'interférométrie (voir les détails dans le Chapitre~\ref{chap:GRAVITY}). Pour simplifier, au lieu de faire un télescope de, par exemple, 500 m de diamètre, on va faire plusieurs télescopes de taille plus raisonnable, mais espacés de 500 m. En combinant les signaux reçus et en les faisant interférer les uns avec les autres, on crée un télescope dont la taille effective (pour la résolution spatiale) correspond à la distance maximale entre ces télescopes. Ce principe peut être étendu à l'infini (y compris dans l'espace) avec néanmoins des limitations que nous aborderons plus tard. C'est le principe de l'EHT qui observe à 1,3 mm, mais à l'échelle de la Terre entière, à savoir un peu moins de 12 000 km. Ainsi, l'EHT a une résolution suffisante pour imager (encore une fois, voir les détails dans le Chapitre~\ref{chap:GRAVITY}) Sagittarius A*. L'EHT a ainsi publié la première image de l'ombre et du disque d'accrétion d'un trou noir, tout d'abord M87* le 10 Avril 2019 \cite{EHT2019}, puis Sagittarius A* le 12 Mai 2022 \cite{EHT2022a}.

\begin{figure}
    \centering
    \resizebox{0.5\hsize}{!}{\includegraphics{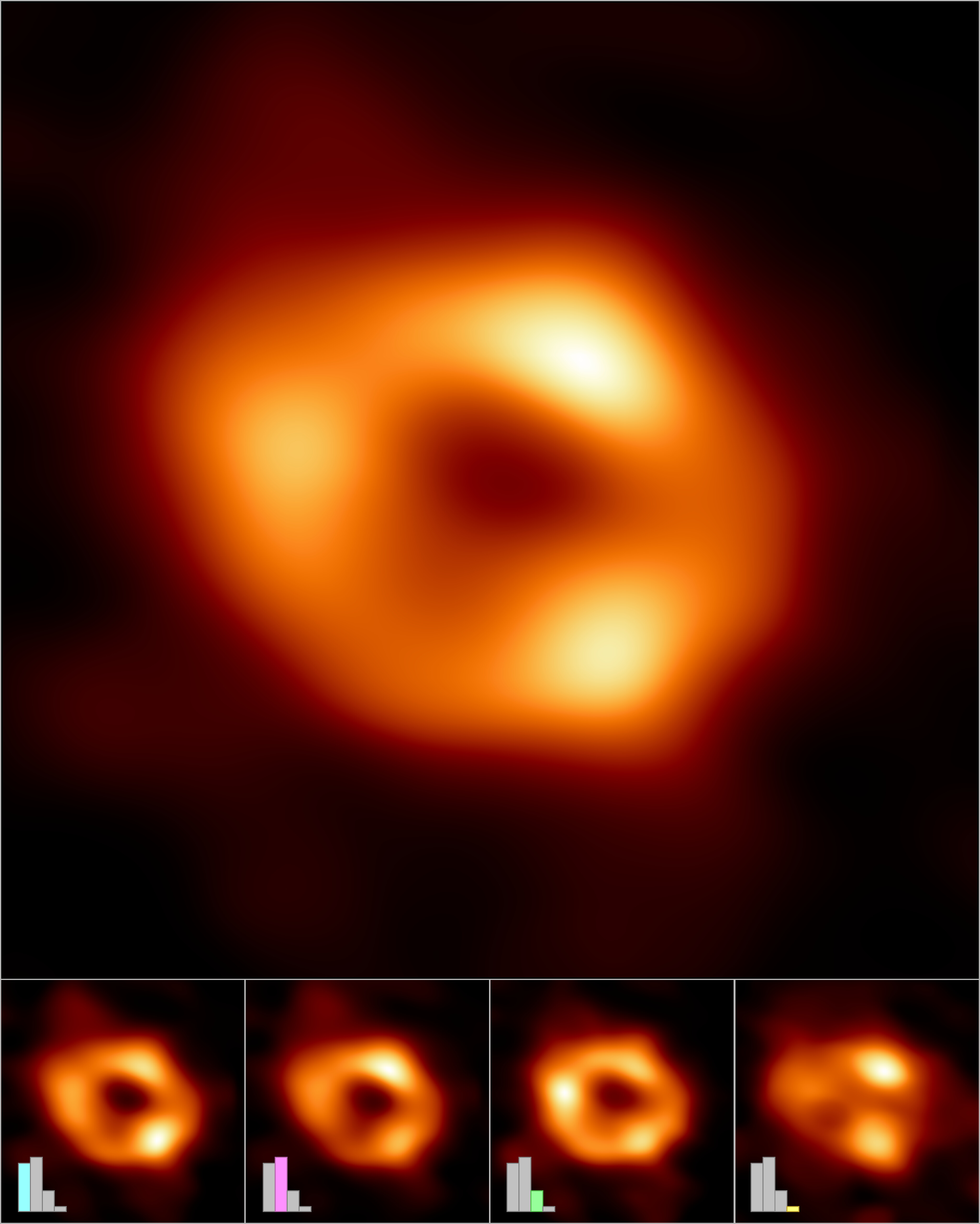}}
    \caption{Images reconstruites de Sgr~A* crées par la collaboration EHT en combinant les images obtenues par les observations de l'EHT. L'image principale (\textbf{haut}) a été produite en moyennant plusieurs milliers d'images reconstruites, en utilisant plusieurs méthodes, qui ajustent les données observées. L'image moyennée montre les caractéristiques qui apparaissent le plus et réduit les particularités peu fréquentes. Les milliers d'images produites ont été regroupées en quatre catégories présentant les mêmes caractéristiques. Les quatre images en \textbf{bas} montrent l'image moyennée ainsi que la proportion d'image de chaque catégorie. Les trois premières montrent une forme d'anneau, mais avec une distribution différente de luminosité. Ces catégories représentent la majeure partie des images observées. La hauteur des barres indique le poids de chaque image dans l'image principale.
    Crédit : EHT Collaboration}
    \label{fig:EHT_img}
\end{figure}

\begin{figure}
    \centering
    \resizebox{0.7\hsize}{!}{\includegraphics{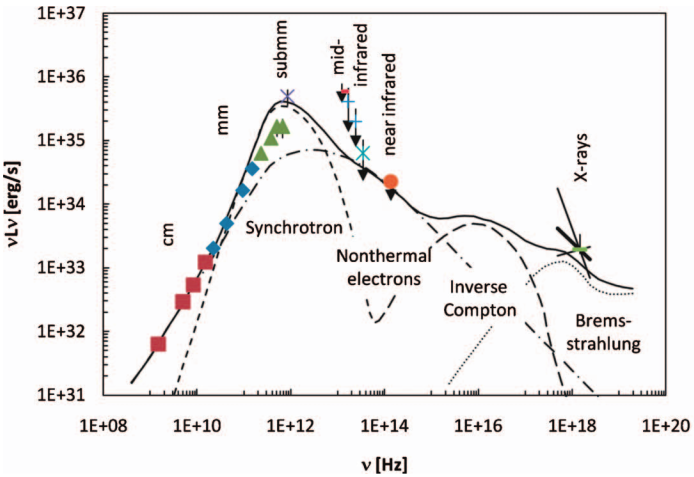}}
    \caption{Spectre de Sgr~A* à l’état quiescent (sans sursaut de rayonnement). Les données, en couleur, proviennent de divers auteurs dont on trouvera les références dans \cite{Genzel2010}. Les spectres modélisés proviennent du modèle de \cite{Yuan2003} : en tirets courts, l’émission synchrotron d’électrons thermiques ; en tirets longs, la diffusion Compton inverse des photons émis par synchrotron sur les électrons thermiques ; en tirets points, l’émission synchrotron d’une population non thermique d’électrons ; en pointillés, l’émission Bremsstrahlung des régions externes de la structure d’accrétion. Crédit : \cite{Genzel2010}.}
    \label{fig:quiescent_spectrum}
\end{figure}

L'étude de Sgr~A* dans le domaine radio présente plusieurs intérêts majeurs. Le spectre quiescent de Sgr~A*, obtenu à partir d'observations multi-longueurs d'onde, y compris en radio, et présenté dans la Fig.~\ref{fig:quiescent_spectrum}, montre une bosse dans le domaine submillimétrique (sub-mm) correspondant au changement de régime optique entre les basses fréquences ($\nu \lesssim 10^{11}$ Hz), où le plasma est optiquement épais ; et les hautes fréquences ($\nu \gtrsim 10^{11}$ Hz), où il est optiquement mince. Le spectre a été modélisé par~\cite{Yuan2003} avec du rayonnement synchrotron (voir Chap.~\ref{chap:modele_hotspot+jet}) thermique pour le domaine radio, du synchrotron non thermique en IR et du rayonnement Inverse Compton et Bremstrahlung pour les hautes énergies (rayons~X et rayons $\gamma$). L'étude de la transition entre les régimes optiquement épais et mince dans le domaine mm, aux alentours de 230 GHz, est donc crucial pour contraindre la physique du flot d'accrétion de Sgr~A*, ce qui a mené la collaboration EHT à créer l'image de la Fig.~\ref{fig:EHT_img}.

L'EHT n'est pas le seul interféromètre radio à observer Sgr~A* : l'interféromètre ALMA l'observe aussi fréquemment de manière plus régulière afin d'étudier sa dynamique. En effet, Sagittarius~A* est une source variable dans le domaine radio, ce qui a grandement complexifié la construction de l'image de l'EHT. Cependant, contrairement au domaine IR (voir plus bas) où les pics d'émission, appelés sursauts, peuvent atteindre 100 fois l'émission continue (voir Chap.~\ref{chap:Sgr~A* flares}), le flux lors des sursauts dans le domaine radio n'augmente que de quelques dizaines de pourcents par rapport au flux continu~\cite{Wielgus2022} (plus de détails dans le Chap.~\ref{chap:Sgr~A* flares}).

\subsubsection{Observations en infrarouge proche}
Bien que le pic d'émission de Sagittarius A* soit dans le domaine radio, il est tout de même observable en IR. Il a fallu néanmoins attendre la construction de grands télescopes comme le \textit{W. M. Keck Observatory}, ci-après Keck, et les quatre grands télescopes du \textit{Very Large Telescope} (VLT), dans les années 1990, ainsi que les techniques d'Optique Adaptative (OA) et d'interférométrie optique en infrarouge proche (NIR) avec l'instrument GRAVITY (voir Chapitre~\ref{chap:GRAVITY}) pour distinguer Sgr~A* des étoiles environnantes (étoiles S). 

La magnitude médiane de Sgr~A* dans la bande K, centrée autour de 2,2 $\mu m$, est de $m_K\approx 17$ (calculé à partir du flux médian dérougi de 1,1 mJy et $A_K=2,5$ \cite{Gravity2020b}). Tout comme en radio, Sgr~A* présente une variabilité en NIR avec cependant une plus grande amplitude et avec une plus grande diversité de temps caractéristique. La distribution du flux de Sgr~A* a fait l'objet de nombreuses études afin de mieux contraindre cette source. En utilisant les données du télescope spatial \textit{Spitzer}-IRAC et du Keck-NIRC2 + VLT-NACO avec une approche Bayésienne, \cite{Witzel2018} modélisent la distribution du flux de Sgr~A* à 4,5~$\mu m$ et 2,2~$\mu m$ respectivement avec des distributions log-normales dont la fonction de densité de probabilité (PDF) est
\begin{equation}
    f(x) = LN(x) = \frac{1}{x \sigma \sqrt{2 \pi}} \mathrm{exp} \left( -\frac{(\mathrm{ln} x -\mu)^2}{2 \sigma^2} \right)
\end{equation}
avec $\mu_M=1,01^{+0,47}_{-0,44}$, $\sigma_M=0,39^{+0,15}_{-0,13}$ et $\mu_K=-1,35^{+0,62}_{-0,60}$, $\sigma_K=0,56^{+0,24}_{-0,21}$ pour les bandes M (\textit{Spitzer}) et K (Keck et VLT) respectivement.

\begin{figure}
    \centering
    \resizebox{0.7\hsize}{!}{\includegraphics{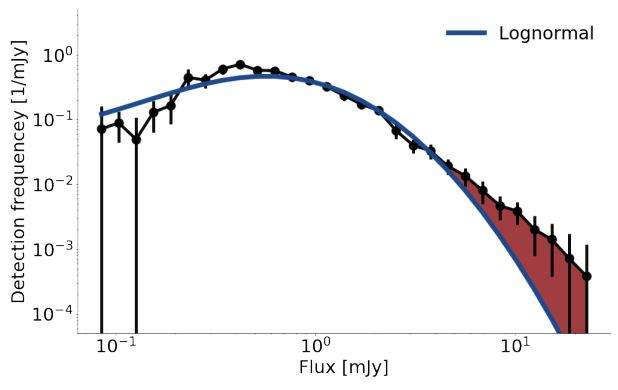}}
    \caption{Distribution du flux dérougi de Sgr~A* observé en bande K par GRAVITY avec le meilleur ajustement d'une loi log-normale en trait bleu. On constate que cette dernière n'ajuste pas correctement les données pour les flux supérieurs à 5 mJy avec l'écart entre la loi log-normale et les données en rouge. Crédit : \cite{Gravity2020b}.}
    \label{fig:distribution_SgrA}
\end{figure}

Depuis 2016, l'instrument GRAVITY, qui utilise les quatre grands télescopes du VLT en mode interférométrique (voir Chapitre~\ref{chap:GRAVITY} pour plus de détails), permet d'observer Sgr~A* et les étoiles S passant proche du trou noir avec une très grande précision. Comme il s'agit d'un instrument basé sur l'interférométrie, les données mesurées sont des visibilités complexes et des phases. Lorsqu'il y a deux sources dans le champ de vue de l'instrument, on utilise un modèle pour ajuster les visibilités et phases observées (voir Chap.~\ref{chap:GRAVITY}). Parmi les paramètres du modèle, le rapport de flux entre les deux sources influence principalement la visibilité observée. Ainsi, lors du passage de l'étoile S2 proche de Sgr~A* entre 2017 et 2019, connaissant la valeur du flux de cette dernière, le flux de Sgr~A* a pu être mesuré avec une grande précision. En utilisant cette technique, \cite{Gravity2020a} a étudié la distribution du flux de Sgr~A* entre 2017 et 2019. Cependant, la distribution du flux observé sur cette période ne correspond pas à une loi log-normale. En effet, Sgr~A* présentait plus de flux élevés ($> 5$ mJy) que la prédiction d'une loi log-normale (voir Figure~\ref{fig:distribution_SgrA}). Cependant, les données GRAVITY peuvent être expliquées en utilisant une loi log-normale avec une loi de puissance pour les flux $x > x_{min} = 5$~mJy tel que
\begin{equation}
    f(x) =
    \begin{cases}
        LN(x) & x \leq x_{min} \\
        c LN(x_{min}) F^{-\alpha}/x_{min}^{-\alpha} & x > x_{min}
    \end{cases}
\end{equation}

Ainsi, on peut définir deux états pour Sgr~A*, un état dit quiescent avec un flux faible dont la distribution du flux est une loi log-normale, liée à des processus stochastiques avec une faible variabilité de l'ordre de quelques jours voire semaines, et un état de sursaut avec un flux minimal de $5$ mJy avec une distribution en loi de puissance. L'année 2019 a encore plus marqué cette distinction entre les deux états avec plus de flux élevés recensés comparés aux années précédentes~\citep{Do2019, Gravity2020a} (voir Figure~\ref{fig:distribution_SgrA}). De plus, l'indice spectral des sursauts, qui est compris entre $-0,5$ et $0,5$, est significativement supérieur à celui de l'état quiescent qui est compris entre $-3$ et $-1$ \cite{Genzel2010} (voir Figure~\ref{fig:quiescent_spectrum}).

\subsubsection{Observations en rayons~X}
L'observation en rayons~X est plus délicate, car elle nécessite des télescopes spatiaux à cause de l'atmosphère qui absorbe les photons de ces énergies. De plus, la technique classique des télescopes optiques, à savoir une succession de miroirs, ne fonctionne plus, car les photons traversent simplement les matériaux avec une incidence normale. Cependant, avec une incidence rasante, les photons peuvent être guidés vers un détecteur. Cette technique fonctionne pour les rayons~X dits mous ($< 5$ keV) mais pas pour les rayons~X durs (5-10 keV). Dans ce cas-là, on utilise des calorimètres qui vont absorber l'énergie du photon au fur et à mesure qu'il traverse le matériau ou utiliser un masque ayant une signature unique sur le détecteur selon la direction d'incidence du rayon X. On peut ainsi connaître sa trajectoire et donc son point d'origine avec une précision modérée (nettement plus faible qu'aux longueurs d'ondes plus grandes).

Sagittarius~A* est aussi détectable en rayons~X par les observatoires \textit{Chandra} et \textit{XMM-Newton}. \cite{Baganoff2003} ont été les premiers à déterminer la luminosité de l'état quiescent de Sgr~A* avec \textit{Chandra} à 2-10 keV à $L_{2-10\mathrm{keV}} \sim 2,4 \times 10^{33}$ erg/s (très inférieur à la limite d'Eddington) avec un indice spectral de -2,7. En plus d'observer l'état quiescent de Sgr~A*, \textit{Chandra} et \textit{XMM-Newton} observent aussi des sursauts, comme en NIR. Chaque sursaut en rayons~X est accompagné d'un sursaut en IR, cependant, certains sursauts observés en IR n'ont pas de contrepartie en rayons~X. \textit{Chandra} a permis d'estimer la fréquence des sursauts de Sgr~A* en rayon X à environ un par jour contre environ quatre par jour en IR.

\newpage
\thispagestyle{plain}
\mbox{}
\newpage

\chapter{GRAVITY, un instrument pour sonder Sagittarius~A*}\label{chap:GRAVITY}
\markboth{GRAVITY, un instrument pour sonder Sagittarius~A*}{GRAVITY, un instrument pour sonder Sagittarius~A*}
{
\hypersetup{linkcolor=black}
\minitoc 
}

\section{L'interférométrie comme outil pour gagner en résolution angulaire} \label{sec:interfero}
\subsection{Principe de base}
\begin{figure}
    \centering
    \resizebox{0.7\hsize}{!}{\includegraphics{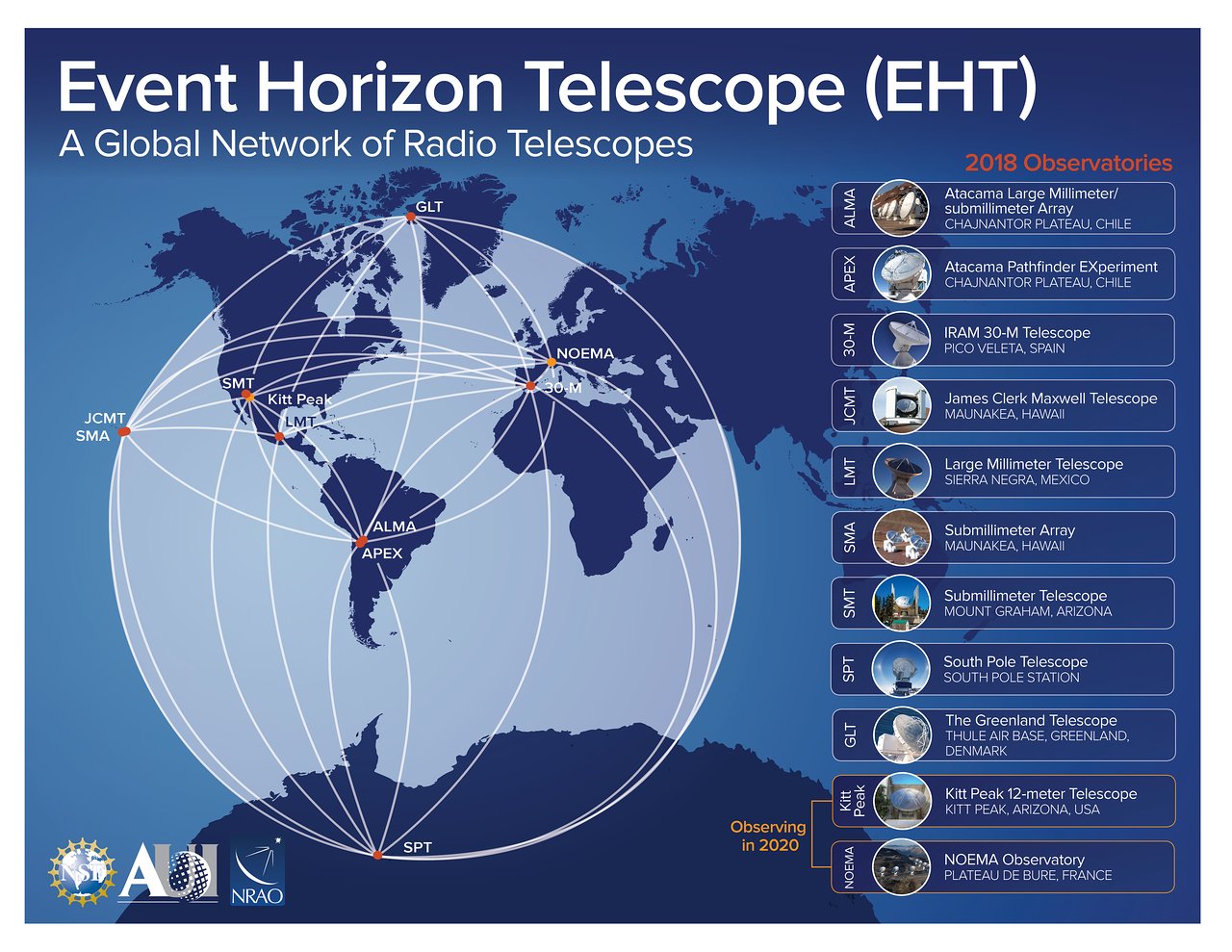}}
    \caption{Radiotélescopes utilisés lors des campagnes d'observation de M87 et Sgr~A* par la Collaboration EHT. Crédit: \href{https://www.eso.org/public/france/images/eso1907p/}{NRAO}.}
    \label{fig:EHT_telescope}
\end{figure}

Sagittarius~A* est une source très compacte avec une taille angulaire de l'ordre de $20\, \mu$as \cite{EHT2022a}. La formule définissant la résolution d'un instrument à partir de sa taille est :
\begin{equation} \label{eq:resolution_angulaire}
    \theta = \frac{\lambda}{D}
\end{equation}
avec $\theta$ la résolution angulaire, correspondant à la largeur à mi-hauteur caractéristique de la PSF, $\lambda$ la longueur d'onde d'observation et $D$ l'ouverture du télescope. Lorsque l'on utilise un seul télescope, $D$ correspond au diamètre du miroir primaire. Ainsi pour un télescope de 8m de diamètre comme un des \textit{Unit Telescope} (UT) du \textit{Very Large Telescope} (VLT, ESO) à Paranal au Chili observant à $2,2\, \mu$m, on a une résolution angulaire de $\sim 55$ mas, 3 ordres de grandeurs au-dessus de la résolution nécessaire pour imager Sgr~A*. En passant dans le domaine radio à 230 GHz (1,3 mm), à plus grande longueur d'onde, la résolution se dégrade et atteint $\sim 33$ secondes d'arc pour le même diamètre $D$. La taille des télescopes (radio, IR ou optique) étant limitée par des contraintes mécaniques notamment, il existe une limite de résolution maximale qui malheureusement, en utilisant un unique télescope, est bien inférieure à la résolution nécessaire pour étudier l'environnement proche de Sgr~A*. Cependant, en combinant la lumière reçue par deux télescopes distants, en les faisant interférer, on échantillonne le signal que recevrait un télescope avec une ouverture effective (pour le calcul de la résolution) égal à la distance entre les deux télescopes. Ainsi, la Collaboration EHT a combiné la lumière de nombreux radiotélescopes à travers le monde entier afin de créer un télescope équivalent de la taille de la Terre (voir Fig.~\ref{fig:EHT_telescope}).

\begin{figure}
    \centering
    \resizebox{0.5\hsize}{!}{\includegraphics{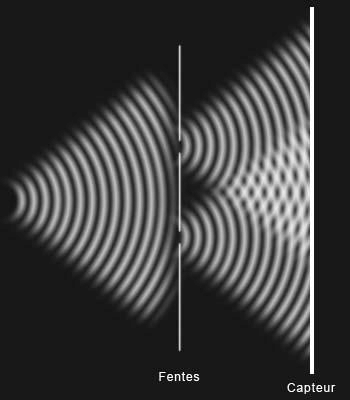}}
    \caption{Illustration de l'expérience des fentes de Young et du phénomène d'interférence entre deux ondes cohérentes. Crédit : \href{https://www.researchgate.net/publication/313435860_Analyses_de_lichens_par_spectrometrie_de_masse_dereplication_et_histolocalisation}{Le Pogam, Pierre. (2016)}.}
    \label{fig:fente_de_Young}
\end{figure}

\begin{figure}
    \centering
    \resizebox{0.7\hsize}{!}{\includegraphics{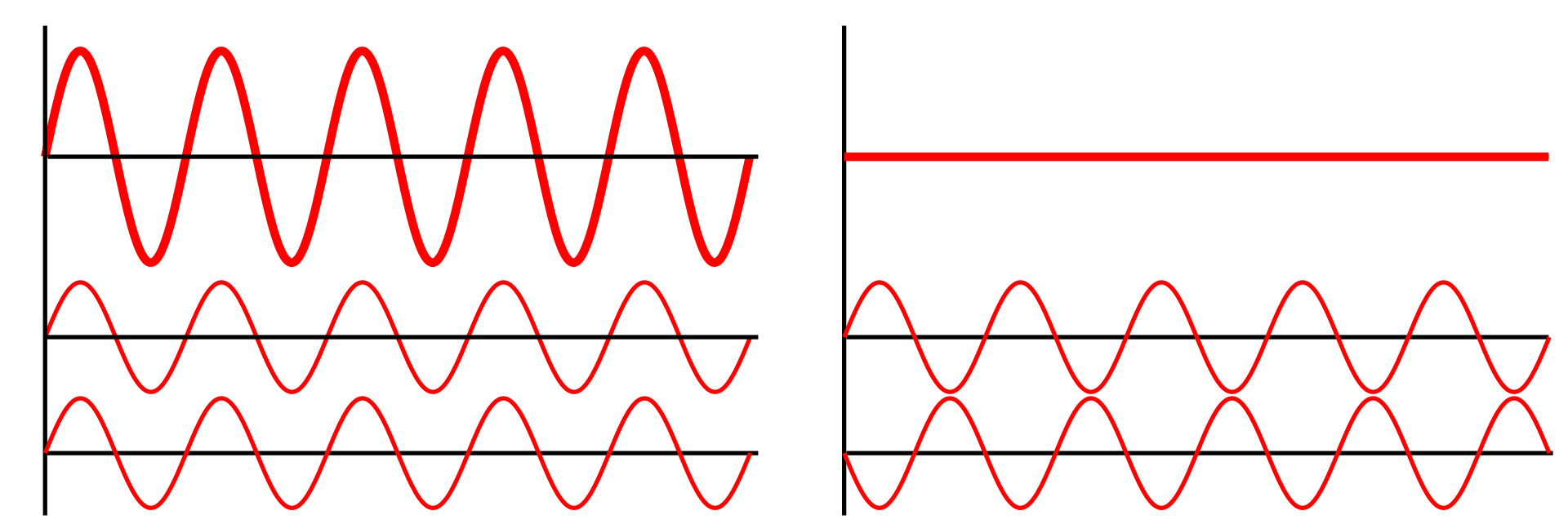}}
    \caption{Interférence de deux ondes. Lorsqu'elles sont en phase, les deux ondes inférieures créent une interférence constructive (à gauche), ce qui donne une onde de plus grande amplitude. Lorsqu'elles sont déphasées de 180°, elles créent une interférence destructive (à droite). Crédit : \href{https://commons.wikimedia.org/wiki/File:Interference_of_two_waves.svg}{Wikimedia}.}
    \label{fig:interference}
\end{figure}

La technique de l'interférométrie repose, comme son nom l'indique, sur le phénomène d'interférence entre deux (ou plus) ondes nécessitant qu'elles soient cohérentes spatialement et temporellement, c'est-à-dire qu'elles proviennent d'une même source ponctuelle et qu'elles soient monochromatiques respectivement\footnote{On peut obtenir des sources ayant une cohérence spatiale ou temporelle d'autres manières qui vont au-delà des limites de cette thèse.}. Dans un cas d'étude bien connu, l'expérience des fentes de Young (voir Fig.~\ref{fig:fente_de_Young}), on utilise une source lumineuse unique que l'on fait passer dans deux fentes (ou trous) espacées afin de créer deux nouvelles sources. Les ondes issues de ces dernières, qui peuvent être décrites par un nombre complexe avec une amplitude $A_i$ et une phase $\Phi_i$, vont parcourir un chemin optique différent $\Delta l$ en fonction du point de l'écran considéré $x$, positionné à une distance $d$ des fentes, et de l'espace entre les deux fentes $a$ tel que 
\begin{equation}
    \Delta l = \frac{a x}{d}.
\end{equation}
On obtient alors soit une frange brillante lorsque les ondes sont en phase, on dit que leur interférence est constructive, soit une frange sombre lorsque les ondes s'annihilent, on dit que leur interférence est destructive. Le caractère constructif ou destructif de la superposition des deux ondes dépend de la différence de marche $\Delta l$ par rapport à la longueur d'onde $\lambda$ 
\begin{equation}
\left\{ \begin{aligned}
        \Delta l &=n \times \lambda, \text{constructives} \\
        \Delta l &=(2n+1) \times \frac{\lambda}{2}, \text{destructives}
    \end{aligned} \right.
\end{equation}
avec $n$ un entier (voir Fig.~\ref{fig:interference}). L'image d'une source ponctuelle unique à travers deux fentes est une succession de franges sombres et brillantes où l'amplitude de l'onde résultante $A$ est définie en tout point comme :
\begin{equation}\label{eq:amplitude_interfero}
    A^2 = A_1^2 + A_2^2 + 2 A_1 A_2 \cos{\Delta \Phi}
\end{equation}
avec $A_1$ l'amplitude de l'onde au niveau de la fente 1, $A_2$ au niveau de la fente 2 et $\Delta \Phi$ le déphasage entre les deux ondes au point considéré qui vaut 
\begin{equation} \label{eq:dephasage}
    \Delta \Phi = \Delta l /\lambda \quad [2 \pi].
\end{equation}

\begin{figure}
    \centering
    \resizebox{0.7\hsize}{!}{\includegraphics{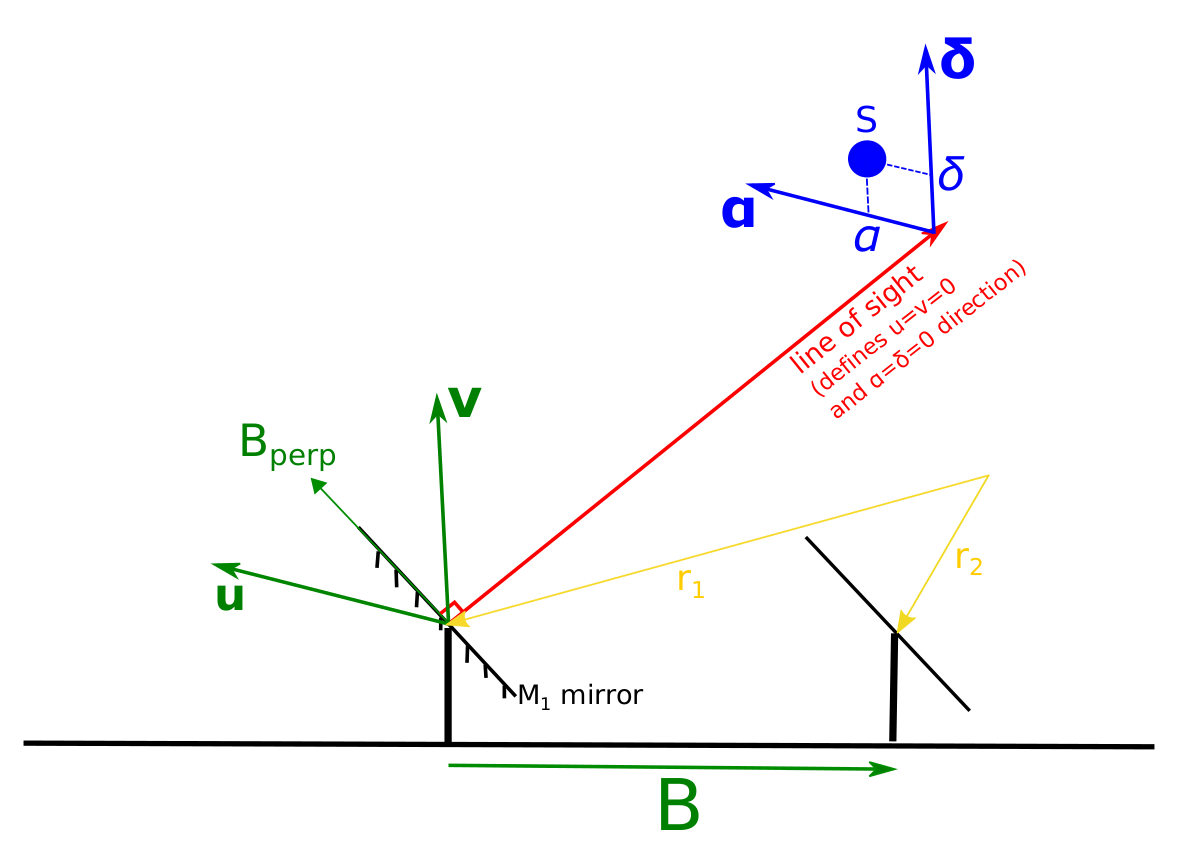}}
    \caption{Géométrie du système : la ligne de visée (flèche rouge) est la direction orthogonale au miroir M1 du télescope (cette direction est supposée identique pour les deux télescopes représentés ici, car la source est très éloignée). Le plan du ciel (en bleu) est étiqueté par les coordonnées angulaires ($\alpha$, $\delta$). La ligne de visée désigne le centre du champ de vision $(\alpha,\delta)= (0,0)$. Le plan d'ouverture (en vert) est étiqueté par les coordonnées $(u, v)$ en rad$^{-1}$. Les deux télescopes sont séparés par la ligne de base $B$. La projection de cette ligne de base orthogonale à la ligne de visée est représentée dans le plan $(u, v)$. Le champ électrique émis par la source S est échantillonné aux deux endroits $r_1$ et $r_2$ (en jaune), aux instants $t_1$ et $t_2$. La ligne de base est $B = r_2 - r_1$. Crédit : Frédéric Vincent.}
    \label{fig:interfero}
\end{figure}

L'expérience des fentes de Young est un cas particulier de l'interférométrie que l'on peut généraliser dans le cas d'observations astronomiques optique\footnote{Le traitement dans le domaine radio est légèrement différent, car on mesure directement l'intensité du champ électrique, on peut donc faire interférer les signaux a posteriori contrairement à plus haute fréquence (NIR ou optique) mais le principe reste le même.}. Prenons le cas de deux télescopes espacés d'une distance $B$ et d'une source étendue aux coordonnées $(\alpha,\delta)$ située à grande distance de l'observateur, comme illustré dans la Fig.~\ref{fig:interfero}. La source émet de la lumière, décrite ici par son champ électrique, mesuré par le télescope 1, situé au point $r_1$, au temps $t_1$ et par le télescope 2, situé au point $r_2$, au temps $t_2$. La lumière collectée par les deux télescopes est envoyée vers un recombinateur où les deux signaux seront combinés. Cependant, pour détecter la figure d'interférence, il faut que le déphasage, autrement dit la différence de marche, soit nulle ou proche d'être nulle du fait du nombre de franges limité par la cohérence temporelle. Il faut donc installer des retards de chemin optique afin de resynchroniser les signaux lorsqu'on les fait interférer\footnote{C'est un des gros challenges de l'interférométrie optique, car il faut une précision sur le décalage de marche inférieure à la longueur d'onde, donc $< 1 \, \mu m$ pour l'instrument GRAVITY qui observe à $2,2 \mu m$.}. Le flux obtenu s'exprime de la manière suivante
\begin{equation}
    \begin{aligned}
        F &= \iint_\mathrm{fov} \vert E(\alpha, \delta, r_1, t_1) + E(\alpha, \delta, r_2, t_2) \vert^2 \, \dd \alpha \dd \delta \\
        &= \iint_\mathrm{fov} I_1 + I_2 + 2 \mathrm{Re} \left[ E(\alpha, \delta, r_1, t_1) E^\star(\alpha, \delta, r_2, t_2) \right] \, \dd \alpha \dd \delta
    \end{aligned}
\end{equation}
avec $I_1$ et $I_2$ les intensités (norme du champ électrique au carré : $I=EE^\star=\vert E \vert^2$) reçues par chaque télescope. Pour la suite, on suppose que chaque télescope reçoit la même intensité, c-à-d, $E_1 = E_2 = E$ ou encore $I_1=I_2=I$. La double intégrale sur le champ de vue (fov) prend en compte l'extension de la source. L'étoile, quant à elle, marque la conjugaison complexe du champ électrique. Le terme le plus important qui définit l'interférométrie est le produit des champs électriques complexes. On effectue une moyenne temporelle sur un intervalle de temps très supérieur à la période d'oscillation de l'onde afin de s'affranchir de la dépendance à cette dernière. En supposant une source stationnaire (indépendante de $t$), on obtient donc la fonction de cohérence mutuelle, décrivant la corrélation spatiale et temporelle, définie comme
\begin{equation}
    \Gamma(r_1,r_2,\tau) = \frac{\iint_\mathrm{fov} \langle E(\alpha, \delta, r_1, t) E^\star(\alpha, \delta, r_2, t+\tau) \rangle \, \dd \alpha \dd \delta}{\iint_\mathrm{fov} \vert E \vert^2(\alpha, \delta) \, \dd \alpha \dd \delta}.
\end{equation}

Dans le cas d'observations astronomiques avec GRAVITY (ou l'EHT), on s'intéresse uniquement à la corrélation spatiale, en supposant qu'elle ne dépende que de la géométrie de la source et est donc indépendante du temps. On appelle cette quantité la \textit{visibilité complexe} définie comme
\begin{equation}
    \label{eq:complexe_visibility}
    \mathcal{V}(B=r_2-r_1) = \Gamma(r_1,r_2,\tau=0) = \frac{\iint_\mathrm{fov} \langle E(\alpha, \delta, r_1, t) E^\star(\alpha, \delta, r_2, t) \rangle \, \dd \alpha \dd \delta}{\iint_\mathrm{fov} \vert E \vert^2(\alpha, \delta) \, \dd \alpha \dd \delta}.
\end{equation}
On obtient ainsi le degré de cohérence en prenant la norme de la fonction de cohérence mutuelle. Le flux recombiné mesuré peut donc se réécrire
\begin{equation}
    F = F_1 +F_2 + 2 \sqrt{F_1 F_2} \vert \mathcal{V} \vert \, \cos (\mathrm{arg} \mathcal{V}).
\end{equation}

\subsection{Plan u-v} \label{sec:plan_uv}
Comme on l'a vu, une des grandeurs caractéristiques importantes qui va déterminer le motif des franges observées est la distance projetée par rapport à la ligne de visée entre les télescopes qui vaut, au maximum, la distance réelle entre les deux télescopes. On parle alors de ligne de base projetée ($B_\perp$ dans la Fig.~\ref{fig:interfero}). L'interférométrie permet de mesurer la corrélation entre deux points de la source (qui dépend de la ligne de base). Faire la transformée de Fourier de la source revient à déterminer la corrélation entre chaque point de la source. Lorsque l’on fait des observations en interférométrie, on effectue une mesure d’un point de l'espace de Fourier appelé plan~(u,v)~\cite{Haniff2007a,Haniff2007b} correspond à la corrélation mesurée pour une ligne de base donnée (longueur et orientation dans le ciel). L'axe horizontal du plan~(u,v), c'est-à-dire l'axe u dans l'espace de Fourier, correspond à la direction Est-Ouest dans l'espace réel, positif vers l'Est. L'axe vertical, c'est-à-dire l'axe v dans l'espace de Fourier correspond à la direction Nord-Sud dans l'espace réel, positif vers le Nord. Les coordonnées du point mesuré dans l'espace de Fourier (plan u-v) de la source dépendent de la longueur et l'orientation de la ligne de base ainsi que de la longueur d'onde. On a ainsi
\begin{equation}
    \mathcal{V} (B) \rightarrow \mathcal{V} \left( \frac{B_\perp}{\lambda} \right) = \mathcal{V} (u,v).
\end{equation}

\begin{figure}
    \centering
    \resizebox{\hsize}{!}{\includegraphics{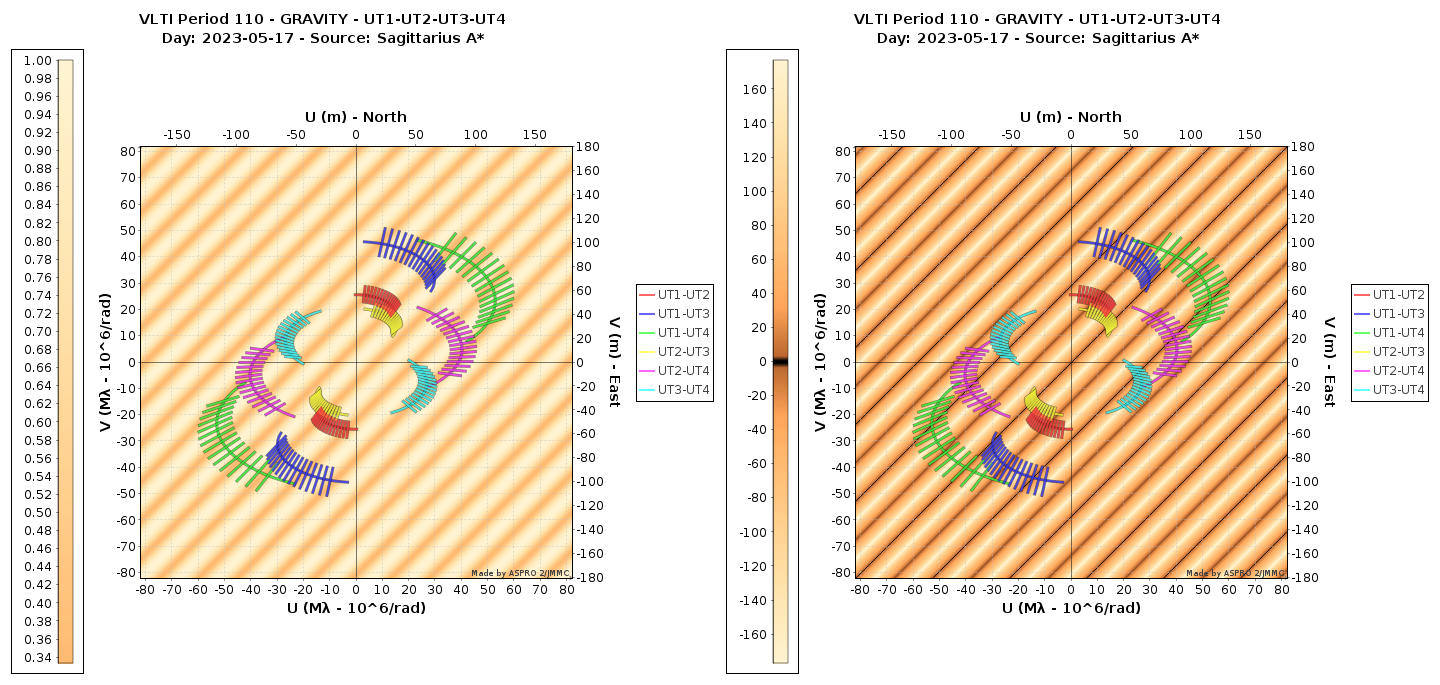}}
    \caption{Échantillonnage du plan~(u,v) lors d'observations simulées du centre galactique centré sur Sagittarius~A* le 17 Mai 2023 par l'instrument GRAVITY avec les 4 UT du VLT. Le temps d'intégration est de 300s avec une observation toutes les 20 min entre 4h15 UT et 9h15 UT (avec un passage au méridien à 6h45 UT à 86° d'élévation). Chaque combinaison de télescopes, correspondant à une couleur, échantillonne l'espace~(u,v) selon un arc. Chaque trait le long de ces arcs montre l'échantillonnage grâce à la décomposition spectrale du signal pour chaque observation. En arrière-plan, on montre à gauche la visibilité et à droite la phase modélisée pour deux points sources, S1 correspondant à Sgr~A* aux coordonnées (0,0) et S2 correspondant à une étoile hypothétique distante de 20 mas de S1 avec un angle de -45° (cet angle se retrouve dans la visibilité et la phase) et un flux deux fois plus grand que celui de S1. Images produites avec \href{http://www.jmmc.fr/apps/beta/Aspro2/}{Aspro2}.}
    \label{fig:uv_coverage}
\end{figure}

Cependant, la source bouge dans le ciel du fait de la rotation de la Terre, ce qui change à la fois la longueur et l'orientation de la ligne de base du point de vue de la source. Ainsi, si l'on fait plusieurs observations d'une même source à différents temps, on mesure différents points de l'espace~(u,v) comme illustré dans la Fig.~\ref{fig:uv_coverage} par les arcs. Ces derniers correspondent à des portions d'ellipses dont le demi grand axe est parallèle à l'axe u et le centre est sur l'axe v, de différentes couleurs, correspondant à différentes combinaisons entre les quatre grands télescopes du VLT. On parle alors de synthèse d'ouverture. 

Le motif de franges obtenu peut être caractérisé par un contraste, aussi appelé \textit{Visibilité (de Michelson)} défini comme
\begin{equation} \label{eq:visibilité}
    \begin{aligned}
        V &= \frac{F_\mathrm{max} - F_\mathrm{min}}{F_\mathrm{max} + F_\mathrm{min}}, \\
          &= \vert \mathcal{V} \vert,
    \end{aligned}
\end{equation}
où $F_\mathrm{min}$ et $F_\mathrm{max}$ sont les flux minimum et maximum resp. dans le plan u-v, et une \textit{phase} $\varphi$, définie comme la position de la frange brillante par rapport à un point de référence, souvent l'axe optique, défini comme la droite où la différence de marche est nulle et dans le plan médiateur de la ligne de base. Ce sont ces deux quantités qui vont être mesurées et ajustées avec des modèles (voir section~\ref{sec:fitting}). Dans le cas d'une source unique au niveau du plan médiateur, on a $F_1=F_2$ et donc $F_\mathrm{min}=F(\mathrm{arg}(\mathcal{\gamma})=\pi)=0$, on obtient donc $V=1$ et $\varphi=0$ puisque la frange brillante est sur l'axe optique comme dans la Fig.~\ref{fig:fente_de_Young}. Lorsque la source n'est pas sur l'axe optique, la visibilité reste égale à un mais la phase sera différente de zéro. On note que ceci est vrai dans le cas d'une onde monochromatique, mais pas en bande large dont le traitement est plus complexe.

Le déphasage $\Delta \Phi$ entre les deux ondes issues des deux ouvertures dépend de la distance $a$ entre ces dernières, mais aussi de longueur d'onde $\lambda$ (Eq.~\ref{eq:dephasage}). Ainsi, mesurer la visibilité et la phase à deux longueurs d'ondes différentes revient à échantillonner deux points différents de l'espace~(u,v). Cela est clairement visible dans la Fig.~\ref{fig:uv_coverage} avec les traits radiaux le long des arcs. En effet, l'instrument GRAVITY mesure la visibilité et la phase dans différents canaux spectraux entre $2,0$ et $2,4$ $\mu m$.

\subsection{Reconstruction d'image}
Le module au carré de la transformée de Fourier d'un signal quelconque dans le temps est un spectre de puissance, c'est-à-dire la puissance par unité de fréquence. Ainsi, la transformée de Fourier d'un signal sinusoïdal est un Dirac à la fréquence de la sinusoïde. Dans notre cas, ce n'est pas un signal temporel (1D) mais un signal spatial à deux dimensions (d'où une surface au lieu d'une courbe). De manière similaire au cas du signal temporel, on parle de fréquence spatiale pour le plan~(u,v). Le théorème de Zernike-van Cittert dit que la corrélation spatiale du champ électrique (décrite précédemment par la visibilité complexe Eq.~\eqref{eq:complexe_visibility}) est directement reliée à la distribution spatiale d'intensité de la source telle que
\begin{equation}
    \frac{\mathcal{V} (u,v)}{\mathcal{V} (0,0)} = \frac{\iint I(\alpha,\delta) \exp (-2i\pi(u\alpha +v\delta))\, \dd \alpha \dd \delta}{\iint I(\alpha,\delta)\, \dd \alpha \dd \delta}.
\end{equation}
Les courtes fréquences spatiales (u et v) correspondent aux variations à grande échelle et inversement, les grandes fréquences spatiales correspondent aux plus petites échelles dans le ciel. Ainsi, si l'on veut mesurer des détails à de petites échelles (spatiales dans le ciel) il faut échantillonner les grandes fréquences spatiales et donc avoir de longues lignes de base. On retrouve ainsi l'équivalence entre un télescope unique d'un diamètre $D$ et un interféromètre dont la ligne de base la plus longue est $D$ en termes de résolution angulaire.

Lorsque que l'on a un bon échantillonnage, c'est-à-dire un grand nombre de points, idéalement répartis uniformément dans l'espace~(u,v), on peut reconstruire l'image de la source en faisant une transformée de Fourier inverse. Cependant, il n'est pas possible d'échantillonner de manière uniforme l'espace~(u,v) du fait de la rotation de la Terre qui forme des arcs dans l'espace~(u,v). De plus, échantillonner aussi bien les basses que les hautes fréquences spatiales requiert un grand nombre de lignes de base avec des distances couvrant plusieurs ordres de grandeurs. Les "images" de M87 et Sgr~A*~\cite{EHT2019, EHT2022a} obtenues par la collaboration EHT sont basées sur ce principe sans avoir un échantillonnage uniforme (impossible à obtenir en pratique). Les images obtenues sont donc des approximations de la réalité, construites à partir d'un échantillonnage limité (mais néanmoins conséquent) du plan~(u,v) et basées sur de nombreuses hypothèses pour résoudre le problème inverse et les dégénérescences. La plupart des lignes de base de la collaboration EHT ont des longueurs de plusieurs dizaines voire milliers de kilomètres permettant d'atteindre une résolution suffisante pour reconstruire l'image de l'ombre et l'environnement proche de M87 et Sgr~A*. Néanmoins, le manque de courtes lignes de base (les réseaux interférométriques comme ALMA sont utilisés comme une antenne unique) entraîne un manque d'information à grande échelle. Les images reconstruites à partir de données interférométriques sont toujours des approximations de la réalité du fait de l'échantillonnage imparfait du plan~(u,v).

\section{Performances de GRAVITY}
On a évoqué brièvement l'instrument GRAVITY précédemment, on rappelle ici ses caractéristiques techniques ainsi que ses performances.

\subsection{Caractéristiques de l'instrument}
GRAVITY est un interféromètre qui combine la lumière de quatre télescopes du VLT, soit les UT de 8m de diamètre, soit les \textit{Auxilary Telescopes} (AT) de 1,8m de diamètre. Chaque télescope est équipé d'optique adaptative (AO) qui permet de corriger les effets de la turbulence atmosphérique et ainsi atteindre leur plein potentiel. Pour les observations du centre galactique, il est nécessaire d'utiliser les UT afin d'atteindre une sensibilité photométrique suffisante pour étudier cet environnement (jusqu'à magnitude 19-20 en bande K). 

\begin{figure}
    \centering
    \resizebox{\hsize}{!}{\includegraphics{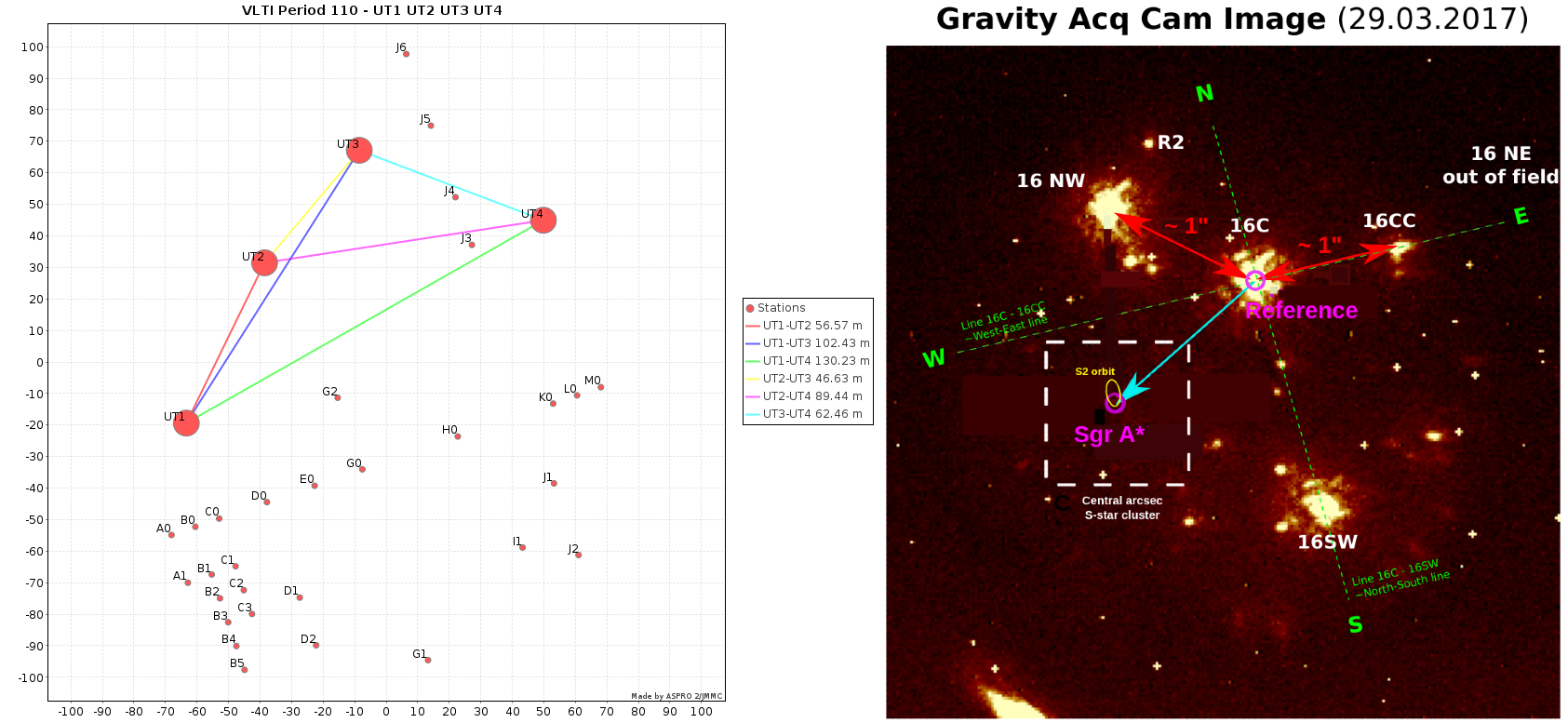}}
    \caption{\textbf{(Gauche)} Configuration des \textit{Unit Telescopes} du VLT Interferometer à Paranal, Chili. Image produite avec \href{http://www.jmmc.fr/apps/beta/Aspro2/}{Aspro2}. \textbf{(Droite)} Image du centre galactique observé par la caméra d'acquisition de GRAVITY d'un des UT montrant la source de référence pour le suivi des franges et la cible d'intérêt, à savoir Sgr~A* et les étoiles S.}
    \label{fig:VLTI_config_AcqCam}
\end{figure}

La configuration des UT, pour le calcul des lignes de base, est illustrée dans la Fig.~\ref{fig:VLTI_config_AcqCam} avec une ligne de base maximale de 130 m ce qui conduit à une résolution angulaire maximale, utilisant l'Eq.~\eqref{eq:resolution_angulaire}, de $\sim 4$ mas. GRAVITY observe dans la bande K entre $1,9727$~$\mu m$ et $2,4273$~$\mu m$ et est équipé d'un spectromètre avec différentes résolutions spectrales (LOW : 22, MEDIUM : 500 et HIGH : 4500 ; \href{https://www.eso.org/public/teles-instr/paranal-observatory/vlt/vlt-instr/gravity/}{ESO}).

Pour fonctionner, GRAVITY a besoin d'une référence pour suivre les franges d'interférence. Cette référence doit à la fois être une source brillante et se trouver à maximum $\sim 2$ arcsec de la cible scientifique pour les UT (6 arcsec pour les AT). Dans le cas du centre galactique, cette cible de référence est parmi le groupe d'étoiles IRS 16 comme l'illustre la Fig.~\ref{fig:VLTI_config_AcqCam}, qui montre aussi la localisation de Sagittarius~A* dans les quelques parsecs centraux de la Voie Lactée, vue par les télescopes du VLT.


\subsection{Ajustement des données}\label{sec:fitting}
\begin{figure}
    \centering
    \resizebox{0.7\hsize}{!}{\includegraphics{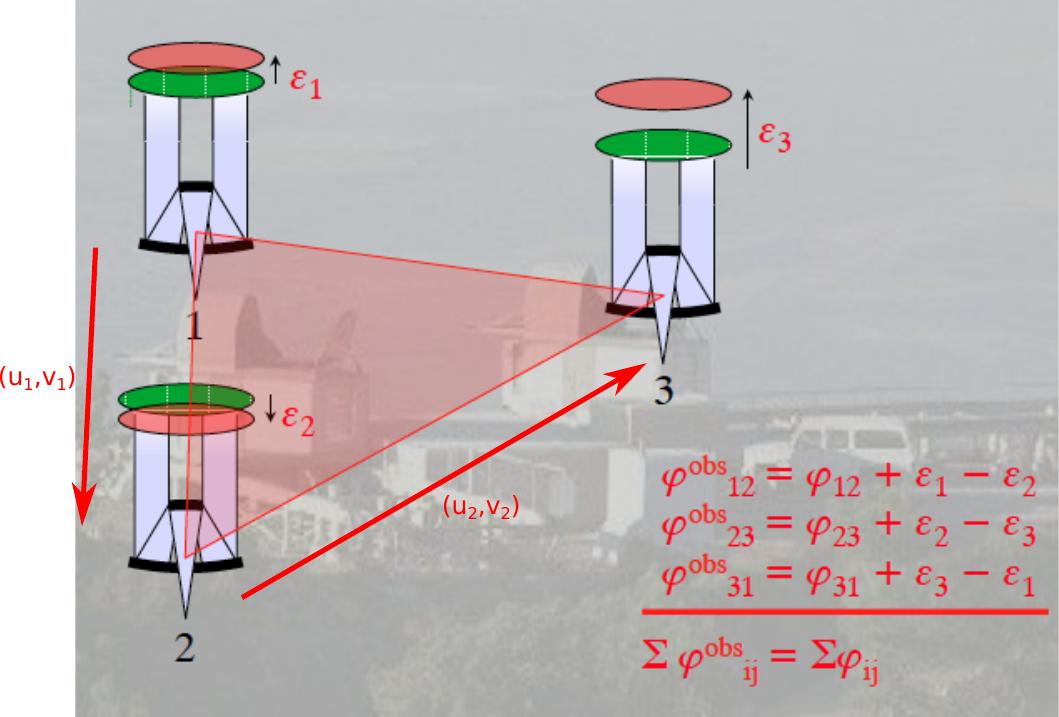}}
    \caption{Schéma illustrant le décalage d'un même plan d'onde dû à l'atmosphère entre différents télescopes. En faisant la somme des phases de trois télescopes (combinés deux à deux), ces décalages s'annulent. Crédit : Guy Perrin}
    \label{fig:T3Phi}
\end{figure}

Comme on l'a vu précédemment, les données mesurées sont la visibilité (Vis) ou le module de la visibilité au carré (Vis2) et la phase des franges d'interférence par rapport à un point de référence (VisPhi). Cependant, la phase est sensible aux perturbations atmosphériques. En effet, la lumière de la source atteignant chaque télescope a traversé une couche d'atmosphère légèrement différente. L'optique adaptative permet de retrouver une onde quasi plane en déformant un miroir en opposition à la perturbation de l'atmosphère. Cependant, les masses d'air étant différentes, ce plan d'onde peut être décalé entre chaque télescope comme illustré dans la Fig.~\ref{fig:T3Phi}. Ce décalage du même plan d'onde par rapport à chaque télescope, appelé \textit{piston différentiel}, se traduit en une différence de marche et donc une différence de phase lorsque l'on fait interférer les signaux qui ne sont pas liés à la source, mais uniquement à l'atmosphère. Afin de compenser ce décalage, on fait la somme des phases de trois télescopes (combiné deux à deux) afin que ces décalages s'annulent entre eux. On appelle cette quantité corrigée des biais atmosphérique la clôture de phase T3Phi.

Le signal obtenu pour chaque combinaison de télescopes et aux différentes longueurs d'ondes dépend de la composition de l'objet observé. Pour un objet composé d'un unique point source, on obtient une visibilité de 1 partout dans le plan u-v. La phase, cependant, dépend de la position du point source par rapport au point de référence (le plus simple étant le centre du champ de vue) et varie linéairement dans la direction orthogonale du vecteur position du point source dans le champ de vue (comme dans l'expérience des fentes de Young, où la figure d'interférence s'étale dans la direction perpendiculaire de l'orientation des fentes\footnote{Attention, il s'agit ici d'une comparaison afin de se représenter le signal de phase dans le plan u-v, qui est de nature complètement différente de la figure d'interférence générée par les fentes de Young !}).

\begin{figure}
    \centering
    \resizebox{0.7\hsize}{!}{\includegraphics{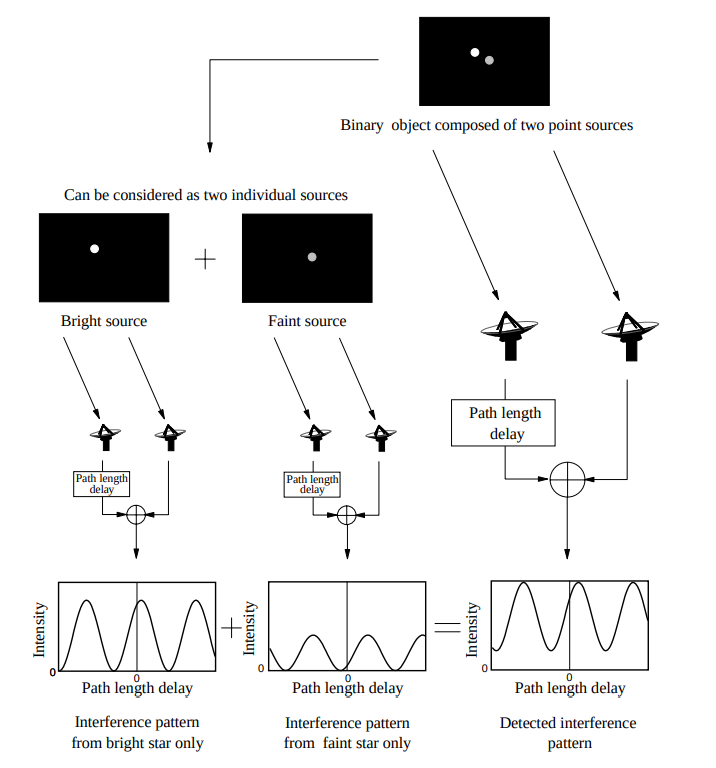}}
    \caption{Relation entre la distribution de la luminosité d'une source particulière et les franges observées dans un interféromètre à deux éléments. Dans le cas illustré ici, l'objet est une étoile binaire composée de deux éléments inégaux et non résolus. Chaque composante produit sa propre figure d'interférence avec une amplitude qui dépend de sa luminosité et une phase, $\varphi$, liée à sa position dans le ciel. Il est à noter que, puisque les deux composantes ne sont pas résolues, chacune d'entre elles produit une figure de frange avec une visibilité $V = 1$. La configuration globale des franges observées est la superposition des schémas d'intensité de chaque composante de la source. L'information sur la luminosité relative et la localisation des deux composantes est contenue dans la visibilité et la phase de cette image résultante. Crédit : \cite{Haniff2003}.}
    \label{fig:two_source_interfero}
\end{figure}

En pratique, le champ de vue d'un télescope contient, la plupart du temps, plusieurs sources qui peuvent être ponctuelles, par exemple plusieurs étoiles, ou étendues comme une nébuleuse, une galaxie, ou du gaz interstellaire faisant office de fond de ciel. D'un point de vue interférométrique, chaque point source va créer une figure d'interférence. La visibilité complexe ainsi mesurée est la somme de la figure d'interférence de chaque source incohérente entre elles, pondérée par leurs flux respectifs~\cite{Waisberg2019} comme l'illustrent les Fig.~\ref{fig:uv_coverage} et \ref{fig:two_source_interfero} (par le théorème de Zernicke-van Cittert). Une source étendue est vue comme un ensemble (plus ou moins important) de points sources rapprochés. Chaque point source a, individuellement, une visibilité égale à 1, mais une phase qui dépend de sa position sur le ciel.

\begin{figure}
    \centering
    \resizebox{0.7\hsize}{!}{\includegraphics{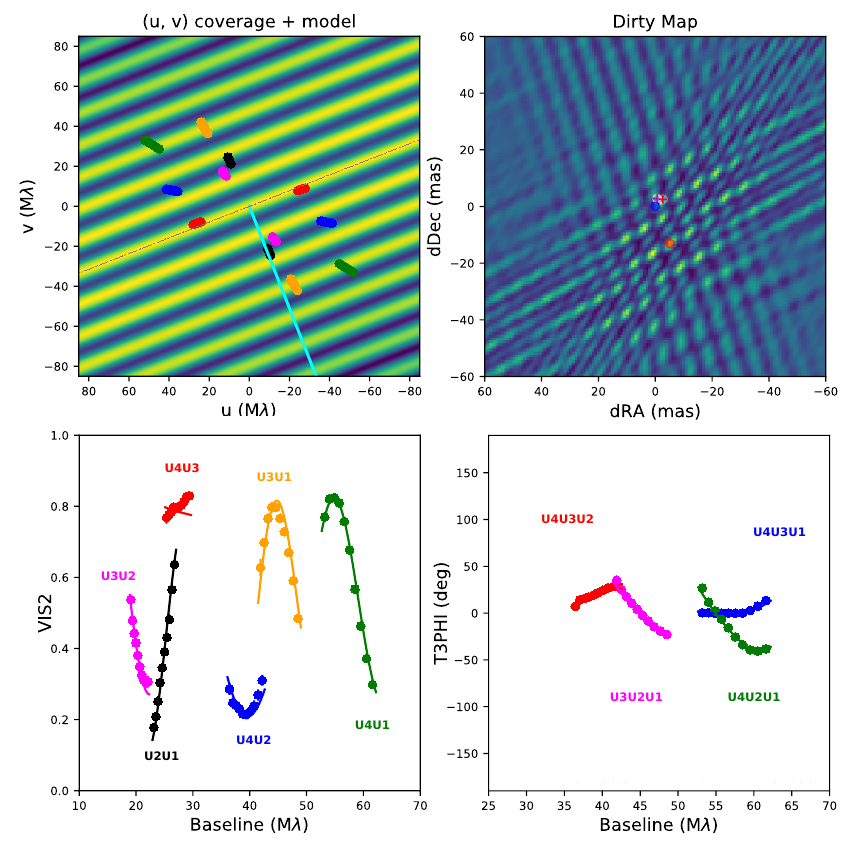}}
    \caption{Exemple d'ajustement des données GRAVITY du 05 Mai 2018 centré sur Sagittarius~A*. Le modèle est composé de deux points sources Sgr~A* et S2 dont les positions sont représentées dans le cadre en haut à droite par des points de couleurs bleu et rouge respectivement. Le cadre en haut à gauche montre la couverture u-v de cette observation ainsi que l'amplitude modèle en arrière-plan. Les deux cadres du bas montrent respectivement les quantités VIS2 et T3Phi observées ainsi que le meilleur ajustement du modèle. On note que la ligne de base rouge est mal ajustée par le modèle, car elle est perpendiculaire à la direction de séparation des deux sources et ne contient donc aucune signature de cette modulation.}
    \label{fig:GRAVITY_data}
\end{figure}

Les quantités observées sont donc ajustées à partir de modèles analytiques qui peuvent être composés, entre autres, d'un nombre arbitraire de points sources\footnote{Dans le cas du centre galactique, pour d'autres cibles, il existe d'autres modèles.} positionnés à des coordonnées particulières et avec un certain rapport de flux par rapport à une source de référence. Ces derniers (position et flux relatif) forment un ensemble de paramètres du modèle qui sont déterminés grâce à un algorithme d'ajustement du type de l'algorithme python \textit{curvefit} ou \textit{Markov Chain Monte Carlo} (MCMC). Il existe d'autres paramètres comme l'indice spectral de la source et l'arrière-plan qui peuvent être fixés afin de limiter le nombre de paramètres à ajuster. La Fig.~\ref{fig:GRAVITY_data} montre des données prises durant la nuit du 05 Mai 2018 centrées sur Sagittarius~A*. On peut notamment y voir la couverture u-v ainsi que la visibilité du modèle en haut à gauche, et les points de données pour chaque combinaison de télescope avec leurs incertitudes et le modèle en trait plein pour la visibilité au carré (VIS2), et T3PHI. Le modèle utilisé est composé de deux sources ponctuelles dont les positions issues du meilleur ajustement sont marquées dans le cadre en haut à droite nommé \textit{Dirty Map}, correspondant à la transformée de Fourier des visibilités (V2 et T3Phi), représentées par les points bleus et rouges.

On note que l'on n'ajuste pas toujours l'ensemble des quantités mesurées (VISAMP, VIS2, VISPHI, T3PHI et T3AMP) comme c'est le cas dans l'exemple ci-dessus. En effet, si l'information qui nous intéresse est de connaître la séparation entre les sources sans se préoccuper de leur position absolue dans le champ de vue, il n'est pas nécessaire d'ajuster VISPhi puisqu'elle est définie par rapport à un point de référence. Dans ce cas de figure, seules VIS2 et T3Phi peuvent être ajustées. Cependant, si l'information de la position des étoiles dans le champ est nécessaire, pour de l'astrométrie à deux champs (\textit{dual Field astrometry} en anglais), comme on le verra dans la section suivante, il faut aussi ajuster la phase VISPhi.

\subsection{Précision astrométrique}
La résolution angulaire maximale de GRAVITY avec les quatre UT soit $\sim 4$ mas, les signaux mesurés sont des quantités interférométriques (VIS2, T3Phi, VISPhi) qui permettent de déterminer la position des points sources (ponctuels) les uns par rapport aux autres et par rapport au centre de phase. La phase étant très sensible à la position des sources les unes par rapport aux autres, la valeur obtenue sur les positions est nettement plus précise que la résolution angulaire et dépend des incertitudes de mesure et de la couverture u-v. En effet, dans le cas d'incertitudes trop grandes ou de mauvaise couverture u-v, il y a un grand nombre de modèles qui peuvent expliquer les données. Au contraire, plus les incertitudes sur les données sont faibles et meilleur est la couverture u-v, plus le modèle nécessaire pour expliquer les données sera précis (voire complexe). Cela se traduit par les barres d'erreurs faibles sur chaque paramètre ajusté. En moyenne, la précision astrométrique de GRAVITY, c'est-à-dire l'incertitude sur les positions des sources modélisées $\delta$RA et $\delta$DEC, est de l'ordre de $\sim 30-50\, \mu$as.

\begin{figure}
    \centering
    \resizebox{0.8\hsize}{!}{\includegraphics{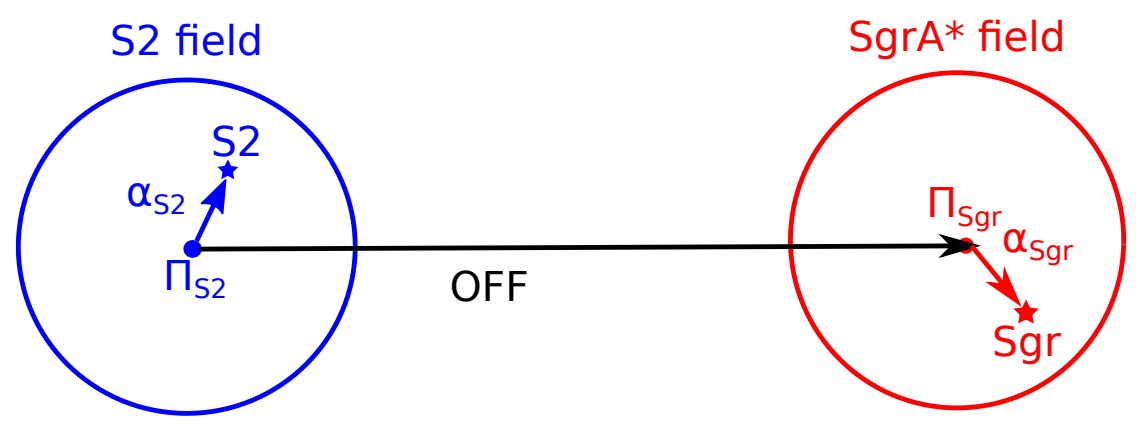}}
    \caption{Illustration de la détermination de la séparation angulaire entre deux sources (Sgr~A* et S2) ne pouvant pas être dans un unique champ de vue. On fait alors de l'astrométrie à deux champs séparé d'une distance angulaire OFF.}
    \label{fig:dual_field_astrometry}
\end{figure}

Ainsi, l'incertitude de mesure de position des étoiles~S pour déterminer leur orbite autour de Sgr~A* est de cet ordre de grandeur lors de leur passage au péri-astre. Lorsqu'une étoile~S est en dehors du champ de vue de Sgr~A*, on fait de l'astrométrie à deux champs, c'est-à-dire un champ "centré"\footnote{On place le champ aux coordonnées prédites qui peuvent être différentes de la position réelle.} sur Sgr~A* et l'autre "centré" sur l'étoile qui nous intéresse. Comme les deux sources ne sont pas dans le même champ, on ne peut pas se contenter d'ajuster T3Phi pour avoir leur séparation. On ajuste donc VISPhi dans les deux champs de vue afin de connaître la position des deux sources dans leurs champs de vue respectifs. Comme on connaît l'écart entre les deux champs de vue, nommé OFF (ce n'est pas une mesure, mais une grandeur déterminée), on peut en déduire la séparation entre les deux sources comme illustré par la Fig.~\ref{fig:dual_field_astrometry}.

Sagittarius~A* est une source variable qui présente parfois de puissants sursauts de rayonnement. La source de ces sursauts ne fait pas encore l'unanimité bien qu'un scénario commence à se démarquer. C'est l'objet de cette thèse. Un des objectifs de GRAVITY est d'étudier ces sursauts de rayonnement en mesurant l'évolution de la position de ces sursauts au cours du temps. En effet, en mesurant la séparation angulaire entre une étoile de référence et Sgr~A* (en état de sursaut) à différents temps, on peut déterminer la trajectoire dans le ciel de la source de ce rayonnement en soustrayant le mouvement de l'étoile (plus de détails dans le Chap~\ref{chap:Sgr~A* flares}) avec la même précision astrométrique que précédemment, à savoir $\sim 30-50$ $\mu$as. Dans la plupart des scénarios envisagés, la source des sursauts se situe proche de l'horizon du trou noir, à quelques rayons gravitationnels (voir définitions dans le Chap~\ref{chap:GYOTO}). En termes de distance angulaire, cela correspond de quelques dizaines à une centaine de $\mu$as. La précision astrométrique de GRAVITY est donc suffisante, mais surtout nécessaire pour étudier ce phénomène.

\chapter{Notions de Relativité Générale et le code de tracé de rayon GYOTO}\label{chap:GYOTO}
\markboth{Notions de Relativité Générale et le code de tracé de rayon GYOTO}{Notions de Relativité Générale et le code de tracé de rayon GYOTO}
{
\hypersetup{linkcolor=black}
\minitoc 
}

\section{Notions de Relativité Générale}
Cette section est inspirée par le cours de 
 \href{http://luth.obspm.fr/~luthier/gourgoulhon/fr/master/relat.html}{Relativité Générale d'Éric Gourgoulhon}, on résume ici les équations principales qui nous intéressent pour le reste de ce manuscrit. Le lecteur peut se référer à ce cours pour les étapes intermédiaires de calcul.

\subsection{Fondements de la Relativité Générale}
La publication de la Relativité Générale en 1915 par Albert Einstein a totalement bouleversé notre compréhension de la gravitation. Précédemment, la gravitation était considérée comme une force fondamentale comme l'électromagnétisme, les forces nucléaires forte et faible. Ainsi, la trajectoire d'une particule (dans le sens général du terme) dans le vide, proche d'un objet massif, était définie par la force gravitationnelle de ce dernier. En Relativité Générale, la gravitation n'est plus une force fondamentale, mais une conséquence de la géométrie de l'espace-temps qui est dictée par son contenu en matière ou énergie.

Pour construire la Relativité Générale, Einstein a formulé des postulats qui sont les fondements de la RG résumés ci-après.

\subsubsection{Le principe de relativité}
Ce principe, exprimé en premier lieu par Galilée, est le fondement des théories d'Einstein au point qu'il sera inclus dans leurs noms. Il stipule que les Lois de la physique sont invariantes quel que soit le référentiel choisi, à condition que ce soit un référentiel inertiel, c'est-à-dire en mouvement rectiligne uniforme dans le cas de la Relativité Restreinte, et quelconque en Relativité Générale. Un exemple classique de ce principe fait intervenir deux personnes, une sur Terre, immobile par rapport au sol et l'autre dans le nid de pie d'un galion en mouvement rectiligne à vitesse constante. Ainsi, les deux référentiels sont inertiels (en négligeant le mouvement de la Terre, hypothèse valable si la durée de l'expérience est très inférieure à une année). Ces deux personnes lâchent un boulet de canon sans vitesse initiale. Dans leur référentiel respectif, le boulet tombe verticalement, car les lois de la physique, en l'occurrence, la gravitation dans cet exemple, s'appliquent de la même manière. Cependant, vu depuis l'observateur sur Terre, le boulet lâché depuis le haut du mât du galion a une trajectoire parabolique (loi de composition des vitesses). Par ce principe, les lois de la physique sont les mêmes partout dans l'univers, que ce soit sur Terre, en laboratoire, dans une galaxie lointaine (même très lointaine) ou proche d'un trou noir, à condition que le référentiel soit inertiel. Une des conséquences est qu'il n'existe pas de référentiel privilégié. Il est à noter cependant que ce principe n'a pas été vérifié et que certaines théories en cosmologie rompent ce principe.

\subsubsection{Universalité de la vitesse de la lumière}
Afin de créer la Relativité Restreinte (RR), Einstein a établi un postulat impliquant une modification importante d'une loi fondamentale de la mécanique, la loi de composition des vitesses. En effet, si l'on reprend l'exemple ci-dessus, du point de vue de l'observateur sur Terre, le boulet lâché depuis le haut du mat du galion suit une trajectoire parabolique, car à tout instant sa vitesse est la somme vectorielle de sa vitesse verticale, déterminé par la pesanteur, et sa vitesse horizontale définie par la vitesse du référentiel inertiel dans lequel il chute, à savoir le galion. Einstein postula que la vitesse de la lumière (dans le vide) est constante quel que soit le référentiel, ce qui modifie sensiblement la loi de composition des vitesses en relativité (Restreinte et Générale).

Ceci mène à des effets décrits par la Relativité Restreinte. Afin de les illustrer, prenons un observateur A quittant la galaxie de Pégase (à bord d'un vaisseau Lantien de classe Aurore) en direction de la Voie Lactée à 90\% de la vitesse de la lumière dans le référentiel de la Terre (environ 3 millions d'années-lumière de distance) ainsi qu'un observateur B immobile par rapport à cette dernière, positionné à mi-parcours du trajet du vaisseau. A l'instant $t_0=0$, les deux observateurs sont confondu (même positions). L'observateur B émet un photon en direction de la Terre grâce à un laser. L'observateur B et le vaisseau observent la propagation de ce photon à une vitesse d'environ 300 000 km/s. En une seconde, dans le référentiel de l'observateur B, la lumière a parcouru $\sim$300.000 km et le vaisseau $\sim$270.000 km. Ils sont donc distants de 30 000 km. Dans le référentiel du vaisseau, en une seconde le photon a parcouru la même distance, donc $\sim$300.000 km, bien plus que les 30.000 km précédents, ce qui parait paradoxal. La solution à ce problème est que le temps et l'espace mesurés par l'observateur A dans le vaisseau, ne sont pas égaux à ceux mesurés par l'observateur B dans leurs référentiels respectifs. Une seconde mesurée dans le référentiel du vaisseau correspond à 2,3 secondes (voir calcul plus bas) dans le référentiel de l'observateur B. C'est la dilatation du temps en Relativité Restreinte due à la vitesse relative entre deux référentiels inertiels. On note que la définition de la seconde (et du mètre) est la même pour tous les référentiels propres (en RR), la dilatation du temps concerne le temps mesuré dans un référentiel en mouvement comparé au temps propre (mesuré dans le référentiel propre, c'est-à-dire comobile avec le phénomène mesuré). Un autre effet comparable s'applique sur les longueurs qui sont, à l'inverse du temps, contractées par rapport aux longueurs propres mesurées dans un référentiel comobile avec le phénomène considéré. Pour déterminer le facteur de dilatation temporelle et de contraction des longueurs, on peut calculer le facteur de Lorentz
\begin{equation}
    \gamma = (1-\beta^2)^{-1/2}
\end{equation}
avec $\beta=v/c$, $v$ étant la vitesse du vaisseau et $c$ la vitesse de la lumière dans le vide. Ainsi le temps propre $\tau$ de l'observateur A voyageant entre Pégase et la Voie Lactée et le temps mesuré dans le référentiel de l'observateur B $t$ sont liés par
\begin{equation} \label{eq:dilatation_temps}
    t = \gamma \tau
\end{equation}
et les longueurs dans le référentiel de l'observateur A dans le vaisseau $L$ et dans le référentiel de l'observateur B $l$ par
\begin{equation}\label{eq:contraction_longueur}
    l = \frac{L}{\gamma}.
\end{equation}

De manière plus générale, les équations pour passer d'un référentiel à un autre sont regroupées dans les transformations de Lorentz qui expriment les coordonnées ($t$,$x$,$y$,$z$) d'un évènement dans un référentiel fixe en fonction des coordonnées ($t^\prime$,$x^\prime$,$y^\prime$,$z^\prime$) du même évènement dans un référentiel en mouvement selon l'axe $\vec{x}$ du référentiel fixe. Elles s'écrivent :
\begin{equation}\label{eq:Lorentz_transformation}
\left\{ \begin{aligned}
    ct^\prime &= \gamma (ct - \beta x)\\
    x^\prime &= \gamma (x - \beta ct)\\
    y^\prime &= y\\
    z^\prime &= z
    \end{aligned} \right.
\end{equation}

Enfin, un des postulats fondamentaux de la relativité (restreinte et générale) est qu'un objet matériel de masse non nulle a une vitesse maximale asymptotique, c'est-à-dire qu'il ne peut pas égaler, qui est la vitesse de la lumière dans le vide. Ainsi $\beta < 1$ et $\gamma > 1$.

\subsubsection{Le principe d'équivalence}
Ce principe stipule que la masse gravitationnelle $m_g$ (celle qui apparaît dans la force gravitationnelle Newtonienne $\vec{F_g}=m_g \vec{g}$) et la masse inertielle $m_i$ (celle qui apparaît à droite de la troisième loi de Newton $\sum \vec{F} = m_i \vec{a}$) sont égales. Ce principe constitue en réalité le principe d'équivalence faible. Ainsi deux objets soumis uniquement à la gravitation (donc dans le vide) subiront la même accélération, c'est-à-dire tomberont à la même vitesse et atteindront le sol en même temps, quelles que soient leurs masses où leurs propriétés. La vérification de ce principe a été un des objectifs du programme Apollo 15 en faisant tomber un marteau et une plume sans atmosphère. Depuis, des tests expérimentaux tentent de mesurer une différence entre ces deux masses, ou tout du moins une limite supérieure de celle-ci. À ce jour, aucun écart de masse a été mesuré à $\sim 10^{-15}$ près (\href{https://en.wikipedia.org/wiki/MICROSCOPE}{MICROSCOPE}).

Cette équivalence rend la gravitation singulière par rapport aux autres interactions fondamentales, car ces dernières dépendent des propriétés internes de l'objet, comme la charge pour la force électromagnétique, alors que ce n'est pas le cas pour la gravitation. Ainsi, toutes particules (massives ou non) subissent la gravitation, la rendant universelle.

\begin{figure}
    \centering
    \resizebox{0.5\hsize}{!}{\includegraphics{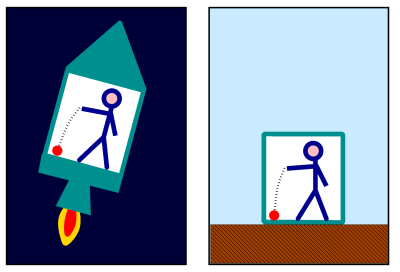}}
    \caption{Illustration du principe d'équivalence en Relativité Générale. La balle tombe vers le sol de la même manière que ce soit dans une fusée ayant une accélération de 1g ou sur Terre du fait de la gravitation. Crédit : \href{https://commons.wikimedia.org/wiki/File:Elevator_gravity2.png}{Wikimedia Commons}}
    \label{fig:principe_equivalence}
\end{figure}

La conséquence est qu'en présence de gravitation, il est impossible de trouver un référentiel inertiel, défini comme un référentiel où une particule ne subissant aucune force\footnote{La gravitation n'est pas une force en Relativité Générale.} suivra un mouvement rectiligne uniforme. Cependant, le principe d'équivalence formulé par Einstein comme postulat de la Relativité Générale, dit que localement, sans aucune force extérieure (donc dans le vide), il est impossible de faire la différence entre un champ de pesanteur gravitationnel et une accélération subie par le référentiel de l'observateur comme illustré par la  Fig.~\ref{fig:principe_equivalence}. Autrement dit, les passagers d'un vaisseau allant vers Alpha Centauri (pour la planète Pandora comme le Venture Star) ayant une accélération constante de 1g ressentiraient le même champ de pesanteur que sur Terre.
La combinaison de ces deux principes mène à la définition moderne du \textit{Principe d'équivalence d'Einstein}, stipulant qu'il existe en tout point de l'espace-temps un référentiel localement inertiel, celui en chute libre dans le champ de gravitation, qui peut être associé à un espace-temps de Minkowski, c'est-à-dire un espace-temps plat où les lois de la physique décrites par la Relativité Restreinte s'appliquent.

\subsection{Définitions et formalisme}
\subsubsection{4-vecteurs}
Comme dit précédemment, en relativité, l'espace et le temps ne sont pas dissociés et forment un espace-temps à quatre dimensions, une temporelle et trois spatiales. On note, dans le cas le plus général, les coordonnées d'un évènement M de l'espace-temps $x_0$, $x_1$, $x_2$ et $x_3$ avec $x_0$ la coordonnée de genre temps et les trois autres de genre espace. Ainsi, dans le cas d'un repère cartésien à quatre dimensions, les coordonnées d'un point de l'espace-temps sont ($ct, x, y, z$)\footnote{On remarque ici que la composante temporelle est $ct$ et non $t$ pour avoir la dimension d'une longueur et être homogène avec les coordonnées spatiales} et dans un repère sphérique ($ct, r, \theta, \varphi$).

De manière similaire à la mécanique classique où l'on définit un vecteur tridimensionnel pour définir une position, vitesse ou accélération, on définit ici des \textit{quadri-vecteurs} ou \textit{4-vecteurs} notés en gras. Ces vecteurs sont définis à travers une base composée de 4 quadri-vecteurs unitaires orthogonaux ($\mathbf{e_0}, \mathbf{e_1}, \mathbf{e_2}, \mathbf{e_3}$), comme la base naturelle cartésienne
\begin{equation}
\left\{ \begin{aligned}
    \mathbf{e_t} &= (1,0,0,0)\\
    \mathbf{e_x} &= (0,1,0,0)\\
    \mathbf{e_y} &= (0,0,1,0)\\
    \mathbf{e_z} &= (0,0,0,1)
    \end{aligned} \right.
\end{equation}
Ainsi, un 4-vecteur quelconque s'exprime de la manière suivante :
\begin{equation}
    \mathbf{v} = v^0 \mathbf{e_0} + v^1 \mathbf{e_1} + v^2 \mathbf{e_2} + v^3 \mathbf{e_3}
\end{equation}
avec ($v^0$, $v^1$, $v^2$, $v^3$) les coordonnées de $\mathbf{v}$ dans la base ($\mathbf{e_0}, \mathbf{e_1}, \mathbf{e_2}, \mathbf{e_3}$).

On note que les chiffres en exposant comme le $0$ de $v^0$ correspondent ici à l'indice des coordonnées d'un 4-vecteur allant de 0 à 3 (et non de 1 à 4). Il ne faut pas les confondre avec une puissance ! On écrirait dans ce cas $(v^0)^2$ pour le carré de la composante de temps du 4-vecteur $\mathbf{v}$.

On peut donc exprimer un 4-vecteur comme la somme de chaque composante multipliée par leur vecteur de base (comme en physique classique)
\begin{equation}
    \mathbf{v} = \sum^3_{\mu=0} v^\mu \mathbf{e_\mu}
\end{equation}
que l'on simplifie avec la \textit{règle de sommation d'Einstein}
\begin{equation}
    \mathbf{v} = v^\mu e_\mu
\end{equation}
où la sommation sur tous les indices (ici de 0 à 3) est implicite lorsque l'on a un indice en haut et un indice en bas du même coté de l'équation. Par convention, on utilise un caractère Grec lorsque l'on parle de 4-vecteurs (0,1,2,3) et un caractère Latin quand on parle de vecteur tridimensionnel (les trois dimensions de genre espace, c'est-à-dire 1,2,3) que l'on notera avec le symbole de vecteur usuel (pas en gras pour différencier des 4-vecteurs).

La sommation d'Einstein implique d'avoir un terme avec un indice en haut et un autre avec un indice en bas. La position de l'indice a donc une importance cruciale et est loin d'être arbitraire. Lorsque l'indice des coordonnées est en haut, cela signifie que les coordonnées sont dites \textit{contravariantes}, c'est-à-dire les composantes d'un 4-vecteur. C'est le cas des coordonnées de position, vitesse, accélération, force, etc. À l'inverse, lorsque l'indice est en bas, on traite des quantités dites covariantes, qui correspondent aux coordonnées de formes linéaires, des objets mathématiques permettant d'associer un nombre réel à un vecteur. On note que les vecteurs de base d'un système de coordonnées arbitraire $\mathbf{e_\alpha}$ sont des 4-vecteurs (il y a quatre 4-vecteurs) avec un indice en bas pour la notation d'Einstein. Le lien entre coordonnées contravariantes et covariantes se fait grâce à la métrique (voir section~\ref{sec:métrique}). Certaines quantités s'expriment plus naturellement en coordonnées covariantes, qu'en coordonnées contravariantes, comme les constantes du mouvement à partir des coordonnées covariantes de la 4-impulsion (voir section~\ref{sec:gyoto_géodésique}).

\subsubsection{Produit scalaire et tenseur métrique} \label{sec:métrique}
On se place pour le moment en espace-temps plat. En physique classique, dans un espace tridimensionnel, le produit scalaire est notamment utilisé pour calculer la norme d'un vecteur, dans le cas du vecteur position, la distance entre deux points A et B. Il s'exprime, en coordonnées cartésiennes, tel que
\begin{equation}
    \vec{AB} \cdot \vec{AB} = (x_B - x_A)^2 + (y_B - y_A)^2 + (z_B - z_A)^2.
\end{equation}
Or, avec la contraction des longueurs (Eq.~\eqref{eq:contraction_longueur}) déduite des transformées de Lorentz (Eq.~\eqref{eq:Lorentz_transformation}, en espace-temps plat), il est aisé de constater que le produit scalaire classique ne se conserve pas par changement de référentiel, ce qui est embêtant pour faire de la physique. Afin d'obtenir un produit scalaire invariant par changement de référentiel, on définit d'abord l'intervalle d'espace-temps infinitésimal $\dd s$ entre deux points d'espace-temps infiniment proches, observés par un observateur quelconque, qui lie les coordonnées d'espace et de temps en une quantité invariante (indépendante du référentiel) tel que
\begin{equation} \label{eq:ds2}
    \dd s^2 = -(c\dd t)^2 + \dd x^2 + \dd y^2 + \dd z^2.
\end{equation}
Pour se convaincre de l'invariance de $\dd s^2$, on utilise le changement de référentiel par transformée de Lorentz, Eq.~\eqref{eq:Lorentz_transformation}, comme suit :
\begin{equation}
    \begin{aligned}
        \dd s^2 &= -(c\dd t)^2 + \dd x^{2} + \dd y^{2} + \dd z^{2}\\
        &= -\left( \gamma (c\dd t^\prime + \beta \dd x^\prime) \right)^2 + \left( \gamma (\dd x^\prime + \beta c\dd t^\prime) \right)^2 + \dd y^{\prime 2} + \dd z^{\prime 2}\\
        &= \gamma^2 \left(-c^2\dd t^{\prime 2} - \beta^2 \dd x^{\prime 2} - 2 \beta c \dd t^\prime \dd x^\prime + \dd x^{\prime 2} + \beta^2 c^2 \dd t^{\prime 2} + 2 \beta c \dd t^\prime \dd x^\prime \right) + \dd y^{\prime 2} + \dd z^{\prime 2}\\
        &= \gamma^2 \left( 1 - \beta^2 \right) \left( -c^2\dd t^{\prime 2} + \dd x^{\prime 2} \right) + \dd y^{\prime 2} + \dd z^{\prime 2}\\
        &= -c^2\dd t^{\prime 2} + \dd x^{\prime 2} + \dd y^{\prime 2} + \dd z^{\prime 2}.
    \end{aligned}
\end{equation}
$\dd s^2$ est donc bien invariant par changement de référentiel.

Soient deux points d'espace-temps A et B de coordonnées $(t_A,x_A,y_A,z_A)$ et $(t_B,x_B,y_B,z_B)$ respectivement. On définit le vecteur $\mathbf{AB} = (c(t_B-t_A),x_B-x_A,y_B-y_A,z_B-z_A)$ et la distance au carré entre A et B via le produit scalaire de la manière suivante :
\begin{equation} \label{eq:produit_scalaire}
    \mathbf{AB} \cdot \mathbf{AB} = -c^2(t_B-t_A)^2 + (x_B-x_A)^2 + (y_B-y_A)^2 + (z_B-z_A)^2.
\end{equation}
De manière similaire à l'intervalle d'espace-temps, le produit scalaire, tel que défini dans l'Eq.~\eqref{eq:produit_scalaire}, entre deux 4-vecteurs $\mathbf{u}$ et $\mathbf{v}$ s'écrivant
\begin{equation} \label{eq:scalar_objectif}
    \mathbf{u} \cdot \mathbf{v} = -u^0 v^0 + u^1 v^1 + u^2 v^2 + u^3 v^3
\end{equation}
est invariant par changement de référentiel. On a donc défini un produit scalaire permettant de définir une longueur invariante par changement de référentiel, ce qui est nécessaire pour faire de la physique.

En exprimant les vecteurs $\mathbf{u}$ et $\mathbf{v}$ à partir de leurs composantes dans une base orthonormée $(\mathbf{e_\alpha})$, le produit scalaire s'écrit
\begin{equation}
    \begin{aligned}\label{eq:scalar_vector}
        \mathbf{u} \cdot \mathbf{v} &= \left( u^0 \mathbf{e_0} + u^1 \mathbf{e_1} + u^2 \mathbf{e_2} + u^3 \mathbf{e_3} \right) \cdot \left( v^0 \mathbf{e_0} + v^1 \mathbf{e_1} + v^2 \mathbf{e_2} + v^3 \mathbf{e_3} \right) \cdot \\
        &= u^\alpha v^\beta \, \mathbf{e_\alpha} \cdot \mathbf{e_\beta}.
    \end{aligned}
\end{equation}
On utilise ici la propriété de linéarité du produit scalaire. On note qu'ici, il y a une double sommation, car les deux indices $\alpha$ et $\beta$ sont présents en haut et en bas. Par identification avec l'Eq.~\eqref{eq:scalar_objectif}, on en déduit les propriétés suivantes pour les vecteurs de base ($\mathbf{e_\alpha}$) :
\begin{equation}
    \begin{aligned}
        \mathbf{e_0} \cdot \mathbf{e_0} &= -1, \\
        \mathbf{e_i} \cdot \mathbf{e_i} &= 1, \\
        \mathbf{e_\alpha} \cdot \mathbf{e_\beta} &= 0, \mathrm{ pour  }\, \alpha \neq \beta\\
    \end{aligned}
\end{equation}
L'ensemble des produits scalaires des vecteurs de base tel que défini ci-dessus forment la métrique de notre espace-temps sous la forme d'un tenseur.
\begin{definition}
    \textit{En évitant tout détail mathématique, on définit ici un tenseur comme un objet mathématique agissant sur un ensemble de vecteurs pour donner un nombre réel. Le tenseur métrique est un bon exemple, qui prend en entrée deux vecteurs, et ressort un nombre.}
\end{definition}
Les tenseurs sont caractérisés par leur dimension et sont indépendants du choix de base\footnote{Dit autrement, un tenseur est un objet "abstrait" à l’image d’un épouvantar dont on ne connaît pas la forme réelle. Ce n’est que lorsqu’on utilise un système de coordonnées (un humain se présente devant l’épouvantar) qu’il se matérialise en quelque chose de concret (tangible).}. Ils peuvent être représentés par des "tableaux" de leur dimension (0 pour un scalaire, 1 pour un vecteur, 2 pour une matrice, etc.) en choisissant une base spécifique. Lorsqu'on change de base, leur représentation change aussi. Par exemple, les coordonnées d'un point vont changer si on les exprime en coordonnées cartésiennes ou sphériques. Les tenseurs ont deux opérations importantes, à savoir un produit tensoriel, qui est la généralisation du produit vectoriel, et un produit de contraction qui est la généralisation du produit scalaire pour des vecteurs (tenseurs d'ordre 1). Les tenseurs sont donc la généralisation, indépendante du choix de base, de quantités usuelles comme les scalaires, vecteurs et matrices.

\begin{definition}
    \textit{Le tenseur métrique est un tenseur covariant d'ordre 2, permettant de définir le produit scalaire de deux vecteurs en chaque point de l'espace-temps, ainsi que les longueurs et les angles. Appliqué à un système de coordonnées (une base), il se représente par une matrice carrée symétrique de dimension quatre en Relativité Générale et Restreinte.}
\end{definition}

Une notion importante est la signature de la métrique. Il existe deux conventions $(-,+,+,+)$ donc -1 pour la coordonnée temporelle et +1 pour les coordonnées spatiales et $(+,-,-,-)$. La première étant généralement utilisée en Relativité Générale et la seconde en Relativité Restreinte. Pour toute la suite de cette thèse, on utilisera la plus répandue en RG, à savoir la première $(-,+,+,+)$.

Dans le cas précédent d'un espace-temps plat, le tenseur métrique dit de Minkowski, souvent noté $\eta_{\mu \nu}$, s'exprime de la manière suivante dans la base ($\mathbf{e_\mu}$) :
\begin{equation}
      \mathbf{g}(\mathbf{e_\mu}, \mathbf{e_\nu}) = g_{\mu \nu} =  \begin{pmatrix}
        -1 & 0 & 0 & 0 \\
        0 & 1 & 0 & 0 \\
        0 & 0 & 1 & 0 \\
        0 & 0 & 0 & 1
    \end{pmatrix} = \eta_{\mu \nu}.
\end{equation}
On retrouve la définition de l'intervalle d'espace-temps temps à partir de la métrique via
\begin{equation} \label{eq:metric_Minkowski}
    ds^2 = \eta_{\mu \nu} dx^\mu dx^\nu = -(c dt)^2 + dx^2 + dy^2 + dz^2.
\end{equation}

Comme le précise la définition, la fonction principale du tenseur métrique $\mathbf{g}$ est de définir le produit scalaire de deux 4-vecteurs $\mathbf{u}$ et $\mathbf{v}$. Ainsi, le produit scalaire peut se réécrire de la manière suivante
\begin{equation}
    \mathbf{u} \cdot \mathbf{v} = \mathbf{g}(\mathbf{u}, \mathbf{v}) = g_{\mu \nu} u^\mu v^\nu,
\end{equation}
et la norme au carré d'un 4-vecteur $\mathbf{v}$ est
\begin{equation}
    \vert \mathbf{v} \vert^2 = \mathbf{v} \cdot \mathbf{v} = g_{\mu \nu} v^\mu v^\nu.
\end{equation}

On peut aussi définir la métrique contravariante qui est la métrique conjuguée de la précédente appelée \textit{métrique inverse} et définie comme suit
\begin{equation} \label{eq:relation_metrique}
    g^{\mu \rho} g_{\rho \nu} = \delta^\mu_\nu,
\end{equation}
$\delta^\mu_\nu$ étant le symbole de Kronecker qui vaut un lorsque $\mu=\nu$ et zéro sinon. Ces deux tenseurs sont très utiles pour monter ou descendre des indices d'un 4-vecteur. En effet, grâce à la règle de sommation de la notation d'Einstein, pour un 4-vecteur quelconque $\mathbf{v}$, le lien entre ses composantes covariantes et contravariantes est fait par les tenseurs métriques tels que
\begin{equation}
    v_\alpha = g_{\alpha \beta} v^\beta
\end{equation}
et réciproquement
\begin{equation}
    v^\alpha = g^{\alpha \beta} v_\beta.
\end{equation}

\subsubsection{Classification des 4-vecteurs et cônes de lumière}
Contrairement au cas 3D où le produit scalaire d'un vecteur avec lui-même, est défini positif ou nul, en Relativité Générale, la norme d'un vecteur $\mathbf{v}$ peut avoir n'importe quel signe à cause de la signature de la métrique $(-,+,+,+)$. Les vecteurs sont regroupés en trois catégories selon la valeur de leur norme :
\begin{itemize}
    \item du genre temps si et seulement si $\mathbf{v} \cdot \mathbf{v} < 0$;
    \item du genre espace si et seulement si $\mathbf{v} \cdot \mathbf{v} > 0$;
    \item du genre lumière si et seulement si $\mathbf{v} \cdot \mathbf{v} = 0$ aussi appelé \textit{vecteur isotrope}.
\end{itemize}
Dans le cas d'un vecteur position entre deux points de l'espace-temps (qui correspond à un intervalle d'espace-temps $\Delta s$), si ce vecteur est de genre temps, cela signifie que la distance séparant les deux points est inférieure à la distance parcourue par la lumière sur l'intervalle de temps. Ainsi, il peut y avoir un lien de causalité entre ces deux points d'espace-temps. À l'inverse, s'il est de genre espace, alors la distance entre les deux points est supérieure à la distance parcourue par la lumière sur l'intervalle de temps. Il ne peut donc pas y avoir de lien de causalité entre ces deux points. Le troisième genre correspond donc à la distance (spatiale) parcourue par un photon (pour être plus exact une particule de masse nulle) en un certain temps dans le référentiel considéré (autre que le référentiel propre du photon\footnote{La notion de temps n'est pas définie pour un photon puisque l'on a $\Delta s = 0$}), d'où son appellation.

Par exemple, la 4-vitesse qui s'exprime
\begin{equation}
    \mathbf{u} = u^\alpha \mathbf{e_\alpha} = u^0 \mathbf{e_0} + u^1 \mathbf{e_1} + u^2 \mathbf{e_2} + u^3 \mathbf{e_3}
\end{equation}
est définie de telle sorte que $\mathbf{u} \cdot \mathbf{u} = -1$ via le produit scalaire, ce qui en fait un vecteur de genre temps. Contrairement à la vitesse « ordinaire », la 4-vitesse d’une particule matérielle est définie sans référence à un observateur ou à un référentiel. Il s’agit d’une quantité absolue, qui ne dépend que de la particule considérée. Ainsi, tout n’est pas relatif dans la théorie de la relativité...

\begin{figure}
    \centering
    \resizebox{0.5\hsize}{!}{\includegraphics{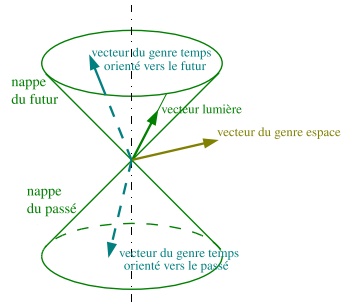}}
    \caption{Cône isotrope de la métrique $\mathbf{g}$ en un point de l'espace-temps. Crédit : \href{http://luth.obspm.fr/~luthier/gourgoulhon/fr/master/relat.html}{Cours RG Éric Gourgoulhon 2013-2014}.}
    \label{fig:cone_de_lumière}
\end{figure}

On peut ainsi créer une représentation graphique de ces genres de vecteurs pour un point de l'espace-temps via les \textit{cônes de lumière} (voir figure~\ref{fig:cone_de_lumière}). Dans cette représentation, on remplace une dimension d'espace par la dimension temporelle (ici la dimension verticale). On appelle cône de lumière la limite de causalité entre le point d'origine de l'espace-temps considéré (sommet du cône) et le reste des points de l'espace-temps. Cela correspond à l'ensemble des points pour lesquels l'intervalle d'espace-temps est nul, donc à des vecteurs de genre lumière. Dans un cas à deux dimensions, une spatiale et une temporelle, les vecteurs de genre lumière ont pour origine le point d'espace-temps considéré et sont colinéaires à une des deux droites dont la pente est $\pm c$. Le cône de lumière est la représentation à trois dimensions de ces droites (par symétrie de révolution). Les vecteurs de genre temps sont à l'intérieur du cône et ceux de genre espace sont à l'extérieur. De plus, comme on a la dimension temporelle qui est représentée par l'axe vertical, on peut définir le passé et le futur du point d'espace-temps considéré et donc l'orientation des vecteurs (vers le passé ou vers le futur). On note que cette construction est indépendante du référentiel, car construite uniquement à partir du tenseur métrique. C'est une conséquence directe du postulat de l'universalité de la vitesse de la lumière.

\subsection{L'équation d'Einstein}
L'équation fondamentale de la Relativité Générale appelée l'équation d'Einstein est l'équation tensorielle reliant la géométrie de l'espace-temps au contenu en matière et énergie de ce dernier. Elle s'écrit de la manière suivante avec la notation d'Einstein :
\begin{equation} \label{eq:Einstein}
    \boxed{
    R_{\mu \nu} - \frac{1}{2} g_{\mu \nu}R = \frac{8 \pi \mathcal{G}}{c^4} T_{\mu \nu}}
\end{equation}
avec $R_{\mu \nu}$ le tenseur de Ricci (symétrique) qui exprime la déformation de l'espace-temps, $g_{\mu \nu}$ le tenseur métrique (voir~\ref{sec:métrique}), $R$ la courbure scalaire qui est la trace du tenseur de Ricci par rapport à $g_{\mu \nu}$ et $T_{\mu \nu}$ le tenseur énergie-impulsion qui représente la répartition de masse et d'énergie dans l'espace-temps.

Il s'agit ici d'une équation non-linéaire faisant intervenir $g_{\mu \nu}$ et ses dérivées, qui peut être décomposée en 6 équations indépendantes. De plus, le tenseur de Ricci $R_{\mu \nu}$, et donc la courbure scalaire $R$, dépendent de la métrique $g_{\mu \nu}$. Un terme supplémentaire peut être ajouté dans le membre de gauche de cette équation dans le cas de la cosmologie, à savoir $\Lambda g_{\mu \nu}$ avec $\Lambda$ la constante cosmologique qui correspond à l'énergie du vide ou énergie noire.

On peut encore compacter l'équation~\eqref{eq:Einstein} en définissant le tenseur d'Einstein
\begin{equation}
    G_{\mu \nu} =  R_{\mu \nu} - \frac{1}{2} g_{\mu \nu}R,
\end{equation}
en prenant $\Lambda=0$, en utilisant le système d'unité géométrique où $\mathcal{G}=c=1$, ce qui donne
\begin{equation}
    G_{\mu \nu} = 8 \pi T_{\mu \nu}.
\end{equation}

Cette équation montre bien le lien entre le contenu en matière/énergie à droite et la géométrie de l'espace-temps à gauche. Cette dernière est donc exprimée par le tenseur métrique qui est donc le tenseur le plus important de la Relativité Générale.

\subsection{Équation des géodésiques}
\begin{figure}
    \centering
    \resizebox{0.5\hsize}{!}{\includegraphics{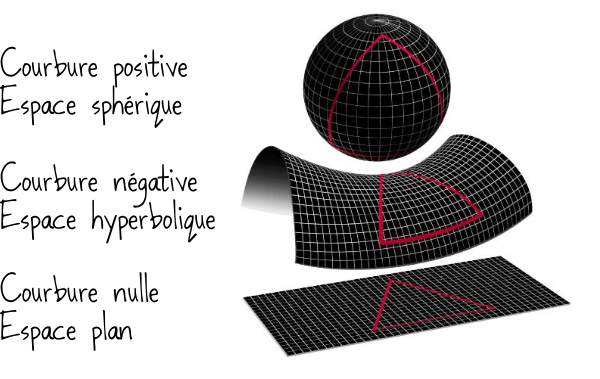}}
    \caption{Schéma de trois espaces-temps ayant resp. une courbure positive, négative et nulle, représentées dans un espace-temps plat (classique). Dans un espace-temps plat (courbure nulle), la somme des angles d'un triangle est égale à 180°. Dans un espace-temps sphérique (courbure positive) cette somme est supérieure à 180° et dans un espace-temps hyperbolique (courbure négative) elle est inférieure à 180°. Crédit : \href{https://commons.wikimedia.org/wiki/File:End_of_universe.jpg}{Wikimedia commons}.}
    \label{fig:courbure}
\end{figure}

\begin{definition}
    \textit{En géométrie, une géodésique en espace-temps courbe généralise la ligne droite en espace-temps plat. Elle correspond à une distance minimale (il peut en exister plusieurs) entre deux points de l'espace définis par une métrique.}
\end{definition}
Dans un espace plat (de Minkowski) tridimensionnel, la géodésique entre deux points est une droite reliant ces deux points. Dans un espace ayant une courbure non nulle (positive ou négative, voir Fig.~\ref{fig:courbure}), la représentation de la géodésique d'un point de vue "extérieur" (en espace plat) sera une courbe. C'est là tout le principe de la Relativité Générale, le mouvement de la Terre dans l'espace-temps du système solaire, qui peut être approximé par la métrique de Schwarzschild (voir plus de détails dans la partie~\ref{sec:metric_Schwarzschild}), est circulaire (on néglige l'excentricité dans cet exemple), ce qui est dû à la courbure de l'espace-temps par la présence d'une masse (le Soleil).

Prenons un système de coordonnées $x^\mu$ et un tenseur métrique $g_{\mu \nu}$, on a donc
\begin{equation}
    ds^2 = g_{\mu \nu} dx^\mu dx^\nu.
\end{equation}
On considère que la courbe est paramétrée par rapport à une variable $\lambda$ telle que $x^\mu = X^\mu (\lambda)$, on a donc $dx^\mu = \frac{dX^\mu}{d\lambda} d\lambda = \Dot{X}^\mu d\lambda$. On obtient donc 
\begin{equation}
    \Dot{s} = \frac{ds}{d\lambda} = \sqrt{-g_{\mu \nu} \Dot{X}^\mu \Dot{X}^\nu}
\end{equation}
où le point marque la dérivée totale par rapport au paramètre $\lambda$. On obtient la longueur de la trajectoire en intégrant $ds$ par rapport à $\lambda$
\begin{equation}
    s = \int d\lambda \sqrt{-g_{\mu \nu} \Dot{X}^\mu \Dot{X}^\nu}.
\end{equation}
Comme dit dans la définition, la géodésique correspond à la longueur minimale entre deux points de l'espace-temps, il faut donc minimiser $s$. On note qu'il est équivalent de maximiser le temps propre comme c'est le cas dans la démonstration de \href{http://luth.obspm.fr/~luthier/gourgoulhon/fr/master/relat.html}{Cours RG Éric Gourgoulhon 2013-2014}. Pour cela, on utilise les équations d'Euler-Lagrange qui stipulent que $s$ est minimal si et seulement si
\begin{equation}\label{eq:Euler-Lagrange}
    \frac{\partial \Dot{s}}{\partial X^\alpha} - \frac{d}{d \lambda} \left( \frac{\partial \Dot{s}}{\partial \Dot{X}^\alpha} \right) = 0.
\end{equation}


Pour simplifier la notation, on note les dérivées partielles de la métrique par rapport à une composante de la manière suivante 
\begin{equation} \label{eq:derive_metric}
    \frac{\partial g_{\alpha \beta}}{\partial x^\rho} = \partial_\rho g_{\alpha \beta}.
\end{equation}


Après plusieurs étapes de calcul et de simplification détaillées dans le cours de \href{http://luth.obspm.fr/~luthier/gourgoulhon/fr/master/relat.html}{Relativité Générale d'Éric Gourgoulhon}, on obtient l'équation des géodésiques générales pour un paramètre quelconque $\lambda$
\begin{equation}\label{eq:geodesic_general}
\boxed{
    \ddot{X}^\alpha + \Gamma^\alpha_{\ \mu \nu} \Dot{X}^\mu \Dot{X}^\nu = \kappa(\lambda) \Dot{X}^\alpha,
    }
\end{equation}
avec
\begin{equation} \label{eq:Christoffel}
    \boxed{
    \Gamma^\beta_{\ \mu \nu} = \frac{1}{2} g^{\beta \sigma} (\partial_\mu g_{\sigma \nu} + \partial_\nu g_{\mu \sigma} - \partial_\sigma g_{\nu \mu})
    }
\end{equation}
les \textit{symboles de Christoffel} qui ne dépendent que de la métrique et du système de coordonnées utilisées. Bien que le calcul de ces quantités semble compliqué, la difficulté réside dans la gestion des indices et des sommations (ce qui peut néanmoins facilement mener à des erreurs) car la métrique est symétrique et contient beaucoup de termes nuls.

En choisissant judicieusement le paramètre $\lambda$, selon si l'on considère des vecteurs de genre temps en utilisant le temps propre $\tau$ (resp. de genre lumière en utilisant un paramètre dit \textit{affine}), et en définissant des conditions initiales, on peut donc résoudre ce système d'équations différentielles du second ordre et ainsi définir la trajectoire d'une particule massive (resp. de masse nulle) dans notre espace-temps. L'Eq.~\eqref{eq:geodesic_general} est l'équivalent relativiste du principe fondamental de la dynamique classique (Newtonien) avec uniquement la force de gravité (dans le vide).

\subsection{Objets compacts et flot d'accrétion}
Bien que la Relativité Générale soit valable dans tous les systèmes faisant appel à la gravitation, y compris pour le bon fonctionnement des satellites GPS autour de la Terre, elle est particulièrement nécessaire pour décrire l'espace-temps dans un champ gravitationnel fort. C'est le cas pour des objets compacts comme les étoiles à neutrons, les naines blanches et les trous noirs. On définit la compacité d'un objet comme 
\begin{equation}
    \Xi = \frac{\mathcal{G}M}{Rc^2}
\end{equation}
où $M$ et $R$ sont resp. la masse et le rayon de l'objet. On peut ainsi comparer la compacité d'une planète comme la Terre au Soleil, à une étoile à neutrons et un trou noir d'une masse solaire (Table~\ref{tab:compacité}), le rayon que l'on utilise pour le trou noir étant son horizon des évènements (voir~\ref{sec:metric_Schwarzschild}).

\begin{table}[ht]
    \centering
    \begin{tabular}{|c|c|c|c|}
    \hline
      Objet & Masse & Rayon & $\Xi$ \\
      \hline
      Terre & $6 \times 10^{24}$ kg & 6.378 km & $7 \times 10^{-10}$ \\
      Soleil & $2 \times 10^{30}$ kg = $1 M_\odot$ & 700.000 km & $5 \times 10^{-6}$ \\
      Naine blanche & $1 M_\odot$ & 6.000 km & $5 \times 10^{-4}$ \\
      Étoile à neutrons & $1,4 M_\odot$ & 30 km & $0,15$ \\
      Trou noir de Schwarzschild & $2 M_\odot$ & $\sim 6$ km & $0,5$ \\
      \hline
    \end{tabular}
    \caption{Comparaison de la compacité de plusieurs objets astrophysiques.}
    \label{tab:compacité}
\end{table}

Pour cette thèse, nous nous intéressons aux trous noirs. Il en existe trois types, classés selon leur masse. Les trous noirs dits \textit{stellaires} ont une masse comparable à celle des étoiles avec un minimum de $\sim 2-3\, M_\odot$ et sont le résultat de l'effondrement gravitationnel d'étoiles massives (ayant une masse initiale d'au moins $8\, M_\odot$). La seconde classe de trou noir regroupe les trous noirs dits \textit{supermassifs}. Comme le nom l'indique, la masse de ces trous noirs est très grande comparée à celle des étoiles et peut atteindre plusieurs milliards de masses solaires. Le processus de formation de ces trous noirs est encore débattu. On suppose que chaque galaxie, hormis les galaxies irrégulières, a un trou noir supermassif en son cœur. Ainsi, lors de la fusion de deux galaxies, les deux trous noirs vont aussi fusionner pour en former un plus massif. La dernière classe est le chaînon manquant entre ces deux extrêmes. Les trous noirs de ce type sont dits \textit{intermédiaires} (quelques milliers de masses solaires) et sont le résultat de la fusion de trous noirs stellaires. La première détection de la coalescence de deux trou noirs stellaires a eu lieu le 14 Septembre 2015 par le détecteur d'ondes gravitationnelles \href{https://www.ligo.caltech.edu/}{LIGO} (\textit{Laser Interferometer Gravitational-wave Observatory}) qui a observé le passage d'ondes gravitationnelles (OG) dues à la coalescence de deux trous noirs de $M_1=36^{+5}_{-4}\, M_\odot$ et $M_2=29^{+4}_{-4}\, M_\odot$ donnant "naissance" à un trou noir de $M_f=62^{+4}_{-4} \, M_\odot$, les $3^{+0,5}_{-0,5}\, M_\odot$ restant étant rayonnés sous forme d'ondes gravitationnelles.

La compacité est à différencier de la masse volumique $\rho = \frac{3M}{4 \pi R^3}$. Bien que tous les trous noirs de Schwarzschild (sans rotation) ont une compacité de $0,5$ par définition, la masse volumique ne sera pas la même pour un trou noir stellaire de $1\, M_\odot$ et un trou noir supermassif de $10^6\, M_\odot$. En effet, comme on le verra dans la section~\ref{sec:metric_Schwarzschild}, le rayon de l'horizon des évènements d'un trou noir est proportionnel à sa masse, or la masse volumique est $\propto M/R^3$ soit $M^{-2}$. Ainsi, les trous noirs supermassifs ont une densité relativement faible, pouvent être comparable à celle de l'eau, comme Gargantua qui a une masse de 100 millions de masses solaires et donc une masse volumique de $\sim 1.800$~$kg/m^3$ (l'eau ayant une masse volumique de $991$~$kg/m^3$). Gargantua est ainsi qualifié de "singularité douce".

\begin{figure}
    \centering
    \resizebox{\hsize}{!}{\includegraphics{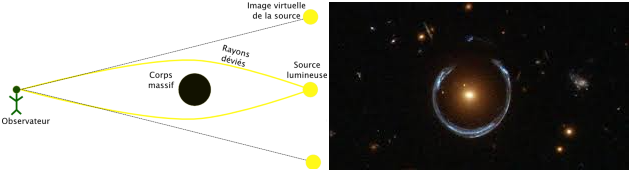}}
    \caption{(\textbf{Gauche}) Schéma illustrant la déviation de rayons lumineux par un objet massif dû à la courbure de l'espace-temps par ce dernier. Crédit : \href{https://scienceetonnante.com/2010/12/17/les-lentilles-gravitatio-noel/}{Science étonnante}. (\textbf{Droite}) Photo d'un anneau d'Einstein presque complet prise par le télescope spatial Hubble. Crédit : ESA/Hubble \& NASA}
    \label{fig:deviation_lumière}
\end{figure}

De par leur côté obscur, les trous noirs sont difficiles à détecter puisqu'ils n'émettent pas de lumière (qui n'arriverait de toute façon pas à s'échapper de l'horizon). Il existe quatre types de méthode détection : 
\begin{itemize}
    \item[$\bullet$] \textbf{Méthode 1 :} par le passage \textbf{d'ondes gravitationnelles}. Comme dit plus haut, l'observation d'ondes gravitationnelles est une méthode de détection de la coalescence de trous noirs. Cependant, ce genre d'évènement n'est pas exclusif aux trous noirs, il y a aussi des coalescences d'étoiles à neutrons dont le résultat peut être une plus grosse étoile à neutrons ou un trou noir. Avec les instruments actuels comme LIGO, \href{https://www.virgo-gw.eu/}{Virgo} (l'interféromètre de Michelson Européen) ou \href{https://gwcenter.icrr.u-tokyo.ac.jp/en/}{KAGRA} (l'observatoire d'OG japonais), seule la fusion de trous noirs stellaires et intermédiaires est détectable. Néanmoins, avec l'arrivée de \href{https://www.elisascience.org/}{LISA} (Laser Interferometer Space Antenna), la détection de la coalescence de trous noirs supermassifs sera possible. Une quasi-détection a été rapportée par la technique de chronométrage des pulsars (PTA pour \textit{Pulsar Timing Array} en anglais)~\cite{PTA2023}.

    \item[$\bullet$] \textbf{Méthode 2 :} par effet de \textbf{lentille gravitationnelle}. En effet, la trajectoire des objets matériels va être affectée par la courbure de l'espace-temps, mais c'est aussi le cas de la lumière ! En 1919, lors de l'éclipse totale du 29 Mai, Eddington mesure le décalage de la position de plusieurs étoiles proches du disque solaire par rapport à leur position de référence. Ce décalage est dû à la courbure de l'espace-temps par le Soleil (voir panneau de gauche de la Fig.~\ref{fig:deviation_lumière}), c'est l'effet de lentille gravitationnelle. Cet effet est d'autant plus marqué que la masse est importante (et compacte), faisant des trous noirs des candidats idéaux pour observer cet effet qui, lorsque l'alignement entre la source, l'objet massif et l'observateur est parfait, crée ainsi un \textit{anneau d'Einstein} (panneau de droite de la Fig.~\ref{fig:deviation_lumière}).

    \item[$\bullet$] \textbf{Méthode 3 :} par la lumière émise par la \textbf{matière environnant le trou noir}. En effet, que ce soit pour des trous noirs stellaires ou supermassifs, les trous noirs peuvent être "entourés" de matière, ce qui est dû à une étoile compagnon pour les trous noirs stellaires, ou au gaz interstellaire pour les supermassifs. On parle alors de flot d'accrétion qui, par conservation du moment cinétique, va former un disque plus ou moins épais. Cette méthode est très dépendante de la physique du flot d'accrétion, qui est le centre d'intérêt de cette thèse.
    
    \item[$\bullet$] \textbf{Méthode 4 :} par influence \textbf{gravitationnelle}. Lorsqu'un trou noir est accompagné d'une ou plusieurs étoiles, comme c'est le cas des étoiles~S autour de Sgr~A*, l'influence gravitationnelle de ce dernier affecte l'orbite (précession) et la lumière reçue (rougissement) \cite{Gravity2020a} de ces étoiles.
    
\end{itemize}

\subsubsection{Métrique de Schwarzschild}\label{sec:metric_Schwarzschild}
La métrique de Schwarzschild est la première solution exacte contenant une masse trouvée pour l'équation d'Einstein (Eq.~\eqref{eq:Einstein}) par Karl Schwarzschild, 6 mois après la publication de la Relativité Générale par Einstein en pleine guerre mondiale.

Cette métrique décrit l'espace-temps en considérant un objet de masse $M$ ponctuel, statique, sans rotation et entouré de vide. C'est l'hypothèse de base de la mécanique du point enseignée jusqu'en Master, par exemple pour calculer l'orbite des planètes autour du Soleil (on réutilisera cet exemple plus tard). En utilisant un système de coordonnées sphériques $(x^\alpha) = (ct, r, \theta, \varphi)$ dont l'origine est le point où toute la masse est contenue, la métrique scalaire de Schwarzschild est
\begin{equation} \label{eq:metric_Schwarzschild}
    ds^2 = -\left( 1-\frac{2 \mathcal{G}M}{c^2 r} \right) c^2 dt^2 + \frac{dr^2}{1-\frac{2 \mathcal{G}M}{c^2 r}} + r^2(d\theta^2 + \sin^2 \theta d\varphi^2)
\end{equation}
ou de manière matricielle
\begin{equation}
    g_{\mu \nu} = \begin{pmatrix}
        -\left( 1-\frac{2 \mathcal{G}M}{c^2 r} \right) & 0 & 0 & 0 \\
        0 & \frac{1}{1-\frac{2 \mathcal{G}M}{c^2 r}} & 0 & 0 \\
        0 & 0 & r^2 & 0 \\
        0 & 0 & 0 & r^2 \sin^2 \theta
    \end{pmatrix}
\end{equation}
où $r$ est la coordonnée radiale, $\theta$ la colatitude (angle entre l'axe vertical et le point) et $\varphi$ la longitude. On constate que cette matrice est diagonale, sans terme croisé, car on considère une masse statique sans rotation.

L'observation des termes de cette métrique montre deux singularités pour deux valeurs particulières du rayon (où un terme devient infini), la première est pour $r=0$ qui correspond à la singularité centrale et la seconde, qui est due au système de coordonnées (mais n'est pas physique), est appelée rayon de Schwarzschild 
\begin{equation}
    r_s = \frac{2 \mathcal{G}M}{c^2} = 2 r_g
\end{equation}
avec
\begin{equation}
    r_g = \frac{\mathcal{G}M}{c^2}.
\end{equation}
le rayon gravitationnelle.
Le rayon de Schwarzschild définit dans le cas des trous noirs, dits de Schwarzschild, l'horizon des évènements. Cette métrique est valide en dehors de cette limite ($r>r_s$) donc en dehors de l'horizon des évènements.

Ainsi, pour tout objet massif sans rotation dont le rayon est supérieur ou égal à $r_s$, on peut utiliser cette métrique pour décrire l'espace-temps autour de cet objet. Reprenons l'exemple du Soleil dont le rayon est de $R_\odot \sim 700.000$~km très supérieur au rayon de Schwarzschild du Soleil $r_{s,\odot}\sim 3$~km. Bien que le Soleil soit en rotation, on va la négliger pour cet exemple. Par conséquent, pour tout $r>R_\odot$, la métrique de Schwarzschild s'applique. On peut ainsi déterminer la trajectoire d'une planète autour du Soleil en utilisant cette métrique.


Du fait de la symétrie sphérique, on peut supposer que $\theta_0 = \frac{\pi}{2}$, $\varphi_0=0$ et que la composante en $\theta$ du vecteur vitesse $u^\theta=0$. Cela correspond à une trajectoire dans le plan équatorial $\theta (\tau) = \theta_0$. Même si ce n'est pas le cas, on peut faire un changement de référentiel en utilisant des matrices de rotation pour se ramener à ce cas. Bien que l'on puisse déterminer la trajectoire à partir de l'équation des géodésiques~\eqref{eq:geodesic_general} de genre temps (avec $\lambda=\tau$), il est plus judicieux d'utiliser des lois de conservations.

En effet, $\mathbf{\xi}  \cdot \mathbf{p}$ est conservé tout le long de la géodésique, avec $\mathbf{p}=mc\mathbf{u}$ la 4-impulsion et $\mathbf{\xi}$ un \textit{vecteur de Killing} dont la propriété est que le tenseur métrique ne varie pas le long des lignes de champ de ce vecteur. Par exemple, les 4-vecteurs $\mathbf{\xi}_{(0)} = c^{-1} \mathbf{\partial_t}$ et $\mathbf{\xi}_{(z)} = \mathbf{\partial_\varphi}$ sont des vecteurs de Killing (car la métrique est indépendante du temps et de l'angle $\varphi$) que l'on va utiliser pour déterminer notre trajectoire à travers la conservation de
\begin{equation}
    \epsilon = - \frac{c}{m} \mathbf{\xi}_{(0)} \cdot \mathbf{p} = -c^2 \mathbf{\xi}_{(0)} \cdot \mathbf{u}
\end{equation}
qui correspond, lorsque $r \gg r_s$, à l'énergie par unité de masse et
\begin{equation}
    l = \frac{1}{m} \mathbf{\xi}_{(z)} \cdot \mathbf{p} = c \mathbf{\xi}_{(z)} \cdot \mathbf{u}
\end{equation}
qui correspond lorsque $\frac{dt}{d\tau} \approx 1$ au moment cinétique par rapport à l'axe z par unité de masse.
En utilisant la 4-vitesse avec le système de coordonnées et la métrique définie précédemment, ces quantités deviennent
\begin{equation}
    \epsilon = -c^2 g_{\mu \nu} (\partial_0)^\mu u^\nu = c^2 \left( 1 - \frac{r_s}{r} \right) \frac{dt}{d\tau},
\end{equation}
\begin{equation}
    l = c g_{\mu \nu} (\partial_0)^\mu u^\nu = r^2 \sin^2 \theta \frac{d\varphi}{d\tau}.
\end{equation}

Ces quantités permettent de calculer les composantes $u^0$ et $u^3$ du 4-vecteur vitesse en fonction du temps et des conditions initiales. On a donc pour le moment 
\begin{equation}
    \begin{aligned}
        u^0 &= \left( 1 - \frac{r_s}{r} \right)^{-1} \frac{\epsilon}{c^2}\\
        u^\theta &= 0\\
        u^\varphi &= \frac{l}{cr^2}.
    \end{aligned}
\end{equation}
en utilisant $sin\ \theta = 1$ puisque $\theta = \frac{\pi}{2}$. La dernière composante de $\mathbf{u}$ peut-être déterminée grâce à sa normalisation $\mathbf{u} \cdot \mathbf{u}=g_{\mu \mu}(u^\mu)^2 = -1$ (puisque $\mathbf{g}$ est diagonale), ce qui donne 
\begin{equation}
    - \left( 1 - \frac{r_s}{r} \right)^{-1} \frac{\epsilon^2}{c^4} + \left( 1 - \frac{r_s}{r} \right)^{-1} \left( \frac{1}{c} \frac{dr}{d\tau} \right)^2 + \frac{l^2}{c^2 r^2} = -1
\end{equation}
On en déduit, en remplaçant $r_s$ par $2 \mathcal{G} M /c ^2$ et après quelques étapes intermédiaires
\begin{equation}\label{eq:conservation_energie_mecanique}
    \frac{1}{2}\left(\frac{dr}{d\tau} \right)^2 + V_{\mathrm{eff}}(r) = \frac{\epsilon^2 - c^4}{2c^2} = cste
\end{equation}
avec
\begin{equation}\label{eq:potentiel_effectif}
\boxed{
    V_{\mathrm{eff}}(r) = \frac{c^2}{2} \left( -\frac{r_s}{r} + \frac{l^2}{c^2 r^2} - \frac{l^2 r_s}{c^2 r^3} \right)}.
\end{equation}
L'Eq.~\eqref{eq:conservation_energie_mecanique} est l'analogie de la conservation de l'énergie mécanique dans un potentiel qui ici est un potentiel effectif, différent du potentiel Newtonien, mais similaire.
On définit le paramètre sans dimension $\bar{l} = \frac{l}{c r_s}$. Le potentiel effectif $V_{\mathrm{eff}}(r)$ s'écrit avec ce paramètre
\begin{equation}
    V_{\mathrm{eff}}(r) = \frac{c^2}{2} \left( -\frac{r_s}{r} + \bar{l^2} \frac{r_s}{r} - \bar{l^2}\frac{r_s^3}{r^3} \right).
\end{equation}

Tout comme en mécanique classique, on cherche à déterminer les extrema de $V_{\mathrm{eff}}$ en fonction de $r$ qui sont définis comme
\begin{equation}
    \frac{dV_{\mathrm{eff}}}{dr} = \frac{c^2 r_s}{2r^2} \left( 1- 2 \bar{l^2} \frac{r_s}{r} +3 \bar{l^2} \frac{r_s^2}{r^2} \right) = 0,
\end{equation}
soit
\begin{equation}\label{eq:potentiel_polynome}
    \left( \frac{r}{r_s} \right)^2 - 2 \bar{l^2} \left( \frac{r}{r_s} \right) + 3 \bar{l^2} = 0.
\end{equation}

On reconnaît ici un polynôme du second degré avec un déterminant $\Delta=4 \bar{l^4} - 4\cdot3\bar{l^2}$, l'Eq.~\eqref{eq:potentiel_polynome} admet des racines si et seulement si $\bar{l^2} > 3$, on définit donc $\bar{l}_{crit} = \sqrt{3}$ ce qui correspond à 
\begin{equation}
    l_{crit} = 2\sqrt{3} \frac{\mathcal{G}M}{c}.
\end{equation}

Ainsi pour $| l | > l_{crit}$, $V_{\mathrm{eff}}$ admet un maximum et un minimum en
\begin{equation}
    \frac{r_{max}}{r_s} = \bar{l^2} \left( 1 - \sqrt{1-\frac{3}{\bar{l^2}}} \right) \quad \mathrm{et} \quad \frac{r_{min}}{r_s} = \bar{l^2} \left( 1 + \sqrt{1-\frac{3}{\bar{l^2}}} \right),
\end{equation}
$r_{min}$ est donc un point d'équilibre correspondant à une orbite circulaire et ne dépend que de la valeur de $l$ qui, on le rappelle, est une constante dépendant de $u^\phi$. Ainsi, on peut définir une vitesse orbitale $\Omega$ pour un rayon $r$ donné correspondant à une orbite circulaire. Après plusieurs étapes de calcul, on obtient
\begin{equation} \label{eq:vitesse_keplerienne}
    \boxed{
    \Omega = \sqrt{\frac{\mathcal{G}M}{r^3}}}.
\end{equation}
Cela correspond exactement à la formule Newtonienne pour une vitesse dite Képlérienne.

Pour $l=l_{crit}$, il n'y a qu'un seul extremum en $r = \bar{l^2} r_s$ qui correspond donc au rayon minimal pour lequel une orbite circulaire est possible, appelé \textit{rayon de la dernière orbite circulaire stable}, ISCO en anglais pour \textit{Innermost Stable Circular Orbit}. Ce rayon particulier vaut
\begin{equation}\label{eq:ISCO}
    r_{ISCO}= 3\, r_s
\end{equation}
dans la métrique de Schwarzschild. On peut aussi définir un rayon particulier pour les particules sans masse, par exemple des photons, de manière similaire. On obtient dans ce cas le rayon de la sphère de photon qui correspond à la dernière orbite circulaire, mais instable, des photons $r_{ph}$
\begin{equation}
    r_{ph} = 1.5\, r_s.
\end{equation}

\subsubsection{Métrique de Kerr}
La métrique de Schwarzschild décrit un espace-temps à symétrie sphérique et statique défini par une masse ponctuelle sans rotation. Cependant, quasiment tous les objets astrophysiques ont une rotation ou plus généralement un moment angulaire non nul. Or, le moment angulaire d'un système fermé se conserve. Ainsi, lors de l'effondrement d'une étoile massive en trou noir, ce dernier a une rotation non nulle\footnote{On note que lors d'une supernova avec création d'un trou noir, une partie de la matière est éjectée. Le système n'est donc pas entièrement fermé.}. De plus, la matière accrétée par un trou noir a aussi un moment cinétique a priori non nul ce qui peut augmenter celui du trou noir lorsque la matière traverse l'horizon. On définit le paramètre de spin ayant la dimension d'une masse $a$ par
\begin{equation}\label{eq:spin}
    |a|=\frac{c|J|}{\mathcal{G}M}
\end{equation}
avec $J$ le moment cinétique. On définit aussi souvent le spin sans dimension $a_\star=a/M$.

\begin{figure}
    \centering
    \resizebox{0.5\hsize}{!}{\includegraphics{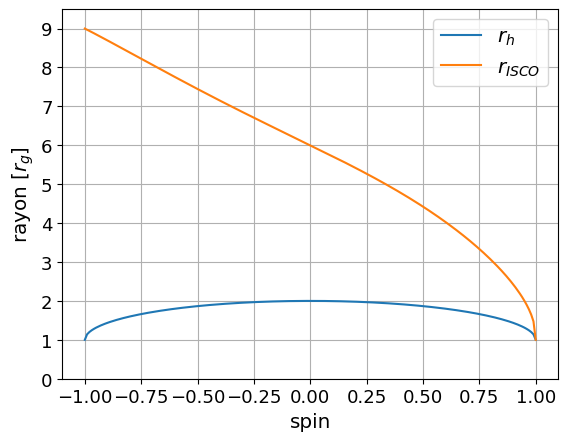}}
    \caption{Évolution du rayon de l'horizon des évènements (en bleu) et du rayon de l'ISCO (en orange) en fonction du spin. On constate que pour un trou noir ayant une rotation maximale, le rayon de l'ISCO est confondu avec l'horizon des évènements.}
    \label{fig:rh_spin}
\end{figure}

Les trous noirs ayant un spin non nul (avec rotation) sont appelés \textit{trous noirs de Kerr}. Du fait de la rotation, le rayon de l'horizon des évènements d'un trou noir de Kerr $r_h$ est plus petit que celui de Schwarzschild et dépend de la valeur du spin $|a_\star | \leq 1$, il s'écrit
\begin{equation}
    r_h = \frac{r_s}{2} \left( 1 + \sqrt{1-a_\star^2} \right).
\end{equation}
L'horizon des évènements $r_h$ d'un trou noir de Kerr est donc compris entre $r_s/2$ et $r_s$, ces deux limites correspondant à un trou noir avec une rotation maximale, on parle alors de singularité nue, et d'un trou noir sans rotation, donc de Schwarzschild (voir Fig.~\ref{fig:rh_spin}). Une autre différence avec les trous noirs de Schwarzschild est que, dans le cas de Kerr, la singularité a la géométrie d'un anneau.

\begin{figure}
    \centering
    \resizebox{0.5\hsize}{!}{\includegraphics{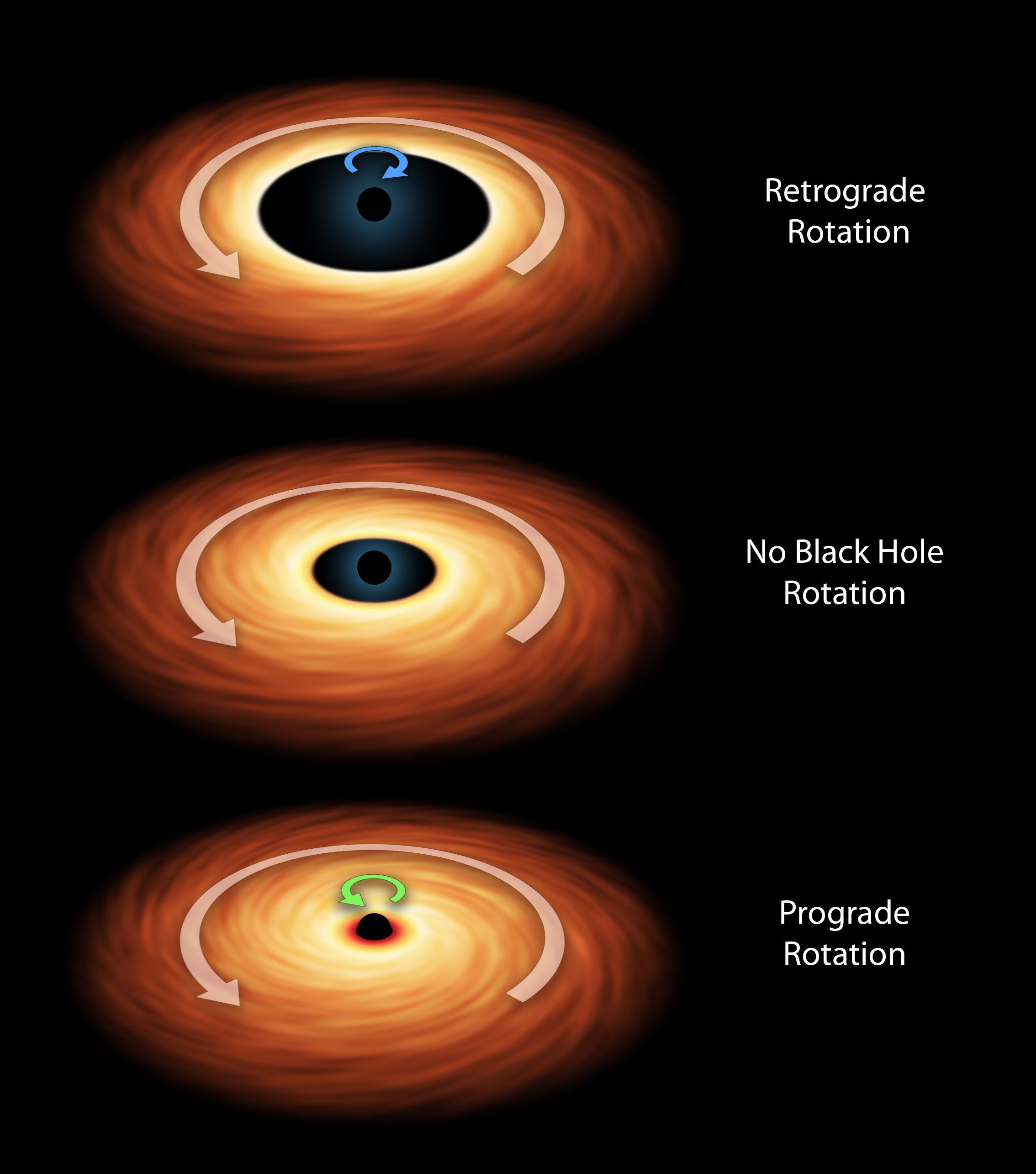}}
    \caption{Illustration de la dépendance de l'ISCO qui définit le bord interne du disque d'accrétion par rapport au spin du trou noir (rétrograde, nul ou prograde comparé à celui du disque). Crédits : \href{https://www.nasa.gov/mission_pages/nustar/multimedia/pia16696.html}{NASA/JPL-Caltech}}
    \label{fig:disque_spin}
\end{figure}

Le rayon de l'horizon des évènements n'est pas le seul rayon particulier à être affecté par la rotation. En effet, le rayon de l'ISCO va aussi dépendre du paramètre de spin. Cependant, cette valeur dépend aussi du type d'orbite, selon que l'on considère une orbite dans le même sens que la rotation du trou noir (\textit{prograde}, signe négatif dans $r_{ISCO}$) ou dans le sens opposé (\textit{rétrograde}, signe positif dans $r_{ISCO}$). L'expression est plus complexe et s'écrit
\begin{equation}
    r_{ISCO} = \frac{r_s}{2} \left( 3 + Z_2 \pm \sqrt{(3-Z_1)(3+Z_1+2Z_2)} \right) \leq \, 4.5 r_s
\end{equation}
avec
\begin{equation}
    \begin{aligned}
        Z_1 &= 1 + \sqrt[3]{1-a_\star^2}\ \left(\sqrt[3]{1+a_\star}+\sqrt[3]{1-a_\star}\right)\\
        Z_2 &= \sqrt{3a_\star^2 + Z_1^2}.
    \end{aligned}
\end{equation}
La Fig.~\ref{fig:rh_spin} montre l'évolution du rayon de l'ISCO par rapport au paramètre de spin. La plupart du temps, on considère que la matière en-dessous de l'ISCO est accrétée avec un temps court comparé au temps dynamique en dehors de cette limite. Ainsi, dans la plupart des simulations numériques et des modèles d'accrétion, le rayon de l'ISCO peut être utilisé comme bord interne du disque (ou plus généralement du flot) d'accrétion, comme l'illustre la Fig.~\ref{fig:disque_spin}.

\begin{figure}
    \centering
    \resizebox{0.5\hsize}{!}{\includegraphics{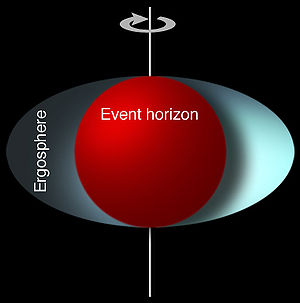}}
    \caption{Illustration de l'ergosphère et de l'horizon des évènements d'un trou noir de Kerr. Crédits : \href{https://fr.wikipedia.org/wiki/Fichier:Ergosphere_of_a_rotating_black_hole.svg}{MesserWoland}}
    \label{fig:ergosphere}
\end{figure}

Un effet important du spin du trou noir est l'entraînement de l'espace-temps et donc de la matière dû à la rotation de l'espace-temps le long de la composante azimutale. Ce phénomène est appelé \textit{effet Lense-Thirring}. On définit la région de l'espace-temps (proche du trou noir) pour laquelle il est impossible pour un observateur de rester immobile par rapport à des étoiles lointaines considérées comme fixes, comme l'ergorégion qui est délimitée par l'ergosphère définie comme la surface
\begin{equation}
    r_{ergo} = \frac{r_s}{2} \left( 1 + \sqrt{1-a_\star^2 \cos^2 \theta} \right).
\end{equation}
Ainsi, l'ergosphère est confondue avec l'horizon des évènements au niveau des pôles et un rayon égal à $r_s$ au niveau de l'équateur. On note que l'ergosphère et l'horizon des évènements sont confondus partout dans le cas d'un trou noir de Schwarzschild ($a=0$). La Fig.~\ref{fig:ergosphere} représente l'ergosphère et l'horizon des évènements d'un trou noir de Kerr. La particularité de l'ergosphère est que l'on peut extraire de l'énergie de rotation du trou noir en envoyant dans cette région des particules massives (processus de \textit{Penrose}) ou des photons (\textit{supperadiance}). Le détail du fonctionnement de ces mécanismes est au-delà des limites de cette thèse (néanmoins, il s'agit de la meilleure source d'énergie possible pour alimenter une civilisation).

Ces rayons caractéristiques peuvent être déterminés à partir de la métrique et de l'équation des géodésiques~\eqref{eq:geodesic_general}. Dans cette thèse, on utilisera deux systèmes de coordonnées pour décrire un trou noir de Kerr, le système de coordonnées de \textit{Boyer-Lindquist} que l'on nommera par la suite Kerr BL et le système de \textit{Kerr-Schild}.

\begin{itemize}
    \item[$\bullet$] \textbf{Métrique Kerr BL} :
Cette métrique utilise un système de coordonnées sphériques similaire à la métrique de Schwarzschild ($ct$,$r$,$\theta$,$\varphi$) avec, on le rappelle, $\mathcal{G}=c=1$. Bien que surprenante au premier abord, cette convention permet de travailler avec un système d'unité géométrique. En effet, les distances et les temps (entre autres) sont directement mis à l'échelle par la masse du trou noir $M$. Ainsi, plutôt que d'utiliser des rayons ou distances en mètres et des temps en secondes qui dépendent de la masse du trou noir que l'on considère, on va utiliser ces unités géométriques qui expriment ces grandeurs en unités de $M$ et sont donc valables quelle que soit la masse du trou noir considéré. Ce n'est qu'à la fin, lorsque l'on veut connaître le temps caractéristique en secondes d'un phénomène dont on connaît la valeur en unité géométrique, que l'on fait la conversion avec les vraies valeurs de $\mathcal{G}$ et $c$. En unités géométriques, le rayon de Schwarzschild vaut
\begin{equation*}
    r_s = 2 M = 2\, r_g
\end{equation*}
avec $r_g = 1 M$ (sous-entendu $r_g = \mathcal{G}M/c^2$). Cette distance est parcourue par la lumière en un temps $t_g = r_g/c = 1 M$ sous-entendu ($t_g = \mathcal{G}M/c^3$). On note que le système de coordonnées devient avec cette convention ($t$,$r$,$\theta$,$\varphi$).

La métrique de Kerr en coordonnées Boyer-Lindquist s'écrit, en prenant l'axe du spin comme axe vertical servant de référence pour la colatitude $\theta$,
\begin{equation} \label{eq:metric_KerrBL}
    ds^2 = -\frac{\Delta}{\rho^2}(dt-a \sin^2 \theta d\varphi)^2 + \frac{\sin^2 \theta}{\rho^2} \left( (r^2+a^2)d\varphi - a dt \right)^2 + \frac{\rho^2}{\Delta}dr^2 + \rho^2 d\theta^2
\end{equation}
avec $\Delta = r^2 - 2Mr + a^2$ que l'on nomme le discriminant et $\rho = r^2 + a^2 \cos^2 \theta$. On note qu'avec la convention décrite précédemment, $M$ et $a$ ont une dimension de longueur.

La matrice de cette métrique n'est pas diagonale, contrairement à celle de Schwarzschild, puisqu'il y a des termes croisés $dt d\varphi$ non nuls. Ce sont ces termes qui marquent l'entraînement de l'espace-temps le long de $\mathbf{\partial \varphi}$ du fait de la rotation (effet Lense-Thirring). Bien que l'utilisation d'un système de coordonnées sphériques soit plus naturel et intuitif pour décrire cet espace-temps, la métrique de Kerr dans ce système de coordonnées présente une singularité mathématique (autrement dit, des infinis pour des valeurs spécifiques) au passage de l'horizon, limitant parfois son utilisation.

    \item[$\bullet$] \textbf{Métrique Kerr KS} :
Le système de coordonnées de \textit{Kerr-Schild} basé sur un système de coordonnées cartésiennes, contrairement au système de coordonnées Kerr BL, a été conçu pour s'affranchir de la singularité mathématique de l'horizon. Elle s'exprime, dans le système de coordonnées ($t$,$x$,$y$,$z$), de la manière suivante
\begin{equation}
    g_{\mu \nu} = \eta_{\mu \nu} + f k_\mu k_\nu
\end{equation}
où
\begin{equation}
    f = \frac{2 M r^3}{r^4 + a^2 z^2},
\end{equation}
\begin{equation}
    \vec{k} = (k_x,k_y,k_z) = \left( \frac{rx+ay}{r^2+a^2}, \frac{ry-ax}{r^2+a^2}, \frac{z}{r} \right)
\end{equation}
et $k_0=1$. $\eta_{\mu \nu}$ étant la métrique de Minkowski, $\vec{k}$ un 3-vecteur unitaire, ce qui fait que le 4-vecteur $\mathbf{k}$ est un vecteur nul à la fois par rapport à $g_{\mu \nu}$ et $\eta_{\mu \nu}$. Il faut aussi noter que $r$ n'est pas le rayon classique, mais est défini par
\begin{equation}
    \frac{x^2+y^2}{r^2+a^2} + \frac{z^2}{r^2} = 1
\end{equation}
ce qui correspond à la définition du rayon classique pour $a \to 0$.

\end{itemize}

\section{Le code de tracé de rayon \textsc{Gyoto}}

\subsection{Principe de fonctionnement}
La courbure de l'espace-temps par un objet massif affecte la trajectoire des photons et, dans le cas des objets compacts, en particulier les trous noirs, de manière extrême, comme on l'a vu avec la Fig.~\ref{fig:deviation_lumière}. Ainsi, générer des images d'un trou noir (il s'agit d'un abus de langage vu que les trous noirs n'émettent pas de rayonnement contrairement à la matière qui les entoure) est une tâche beaucoup moins triviale que pour des objets non relativistes (ici dans le sens de la Relativité Générale). Pour cela, il faut tracer la trajectoire des photons observés à l'aide d'un code de tracé de rayon comme \textsc{Gyoto}. 

\begin{figure}
    \centering
    \resizebox{0.5\hsize}{!}{\includegraphics{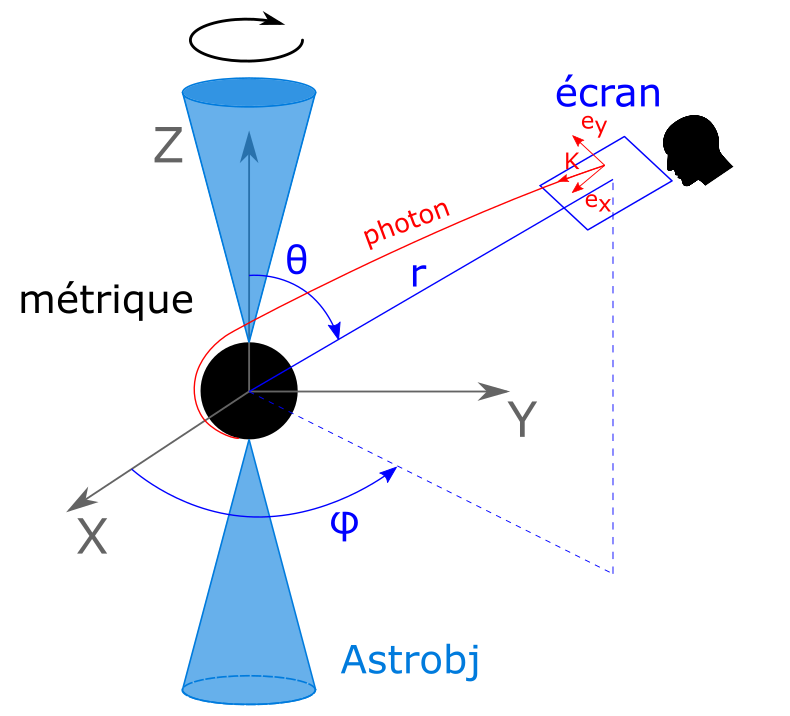}}
    \caption{Illustration d'une scène dans \textsc{GYOTO} composée d'un écran, d'un objet compact défini par la métrique et d'une source d'émission représentée ici par un jet. La géodésique des photons est intégrée depuis l'écran jusqu'à la source d'émission. Dans cette illustration, le PALN (\textit{Position Angle of the Line of Nodes}, égal à 180°) est choisi tel que l'axe vertical de l'écran $\vec{e_y}$ soit le long de l'axe $\vec{Z}$ qui est aligné sur le spin, autrement dit selon $-\vec{e_\theta}$.}
    \label{fig:scenery_gyoto}
\end{figure}

Prenons la situation, que l'on nommera \textit{Scène}, suivante, illustrée par la Fig.~\ref{fig:scenery_gyoto}: on veut faire une image d'un jet lancé par un trou noir de Kerr ayant un spin $a$. L'observateur se situe à une distance $d$ et à une inclinaison (par rapport à l'axe de spin) $i$. Deux directions de propagation du photon sont possibles pour générer une (ou plusieurs) image(s), soit depuis la source vers l'observateur (dans le sens du temps, \textit{forward ray-tracing} en anglais), soit depuis l'observateur vers la source (dans le sens inverse du temps, \textit{backward ray-tracing} en anglais). Chaque point de notre source va émettre des photons dans des directions différentes, en faisant ainsi du tracé de rayon vers l'avant (dans le sens du temps), on ne sait pas à l'avance si le photon émis atteindra l'observateur, les trajectoires des photons pouvant être très complexes en faisant plusieurs tours du trou noir avant d'atteindre l'observateur. Il y a donc une grande quantité de photons qui n'atteindront pas l'observateur, demandant du temps de calcul pour... rien. Il est plus judicieux de calculer la trajectoire des photons dont on sait qu'ils atteindront l'observateur, mais ne sachant pas a priori s'il y a une source sur cette trajectoire (donc dans le sens opposé au temps). 

On commence donc par définir la position ($x^0$,$x^1$,$x^2$,$x^3$) d'un \textit{écran} (l'observateur) dans un système de coordonnées centré sur l'objet compact. Ce dernier est défini par la \textit{métrique} que l'on considère, qui peut être celle d'un trou noir de Kerr ou d'objets plus exotiques comme des trous de vers vers d'autres planètes, ou encore des étoiles bosoniques. On définit la position et la dynamique de notre source de rayonnement, nommée \textit{Astrobj}. Afin de créer notre image, on "lance" un photon (pour être plus exact un paquet de photons) depuis chaque pixel de notre écran dont on va suivre la trajectoire jusqu'à la région d'émission aux abords de l'objet compact via l'équation des géodésiques. Une fois la région d'intérêt atteinte, on détermine si la géodésique passe par la source d'émission (dans le cas d'une "étoile" en orbite, on calcule sa position au temps $t$ du photon). Si c'est le cas, on intègre l'équation du transfert radiatif le long de la partie de la géodésique traversant la source.

Comme on intègre les géodésiques des photons dans le sens opposé du temps, il faut correctement définir les conditions d'arrêt de l'intégration pour éviter une intégration infinie. On définit donc trois types de conditions d'arrêt :
\begin{itemize}
    \item[$\bullet$] par la métrique : la géodésique s'approche de l'horizon des évènements d'un trou noir. En effet, comme aucune lumière ne peut s'en échapper, il est inutile de continuer à intégrer la géodésique. On note que toutes les métriques n'ont pas d'horizon, comme les étoiles bosoniques;
    \item[$\bullet$] par transfert radiatif : lorsque la transmission de la source de rayonnement déjà intégrée atteint zéro, c'est-à-dire que l'on devient optiquement épais. Le rayonnement émis en amont dans le temps, donc plus loin dans l'intégration de la géodésique, ne contribue pas, car celui-ci serait absorbé;
    \item[$\bullet$] par la trajectoire : lorsqu'un photon atteint une grande distance et s'éloigne de la zone d'intérêt. La définition de cette zone est donc assez importante : trop grande, on va intégrer plus que nécessaire et prendre du temps de calcul inutile, trop petite, on risque de stopper l'intégration alors qu'il aurait pu revenir dans la zone d'intérêt et avoir une contribution supplémentaire. Cependant, ce dernier cas ne se produit que très près de l'objet compact ($\sim r_g$).
\end{itemize}

\subsection{Conditions initiales}
L'intégration des géodésiques de genre lumière se fait depuis l'écran que l'on suppose statique aux coordonnées ($t$,$r$,$\theta$,$\varphi$). On se place ici dans une métrique d'un trou noir de Kerr en coordonnées Boyer-Lindquist pour simplifier les notations, mais cela reste valable dans le cas général. Ces coordonnées peuvent aussi être déterminées à partir de la distance, du temps, de l'inclinaison de l'écran par rapport à l'objet compact, l'inclinaison étant définie par rapport à l'axe de spin (l'axe vertical en coordonnées Cartésiennes) et correspondant donc ici à $\theta$. La composante azimutale (dans le plan équatorial) de l'écran par rapport à l'axe de référence $\varphi$ complète les coordonnées comme illustré Fig.~\ref{fig:scenery_gyoto}. On définit un dernier angle, nommé PALN (\textit{Position Angle of the Line of Nodes} en anglais), qui correspond à l'angle de position (à partir du Nord compté positivement vers l'Est) de l'intersection entre le plan équatorial de l'objet compact et le plan de la caméra, autrement dit l'orientation de la tête de l'observateur/écran dans la Fig.~\ref{fig:scenery_gyoto}.
Pour finir, on définit le champ de vue de l'écran que l'on note $FOV$ (pour \textit{field of view} en anglais) en radian (ou $\mu as$), et sa résolution avec un certain nombre de pixels $n_{pix}$ sur les axes vertical et horizontal (exemple 100x100 pixels)

On intègre les géodésiques des photons pour chaque pixel de la caméra, on définit donc le vecteur tangent au photon $\mathbf{k}$ pour chaque pixel de telle sorte que le 4-vecteur $\mathbf{K}$ défini par
\begin{equation}\label{eq:decomposition_vitesse}
    \mathbf{k} = \alpha \mathbf{u} + \mathbf{K}
\end{equation}
avec $\mathbf{u}$ la 4-vitesse de l'observateur qui est supposée statique ($u^r=u^\theta=u^\varphi=0$ et $u^t=-1$) et $\alpha=1$ une constante que l'on impose égale à un pour normaliser la fréquence observée $\nu_{obs}=1$\footnote{On note que la valeur physique de la fréquence est restaurée pour les calculs de transferts radiatifs.}. On calcule ainsi le décalage spectral dû au potentiel gravitationnel, l'effet Doppler relativiste ou encore le beaming (voir Chap.~\ref{chap:modele_hotspot+jet}) par rapport à cette fréquence observée normalisée que l'on applique à la fin de l'intégration à la fréquence réelle. Cela est particulièrement pratique si l'on veut faire un spectre par exemple, puisqu'il suffit d'appliquer ce facteur pour chaque fréquence sans refaire l'intégration, ou obtenir une carte (par pixel) du décalage spectral. $\mathbf{K}$ est donc un 4-vecteur dont la composante temporelle est nulle puisque $k^t=g^{tt}k_t=1$, et dont les trois composantes spatiales correspondent au 3-vecteur vitesse de la lumière qui est perpendiculaire à l'écran dans la direction de l'objet compact (voir Fig.~\ref{fig:scenery_gyoto}). Les vecteurs $\mathbf{N}$ et $\mathbf{W}$ définissent l'axe vertical (direction Nord) et l'axe horizontal (direction Ouest) de l'écran. Ils forment avec $\mathbf{K}$ une base orthonormale directe. On note que la décomposition précédente~\eqref{eq:decomposition_vitesse} est généralisable pour tout 4-vecteur de genre temps pour n'importe quel point de l'espace-temps.

\subsection{Intégration des géodésiques}\label{sec:gyoto_géodésique}
Une fois nos conditions initiales établies, on peut remonter la trajectoire des photons en intégrant l'équation des géodésiques \eqref{eq:geodesic_general} avec un vecteur de genre lumière. Comme dans le cas de particules massives dans la partie~\ref{sec:metric_Schwarzschild} on peut utiliser la conservation de certaines quantités de la 4-impulsion d'un photon $\mathbf{k}$, définie comme 
\begin{equation}
    k^\alpha = g^{\alpha \beta} k_\beta = \frac{dx^\alpha}{d\lambda}
\end{equation}
et qui satisfait
\begin{equation} \label{eq:impulsion_photon}
    -\mathbf{k} \cdot \mathbf{k}=m^2=0,
\end{equation}
ce qui fait de la 4-impulsion d'un photon un vecteur de genre lumière idéal pour l'équation des géodésiques. Dans le cas de la métrique de Kerr, le calcul de la trajectoire des photons a une solution simple et usuelle. L'intégration se fait en conservant la masse via l'Eq.~\eqref{eq:impulsion_photon}, l'énergie de la particule mesurée par un observateur à l'infini E, la composante axiale de son moment cinétique L (comme dans la section \ref{sec:metric_Schwarzschild}) et une quatrième constante (moins directe) que l'on nomme la constante de Carter Q \citep{Carter1968} qui s'exprime directement via les composantes covariantes de la 4-impulsion comme
\begin{equation}
    \begin{aligned}
        E \quad &= \quad -p_t,\\
        L \quad &= \quad p_\varphi,\\
        Q \quad &= \quad p_\theta^2+\cos^2 \theta \left( -a^2 E^2 + \frac{p_\varphi}{\sin^2 \theta} \right),\\
        0 \quad &= \quad g^{\alpha \beta} p_\alpha p_\beta.
    \end{aligned}
\end{equation}

Cependant, \textsc{Gyoto} intègre directement l'équation des géodésiques pour n'importe quel métrique à partir du moment où l'on connait les symboles de Christoffel. On peut ainsi utiliser \textsc{Gyoto} pour générer des images pour une métrique d'étoile bosonique, de champ scalaire et même pour une binaire de trou noir (en lui fournissant les symboles de Christoffel dont il n'existe pas de formule analytique).

\subsection{Intégration du transfert radiatif}
Le but principal de \textsc{GYOTO} est de produire des images, en calculant pour chaque pixel l'intensité spécifique $I_\nu$ dont l'unité est J m$^{-2}$ s$^{-1}$ str$^{-1}$ Hz$^{-1}$ dans le système SI à la fréquence observée $\nu_{obs}$. Lorsqu'un photon traverse un milieu matériel, l'intensité transmise dépend de l'absorption $\alpha_\nu$ [m$^{-1}$] et de l'émission $j_\nu$ [J m$^{-3}$ s$^{-1}$ str$^{-1}$ Hz$^{-1}$] de ce milieu. On distingue ainsi deux types de milieux selon leur transmission optique $T_\nu = \exp(- \int \alpha_\nu ds)$ avec $ds$ l'élément de longueur propre infinitésimale :
\begin{itemize}
    \item[$\bullet$] milieu optiquement \textbf{épais}, caractérisé par une transmission nulle $T_\nu = 0$ qui correspond à une forte absorption. L'intensité reçue dans ce cas a été émise à la surface de la source et marque la fin de l'intégration de la géodésique;
    
    \item[$\bullet$] milieu optiquement \textbf{mince}, où par opposition au cas optiquement épais, la transmission est non nulle. Chaque élément de longueur de la source va contribuer à l'intensité totale reçue par l'observateur. 
\end{itemize}

Pour chaque élément de longueur $ds$, dans le référentiel de la source, le milieu va absorber une partie de l'intensité spécifique $I_\nu$, la variation d'intensité spécifique résultante est la suivante
\begin{equation} \label{eq:absorption}
    dI_\nu = - \alpha_\nu I_\nu ds.
\end{equation}
Le milieu va aussi émettre du rayonnement. Que ce soit du rayonnement synchrotron, de corps noir ou autre, la variation d'intensité spécifique liée est
\begin{equation} \label{eq:emission}
    dI_\nu = j_\nu ds.
\end{equation}

Ainsi, la variation d'intensité spécifique totale prenant en compte l'absorption et l'émission, correspondant à l'équation du transfert radiatif (dans le référentiel de l'émetteur !) s'écrit
\begin{equation} \label{eq:transfert_rad}
    \frac{dI_\nu}{ds} = j_\nu - \alpha_\nu I_\nu.
\end{equation}

On cherche à déterminer l'intensité spécifique observée et non celle émise. En effet, les effets de la Relativité Restreinte et Générale comme l'effet Doppler relativiste et gravitationnel vont affecter la fréquence du photon le long de la géodésique. Cependant, on peut définir l'intensité spécifique invariante le long des géodésiques (dans le sens qui ne dépend pas du référentiel) comme  
\begin{equation}
    \mathcal{I} = \frac{I_\nu}{\nu^3}.
\end{equation}
On peut ainsi intégrer l'équation du transfert radiatif~\eqref{eq:transfert_rad} dans le référentiel de l'émetteur puis calculer l'intensité spécifique dans le référentiel de l'observateur $I_{\nu,obs}$ grâce à l'invariance précédente  en connaissant le rapport de fréquence $\nu_{obs}/\nu_{em}$
\begin{equation}
    I_{\nu,obs}= \left( \frac{\nu_{obs}}{\nu_{em}} \right)^3 I_{\nu,em}.
\end{equation}

D'un point de vue numérique, pour chaque pas d'intégration de l'équation des géodésiques, et à condition que le photon soit dans la région d'intérêt (que l'on définit pour la suite, sauf indication contraire, comme $r_{ph}<50 M$), on teste si le photon se situe à l'intérieur d'une source d'émission astrophysique. Le résultat de cette condition dépend des coordonnées (spatiales et temporelles) du photon, ainsi que de la géométrie et de la dynamique de la source (un disque d'accrétion mince ou épais, ou une étoile en orbite). On détermine le rapport de fréquence (le facteur de décalage spectral) pour ce pas d'intégration à la position $x^\alpha$ du photon comme
\begin{equation}
    \frac{\nu_{em}}{\nu_{obs}} = - \mathbf{u}_{obj} \cdot \mathbf{u}_{ph}
\end{equation}
où $\mathbf{u}_{obj}$ est la 4-vitesse de l'objet et $\mathbf{u}_{ph}$ est la 4-vitesse du photon.

\begin{figure}
    \centering
    \resizebox{0.8\hsize}{!}{\includegraphics{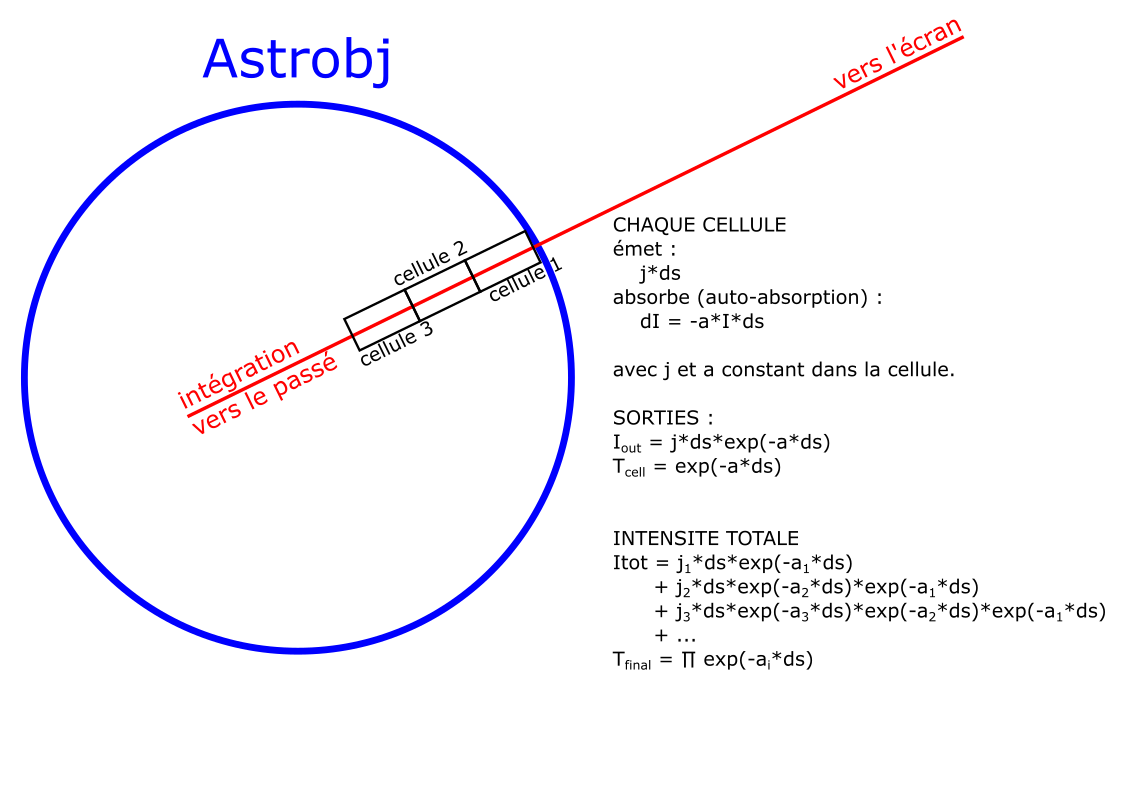}}
    \caption{Schéma illustrant l'intégration du transfert radiatif le long d'une géodésique traversant un objet astrophysique nommé \textit{Astrobj}. L'Astrobj est divisé en différentes cellules dans lesquelles l'incrément d'intensité spécifique et la transmission sont calculés à partir des coefficients d'émission et d'absorption. Crédit : Frédéric Vincent.}
    \label{fig:schema_intreg_transfert_rad}
\end{figure}

Puis on calcule, dans le référentiel de l'émetteur, l'intensité spécifique $I_{\nu,em}$ pour la cellule $i$ (l'indice du pas d'intégration de la géodésique à l'intérieur de la source, voir Fig.~\ref{fig:schema_intreg_transfert_rad}) avec un incrément de distance $\Delta s$ ainsi que les coefficients d'émission $j_{\nu,em}$ et d'absorption $\alpha_{\nu,em}$ supposés constant le long de $\Delta s$ qui dépendent des paramètres physiques du milieu et du processus d'émission de la manière suivante
\begin{equation} \label{eq:increment_intensité_cell}
    I_{\nu,em}^i = \exp (-\alpha_{\nu,em} \Delta s) j_{\nu,em} \Delta s
\end{equation}
et la transmission sur cet incrément de distance $T_{\nu,em}^i$
\begin{equation}
    T_{\nu,em}^i = \exp (-\alpha_{\nu,em} \Delta s).
\end{equation}

Cependant, la totalité de cette intensité émise ne va pas forcément atteindre l'observateur. En effet, une partie peut être absorbée par les cellules précédentes. La contribution réelle de la cellule $i$ à l'intensité totale, c-à-d l'incrément d'intensité spécifique de la cellule $i$ est 
\begin{equation} \label{eq:increment_intensité}
    \Delta I_{\nu,em}^i = I_{\nu,em}^i \times T^i_{prev}
\end{equation}
où
\begin{equation}
    T^i_\mathrm{prev} = \prod_{j=0}^{i-1} T_{\nu,em}^j.
\end{equation}
est la transmission de l'ensemble des cellules précédentes. L'intensité spécifique totale émise est donc la somme des incréments calculés par l'équation~\eqref{eq:increment_intensité}\footnote{L'intensité spécifique est initialisée à zéro au niveau de l'écran et la transmission à 1.} comme illustré dans la Fig.~\ref{fig:schema_intreg_transfert_rad}. On a donc les expressions suivantes pour l'intensité spécifique observée et la transmission à la cellule $N$ :
\begin{equation}
    I_{\nu,obs}^N = \sum_{i=0}^N \Delta I_{\nu,em}^i \left( \frac{\nu_{obs}}{\nu_{em}} \right)^3
\end{equation}
\begin{equation}
    T^N = \prod_{i=0}^N T_{\nu,em}^i.
\end{equation}

\part{Les sursauts de Sagittarius A*}
\chapter{Propriétés observationnelles des sursauts de Sagittarius A*}\label{chap:Sgr~A* flares}
\markboth{Propriétés observationnelles des sursauts de Sagittarius A*}{Propriétés observationnelles des sursauts de Sagittarius A*}
{
\hypersetup{linkcolor=black}
\minitoc 
}

\section{Les sursauts en IR}
\subsection{Statistiques}
\begin{figure}
    \centering
    \resizebox{\hsize}{!}{\includegraphics{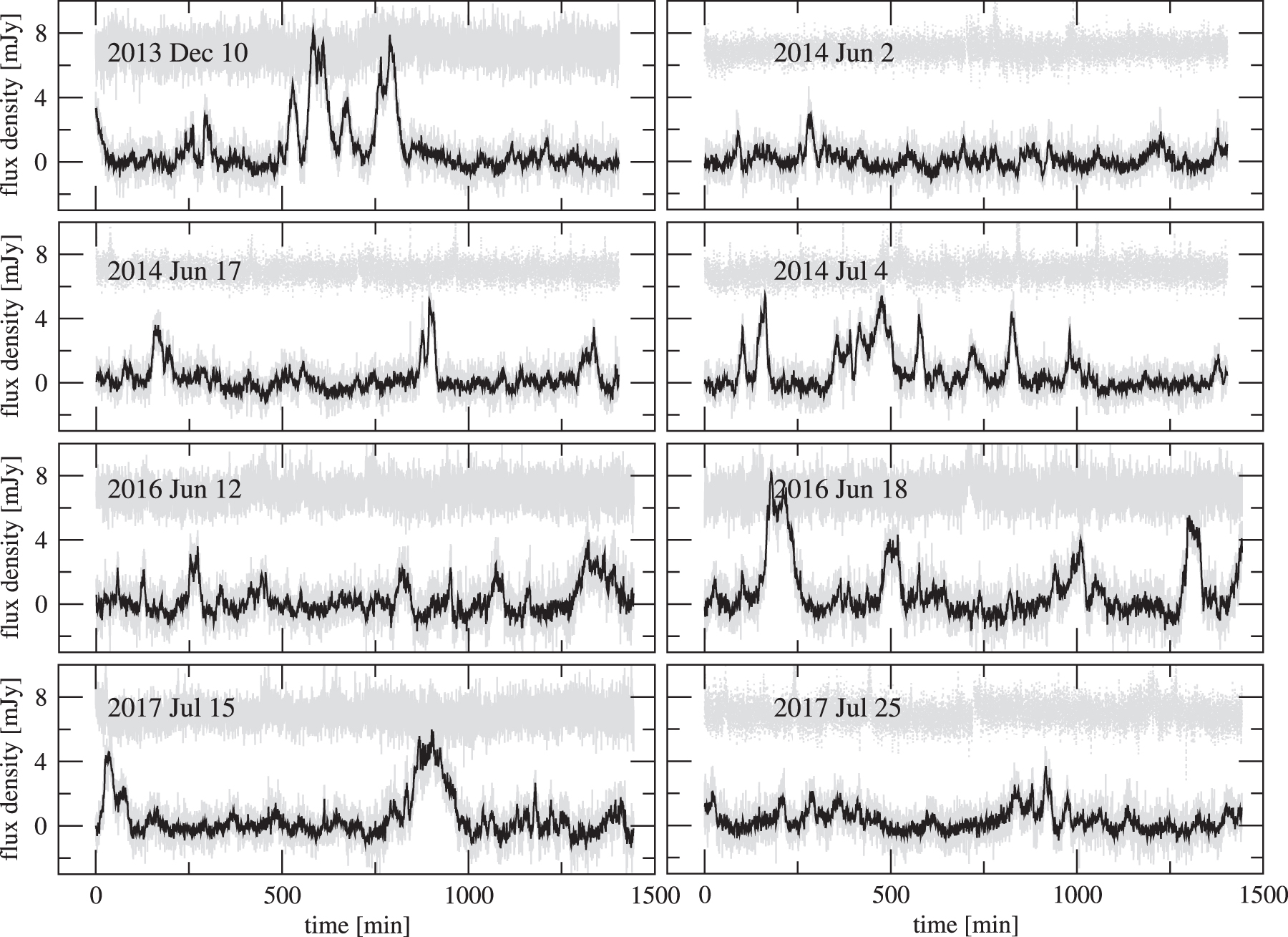}}
    \caption{Densité de flux en fonction du temps pour 8 époques (de 24h) d'observations avec \textit{Spitzer}. Les courbes grises sont les courbes de lumière construites avec chacune des images (toutes les $6.4$s), celles en noir correspondent à une moyenne de flux sur une minute. Crédit : \cite{Witzel2018}.}
    \label{fig:Spitzer_LC}
\end{figure}

Peu après la première détection d'un sursaut de Sagittarius A* en rayons~X \cite{Baganoff2001}, les sursauts de Sgr~A* ont été détectés en IR~\cite{Genzel2003} et sont régulièrement observés depuis 20 ans avec différents instruments installés sur de grands télescopes terrestres comme le \href{https://www.keckobservatory.org/}{Keck Observatory} et le \href{https://www.eso.org/public/france/teles-instr/paranal-observatory/vlt/}{Very Large Telescope} ou avec des télescopes spatiaux comme \href{https://www.spitzer.caltech.edu/}{\textit{Spitzer}}. Ces observations ont notamment permis de distinguer l'état de sursaut de l'état quiescent de Sgr~A* (voir Chap.~\ref{chap:GC}). L'observation continue sur de longues périodes de Sgr~A*, notamment par \textit{Sptizer}, a permis de déterminer que la fréquence d'occurrence des sursauts en IR est de l'ordre de $4\pm2$ par jour~\cite{Dodds-Eden2009, Eckart2006, Yusef-Zadeh2006, Gravity2020b} correspondant donc à une période d'inactivité (quiescente) de l'ordre de $\sim6$h. La durée totale des sursauts observés est, quant à elle, de l'ordre de $\sim 30$ min, pouvant aller jusqu'à 2h~\cite{Genzel2003, Meyer2006, Eckart2006, Trippe2007} comme le montrent les courbes de lumière obtenues avec \textit{Sptizer} dans la Fig.~\ref{fig:Spitzer_LC}.

\begin{figure}
    \centering
    \resizebox{0.8\hsize}{!}{\includegraphics{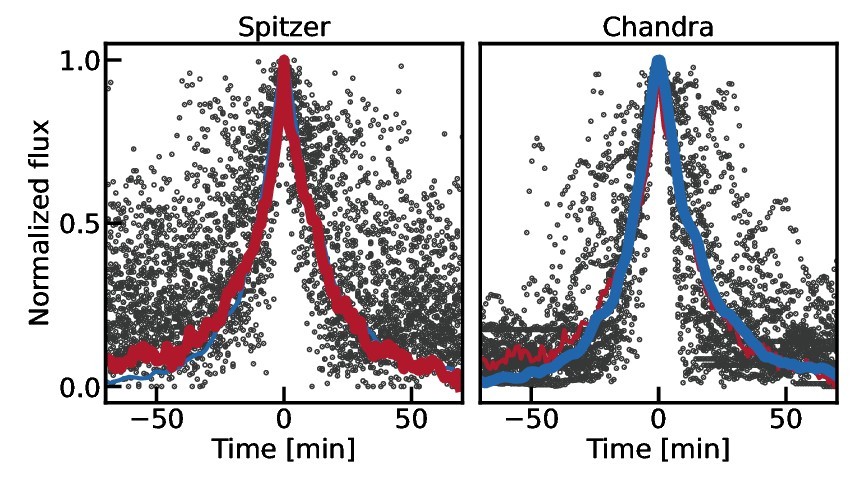}}
    \caption{Formes caractéristiques des sursauts en IR et en rayons~X. \textbf{À gauche} : segments de données normalisées pour 25 sursauts identifiés dans la Fig.~\ref{fig:Spitzer_LC}. La ligne rouge épaisse indique la première composante principale dérivée des données et la fine ligne bleue la première composante principale dérivée des données en rayons~X \textit{Chandra} (voir le panneau de droite). \textbf{À droite} : même chose que le panneau de gauche pour 26 sursauts en rayons~X de \textit{Chandra}. Crédit : \cite{von_Fellenberg2023}.}
    \label{fig:von_Fellenberg2023}
\end{figure}

Après avoir normalisé et aligné les pics de ces courbes de lumière, \cite{von_Fellenberg2023} les ont regroupés pour former un jeu de données sur lequel une analyse statistique peut être faite (voir l'article pour les détails mathématiques). À partir de ces données, ils en déduisent un profil de courbe de lumière typique illustré dans la Fig.~\ref{fig:von_Fellenberg2023} pouvant être décrit par deux exponentielles (croissante puis décroissante) avec des temps caractéristiques en IR $\tau_{rise}=15,3 \pm 4,2$ min et $\tau_{decay}=15,6 \pm 4,1$ min.

\subsection{Observations des sursauts avec GRAVITY}
\subsubsection{Mesure du mouvement orbital}
\begin{figure}
    \centering
    \resizebox{0.8\hsize}{!}{\includegraphics{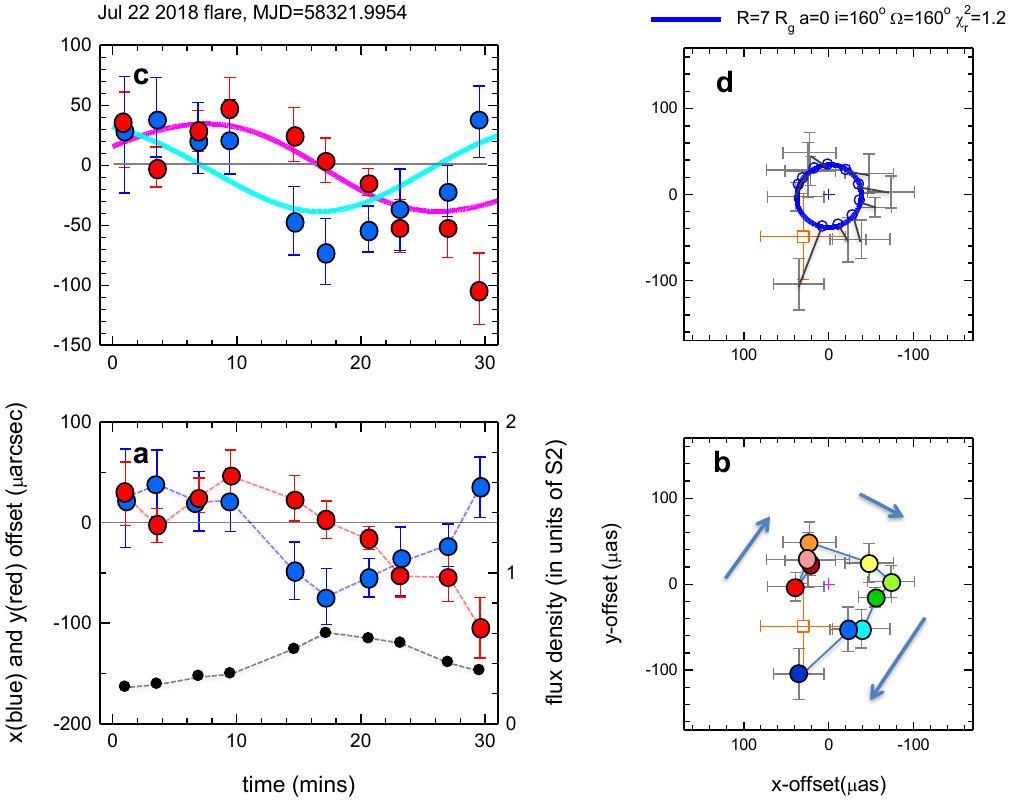}}
    \caption{\textbf{En bas à gauche} (panneau a) : évolution temporelle de la position Est-Ouest (positif vers l'Est, bleu) et Nord-Sud (rouge) des centroïdes du sursaut du 22 juillet (MJD = 58321,9954) par rapport à leur médiane, ainsi que l'évolution de la densité de flux (axe des ordonnées de droite, noir) en unités du flux de S2 (magnitude $m_K = 14,0$). Les barres d'erreur sont de 1$\sigma$. L'intensité totale a été calculée à partir de la somme des deux directions de polarisation. \textbf{En bas à droite} (panneau b) : projection de l'orbite du centroïde du sursaut sur le ciel (couleur allant du brun au bleu foncé comme marqueur qualitatif du temps au cours des 30 minutes d'observation, par rapport à leur médiane (petite croix rose) et après élimination du mouvement de S2). Le carré orange et l'incertitude de 1$\sigma$ représentent la position astrométrique à long terme du centre de masse de l'orbite de S2. \textbf{En haut à gauche} (panneau c) et \textbf{en haut à droite} (panneau d) : comparaison des données des deux panneaux inférieurs avec une réalisation d'un modèle simple de point chaud dans la métrique de Schwarzschild. Les courbes continues violette et cyan en (c) montrent la même orbite en x(t) et y(t), comparées aux données en bleu et rouge. La courbe continue bleue en (d) désigne un point chaud sur une orbite circulaire avec $R = 1,17 \times R(ISCO, a = 0, M = 4,14 \times 10^6 M_\odot)$, vu à l'inclinaison $160\degree$ (dans le sens des aiguilles d'une montre sur le ciel, comme pour les données en (d)) et avec la ligne des nœuds à $\Omega = 160\degree$ $(\chi_r^2 = 1,2)$. Les cercles bleus ouverts et les barres grises relient les points de données à leurs emplacements sur l'orbite la mieux ajustée. Crédit : \cite{Gravity2018}.}
    \label{fig:Gravity2018}
\end{figure}

L'intérêt principal de cette thèse est l'étude des sursauts observés par GRAVITY. L'extrême précision astrométrique de l'instrument, de l'ordre de $30-50$ $\mu$as, a permis à la collaboration GRAVITY de détecter le mouvement de la source de flux autour de Sgr~A* durant trois sursauts brillants en 2018~\cite{Gravity2018}. Les astrométries de ces sursauts, c'est-à-dire la position en fonction du temps du centroïde du flux de la source, présentant un mouvement elliptique, comme illustré dans la Fig.~\ref{fig:Gravity2018}, ont été associées à l'orbite d'un point chaud autour du trou noir. Le modèle utilisé par \cite{Gravity2018} pour ajuster les données est un modèle de point chaud simple similaire à \cite{Hamaus2009, Vincent2011, Vincent2014}, correspondant à l'orbite circulaire dans le plan équatorial du trou noir d'une sphère émettant un rayonnement constant. Plusieurs codes de tracé de rayons ont été utilisés pour ajuster les données, à savoir \texttt{grtrans}\footnote{\href{https://github.com/jadexter/grtrans}{https://github.com/jadexter/grtrans}} \cite{Dexter2016, Dexter2009} où l'émission considérée est du rayonnement synchrotron provenant d'une distribution des électrons non thermiques en loi de puissance (voir Chap.~\ref{chap:modele_hotspot+jet}) \cite{Broderick2006}, \textsc{Gyoto}\footnote{\href{https://gyoto.obspm.fr}{https://gyoto.obspm.fr}} (voir Chap~\ref{chap:GYOTO}) où la source est optiquement épaisse avec une émission $I_\nu (\nu) =$ constante et \texttt{NERO} où l'émission est calculée à partir de la densité intégrée le long de la géodésique traversant le point chaud dont la densité est gaussienne. Les différents codes et approches comme Markov Chain Monte Carlo (MCMC), ajustement de courbe ou technique des moindres carrés, convergent vers les mêmes résultats, à savoir un rayon orbital $r \approx 7 \pm 0,5$ $r_g$, donc une période orbitale de $P=40 \pm 8$ min pour un spin $a=0$, une inclinaison $i \approx 160 \degree \pm 10 \degree$ et l'angle de position\footnote{\href{https://en.wikipedia.org/wiki/Position_angle}{https://en.wikipedia.org/wiki/Position\_angle}} de la ligne des nœuds (PALN) $\Omega \approx 115 \degree - 160 \degree$ (avec une valeur de $160 \degree$ pour le modèle de la Fig.~\ref{fig:Gravity2018}). La forme quasi-circulaire de l'astrométrie, et le faible contraste dans la courbe de lumière et la polarisation nécessitent une faible inclinaison~\cite{Hamaus2009}. Cette contrainte a été plus tard confirmée par les observations radio de l'EHT~\cite{EHT2022a} et \cite{Wielgus2022} suggérant aussi une faible inclinaison ($i \leq 30 \degree$).

Dans le modèle utilisé par \cite{Gravity2018}, quel que soit le code utilisé, l'orbite du point chaud est circulaire dans le plan équatorial, la vitesse est supposée Képlérienne et est liée au rayon orbital (Eq.~\eqref{eq:vitesse_keplerienne}). De ce fait, les propriétés spatiales et temporelles du modèle de sursaut sont liées. Il en résulte, comme on peut le constater dans le panneau (d) de la Fig.~\ref{fig:Gravity2018}, que l'ajustement des données ne permet pas d'ajuster à la fois le rayon et la période orbitale de manière indépendante. Afin d'avoir une vitesse orbitale suffisante pour correspondre aux données, le rayon obtenu par l'ajustement est plus faible que ce que suggèrent les données. En effet, l'ensemble des données astrométrique dans le panneau (d) de la Fig.~\ref{fig:Gravity2018} sont à l'extérieur du meilleur ajustement du modèle illustré par la courbe bleue. En levant la corrélation entre la vitesse orbitale et le rayon, un meilleur ajustement serait possible avec un rayon orbital plus élevé, mais une vitesse égale (ou très proche). Ainsi, les données du sursaut du 22 juillet 2018 suggèrent une vitesse orbitale super-Képlérienne. Cette propriété est très importante et est un élément essentiel du modèle présenté dans le Chap.~\ref{chap:Plasmoid Flare model}.

\begin{figure}
    \centering
    \resizebox{0.5\hsize}{!}{\includegraphics{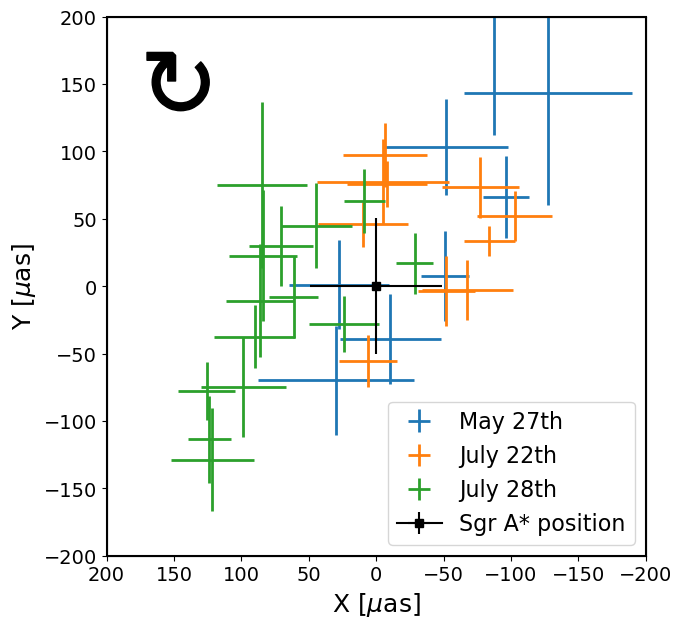}}
    \caption{Positions des centroïdes des trois sursauts observés avec GRAVITY, centrés sur la position de Sgr~A* (carré noir). Les sursauts du 27 Mai et du 22 Juillet semblent similaires dans leur orientation orbitale, tandis que le centroïde du sursaut du 28 Juillet pointe largement dans la direction opposée. Aucun des centres du mouvement projeté n'est centré sur la position de Sgr A*. La direction du mouvement orbital est représentée par une flèche noire dans le coin supérieur gauche.}
    \label{fig:Ball2021}
\end{figure}

De plus, la position de Sgr~A*, ou plus précisément la position du centre de masse dérivé à partir de l'orbite à long terme de l'étoile S2, est décalé par rapport au centre de l'orbite modélisée. Bien que ce dernier soit dans la barre d'erreur de $1\sigma$ sur la position de Sgr~A*, pour le sursaut du 22 juillet 2018, ce décalage n'est pas le même pour les deux autres sursauts observés en 2018 comme l'illustre la Fig.~\ref{fig:Ball2021}. Les sursauts du 27 mai et du 22 juillet semblent similaires dans leurs orientations orbitales, tandis que le centroïde du sursaut du 28 juillet pointe largement dans la direction opposée. Ce décalage sera aussi un élément important des modèles présentés dans les prochains chapitres.

\subsubsection{Courbe de lumière et polarisation}
\begin{figure}
    \centering
    \resizebox{0.4\hsize}{!}{\includegraphics{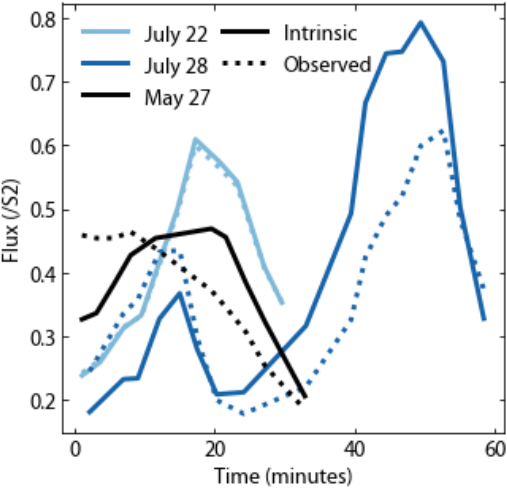}}
    \caption{Courbes de lumière observées et intrinsèques déduites pour les trois sursauts observés par GRAVITY en 2018. Le flux est mesuré par rapport à l'étoile voisine S2. Chaque ligne pointillée correspond à la courbe de lumière observée par GRAVITY pendant le sursaut en question. La ligne continue montre la courbe de lumière intrinsèque une fois que tous les effets relativistes prédits par le modèle orbital le mieux adapté ont été supprimés. Crédit : \cite{Gravity2020c}.}
    \label{fig:Gravity2020_model}

    \resizebox{0.9\hsize}{!}{\includegraphics{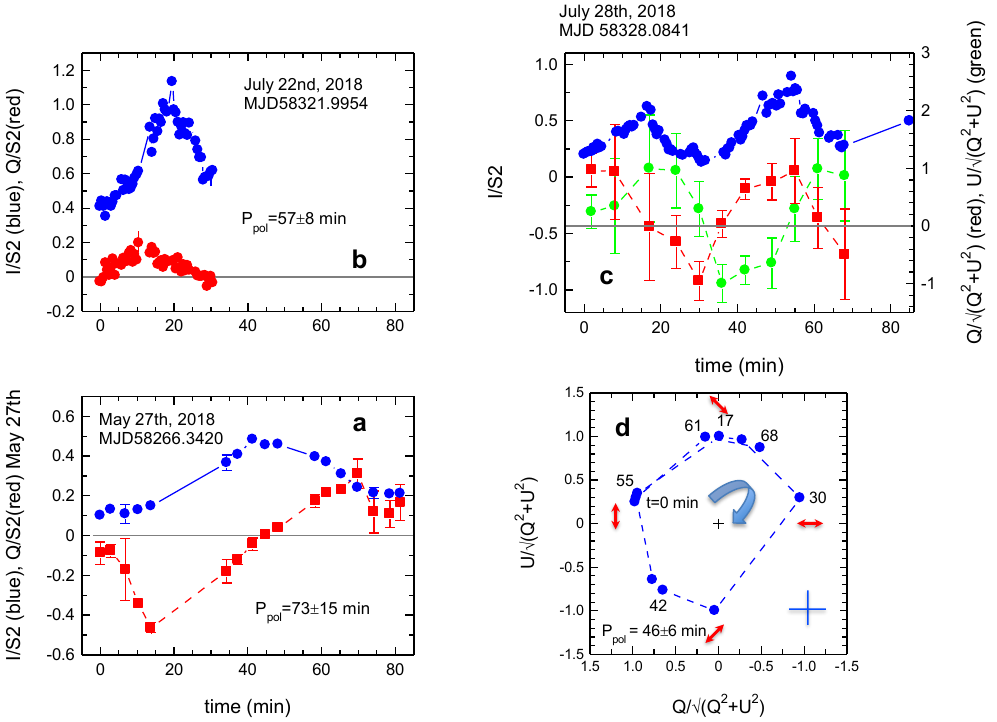}}
    \caption{\textbf{En bas à gauche} (panneau a) : flux total I/S2 (bleu, par rapport à S2) et rapport de flux Q/I en fonction du temps pour le sursaut du 27 mai. \textbf{En haut à gauche} (panneau b) : comme en (a) mais pour le sursaut du 22 juillet. \textbf{En haut à droite} (panneau c) : Évolution de I/S2 (bleu), Q/(Q$^2$ + U$^2$)$^{1/2}$ (rouge) et U/(Q$^2$ + U$^2$)$^{1/2}$ (vert) pendant le sursaut du 28 juillet. \textbf{En bas à droite} (panneau d) : évolution du sursaut du 28 juillet dans le plan des paramètres de Stokes normalisés Q/(Q$^2$ + U$^2$)$^{1/2}$ (horizontal) et U/(Q$^2$ + U$^2$)$^{1/2}$ (vertical). Les flèches rouges indiquent les directions de polarisation sur le ciel. La croix bleue en bas à droite indique une barre d'erreur typique. Crédit : \cite{Gravity2018}.}
    \label{fig:Gravity_polar}
\end{figure}

En plus de l'astrométrie, GRAVITY a mesuré le flux des sursauts de 2018, mais celui-ci n'a cependant pas été ajusté par les modèles de~\cite{Gravity2018}. Les sursauts du 22 juillet et du 27 mai sont assez similaires avec un pic unique et une pente de décroissance comparables avec néanmoins une phase de croissance significativement différente (voir courbes cyan et noir en pointillés de la Fig.~\ref{fig:Gravity2020_model}). La courbe de lumière du sursaut du 28 juillet (en pointillés bleus) est assez différente des deux premières et présente deux pics avec néanmoins une fenêtre d'observation de 60 min au lieu de 30. \cite{Gravity2020c} a modélisé ces sursauts avec un modèle de point chaud en prenant en compte un mouvement hors du plan équatorial (une orbite inclinée) ainsi que le déchirement par la rotation différentielle autour du trou noir. Ce modèle permet notamment de contraindre la taille maximale du point chaud à $R_\mathrm{max} \sim 5 \, r_g$. En effet, au-delà d'une certaine taille, la rotation différentielle va étendre fortement le point chaud autour du trou noir. Le centroïde résultant présentera un mouvement orbital beaucoup plus réduit, car un objet plus axisymmétrique a un centroïde plus près du centre du champ. De plus, l'utilisation de code de tracé de rayons permet de calculer la courbe de lumière obtenue. Ainsi, il est possible de calculer les effets relativistes (voir Chap.~\ref{chap:modele_hotspot+jet}) qui influencent la courbe de lumière observée à partir des paramètres orbitaux ajustés. Après avoir soustrait ces effets de la courbe de lumière mesurée, \cite{Gravity2020c} obtiennent la courbe de lumière intrinsèque que doivent avoir ces sursauts avec les paramètres orbitaux ajustés (voir Fig.~\ref{fig:Gravity2020_model}). On note ici qu'il ne s'agit pas d'un ajustement de la courbe de lumière, qui nécessite une modélisation de l'émission du point chaud, mais d'une déduction de la courbe de lumière intrinsèque que doivent avoir les sursauts en soustrayant les effets relativistes prédits par le modèle orbital le mieux adapté.

GRAVITY a aussi mesuré le flux des sursauts dans deux directions de polarisation, permettant ainsi de déterminer la direction du vecteur polarisation en fonction du temps\footnote{Pour certains sursauts, seule une direction de polarisation a été mesurée, ne permettant pas de déterminer l'orientation du vecteur polarisation, mais présentant un signal périodique.}, le long de l'orbite, ainsi que la fraction de polarisation. De plus, en exprimant ces quantités avec les paramètres de \href{https://en.wikipedia.org/wiki/Stokes_parameters}{Stokes}, on obtient des boucles de polarisation dans le plan Q-U (plus de détails dans le Chap.~\ref{chap:Polarization}; voir la Fig.~\ref{fig:Gravity_polar}). Les propriétés des boucles dans le plan Q-U dépendent fortement de la configuration magnétique de la zone d'émission ainsi que de la géométrie du système~\cite{Wielgus2022} permettant de donner des contraintes observationnelles. Cependant, la précision et l'échantillonnage des sursauts de 2018 ne permettent pas de conclure de manière certaine. Cependant, comme la période de polarisation mesurée est la même que la période orbitale et que l'on observe l'orbite avec une faible inclinaison, cela exclut une configuration magnétique toroïdale \cite{Gravity2018}. L'angle de polarisation correspond à la direction du vecteur électrique du photon, mesuré du Nord vers l'Est, et est compris dans l'intervalle [$-\pi /2, \pi / 2 $[. Dit autrement, la direction du vecteur polarisation n'a pas d'importance, seule son orientation dans le plan du ciel est pertinente.

\section{Propriétés multi-longueurs d'onde}
\subsection{Délais entre les différentes longueurs d'onde}
Avant de rentrer dans les détails de la modélisation des sursauts de Sgr~A* observés par GRAVITY, il est intéressant de voir les propriétés multi-longueurs d'onde des sursauts de Sgr~A*. Ces derniers étant observés en NIR, en rayons~X et en radio, plusieurs campagnes d'observations à ces longueurs d'ondes utilisant simultanément des télescopes terrestres et spatiaux ont été menées. La première détection d'un sursaut à la fois en rayons~X et en NIR a été observée le 19 juin 2003 par les instruments NACO (NIR; VLT) et ACIS-I (rayons~X, \textit{Chandra})~\cite{Eckart2004}. Le sursaut ayant eu lieu au début, voire un peu avant le chevauchement des observations, seule une valeur limite maximale du décalage temporel entre la fin du sursaut en rayons~X et NIR a pu être estimée à $\sim 15$ min. Les campagnes d'observation suivantes~\cite{Eckart2006a}, avec en plus le réseau submillimétrique \href{https://lweb.cfa.harvard.edu/sma/}{SMA} et le \href{https://public.nrao.edu/telescopes/vla/}{VLA} (\textit{Very Large Array} en anglais), ont permis d'observer quatre sursauts simultanément en NIR et rayons~X. Cependant, un seul de ces sursauts était visible dans les deux domaines de longueurs d'onde, les autres ayant été observés uniquement en NIR. D'une manière générale, les sursauts en rayons~X ont systématiquement une contrepartie en NIR (à condition d'avoir des moyens d'observation disponibles) mais pas l'inverse.

\begin{figure}
    \centering
    \resizebox{\hsize}{!}{\includegraphics{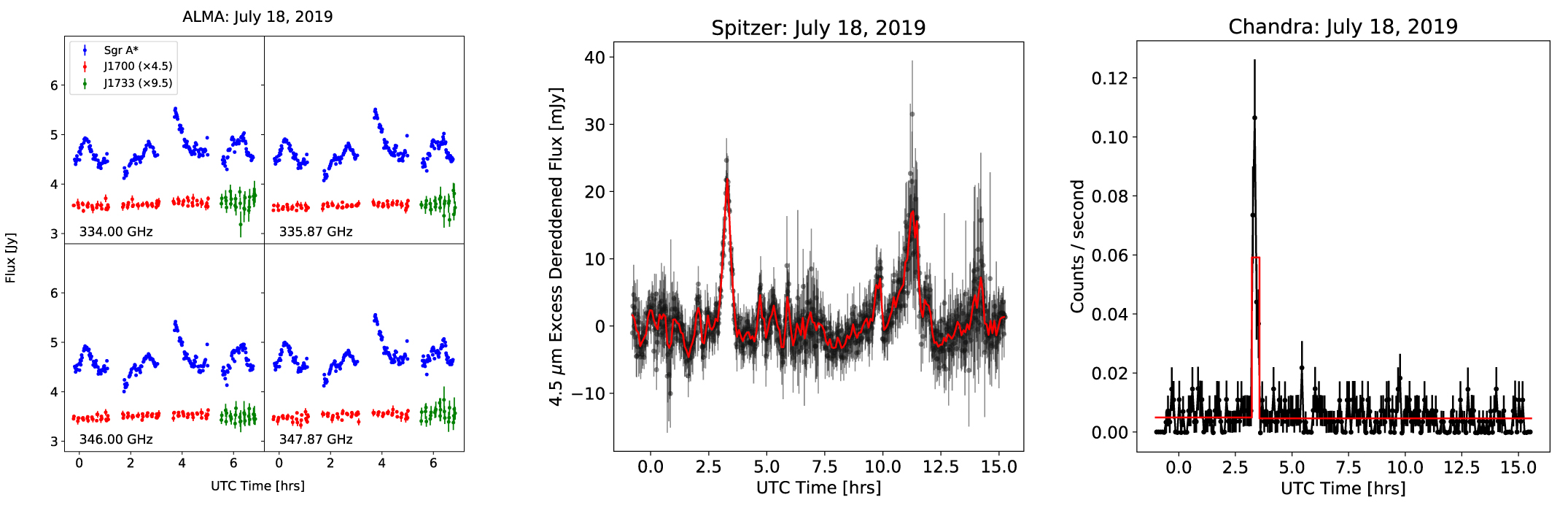}}
    \caption{Courbes de lumière multi-longueurs d'onde de Sgr A* le 18 juillet 2019. \textbf{À gauche} : Sgr A* (bleu) tel qu'observé par ALMA ; les calibrateurs de phase J1700 et J1733 (rouge et vert, respectivement) sont montrés. Un temps de binning de 60 s est utilisé. \textbf{Au centre} : courbe de lumière \textit{Spitzer} de Sgr A* avec un binning de 60 s (noir) et un binning de cinq minutes (rouge). \textbf{À droite} : Courbe de lumière \textit{Chandra} 2-8 keV soustraite du bruit de fond de Sgr A* avec un binning de cinq minutes (noir). Crédit : \cite{Michail2021}.}
    \label{fig:Michail2021a}
    
    \vspace{1cm}
    
    \resizebox{\hsize}{!}{\includegraphics{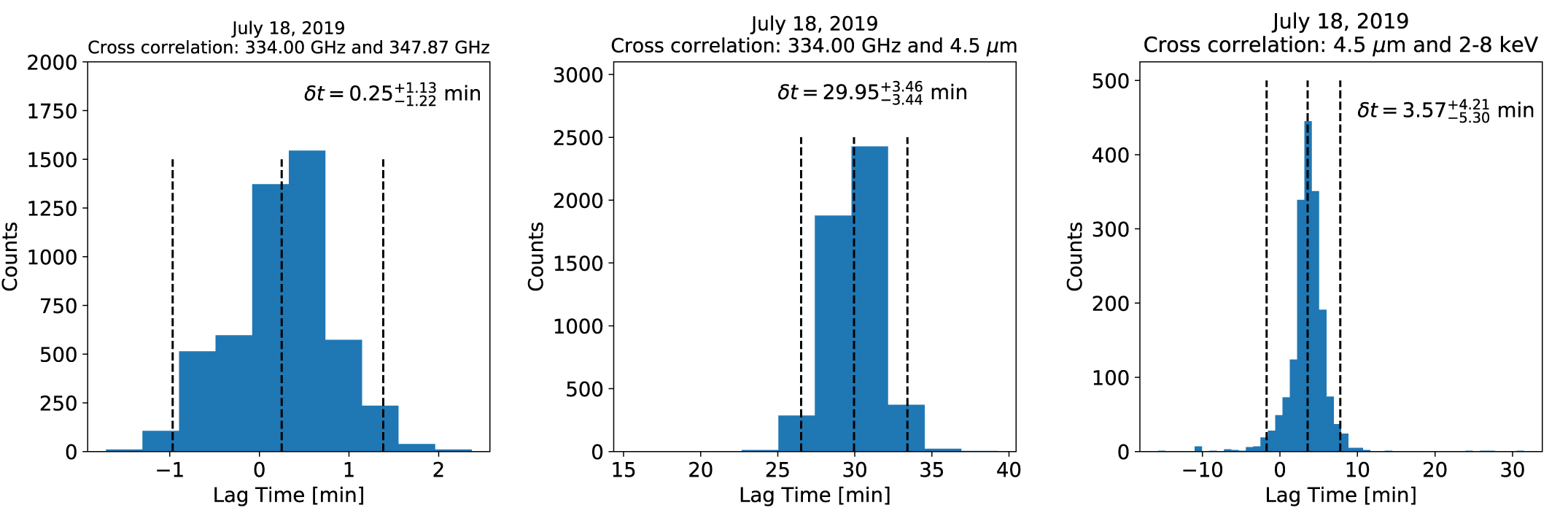}}
    \captionof{figure}{Histogrammes des corrélations entre les paires de courbes de lumière à l'aide du package Python \texttt{pyCCF} \cite{Sun2018}. L'intervalle de confiance (IC) à 95$\%$ (2$\sigma$) est indiqué. \textbf{À gauche} : les corrélations entre les fréquences ALMA les plus largement séparées. \textbf{Au centre} : l'histogramme de corrélation entre la fréquence ALMA la plus basse et la courbe de lumière \textit{Spitzer}; la courbe de lumière ALMA atteint son maximum moins de 30 minutes après les données Spitzer. \textbf{À droite} : l'histogramme de corrélation entre \textit{Spitzer} et \textit{Chandra}, où les pics des sursauts se produisent simultanément. Crédit : \cite{Michail2021}.}
    \label{fig:Michail2021b}
\end{figure}

\begin{figure}
    \centering
    \resizebox{0.8\hsize}{!}{\includegraphics{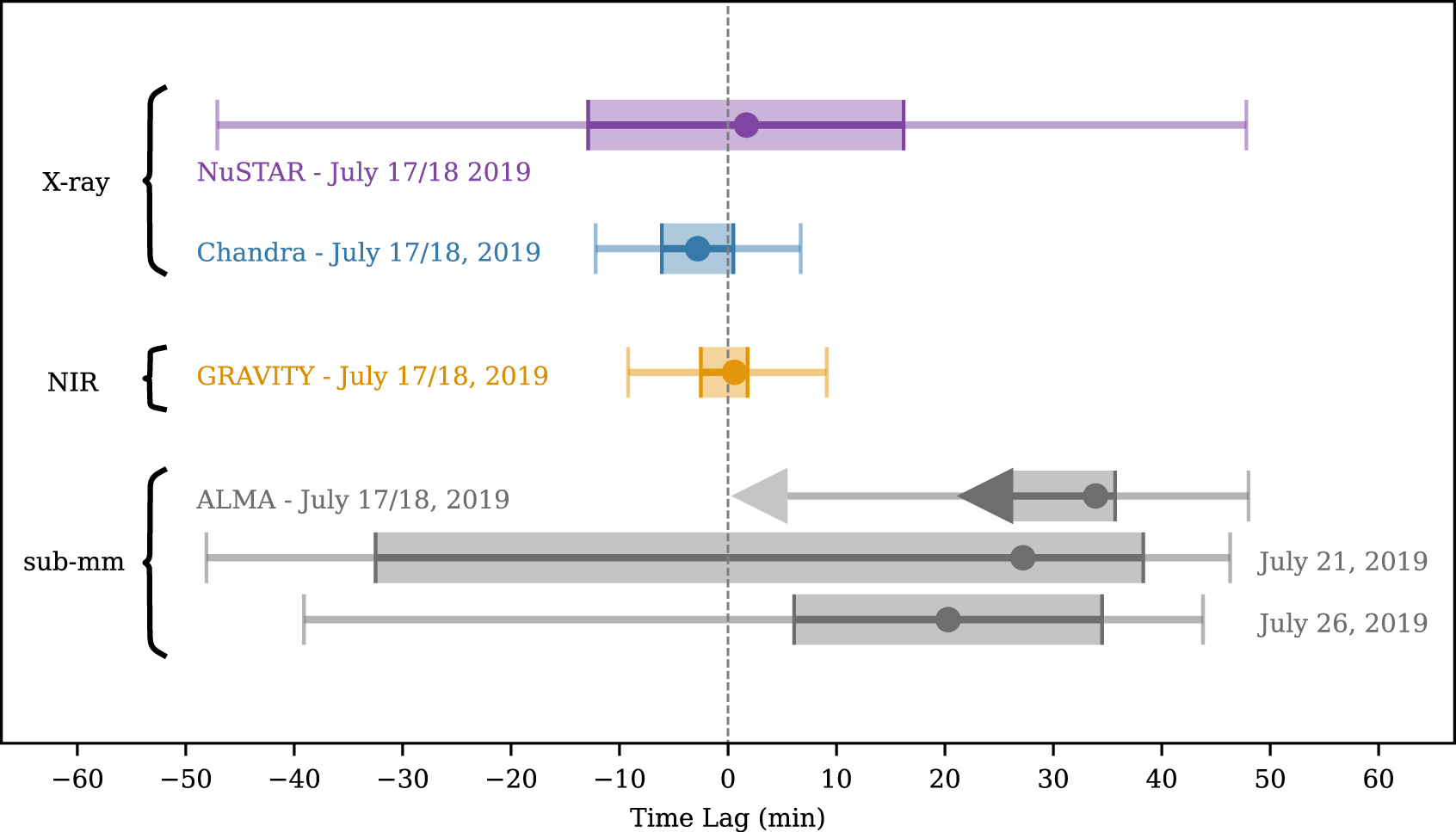}}
    \caption{Délais entre toutes les observations multi-longueurs d'onde et les courbes de lumière de \textit{Spitzer} $4,5\, \mu $m pour la campagne d'observation de Sgr~A* de juillet 2019. Les points violets, bleus, orange et gris représentent respectivement les décalages \textit{NuSTAR} 3-10 keV, \textit{Chandra} 2-8 keV, GRAVITY bande K ($2,2\, \mu $m) et ALMA 340 GHz. Les intervalles de confiance à 68$\%$ sont représentés par les cases ombrées et les intervalles à 99,7$\%$ sont représentés par les fines barres d'erreur. Comme le retard submillimétrique mesuré le 18 juillet est une limite supérieure, le pic du sursaut n'a pas été capturé. Crédit : \cite{Boyce2022}.}
    \label{fig:Boyce2022}
\end{figure}

L'aspect temporel entre les différentes longueurs d'onde n'est pas trivial. En effet, des études récentes~\cite{Michail2021, Boyce2022} utilisant des données ALMA entre 334,00 GHz et 347,87 GHz (radio), \textit{Spitzer} à $4,5 \, \mu$m (IR), GRAVITY à $2,2 \, \mu$m, \textit{Chandra} entre 2 et 8 keV (rayons~X mou) et \textit{NuSTAR} entre 3 et 10 keV (rayons~X mou) obtenues le 18 juillet 2019 (voir Fig.~\ref{fig:Michail2021a}) suggèrent une quasi-simultanéité entre le sursaut en rayons~X et celui en IR avec un décalage en temps $\delta_{2-8\, keV}^{4,5\, \mu m} = 4,67^{+4,07}_{-5,31}$ min (comme~\cite{Eckart2006a}) mais un décalage significatif entre le sursaut observé en IR et celui observé en radio avec un décalage de $\delta_{4,5\, \mu m}^{334\, GHz} = 21,48^{+3,44}_{-3,57}$ min comme illustré dans la Fig.~\ref{fig:Michail2021b}. Cependant, comme on peut le constater avec les panneaux de gauche de la Fig.~\ref{fig:Michail2021a}, présentant les données ALMA, le moment du pic d'émission IR n'est pas observé en radio, ce qui soulève la possibilité que le délai soit la conséquence d'un échantillonnage incomplet de la courbe de lumière. D'autres observations simultanées réalisées avec \textit{Spitzer} et ALMA ont des résultats similaires (voir Fig.~\ref{fig:Boyce2022}).

\begin{figure}
    \centering
    \resizebox{\hsize}{!}{\includegraphics{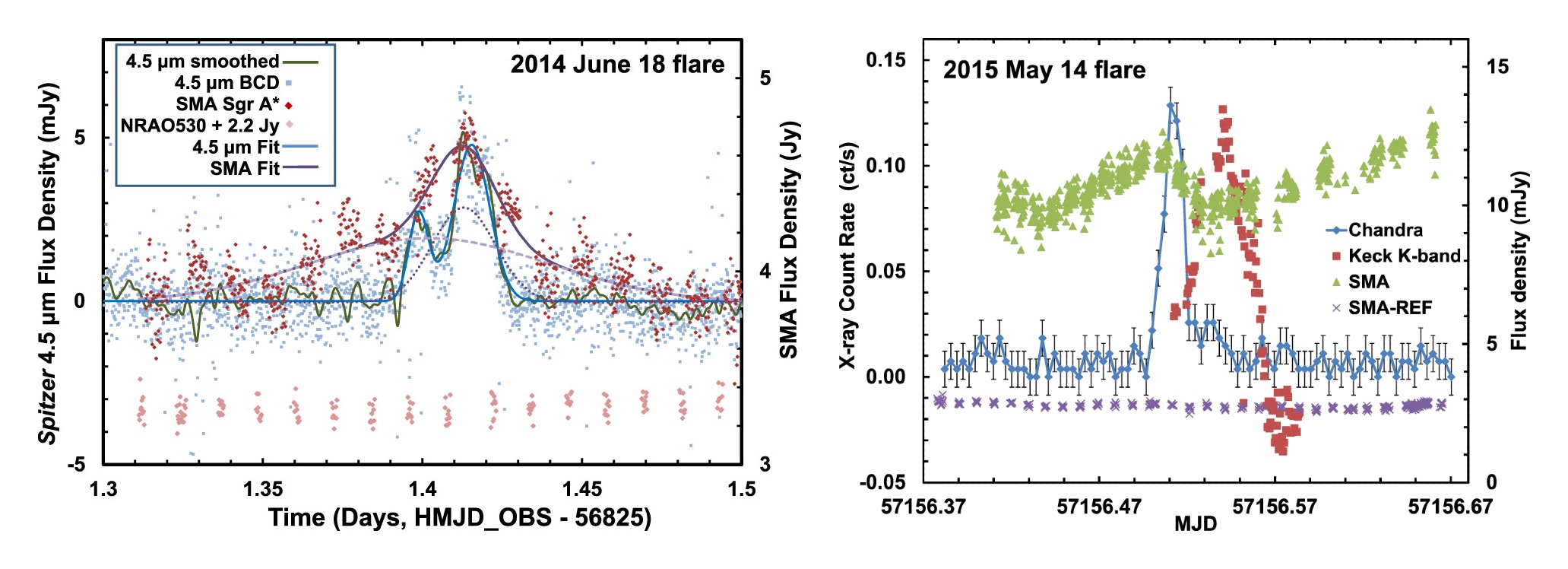}}
    \caption{\textbf{À gauche} : Observations conjointes d'un sursaut à double pic de Sgr~A* le 18 juin 2014 par \textit{Spitzer}/IRAC à $4,5\, \mu$m (points bleus et ligne lissée verte, échelle en ordonnée à gauche) et le SMA à 875 $\mu$m (points rouges, échelle en ordonnée à droite). La densité de flux du calibrateur du SMA (NRAO 530) (en rouge clair en bas) est de $\sim$1~Jy (une constante de 2,2~Jy a été ajoutée pour placer les données sur l'échelle de l'ordonnée de droite). La ligne bleue lissée est l'ajustement de la courbe à deux gaussiennes aux données de $4,5\, \mu$m. Les lignes violettes en tirets et en pointillés montrent deux courbes gaussiennes ajustant les données submillimétriques du SMA, et la ligne violette continue montre leur somme. \textbf{À droite} : Observations du 14 mai 2015 du sursaut à un seul pic de Sgr~A*. Les carrés rouges montrent les données du \textit{Keck} à 2,12 $\mu$m (échelle en ordonnée à droite), les triangles verts, les données du SMA à 1,32 mm (l'échelle est de 500$\times$ l'ordonnée à droite, c'est-à-dire que la densité de flux maximale est de $\sim$6~Jy), et les points bleus, les données de \textit{Chandra} à 2-8 keV (ordonnée à gauche). La courbe de lumière des rayons~X est additionnée sur des bins de 300~s, et les barres d'erreur de Poisson sur le taux de comptage des rayons~X sont représentés par des barres d'erreur noires. La densité de flux du calibrateur SMA (NRAO 530) ($\sim$1,4~Jy), dont l'échelle est la même que pour Sgr A*, est représentée par des croix violettes. Crédit : \cite{Fazio2018}.}
    \label{fig:Fazio2018}
\end{figure}

Cependant, la campagne d'observation de 2014-2015 réalisée avec \textit{Spitzer}, l'observatoire \textit{Keck}, \textit{Chandra} et le réseau SMA montre des résultats très différents et bouscule le schéma bien établi décrit précédemment~\cite{Fazio2018}. Durant cette campagne, deux sursauts ont été observés à plusieurs longueurs d'onde montrant chacun des propriétés temporelles inhabituelles. Le premier sursaut, illustré par le panneau de gauche dans la Fig.~\ref{fig:Fazio2018}, observé en radio avec SMA et en IR par \textit{Spitzer}, montre une simultanéité entre ces deux longueurs d'ondes. On note que la courbe de lumière IR est ajustée à l'aide de deux gaussiennes, tout comme la courbe de lumière en radio, mais avec des largeurs à mi-hauteur (FWHM) plus importantes en radio. Le second sursaut, illustré par le panneau de droite dans la Fig.~\ref{fig:Fazio2018}, quant à lui a été observé avec \textit{Chandra}, le \textit{Keck} et SMA. Les formes des courbes de lumière IR et~X sont similaires aux courbes de lumière habituelles, mais présentent un délai significativement plus important que dans les études précédentes de l'ordre de $\sim 38$ min. De plus, la contrepartie radio est aussi intrigante, avec une diminution du flux au moment du pic d'émission en rayons~X jusqu'au moment du pic d'émission IR où le flux radio recommence à monter. Plusieurs scénarios sont évoqués concernant la diminution radio. Dans le premier, cette diminution est associée à un sursaut radio précédent n'ayant aucun lien avec les sursauts IR et~X. Dans le second, la diminution du flux précédant l'augmentation du sursaut est le résultat d'un changement significatif dans le flot d'accrétion\footnote{Je remercie Mr Bart Ripperda pour les discussions sur ce sujet.} (une éjection d'une partie du disque ? Plus de détails dans le Chap.~\ref{chap:Plasmoid Flare model}).

\subsection{Polarisation en radio}
Bien que la campagne d'observation de la collaboration EHT pour imager Sgr~A* ne soit pas multi-longueurs d'onde, la variabilité de Sgr~A* a été un vrai défi pour produire l'image de la Fig.~\ref{fig:EHT_img}. Durant la campagne, un sursaut en rayons~X a été détecté en simultané des observations. Les sursauts étant un phénomène transitoire pouvant potentiellement altérer la dynamique du flot d'accrétion, la collaboration EHT a décidé d'exclure les données de cette période pour la construction de l'image de Sgr~A*. Cependant, avec le réseau d'antennes radio ALMA, le flux et la polarisation radio ont pu être mesurés au moment du sursaut en rayons~X~\cite{Wielgus2022}. La courbe de lumière et plus particulièrement les boucles de polarisation dans l'espace Q-U ont ainsi pu être modélisées par un point chaud ayant une configuration magnétique verticale, observé à faible inclinaison, orbitant autour du trou noir avec une vitesse Képlerienne, un rayon orbital de $\sim 10$ $r_g$, similaire aux résultats \cite{Gravity2018}. Plus de détails dans le Chap.~\ref{chap:ouvertures}.

\subsection{Indices spectraux et Densité Spectrale d'Énergie}
\begin{figure}
    \centering
    \resizebox{\hsize}{!}{\includegraphics{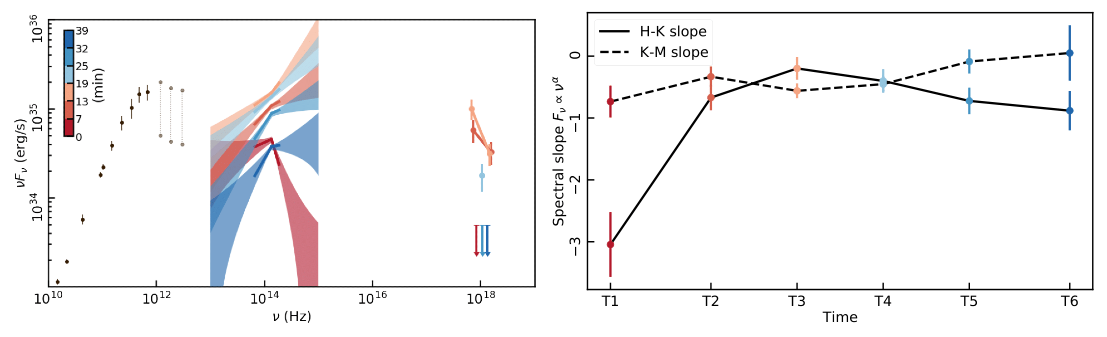}}
    \caption{\textbf{À gauche} : Évolution temporelle de la SED. La couleur indique le temps : du rouge foncé au bleu foncé, comme indiqué dans la barre de couleur. Pour deux pas de temps, le spectre des rayons~X peut être divisé en deux points (T2 et T3). Pour T4, une seule mesure du flux de rayons~X est possible. Les limites supérieures sont représentées pour T1, T5 et T6. Les mesures dans le proche infrarouge sont indiquées par des lignes épaisses, avec les incertitudes indiquées et extrapolées par la zone ombrée. La SED submillimétrique est tracée pour comparaison, les données radios et submillimétriques proviennent de \cite{Falcke1998, Bower2015, Bower2019, Brinkerink2015, Liu2016}. \textbf{À droite} : Pentes spectrales infrarouges $\alpha$ pour les six temps T1 à T6. La couleur indique le temps, du rouge foncé au bleu foncé. La ligne noire continue montre la pente H-K ; la ligne noire en pointillés montre la pente K-M. Crédit : \cite{Gravity2021}.}
    \label{fig:Gravity2021}
\end{figure}

Les observations multi-longueurs d'onde des sursauts permettent in-fine de déterminer leurs densités spectrales d'énergie (SED pour \textit{Spectral Energy Density)}. Cependant, les points de données sont très espacés et ne forment pas, comme pour l'état quiescent, un échantillon régulier. La SED des sursauts se construit à partir des mesures de flux dans une bande de fréquence, au mieux radio + NIR + rayons~X, qui eux-mêmes dépendent du temps. Le SED du sursaut du 18 juillet 2019 observé avec \textit{Spitzer}, GRAVITY en bande K et en bande H, \textit{Chandra} et \textit{NuSTAR}~\cite{Gravity2021} à différents temps est illustré dans le panneau de gauche de la Fig.~\ref{fig:Gravity2021}. Comme on peut le voir, les sursauts sont variables dans le temps, mais aussi du point de vue spectral, avec un indice spectral entre les bandes H-K et K-M qui dépendent du temps (voir panneau de droite de la Fig.~\ref{fig:Gravity2021}). Ces caractéristiques fréquentielles permettent d'envisager deux scénarios d'émission. Le premier est du rayonnement synchrotron non thermique en loi de puissance (voir Chap.~\ref{chap:modele_hotspot+jet}) avec une fréquence de coupure par refroidissement synchrotron (dépendante du temps) et le second est un rayonnement synchrotron combiné à l'effet Compton inverse, où les photons submillimétriques du rayonnement synchrotron (de basse énergie) extraient de l'énergie des électrons à haute température qui ont produit le rayonnement synchrotron, on parle alors de rayonnement \textit{Synchrotron Self Compton} (SSC). Les données du sursaut du 18 juillet 2019 ne permettent pas de distinguer les deux processus de rayonnement, cependant, la densité d'électrons nécessaire pour du SSC est de plusieurs ordres de grandeurs au-dessus de la densité estimée pour l'état quiescent de Sgr~A*, favorisant ainsi le scénario de synchrotron en loi de puissance (\textit{PLcool}).

\chapter{Un modèle analytique simple de Sagittarius~A*}\label{chap:modele_hotspot+jet}
\markboth{Un modèle analytique simple de Sagittarius~A*}{Un modèle analytique simple de Sagittarius~A*}
{
\hypersetup{linkcolor=black}
\minitoc 
}

\section{Historique des modèles de sursauts}
La première détection de sursauts de Sagittarius A* a eu lieu en 2001 \cite{Baganoff2001} en rayons~X puis en 2003 \citep{Genzel2003} en NIR. Depuis, de nombreux modèles ont été évoqués afin d'expliquer ces sursauts. Nous allons faire dans cette partie une petite revue, non exhaustive, des modèles qui ont été envisagés pour les sursauts de Sgr~A* ainsi que leurs prédictions comparées aux données récentes (principalement les résultats de GRAVITY, voir Chap.~\ref{chap:Sgr~A* flares}).
\subsection{Bruit rouge}
Dans ce modèle, le haut niveau de flux mesuré par un observateur lors d'un "sursaut" n'est pas lié à un évènement particulier distinct du processus d'émission de l'état quiescent, mais représente un cas extrême (potentiellement amplifié par les effets de la relativité) de la variabilité de l'état quiescent~\citep{Do2009}. En effet, en pratique, la plupart des processus astrophysiques ne sont pas complètement statiques et présentent une variabilité plus ou moins importante, même autour d'un état d'équilibre. Ainsi, dans un disque d'accrétion ayant atteint un état d'équilibre, il y a toujours des fluctuations de densité ou de température (pour ne citer que ces paramètres). Ces fluctuations vont se transmettre au flux émis par cette région du disque. Ces fluctuations, qui sont dues à un processus stochastique, sont souvent décrites par du bruit statistique qui peut être soit du bruit blanc, c'est-à-dire que le spectre de ce bruit est constant et ne dépend pas de la fréquence\footnote{Ici, la fréquence correspond à l'inverse de la période de la variation. Pour obtenir le spectre, on effectue la transformée de Fourrier du signal.}, soit, dans le cas qui nous intéresse ici, un bruit rouge, c'est-à-dire que la densité spectrale de puissance diminue lorsque la fréquence augmente ($P(f) \propto f^{-\alpha}$, avec $\alpha > 0$). 

\begin{figure}
    \centering
    \resizebox{0.6\hsize}{!}{\includegraphics{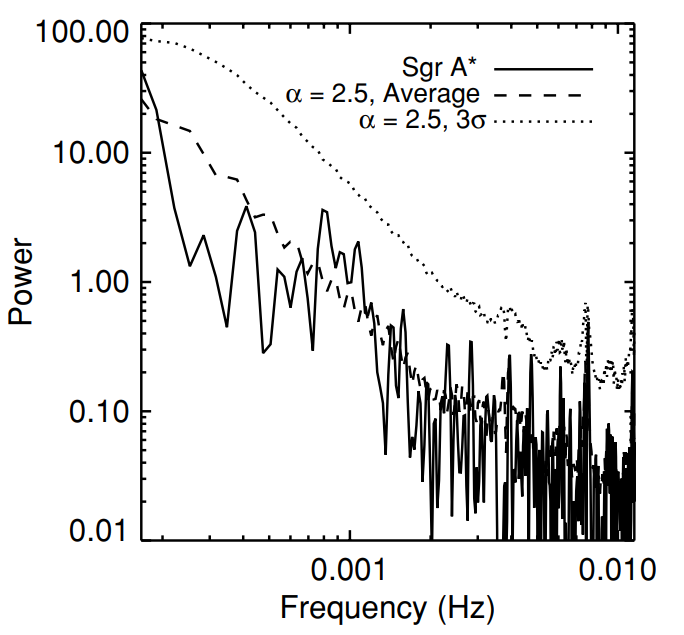}}
    \caption{Périodigramme de la courbe de lumière de Sagittarius A* (trait plein) ainsi que la puissance moyenne (tirets) et le seuil de 3$\sigma$ (pointillé) des simulations Monte Carlo d'un bruit rouge avec une pente de $\alpha=2,5$. Crédit: \cite{Do2019}.}
    \label{fig:périodigram_SgrA}
\end{figure}

\citet{Do2009} ont observé Sgr~A* durant 3 heures par nuit, durant 7 nuits entre Juillet 2005 et Août 2007 avec une cadence d'une minute, leur permettant d'obtenir le spectre de Sgr~A* pour des périodes entre deux minutes et une heure\footnote{Critères de Nyquist.}. Le spectre obtenu (voir Fig.~\ref{fig:périodigram_SgrA}) correspond à un spectre de type bruit rouge avec une pente pour la loi de puissance de $\alpha \approx 2,5$. Ce type de modèle et cette étude excluent la notion de sursauts puisque si la source des sursauts était liée à un phénomène physique distinct de l'état quiescent, un pic serait observé dans la Fig.~\ref{fig:périodigram_SgrA} à la fréquence correspondant à l'inverse de la période du sursaut qui selon \cite{Gravity2018} est de l'ordre de 30 min-1h ($2-5. 10^{-4}$ Hz). Bien que les paramètres de leurs observations permettent un critère de Nyquist compatible avec la détection de ce signal dans le périodigramme, la faible fréquence d'occurrence\footnote{À ne pas confondre avec la fréquence spectrale.} (le nombre moyen de sursauts observés, ayant un flux supérieur à $5$~mJy, par jour qui est de l'ordre de 4), et l'intervalle de temps d'observation par nuit, ne permettent pas de conclure de manière certaine. Une observation continue sur une large période comprenant un nombre significatif ($> 1$) de potentiels sursauts, typiquement de l'ordre de 24h, est nécessaire pour revendiquer la détection (ou la non détection) de ce signal dans le périodigramme. De telles observations sont impossibles avec des télescopes terrestres, mais largement dans les capacités du télescope spatial James Webb. Cependant, le mouvement orbital observé par \cite{Gravity2018} est difficilement compatible avec un processus stochastique comme le modèle de bruit rouge.

\subsection{Interaction étoile disque}
\begin{figure}
    \centering
    \resizebox{0.6\hsize}{!}{\includegraphics{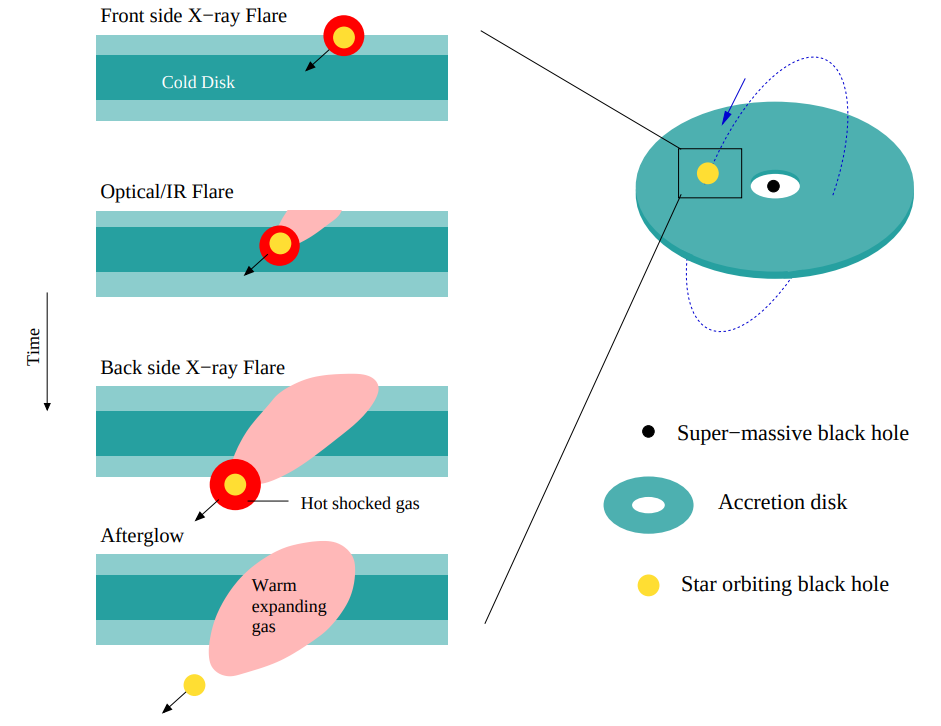}}
    \caption{Schéma d'une étoile en orbite autour d'un trou noir supermassif traversant le disque d'accrétion, chauffant le gaz et générant une onde de choc. Le résultat de cette interaction se manifestant par un sursaut en visible, IR et rayons~X. Crédit : \cite{Nayakshin2004}.}
    \label{fig:star_disk_interaction}
\end{figure}

Un mécanisme possible comme source pour les sursauts de Sgr~A* est une interaction entre le disque d'accrétion et une étoile comme décrit par \cite{Nayakshin2004} et schématisé dans la Fig.~\ref{fig:star_disk_interaction}. Le scénario est le suivant : le flot d'accrétion de Sgr~A* est modélisé par un disque très étendu ($R_{max} > 10^4\, r_s$) alimenté par le vent des (nombreuses) étoiles environnantes. On distingue deux régions pour le disque, la région interne (bleu foncé dans la Fig.~\ref{fig:star_disk_interaction}) optiquement épaisse et la région externe (bleu clair dans la Fig.~\ref{fig:star_disk_interaction}) optiquement mince. On considère qu'il existe un groupe d'étoiles de faibles masses (peu lumineuses) en orbite autour du trou noir dont la trajectoire traverse le disque (à deux reprises). Lors du passage d'une étoile à travers le disque, l'étoile va chauffer le gaz en créant une onde de choc. Le gaz "choqué" va donc rayonner et émettre un sursaut en rayons~X (dû principalement au choc). Lorsque l'étoile, ou plus exactement l'onde de choc, est dans la zone optiquement épaisse ou derrière (par rapport à l'observateur), le sursaut en rayons~X est absorbé, ce qui a pour effet de chauffer le gaz environnant qui va rayonner en visible et IR, créant aussi un sursaut dans ces longueurs d'ondes. 

\cite{Nayakshin2004} a estimé le nombre d'interactions étoile-disque par jour à $\sim 3$ à partir du profil de densité stellaire estimé dans \cite{Genzel2003} et de la taille du disque \citep[aussi contraint par la population stellaire du parsec central du centre galactique;][]{Nayakshin2004}. On note que cette estimation de la fréquence des sursauts est comparable aux observations. De plus, comme on l'a vu au Chap.~\ref{chap:Sgr~A* flares}, les sursauts observés à plusieurs longueurs d'ondes ont des propriétés temporelles très variées avec, pour certains sursauts, une simultanéité entre rayons~X et IR, et d'autres avec un délai entre ces deux longueurs d'ondes. Ce modèle d'interaction étoile-disque comme origine des sursauts permet naturellement d'expliquer les délais (positifs ou négatifs selon la position de l'observateur par rapport à la direction de propagation de l'étoile) entre rayons~X et les autres longueurs d'onde à travers le régime optique du disque à une longueur d'onde donnée (optiquement épais ou optiquement mince).

Ces interactions, par principe, peuvent avoir lieu n'importe où dans le disque qui s'étend jusqu'à une très grande distance du trou noir ($R_{max} > 10^4\, r_s$), ce qui permet d'avoir une fréquence de sursauts comparable aux observations. En effet, si le disque est moins étendu que la limite estimée précédemment, il y a moins d'étoiles dont l'orbite traverse le disque et donc moins de sursauts. Il existe aussi une limite inférieure sur le rayon pour l'existence d'étoiles capables de générer ces interactions. En effet, en dessous d'une certaine limite $R_{TDE}$, les forces\footnote{Il s'agit d'un abus de langage puisque la gravité n'est pas une force.} de marée du trou noir seront supérieures à la force de cohésion de l'étoile (sa propre gravité) ce qui a pour effet de "déchirer" l'étoile, une partie de sa masse (voire la totalité) étant accrétée par le trou noir dans son disque d'accrétion. On parle alors \textit{d'évènement de rupture par effet de marée} ou \textit{Tidal Disruption Event} (TDE) en anglais. Ce rayon limite dépend à la fois de la masse du trou noir $M_{BH}$ mais aussi de celle de l'étoile $M_\star$ et de son rayon $R_\star$ \citep{Rees1988,Gezari2014}
\begin{equation}
    R_{TDE} \approx R_\star \left( \frac{M_{BH}}{M_\star} \right)^{1/3}.
\end{equation}
Pour une étoile comme le soleil autour de Sgr~A*, cette limite est de $R_T\sim 9\, r_s = 18\, r_g$, du même ordre de grandeur que le rayon orbital des sursauts observés par GRAVITY. Le rayon limite d'étoiles moins massives que le Soleil (donc plus petites) est plus proche de l'horizon des évènements et donc comparable aux observations, rendant ce scénario tout à fait crédible. Le mouvement orbital observé est l'orbite du point d'impact dans le disque autour du trou noir avec le reste du flot d'accrétion. Cependant, pour un disque d'accrétion, on s'attend à une vitesse orbitale Képlérienne ou sub-Keplérienne or les observations GRAVITY requièrent une vitesse super-Képlérienne comme discuté dans le Chap.~\ref{chap:Sgr~A* flares}.

\subsection{Instabilités dans le disque}
\begin{figure}
    \centering
    \resizebox{0.6\hsize}{!}{\includegraphics{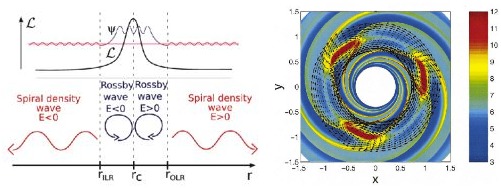}}
    \caption{(\textbf{Droite}) Vue schématique d'une RWI avec les deux régions de propagation pour les ondes de Rossby, et la formation de vortex au centre. (\textbf{Gauche}) Simulations hydrodynamiques 2D de vortex de Rossby dans un disque, adaptées de Li et al (2001) pour $m = 3$. La couleur code la pression (en unités de $10^{-3}p_0$). Les flèches indiquent le modèle d'écoulement près de $r_0$ dans un cadre mobile se déplaçant à la vitesse $u_\varphi(r_0)$. Les vortex sont anticycloniques et entourent des régions de hautes pressions. Des spirales à grande échelle sont également produites en liaison avec les vortex. Crédit : \cite{Lovelace2014}.}
    \label{fig:RWI_Lbump}
\end{figure}

Bien que difficilement détectable en IR, Sgr~A* ou plus précisément son flot d'accrétion a un pic d'émission dans le domaine submillimétrique (Fig.~\ref{fig:quiescent_spectrum}). Comme dit pour le précédent modèle, le disque d'accrétion est alimenté par le vent stellaire des étoiles massives en orbite autour du trou noir (voir panneau du milieu à gauche de la Fig.\ref{fig:GC_scales}). Même en état d'équilibre global, un certain nombre d'instabilités peuvent apparaître dans le disque d'accrétion. On définit le profil de vorticité potentielle inverse $\mathcal{L}$ tel que
\begin{equation}
    \mathcal{L} = \frac{k^2}{2 \Omega \Sigma}
\end{equation}
où $\Omega$ est la fréquence de rotation, $\Sigma$ la densité surfacique du disque et $k$ la fréquence épi-cyclique. Si $\mathcal{L}$ admet un extremum (une bosse comme illustré Fig.~\ref{fig:RWI_Lbump}), une instabilité comparable à l'instabilité de Kelvin-Helmholtz, appliquée aux disques d'accrétion, nommée \textit{instabilité par ondes de Rossby} (RWI) en anglais, se développe en $r_0$ le plus souvent proche de l'ISCO. Une fois déclenchée, cette instabilité se manifeste par des spirales de surdensité et des vortex de Rossby où la densité augmente exponentiellement. Le nombre de vortex formés $m$ dépend des conditions du disque~\citep{Tagger2006}.

Dans le cas d'un rayonnement synchrotron (voir Section~\ref{sec:synchrotron} pour plus de détails), le flux émis est proportionnel à la densité. Donc, dans ce type de modèle, les sursauts observés sont le résultat de la formation de ces surdensités dans le disque.

\subsection{Point chaud}
Contrairement aux modèles précédents basés sur des phénomènes physiques particuliers, le modèle dit de point chaud est, dans la grande majorité des cas, analytique. Certains modèles liés à un phénomène physique, comme les modèles de plasmoïdes issus de la reconnexion magnétique (voir Chapitre~\ref{chap:Plasmoid Flare model}) peuvent néanmoins être assimilés à des points chauds, car très similaires en première approximation. Le principe de base est relativement simple, on considère une région sphérique, potentiellement déformable par rotation différentielle comme dans \cite{Gravity2020b}, ayant un mouvement orbital autour du trou noir. On distingue deux types de mouvements, une orbite, la plupart du temps Képlérienne, dans le plan équatorial \citep{Hamaus2009,Gravity2018} ou une éjection dans la magnétosphère (hors du disque et du jet s'il existe) comme dans \cite{Vincent2014} ou \cite{Ball2021}. Comme il s'agit de modèles (la plupart du temps) agnostiques, la vitesse orbitale peut être un paramètre libre du modèle. L'objectif de ce genre de modèle est d'ajuster les observations avec un nombre limité de paramètres libres afin d'explorer l'espace des paramètres avec un temps de calcul raisonnable.

Les premiers modèles de point chaud ont une émission intrinsèque (dans le référentiel de l'émetteur) constante \cite{Genzel2003, Broderick2006, Vincent2014, Gravity2018, Gravity2020b}. La variabilité observée étant due aux effets relativistes comme le beaming (plus de détails dans \ref{sec:synchrotron}), l'effet Doppler relativiste (voir \ref{sec:effet_relat_LC}) et du processus d'émission (corps noir, synchrotron ou loi de puissance). Cependant, les modèles de point chaud récents~\cite{Hamaus2009,von_Fellenberg2023, Aimar2023} incluent une variabilité intrinsèque, c'est le cas du premier modèle considéré dans cette thèse, détaillé dans la section suivante.

\section{Modèle de point chaud variable}
\subsection{Motivations physiques} \label{sec:motivation}
Dans cette partie, on présente un modèle dit de point chaud reprenant l'idée de base de \cite{Hamaus2009}, à savoir une sphère en orbite autour du trou noir émettant du rayonnement. Comme on l'a vu au Chap.~\ref{chap:Sgr~A* flares}, les sursauts de Sgr~A* observés par GRAVITY à $2,2$ $\mu$m présentent un mouvement elliptique, quasi circulaire. \cite{Gravity2018} ont utilisé un modèle de point chaud avec une émission constante, en orbite circulaire dans le plan équatorial du trou noir, pour modéliser l'astrométrie. 
L'aspect très circulaire de l'orbite, ainsi que l'indépendance de la position dans le ciel des sursauts au moment de leur maximum d'émission, permettent de contraindre l'inclinaison à une valeur inférieure à 25°. En effet, à faible inclinaison, les effets relativistes comme le beaming et l'effet de lentille vont moins impacter la courbe de lumière observée (d'un facteur inférieur à 1,5 \citep{Gravity2018}) qu'à forte inclinaison où ces effets dominent entièrement la courbe de lumière \citep{Hamaus2009} (voir Fig.~\ref{fig:i_var_star}). La variabilité observée, bien qu'affectée par les effets relativistes, est dominée par la variabilité intrinsèque. Le temps caractéristique de refroidissement par rayonnement synchrotron $t_\mathrm{cool}$
\begin{equation}
    t_\mathrm{cool} = 15 \times \left(\frac{B}{20 \mathrm{G}} \right)^{-1,5} \left(\frac{\lambda}{2,2 \mu\mathrm{m}} \right)^{0,5} \mathrm{min}
\end{equation}
est de l'ordre de 15 min à $2,2$ $\mu$m, pour un champ magnétique de 20 G, qui est l'ordre de grandeur attendu pour Sgr~A*. Il est très comparable au temps caractéristique de la variabilité des sursauts qui est de l'ordre de $\sim 20-30$ min. De plus, la période de la boucle de rotation de la polarisation observée est compatible avec une émission synchrotron avec un champ magnétique poloidal (plus de détails sur la polarisation dans le Chap.~\ref{chap:Polarization}). Ainsi, l'étude d'un modèle de point chaud orbitant autour du trou noir et émettant du rayonnement synchrotron avec une variabilité intrinsèque apparaît comme naturelle au vu des résultats de~\cite{Gravity2018}.

Les objectifs d'un tel modèle sont les suivants, premièrement, améliorer le modèle de point chaud de \cite{Gravity2018} en ajoutant une émission réaliste, le rayonnement synchrotron, et en considérant une variabilité intrinsèque tout en gardant la simplicité d'une orbite Képlérienne dans le plan équatorial. Deuxièmement, étudier l'impact sur les observables, à savoir l'astrométrie et la courbe de lumière\footnote{Jusqu'à présent seule l'astrométrie des sursauts de Sgr~A* observée par GRAVITY a été ajustée.}, de la variabilité intrinsèque. Pour cela, on génère grâce à \textsc{GYOTO} une série d'images à plusieurs temps d'observation avec un certain nombre de pixels $N$ de l'environnement proche de Sgr~A* composées de différentes sources (dont le point chaud qui nous intéresse ici). Pour construire la courbe de lumière, on calcule, pour chaque temps d'observation $t_{obs}$, le flux total observé comme suit
\begin{equation}\label{eq:calcul_luminosité}
    F_\nu (t_{obs}) = \sum_j^{N} \sum_i^{N} I_\nu (i,j,t_{obs}) \, \delta
\end{equation}
avec $I_\nu (i,j,t_{obs})$ l'intensité spécifique calculée par \textsc{GYOTO} au pixel $(i,j)$ et $\delta=(\theta/N))^2$ l'élément d'angle solide par pixel où $\theta$ est le champ de vue en radian. Les coordonnées du centroïde de l'image ($X(t_{obs})$,$Y(t_{obs})$) correspondant au point astrométrique au temps $t_{obs}$ sont calculées de la manière suivante
\begin{subequations}\label{eq:calcul_centroïde}
\begin{align}
    X(t_{obs}) = \sum_j^{N} \sum_i^{N} \left(i-\frac{N-1}{2} \right) \ \zeta \times \frac{I_\nu (i,j,t_{obs}) \, \delta}{F_\nu} \\
    Y(t_{obs}) = \sum_j^{N} \sum_i^{N} \left(j-\frac{N-1}{2} \right) \ \zeta \times \frac{I_\nu (i,j,t_{obs}) \, \delta}{F_\nu}
\end{align}
\end{subequations}
où $\zeta=\theta/N$ est l'angle (ascension droite ou déclinaison) par pixel que l'on convertit en $\mu \mathrm{as} /\text{pix}$ pour comparer aux données GRAVITY.

%

\subsection{Processus d'émission : le rayonnement synchrotron}\label{sec:synchrotron}
\subsubsection{Formules générales}
\begin{figure}
    \centering
    \resizebox{0.5\hsize}{!}{\includegraphics{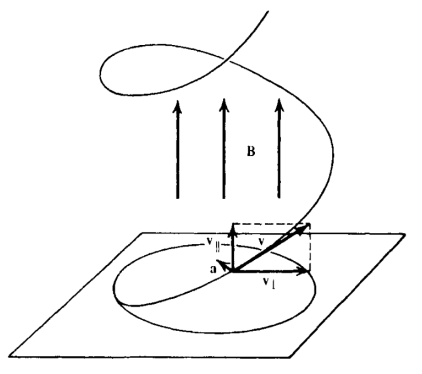}}
    \caption{Illustration de la trajectoire hélicoïdale d'un électron autour d'une ligne de champ magnétique. Crédit : \cite{RL86}}
    \label{fig:electron_trajectory}
\end{figure}

Pour toute la suite de cette thèse, sauf indication contraire, on considère un rayonnement d'origine synchrotron. Les expressions de cette partie seront exprimées dans le système d'unité CGS (voir Annexe~\ref{ap:Units} pour la conversion en unité SI). Ce rayonnement est un cas particulier du rayonnement bremsstrahlung qui décrit le rayonnement issu d'une particule chargée subissant une accélération. Dans le cas du synchrotron, il s'agit la plupart du temps d'électrons\footnote{Le rayonnement émis étant inversement proportionnel à la masse de la particule au carré, le rayonnement d'un proton est donc $\sim 2000^2$ fois plus faible que celui d'un électron dans les mêmes conditions.} (ou positrons) subissant une accélération due à la présence d'un champ magnétique comme illustré dans la Fig.~\ref{fig:electron_trajectory}. La quantité infinitésimale d'énergie $dE$ émise par un électron (ou positron\footnote{La charge est toujours au carré donc les deux ont la même influence.}) de charge $e$ durant l'intervalle de temps $dt$ dans l'angle solide $d\Omega$ autour de la direction d'émission $\vec{k}$ s'exprime grâce à la \textit{formule de Larmor} qui, dans le cas non relativiste, s'écrit~\citep{RL86}
\begin{equation}\label{eq:Larmor}
    \frac{dE}{d\Omega dt} = \frac{e^2 a^{\prime 2}}{4 \pi c^3} \sin^2 \Theta
\end{equation}
avec $a^\prime$ l'accélération subie par l'électron dans son référentiel propre à l'instant $t$\footnote{L'accélération dans le référentiel propre de l'électron et dans le référentiel de l'observateur $a$ sont liée tel que $a^\prime = \gamma^2 a$~\citep{RL86}.} et $\Theta$ l'angle entre cette dernière et la direction de l'émission ($\Theta=\frac{\vec{k} \cdot \vec{a}}{\vert \vert \vec{k} \vert \vert \cdot \vert \vert \vec{a} \vert \vert}$). On constate ainsi que ce rayonnement appelé rayonnement cyclotron (car vitesse non relativiste\footnote{Le rayonnement synchrotron est le cas relativiste du rayonnement cyclotron.}) n'est pas isotrope et dépend fortement de la direction de propagation via $\sin^2 \Theta$. On constate aussi que la puissance rayonnée est proportionnelle au carré de l'accélération, laquelle dans le cas du rayonnement cyclotron est liée à la force de Lorentz $m \vec{a} = q(\vec{v} \times \vec{B})$ (dans le cas non relativiste).

\begin{figure}
    \centering
    \resizebox{0.4\hsize}{!}{\includegraphics{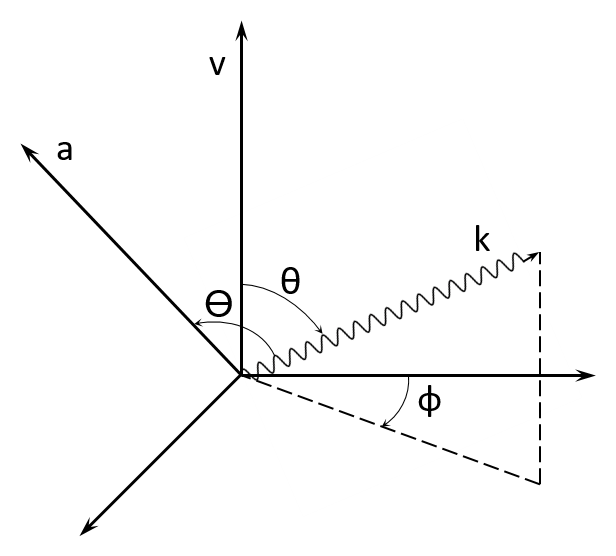}}
    \caption{Schéma détaillant la définition des angles dans les formules de Larmor à partir des vecteurs vitesse, accélération et direction d'émission.}
    \label{fig:angle_emission}
\end{figure}

\begin{figure}
    \centering
    \resizebox{0.7\hsize}{!}{\includegraphics{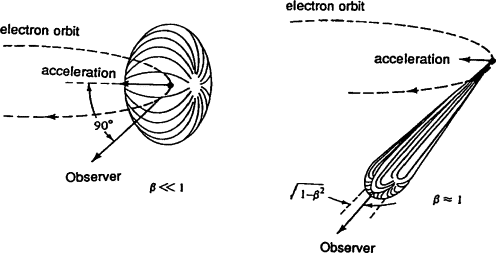}}
    \caption{Illustration de l'effet de beaming, c'est-à-dire la focalisation du rayonnement émis par une particule relativiste dans sa direction de propagation. Crédit: \cite{Johnson1997}.}
    \label{fig:beaming}
\end{figure}

Cependant, les sources qui nous intéressent pour cette thèse sont des sources thermiques, avec une température très élevée, c'est-à-dire que la vitesse thermodynamique des particules est relativiste ($\beta \sim 1$, $\gamma \gg 1$), ou des sources qui font appel à des processus d'accélération générant une population de particules dite non thermique (voir section~\ref{sec:distributions}), à des énergies très élevées ($\gamma > 100$). Ainsi, on va s'intéresser à la version relativiste de la formule de Larmor avec une accélération perpendiculaire à la vitesse, c'est à dire du rayonnement synchrotron (version relativiste du rayonnement cyclotron). La quantité infinitésimale d'énergie émise par une charge ayant une vitesse relativiste dans l'angle solide autour de la direction d'émission $\vec{k}$ s'écrit
\begin{equation} \label{eq:Larmor_relat}
    \frac{dE}{d\Omega dt} = \frac{e^2 a^2}{4 \pi c^3} \frac{1}{(1 - \beta \cos \theta)^4} \left( 1 - \frac{\sin^2 \theta \cos^2 \phi}{\gamma^2(1-\beta \cos \theta)^2} \right)
\end{equation}
avec $a$ l'accélération orthogonale à la vitesse, $\theta$ l'angle entre le vecteur vitesse de la particule et la direction d'émission et $\phi$ l'angle entre cette dernière et la normale du plan ($\vec{a}$,$\vec{v}$) comme illustré dans la Fig.~\ref{fig:angle_emission}. On constate à partir de l'Eq.~\eqref{eq:Larmor_relat} que l'émission est maximale dans la direction de la vitesse de la particule. En effet, pour $\theta$ proche de zéro, $\sin^2 \theta \to 0$ et donc la puissance émise est maximale. Cet effet, à savoir que l'émission d'une particule chargée relativiste est focalisée dans la direction de la vitesse de la particule, est connu sous le nom de \textit{beaming}, et illustré par la Fig.~\ref{fig:beaming}. Cette focalisation/concentration est d'autant plus importante que le facteur de Lorentz de la particule est élevé.

En résolvant l'équation du mouvement (relativiste) pour un électron/positron se déplaçant dans un champ magnétique ambiant $B$, on obtient
\begin{equation}\label{eq:acceleration}
    \begin{aligned}
        a_\perp &= \frac{e v_\perp B}{\gamma m_e c}, \\
        a_\parallel &= 0
    \end{aligned}
\end{equation}
où $\parallel$ et $\perp$ marquent respectivement les composantes parallèle et perpendiculaire à la direction du champ magnétique ambiant. En intégrant l'Eq.~\eqref{eq:Larmor_relat} et en injectant l'Eq.~\eqref{eq:acceleration}, on obtient la puissance totale émise par un électron/positron dans un champ magnétique ambiant $B$
\begin{equation}\label{eq:puissance_totale}
    \frac{dE}{dt} = \frac{2}{3} r_0^2 c \beta^2_\perp \gamma^2 B^2
\end{equation}
avec $r_0 = e^2/mc^2$ le rayon classique de l'électron. Il est intéressant de noter que la puissance émise est proportionnelle au carré du facteur de Lorentz de la particule. Ainsi, les électrons les plus énergétiques émettront le plus de rayonnement. Dans la plupart des systèmes astrophysiques, on considère des distributions d'électrons isotropes, c'est-à-dire avec aucune direction privilégiée. Par conséquent, la valeur de $\beta_\perp$ va fortement varier en fonction des électrons de la distribution considérée. En définissant l'angle $\alpha$ appelé \textit{pitch angle} en anglais, entre la direction du champ magnétique B et le vecteur vitesse de l'électron, on peut calculer la puissance moyenne émise d'un électron de ce genre de distribution comme
\begin{equation}\label{eq:puissance_synchrotron_moy}
\begin{aligned}
    \langle \frac{dE}{dt} \rangle &= \frac{2}{3} r_0^2 c \langle \beta^2_\perp \rangle \gamma^2 B^2 \\
    &= \frac{2}{3} r_0^2 c \gamma^2 B^2 \frac{\beta^2}{4 \pi} \int \sin^2 \alpha d\Omega\text{'}\\
    &= \frac{4}{9} r_0^2 c \beta^2 \gamma^2 B^2 \\
    & = \frac{4}{3} \sigma_T c \beta^2 \gamma^2 U_B
\end{aligned}
\end{equation}
où $\sigma_T=8 \pi r_0^2 /3$ est la section efficace de Thomson et $U_B=B^2/8 \pi$ la densité d'énergie magnétique.
\textbf{Attention, il ne faut pas confondre l'angle solide du vecteur vitesse par rapport à la direction d'émission $d\Omega$ précédent, comme dans l'Eq.~\eqref{eq:Larmor_relat}, et celui par rapport au champ magnétique $d\Omega^\prime$ dans l'Eq.~\eqref{eq:puissance_synchrotron_moy}}.

\begin{figure}
    \centering
    \resizebox{0.6\hsize}{!}{\includegraphics{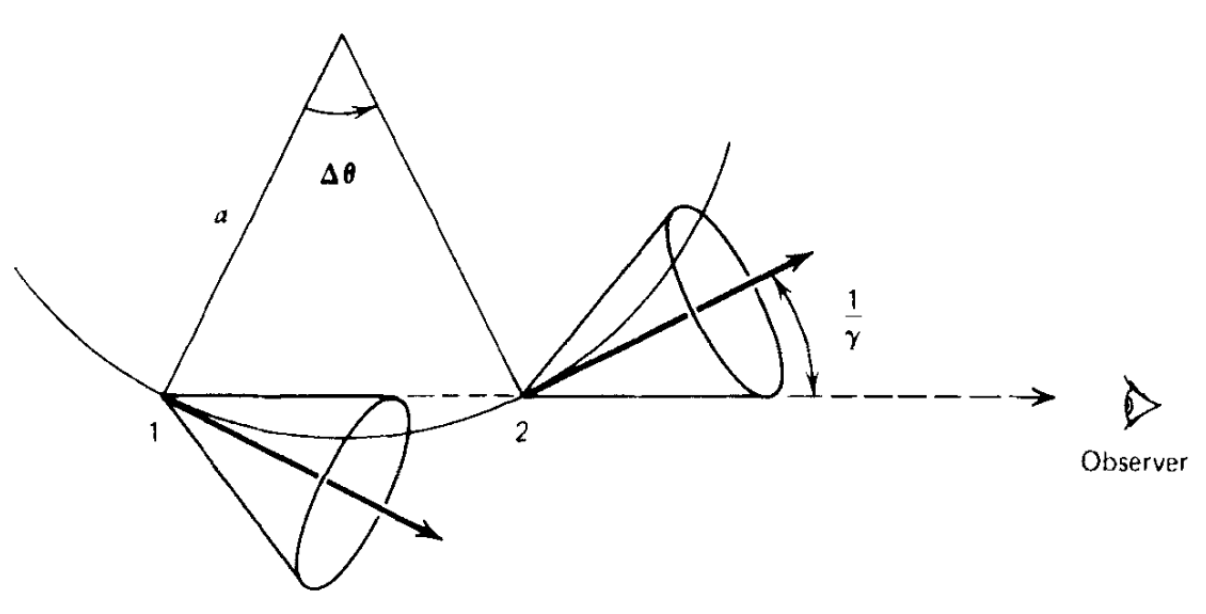}}
    \caption{L'observateur reçoit du rayonnement synchrotron uniquement lorsque l'électron est entre les positions 1 et 2, dû à la composante circulaire de son mouvement et de l'effet du beaming. Crédit : \cite{RL86}.}
    \label{fig:pulse_synchrotron}
\end{figure}

\begin{figure}
    \centering
    \resizebox{0.5\hsize}{!}{\includegraphics{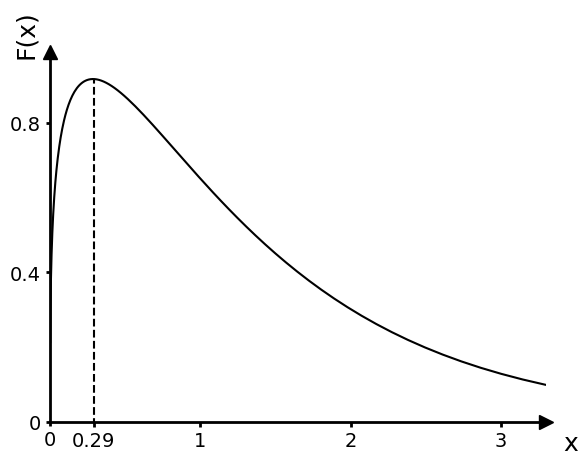}}
    \caption{La fonction F(x) telle que définie par l'Eq.~\eqref{eq:F(x)}.}
    \label{fig:F(x)}
\end{figure}

On a ainsi déterminé la puissance totale moyenne émise par un électron en fonction de son énergie $\gamma$ et de la densité d'énergie magnétique $U_B$. Il s'agit ici d'une quantité bolométrique, or les observations sont restreintes à une bande de fréquences limitée. Il est donc plus intéressant d'exprimer la puissance (moyenne) émise par unité de fréquence. Le mouvement hélicoïdal combiné à l'effet de beaming, illustré dans les Fig.~\ref{fig:electron_trajectory} et \ref{fig:beaming}, a pour conséquence qu'un observateur statique va observer une série de pulsations lorsqu'il se trouve (approximativement) dans la direction de propagation de l'électron (voir Fig.~\ref{fig:pulse_synchrotron}). Lorsque l'observateur est en dehors de celle-ci, il ne reçoit aucun flux. La durée des pulsations dépend de l'angle d'ouverture de l'émission qui ne dépend que de $\gamma$. La fréquence de répétition de ces pulsations correspond à la fréquence "orbitale" de l'électron autour de la ligne de champ magnétique qui dépend de l'intensité de ce dernier $B$ et du facteur de Lorentz $\gamma$ de l'électron. On appelle cette fréquence la \textit{fréquence critique} qui s'exprime de la manière suivante
\begin{equation}\label{eq:critical_freq}
    \nu_{crit} = \frac{3}{2} \gamma^2 \nu_{cyclo} \sin \alpha
\end{equation}
où
\begin{equation}
    \label{eq:cyclotron_freq}
    \nu_{cyclo} = \frac{eB}{2 \pi m_e c}.
\end{equation}
Cependant, tout le rayonnement observé n'est pas à cette unique fréquence $\nu_{crit}$, \cite{RL86} ont démontré qu'à partir de ce principe et de l'Eq.~\eqref{eq:puissance_totale}, la puissance émise par unité de fréquence s'écrit
\begin{equation}\label{eq:synchrotron_exact}
    \frac{dE}{dt d\nu} = \sqrt{3}\frac{e^3 B \sin \alpha}{m_e c^2} F \left( \frac{\nu}{\nu_{crit}} \right)
\end{equation}
avec
\begin{equation}\label{eq:F(x)}
    F(x) = x \int_x^\infty K_{5/3}(u) du
\end{equation}
avec $K_{5/3}$ une fonction de Bessel modifiée. La courbe $F(x)$ est présentée dans la Fig.~\ref{fig:F(x)}, on constate bien que la puissance émise dépend fortement de la fréquence avec un pic d'émission à $\nu_{peak} = 0,29\ \nu_{crit}$. La fonction, donc la puissance émise, chute rapidement et tend vers zéro au-delà de $x=1$, c'est-à-dire au-delà de la fréquence critique. 

On a ainsi déterminé la puissance du rayonnement synchrotron émis par unité de fréquence d'un électron unique plongé dans un champ magnétique. De toute évidence, dans n'importe quelle source astrophysique, il y a un grand nombre d'électrons avec une certaine distribution de vitesse qui, si on la suppose isotrope, se résume à une distribution à une dimension, à savoir la vitesse que l'on exprime le plus souvent via le facteur de Lorentz $N_e(\gamma) = d n_e/d\gamma$. Le coefficient d'émission synchrotron $j_\nu$ [erg s$^{-1}$ cm$^{-3}$ sr$^{-1}$ Hz$^{-1}$] à la fréquence $\nu$ correspond à l'intégrale sur toute la distribution de la puissance émise par unité de fréquence de l'Eq.~\eqref{eq:synchrotron_exact} par unité d'angle solide $d\Omega$ et s'écrit
\begin{equation}\label{eq:emission_synchrotron_exact}
    j_\nu = \frac{dE}{dt d\nu dV d\Omega}  = \frac{\sqrt{3}}{4 \pi} \frac{e^3 B sin(\alpha)}{m_e c^2} \int_0^\infty N_e(\gamma) F \left( \frac{\nu}{\nu_{crit}} \right) d\gamma.
\end{equation}
Le facteur $1/4\pi$ vient de notre hypothèse d'une distribution isotrope de vitesse.

On définit aussi le coefficient d'absorption synchrotron $\alpha_\nu$ [cm$^{-1}$] qui englobe les effets d'absorption et diffusion du rayonnement par cette même distribution d'électrons par \cite{Ghisellini1991}
\begin{equation}\label{eq:absorption_synchrotron_exact}
    \alpha_\nu = -\frac{\sqrt{3}}{8\pi m_e \nu^2} \frac{e^3 B sin(\alpha)}{m_e c^2} \int_0^\infty \frac{N_e(\gamma)}{\gamma (\gamma^2-1)^{1/2}} \frac{d}{d\gamma} \left[\gamma (\gamma^2-1)^{1/2} F \left( \frac{\nu}{\nu_{crit}} \right) \right].
\end{equation}

\subsubsection{Distributions classiques d'électrons}\label{sec:distributions}
\begin{figure}
    \centering
    \resizebox{0.5\hsize}{!}{\includegraphics{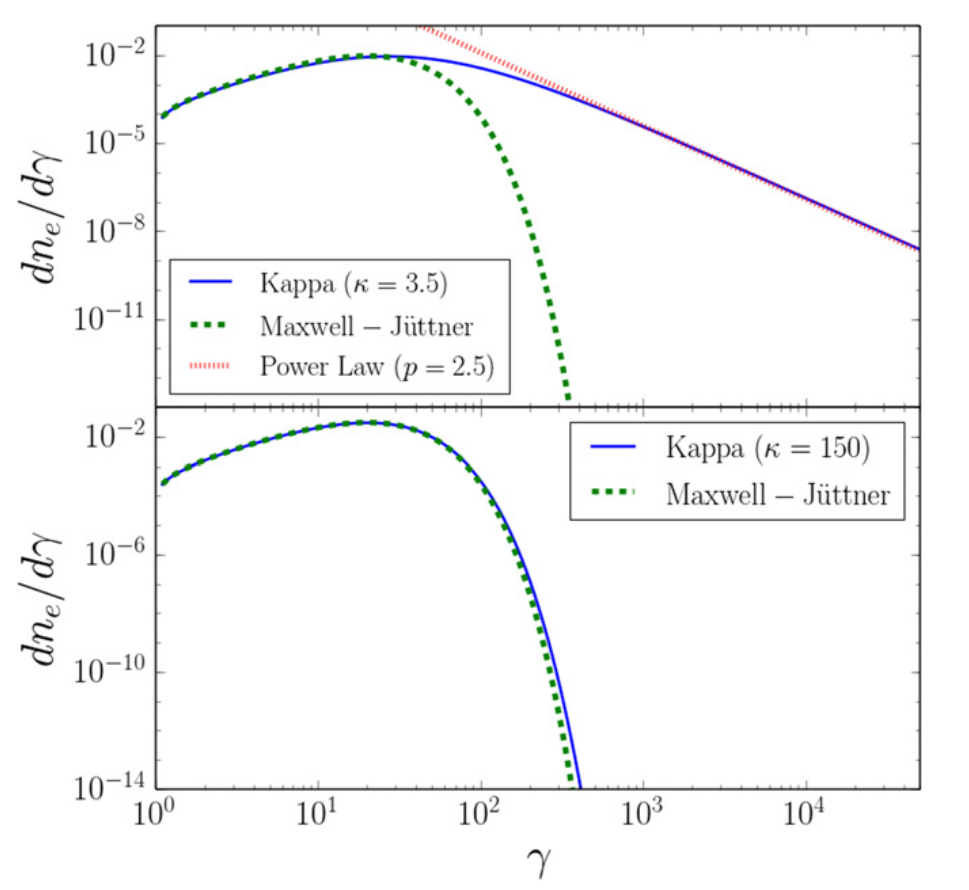}}
    \caption{Illustration des trois types de distributions usuelles (thermique, loi de puissance, et Kappa). Crédit : \cite{Pandya2016}}
    \label{fig:distributions}
\end{figure}

Il existe trois types de distributions bien définies par des formules analytiques illustrées dans la Fig.~\ref{fig:distributions}:
\begin{itemize}
    \item[$\bullet$] \textbf{Distribution de Maxwell-Jüttner} : cette distribution correspond à l'équivalent relativiste d'une distribution de vitesse Maxwellienne en équilibre thermodynamique, donc liée à la température. Pour la suite, on simplifiera le nom de cette distribution en \textit{distribution thermique}. Elle s'exprime de la manière suivante
    \begin{equation} \label{eq:distrib_thermique}
        \frac{dN_e}{d\gamma} = \frac{N_e}{\Theta_e} \frac{\gamma (\gamma^2-1)^{1/2}}{K_2(1/\Theta_e)} \exp \left( -\frac{\gamma}{\Theta_e} \right)
    \end{equation}
    avec $\Theta_e = k_B T / m_e c^2$ la température sans dimension et $K_2$ une fonction de Bessel modifiée du second ordre. Il est toutefois important de noter que cette distribution suppose un équilibre thermodynamique, ce qui est le cas lorsque le plasma est collisionel, c'est-à-dire qu'il y a suffisamment de collisions dues à la densité pour que les particules s'échangent de l'énergie cinétique jusqu'à atteindre un équilibre. Or, beaucoup de sources astrophysiques de rayonnement synchrotron, y compris le flot d'accrétion de Sgr~A*, sont non collisionelles. Il n'y a donc, a priori, pas d'équilibre thermodynamique, rendant l'utilisation de cette distribution caduque. Cependant, sa simplicité (un seul paramètre $\Theta_e$) en fait une distribution très largement utilisée, même pour des plasmas non collisionels notamment dans un contexte de simulation Magnéto-HydroDynamique (MHD).

    \item[$\bullet$] \textbf{Distribution en loi de puissance (\textit{Power Law})} : par opposition à la distribution précédente, on qualifie cette distribution de non thermique, avec une densité qui décroit lorsque que $\gamma$ augmente en une loi de puissance. Il faut néanmoins trois paramètres pour construire cette distribution, l'énergie minimale $\gamma_{min}$\footnote{Même si l'énergie d'une particule est $E=\gamma m c^2$, on simplifie la formulation en confondant $E$ et $\gamma$ puisque $mc^2$ est une constante.}, l'énergie maximale $\gamma_{max}$ et l'indice de la loi de puissance $p$. La fonction de la distribution en loi de puissance s'écrit
    \begin{equation} \label{eq:distrib_PL}
        \frac{dN_e}{d\gamma} = \frac{N_e (p-1)}{\gamma_{min}^{1-p}-\gamma_{max}^{1-p}} \gamma^{-p} , \text{pour } \gamma_{min} \leq \gamma \leq \gamma_{max}
    \end{equation}
    et vaut zéro partout ailleurs. Ce type de distribution est motivé par les observations notamment dans le vent solaire ou dans des sources très énergétiques comme les Noyaux Actifs de Galaxie (AGN en anglais) mais aussi par des simulations de type Particule-In-Cells (PIC) de choc et de reconnexion magnétique (voir Chap.~\ref{chap:Plasmoid Flare model}).

    \item[$\bullet$] \textbf{Distribution Kappa} : cette distribution est une fusion des deux précédentes. On considère ici un coeur thermique à faible énergie et une loi de puissance à haute énergie. L'intérêt de ce genre de distribution est de prendre en compte à la fois une partie en équilibre thermodynamique (ou un mouvement d'ensemble) et une partie hors équilibre non thermique. La distribution est construite de telle sorte qu'il n'y ait pas de discontinuité et nécessite deux paramètres, la température sans dimension $\Theta_e=w$ et un indice $\kappa$ lié à l'indice de la loi de puissance $p$ tel que $\kappa = p + 1$. La fonction de distribution Kappa s'écrit
    \begin{equation}\label{eq:kappa_distri}
        \frac{dN_e}{d\gamma} = N \gamma (\gamma^2-1)^{1/2} \left( 1+ \frac{\gamma -1}{w\ \kappa} \right)^{-(\kappa + 1)}
    \end{equation}
    où $N$ est un facteur de normalisation impliquant la fonction spéciale Gamma\footnote{\url{https://fr.wikipedia.org/wiki/Fonction_gamma}} $\Gamma(n)$ et s'exprime de la manière suivante
    \begin{equation}
    N (\kappa, w) = 
        \begin{cases}
            N_e \left( \frac{2}{\pi \kappa^3 w^3} \right)^{1/2} \frac{\Gamma(\kappa+1)}{\Gamma(\kappa - 1/2)}, & \text{si } \kappa w \ll 1 \\
            N_e \frac{(\kappa-2)(\kappa-1)}{2 \kappa^2 w^3} , & \text{si } \kappa w \gg 1.
        \end{cases}
    \end{equation}
    On se limitera quasiment exclusivement au cas relativiste $\kappa w \gg 1$.
\end{itemize}

\subsubsection{Formules d'approximations analytiques}
Le calcul exact des coefficients d'émission et d'absorption via les Eq.~\eqref{eq:emission_synchrotron_exact} et \eqref{eq:absorption_synchrotron_exact} respectivement requièrent une double intégration en comptant celle dans l'Eq.~\eqref{eq:F(x)} ce qui est très lourd en termes de temps de calcul. Fort heureusement, \cite{Pandya2016} ont dérivé des formules approximées, valides dans certaines zones de l'espace des paramètres, qui ne font plus appel à aucune intégrale, pour calculer ces coefficients à partir des paramètres de la distribution (thermique, loi de puissance ou kappa), de la fréquence et du champ magnétique (pris en compte dans $\nu_{crit}$).

Pour chacune des trois distributions, le coefficient d'émission synchrotron peut s'écrire \cite{Pandya2016}
\begin{equation}
    j_\nu = \frac{N_e e^2 \nu_{crit}}{c} \mathbf{J_S} \left( \frac{\nu}{\nu_{crit}}, \alpha \right),
\end{equation}
et le coefficient d'absorption
\begin{equation}
    \alpha_\nu = \frac{N_e e^2}{\nu m_e c} \mathbf{A_S} \left( \frac{\nu}{\nu_{crit}}, \alpha \right),
\end{equation}
où $J_S$ et $A_S$ sont respectivement les coefficients d'émission et d'absorption sans dimension qui dépendent de la distribution considérée et $\alpha$ l'angle entre la direction d'émission et le champ magnétique. Ainsi, toute la complexité de la double intégration et de la dépendance à la distribution
est contenue dans les coefficients sans dimension qui sont ici des formules d’ajustement~\cite{Pandya2016}.

Les formules de \cite{Pandya2016}, réécrites dans l'annexe~\ref{ap:coefs_synchrotron}, permettent d'obtenir une approximation des coefficients synchrotron avec une erreur relative inférieure ou égale à $40\%$ pour $10 < \nu / \nu_{crit} < 3 \times 10^{10}$, et $15\degree < \alpha < 85\degree$. De plus, le domaine de validité sur les paramètres de chacune des distributions est aussi limité à $3 < \Theta_e < 40$ pour une distribution thermique, $1,5 < p < 6,5$ pour la distribution en loi de puissance et $3 < w < 40$, $2,5 < \kappa < 7,5$ pour la distribution kappa.

\subsection{Effets relativistes sur la courbe de lumière observée}\label{sec:effet_relat_LC}
Avant d'introduire la variabilité intrinsèque de notre source, on rappelle ici les différents effets relativistes qui vont affecter la courbe de lumière observée. En effet, même lorsque la source de rayonnement est constante dans le référentiel de l'émetteur comme dans \cite{Hamaus2009}, le flux observé sera variable. 

Comme on l'a vu au Chap.~\ref{chap:GYOTO}, la quantité (radiative) qui se conserve le long d'une géodésique, autrement dit, qui est indépendante du référentiel, est
\begin{equation}\label{eq:conserv_Inu}
    \frac{I_\nu}{\nu^3} = \text{constante}.
\end{equation}

Ainsi, une première source de variabilité évidente est l'effet Doppler (dans le cas relativiste). En effet, durant une orbite autour du trou noir et selon l'inclinaison de l'observateur, la vitesse radiale du point chaud sera tantôt positive et tantôt négative selon sa phase orbitale. Il en résulte un \textbf{effet Doppler longitudinal} (en vitesse radiale) périodique tantôt décalé vers le rouge et tantôt décalé vers le bleu dont la période est la période orbitale. Il s'exprime de la manière suivante
\begin{equation}
    \frac{\nu_{obs}}{\nu_{ém}} = \sqrt{\frac{1-\beta_\parallel}{1+\beta_\parallel}}
\end{equation}
où $\beta_\parallel=v_\parallel/c$ avec $v_\parallel$ la composante radiale de la vitesse par rapport à l'observateur\footnote{À ne pas confondre avec la composante radiale de la vitesse dans le système de coordonnées $v_r$.} et peut donc être positive ou négative selon que la source s'éloigne ou se déplace vers l'observateur respectivement.

À cela s'ajoute \textbf{l'effet Doppler transversal} qui n'a pas d'équivalent en mécanique classique. Ce décalage de fréquence est présent lorsque le mouvement de la source est orthogonal à la ligne de visée, et est dû à la dilatation temporelle qui est liée à la norme de la vitesse de la source dans le référentiel de l'observateur (voir \eqref{eq:dilatation_temps}).

Les deux décalages spectraux précédents sont uniquement des effets de Relativité Restreinte (mouvement d'une source par rapport à un observateur). Il faut aussi prendre en compte le décalage spectral gravitationnel, aussi appelé \textbf{décalage d'Einstein} dû, comme précédemment, à la dilatation temporelle de la source par rapport à l'observateur, générée par la métrique (et non la vitesse de la source). Il s'exprime de la manière suivante
\begin{equation}
    \frac{\nu_{obs}}{\nu_{ém}} = \sqrt{1-\frac{r_s}{r}}
\end{equation}
avec $r_s$ le rayon de Schwarzschild et $r$ le rayon orbital de la source. On constate aisément que ce rapport est strictement inférieur à un. Ainsi, la fréquence observée est toujours plus petite que la fréquence émise\footnote{En ne prenant en compte que ce décalage.} et ce décalage est d'autant plus grand que le rayon orbital de la source est proche de l'horizon (dans le cas d'un trou noir sans spin).

\begin{figure}
    \centering
    \resizebox{0.8\hsize}{!}{\includegraphics{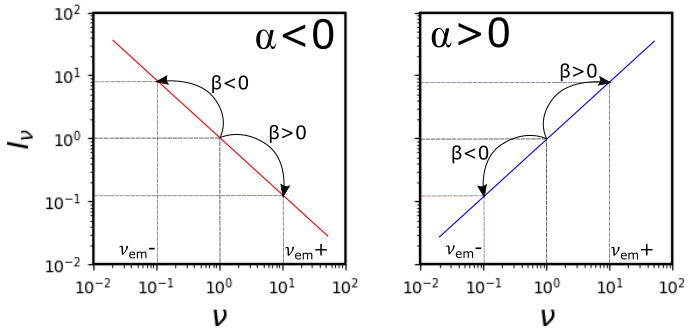}}
    \caption{Schéma illustrant l'influence du spectre de la source sur l'intensité émise puis observée avec un indice spectral négatif, c'est-à-dire que le flux diminue lorsque la fréquence augmente, à \textbf{gauche}, et un indice spectral positif, c'est-à-dire que le flux augmente avec la fréquence, à \textbf{droite}.}
    \label{fig:spectral_Inu}
\end{figure}

\begin{figure}
    \centering
    \resizebox{0.8\hsize}{!}{\includegraphics{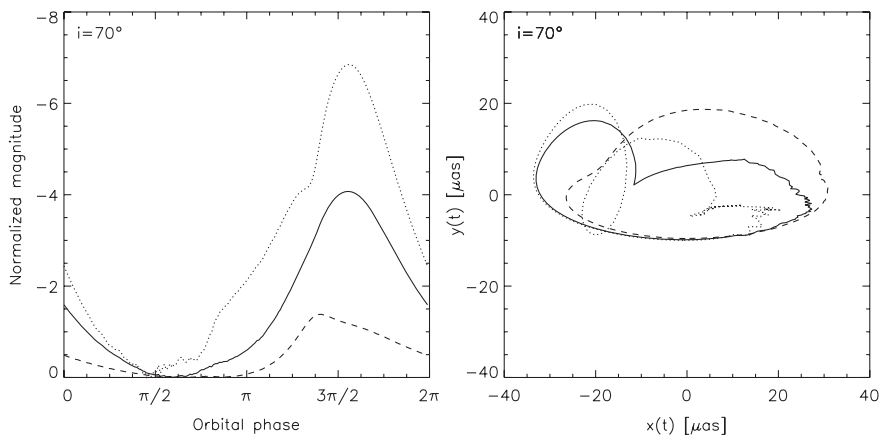}}
    \caption{Courbe de lumière \textbf{(gauche)} et astrométrie \textbf{(droite)} d'un point chaud orbitant à l'ISCO d'un trou noir de Schwarzschild vu avec une inclinaison de $70$° avec trois indices spectraux ($\alpha = -3$ en pointillés, $\alpha = 0$ en trait plein et $\alpha = 3$ en tirets). Crédit : \cite{Hamaus2009}.}
    \label{fig:Hamaus_spectral}
\end{figure}

L'Eq.~\eqref{eq:conserv_Inu} montre que les décalages spectraux précédents ont une grande importance dans l'intensité spécifique mesurée à travers la dépendance en $\nu^3$. Cependant, la dépendance spectrale de l'émission de la source apparaît aussi comme cruciale. En effet, pour une source spectralement plate $I_\nu=\text{cste}$ (spectre blanc), la variabilité provient uniquement du facteur $\nu^3$. Cependant, ce genre de source est rare ou n'existe que localement, sur une gamme de fréquences limitée, par exemple autour de la fréquence du maximum d'émission. La plupart des sources ont une émission qui dépend de la fréquence. On peut définir localement un indice spectral $\alpha$ tel que $I_\nu \propto \nu^\alpha$. Cet indice spectral est important à la fois pour la courbe de lumière, mais aussi pour l'astrométrie (voir Section~\ref{sec:effet_relat_astrometrie}). En effet, lorsque le point chaud se déplace en direction de l'observateur, on a vu que le rayonnement est décalé vers le bleu et donc amplifié (Eq.~\ref{eq:conserv_Inu}). Selon le signe de l'indice spectral $\alpha$, le flux observé va être encore plus amplifié si $\alpha < 0$ (spectre bleu) ou réduit si $\alpha < 0$ (spectre rouge). On constate aisément cet effet dans les Figs~\ref{fig:spectral_Inu} et \ref{fig:Hamaus_spectral} qui montre la courbe de lumière (gauche) d'un modèle de point chaud orbitant à l'ISCO d'un trou noir de Schwarzschild pour les trois types de spectres discuté.

\begin{figure}
    \centering
    \resizebox{0.8\hsize}{!}{\includegraphics{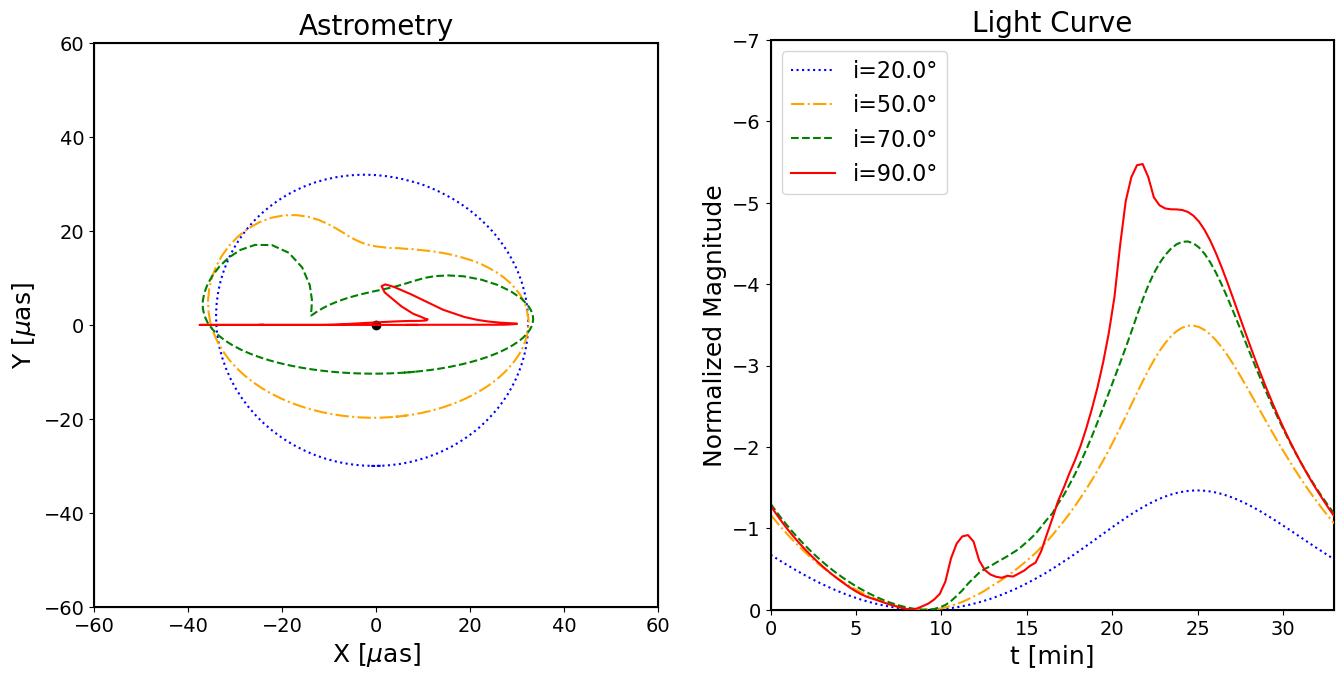}}
    \caption{Astrométries \textbf{(gauche)} et courbes de lumières \textbf{(droite)} d'un point chaud sphérique de rayon $R=1\, r_g$, orbitant avec une vitesse Képléreinne à $r=6\, r_g$ d'un trou noir de Schwarzschild pour quatre inclinaisons : $20\degree$ (bleu), $50\degree$ (orange), $70\degree$ (vert) et $90\degree$ (rouge). L'émission est du synchrotron avec une distribution kappa ($\kappa=5$) équivalente à une émission en loi de puissance $I_\nu \propto \nu^\alpha$ de coefficient $\alpha=-1,5$. Le point noir dans le panneau de gauche correspond à la position du trou noir. L'astrométrie à $90\degree$ est fortement bruité due à des effets de résolution (elle devrait être parfaitement horizontale).}
    \label{fig:i_var_star}
\end{figure}

Contrairement à la physique classique où les rayons lumineux ont une trajectoire rectiligne, en relativité générale, on a vu au Chap.~\ref{chap:GYOTO} qu'ils suivent les géodésiques de genre lumière qui, proches des trous noirs, sont courbées. Ainsi certains rayons lumineux émis dans une certaine direction, autre que celle de l'observateur, vont être déviés et renvoyés à ce dernier (voir Fig.~\ref{fig:deviation_lumière}). On va donc observer plusieurs images d'un même objet qui vont chacune contribuer à la courbe de lumière. On définit l'ordre de cette image $n$ par le nombre de demi-orbites autour du trou noir des photons depuis la source jusqu'à l'observateur en partant de 1. Ainsi, l'image dite "primaire" ($n=0$) correspond aux photons ayant une trajectoire directe entre la source et l'observateur et l'image secondaire ($n=1$), ceux ayant fait un demi-tour autour du trou noir. La position de l'image secondaire dans le ciel est décalée par rapport à celle de l'image primaire (voir panel de gauche Fig.~\ref{fig:images_exemple}). Les photons $n \geq 2$, qui font au moins un tour autour du trou noir, sont regroupées dans l'anneau de photons, issues d'une coquille fine de l'espace-temps proche du trou noir où les géodésiques de genre lumière sont instables, c'est-à-dire qu'un photon entrant dans cette coquille peut s'en échapper après un certain nombre d'orbites. Sauf indication contraire, on se limitera aux images primaire et secondaire à cause d'effets de résolution, les images d'ordre élevé sont très petites, et demandent un grand nombre de pixels pour correctement les résoudre avec les codes de tracé de rayons, ce qui requiert un plus grand temps de calcul et d'un point de vue observationnel, une résolution instrumentale hors d'atteinte actuellement. Lorsqu'il y a un alignement entre la source, le trou noir et l'observateur, on a alors un effet de lentille gravitationnelle se traduisant par une forte augmentation du flux dans la courbe de lumière puisque la surface émettrice est plus grande. Pour un point chaud en orbite équatoriale, cet effet est visible à forte inclinaison ($i \sim 90 \degree$, voir Fig.~\ref{fig:i_var_star}) lorsque le point chaud passe derrière le trou noir.

\subsection{Effets relativistes sur l'astrométrie}\label{sec:effet_relat_astrometrie}
L'observable principale de GRAVITY est l'astrométrie, on le rappelle, le suivi du centroïde du flux issu de (l'environnement proche de) Sgr~A* puisque la source est non résolue. Les effets cités à la section précédente ont aussi un impact sur l'astrométrie observée. Même si l'on considère une seule source de rayonnement, on observe plusieurs images de cet objet. Ainsi, le centroïde de l'image correspond au barycentre entre les différents ordres d'images pondérées par leurs flux respectifs. On rappelle que les images primaire et secondaire sont fortement décalées l'une par rapport à l'autre (environ une demi-orbite à faible inclinaison). 

Prenons le cas où le point chaud se situe dans la zone de beaming positif, c'est-à-dire la zone du plan du ciel où le point chaud se déplace vers l'observateur, son flux est alors amplifié (voir section~\ref{sec:effet_relat_LC}). Dans ce cas, c'est l'image primaire qui domine le centroïde final puisque son flux est amplifié alors que celui de l'image secondaire est soit moins amplifié, soit au contraire diminué. En effet, du fait de la différence de chemin parcouru par les photons, plus les effets de dilatation temporelle gravitationnelle, le temps d'émission de l'image secondaire est toujours antérieur à celui de l'image primaire. Ainsi, l'image secondaire du point chaud correspond à une phase antérieure de son orbite et donc un beaming différent\footnote{Sans prendre en compte une variabilité intrinsèque pour le moment.}. À l'inverse, lorsque l'image primaire se situe dans la zone de beaming négatif, c'est-à-dire la zone du plan du ciel où le point chaud s'éloigne de l'observateur, le flux de l'image secondaire est quant à lui amplifié, car au temps d'émission de cette image, le point chaud s'approche de l'observateur (est à une phase différente de son orbite).

\begin{figure}
    \centering
    \resizebox{0.8\hsize}{!}{\includegraphics{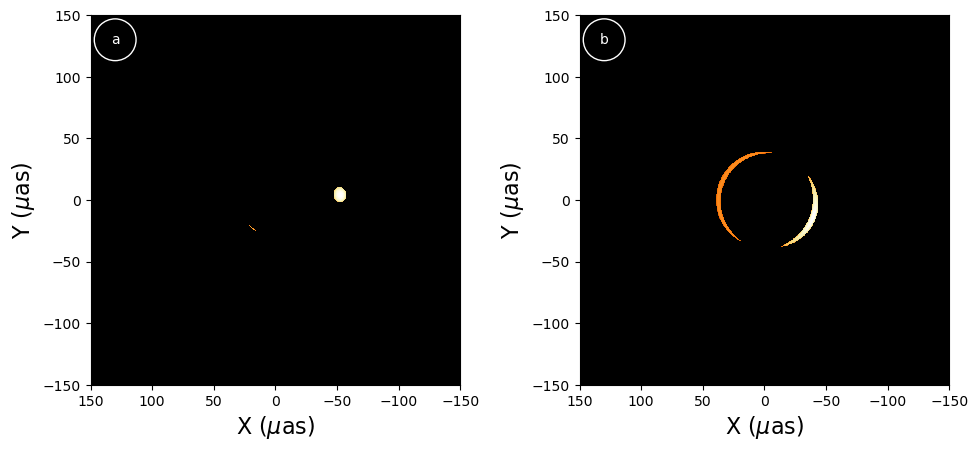}}
    \caption{Images générées par \textsc{GYOTO} d'un point chaud orbitant un trou noir de Schwarzschild à $r=9$ $r_g$ avec une inclinaison de \textbf{(a)} 20° et \textbf{(b)} 90° illustrant la taille relative entre l'image primaire (blanche) et secondaire (rouge) pour ces deux inclinaisons (à un temps d'observation particulier). Note : il est possible que l'image secondaire de l'image de gauche ne soit pas visible dans un format imprimé.}
    \label{fig:images_exemple}
\end{figure}

Cependant, comme on peut le voir sur le panel de gauche de la Fig.~\ref{fig:images_exemple}, la taille apparente de l'image secondaire est nettement plus petite que celle de l'image primaire lorsque l'inclinaison est faible ou modérée\footnote{On rappelle que l'on se place ici dans le cas d'une source dans le plan équatorial.}. Cela n'est plus vrai à forte inclinaison lors d'un effet de lentille gravitationnelle comme on peut le voir dans le panel de droite de la Fig.~\ref{fig:images_exemple} où la taille des deux images est comparable et elles vont donc avoir une contribution similaire dans le calcul du centroïde. Ce sont ces deux effets combinés qui sont responsables du rebroussement de la courbe verte ($i=70$°) du panel de gauche de la Fig.~\ref{fig:i_var_star} en $X\sim -17\, \mu$as, $Y \sim 2\, \mu$as.

\subsection{Dynamique et évolution temporelle du point chaud}
Étant donné la rareté et l'absence observées de répétitions des sursauts de Sgr~A*, ainsi que les arguments présentés dans la section~\ref{sec:motivation}, il est naturel de considérer un modèle avec une variabilité intrinsèque. On cherche dans un premier temps un modèle analytique simple incluant une variation temporelle du flux émis avec une phase de croissance et une phase de décroissance. Pour cela, on applique une modulation Gaussienne sur la densité et la température, que l'on considère comme uniformes, imitant une sur-densité grandissante ayant été chauffée (par un mécanisme quelconque) et se refroidissant et qui finit par s'estomper (en se mélangeant avec le reste du flot d'accrétion par exemple). On note que le processus de refroidissement considéré ici n'est pas du refroidissement synchrotron, mais un processus quelconque là aussi. On considère ici un seul temps caractéristique $t_\sigma$ valable à la fois pour la phase de chauffage et de refroidissement afin de limiter le nombre de paramètres. Comme les coefficients d'émission et d'absorption dépendent de la densité et de la température du plasma, la courbe de lumière intrinsèque résultante est aussi une Gaussienne. La densité et la température en fonction du temps s'écrivent
\begin{equation}\label{eq:ne(t)}
    n_e(t)=n_e^0\ \exp \left( -0,5 \times \left(\frac{t-t_{ref}}{t_{\sigma}}\right)^2\right),
\end{equation}
\begin{equation}\label{eq:Te(t)}
    T_e(t)=T_e^0\ \exp \left( -0,5 \times \left(\frac{t-t_{ref}}{t_{\sigma}}\right)^2\right),
\end{equation}
avec $n_e^0$ et $T_e^0$ les valeurs maximales de densité et température, et $t_{ref}$ le temps coordonné de référence du maximum d'émission.

Pour calculer les coefficients synchrotron, il faut définir la valeur du champ magnétique ainsi que, en principe, sa configuration afin de calculer l'angle $\alpha$ entre le vecteur champ magnétique et le vecteur tangent au photon, les deux projetés dans le référentiel de l'émetteur. Cependant, on ne va pas considérer une configuration magnétique particulière, mais un champ magnétique isotrope, c'est-à-dire que l'on va moyenner les coefficients par rapport à $\alpha$. Cela permet d'être agnostique sur la configuration du champ magnétique dans un premier temps. La norme du champ magnétique, quant à elle, est déterminée à partir de la magnétisation
\begin{equation} \label{eq:magnetisation}
    \sigma =  \frac{v_A^2}{c^2}
\end{equation}
où $v_A = B/\sqrt{4 \pi m_p n_e}$ est la vitesse d'Alvèn du plasma, que l'on fixe à une valeur donnée. On remarque que la magnétisation dépend de la densité qui est elle-même une fonction du temps, ainsi si l'on impose une valeur fixe pour $\sigma$, la valeur du champ magnétique sera, elle aussi, fonction du temps (évolution identique à la densité).

Pour le calcul des coefficients synchrotron, on utilise une distribution Kappa (coeur thermique + loi de puissance à haute énergie) paramétrée par la température sans dimension $\Theta_e$ et l'indice $\kappa = p+1$. Le choix de cette distribution est motivé par deux choses : (1) on suppose une sphère de plasma de rayon $R = 1\, r_g$ faisant partie du flot d'accrétion de Sgr~A* (dans le disque) qui est chauffé puis refroidi, nécessitant donc une composante thermique. On prend en compte en plus une composante non thermique à haute énergie afin d'atteindre des niveaux de flux suffisamment élevés sans avoir recours à une température extrême. (2) En NIR, on peut relier l'indice de la loi de puissance de la distribution des électrons $p$ à l'indice spectral observé $\alpha$ ($\nu F_\nu \propto \nu^\alpha$)\footnote{Attention, cet indice spectral $\alpha$ n'a pas la même définition que celui défini plus haut et il ne faut pas le confondre avec l'angle entre le vecteur tangent au photon et le vecteur champ magnétique.} par
\begin{equation} \label{eq:spectral_index}
    \alpha = \frac{3-p}{2}.
\end{equation}
Pour les sursauts de Sgr~A*, on observe un indice spectral compris entre -0.5 et 0.5 \citep{Genzel2010} qui se traduit pour l'indice de la loi de puissance $2 < p < 4$, soit $3 < \kappa < 5$. On rappelle que plus l'indice est élevé, moins il y a d'électrons à très hautes énergies (voir Fig.~\ref{fig:distributions}). On va fixer $\kappa=5$ pour ce modèle afin de tenir compte de la composante non thermique sans que celle-ci soit dominante.

Enfin, pour la dynamique du point chaud, on garde le même mouvement orbital que dans \cite{Gravity2018}, à savoir une orbite circulaire à vitesse Képlérienne dans le plan équatorial. La vitesse est donc déterminée à partir du rayon par
\begin{equation}
    v_{orb} = (r^{1.5} + a_\star)^{-1}
\end{equation}
où $r$ est le rayon orbital en unité de $r_g$ et $a_\star$ le spin sans dimension du trou noir.

\subsection{Influence de la variabilité intrinsèque}\label{sec:variabilité_intrinsèque}
Le fait de prendre une source variable a plusieurs conséquences. En effet, comme on l'a vu plus haut, lorsque l'on fait une image à un temps d'observation donné, on constate plusieurs images de notre source. Cependant, dû à la différence de chemin parcouru par les photons entre l'image primaire et secondaire, le temps d'émission n'est pas le même. Pour une source constante, seule la phase de l'orbite, se traduisant par un décalage spectral différent (dû au beaming), va créer une différence. Or dans le cas d'une source variable comme c'est le cas de notre point chaud, le flux émis sera aussi différent et dépend de l'écart entre le temps coordonné d'émission $t_{ém}$ et le temps coordonné de référence de la Gaussienne $t_{ref}$ ainsi que du temps caractéristique de la Gaussienne $t_\sigma$. 

Prenons deux exemples avec un temps caractéristique $t_\sigma = 15$ minutes :
\begin{itemize}
    \item[$\bullet$] \textbf{Exemple 1} : prenons le cas où le temps d'émission de l'image primaire $t_{ém}^1$ est égal à $t_{ref}$, autrement dit l'image primaire est à son maximum d'émission (peu importe la phase orbitale pour le moment). Le temps d'émission de l'image secondaire $t_{ém}^2$ est toujours antérieur à celui de l'image primaire $t_{ém}^2 < t_{ém}^1$. Le décalage de temps d'émission entre l'image primaire et secondaire dépend de l'écart de distance parcourue entre les deux photons et dépend donc du rayon orbital du point chaud. Pour un rayon orbital $r= 9\, r_g$, on a $\Delta t_{ém} = t_{ém}^2 - t_{ém}^1 \approx -7,6$ minutes. Ce qui signifie que la densité et la température sont $\approx~25 \%$ plus faibles que pour l'image primaire résultant en une intensité spécifique, à la même fréquence, c'est-à-dire sans prendre en compte les effets relativistes des sections précédentes (section~\ref{sec:effet_relat_LC} et \ref{sec:effet_relat_astrometrie}), environ $5$ fois plus faible. La détermination de ce facteur via une formule analytique est loin d'être simple à cause de la dépendance de l'intensité spécifique aux différents paramètres et de leur interdépendance (voir Chap.~\ref{chap:Plasmoid Flare model}). L'importance de l'image secondaire dans la courbe de lumière est alors négligeable, tout comme sa contribution dans le calcul du centroïde qui correspond approximativement à la position de l'image primaire.

    \item[$\bullet$] \textbf{Exemple 2} : prenons maintenant le cas inverse, le temps d'émission de l'image secondaire $t_{ém}^2$ est égal à $t_{ref}$, alors le temps d'émission de l'image primaire $t_{ém}^1$ est $7,6$ minutes après le maximum d'émission résultant en une intensité spécifique $5$ fois plus faible. Cependant, la taille angulaire de l'image primaire étant significativement plus grande, dans la majorité des cas, que celle de l'image secondaire, le flux issu de l'image primaire reste majoritaire ou tout du moins comparable à celui de l'image secondaire. Néanmoins, dans cette situation, la contribution de l'image secondaire à la courbe de lumière n'est plus négligeable et le centroïde observé ne correspond plus à la position de l'image primaire, mais se situe entre les deux images.
\end{itemize}

\begin{figure}
    \centering
    \resizebox{\hsize}{!}{\includegraphics{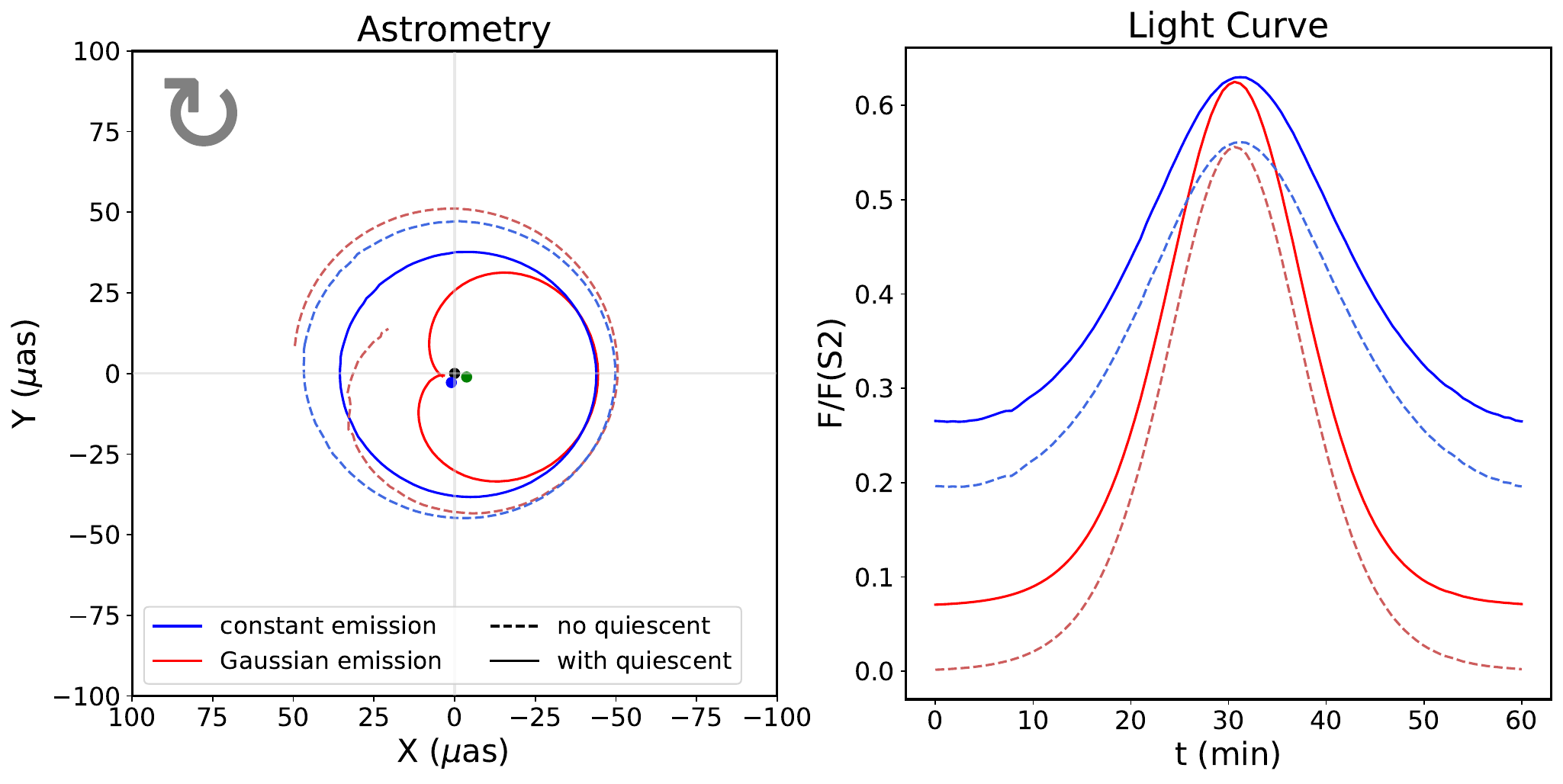}}
    \caption{Astrométrie (\textbf{gauche}) et courbes de lumière (\textbf{droite}) du modèle point chaud + jet avec deux valeurs pour l'état quiescent correspondant à sans quiescent (tirets) et avec quiescent (lignes pleines). En bleu, le point chaud a une émission presque constante ($t_\sigma >> t_{orbit}$). L'effet du beaming est reflété dans les courbes de lumière. En rouge, le point chaud a une émission gaussienne avec $t_\sigma=30$ min. Les paramètres du point chaud sont énumérés dans la Table~\ref{tab:hotspot table params}. Nous avons synchronisé le beaming et le maximum intrinsèque de la modulation gaussienne. Les points, noir, bleu et vert dans le panneau de gauche, représentent respectivement la position de Sgr~A*, le centroïde du jet et le centroïde du disque.}
    \label{fig:influence_variabilité}
\end{figure}

Nous avons vu dans les sections~\ref{sec:effet_relat_LC} et \ref{sec:effet_relat_astrometrie} les effets relativistes pour une source constante et, au-dessus, l'influence de la variabilité intrinsèque sans ces effets. La courbe de lumière et l'astrométrie observées sont une combinaison de tous ces effets, dans toutes les configurations possibles, que l'on ne va pas détailler de manière exhaustive. La Fig.~\ref{fig:influence_variabilité} permet d'illustrer la plupart des configurations importantes. Elle montre l'astrométrie et la courbe de lumière d'une orbite complète à $r=9\, r_g$ d'un modèle de point chaud avec une émission constante en bleu ($t_\sigma >> t_{orbite}$) et notre modèle de point chaud en rouge avec un temps caractéristique $t_\sigma = 30$ min vu avec une inclinaison de $20\degree$. On ne s'intéresse ici qu'aux courbes en tirets où il n'y a que le point chaud. On constate que dans le cas d'une émission constante, l'astrométrie est fermée puisque l'état du système final est identique au système initial (à la même phase, mais une orbite plus tard). Cependant, ce n'est pas le cas lorsqu'on a de la variabilité intrinsèque. On se place dans le cas où le maximum d'émission est synchronisé avec le maximum du beaming au milieu de la fenêtre d'observation ($t_{obs} \approx 30$ min $\sim P_{\text{orb}}/2$).

Au début de l'observation, les temps d'émission de l'image primaire et secondaire sont antérieurs à $t_{ref}$ avec $t_{ém}^2 < t_{ém}^1 < t_{ref}$. Le flux de l'image primaire est certes faible, car encore plus atténué par l'effet de beaming "négatif" mais reste significativement supérieur à celui de l'image secondaire qui est totalement négligeable malgré un effet de beaming positif (tant dû à l'intensité spécifique proche de zéro que sa taille relative). De ce fait, la contribution de l'image secondaire dans l'astrométrie est aussi négligeable, contrairement au cas de l'émission constante dont les points astrométriques sont par conséquent plus resserrés (l'image secondaire étant approximativement en opposition par rapport au trou noir, donc du côté des $X<0$).

Lorsqu'on atteint le milieu de l'orbite à $t_{obs}\sim 30$~min, c'est-à-dire le moment du maximum de beaming et d'émission de l'image primaire, le flux de l'image secondaire est négligeable dans les deux cas, car il subit un beaming négatif et a une taille apparente, encore une fois, très faible. De plus, dans le cas de l'émission variable, le flux émis par l'image secondaire est encore plus faible que dans le cas de l'émission constante, car elle n'a pas encore atteint le maximum d'émission. Cependant, dans ce cas particulier, l'image primaire étant fortement amplifiée, cet effet est négligeable et les deux courbes en tirets se confondent en $X \sim -30\, \mu$as et $Y \sim -20\, \mu$as. 

Les choses intéressantes commencent maintenant, à partir du moment où les deux astrométries en tirets divergent. Pour $t_{obs} > P_{\text{orb}}/2$, le temps d'émission de l'image primaire a dépassé le temps du maximum d'émission $t_{ém}^1 > t_{ref}$ mais ce n'est pas encore le cas pour l'image secondaire. Ainsi, le flux de l'image secondaire va augmenter à la fois via la modulation Gaussienne jusqu'à atteindre le maximum et l'effet de beaming positif. À l'inverse, le flux de l'image primaire va diminuer à la fois dû à la décroissance du flux intrinsèque ($t_{ém}^1 > t_{ref}$) et de l'effet de beaming négatif. Ainsi le rapport des flux entre l'image primaire et secondaire qui était penché fortement du côté de l'image primaire précédemment, est plus équilibré dans cette partie de l'orbite et dans ce cas particulier. Le rapport de flux étant de l'ordre de grandeur de 1 (au lieu de $10^{5}$), l'astrométrie est plus resserrée et surtout ne forme pas une boucle fermée. On note que le flux des deux images est très faible comparé au maximum, mais sont comparables l'un à l'autre à ce moment d'observation.

Pour résumer, ce qui détermine l'astrométrie dans un modèle avec uniquement un point chaud est le rapport de flux entre les deux images qui sont en opposition par rapport au trou noir (origine des coordonnées). La question à se poser est quelle image domine l'astrométrie au temps d'observation considéré et pourquoi. La réponse étant la convolution entre le beaming et le flux émis à ce temps d'émission qui est différent pour les deux images. La courbe de lumière observée étant le résultat de cette convolution, plus la contribution relative des deux images, la valeur du maximum ainsi que la largeur de la Gaussienne observée dépendent de la phase orbitale, comme illustré par le panel de droite de la Fig.~\ref{fig:multi_phi0}.

\section{Modélisation de l'état quiescent}
En plus de la variabilité temporelle, on va rajouter ici un modèle de l'état quiescent de Sgr~A* afin d'étudier l'impact de ce dernier sur les observables et en particulier sur l'astrométrie.
\subsection{Modèle de Tore+Jet}
La nature exacte de l'état quiescent de Sgr~A* n'est pas connu. Sgr~A* est un flot d'accrétion de faible luminosité avec un taux d'accrétion de $(5,2-9,5) \times 10^{-9} M_\odot$.yr$^{-1}$ et une luminosité bolométrique de $(6,8-9,2) \times 10^{35}$ erg.s$^{-1}$ \citep{Bower2019, EHT2022b} et accrété donc à un taux très inférieur au taux d'Eddington. La présence d'étoiles massives dans le voisinage du trou noir alimente le flot d'accrétion de Sgr~A* à travers leur vent stellaire. Il est donc vraisemblable qu'un disque d'accrétion se forme. Étant donné la faible luminosité de ce disque, on est dans une configuration RIAF (\textit{Radiatively Inefficient Accretion Flow}), c'est-à-dire un disque géométriquement épais et optiquement mince. 

\begin{figure}
    \centering
    \resizebox{0.8\hsize}{!}{\includegraphics{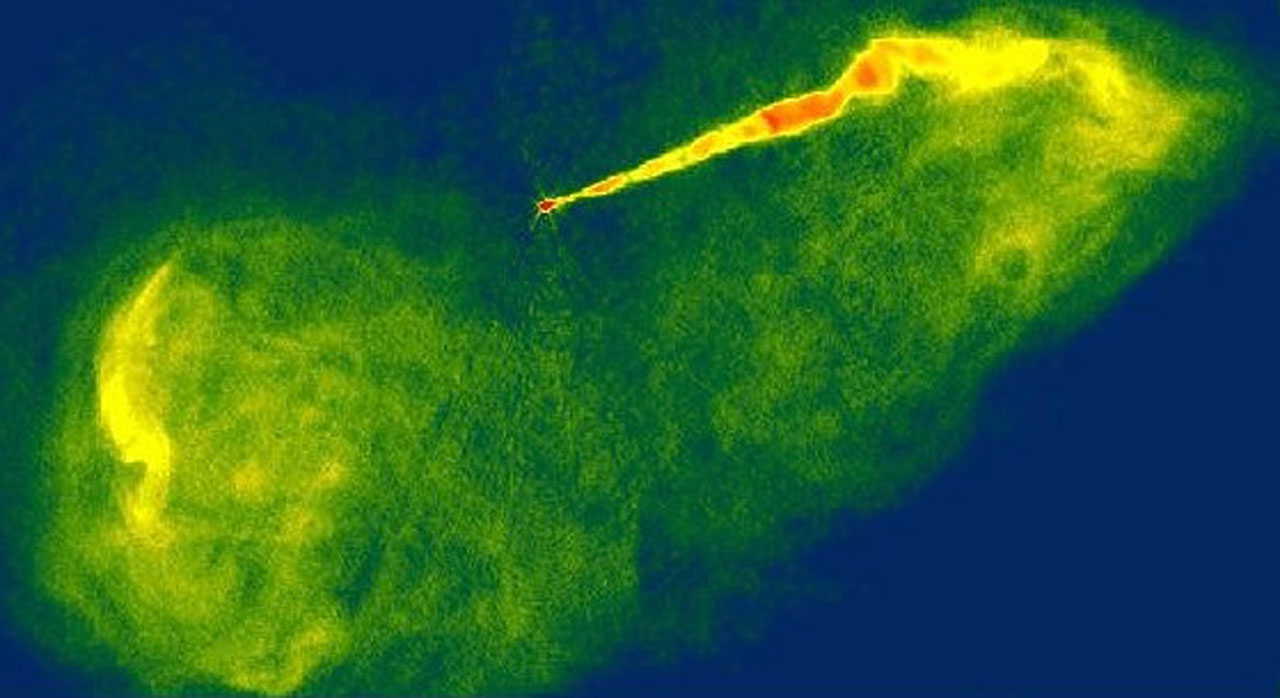}}
    \caption{Cette image radio de la galaxie M 87, prise avec le radiotélescope Very Large Array (VLA) en février 1989, montre des structures géantes ressemblant à des bulles où l'on pense que l'émission radio est alimentée par les jets de particules subatomiques provenant du trou noir central de la galaxie. La fausse couleur correspond à l'intensité de l'énergie radio émise par le jet. Crédit : \href{https://esahubble.org/images/opo9943b/}{National Radio Astronomy Observatory/National Science Foundation}.}
    \label{fig:M87}
\end{figure}

\begin{figure}
    \centering
    \resizebox{0.8\hsize}{!}{\includegraphics{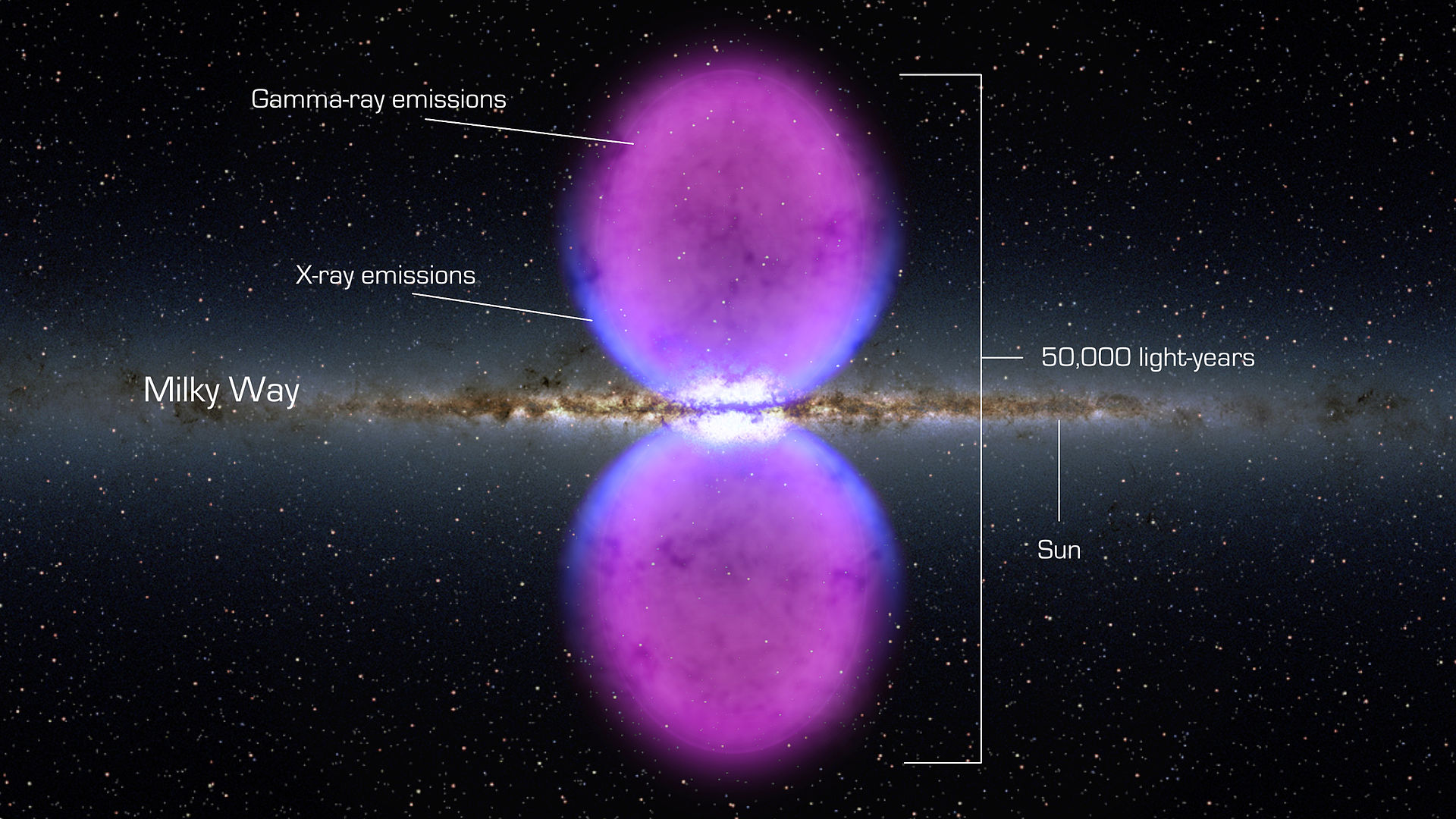}}
    \caption{Vue d'artiste des bulles de Fermi. Crédit : \href{http://www.nasa.gov/mission_pages/GLAST/news/new-structure.html}{NASA's Goddard Space Flight Center}}
    \label{fig:bulle de Fermi}
\end{figure}

La présence d'un disque apparaît donc comme évidente, cependant, d'autres systèmes astrophysiques similaires comme les AGN ou les micro-quasars qui sont tous deux des trous noirs avec un flot d'accrétion, présentent aussi un jet dont l'un des plus connus est celui de M87* illustré dans la Fig.~\ref{fig:M87}. D'un point de vue théorique, il existe deux mécanismes capables de générer des jets à partir d'un disque d'accrétion :
\begin{itemize}
    \item[$\bullet$] \textbf{Le processus de Penrose} : ce processus décrit comment extraire de l'énergie ou du moment cinétique depuis l'ergorégion (voir Chap.~\ref{chap:GYOTO}) d'un trou noir en rotation (spin non-nul) et se base sur l'effet Lense-Thiring \citep{Penrose1969}. Pour cela, on lance une particule suivant une trajectoire passant par l'ergorégion où l'on scinde cette particule en deux de telle manière qu'une partie soit absorbée par le trou noir en lui faisant perdre du moment cinétique. L'autre partie est alors éjectée avec une vitesse supérieure à celle de la particule incidente et même avec une énergie totale supérieure à l'énergie de masse de la particule incidente. Dans le cas où la particule initiale est un photon de haute énergie capable de générer une paire d'électron-positron, ces derniers sont donc les fragments qui vont, pour l'un des deux, être absorbé par le trou noir et pour l'autre être éjecté et contribuer à former le jet. 
    
    \item[$\bullet$] \textbf{Le processus de Blandford-Znajek} : comme pour le processus précédent, il décrit comment extraire de l'énergie de rotation d'un trou noir, mais cette fois-ci via les lignes de champ magnétique du flot d'accrétion \citep{Blandford-Znajek1977}. Il nécessite comme le précédent un trou noir avec rotation et un disque fortement magnétisé avec des lignes de champ magnétique poloïdal. Les lignes de champ magnétique traversant l'ergosphère ne peuvent rester statiques par rapport à l'infini et s'enroulent autour de l'axe de rotation du trou noir et gagnent une forte composante toroïdale. Les particules chargées suivant ces lignes de champ magnétique vont ainsi gagner de l'énergie et être éjectées en un jet collimaté.
\end{itemize}
De plus, de nombreuses simulations numériques GRMHD de flot d'accrétion RIAF en configuration MAD (\textit{Magnetically Arrested Disk}) ou SANE (\textit{Standard and Normal Evolution})\footnote{On reviendra sur ces deux configurations dans le Chap.~\ref{chap:Plasmoid Flare model}.} génèrent un jet dans la région polaire du trou noir \citep{Dexter2020, Jiang2023, Nathanail2022a, Porth2021, Ripperda2020, Ripperda2022, Yoon2020}. Ainsi, la présence d'un jet est possible d'un point de vue théorique si l'on considère que Sgr~A* a un spin non-nul ce qui  est le plus probable et d'un point de vue numérique. Toutefois, l'existence d'un jet pour Sgr~A* est encore débattu. En effet, le spectre de l'état quiescent de Sgr~A* peut-être modélisé par uniquement un disque d'accrétion avec néanmoins une température très élevée \citep{Narayan1995}. D'un autre côté, la présence des bulles de Fermi, de très large zone ($\sim$ 25.000 années-lumière) émettant des rayons gamma, des rayons~X et des ondes radio en dehors du plan galactique avec un point de convergence au niveau du centre galactique (voir Fig.~\ref{fig:bulle de Fermi}) peut être associé à un ancien jet de Sgr~A* aujourd'hui (ou plutôt il y $\sim 24.000$ ans) disparu \citep{Cheng2011}. De plus, de récentes observations en radio \citep{Yusef-Zadeh2020} et en rayons~X \citep{Zhu2019} tendent à suggérer l'existence d'un jet lancé par Sgr~A* à travers l'analyse des propriétés temporelles et spectrales dans ces deux domaines de longueurs d'ondes. On va donc considérer, en plus d'un disque d'accrétion, la présence d'un jet pour l'état quiescent de Sgr~A*.

\begin{figure}
    \centering
    \resizebox{0.4\hsize}{!}{\includegraphics{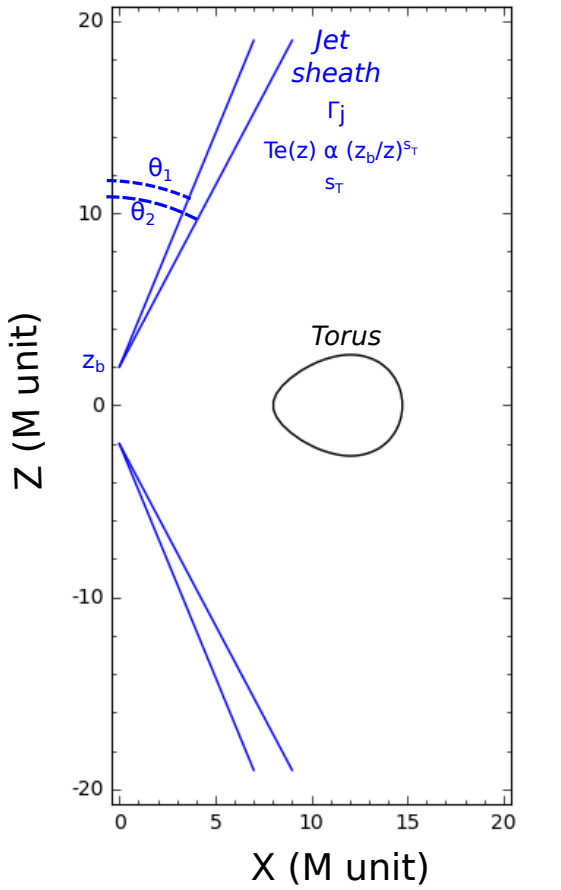}}
    \caption{Schéma du modèle de tore+jet pour l'état quiescent de Sgr~A*. Le jet est axisymétrique et symétrique par rapport au plan équatorial. Crédit : \cite{Vincent2019}.}
    \label{fig:modèle_quiescent}
\end{figure}

On modélise donc l'état quiescent de Sgr~A* par deux objets distincts :
\begin{itemize}
    \item[$\bullet$] Un \textbf{tore} modélisant le disque d'accrétion géométriquement épais. La géométrie exacte du tore est définie par le bord interne $r_{in}$ et le moment angulaire $l = - u_\varphi/u_t$ où $\va{u}$ est la 4-vitesse du fluide. Pour tous les détails, on encourage le lecteur à se référer à \cite{Vincent2019}. On considère une émission synchrotron d'une distribution thermique à la température $T_e^T$ et avec une densité $n_e^T$ (l'exposant $^T$ indique que la quantité est celle du tore). Ces deux quantités dépendent de leur valeur au centre du tore défini par les paramètres $T_e^{T\ \text{cen}}$ et $n_e^{T\ \text{cen}}$ ainsi que leur profil d'évolution dans le tore défini par l'indice polytropique $k$. Enfin, comme pour le point chaud, le champ magnétique est considéré comme mélangé et la norme est définie par la magnétisation $\sigma^T$.
    
    \item[$\bullet$] Un \textbf{jet} relativiste, les simulations GRMHD montrent que la quasi-totalité de l'émission issue du jet provient de sa gaine, c'est-à-dire du bord extérieur \citep{Moscibrodzka2013, Ressler2017, Davelaar2018}. En effet, la densité dans le cœur du jet est extrêmement faible et atteint la limite inférieure des simulations. Les fortes lignes de champ magnétique agissent comme une barrière empêchant le plasma d'entrer dans le jet. Il est alimenté en matière par un des deux processus décrits précédemment. Dans le cas de disque d'accrétion magnétisé, comme c'est le cas dans les simulations citées précédemment, c'est le processus Blandford-Znayek qui génère le jet. La gaine du jet se caractérise par une magnétisation proche de l'unité, alors que le cœur de ce dernier est fortement magnétisé (à cause de la faible densité) et que la magnétisation du disque est le plus souvent inférieure à 1. Il s'agit donc d'une zone de transition entre le disque (ou la magnétosphère) et le cœur du jet. Pour notre modèle, on se limite donc à cette gaine dont on définit la géométrie à travers trois paramètres : l'angle d'ouverture (bord interne) $\theta_1$ et de fermeture (bord externe) $\theta_2$ et la position de la base du jet $z_b$ comme illustré dans la Fig.~\ref{fig:modèle_quiescent}.
    
    La distribution des électrons dans la gaine du jet (que l'on va à partir de maintenant simplifier par "le jet") est une distribution kappa, en effet, on considère que lors de leur injection dans le jet avec un facteur de Lorentz de groupe $\Gamma^J$\footnote{Permettant de déterminer la vitesse du plasma dans le jet.} (l'exposant $^J$ indique que la quantité est celle du jet), les électrons ont été accélérés, résultant en deux composantes, l'une thermique et l'autre non thermique. Ce choix est motivé par les observations dans le domaine IR de l'état quiescent de Sgr~A* qui pointent vers une source non thermique. Le rayonnement synchrotron a pour effet de diminuer l'énergie des électrons. Ainsi, en principe, après un certain temps, les électrons non thermiques devraient avoir refroidi, tout comme la température doit diminuer au cours du temps. Cependant, on considère ici une distribution constante au cours du temps, ce qui nécessite un processus de chauffage afin de contrer le refroidissement synchrotron. La reconnexion magnétique turbulente comme décrite dans certaines simulations PIC \citep{Nättilä2021} est un processus possible permettant de maintenir une distribution kappa dans la gaine du jet.
    
    La température et la densité sont calculées à partir de leur valeur à la base du jet $T_e^J$ et $n_e^J$ respectivement et de leur profil le long du jet. La densité évolue en fonction du rayon cylindrique $n_e \propto r_{cyl}^{-2}$ alors que la température évolue en fonction de la distance verticale $T_e \propto z^{-s_T}$ avec $s_T$ un paramètre du modèle.
\end{itemize}

\begin{figure}
    \centering
    \resizebox{\hsize}{!}{\includegraphics{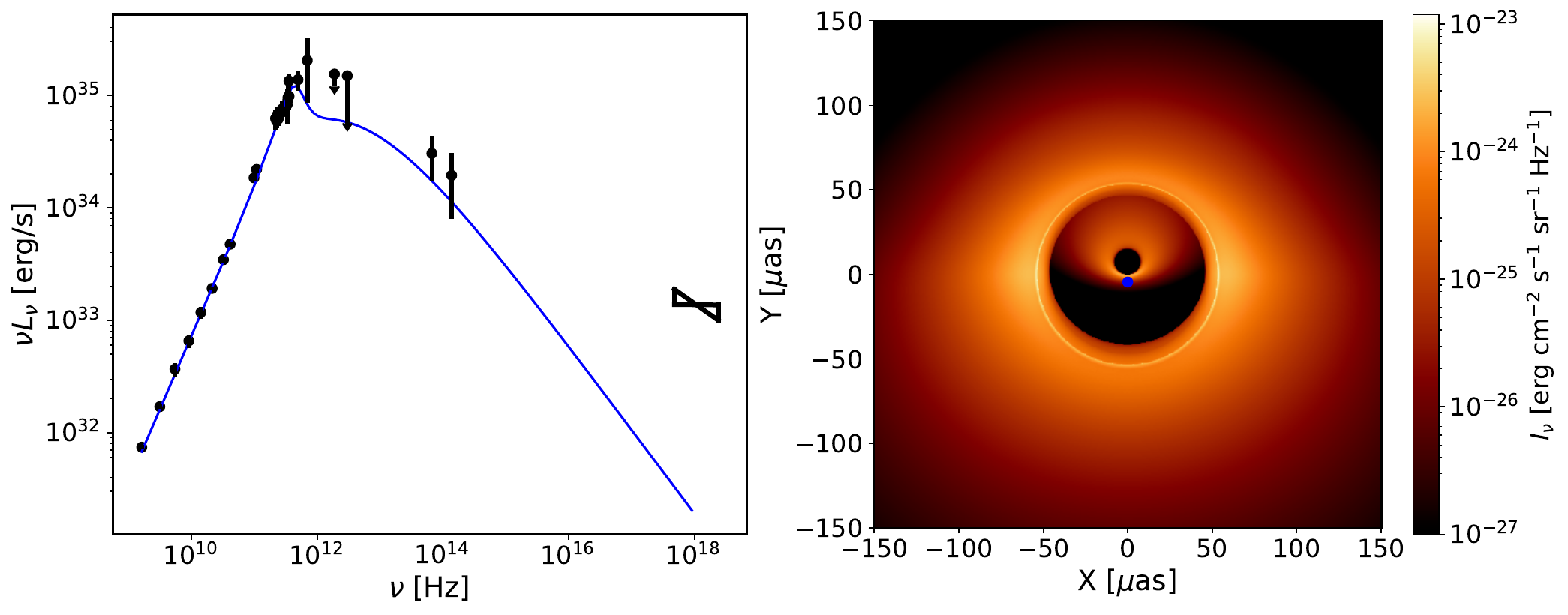}}
    \caption{\textbf{Gauche :} Spectre associé au meilleur ajustement du modèle de tore+jet (voir Table~\ref{tab:quiescent table params}) pour l'état quiescent de Sgr~A* ($\chi^2_{red}=0,91$ avec ndof=27). Les données proviennent de \cite{Bower2015} pour $\nu < 50$ GHz, \cite{Brinkerink2015} pour les deux points autour de 100 GHz, \cite{Liu2016} pour le point à 492 GHz, \cite{Marrone2006} pour le point 690 GHz, \cite{von_Fellenberg2018} pour les limites supérieures dans l'infrarouge lointain, \cite{Witzel2018} pour les données dans l'infrarouge moyen, et \cite{Baganoff2001} pour le nœud papillon en rayons~X. On note que, comme pour \cite{torus+jet}, les données relatives aux rayons~X n'ont pas été ajustées, car nous n'avons pas pris en compte l'émission Bremsstrahlung ou Compton inverse. \textbf{Droite :} Image du meilleur ajustement à $2,2\, \mu$m du modèle de tore+jet avec un champ de vision de $300\, \mu$as vu avec une inclinaison de $20 \degree$ et un PALN de $\pi$ rad. La barre de couleur donne les valeurs de l'intensité spécifique en unités cgs en échelle logarithmique. L'émission de la région extérieure provient du jet arrière, et l'émission proche du centre provient de la partie avant du jet. Le centroïde du jet est représenté par le point bleu à $\sim$($0,-2,2$).}
    \label{fig:fit_quiescent_spectrum}
\end{figure}

\cite{Vincent2019} ont utilisé ce modèle de tore+jet pour ajuster le spectre quiescent de Sgr~A* allant du domaine radio jusqu'en IR en ajustant la densité et la température centrales du tore et de la base du jet, le rayon interne du tore, l'indice de loi de puissance de la température dans le jet et l'indice kappa de la distribution du jet pour une inclinaison de $20\degree$ et $70\degree$. Les observations de \cite{Gravity2018} et \cite{EHT2022a} favorisant une faible inclinaison, on ne s'intéresse qu'au cas $i=20\degree$. Le but d'un tel modèle est d'ajuster les données radio très basses fréquences et IR majoritairement par le jet et la bosse observée dans le domaine submillimétrique par le tore, voir Fig.~\ref{fig:fit_quiescent_spectrum}. La pente dans le domaine radio est majoritairement dominée par l'indice de la loi de puissance sur la température $s_T$ alors que la pente à "hautes" fréquences, dans le domaine infrarouge, est, elle, déterminée principalement à partir de l'indice $\kappa$ de la distribution des électrons via l'Eq.~\eqref{eq:spectral_index}. Contrairement à \cite{Vincent2019}, on considère une magnétisation du jet (plus exactement sa gaine) $\sigma^J = 1$, plus en accord avec les simulations GRMHD \citep{Moscibrodzka2013, Ressler2017, Davelaar2018}. Après avoir relancé l'algorithme d'ajustement avec cette nouvelle valeur de $\sigma^J$, en laissant la densité et la température de la base du jet ainsi que la pente de température comme paramètres libres et en fixant tous les autres, on obtient les valeurs résumées dans la Table~\ref{tab:quiescent table params} ainsi que le spectre et l'image à $2,2$ $\mu$m faits avec \textsc{GYOTO} du modèle tore+jet dans la Fig.~\ref{fig:fit_quiescent_spectrum}. Comme l'émission du disque est négligeable à $2,2$ $\mu$m, l'image de l'état quiescent à cette longueur d'onde est composée uniquement du jet, on ignore donc le tore pour la suite. Le centroïde du jet, très proche de la position du trou noir (quelques $\mu$as), est marqué par le point bleu dans le panel de droite de la Fig.~\ref{fig:fit_quiescent_spectrum}.

\begin{table}
    \centering
    \begin{tabular}{ l c r }
        \hline
        \hline
        Paramètre & Symbole & Valeur\\
        \hline
        \textbf{Trou Noir} & &\\
        masse [$M_\odot$] & $M$ & $4,297 \times 10^6$ \\
        distance [kpc] & $d$ & $8,277$\\
        spin & $a$ & $0$\\
        inclinaison [$\degree$] & $i$ & $20$ \\
        \hline
        \textbf{Tore} & & \\
        moment angulaire [$r_g/c$] & $l$ & $4$\\
        rayon interne [$r_g$] & $r_{in}$ & $8$ \\
        indice polytropique & $k$ & $5/3$\\
        densité centrale [cm$^{-3}$] & $n_e^T$ & $1,2\times 10^{9}$\\
        température centrale [K] & $T_e^T$ & $7\times 10^9$\\
        magnétisation & $\sigma^T$ & $0,002$\\
        \hline
        \textbf{Jet} & & \\
        angle d'ouverture interne [$\degree$] & $\theta_1$ & $20$\\
        angle d'ouverture externe [$\degree$] & $\theta_2$ & $\theta_1+3,5$\\
        hauteur de la base du jet [$r_g$] & $z_b$ & $2$\\
        facteur de Lorentz de groupe & $\Gamma_j$ & $1,15$\\
        densité de la base [cm$^{-3}$] & $n_e^J$ & $3,5 \times 10^6$\\
        température de la base [K] & $T_e^J$ & $3 \times 10^{10}$\\
        pente de la température & $s_T$ & $0,21$ \\
        indice $\kappa$ & $\kappa^J$ & $5,5$ \\
        magnétisation & $\sigma^J$ & $1$ \\
        \hline
        \hline
    \end{tabular}
    \caption{Paramètres de meilleur ajustement du modèle quiescent tore + jet. On conserve les mêmes paramètres géométriques, facteur de Lorentz et indice $\kappa$ que \citet{Vincent2019}, et l'on ajuste la densité et la température de la base et la pente de température du jet. Les paramètres du tore sont inchangés.}
    \label{tab:quiescent table params}
\end{table}

\begin{table}
    \centering
    \begin{tabular}{l c r}
        \hline
        \hline
        Paramètre & Symbole & Valeur\\
        \hline
        \textbf{Point chaud} & & \\
        densité max [cm$^{-3}$] & $n_e^{hs}$ & $1,05\times 10^7$\\
        température max [K] & $T_e^{hs}$ & $9,03 \times 10^{10}$\\
        temps coordonné du maximum [min] & $t_\mathrm{ref}$ & $-1$\\
        Gaussian sigma [min] & $t_\sigma$ & $30$\\
        magnétisation & $\sigma^{hs}$ & $0,01$\\
        $\kappa$-distribution index & $\kappa^{hs}$ & $5$\\
        rayon orbital [$r_g$] & $R^{hs}$ & $9$\\
        angle initial azimutal [$\degree$] & $\varphi_0^{hs}$ & $90$\\
        PALN [$\degree$] & $\Omega$ & $160$\\
        \hline
    \end{tabular}
    \caption{Résumé des paramètres du modèle de point chaud. On note que nous avons utilisé pour la densité et la température maximales les valeurs ajustées du jet de la Table~\ref{tab:quiescent table params} comme référence et les avons mises à l'échelle pour le point chaud par un facteur $3,01$.}
    \label{tab:hotspot table params}
\end{table}

\subsection{Modèles alternatifs}
Comme discuté précédemment, le choix d'un modèle contenant un jet ne fait pas consensus, par exemple, la collaboration EHT ayant imagé Sgr~A* ne fait pas appel à un jet (ou dont la contribution est négligeable) pour reproduire les observations \cite{EHT2022a,EHT2022b}. De plus, notre modèle quiescent dépend d'un certain nombre d'hypothèses discutables comme le profil de vitesse du plasma dans le jet (uniforme), le choix de la distribution, la géométrie du tore et l'existence d'un jet avec un spin nul. Cette dernière est en contradiction avec les processus de génération des jets. Le choix d'un spin nul vient du précédent ajustement avec ce modèle par \cite{Vincent2019}. En effet, la géométrie du tore dépend du spin à travers le moment angulaire $l$. Le nouvel ajustement réalisé ici concerne uniquement les paramètres du jet, les paramètres du tore sont fixés, car ils ajustent correctement la bosse en sub-mm. Or changer le spin, revient à changer la géométrie du tore et nécessiterait d'ajuster les autres paramètres de ce dernier. Un ajustement sur l'ensemble des paramètres du modèle en imposant un spin non nul lèverait cette contradiction, cependant, le spin de Sgr~A* est très peu contraint, rajoutant un paramètre libre à la dizaine d'autres. Un espace des paramètres aussi grand nécessite des temps de calcul trop importants. Ce modèle étant un modèle analytique simple avec d'autres hypothèses importantes, le gain d'un tel ajustement est relativement faible compte tenu du temps de calcul nécessaire. Cela pourra être réalisé avec une meilleur modélisation à la fois du disque notamment sa géométrie, mais aussi du jet, en particulier le profil de vitesse. 

Afin de tester la fiabilité de la position du centroïde de l'état quiescent de notre modèle de tore+jet, on peut tester des modèles alternatifs. Comme on l'a dit plus haut, la présence d'un disque épais est fortement probable. On calcule donc avec \textsc{GYOTO} l'image d'un disque géométriquement épais émettant un rayonnement synchrotron thermique identique à notre modèle de tore+jet à $2,2$ $\mu $m\footnote{Ce modèle de disque épais n'a pas pour objectif d'ajuster le spectre complet.} afin de déterminer la position de son centroïde. Comme l'inclinaison est faible ($i=20$°) et que le disque est axisymétrique, le centroïde est proche du centre de l'image et donc de la position du trou noir, tout comme notre modèle, qui sont représentés par des points vert et bleu respectivement dans la Fig.~\ref{fig:influence_variabilité}. Cependant, si le disque n'est pas axisymétrique en raison par exemple d'une éjection d'une partie de ce dernier comme le montre~\cite{Ripperda2022} lors d'un évènement de reconnexion magnétique, le centroïde résultant peut-être décalé de la position du trou noir de l'ordre de $\sim 10\, \mu$as.

Malgré les limitations du modèle de tore+jet, comme on s'intéresse principalement à l'astrométrie et à la courbe de lumière, ce modèle est suffisant pour prendre en compte l'état quiescent dans notre modèle de sursaut de Sgr~A*. Néanmoins, un modèle plus évolué permettrait de dériver des valeurs plus réalistes des paramètres physiques comme la densité et la température.

\subsection{Impact du quiescent sur l'astrométrie} \label{sec:impact_quiescent}
L'impact de l'état quiescent sur la courbe de lumière est trivial puisqu'on rajoute un flux constant se manifestant par un simple offset. Cependant, l'impact sur l'astrométrie est beaucoup plus important et complexe. En effet, le flux de l'état quiescent est constant et le centroïde final entre toutes les images de toutes les sources (ici le jet et le point chaud) dépend du rapport de flux et de la position respective de chacune des images. On peut ainsi distinguer deux régimes. La partie de l'astrométrie dominée par le jet, c'est-à-dire que le flux du jet est plus important que le flux des images du point chaud. Le centroïde observé coïncide alors avec celui du jet. À mesure que le flux du point chaud (image primaire ou secondaire) augmente et donc que le rapport de flux entre le point chaud et le jet augmente, l'astrométrie va de plus en plus pencher vers le point chaud\footnote{Plus exactement, le centroïde des deux images du point chaud.}. Le point chaud ayant un mouvement orbital, l'astrométrie va suivre ce mouvement, mais avec un rayon plus petit puisque le jet est toujours présent et contribue au centroïde observé. Inversement, lorsque le flux du point chaud diminue, soit à cause du beaming, soit, car le maximum d'émission est passé (voir la section \ref{sec:variabilité_intrinsèque}), le rapport de flux penche de-nouveau vers le jet et donc l'astrométrie retourne vers le centroïde de ce dernier. C'est ce qui est illustré par la courbe rouge en trait plein de la Fig.~\ref{fig:influence_variabilité}.

\begin{figure}
    \centering
    \resizebox{\hsize}{!}{\includegraphics{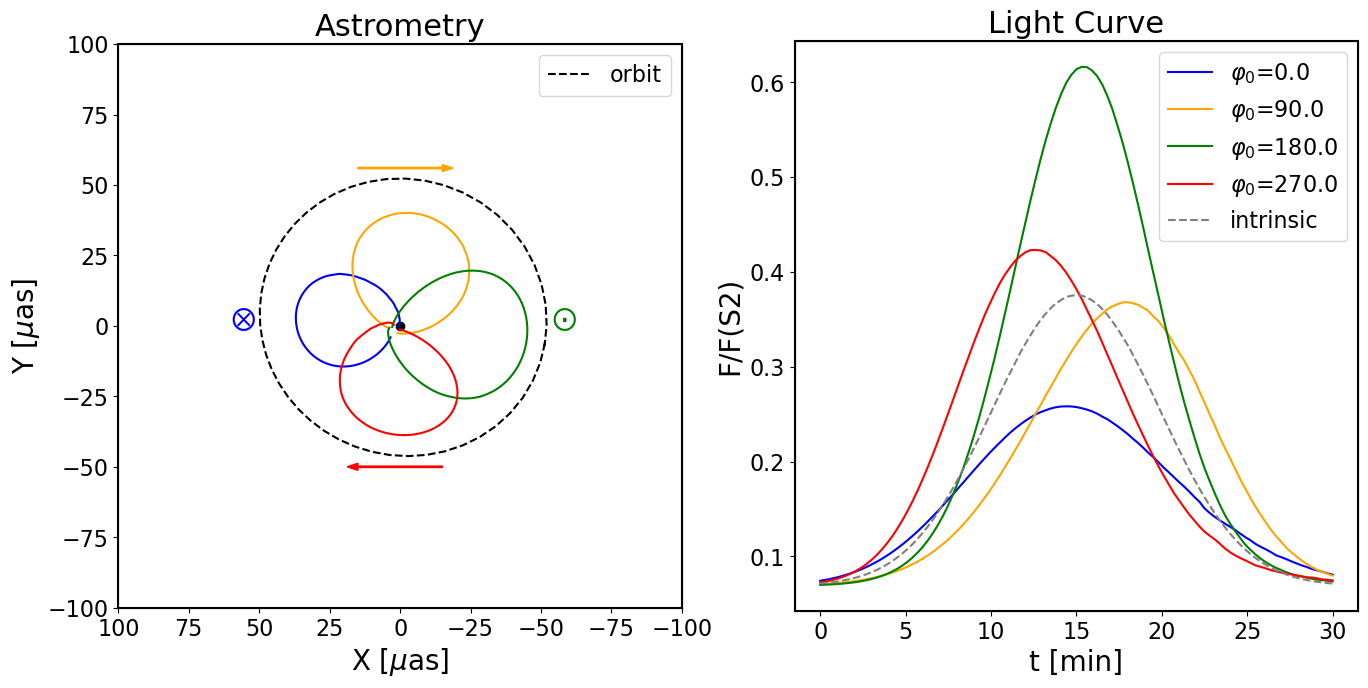}}
    \caption{Astrométrie (\textbf{gauche}) et courbes de lumière (\textbf{droite}) du modèle point chaud + jet pour quatre angles azimutaux initiaux $\varphi_0$ de $0 \degree$ en bleu, $90 \degree$ en orange, $180 \degree$ en vert, et $270 \degree$ en rouge. La ligne noire en pointillés montre la trajectoire du centroïde de l'image primaire sans le jet (dans le sens des aiguilles d'une montre). Le jet domine le début et la fin des sursauts. Les centroïdes observés commencent et finissent donc près du centroïde du jet. Les orbites apparentes tournent autour de ce dernier avec $\varphi_0$ car le maximum d'émission se produit à différents $\varphi$. La modulation Gaussienne, qui a une durée typique de $t_\sigma = 15$ min (ligne grise en pointillés, la même pour les quatre courbes) est affectée par des effets relativistes.}
    \label{fig:multi_phi0}
\end{figure}

Ainsi, pour une fenêtre d'observation plus grande que le temps caractéristique de la modulation Gaussienne du point chaud ($T_{obs} > t_\sigma$), c'est-à-dire qu'au début et à la fin de la fenêtre d'observation, c'est le jet qui domine le flux, l'astrométrie part et finit à la position du centroïde de l'état quiescent. De plus, si la période orbitale $T_{orb}$ est plus grande que $t_\sigma$, on observe alors seulement une partie de l'orbite du point chaud. Étant donné que le début et la fin de l'astrométrie correspondent au centroïde du jet, l'orientation de l'astrométrie n'est pas la même selon la phase de l'orbite observée. C'est ce qui est illustré dans la Fig.~\ref{fig:multi_phi0} où l'on peut voir quatre demi-orbites du point chaud autour du trou noir avec $t_\sigma = 15$ min et un angle azimutal initial différent ($\varphi_0 = 0$° en bleu, $\varphi_0= 90$° en jaune, $\varphi_0=180$° en vert et $\varphi_0=270$° en rouge). Les autres paramètres sont résumés dans la Table~\ref{tab:hotspot table params}. Le centre de l'astrométrie observé est donc décalé par rapport à la position du trou noir et dépend aussi de la partie de l'orbite observée. Ce décalage entre le centre de l'astrométrie et la position du trou noir est une des observations de \cite{Gravity2018}. Les trois sursauts observés ont une orientation différente par rapport à la position de Sgr~A* déduite de l'orbite de l'étoile S2.

La Fig.~\ref{fig:multi_phi0} montre aussi que le rayon de l'orbite observée ne dépend pas uniquement du rayon orbital du point chaud, mais aussi de la convolution entre la variabilité intrinsèque et le beaming. En effet, pour la courbe bleue ($\varphi_0 =0$°), le maximum d'émission survient au moment du beaming négatif comme l'atteste la courbe de lumière. Le flux du point chaud étant plus faible que dans les autres cas, le rapport de flux avec le quiescent est plus faible et donc l'astrométrie est plus proche du centroïde du jet. Par comparaison, l'orbite du point chaud est tracée dans l'astrométrie en tirets noirs.

\begin{figure}
    \centering
    \resizebox{\hsize}{!}{\includegraphics{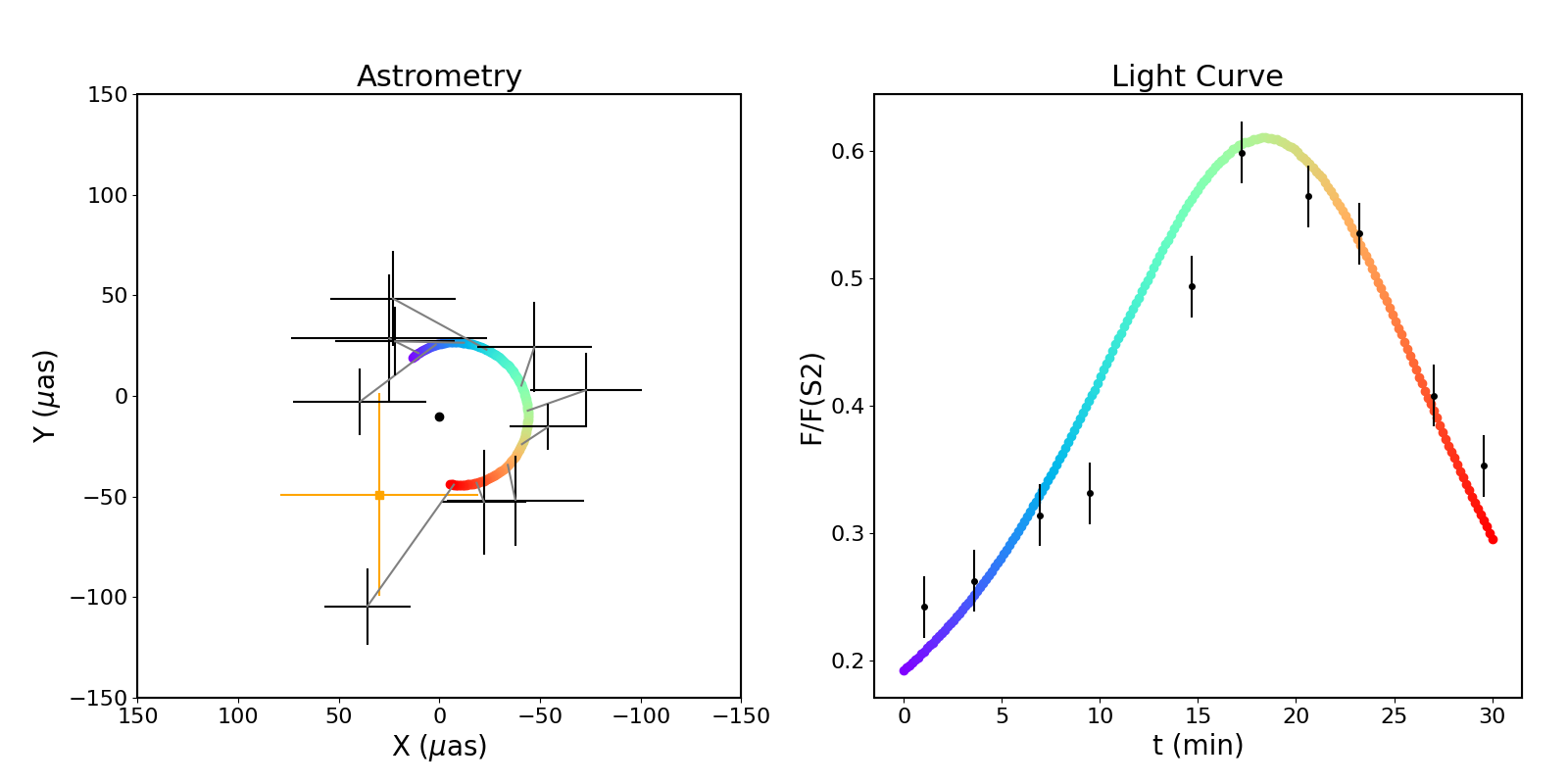}}
    \caption{Astrométrie (\textbf{gauche}) et courbe de lumière (\textbf{droite}) du modèle de point chaud + jet comparées aux données en noir du sursaut du 22 Juillet 2018 \cite{Gravity2018}. La position du trou noir à partir de l'orbite de S2 est marquée par la croix orange dans le panel de gauche. La couleur encode le temps.}
    \label{fig:Flare_22_06_18_hotspot}
\end{figure}

Les cas présentés précédemment dans les Fig.~\ref{fig:influence_variabilité} et \ref{fig:multi_phi0} sont des cas particuliers où les valeurs des paramètres ont été sélectionnées afin de présenter de manière la plus extrême possible les effets de la variabilité intrinsèque et de l'état quiescent sur les observables. Dans la réalité, la fenêtre d'observation commence la plupart du temps une fois que le sursaut a commencé, donc déjà dans la phase de croissance où son flux est supérieur à celui du quiescent. Sgr~A* n'est pas la seule source d'intérêt du centre galactique observé avec GRAVITY. L'observation et la mesure d'astrométrie des étoiles~S sont un sujet d'étude très important ayant déjà abouti à d'importants résultats. Le suivi de ces étoiles est donc tout aussi important que l'observation des sursauts. Cependant, lors d'observation de ces étoiles, Sgr~A* n'est, à quelques exceptions près, pas dans le champ de vue. Donc, bien qu'il soit possible d'observer le sursaut jusqu'à ce que Sgr~A* retourne à son état quiescent, la contrainte du temps d'observation disponible et les différents sujets d'étude de la Collaboration GRAVITY font que l'on change de cible lorsque le flux a fortement diminué (mais pas nécessairement retourné au niveau du quiescent). On n'observe donc pas le départ et l'arrivée de l'astrométrie à la position du centroïde du quiescent. Le temps caractéristique intrinsèque du sursaut est a priori très variable et peu contraint. Néanmoins, le décalage de l'orbite apparente par rapport au trou noir et l'orientation dans le ciel sont toujours présents, bien que moins remarquables. À partir d'un choix particulier de paramètres, et en connaissant les effets décrits précédemment, l'astrométrie et la courbe de lumière du sursaut du 22 Juillet 2018 peuvent être reproduites de manière satisfaisante avec le modèle de point chaud + jet (voir Fig.~\ref{fig:Flare_22_06_18_hotspot}). Les paramètres utilisé résumé dans la Table~\ref{tab:hotspot table params} ne sont pas le résultat d'un algorithme d'ajustement, qui améliorerait sans aucun doute le résultat. Le problème de la vitesse super-Képlérienne n'est pas traité avec ce modèle relativement simple qui permet néanmoins de comprendre les différents effets qui s'appliqueront pour un modèle plus sophistiqué.

\newpage
\thispagestyle{plain}
\mbox{}
\newpage

\chapter{La reconnexion magnétique comme source des sursauts}\label{chap:Plasmoid Flare model}
\markboth{La reconnexion magnétique comme source des sursauts}{La reconnexion magnétique comme source des sursauts}
{
\hypersetup{linkcolor=black}
\minitoc 
}

\section{La reconnexion magnétique}
Les sursauts de Sagittarius A* sont observés à la fois en NIR mais aussi en rayons~X, ce qui démontre des processus de hautes énergies. Parmi les processus physiques capables d'accélérer les particules à de très hautes énergies, la reconnexion magnétique est un candidat très prometteur. Ce type d'événement a été notamment étudié pour les éruptions solaires \cite{Mandrini1996,Zhu2016} et depuis quelques années fait l'objet d'études intensives (voir section~\ref{sec:reconnexion_BH} pour les références) dans les environnements d'objets compacts comme mécanisme d'accélération associé à des émissions de hautes énergies (rayons~X et gamma).

\subsection{Concept général}
\begin{figure}
    \centering
    \resizebox{\hsize}{!}{\includegraphics{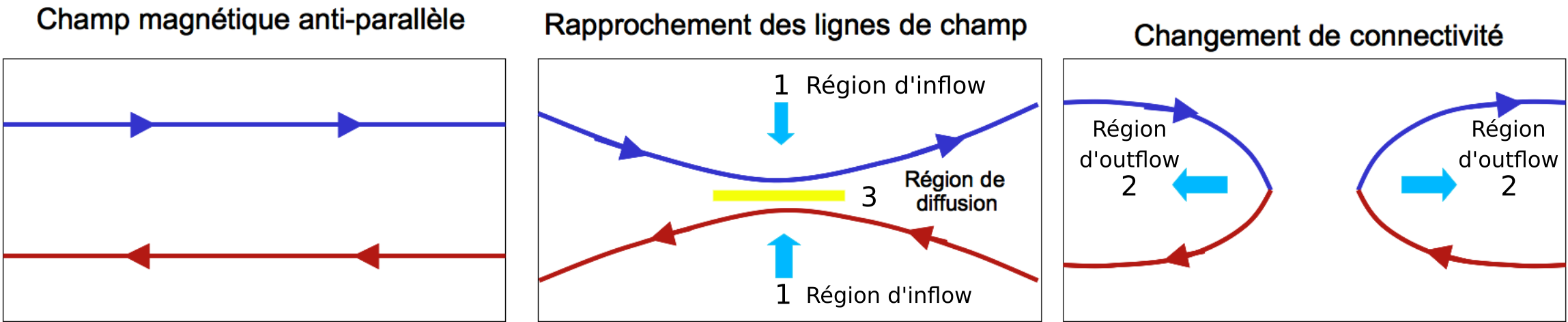}}
    \caption{Vue schématique de la reconnexion magnétique en géométrie 2D. (\textbf{Gauche}) Configuration initiale avec des lignes de champ magnétique horizontales anti-parallèles. Une nappe de courant va se former entre les deux. (\textbf{Centre}) La faible pression magnétique fait rapprocher les lignes de champ entraînant le plasma avec elles. La région de dissipation est marquée en jaune. (\textbf{Droite}) Après reconnexion, les lignes de champ sont éjectées de la zone de dissipation par les forces de tension magnétique. Crédit~:~\href{http://sesp.esep.pro/fr/pages_etoile-planete/html_images/envimage17.html}{Obs. de Paris}.}
    \label{fig:reconnexion}
\end{figure}

La reconnexion magnétique correspond à un changement de topologie des lignes de champ magnétique, ayant pour conséquence d'accélérer les particules dans la région de reconnexion. Cela se produit notamment lorsque deux lignes de champ magnétique antiparallèles se rapprochent l'une de l'autre comme illustré dans la Fig.~\ref{fig:reconnexion} présentant le modèle stationnaire de reconnexion magnétique de Sweet-Parker~\cite{Sweet1958, Parker1957, Zweibel2009}. Dans ce phénomène physique, on distingue trois zones : la première est la zone dite d'\textit{inflow} (flux d'entrée), qui correspond à la zone où les lignes de champ, entraînant le plasma, s'approchent du site de reconnexion ; la seconde est la zone dite d'\textit{outflow} (flux sortant), où les lignes de champ et le plasma sont expulsées du site de reconnexion ; et la dernière est la région où a lieu la reconnexion que l'on nomme pour le moment région 3. Dans les régions d'inflow et d'outflow (régions 1 et 2 respectivement dans les panneaux du milieu et de droite de la Fig.~\ref{fig:reconnexion}), le plasma est fortement conducteur, c'est-à-dire $\sigma_0 \rightarrow \infty$. Au niveau du site de reconnexion, le champ magnétique est nul ($B \rightarrow 0$). Cependant, le champ électrique $E$ est uniforme (en première approximation) quelque soit la région. Pour lier les différentes régions, on peut utiliser la loi d'Ohm
\begin{equation}\label{eq:Ohm}
    J = \sigma_0(E+\frac{1}{c} v \times B)
\end{equation}
où $j$ est le courant électrique et $v$ la vitesse du plasma, en prenant $E=\mathrm{cste}$. Ainsi, dans les régions d'inflow et d'outflow, où $\sigma_0 \rightarrow \infty$, l'Eq.~\eqref{eq:Ohm} devient :
\begin{equation}\label{eq:Ohm_inflow}
    E = -\frac{1}{c} v \times B.
\end{equation}
Dans la région du site de reconnexion (région 3), l'Eq.~\eqref{eq:Ohm} devient :
\begin{equation}\label{eq:Ohm_currentsheet}
    E = \frac{1}{\sigma_0} J.
\end{equation}
Cette région porte le nom de \textit{nappe de courant} du fait du fort courant électrique qui la constitue et de sa finesse. Le courant électrique s'exprime, grâce à la loi de Maxwell-Ampère, en fonction du gradient du champ magnétique tel que
\begin{equation} \label{eq:Maxwell-Ampère}
    \vec{J} = \frac{c}{4 \pi}\vec{\nabla} \times \vec{B}
\end{equation}
où l'on néglige le courant de déplacement $\frac{\partial \vec{E}}{\partial t}$. L'Eq.~\ref{eq:Ohm_currentsheet} devient donc
\begin{equation}\label{eq:Ohm_currentsheet_bis}
    E = \frac{1}{\sigma_0} \frac{c}{4 \pi} \frac{B}{\delta}
\end{equation}
où l'on fait l'approximation que $\vec{\nabla} \times \vec{B} = B/\delta$ avec $\delta$ l'épaisseur de la nappe de courant. Le champ électrique étant uniforme, on peut égaliser les équations~\eqref{eq:Ohm_inflow} et \eqref{eq:Ohm_currentsheet_bis}, ce qui donne, en considérant la région d'inflow avec une vitesse $v_\mathrm{in}$,
\begin{align}\label{eq:v_in}
    \frac{1}{c} v_\mathrm{in} B &= \frac{1}{\sigma_0} \frac{c}{4 \pi} \frac{B}{\delta} \\
    v_\mathrm{in} &= \frac{c^2}{4 \pi \sigma_0} \frac{1}{\delta} \\
    &= \frac{\eta}{\delta}
\end{align}
où $\eta = \frac{c^2}{4 \pi \sigma_0}$ est la diffusivité magnétique.

\begin{figure}
    \centering
    \resizebox{0.6\hsize}{!}{\includegraphics{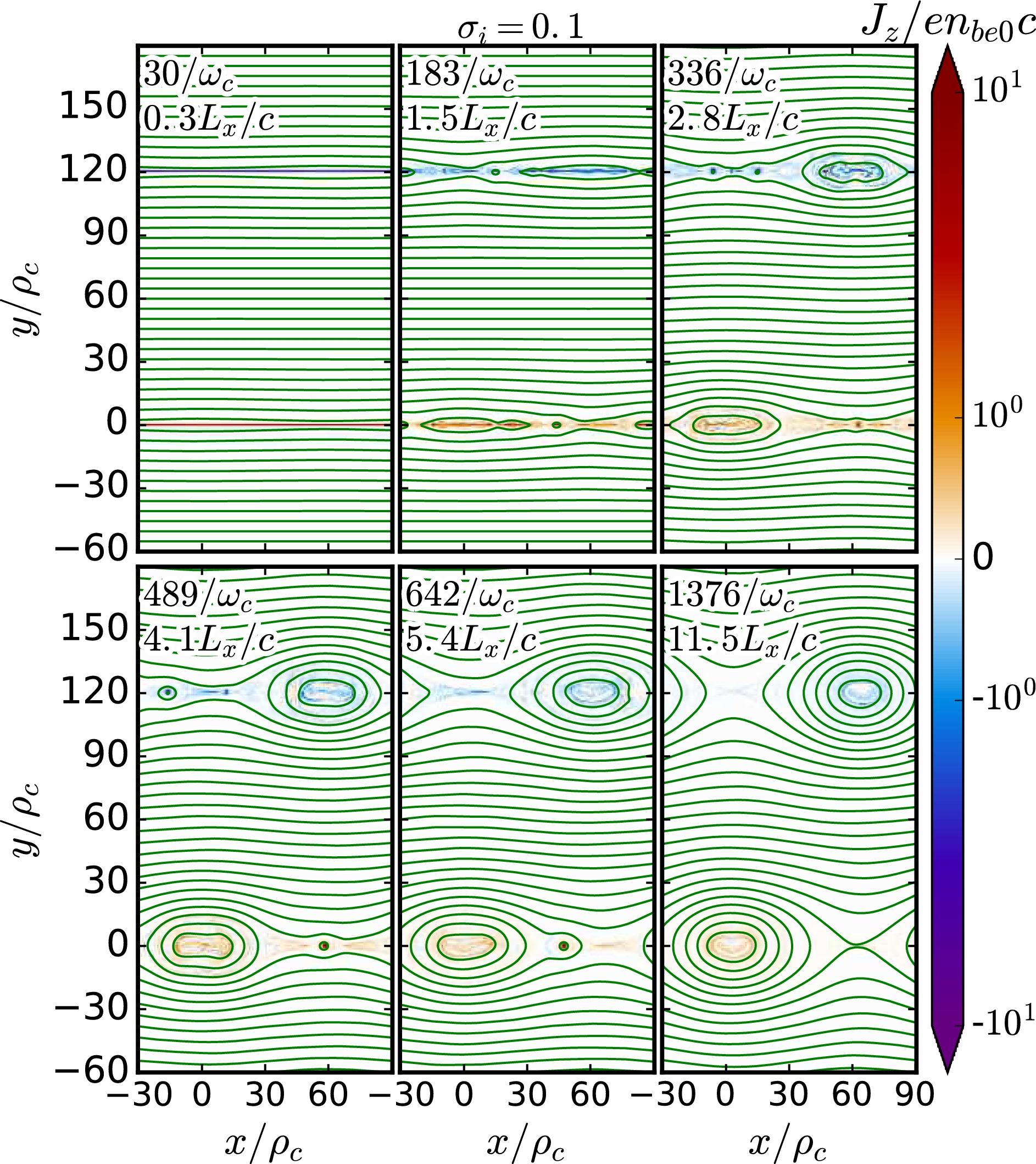}}
    \caption{(Pour $\sigma = 0,1$) Lignes de champ magnétique dans le plan (lignes vertes) et densité de courant hors plan $J_z$ (rouges/bleues) à différents moments, exprimées en $1/\omega_c$ et aussi en termes de temps de propagation de la lumière $L_x/c$ ; Le champ magnétique commence par une petite perturbation (en haut à gauche) de champ magnétique inversé et évolue par reconnexion jusqu'à un état presque stable (en bas à droite). Crédit~:~\cite{Werner2018}.}
    \label{fig:PIC_reconnexion}
\end{figure}

Lors de la reconnexion, la masse se conserve, on peut ainsi relier le flux entrant dans la nappe de courant au flux sortant, c'est-à-dire
\begin{equation}\label{eq:equal_mass_flux}
     v_\mathrm{in} L =  v_\mathrm{out} \delta
\end{equation}
où $v_\mathrm{out}$ est la vitesse des particules dans la région d'outflow et $L$ est la demi-longueur de la nappe de courant. Si la taille de la zone où les lignes de champ magnétique sont antiparallèles est plus grande que $2L$, alors il se formera plusieurs sites de reconnexion. On peut le voir dans le panneau en haut à droite de la Fig.~\ref{fig:PIC_reconnexion}, qui montre l'état d'une simulation \textit{Particule-In-Cells} (PIC) de reconnexion magnétique (voir détails dans la section~\ref{sec:simu_GRPIC}), où se sont formés deux sites de reconnexion appelés points-X en $y/\rho_c=120$, $x/\rho_c=0$ et $x/\rho_c=90$.

La reconnexion magnétique est un moyen efficace pour convertir de l'énergie magnétique en énergie cinétique. Pour déterminer la vitesse d'outflow, on peut relier la pression magnétique en amont de la reconnexion, c'est-à-dire dans la région d'inflow, à la pression dynamique des particules quittant la reconnexion dans la région d'outflow tel que 
\begin{align}\label{eq:v_out}
    \frac{B^2}{8 \pi} &= \rho v_\mathrm{out}^2 \\
    v_\mathrm{out}^2 &= \frac{B^2}{8 \pi \rho} = \frac{1}{2} v_A^2
\end{align}\label{eq:eguale_pression}
avec $\rho=m_p n_e$ la masse volumique du plasma dans la région d'outflow (égale à la densité de la région d'inflow) où $m_p$ est la masse des protons et $n_e$ la densité d'électron. On trouve ainsi que la vitesse du flow sortant du site de reconnexion est\footnote{A un facteur $1/2$ près.} la vitesse d'Alfvén, qui dans un plasma fortement magnétisé est proche de la vitesse de la lumière dans le vide ($v_A \sim c$). La vitesse des particules accélérées est donc relativiste. 

Il est utile de définir le taux de reconnexion sans dimension $R$ (crucial pour la suite) correspondant au rapport entre la vitesse d'entrée $v_\mathrm{in}$ et la vitesse de sortie $v_\mathrm{out}$ qui, à partir de l'Eq.~\eqref{eq:equal_mass_flux} vaut 
\begin{equation}\label{eq:R_1}
    R = \frac{v_\mathrm{in}}{v_\mathrm{out}} = \frac{\delta}{L}.
\end{equation}
Cependant, en utilisant les Eq.~\eqref{eq:v_in} et \eqref{eq:v_out}, on obtient une seconde expression du taux de reconnexion sans dimension
\begin{equation}\label{eq:R_2}
    R = \frac{v_\mathrm{in}}{v_\mathrm{out}} = \frac{\eta}{\delta v_A}.
\end{equation}
En faisant \eqref{eq:R_1} $\times$ \eqref{eq:R_2}, on obtient
\begin{equation}
    R = \sqrt{\frac{\delta}{L} \frac{\eta}{\delta v_A}} = \sqrt{\frac{\eta}{L v_A}} = \frac{1}{\sqrt{S}}
\end{equation}
exprimé en fonction du nombre de Lundquist $S = \frac{L v_A}{\eta}$.

\subsection{Reconnexion magnétique autour des trous noirs} \label{sec:reconnexion_BH}
La question que l'on peut se poser désormais est comment et où ce processus peut se produire autour des trous noirs. La réponse à cette question dépend de la configuration magnétique que l'on considère. L'étude de ce phénomène autour de trous noirs se fait par l'intermédiaire de simulations numériques (voir Sect.~\ref{sec:simulations}).

\begin{figure}
    \centering
    \resizebox{\hsize}{!}{\includegraphics{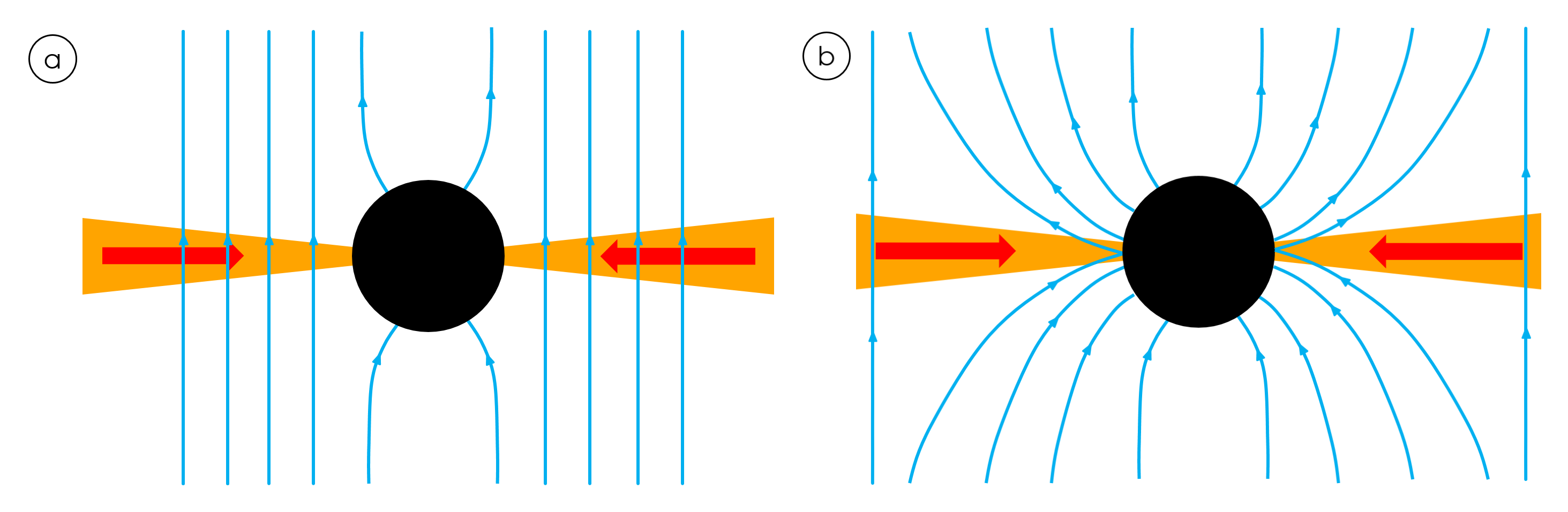}}
    \caption{Schéma de l'évolution des lignes de champ magnétique, en présence d'un disque d'accrétion, dont la configuration initiale est verticale (\textbf{panneau a}). Les lignes de champ sont advectées avec la matière dans le plan équatorial, menant à la création d'une nappe de courant dans ce dernier en raison de l'inversion de la polarité des lignes de champ (\textbf{panneau b}).}
    \label{fig:vertical_B}
\end{figure}

Commençons par une configuration où les lignes de champ magnétique sont initialement verticales, comme illustré par le panneau de gauche de la Fig.~\ref{fig:vertical_B}. En présence d'un disque d'accrétion, ces lignes de champ vont être d'autant plus advectées par le plasma (hypothèse du champ gelé, voir Sect.~\ref{sec:simu_GRMHD}) que la vitesse d'accrétion est importante, donc plus au niveau du plan équatorial que hors de ce dernier et à grande distance (en supposant que la densité décroisse avec le rayon et la hauteur par rapport au plan équatorial). Comme le montre le panneau de droite de la Fig.~\ref{fig:vertical_B}, après un certain temps (typiquement le temps dynamique d'accrétion $\sim 100$~$r_g/c$), on obtient des lignes de champ ancrées au niveau de l'horizon et s'étendant à l'infini avec une polarité opposée de part et d'autre du plan équatorial. Le changement de polarité au plan équatorial à proximité de l'horizon des évènements est la source d'un fort courant électrique (nappe de courant), où aura lieu la reconnexion magnétique et donc la formation de plasmoïdes. À grande distance, on retrouve des lignes de champ verticales, comme dans les conditions initiales, qui vont agir comme barrière. Ne pouvant continuer leur éjection dans le plan équatorial, les plasmoïdes sont éjectés hors du plan équatorial~\cite{Crinquand2022, Ripperda2020, Ripperda2022}.

\begin{figure}
    \centering
    \resizebox{\hsize}{!}{\includegraphics{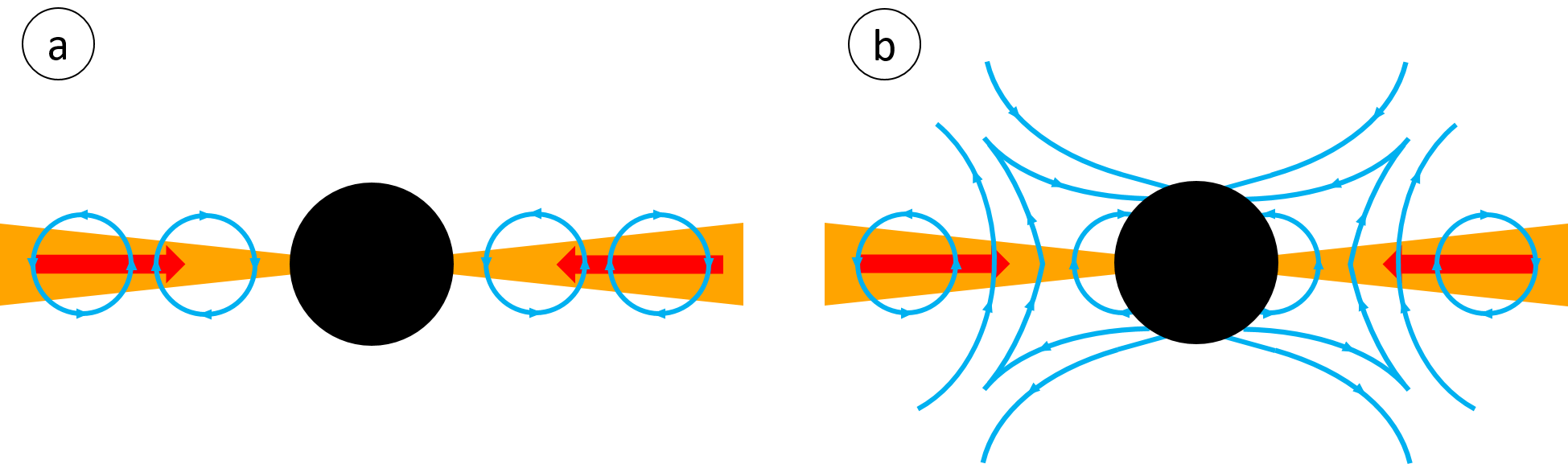}}
    \caption{Même chose qu'à la Fig.~\ref{fig:vertical_B} avec des boucles de lignes de champ magnétique comme conditions initiales (\textbf{panneau a}). Dans ces conditions, les larges boucles s'ouvrent du fait de la rotation différentielle et de l'effet Lense-Thiring. La ligne de champ fermée la plus externe reliant l'horizon et le disque d'accrétion est appelée séparatrice (\textbf{panneau b}).}
    \label{fig:loop_B}
\end{figure}

Une seconde configuration intéressante pour étudier la reconnexion magnétique est l'advection de boucles de champ magnétique poloïdales comme illustré par le panneau de gauche de la Fig.~\ref{fig:loop_B}. Lorsque le point de la boucle le plus proche du trou noir atteint l'ergosphère, la ligne de champ magnétique subit, en ce point, un fort couple dû à l'effet Lense-Thirring. Cependant, au niveau du disque d'accrétion, la vitesse azimutale de la ligne de champ est fixe et correspond à la vitesse locale du fluide (hypothèse du champ gelé), que l'on peut supposer comme Képlérienne. La composante toroïdale de la ligne de champ magnétique va donc croître dans les régions internes (au point d'ancrage au niveau de l'ergosphère) et se propager le long de la ligne de champ. Pour une boucle dépassant une certaine taille initiale, le couple généré par l'effet Lense-Thirring (qui dépend du spin) est suffisamment important pour mener à l'ouverture de la ligne de champ~\cite{Uzdensky2005, de_Gouveia_dal_Pino2005, El_Mellah2022}. Il en résulte deux lignes de champ distinctes : d'une part, celle ancrée au niveau de l'ergosphère et par extension à l'horizon des évènements du trou noir, s'étendant de l'autre côté à l'infini, et d'autre part, celle ancrée au disque d'accrétion et s'étendant également à l'infini, comme illustré dans le panneau de droite de la Fig.~\ref{fig:loop_B}. En plus de ces deux types de lignes de champ, il existe des lignes de champ magnétique liant l'horizon des évènements au disque d'accrétion dont la plus externe est appelée séparatrice. Une nappe de courant se forme entre les deux lignes de champ ouvertes au point~Y de la séparatrice où a lieu la reconnexion magnétique générant des chaines de plasmoïdes qui sont éjectées dans la magnétosphère du trou noir.

\section{Simulations numériques de reconnexion magnétique}\label{sec:simulations}
Afin d'étudier la reconnexion magnétique en général, notament autour d'objets compacts comme les trous noirs, on utilise des simulations numériques. L'objectif de cette section n'est pas de détailler le fonctionnement et toutes les équations des deux types de simulations numériques qui nous intéressent, à savoir la \textit{Magnéto-HydroDynamique en Relativité Générale} (GRMHD) et \textit{Particules-In-Cells} (en Relativité Générale) ((GR)PIC), mais d'introduire brièvement les deux approches ainsi que leurs avantages et inconvénients.

\subsection{Simulations PIC}\label{sec:simu_GRPIC}
\subsubsection{Description cinétique du plasma}
Le plasma est un état de la matière pour lequel les électrons (où tout du moins une partie) d'un atome ont été arrachés à ce dernier, formant ainsi un ensemble de particules chargées composé d'ions et d'électrons. D'une manière plus générale, on parle de plasma lorsque que l'on a un ensemble de particules chargées positivement et négativement. Un ensemble formé de paires électrons-positrons est aussi un plasma. Une caractéristique essentielle des plasmas est la \textit{quasi-neutralité}, c'est-à-dire qu'il doit y avoir autant de charges positives que de charges négatives\footnote{On note qu'à très petite échelle cette condition peut ne plus être vérifiée.}. La présence de charges électriques libres rend le plasma sensible aux champs électrique et magnétique externes. De plus, localement, des champs électrique et magnétique peuvent se former en raison de la présence et la dynamique des charges composant le plasma. Il y a donc une forte rétroaction entre les particules (ions/électrons/positrons) et les champs électrique et magnétique.

\begin{figure}
    \centering
    \resizebox{\hsize}{!}{\includegraphics{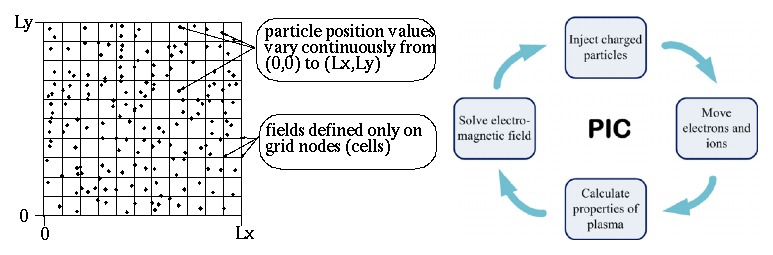}}
    \caption{(\textbf{Gauche}) Schéma d'une grille de simulation PIC. Crédit : \href{https://web.cels.anl.gov/~zippy/publications/buffer/slides/gk_3.html}{CERN WebMaker}. (\textbf{Droite}) Schéma du cycle d'intégration pour chaque pas de temps dans les simulations PIC. Crédit~:~\cite{Li2013}.}
    \label{fig:schema_PIC}
\end{figure}

Une première méthode pour décrire le plasma est une approche dite \textit{cinétique} dans laquelle on suit la dynamique d'un certain nombre de particules représentant un groupe d'électrons ou de positrons ou d'ions\footnote{Avec un traitement particulier, les photons peuvent aussi être considérés comme des particules.}. Numériquement, ces particules sont définies par leur position et leur vitesse (qui sont des variables continues) et se déplacent dans les cellules d'une grille, d'où le nom de ce type de simulation : Particules-In-Cells, illustrée dans le panneau de gauche de la Fig.~\ref{fig:schema_PIC}. Les champs électrique et magnétique sont calculés aux points de la grille et interpolés pour chaque particule afin de déterminer la force qu'elle subit qui permet, en résolvant l'équation du mouvement (Newtonienne)
\begin{equation}
    \frac{d \vec{p}}{dt} = q \left( \vec{E} + \frac{\vec{v}}{c} \times \vec{B} \right)
\end{equation}
de déterminer le déplacement des particules. Pour chaque pas de temps, le déplacement des particules (chargées) va créer un courant 
\begin{equation}
    \vec{J} = \rho \vec{V}
\end{equation}
où $\rho$ est la densité de particules ayant une vitesse $\vec{v}$, calculé à chaque point de la grille. Ce courant est ensuite utilisé pour mettre à jour la valeur des champs électrique et magnétique à travers les équations de Maxwell
\begin{align}
    \vec{\nabla} \cdot \vec{E} &= 4 \pi \rho, \\
    \vec{\nabla} \cdot \vec{B} &= 0, \\
    \frac{\partial \vec{B}}{\partial t} &= - c \vec{\nabla} \times \vec{E}, \\
    \frac{\partial \vec{E}}{\partial t} &= c \vec{\nabla} \times \vec{B} - 4 \pi \vec{J}.
\end{align}
Enfin, ces champs (interpolés) créent une accélération pour chaque particule, recommençant ainsi le cycle d'intégration illustré dans le panneau de droite de la Fig.~\ref{fig:schema_PIC}. On note que les collisions entre particules ne sont pas prises en compte, car très gourmandes en temps de calcul. Cette approche reste valable pour des plasmas dits sans collisions du fait de la très faible densité comme c'est le cas pour Sgr~A*.

\subsubsection{Avantages}
\begin{itemize}
    \item[$\bullet$] \textbf{Modélisation ab-initio} : mis à part l'hypothèse de l'absence de collision entre les particules, restreignant les cas d'utilisation, la modélisation du plasma avec cette méthodologie repose sur les lois physiques de base que sont les équations de Maxwell et l'équation du mouvement.
    \item[$\bullet$] \textbf{Capture de toute la micro-physique} : du fait de la modélisation ab-initio, tous les processus à l'échelle microscopique comme l'accélération de particules par reconnexion, ou chocs, et, dans certain cas, les processus radiatifs comme le rayonnement synchrotron, l'effet Compton inverse, ou encore la création de paires e$^{+/-}$ sont traités. Ce type de simulation est donc crucial pour construire un modèle analytique.
\end{itemize}

\subsubsection{Inconvénients}
\begin{itemize}
    \item[$\bullet$] \textbf{Coût en termes de calcul} : comme on suit la dynamique de chaque particule individuellement, le temps de calcul augmente significativement avec le nombre de particules dans la simulation\footnote{À ne pas confondre avec la densité d'électrons, de positions ou d'ions : une particule peut représenter un grand nombre d'électrons de positrons ou d'ions.}. Or, même pour des plasmas sans collisions, le nombre de particules physiques, c'est-à-dire électrons, positrons ou ions, reste très grand (de l'ordre de $10^6$~cm$^{-3}$ pour Sgr~A*). D'autre part, plus le nombre de particules dans la simulation est élevé, plus leur distribution sera précise (bien échantillonnée).
    \item[$\bullet$] \textbf{Séparation d'échelle} : dans le cas des simulations PIC en relativité générale, un problème supplémentaire apparaît. Les processus cinétiques, avec des valeurs typiques de champ magnétique autour de trous noirs ($\sim 1-100$ Gauss), se produisent à une échelle très petite (rayon de Larmor, $r_L = \frac{m v_\perp}{\vert q \vert B} \approx 170$m pour un électron avec un facteur de Lorentz $\gamma=10^3$ dans un champ magnétique de $100$ G) comparée à la longueur du rayon gravitationnel d'un trou noir de la masse de Sgr~A* ($\sim 1 \, r_g \sim 6,3 \times 10^9$m). Sans RG, il suffit de dimensionner la taille et la résolution (nombre de points de la grille) de la boite de simulation avec le champ magnétique. Lorsqu'on étudie le plasma avec une méthode cinétique autour d'objets compacts, nécessitant l'introduction de la RG, la taille caractéristique est déterminée par la masse de l'objet compact qui est macroscopique. Il y a donc une (beaucoup) trop grande différence d'échelle entre la taille caractéristique du système $r_g$ et l'échelle cinétique (de l'ordre de huit ordres de grandeur). Pour contourner ce problème, le champ magnétique est réduit de telle sorte que l'échelle cinétique puisse être résolue avec un nombre de points de grille raisonnable (par rapport au temps de calcul). Il faut a postériori faire une mise à l'échelle des grandeurs étudiées.
    \item[$\bullet$] \textbf{Évolution à court terme} : du fait du coût en termes de temps de calcul, la durée des simulations GRPIC est souvent limitée à $\sim 10-100 \, r_g/c$, temps inférieur ou proche du temps dynamique d'un flot d'accrétion, alors que les simulations de type GRMHD (voir section~\ref{sec:simu_GRMHD}) suivent l'évolution du système sur des durées de l'ordre de $\sim 10.000 \, r_g/c$.
\end{itemize}

\subsection{Simulations GRMHD}\label{sec:simu_GRMHD}
\subsubsection{Description fluide du plasma}
Une autre méthode pour décrire le plasma est de considérer une approche fluide. Au lieu de s'intéresser à chaque macro particule individuellement, on étudie le comportement d'ensemble du plasma en supposant qu'il est en équilibre thermodynamique. Les grandeurs caractéristiques restantes sont la densité de particules, la température, la vitesse du fluide (correspondant à la vitesse moyenne du groupe des particules) et le champ magnétique. La densité $N$ et la température $T$ peuvent être reliées à une distribution des particules $n(v)$, supposée thermique, de la manière suivante
\begin{equation}
    N = \int n(v) dv,
\end{equation}
\begin{equation}
    k_b T = \frac{1}{N} \int m v^2 n(v) dv.
\end{equation}
où $v$ est la vitesse des particules. 

Comme on s'intéresse au plasma dans un environnement magnétisé, les équations de la mécanique des fluides non magnétisés visqueux, comme l'équation de Navier-Stokes, ne sont pas suffisantes. En effet, pour prendre en compte l'impact du champ magnétique sur les particules de fluide ayant une masse volumique $\rho$ et une pression $p$, ainsi que la rétroaction, il faut introduire de nouveaux termes aux équations d'hydrodynamique ainsi que de nouvelles équations. Les équations de la MHD sont les suivantes : la conservation de la masse
\begin{equation}
    \frac{\partial \rho}{\partial t} + \vec{\nabla} (\rho \vec{v}) = 0,
\end{equation}
la quantité de mouvement
\begin{equation}
    \rho \left( \frac{\partial}{\partial t} + \vec{ \cdot \vec{\nabla}} \right) \vec{v} = \vec{J} \times \vec{B} - \nabla p,
\end{equation}
les équations de Maxwell-Ampere
\begin{equation}
    4 \pi \vec{J} = \vec{\nabla} \times \vec{B},
\end{equation}
et de Maxwell-Faraday
\begin{equation} \label{eq:Maxwell-Faraday}
    \frac{\partial \vec{B}}{\partial t} = - \vec{\nabla} \times \vec{E}
\end{equation}
et la loi d'Ohm Eq.~\eqref{eq:Ohm} qui combiné avec l'Eq. \eqref{eq:Maxwell-Faraday} forment l'équation d'induction
\begin{equation}\label{eq:induction}
    \frac{\partial \vec{B}}{\partial t} = \vec{\nabla} \times (\vec{v} \times \vec{B}) + \eta \nabla^2 \vec{B}.
\end{equation}
Le système d'équations à résoudre est complété par la \textit{relation de fermeture}, souvent supposée comme adiabatique
\begin{equation}
    \frac{d}{dt} \left( \frac{p}{\rho^\alpha} \right) = 0
\end{equation}
où $\alpha$ est l'indice adiabatique, bien que d'autres relations de fermeture existent.

Parmi les simulations MHD, il existe plusieurs approximations. La plus basique est l'approximation de MHD idéale dans laquelle on suppose que le plasma est un conducteur parfait, c'est-à-dire de résistance électrique nulle. La conséquence de cette hypothèse est que les lignes de champ magnétique sont liées au plasma et vice-versa, on parle de lignes de champ "gelées". Ainsi, les lignes de champ magnétique sont advectées avec l'accrétion du plasma sous l'effet de la gravité et inversement, lorsque les lignes de champ sont entraînées (par la rotation du trou noir par exemple), elles vont entraîner le plasma avec elles du fait de l'hypothèse du champ gelé. Cette hypothèse est valide lorsque l'on a un plasma fortement magnétisé avec un nombre de Reynolds magnétique $R_m$, défini comme le rapport entre l'induction magnétique et la diffusion magnétique (provenant de l'équation d'induction), tel que
\begin{equation}
    R_m = \frac{Lu}{\eta} \gg 1
\end{equation}
avec $u$ la norme du vecteur vitesse du fluide. Bien que cette hypothèse soit acceptable pour la majorité du flot d'accrétion, la reconnexion magnétique, qui est un phénomène de diffusion de flux magnétique, ne peut pas exister par construction du fait de la résistivité électrique nulle. La reconnexion magnétique étant un effet non idéal, il faut utiliser la seconde approximation, la MHD résistive (RMHD). Dans cette approximation, la résistivité électrique est non nulle et est un paramètre des simulations servant de terme de dissipation pour la reconnexion. L'ajout de ce terme complexifie significativement les équations, et nécessite une attention particulière lors de la résolution dans les simulations. Il est à noter que bien que la reconnexion soit un effet non idéal, elle peut tout de même être étudiée dans le cas de la MHD idéale. En effet, l'aspect discret des simulations numériques agit comme un terme de diffusion et dépend de la résolution de la simulation. On parle alors de diffusion numérique~\cite{Ripperda2022}.

\subsubsection{Avantages}
\begin{itemize}
    \item[$\bullet$] \textbf{Simulation à grandes échelles spatiale et temporelle} : comme on traite des grandeurs thermodynamiques dans une approche fluide, il est possible d'étudier le flot d'accrétion à très grande échelle, que ce soit à proximité de l'horizon ou à grande distance ($\sim 1000 \, r_g$). De plus, les simulations peuvent faire évoluer le flot d'accrétion(-éjection) durant de longues périodes $\sim 10,000 \, r_g/c$~\cite{Ripperda2022} avec un temps de calcul raisonnable\footnote{Dépendant de la résolution de la simulation} (comparé aux simulations PIC).
    \item[$\bullet$] \textbf{Dynamique réaliste} : comme on simule le flot d'accrétion-éjection sur de grandes échelles spatiale et temporelle, la dynamique globale du plasma est réaliste, ce qui permet de localiser les phénomènes physiques d'intérêt (reconnexion, instabilité, éjection de plasma).
\end{itemize}

\subsubsection{Inconvénients}
\begin{itemize}
    \item[$\bullet$] \textbf{Approche fluide} : ce type d'approche a deux inconvénients pour l'étude de la reconnexion magnétique pour les sursauts de Sgr~A*. Elle repose sur des grandeurs thermodynamiques supposant un équilibre des particules avec, le plus souvent, une distribution thermique. Or, cet équilibre nécessite un grand nombre de collisions entre les particules afin d'échanger de l'énergie. Cette approche s'applique donc aux plasmas avec collisions, ce qui n'est pas le cas de Sgr~A* dont la densité est de l'ordre de $10^6$ cm$^{-3}$. De plus, les processus de chauffage, que ce soit par choc ou par reconnexion, ne sont pas modélisés en MHD idéale. Or, comme on l'a vu avec les simulations PIC, la reconnexion a pour effet d'accélérer les particules à hautes énergies (non thermiques par nature).
    \item[$\bullet$] \textbf{Limite de densité} : du fait de la nature fluide de cette méthode, les zones de faible densité sont source d'erreurs numériques. Pour pallier ce problème, les simulations (GR)MHD ont un seuil de densité, généralement via une valeur maximale de la magnétisation~$\sigma$. Ce problème se pose particulièrement dans les jets qui sont par nature très peu denses.
\end{itemize}

\subsection{Contraintes provenant des simulations}\label{sec:contraintes_simus}
Au vu des avantages et inconvénients des deux approches, il est aisé de constater qu'elles sont complémentaires. Les informations que l'on peut et va extraire des simulations ne seront pas les mêmes selon qu'il s'agisse de simulations GR(R)MHD (GRMHD Résistive) ou (GR)PIC.

\begin{figure}
    \centering
    \resizebox{0.8\hsize}{!}{\includegraphics{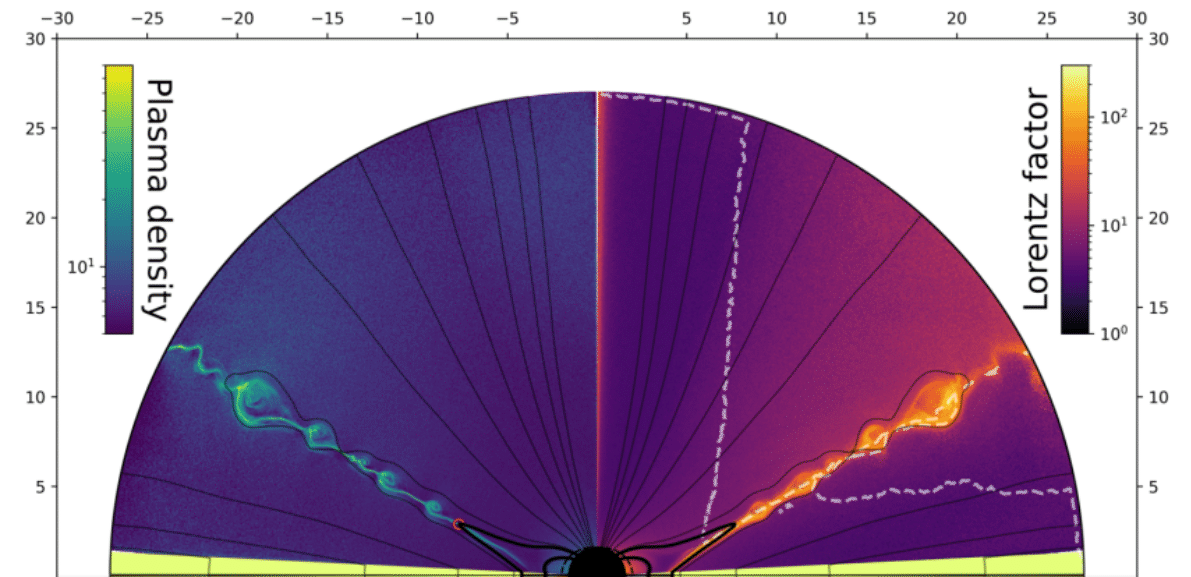}}
    \caption{Lignes de champ magnétique poloïdales autour du trou noir (lignes noires pleines), avec l'horizon des événements (disque noir) et l'ergosphère (ligne noire en pointillés), pour un instantané fiduciaire d'une simulation GRPIC de~\cite{El_Mellah2022} avec $a = 0,8$ et un disque fin (en jaune dans le plan équatorial). Les lignes de champ plus épaisses délimitent la région entre l'ISCO et la séparatrice couplant le trou noir au disque. Les cartes en couleur représentent la densité totale du plasma, $n$ (à gauche), le facteur de Lorentz moyen des particules (à droite). Pour compenser la dilution spatiale, la densité du plasma a été multipliée par $r^2$. Dans la carte de la densité du plasma, le cercle rouge localise le point Y. Les distances au trou noir sur les axes x et y sont données en unités de $r_g$. Crédit~:~\cite{El_Mellah2022}.}
    \label{fig:El_Mellah2022}
\end{figure}

\begin{figure}
    \centering
    \resizebox{0.7\hsize}{!}{\includegraphics{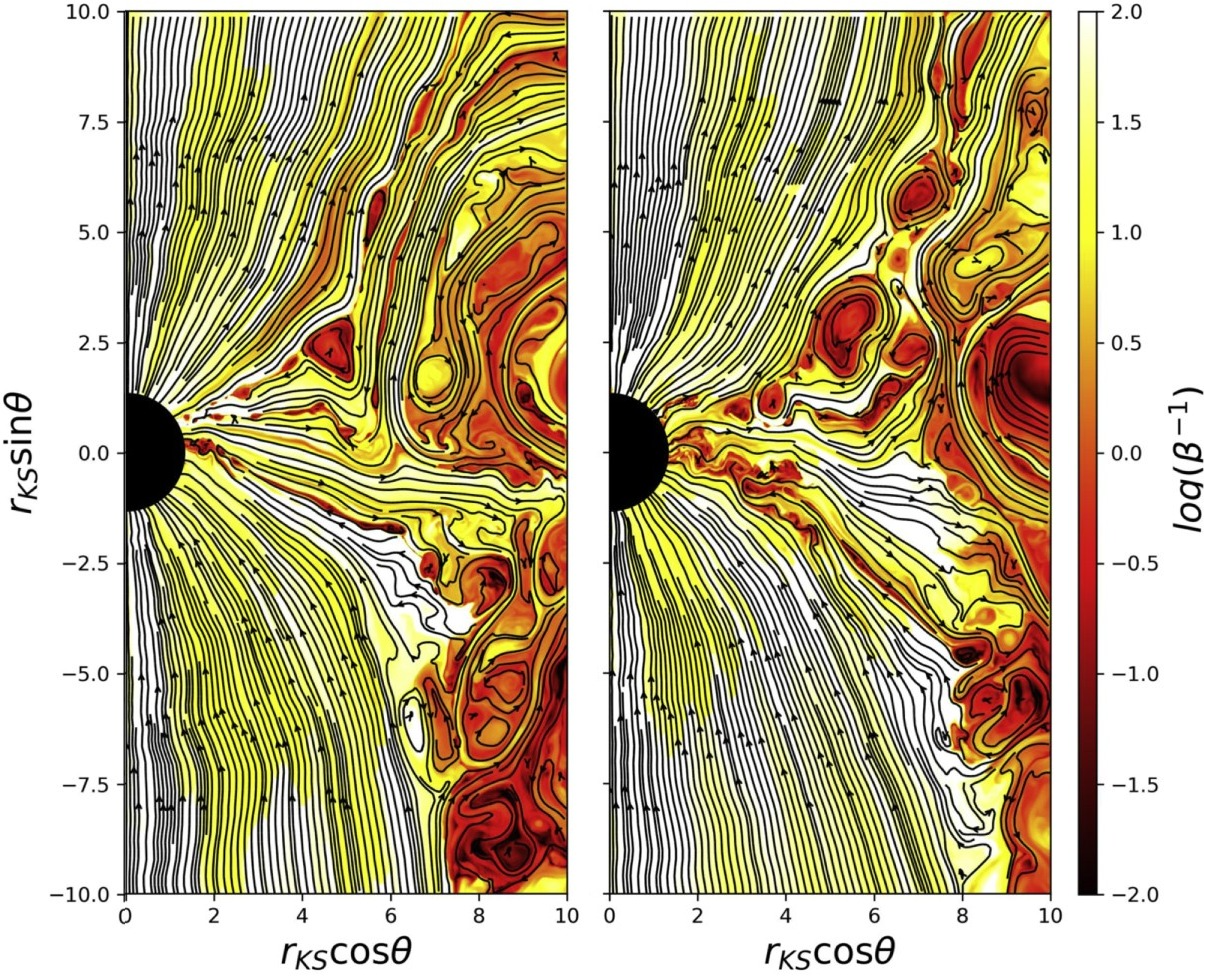}}
    \caption{$\beta^{-1}=b^2/(2p)$ pour deux temps particuliers d'une simulation 2D GRRMHD en configuration \textit{magnetically arrested disk} (MAD). Les lignes de champ magnétique sont tracées en traits pleins noirs. On distingue bien la formation de chaines de plasmoïdes proches de l'horizon qui sont éjectés le long du jet dans la partie supérieure et forment de larges structures dont la taille typique est de $\sim 1 \, r_g$. Crédit~:~\cite{Ripperda2020}.}
    \label{fig:Ripperda2020}
\end{figure}

\begin{figure}
    \centering
    \resizebox{\hsize}{!}{\includegraphics{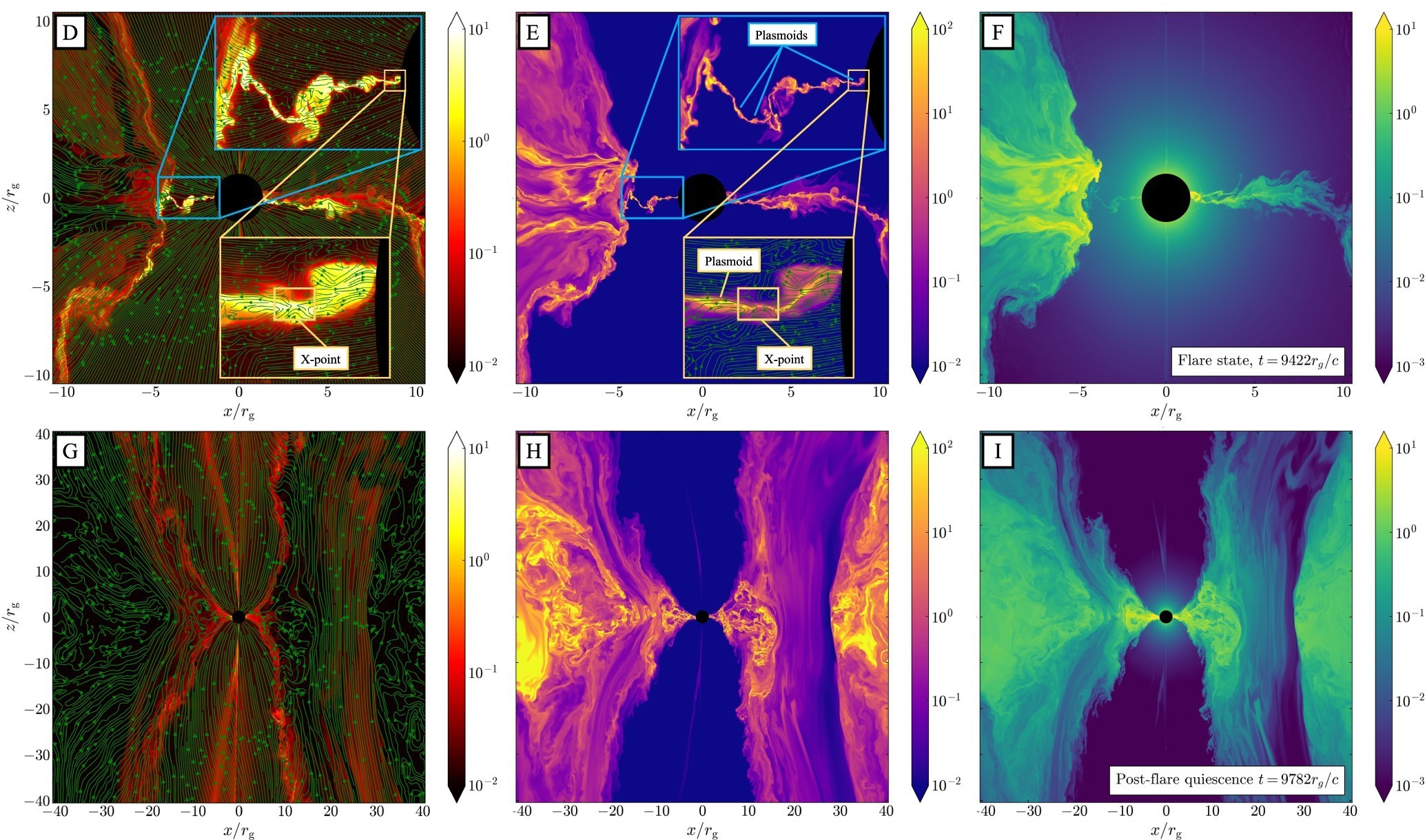}}
    \caption{Instantané du flot d'accrétion-éjection dans le plan méridional à deux temps particuliers d'une simulation 3D GRMHD, identifiés comme un état de sursaut (\textit{Flare} en anglais ; \textbf{première ligne}), et un état post-sursaut, mais pas encore quiescent (\textbf{seconde ligne}). Les colonnes montrent (\textbf{de gauche à droite}) respectivement la température sans dimension $T=p/\rho$, le plasma-$\beta$ et la densité $\rho$. On note que l'échelle n'est pas la même dans les deux lignes. Crédit~:~\cite{Ripperda2022}.}
    \label{fig:Ripperda2022}
\end{figure}

La première information qui nous intéresse est le lieu de la reconnexion, ou plus précisément, où se forment les plasmoïdes qui sont la source du rayonnement observé. Les deux types de simulations montrent, quelle que soit la configuration du champ magnétique, que la reconnexion a lieu à proximité de l'horizon à un rayon inférieur à $\sim 20 \, r_g$ qui dépend de la valeur du spin~\cite{Ripperda2020, Ripperda2022, Cemeljic2022, El_Mellah2022, Crinquand2022}. Cependant, la dynamique des plasmoïdes, appelés tubes de flux en 3D du fait de leur forme, diffère selon la configuration magnétique, comme décrit plus haut. Dans le cas de boucles de champ magnétique, les plasmoïdes se forment directement dans la magnétosphère et sont éjectés le long de la gaine du jet~\cite{El_Mellah2022} (voir Fig.~\ref{fig:El_Mellah2022}). Dans le cas de lignes de champ magnétique initialement verticales, la nappe de courant, et donc les plasmoïdes, se forment dans le plan équatorial où ils peuvent être soit advectés~\cite{Crinquand2021}, soit éjectés. L'éjection peut être directement le long du jet comme on peut le voir dans les simulations de \cite{Ripperda2020, Cemeljic2022} (voir Fig.~\ref{fig:Ripperda2020}) ou dans le plan équatorial comme dans \cite{Ripperda2022}. Cependant, dans ce dernier, un second effet de la reconnexion est l'éjection d'une partie du disque d'accrétion et la formation d'une barrière magnétique verticale (en $X\sim 22$~$r_g$ du panneau en bas à gauche de la Fig.~\ref{fig:Ripperda2022}) contre laquelle les plasmoïdes vont être déviés. Il en résulte une éjection dans la magnétosphère similaire aux autres cas décrits précédemment. Ainsi, la dynamique des plasmoïdes peut être modélisée par une éjection dans la magnétosphère en plus du mouvement orbital.

La seconde contrainte que l'on peut dériver à partir des simulations est la durée de la reconnexion. Il est néanmoins important de ne pas confondre la durée de décroissance du flux magnétique par reconnexion (qui est la quantité qui nous intéresse) avec l'efficacité de reconnexion magnétique, appelé dans la littérature taux de reconnexion (\textit{reconnection rate} en anglais), qui caractérise la conversion de l'énergie magnétique en énergie cinétique dans les particules composant le plasma. Le taux de reconnexion $\beta_\mathrm{rec}$, correspondant au rapport entre la vitesse du plasma entrant dans le site de reconnexion $v_\mathrm{in}$ et la vitesse de sortie $v_\mathrm{out} = v_A \sim c$ ($\beta_\mathrm{rec} \sim v_\mathrm{in} / c$), est très différent dans les simulations GRMHD avec une valeur de l'ordre de $\sim 0.01$ alors que les simulations PIC obtiennent une valeur d'un ordre de grandeur plus grand (donc plus rapide) $\sim 0.1$. Il est important de noter qu'il y a deux taux qui nous intéresse ici. Le taux de conversion d'énergie magnétique en énergie cinétique et le taux d'apport de flux magnétique dans la nappe de courant par le flot d'accrétion. Comme décrites dans les sections précédentes, les simulations PIC opèrent sur de courtes durées et ne prennent pas en compte, en général, la dynamique du flot d'accrétion\footnote{En tout cas pour les simulations PIC qui nous intéressent.}. Le taux de d'apport de flux magnétique dans la nappe de courant n'est donc pas pris en compte puisque \cite{El_Mellah2022,El_Mellah2023} apportent continuellement du flux magnétique à diffuser dans la nappe de courant. Le taux de conversion d'énergie magnétique en énergie cinétique, quant à lui, est réaliste dans ces simulations par construction. A contrario, les simulations GRMHD ne prennent pas en compte toute la micro-physique, le taux de conversion d'énergie magnétique en énergie cinétique n'est alors pas réaliste, contrairement au taux de d'apport de flux magnétique puisqu'il n'est pas uniquement prescrit, mais calculé depuis une condition initiale imposée et la dynamique du flot d'accrétion.
Pour résumer, la reconnexion magnétique est caractérisée par deux taux caractéristiques : le taux de conversion d'énergie magnétique en énergie cinétique (réaliste dans les simulations PIC) ; et le taux d'apport de flux magnétique (réaliste dans les simulations GRMHD), qui est gouverné par la dynamique du flot d'accrétion. C'est la durée\footnote{Qui est différente du taux} de ce dernier qui nous intéresse comme temps caractéristique.

La reconnexion magnétique est un phénomène efficace pour accélérer des particules à très hautes énergies. La distribution en énergie (ou de manière équivalente en facteur de Lorentz) des particules est un élément crucial de l'étude de la reconnexion magnétique qui peut être étudié à partir des simulations PIC (grâce à leur nature cinétique). Ces dernières suggèrent une distribution des électrons en loi de puissance à haute énergie avec un indice $p$ qui dépend de la magnétisation du plasma (voir~\ref{sec:distrib_evol}). Comme on l'a vu précédemment, l'énergie maximale atteinte par les électrons dépend aussi de la magnétisation. Or ce paramètre peut être difficile à estimer. En effet, les simulations GRPIC ont un champ magnétique très faible pour résoudre le problème de la séparation d'échelle et les simulations GRMHD souffrent d'un seuil sur la densité pour maintenir les lois de conservation. Dans les deux cas, ces limitations affectent significativement la valeur de la magnétisation. Cette dernière est donc très peu contrainte.

\section{Modèle de point chaud basé sur la reconnexion magnétique}
L'idée principale de ce modèle, et de cette thèse, est de regrouper les informations essentielles issues des simulations numériques fluides et cinétiques ainsi que les contraintes observationnelles (voir Chap~\ref{chap:Sgr~A* flares}) dans un modèle semi-analytique simple afin d'ajuster les données des sursauts observés par GRAVITY avec des temps de calcul raisonnables par rapport aux simulations GRPIC et GRMHD.

\subsection{Description générale}
\begin{figure}
    \centering
    \resizebox{0.6\hsize}{!}{\includegraphics{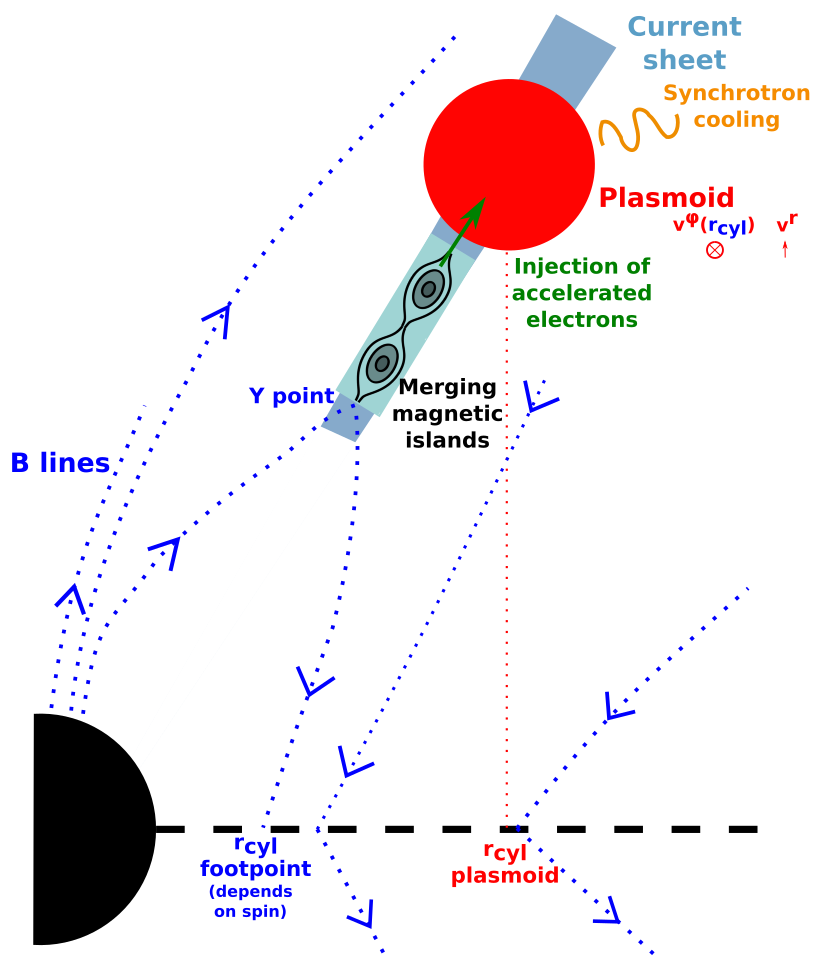}}
    \caption{Croquis de la reconnexion magnétique dans la magnétosphère du trou noir, tel qu'illustré par \citep{El_Mellah2022}, sur lequel notre modèle de plasmoïde est construit. Trois types de lignes de champ magnétique sont représentés : les lignes qui traversent l'horizon des événements, qui s'étendent à l'infini, les lignes ancrées sur le disque, qui s'étendent également à l'infini, et la séparatrice, qui relie le disque et l'horizon des événements du trou noir. Cette dernière forme un point Y et une nappe de courant dans laquelle se forment les chaînes de plasmoïdes. Nous modélisons un plasmoïde unique comme le résultat de fusions multiples.}
    \label{fig:schema_plasmoid}
\end{figure}

\textbf{Il est tout d'abord important de noter que l'on ne modélise pas la reconnexion magnétique elle-même, mais le produit final de la reconnexion, à savoir les plasmoïdes.} Pour simplifier, on ne considère qu'un seul plasmoïde macroscopique correspondant à un point chaud. De même, pour simplifier et limiter le nombre de paramètres du modèle, on considère une sphère de plasma, et non une forme cylindrique, de rayon constant, alors que les simulations montrent une augmentation de la taille à travers les fusions successives des plasmoïdes. Ainsi, aucune expansion adiabatique n'est prise en compte dans ce modèle, contrairement à d'autres modèles de sursauts \cite{Boyce2022}. Néanmoins, les fusions successives des plasmoïdes dans notre point chaud sont prises en compte sous la forme d'un apport de particules accélérées dans le point chaud. L'hypothèse d'un rayon constant peut tout de même être justifiée en considérant que le temps de formation du plasmoïde macroscopique ($\sim (1-10)\, L/c$ avec $L=1\, r_g$) \cite{Nalewajko2015, Rowan2017, Werner2018, Klion2023} est très inférieur à la durée de la reconnexion ($\sim 100 \, r_g/c$). Tout comme dans les simulations GRPIC de \cite{El_Mellah2022} et GRMHD de \cite{Ripperda2020, Ripperda2022}, où les plasmoïdes sont éjectés le long du jet, la trajectoire du point chaud est une éjection en-dehors du plan équatorial, comme illustré dans la Fig.~\ref{fig:schema_plasmoid}.

Contrairement aux simulations de GRMHD, parfois utilisées en tant que point de comparaison des données, où les quantités physiques comme la densité, la température et le champ magnétique sont évoluées à partir de conditions initiales et la distribution des électrons supposée comme thermique~\cite{Dexter2020}, pour ce modèle, on considère un schéma inverse. L'évolution des quantités physiques est analytique, mais on traite l'évolution de la distribution des électrons, responsable de l'émission du rayonnement observé, en prenant en compte l'injection de particules et leur refroidissement synchrotron à travers l'utilisation du code \texttt{EMBLEM}~\cite{dmytriiev2021}.

\subsection{Dynamique du point chaud}
On se place dans le système de coordonnées de Boyer-Lindquist (sphérique). Comme dit précédemment, on considère pour notre point chaud une éjection le long du jet. Cependant, l'angle polaire d'ouverture du jet $\theta_J$ n'est pas connu et est très dépendant du modèle considéré, comme dans notre modèle quiescent. On va donc considérer que le point chaud a un mouvement conique \cite{Ball2021}, c'est-à-dire avec un angle polaire $\theta=\theta_0$ constant (l'indice zéro indique qu'il s'agit de condition initiale). Ce paramètre est un paramètre libre du modèle, et pour le moment il n'est pas contraint (voir Sect.~\ref{sec:comparaison_obs}).

Le point chaud est le résultat de la reconnexion magnétique qui a lieu au point~Y (voir Fig.~\ref{fig:schema_plasmoid}) de la dernière (la plus externe) ligne de champ magnétique fermée appelée séparatrice. Or, cette dernière est liée au disque d'accrétion à un certain rayon cylindrique dont la vitesse azimutale peut être supposée comme Képlérienne. Cependant, le rayon projeté du point chaud est supérieur au rayon du point d'ancrage. Cette différence a pour conséquence une vitesse super-Képlérienne (voir Fig.~\ref{fig:schema_plasmoid}) du point chaud qui est entraîné par le disque d'accrétion par la séparatrice, qui agit comme une fronde. On note que le rayon du point~Y ainsi que celui du point d'ancrage de la séparatrice dépendent de la valeur du spin de Sgr A*, comme le montrent les simulations de~\cite{El_Mellah2022}. Cependant, après la reconnexion, les plasmoïdes se détachent de la séparatrice, et ne sont donc plus entraînés par le flot d'accrétion. Ainsi, on considère une vitesse azimutale initiale pour le point chaud libre (a priori super-Keplérienne). Pour déterminer la vitesse azimutale $\dot{\varphi}$\footnote{Le $\dot{}$ marquant la dérivé par rapport au temps.} à l'instant $t$, on applique la conservation du moment cinétique Newtonien
\begin{equation}\label{eq:phi(t)}
    \dot{\varphi}(t) = \dot{\varphi}_0\left( \frac{r_0}{r(t)} \right)^2
\end{equation}
où $\dot{\varphi}(t)$ est la vitesse azimutale du point chaud (en rad.s$^{-1}$) au temps coordonné $t$, $\dot{\varphi}_0$ et $r_0$ la vitesse azimutale et le rayon initial (au temps coordonné $t_0$), et $r(t)$ le rayon au temps $t$. Comme on considère une éjection, on définit une vitesse radiale positive et constante par simplicité $\dot{r}(t)=\dot{r}_0 > 0$. Cette dernière est aussi un paramètre libre du modèle. Le rayon orbital à l'instant $t>t_0$ est défini comme
\begin{equation} \label{eq:r(t)}
    r(t) = r_0 + \dot{r}_0 (t - t_0).
\end{equation}

Ainsi, toute la dynamique du point chaud, correspondant à une éjection conique, est définie par un ensemble de six paramètres : la position initiale ($t_0$, $r_0$, $\theta_0$, $\varphi_0$) et les vitesses radiale et azimutale initiales ($\dot{r}_0$, $\dot{\varphi_0}$). L'évolution des coordonnées en fonction du temps est résumée ci-dessous
\begin{equation}
    \begin{aligned}
        r(t) &= r_0 + \dot{r}_0 (t - t_0), \\
        \theta(t) &= \theta_0, \\
        \varphi(t) &= \varphi_0 + r_0^2 \frac{\dot{\varphi}_0}{\dot{r}} \left( \frac{1}{r_0} - \frac{1}{r(t)} \right), \\
        \dot{r}(t) &= \dot{r}_0, \\
        \dot{\theta}(t) &= 0, \\
        \dot{\varphi}(t) &= \dot{\varphi}_0 \left( \frac{r_0}{r(t)} \right)^2.
    \end{aligned}
\end{equation}
L'équation de $\varphi(t)$ étant le résultat de l'intégration de l'Eq~\eqref{eq:phi(t)} avec l'Eq~\eqref{eq:r(t)}.

\subsection{Évolution cinétique des particules}
La particularité du modèle de point chaud présenté ici, issu de la reconnexion magnétique, est le traitement particulier de la distribution des électrons responsables du rayonnement synchrotron observé.
\subsubsection{Phases de vie du point chaud}
Le cycle de vie du point chaud est divisé en deux phases distinctes liées à la reconnexion, une \textit{phase de croissance} et une \textit{phase de refroidissement}.

\begin{figure}
    \centering
    \resizebox{0.6\hsize}{!}{\includegraphics{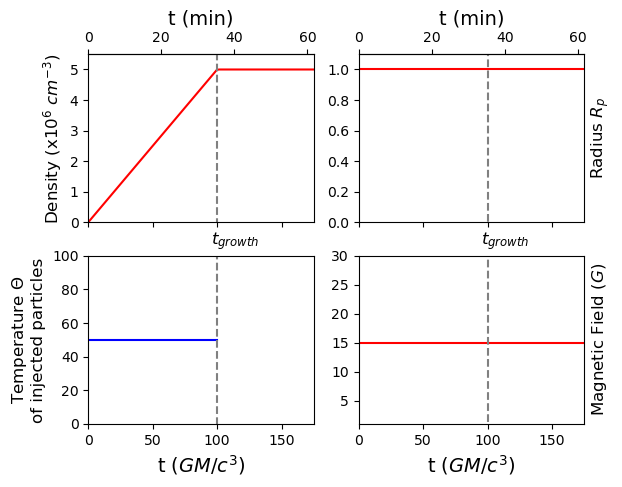}}
    \caption{Évolution temporelle de la densité (\textbf{en haut à gauche}), du rayon (\textbf{en haut à droite}), de la température sans dimension des particules injectées (\textbf{en bas à gauche}) et du champ magnétique (\textbf{en bas à droite}). Seule la densité évolue au cours du temps de façon linéaire jusqu'à atteindre son maximum $n_{e,max}$. La courbe de température sans dimension s'arrête à la fin de la phase de croissance $t = t_\mathrm{growth}$ puisqu'il n'y a plus d'injection d'électrons. Les valeurs des paramètres utilisés dans cette figure correspondent à celle de la Table~\ref{tab:emblem} pour le sursaut du 22 juillet 2018.}
    \label{fig:evol_quantité}
\end{figure}

\begin{itemize}
    \item[$\bullet$] \textbf{Phase de croissance} : Durant cette phase, on injecte à un taux constant des particules suivant une distribution kappa (cœur thermique avec une loi de puissance à haute énergie) dans le point chaud. Une fois dans le point chaud, elles refroidissent en émettant du rayonnement synchrotron. Les autres quantités physiques, la température sans dimension $\Theta_e$, l'indice kappa $\kappa$ de la distribution injectée et l'intensité du champ magnétique $B$ dans le point chaud sont des paramètres constants au cours du temps. Les deux derniers paramètres physiques principaux du modèle sont la densité maximale atteinte dans le point chaud à la fin de la phase de croissance $n_{e,max}$, qui survient après un temps de croissance $t_\mathrm{growth}$. L'évolution de ces paramètres physiques en fonction du temps est résumée dans la Fig.~\ref{fig:evol_quantité}.
    
    Ce choix d'évolution des paramètres est fait pour modéliser l'apport de particules accélérées par la nappe de courant, au point de reconnexion, dans le point chaud, via la fusion des petits plasmoïdes avec ce dernier. Le choix d'une distribution kappa comme distribution injectée dans le point chaud résulte du chauffage du plasma par la reconnexion et de l'accélération des électrons par cette dernière à très hautes énergies comme le montrent les simulations PIC de reconnexion magnétique~\cite{Selvi2022, Chernoglazov2023, Schoeffler2023, Ball2018, Hoshino2022, Johnson2022, Klion2023, Werner2018, Zhang2021}.
    
    Un paramètre important est la durée de la phase de croissance $t_\mathrm{growth}$ (voir section~\ref{sec:distrib_evol}). Ce temps caractéristique correspond à la durée de survie de la reconnexion, c'est-à-dire, la durée pendant laquelle la reconnexion produit des plasmoïdes qui vont fusionner avec le point chaud. En effet, la reconnexion correspond à de la dissipation de l'énergie magnétique dans le plasma. Cette dissipation n'est évidemment pas infinie et a un temps caractéristique qui dépend des conditions du flot d'accrétion (section~\ref{sec:contraintes_simus}). La récente simulation GRMHD en 3D de \cite{Ripperda2022} montre une durée de décroissance du flux magnétique, donc la durée de la reconnexion, de l'ordre de 100 $r_g/c$, similaire au temps caractéristique rapporté par \cite{Parfrey2015}, qui se traduit par $\sim 30$~min pour Sgr~A*.
    
    \item[$\bullet$] \textbf{Phase de refroidissement} : Bien que le refroidissement par rayonnement synchrotron soit aussi présent durant la phase de croissance, la fin de cette dernière marque la fin de tout processus de chauffage. Les électrons présents dans le point chaud ne font plus que refroidir par rayonnement synchrotron. On note que l'utilisation d'un code d'évolution cinétique de la distribution des particules implique une limite en temps de cette évolution. Pour limiter le temps de calcul, tout en ayant un intervalle de simulation suffisamment grand pour que lors de l'intégration des géodésique, on ne dépasse pas ce temps maximum, on prend un temps maximal de simulation de $400\, r_g/c$ ($\sim$~2h pour Sgr~A*).
\end{itemize}

\subsubsection{Distribution des particules en fonction du temps}\label{sec:distrib_evol}
Un aspect fondamental du modèle est le traitement de la distribution des électrons dont l'évolution est gouvernée par l'équation cinétique
\begin{equation} \label{eq:kineticeq}
\dfrac{\partial N_\text{e}(\gamma,t)}{\partial t} = \dfrac{\partial}{\partial \gamma} \left( -\dot{\gamma}_\mathrm{syn}\, N_\text{e}(\gamma,t) \right) + Q_\text{inj}(\gamma,t),
\end{equation}
où $\gamma$ est le facteur de Lorentz des électrons, $Q_\text{inj}(t)$ est le taux d'injection et $N_e = \dd n_e / \dd \gamma$ est la distribution des électrons (avec $n_e$ la densité numérique d'électrons). On distingue dans cette équation deux termes définissant l'évolution dans le temps de la distribution, un terme de refroidissement par rayonnement synchrotron
\begin{equation}
\label{eq:kinetic}
-\dot{\gamma}_\mathrm{syn} N_\text{e} = \frac{4 \sigma_T U_B}{3 m_e c} (\gamma^2 - 1) N_e
\end{equation}
avec $\sigma_T$ la section efficace de Thomson, $m_e$ la masse des électrons et $U_B = B^2/8 \pi$ (en unité CGS) et un terme d'injection $Q_{\text{inj}}(\gamma)$ pouvant être associé à du chauffage. En effet, on injecte des particules suivant une distribution kappa $N_e^\kappa(\gamma)$ (voir Eq.~\eqref{eq:kappa_distri}) tel que
\begin{equation} \label{eq:injterm}
    Q_{\text{inj}}(\gamma,t) = \left\{
    \begin{array}{ll}
        \dfrac{4 \pi N_e^\kappa(\gamma)t^{\alpha-1}}{t_\mathrm{growth}} & \mbox{durant\:la\:phase\:de\:croissance, } \\
        0 & \mbox{durant\:la\:phase\:de\:refroidissement}
    \end{array}
\right.
\end{equation}
où $\alpha$ est l'indice\footnote{Cet indice n'est pas nécessairement un entier.} de la courbe de croissance de la densité en fonction du temps $n_e(t) \propto t^\alpha$ (0~=~constant, 1~=~linéaire, 2~=~quadratique, etc). Pour la suite, on prend $\alpha=1$ compte tenu que l'on considère une évolution linéaire de la densité.

\begin{figure}
    \centering
    \resizebox{0.6\hsize}{!}{\includegraphics{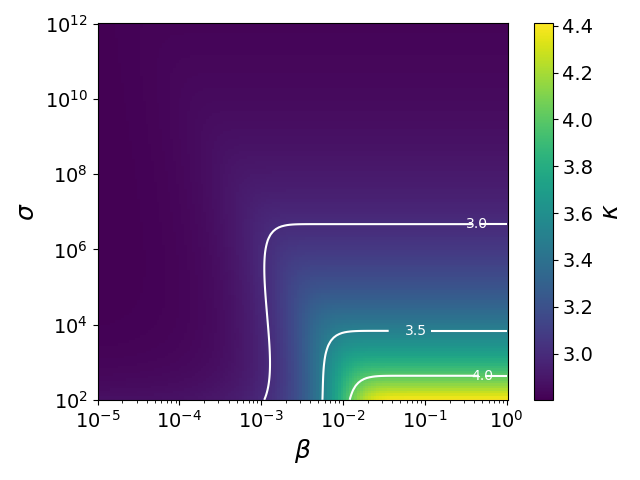}}
    \caption{Carte 2D de l'indice $\kappa$ en fonction de la magnétisation $\sigma$ et du paramètre $\beta$ dérivé à partir des Eq.~\eqref{eq:kappa_index} et~\eqref{eq:kappa_sub}.}
    \label{fig:kappa_value}
\end{figure}

La distribution injectée est définie par deux paramètres : la température sans dimension $\Theta_e$ et l'indice kappa $\kappa$ qui, on le rappelle, est lié à l'indice de la loi de puissance à haute énergie telle que $\kappa = p+1$. Le premier est un paramètre libre du modèle permettant d'ajuster le maximum émis (voir section~\ref{sec:vari_params}). Bien que l'indice kappa soit aussi un paramètre libre du modèle, on peut restreindre l'intervalle de valeurs possibles. En effet, les simulations PIC de reconnexion sans RG montrent que l'indice de la loi de puissance de la distribution des particules accélérées est lié à la magnétisation en amont de la reconnexion~$\sigma_b$ (voir Eq.~\eqref{eq:magnetisation}) tel que
\begin{equation}\label{eq:kappa_index}
\kappa = p+1 = A_p + B_p \tanh{(C_p\ \beta_b)} + 1
,\end{equation}
où
\begin{equation}\label{eq:kappa_sub}
    A_p=1,8 + 0,7/\sqrt{\sigma_b}, B_p=3,7\ \sigma_b^{-0,19}, C_p=23,4\ \sigma_b^{0,26},
\end{equation}
et $\beta_b =  8 \pi p/B^2 \ll 1$ le rapport entre la pression du gaz et la pression magnétique.
De plus, l'énergie maximale atteinte par les électrons accélérés par la reconnexion dépend de la quantité d'énergie emmagasinée par les lignes de champ qui reconnectent, aussi caractérisée par la magnétisation $\gamma_{max} \approx \sqrt{\sigma_b}$~\cite{Ripperda2022}.
À partir de simulations PIC~\cite{Selvi2022, Chernoglazov2023, Schoeffler2023, Ball2018, Hoshino2022, Johnson2022, Klion2023, Werner2018, Zhang2021} ainsi que de la magnétisation attendue autour de Sgr~A*, à savoir a minima $\sigma_b \gg 100$~\cite{Crinquand2022}, on peut restreindre les valeurs de $\kappa$ entre 2,8 et 4,4 (en prenant un ensemble très large pour $\sigma_b$ et $\beta_b$, voir Fig.~\ref{fig:kappa_value}) et fixer le facteur de Lorentz maximal des électrons à $\gamma_{max}=10^6$. On note que l'on peut aller jusqu'à $\sim 10^7$ car la magnétisation réelle de Sgr~A* est peu contrainte. On constate en regardant la Fig.~\ref{fig:kappa_value} que dans le cas d'une magnétisation très importante, comme celle attendue autour de Sgr~A*, la valeur de $\kappa$ est très contrainte autour de 2,8.

\begin{figure}
    \centering
    \resizebox{0.8\hsize}{!}{\includegraphics{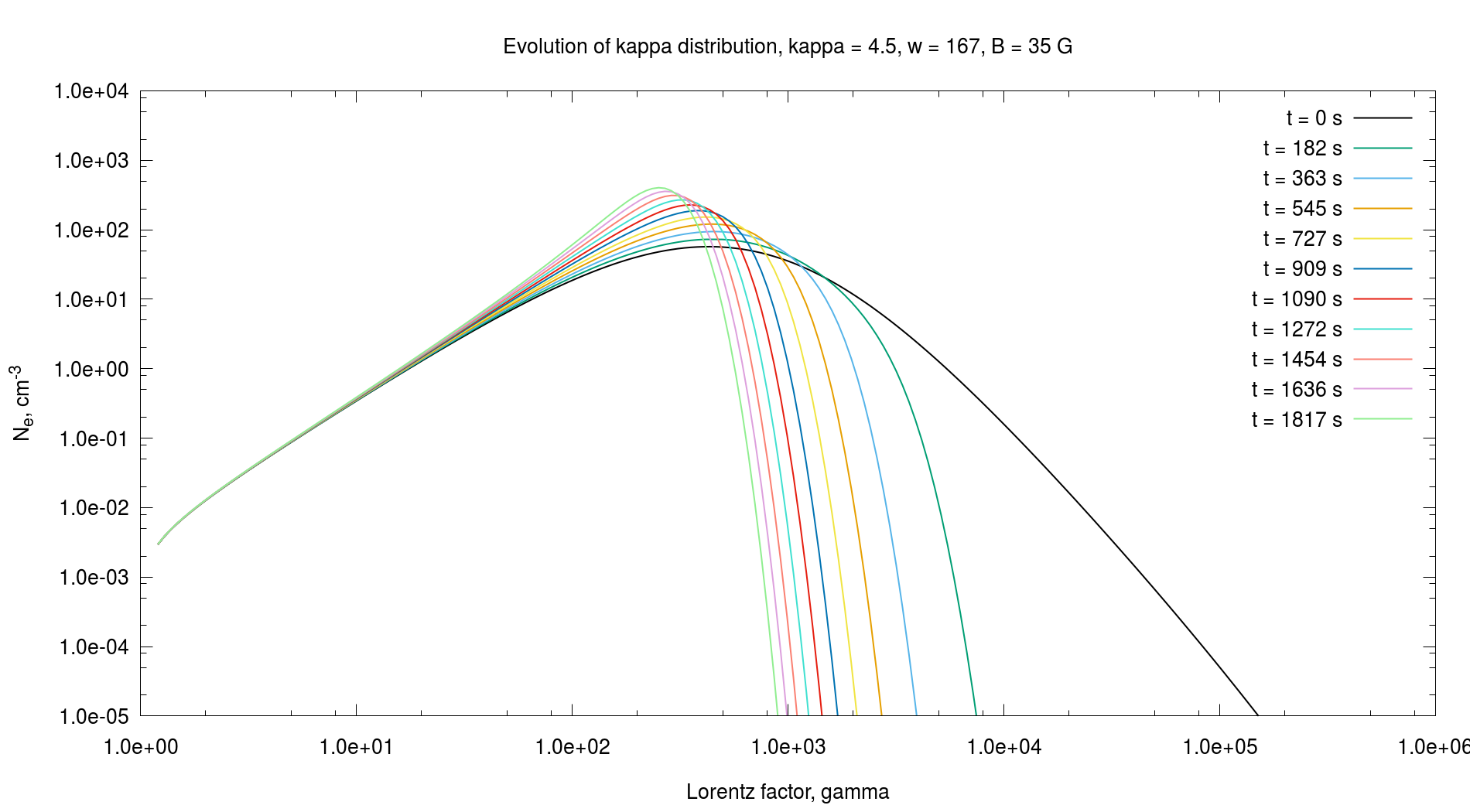}}
    \caption{Évolution d'une distribution des électrons kappa en fonction du temps refroidissant par rayonnement synchrotron. La température sans dimension $\Theta_e$ (noté $w$ dans la figure) est de 167, l'indice kappa de 4,5, et le champ magnétique de 35~G et la densité de $10^6$ cm$^{-3}$.}
    \label{fig:evol_distri_kappa}
\end{figure}

L'évolution de l'énergie d'un électron en fonction du temps émettant du rayonnement synchrotron est \cite{RL86}
\begin{equation} \label{eq:elenevol}
    \gamma(t) = \gamma_0 (1+b_c\gamma_0t)^{-1}
,\end{equation}
\begin{equation}
    \text{avec } b_c=\frac{4}{3} \frac{\sigma_T U_B}{m_e c},
\end{equation}
où $\gamma$ est le facteur de Lorentz de l'électron au temps $t$, et $\gamma_0$ le facteur de Lorentz initial. On retrouve ainsi le terme de refroidissement synchrotron de l'Eq.~\eqref{eq:kinetic}. Le temps caractéristique de refroidissement synchrotron $t_{cool}$ pour un électron avec un facteur de Lorentz $\gamma$ est
\begin{equation}\label{eq:cooling_time}
    t_{cool} =\frac{3}{4} \frac{m_e c}{\sigma_T U_B \gamma}.
\end{equation}
Ainsi, les électrons de hautes énergies refroidissent plus vite que ceux de plus basses énergies. Cela est parfaitement illustré par la Fig.~\ref{fig:evol_distri_kappa} montrant l'évolution d'une distribution d'électrons, correspondant initialement à une distribution kappa, au cours du temps, refroidie par rayonnement synchrotron. On constate aisément que la distribution est fortement déformée et ne correspond plus à une distribution bien définie (thermique, PL ou kappa), montrant l'importance du traitement cinétique de cette dernière. Durant la phase de refroidissement, l'évolution de la distribution est très similaire à la Fig.~\ref{fig:evol_distri_kappa}.

\begin{figure}
    \centering
    \resizebox{\hsize}{!}{\includegraphics{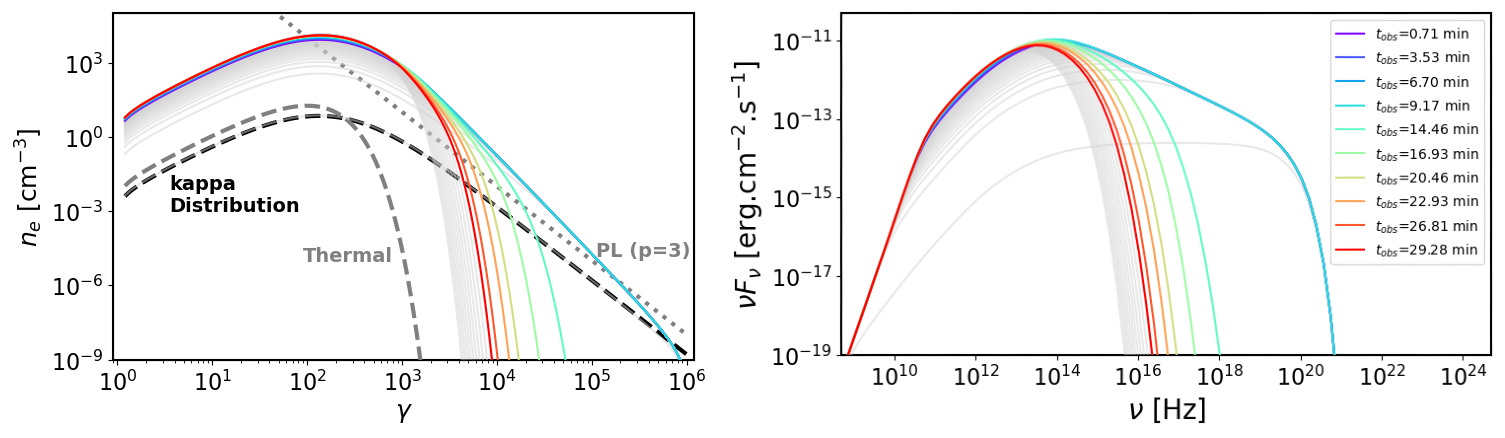}}
    \caption{Évolution de la distribution des électrons (\textbf{à gauche}) et de la densité spectrale d'énergie (SED) (\textbf{à droite}) en fonction du temps. La couleur indique le temps d'observation en lien avec la comparaison au sursaut du 22 juillet 2018. Les courbes hors de la fenêtre d'observation (avant ou après) sont en gris. Dans le panneau de gauche, on indique aussi la distribution kappa injectée durant la phase de croissance en tirets noirs, ainsi que les distributions thermiques (tirets gris) et PL (pointillés gris) permettant de construire cette distribution kappa. La fin de l'injection a lieu pour $t_\mathrm{obs} \sim 12$ min.}
    \label{fig:distri_SED_evol}
\end{figure}

Cependant, durant la phase de croissance, au cours de laquelle on injecte des électrons suivant une distribution kappa, ces derniers vont immédiatement refroidir et peupler des énergies plus basses. Ainsi, à chaque pas de temps, on a une injection de particules et un refroidissement. Pour les hautes énergies ($\gamma > 10^3$), cela résulte en un état d'équilibre, comme illustré dans le panneau de gauche de la Fig.~\ref{fig:distri_SED_evol} où l'on montre l'évolution de la distribution avec les paramètres de la Table~\ref{tab:emblem}. La couleur encode le temps et les courbes en gris correspondent à l'état de la distribution en dehors de la fenêtre d'observation (voir Section~\ref{sec:comparaison_obs}). La fin de la phase de croissance a lieu à $t_{obs} \sim 12$ min. Ainsi, à hautes énergies, les courbes en violet, bleu et cyan (antérieures à cette limite) se superposent en raison de cet état d'équilibre. Dès lors que l'on stoppe l'injection d'électrons, la distribution évolue comme dans la Fig.~\ref{fig:evol_distri_kappa}. 

\begin{table}
    \centering
    \begin{tabular}{lccc}
        \hline
        \hline
        Paramètre & Symbole & Valeur\\
        \hline
        {champ magnétique} [G] & $B_\text{p}$ & 15 \\
        {rayon du point chaud} [$r_g$] & $R_{\text{p}}$  & $1$ \\
        {facteur de Lorentz minimal} & $\gamma_{\text{min}}$ & 1 \\
        {facteur de Lorentz maximal} & $\gamma_{\text{max}}$ & $10^6$ \\
        {indice kappa de la distribution} & $\kappa$ & 4.0 \\
        {température sans dimension de la distribution} & $\Theta_e$ & 50 \\
        {densité d'électrons maximale} [cm$^{-3}$] & $n_{\text{e,max}}$ & $5 \times 10^6$ \\
        {temps de croissance} [$r_g/c$] & $t_\mathrm{growth}$ & 120 \\
        \hline
    \end{tabular}
    \caption{Paramètres du code \texttt{EMBLEM} pour la simulation de l'évolution de la distribution des électrons. Cet ensemble est utilisé pour comparer le modèle au sursaut du 22 Juillet 2018~\citet{Gravity2018}.}
    \label{tab:emblem}
\end{table}

On a donc deux phénomènes qui régissent l'évolution des distributions :
\begin{itemize}
    \item le temps de refroidissement : très lent à bas $\gamma$, donc les distributions évoluent toujours lentement à $\gamma \leq 1000$ ;
    \item l'équilibre injection/refroidissement : qui maintient la distribution constante également à haut $\gamma$ pendant la phase d'injection, avant de laisser la partie à $\gamma \geq 1000$ se refroidir rapidement quand on coupe l'injection.
\end{itemize}

\subsubsection{Calcul des coefficients synchrotron}
La distribution des électrons ne pouvant être décrite par une distribution bien définie (thermique, PL ou kappa), les formules d'approximations analytiques de \cite{Pandya2016} pour les coefficients synchrotron ne peuvent pas être utilisées. Il faut donc résoudre les Eq.~\eqref{eq:emission_synchrotron_exact} et~\eqref{eq:absorption_synchrotron_exact}, or ces dernières contiennent une intégrale de la distribution et l'intégrale de la fonction de Bessel modifiée d'ordre 2 (voir Eq.~\eqref{eq:F(x)}). Afin d'éviter une double intégration coûteuse en temps de calcul, on utilise les formules de \cite{Chiarberge&Guisellini1999}, où les coefficients d'émission et d'absorption sont (avec nos notations)
\begin{equation}
    j_\nu(t) = \frac{1}{4\pi} \int_{\gamma_{min}}^{\gamma_{max}} d\gamma N_e(\gamma,t) P_s(\nu,\gamma),
    \label{eq:j_nu}
\end{equation}
\begin{equation}
    \alpha_\nu(t)=-\frac{1}{8 \pi m_e \nu^2} \int_{\gamma_{min}}^{\gamma_{max}} \frac{N_e(\gamma,t)}{\gamma l} \frac{d}{d\gamma} [\gamma l P_s(\nu,\gamma)]
    \label{eq:a_nu}
\end{equation}
avec 
\begin{equation}
    P_s(\nu,\gamma) = \frac{3\sqrt{3}}{\pi} \frac{\sigma_T c U_B}{\nu_B}x^2 
    \left\{ K_{4/3} (x) K_{1/3} (x)-\frac{3}{5}x [K_{4/3}^2(x)-K_{1/3}^2(x)]  \right\},
    \label{eq:single_elec_emis}
\end{equation}
où $l=(\gamma^2-1)^{1/2}$ est le moment de l'électron en unité de $m_e c$, $x=\nu / (3 \gamma^2 \nu_B)$, $\nu_B=eB/(2\pi m_e c)$\footnote{Il s'agit de $\nu_{cyclo}$ du Chap.~\ref{chap:modele_hotspot+jet}.}, et $K_a(t)$ est la fonction de Bessel modifiée d'ordre $a$. L'intégrale de l'Eq.~\eqref{eq:F(x)} a été remplacée par une somme de fonctions de Bessel modifiées, réduisant ainsi le temps de calcul.

On note une différence dans les termes utilisés entre les Eq.~\eqref{eq:emission_synchrotron_exact} et~\eqref{eq:j_nu}, néanmoins les expressions sont parfaitement équivalentes mise à part l'approximation sur l'intégrale décrite précédemment. On remarque aussi que les Eq.~\eqref{eq:j_nu} et~\eqref{eq:a_nu} sont moyennées sur le \textit{pitch angle}, impliquant un champ magnétique isotrope (non ordonné dans une configuration particulière). Cette hypothèse est raisonnable, car la configuration du champ magnétique autour de Sgr~A* est peu contrainte. Néanmoins, les données \cite{Gravity2018} excluent un champ magnétique toroïdal (voir Chap.~\ref{chap:Sgr~A* flares}). Cette simplification permet donc de ne pas dépendre du choix de la configuration du champ magnétique pour le point chaud et de traiter chaque configuration individuellement.

\subsubsection{Dépendance de la courbe de lumière intrinsèque aux paramètres}\label{sec:vari_params}
À partir du calcul des coefficients synchrotron décrit au-dessus, on peut intégrer l'équation de transfert radiatif afin de déterminer la courbe de lumière intrinsèque, c'est-à-dire dans le référentiel de l'émetteur, sans prendre en compte quelconque effet Doppler, c'est-à-dire $\nu_\mathrm{obs} = \nu_\mathrm{ém} = 2,2\,\mu$m. À cette fréquence, le plasma est optiquement mince, on peut donc négliger l'absorption en première approximation\footnote{Ce ne sera pas le cas lors des calculs avec \textsc{GYOTO}. On fait cette approximation ici pour avoir une intuition sur la courbe de lumière.}. 
La forme et le maximum de la courbe de lumière dépendent de la valeur des paramètres du modèle. Le but de cette section est d'étudier l'espace des paramètres afin de comprendre leur influence sur la courbe de lumière.

Pour commencer, on s'intéresse à la phase de décroissance de la courbe de lumière, c'est-à-dire la phase de refroidissement gouvernée uniquement par refroidissement synchrotron. Comme le montre l'Eq.~\eqref{eq:cooling_time}, le temps caractéristique de refroidissement dépend du champ magnétique et du facteur de Lorentz de l'électron. Dans l'idéal, on voudrait obtenir une expression ne dépendant que des paramètres de notre modèle, à savoir uniquement le champ magnétique dans ce cas. On a vu dans le Chap.~\ref{chap:modele_hotspot+jet} que le rayonnement synchrotron dépend de la fréquence avec la fonction $F(x)$ illustrée par la Fig.~\ref{fig:F(x)}, le maximum d'émission étant à $0,29\, \nu_{crit}$ où $\nu_{crit} = 3 \gamma^2 \nu_B$\footnote{On note que l'on a moyenné sur le \textit{pitch angle}.}. On peut faire l'hypothèse que le rayonnement synchrotron est monochromatique à la fréquence $\nu = 0,29\, \nu_{crit}$. On peut ainsi considérer que tous les électrons émettant à la fréquence $\nu$ ont un facteur de Lorentz donné par
\begin{equation}\label{eq:Dirac_approx}
    \bar{\gamma} = \left( \dfrac{\nu m_e c}{\eta e B} \right)^{1/2}
,\end{equation}
avec $\eta = (0,29 \times 3)/(2\pi)$ un facteur numérique sans dimension. On note que le champ magnétique intervient aussi dans cette équation. En injectant l'Eq.~\eqref{eq:Dirac_approx} dans l'Eq.~\eqref{eq:cooling_time}, le temps caractéristique de refroidissement synchrotron et donc de la phase de refroidissement s'écrit
\begin{equation}
    t_\mathrm{sync} = 19 \times \left(\frac{B}{20 G} \right)^{-1,5}\ \left(\frac{\lambda}{2,2 \mu m} \right)^{0,5} \text{min.}
\end{equation}
Avec cette même hypothèse, et à partir de l'Eq.~\eqref{eq:elenevol}, on en déduit que $\bar{\gamma} \propto t^{-1}$ donc la décroissance du flux est aussi en $\nu F_\nu \propto t^{-1}$ (venant de l'Eq.~\eqref{eq:synsedflareapp}). Bien que l'hypothèse précédente soit une approximation (un électron émet à plusieurs fréquences, voir Fig.~\ref{fig:F(x)}), elle est très utile pour dériver des formules analytiques comme ici.

\begin{figure}
    \centering
    \resizebox{0.8\hsize}{!}{\includegraphics{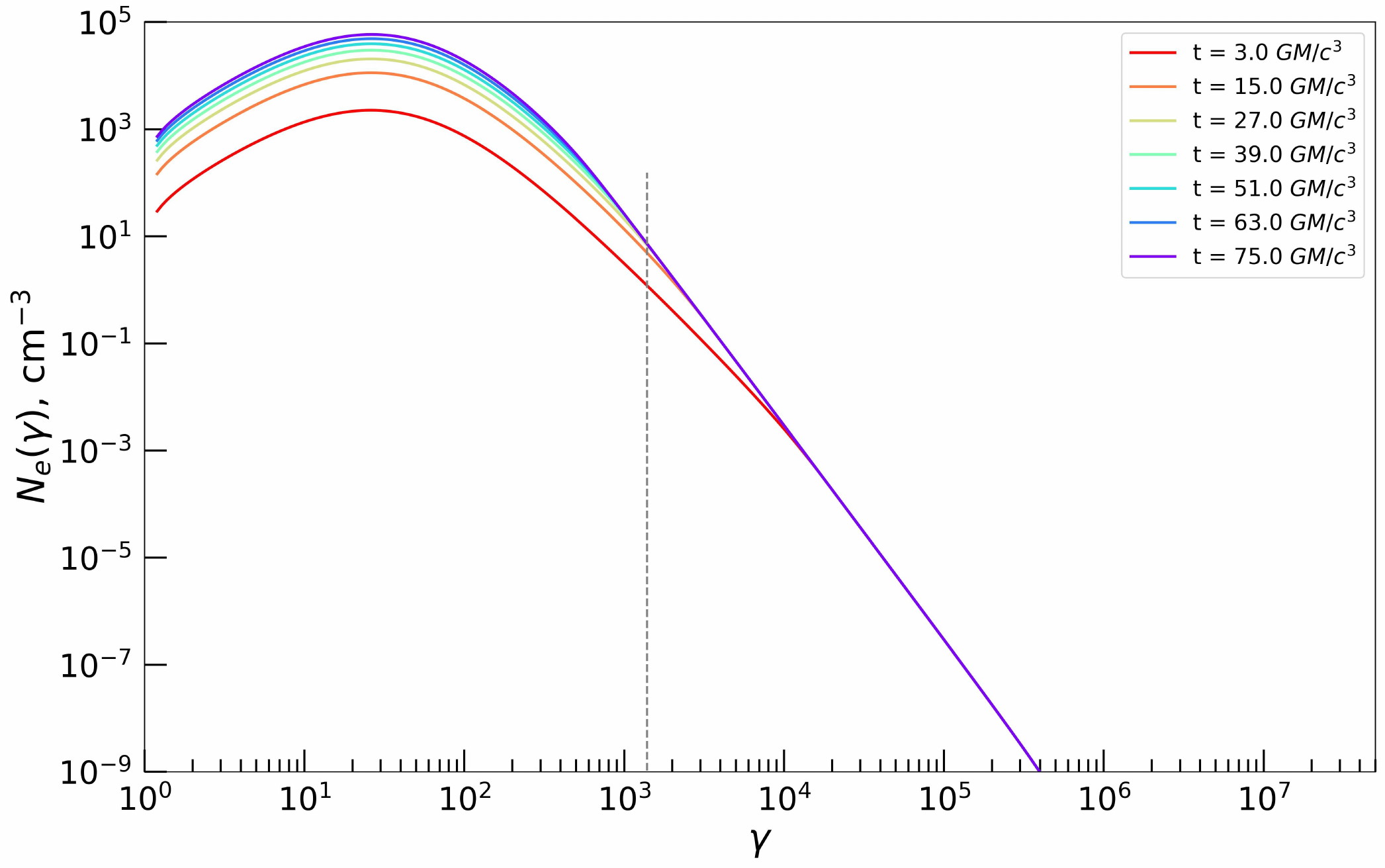}}
    \caption{Évolution de la distribution des électrons avec \texttt{EMBLEM} (traits pleins) de $t=0$ à $t=t_\mathrm{growth}=75 \, r_g/c$ en injectant une distribution kappa avec $\Theta_e=10$ , $\kappa=4,$ et $\dot{n_e}=5,10^6/t_\mathrm{growth}$. Le champ magnétique est fixé à 30 Gauss, résultant en un régime stationnaire pour $\gamma > 10^4$ dès $t=0$. Ce régime s'étend à des valeurs $\gamma$ plus faibles au fur et à mesure que le temps augmente. Pour estimer le flux maximal, nous avons approximé l'ensemble de la distribution (à $t=t_\mathrm{growth}$) par un simple Dirac à $\bar{\gamma}$, représenté par la ligne grise en pointillés.}
    \label{fig:Distrib_approx}
\end{figure}

\begin{figure}
    \centering
    \resizebox{\hsize}{!}{\includegraphics{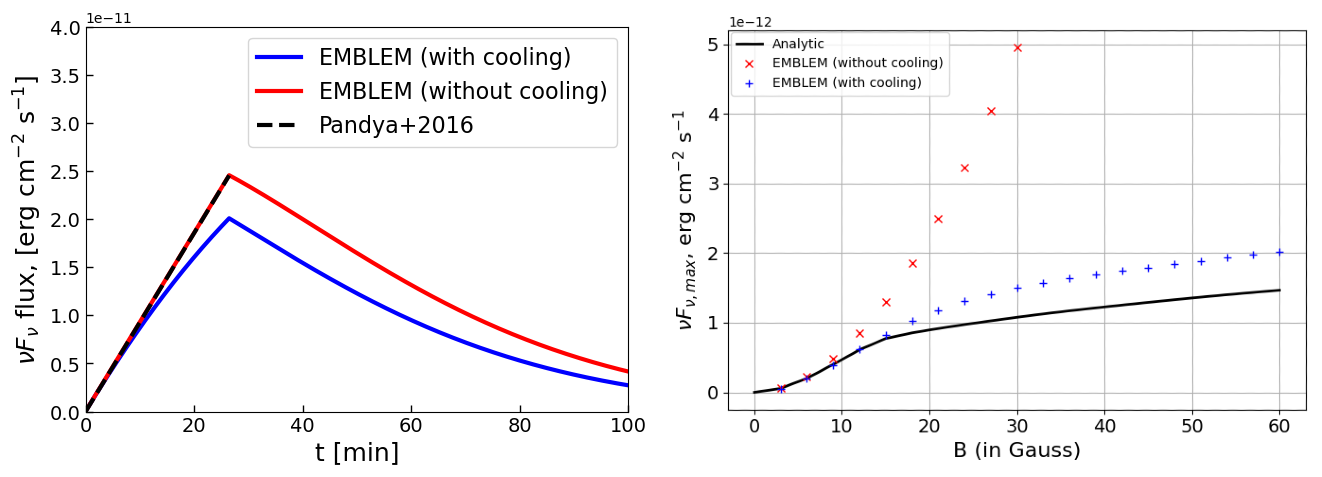}}
    \caption{(\textbf{Gauche}) Courbe de lumière émise par le point chaud dans le référentiel de l'émetteur calculé avec \texttt{EMBLEM}, avec refroidissement en bleu et sans refroidissement en rouge, avec $B=10$G, $\kappa=4$, $\Theta_e=90$, $n_{e,max}=10^6$ cm$^{-3}$ et $t_{inj}=75 \, r_g/c$. La courbe en pointillés noirs correspond au calcul du flux via les formules de \cite{Pandya2016}. (\textbf{Droite}) Évolution du flux maximal $\nu F_\nu (t_\mathrm{growth})$ avec $t_\mathrm{growth}=75 \, r_g/c$ en fonction du champ magnétique. Le refroidissement pendant la phase de croissance se traduit par un flux maximal plus faible (croix bleues) que sans refroidissement (croix rouges). Le flux maximal avec refroidissement peut être estimé à l'aide de l'équation~\eqref{eq:synsedflareapp} (courbe noire).}
    \label{fig:EMBLEM_cooling}
\end{figure}

La forme exacte de la partie croissante de la courbe de lumière ($t < t_\mathrm{growth}$) n'est pas aussi évidente que l'on pourrait le croire à ce stade. En effet, bien que l'on ait choisi une injection constante d'électrons durant la phase de croissance, se traduisant par une augmentation linéaire de la densité, le flux n'augmente pas linéairement comme on peut s'y attendre vu que $I_\nu \propto n_e$ (voir Eq.~\eqref{eq:j_nu}). En effet, en reprenant l'approximation précédente, pour une valeur de champ magnétique fixe, on peut déterminer le facteur de Lorentz $\bar{\gamma}$ des électrons dont le rayonnement est observé à une certaine fréquence $\nu_\mathrm{obs}$. Or, on a vu dans la section~\ref{sec:distrib_evol} qu'à hautes énergies la distribution était dans un état stationnaire pendant la phase d'injection. Pour déterminer si la distribution d'électrons à l'énergie $\gamma_i$ est stationnaire, on compare le temps écoulé, $t$, au temps caractéristique de refroidissement synchrotron $t_\mathrm{cool}$ (qui dépend de $\gamma$). Si $t<t_{cool}(\gamma_i)$, alors peu d'électrons à cette énergie ont refroidi. Cependant, les électrons à hautes énergies ont refroidi et certains atteignent l'énergie $\gamma_i$ à l'instant $t$. Ainsi, la densité à l'énergie $\gamma_i$ augmente (linéairement) pour $t<t_{cool}(\gamma_i)$. En revanche, si $t>t_{cool}(\gamma_i)$\footnote{On se limite toujours à $t < t_\mathrm{growth}$.}, alors la distribution ($\dd n_e/\dd \gamma$) est stationnaire pour $\gamma \geq \gamma_i$ (du fait de l'équilibre entre l'injection et le refroidissement). Ainsi, le facteur de Lorentz limite du régime stationnaire, noté $\gamma_\mathrm{lim}$, diminue avec le temps. Cela est visible dans la Fig.~\ref{fig:Distrib_approx} où l'on trace l'évolution de la distribution des électrons uniquement durant la phase de croissance. On marque aussi la valeur de $\bar{\gamma}$ pour $\nu = 2,2\, \mu$m et $B=30$ G avec une fonction de Dirac en tirets gris. Pour $\bar{\gamma} \ll \gamma_\mathrm{lim}(t)$, le refroidissement est négligeable, l'intensité augmente donc linéairement avec le temps, comme le montre les 10 premières minutes de la courbe bleue dans le panneau de gauche de la Fig.~\ref{fig:EMBLEM_cooling}. Inversement, on a vu que si $\bar{\gamma} > \gamma_\mathrm{lim}(t)$, on est dans le régime stationnaire, avec une intensité constante (non visible dans le panneau de gauche de la Fig.~\ref{fig:EMBLEM_cooling}, car le temps $t_\mathrm{growth}$ n'est pas assez grand pour atteindre le régime stationnaire à $2,2\, \mu$m. Enfin, si $\bar{\gamma} \lesssim \gamma_\mathrm{lim}(t)$, le refroidissement ne peut plus être négligé. En effet, les premiers électrons injectés à cette énergie ont eu suffisamment de temps pour refroidir, atteignant des énergies plus basses (n'émettant pas à $\nu_\mathrm{obs}$). La densité d'électrons à $\bar{\gamma}$, donc l'intensité, n'augmente plus linéairement durant ce régime dit transitoire. On constate bien l'influence de ce dernier en comparant les courbes rouge et bleue du panneau de gauche de la Fig.~\ref{fig:EMBLEM_cooling} présentant les courbes de lumière intrinsèque calculée à partir d'\texttt{EMBLEM}, avec refroidissement en bleue et sans refroidissement en rouge.

La dernière grandeur caractéristique de la courbe de lumière intrinsèque est la valeur maximale du flux. L'effet du refroidissement décrit précédemment complique significativement une description analytique. Cependant, en utilisant la même approximation que précédemment, un rayonnement synchrotron monochromatique, et après de nombreuses étapes de calcul détaillées dans l'Appendix E de \cite{Aimar2023} reproduit en annexe~\ref{ap:Paper}\footnote{Je remercie infiniment Anton Dmytriiev pour ses calculs !}, on obtient une formule analytique pour le flux en fonction du temps pour $t \leq t_\mathrm{growth}$
\begin{equation} \label{eq:synsedflareapp}
    \nu F_{\nu}^{\mathrm{syn}}(\nu,t) = \dfrac{n_e R_b^3 \, \bar{\gamma} m_e c^2}{12 D^2 t_\mathrm{growth} \, \kappa \theta^2}
    \begin{cases}
      \left[ \Psi(\bar{\gamma}) - \Psi(\xi(\bar{\gamma},t)) \right], & \text{pour } \nu < \tilde{\nu}(t) \\
        \Psi(\bar{\gamma}), & \text{pour } \nu \geq \tilde{\nu}(t)
    \end{cases}
,\end{equation}
où $\tilde{\nu}(t) = (\eta e B)/(m_e c b_c^2 t^2)$ est la fréquence correspondant à la condition $\bar{\gamma} = 1/(b_c t)$, avec $\eta$ un facteur sans dimension défini dans l'Eq.~\eqref{eq:Dirac_approx} et 
\begin{equation} \label{eq:psi_function}
\Psi(x) = \left(1 + \dfrac{x - 1}{\kappa \theta} \right)^{-\kappa} \, \left[x^2 (\kappa - 1) + 2x (\kappa\theta - 1) + 2\theta (\kappa \theta - 2) \right].
\end{equation}
Pour déterminer le maximum de flux émis, il suffit d'utiliser les équations précédentes avec $t=t_\mathrm{growth}$. Le panneau de droite de la Fig.~\ref{fig:EMBLEM_cooling} montre la dépendance du flux observé en fonction de l'intensité du champ magnétique en utilisant le code \texttt{EMBLEM} (croix bleues) et avec l'Eq.~\eqref{eq:synsedflareapp} (courbe noire). On distingue bien dans la forme de la courbe noire le régime d'équilibre pour $B \geq 16,2$~G du régime de faible refroidissement pour $B<16,2$~G. L'erreur maximale de l'Eq.~\eqref{eq:synsedflareapp} comparée aux résultats d'\texttt{EMBLEM} est de $\sim 30 \%$ dans le régime stationnaire et de moins de $7 \%$ pour le régime de faible refroidissement, ce qui en fait une bonne approximation pour estimer le flux observé à partir des paramètres du modèle ou à l'inverse estimer la valeur des paramètres à partir d'un maximum de flux.

On constate à partir de la Fig.~\ref{fig:EMBLEM_cooling} que même dans le régime d'équilibre, le flux continue d'augmenter avec le champ magnétique. En effet, bien que la distribution soit stationnaire, le facteur de Lorentz $\bar{\gamma}$ et la puissance synchrotron émise dépendent de $B$. À titre de comparaison, on trace aussi dans la Fig.~\ref{fig:EMBLEM_cooling} le flux maximal en fonction de $B$ en utilisant \texttt{EMBLEM} mais sans terme de refroidissement, ce qui revient à considérer une pure distribution kappa. On constate aisément l'importance du refroidissement de la distribution des électrons pour le calcul de la courbe de lumière. Une comparaison plus exhaustive avec les formules analytiques de \cite{Pandya2016} est présentée dans l'Appendix C de \cite{Aimar2023} en annexe~\ref{ap:Paper}.

À plus haute fréquence, dans le domaine des rayons X ($10^{15}$-$10^{18}$Hz), le temps de refroidissement synchrotron est très court, de l'ordre de $1-100$s. Ainsi, l'état d'équilibre est atteint très rapidement comme illustré dans le panneau de droite de la Fig.~\ref{fig:distri_SED_evol} montrant l'évolution de la densité spectrale d'énergie (SED pour \textit{Spectral Energy Density} en anglais) associée à l'évolution de la distribution du panneau de gauche. Dès lors que l'injection d'électrons s'arrête, le flux à haute fréquence diminue très rapidement avec le même temps caractéristique que mentionné précédemment. Bien que l'étude des sursauts en rayons~X de Sgr~A* ne soit pas l'objet de cette thèse, ce modèle a été construit dans l'optique de modéliser les sursauts de Sgr~A* à toutes les longueurs d'ondes (plus de détails dans le Chap.~\ref{chap:ouvertures}).

\section{Comparaison aux données GRAVITY}\label{sec:comparaison_obs}
L'objectif premier de ce modèle est d'être capable d'expliquer les données des sursauts, à la fois l'astrométrie et la courbe de lumière, observées par GRAVITY, en particulier les propriétés décrites dans le Chap.~\ref{chap:Sgr~A* flares}.

Les sections précédentes décrivent l'évolution intrinsèque des paramètres de notre modèle et de la courbe de lumière résultante dans le référentiel de l'émetteur. Or, comme on l'a vu au Chap.~\ref{chap:modele_hotspot+jet}, plusieurs effets de relativité générale, comme les effets Doppler relativistes et gravitationnels, affectent la courbe de lumière observée. De nombreux cas de figure peuvent se produire comme ceux illustrés dans le Chap.~\ref{chap:modele_hotspot+jet}, mais aussi de nouveaux en raison de la dynamique hors du plan équatorial. On ne présentera pas ici tous les cas de figure possibles, mais seulement certaines configurations présentant un intérêt particulier, notamment au vu des observations GRAVITY. Pour toute la suite, l'inclinaison est fixée à $20$°, tout comme la masse et la distance de Sgr~A* à $4,297 \times 10^6 \, M_\odot$ et 8,277 kpc respectivement. Le sursaut du 22 juillet 2018, présentant l'astrométrie la plus nette en termes de mouvement orbital et une courbe de lumière en forme de cloche, est le sursaut idéal pour comparer les prédictions du modèle. On accorde donc une importance particulière à ce sursaut. 

\begin{figure}
    \centering
    \resizebox{0.6\hsize}{!}{\includegraphics{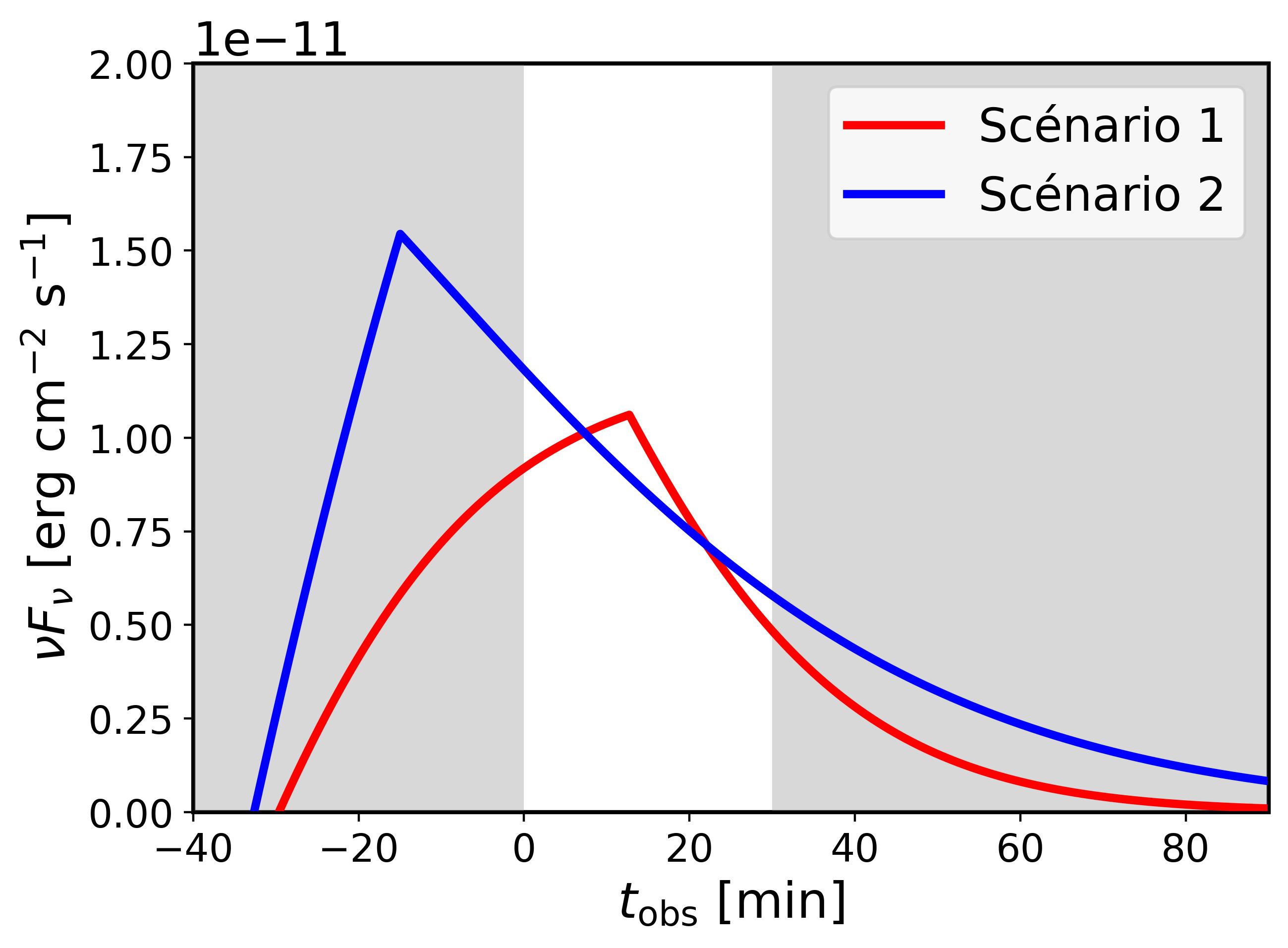}}
    \caption{Courbes de lumières dans le référentiel du point chaud calculé à partir des simulations d'\texttt{EMBLEM} pour les deux configurations envisagées. Pour la première, dont les paramètres sont dans les Tables~\ref{tab:emblem} et \ref{tab:plasmoid_orbital_params}, la fenêtre d'observation, représentée par la partie non grisée, contient à la fois la phase de croissance et de refroidissement. Pour la seconde, on observe uniquement la phase de refroidissement qui va être affecté par le beaming (voir Fig.~\ref{fig:secondary_peak}). Les paramètres utilisés pour cette configuration sont dans la Table~\ref{tab:alternativeJuly22}.}
    \label{fig:CLs_intrinsic}
\end{figure}

\begin{figure}
    \centering
    \resizebox{\hsize}{!}{\includegraphics{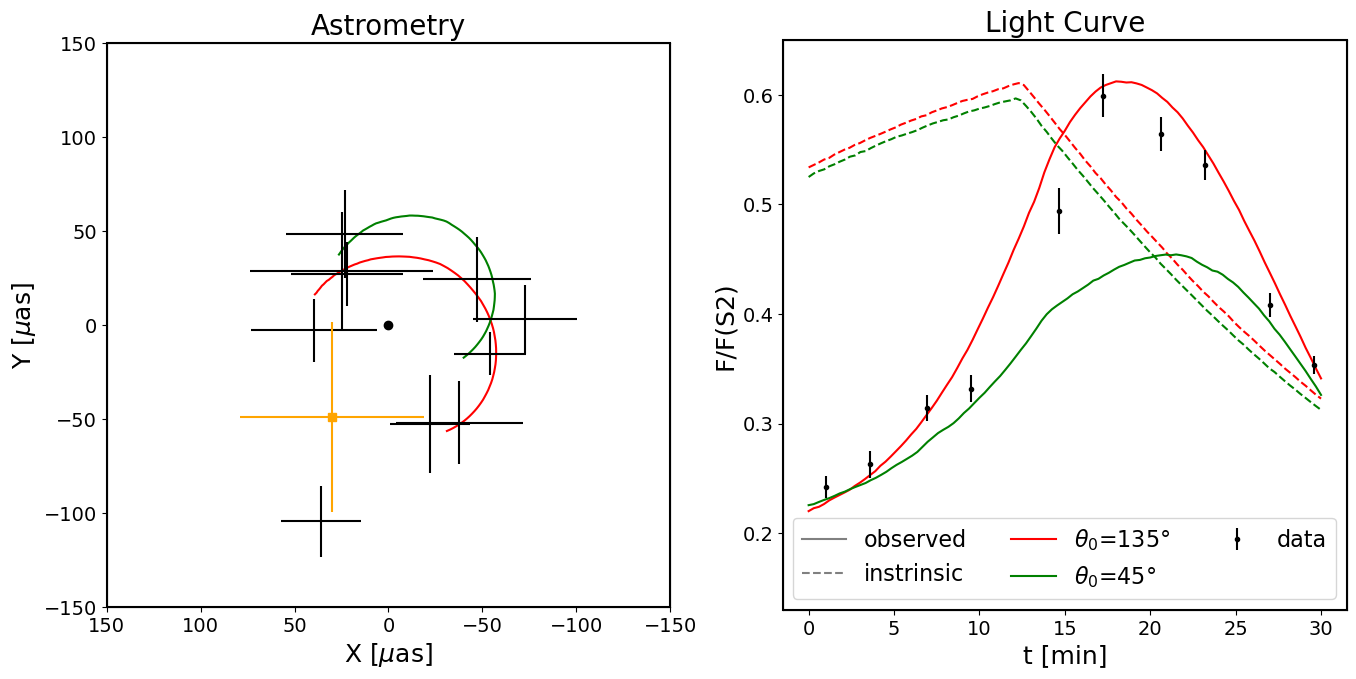}}
    \caption{Données et modèle du sursaut du 22 juillet 2018. Le panneau de gauche montre l'astrométrie du sursaut tandis que le panneau de droite montre les courbes de lumière observées (ligne pleine) et intrinsèques (ligne pointillée). Les courbes en rouge correspondent aux paramètres listés dans les tables~\ref{tab:emblem} et \ref{tab:plasmoid_orbital_params}. On note qu'il ne s'agit pas du résultat d'un ajustement. Les courbes en vert correspondent aux mêmes paramètres que les courbes en rouge avec $\theta_0=45 \degree$. Les courbes de lumière intrinsèques sont légèrement différentes dues à l'effet de la courbure de l'espace-temps sur la forme et la taille de l'image primaire. Le point noir dans le panneau de gauche représente la position de Sgr~A* dans \textsc{GYOTO} et la croix orange représente la position de Sgr~A* mesurée à travers l'orbite de S2.}
    \label{fig:plasmoid_flare}
    
    \vspace{1cm}

    \begin{tabular}{lcc}
        \hline
        \hline
        Paramètre & Symbole & Valeur \\
        \hline
        \textbf{Plasmoïde} & &\\
        temps dans \texttt{EMBLEM} au temps d'observation initial [min] & $t_{obs,0}^{emblem}$ & $29,6$\\
        rayon cylindrique initial [$r_g$] & $r_{cyl,0}$ & $10,6$\\
        angle polaire [$\degree$] & $\theta_0$ & $135$\\
        angle azimutal initial [$\degree$] & $\varphi_0$ & $280$\\
        vitesse radiale initiale [$c$] & $\dot{r}_0$ & $0,01$\\
        vitesse azimutale initiale [rad.s$^{-1}$] & $\dot{\varphi}_0$ & $4,25 \times 10^{-3}$\\
        position X de Sgr A* [$\mu$as] & $x_0$ & $0$\\
        position Y de Sgr A* [$\mu$as] & $y_0$ & $0$\\
        PALN [$\degree$] & $\Omega$ & $160$\\
        \hline
    \end{tabular}
    \captionof{table}{Paramètres orbitaux du modèle de plasmoïde utilisé pour la comparaison au sursaut du 22 juillet 2018 observé par \citet{Gravity2018} (scénario 1).}
    \label{tab:plasmoid_orbital_params}
\end{figure}

Pour cette comparaison, on va considérer deux configurations dont les paramètres sont listés dans les tables~tables~\ref{tab:emblem}, \ref{tab:plasmoid_orbital_params} et table~\ref{tab:alternativeJuly22} respectivement. Pour la première, on observe la fin de la phase de croissance et le début du refroidissement comme illustré par la courbe rouge dans la Fig.~\ref{fig:CLs_intrinsic}, le tout affecté par le beaming, donnant la courbe de lumière observée (courbes en trait plein dans le panneau de droite de la Fig.~\ref{fig:plasmoid_flare}). Le temps de croissance caractéristique est de $t_\mathrm{growth} \sim 100 \, r_g/c$, soit environ 30 min, ce qui est supérieur au temps de montée de la courbe de lumière observé qui est de l'ordre de 15 min, on en déduit que la phase de croissance est partiellement atténuée par le beaming. La phase de décroissance observée est liée au refroidissement combiné au beaming. Ces hypothèses contraignent les valeurs de l'angle azimutal et du temps coordonné initiaux $\varphi_0$ et $t_0 \sim t_\mathrm{growth}$. Comme on l'a vu dans le Chap.~\ref{chap:Sgr~A* flares}, les données du sursaut du 22 juillet 2018 suggèrent un rayon projeté de $r_{cyl,0} \sim 10 \, r_g/c$ et une vitesse super-Képlérienne pour le paramètre $\dot{\varphi}_0$. Le cas de la vitesse radiale sera discuté dans la section~\ref{sec:limitations}. Comme on modélise une éjection le long de la gaine du jet, l'angle polaire $\theta_0$ devrait correspondre à l'angle d'ouverture externe du jet $\theta_2=23,5$°, cependant, afin de ne pas être trop dépendant de notre modèle quiescent, nous considérerons ici une assez large gamme de valeurs de $\theta_0$. De plus, on peut estimer un ensemble de valeurs raisonnables à partir des simulations GRMHD~\cite{Ripperda2020,Ripperda2022} $10\degree \leq \theta_0 \leq 45\degree$. 

À partir de ce raisonnement, on peut déterminer un ensemble de paramètres (adaptables) pour le modèle, listés dans les tables~\ref{tab:emblem} et \ref{tab:plasmoid_orbital_params}, capables de reproduire les caractéristiques du sursaut du 22 juillet 2018 comme illustré par les courbes rouges de la Fig.~\ref{fig:plasmoid_flare}. On note que la position de Sagittarius~A* dans le modèle est aux coordonnées ($0$,$0$) et reste à la limite de la barre d'erreur de 1$\sigma$ sur sa position dérivée du mouvement de l'étoile S2. Cela est dû à notre choix de paramètres, un rayon orbital projeté plus élevé se traduisant par un arc de cercle (avec une plus petite courbure) et une position décalée vers la partie Nord-Ouest dans le plan du ciel. Pour rester compatible avec les données, il faudrait décaler l'ensemble du modèle, c'est-à-dire l'origine des coordonnées du modèle dans la direction opposée, soit au Sud-Est. La position de Sgr~A* dans le modèle sera donc plus proche de la position mesurée à partir de l'orbite de l'étoile S2. L'influence de l'état quiescent est aussi visible, bien que moins marqué que dans la Fig.~\ref{fig:influence_variabilité}. En effet, la distance apparente entre le modèle et la position de Sgr~A* en haut à gauche ($X\sim 30\, \mu$as, $Y \sim 30 \, \mu$as) de l'astrométrie de la Fig.~\ref{fig:plasmoid_flare} est plus courte qu'à droite ($X\sim -70\, \mu$as, $Y \sim-10 \, \mu$as), en raison de la différence de rapport de flux entre le point chaud et l'état quiescent. Les temps de croissance plus courts que ceux prédits par les simulations ($\sim 100 \, r_g/c$) ne sont pas compatibles avec ce scénario. En effet, un temps de croissance court résulterait en une augmentation beaucoup plus rapide du flux observé.

\begin{figure}
    \centering
    \resizebox{0.7\hsize}{!}{\includegraphics{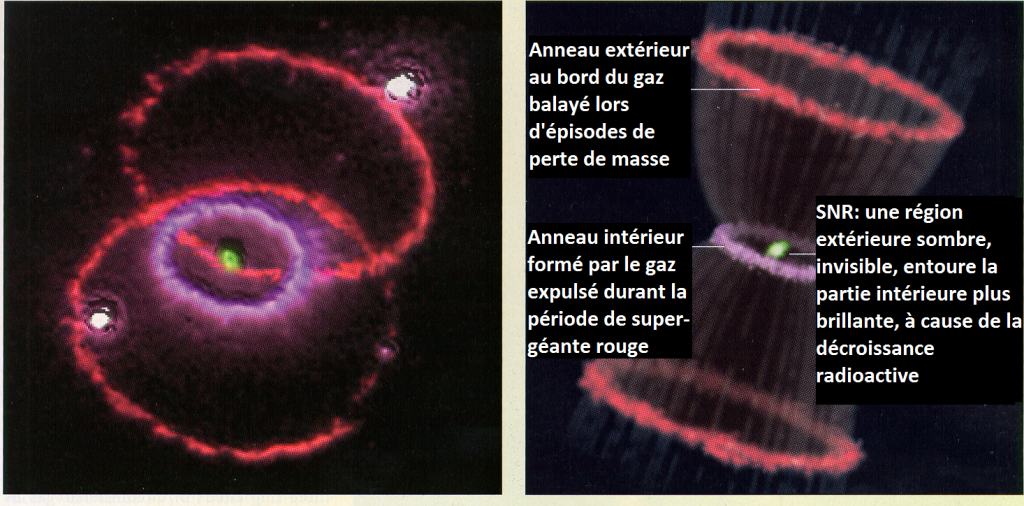}}
    \caption{(\textbf{Gauche}) Image réalisée par \href{https://hubblesite.org/home}{Hubble} du reste de supernova SN1987A. (\textbf{Droite}) Vue d'artiste de SN1987A, vue de côté. Crédit :  \href{https://www.jwst.fr/2017/03/la-supernova-sn-1987a/}{jwst.fr}}
    \label{fig:SN1987A}
\end{figure}

\begin{figure}
    \centering
    \resizebox{\hsize}{!}{\includegraphics{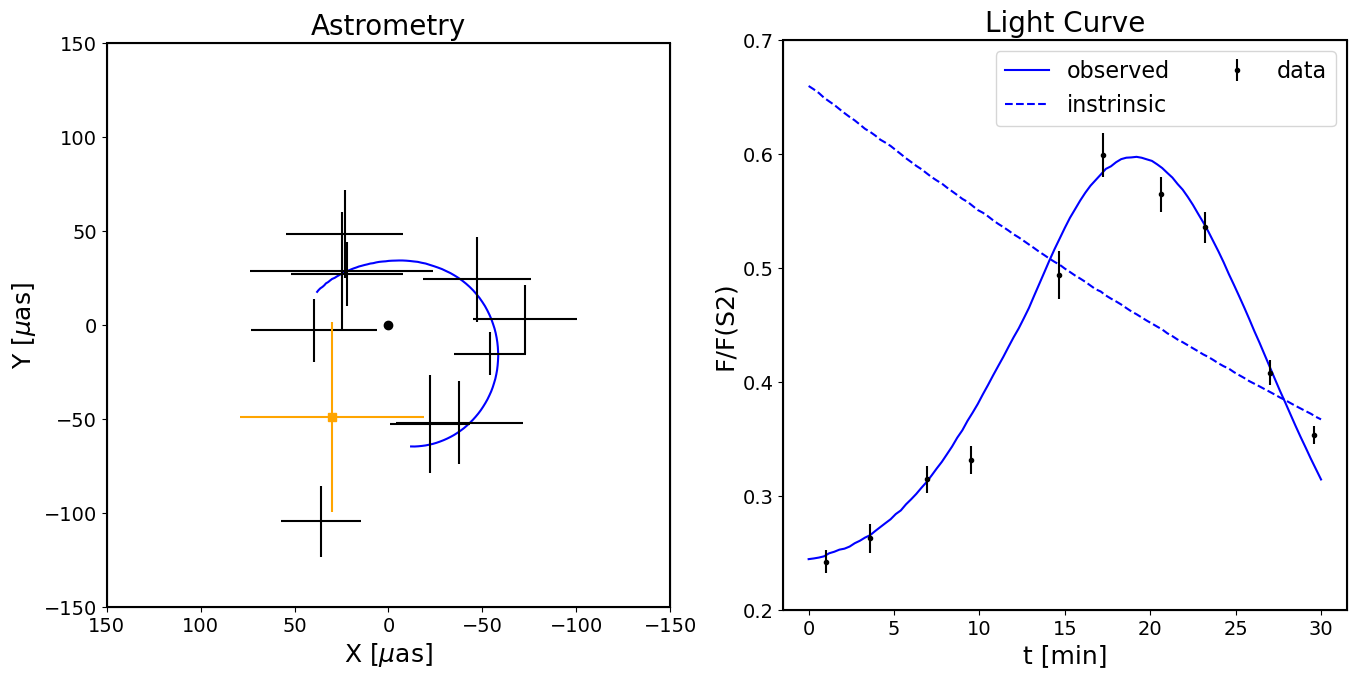}}
    \caption{Même chose qu'à la Fig.~\ref{fig:plasmoid_flare} avec les paramètres listés dans la table~\ref{tab:alternativeJuly22}.}
    \label{fig:secondary_peak}

    \vspace{1cm}
    
    \begin{tabular}{lcc}
        \hline
        \hline
        Paramètre & Symbole & Valeur \\
        \hline
        \textbf{Plasmoïde} & &\\
        temps dans \texttt{EMBLEM} au temps d'observation initial [min] & $t_{obs,0}^{emblem}$ & $32,6$\\
        rayon cylindrique initial [$r_g$] & $r_{cyl,0}$ & $10,6$\\
        angle polaire [$\degree$] & $\theta_0$ & $135$\\
        angle azimutal initial [$\degree$] & $\varphi_0$ & $240$\\
        vitesse radiale initiale [$c$] & $\dot{r}_0$ & $0,01$\\
        vitesse azimutale initiale [rad.s$^{-1}$] & $\dot{\varphi}_0$ & $4,71 \times 10^{-3}$\\
        position X de Sgr A* [$\mu$as] & $x_0$ & $0$\\
        position Y de Sgr A* [$\mu$as] & $y_0$ & $0$\\
        PALN [$\degree$] & $\Omega$ & $160$\\
        champ magnétique [G] & $B_\text{p}$ & 10 \\
        rayon du point chaud [$r_g$] & $R_{\text{p}}$  & $1$ \\
        facteur de Lorentz minimal & $\gamma_{\text{min}}$ & 1 \\
        facteur de Lorentz maximal & $\gamma_{\text{max}}$ & $10^6$ \\
        indice kappa de la distribution & $\kappa$ & 4,0 \\
        température sans dimension de la distribution & $\Theta_e$ & 72 \\
        densité d'électrons maximale [cm$^{-3}$] & $n_{\text{e,max}}$ & $5 \times 10^6$ \\
        temps de croissance [$r_g/c$] & $t_\mathrm{growth}$ & 50 \\
        \hline
    \end{tabular}
    \captionof{table}{Paramètres du modèle de point chaud (plasmoïde) utilisé pour comparer aux données du 22 juillet 2018~\citet{Gravity2018} dans un scénario où la courbe de lumière observée est principalement due à l'effet de beaming (scénario 2).}
    \label{tab:alternativeJuly22}
\end{figure}

Avec les paramètres de la table~\ref{tab:plasmoid_orbital_params}, le point chaud est éjecté derrière le trou noir, permettant un effet de lentille gravitationnelle pouvant générer un second pic (voir plus bas). A priori, les plasmoïdes peuvent être éjectés de chaque côté du trou noir par rapport à l'observateur. On s'intéresse donc à l'influence de la direction de l'éjection, dans la direction opposée comme précédemment (en rouge dans la Fig.~\ref{fig:plasmoid_flare}), ou vers l'observateur (en vert). En effet, la projection du mouvement dans le ciel n'est pas la même selon la position par rapport à l'observateur. Les deux trajectoires sont des ellipses (proches d'un cercle, car l'inclinaison est faible) dont les centres sont décalés entre eux et par rapport au trou noir. Cela est uniquement dû à l'inclinaison de l'observateur à l'image du rémanent de supernoa SN1987A (voir Fig.~\ref{fig:SN1987A}). L'orientation dans le ciel de ces cercles dépend du PALN. On peut constater cet effet dans l'astrométrie de la Fig.~\ref{fig:plasmoid_flare} où le décalage entre le centre de la courbe bleue et la position du trou noir est encore plus marqué dans le cas où l'observateur et le point chaud sont dans le même hémisphère. On constate aussi un effet sur la courbe de lumière. Les temps caractéristiques sont similaires, mais pas le flux maximal. En effet, comme le point chaud est éjecté vers l'observateur, la fréquence reçue par un observateur est décalée vers le bleu (contrairement au cas où le Blob s'éloigne, pour lequel le décalage est vers le rouge) en raison de l'effet Doppler relativiste. Cependant, la fréquence de l'observateur est fixe, à $2,2\, \mu$m. Ainsi, les décalages en fréquences s'appliquent sur la fréquence émise qui est par conséquent, décalée vers le rouge (respectivement vers le bleu). Comme on peut le voir sur le panneau de droite de la Fig.~\ref{fig:distri_SED_evol}, la fréquence d'observation ($\sim 1,3 \times 10^{14}$~Hz) est très proche du maximum de flux. Selon le temps d'émission, qui est différent dans les deux cas du fait de la différence de distance parcourue, le spectre d'émission n'est pas le même. L'intensité émise à la fréquence d'émission, décalée par rapport à celle de l'observateur, peut être supérieure ou inférieure selon le spectre et le décalage spectral. De plus, l'amplitude de l'effet de beaming n'est aussi pas identique (en plus du déphasage dû à la différence de temps d'émission). Tous ces effets combinés résultent en une courbe de lumière moins intense dans le cas où le point chaud est dans le même hémisphère que l'observateur par rapport au cas où ils sont en opposition (en changeant \textbf{uniquement} $\theta_0$).

\begin{figure}
    \centering
    \resizebox{\hsize}{!}{\includegraphics{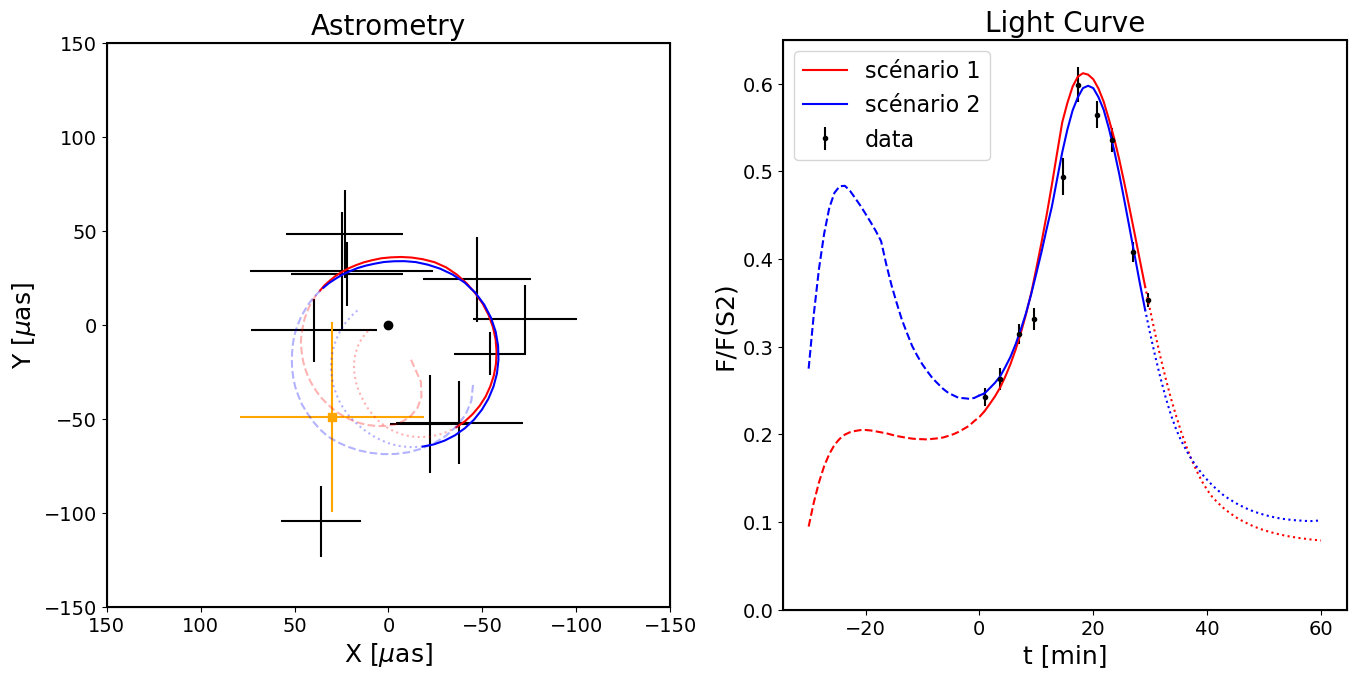}}
    \caption{Même chose qu'aux Fig.~\ref{fig:plasmoid_flare} et~\ref{fig:secondary_peak} avec une fenêtre d'observation plus grande allant de -30 min à 60 min. La courbe rouge correspond au premier scénario, c'est-à-dire le pic d'émission durant l'observation et la bleue au second, dans lequel le sursaut observé est un pic secondaire généré par le beaming. La partie du modèle réellement observée (entre 0 et 30 min) est tracée en trait plein alors que celle en amont des observations est tracée en tirets et celle en aval en pointillés. Les courbes de lumière intrinsèque sont visibles dans la Fig.~\ref{fig:CLs_intrinsic}.}
    \label{fig:peak_extented}

    \vspace{1cm}
    
    \resizebox{\hsize}{!}{\includegraphics{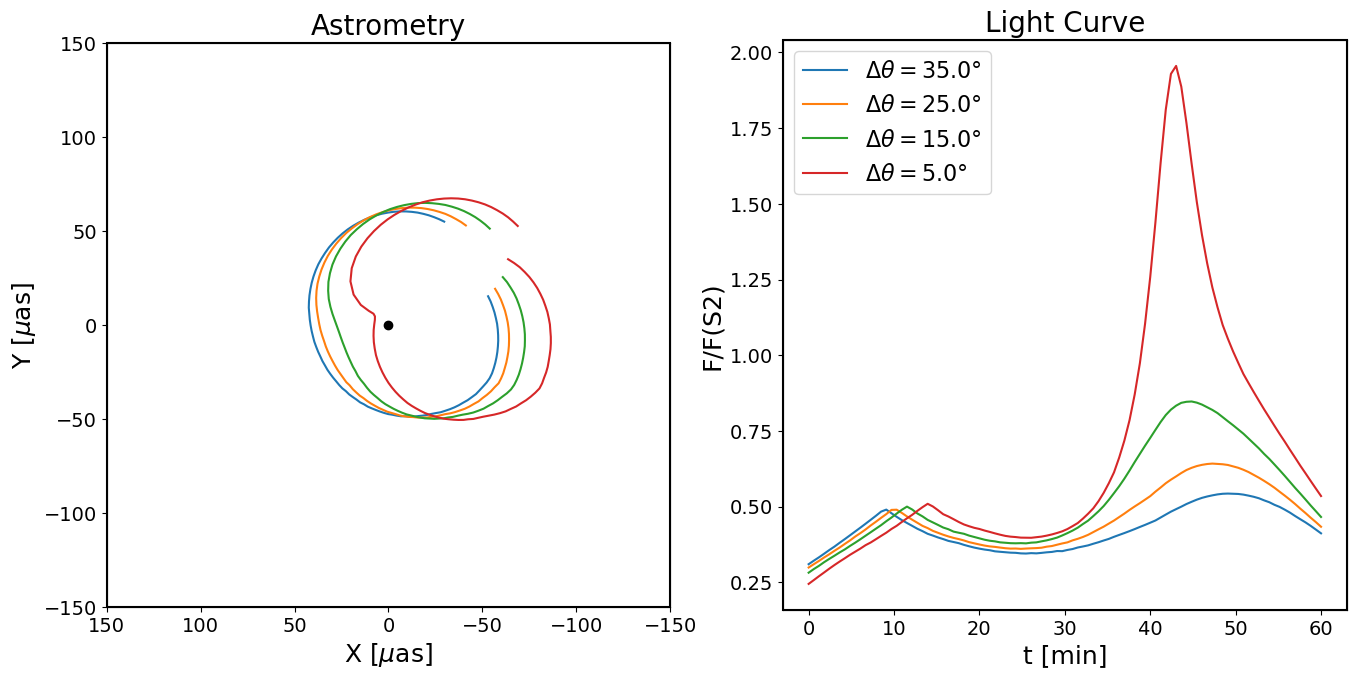}}
    \caption{Astrométrie (\textbf{gauche}) et courbes de lumière (\textbf{droite}) du modèle de plasmoïde~+~jet pour quatre angles polaires $\theta_0$ se traduisant par un écart avec l'inclinaison de  $35 \degree$ en bleu, $25 \degree$ en orange, $15 \degree$ en vert, et $5 \degree$ en rouge. L'effet de lentille gravitationnelle est particulièrement visible sur la courbe rouge. On note que le changement d'angle polaire avec le même rayon résulte en un décalage en temps du premier pic (associé à l'émission intrinsèque) et des astrométries.}
    \label{fig:double_peak_delta_theta}
\end{figure}

Une seconde configuration possible est que la courbe de lumière observée soit principalement générée par l'effet de beaming (positif), potentiellement combiné à un effet de lentille, dont l'amplification survient durant la phase de refroidissement, comme illustré dans la Fig.~\ref{fig:CLs_intrinsic} (courbe bleue). Les paramètres physiques dans ce scénario, listés dans la table~\ref{tab:alternativeJuly22}, sont légèrement différents, mais comparables aux valeurs précédentes, avec néanmoins un temps de croissance $t_\mathrm{growth}=50\, r_g/c$, et donnent un résultat similaire par rapport aux données (voir Fig.~\ref{fig:secondary_peak}). Le pic d'émission intrinsèque, survenant en amont de la fenêtre d'observation, est partiellement atténué, selon la phase exacte du beaming (négatif), mais toujours présent comme illustré par la courbe bleue dans la Fig.~\ref{fig:peak_extented}. Pour être fortement atténué, il faut que le pic d'émission soit synchronisé avec le beaming négatif, et que la phase de croissance soit courte (du même ordre de grandeur qu'une demi-orbite $\sim 15$ min), ce qui n'est pas le cas dans la Fig.~\ref{fig:peak_extented}. Dans ce scénario, comme la phase de croissance n'est pas observée, il est impossible de contraindre les paramètres physiques, comme la densité maximale et la température, qui sont dégénérés puisque seule la phase de décroissance liée au champ magnétique est observée. De plus, la courbe de décroissance dépend de la valeur du champ magnétique, mais aussi du maximum d'émission et du moment d'observation (la phase de la courbe observée), qui ne sont pas contraints. Ainsi, il existe un large ensemble de paramètres qui donneront un résultat similaire. Cependant, la courbe de lumière dans ce scénario présente un pic (plus ou moins important) en amont des observations. L'absence de détection de ce pic peut être due à l'absence de données (observation d'un autre champ de vue par exemple) ou à un niveau de flux faible, confondu avec l'état quiescent.

Dans ces conditions, et uniquement avec les données du sursaut, il est très difficile de distinguer cette configuration de la première. Cependant, les données en amont du sursaut peuvent aider à distinguer les deux scénarios. En effet, dans le premier scénario, où le pic d'émission survient durant la phase d'observation, le flux en amont sera proche de celui de l'état quiescent avec peu de variabilité (courbe rouge dans la Fig.~\ref{fig:peak_extented}), alors que dans le cas d'un pic secondaire, la variabilité en amont du sursaut due à l'émission intrinsèque sera plus importante. La variabilité en aval du sursaut peut aussi être révélatrice du scénario (non illustré dans la Fig.~\ref{fig:peak_extented}). En effet, le rapport de flux et le décalage temporel entre le pic observé et un second pic (s'il est observé) permettraient de faire la distinction (future étude). Bien que peu enviable vis-à-vis de la contrainte des paramètres physiques, le second scénario est tout à fait plausible. 

\begin{figure}
    \centering
    \resizebox{\hsize}{!}{\includegraphics{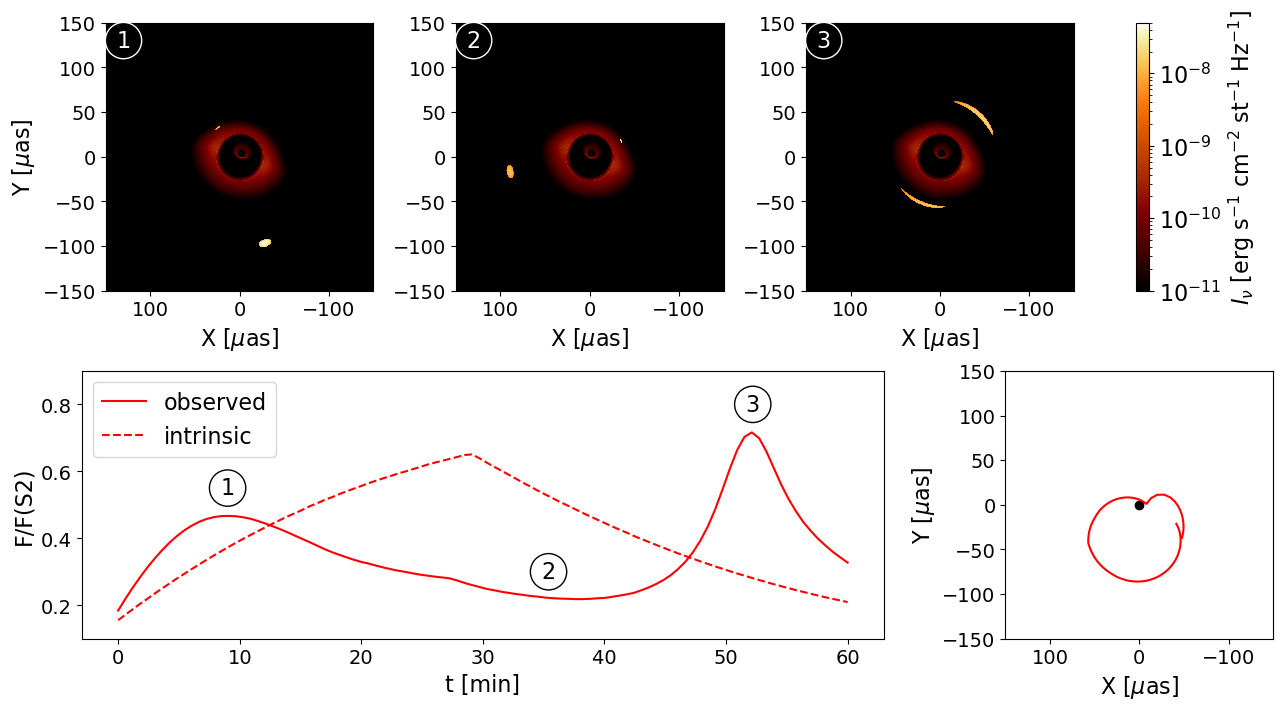}}
    \caption{Courbe de lumière (\textbf{en bas à gauche}), astrométrie (\textbf{en bas à droite}) et 3 images (\textbf{en haut}) du modèle de plasmoïde~+~jet obtenues avec les paramètres de la table~\ref{tab:double_peak}. Les trois images sont obtenues aux temps $t_{obs}=$ [$19$, $35$, $48.7$] min de gauche à droite. La dernière image illustre l'effet de lentille gravitationnelle .}
    \label{fig:double_peak}

    \vspace{1cm}

    \begin{tabular}{lcc}
        \hline
        \hline
        Paramètre & Symbole & Valeur \\
        \hline
        \textbf{Plasmoïde} & &\\
        temps dans \texttt{EMBLEM} au temps d'observation initial [min] & $t_{obs,0}^{emblem}$ & $0$\\
        rayon cylindrique initial [$r_g$] & $r_{cyl,0}$ & $10,6$\\
        angle polaire [$\degree$] & $\theta_0$ & $135$\\
        angle azimutal initial [$\degree$] & $\varphi_0$ & $210$\\
        vitesse radiale initiale [$c$] & $\dot{r}_0$ & $0,01$\\
        vitesse azimutale initiale [rad.s$^{-1}$] & $\dot{\varphi}_0$ & $3,96 \times 10$\\
        position X de Sgr A* [$\mu$as] & $x_0$ & $0$\\
        position Y de Sgr A* [$\mu$as] & $y_0$ & $0$\\
        PALN [$\degree$] & $\Omega$ & $160$\\
        champ magnétique [G] & $B_\text{p}$ & 15 \\
        indice kappa de la distribution & $\kappa$ & 4,0 \\
        température sans dimension de la distribution & $\Theta_e$ & 45 \\
        densité d'électrons maximale [cm$^{-3}$] & $n_{\text{e,max}}$ & $5 \times 10^6$ \\
        temps de croissance [$r_g/c$] & $t_\mathrm{growth}$ & 90 \\
        \hline
    \end{tabular}
    \captionof{table}{Paramètres du modèle de point chaud (plasmoïde) générant une courbe de lumière observée avec deux pics dont le ratio et le timing sont similaires au sursaut du 28 juillet 2018~\cite{Gravity2018}.}
    \label{tab:double_peak}
\end{figure}

À la lumière du scénario précédent, qui nécessite un temps de croissance plus court que les prédictions des simulations\footnote{Mais néanmoins du même ordre de grandeur.}, on peut envisager un dernier scénario d'intérêt. Avec un temps de croissance comparable aux simulations, ou plus précisément égal ou supérieur à la période orbitale, le pic d'émission intrinsèque ne peut plus être entièrement atténué. De plus, avec une fenêtre d'observation supérieure à la période orbitale comme c'est le cas pour le sursaut du 28 juillet 2018~\cite{Gravity2018}, il est possible d'observer deux pics dus au beaming (positif). Le premier survient durant la phase de croissance et le second durant la phase de refroidissement. Le second pic peut en plus être amplifié si le point chaud est éjecté de l'autre côté du trou noir par rapport à l'observateur par l'effet de lentille gravitationnelle (voir Fig.~\ref{fig:deviation_lumière}). Cependant, l'intensité du second pic dépend fortement de l'alignement entre l'observateur, le trou noir et le point chaud avec des paramètres physiques et orbitaux (hormis $\theta_0$) identiques. Dit autrement, le rapport de flux entre les deux pics observés dépend de l'écart d'angle polaire $\Delta \theta$ défini comme
\begin{equation}
    \Delta \theta = \abs{(180 \degree - \theta_0) - i}
\end{equation}
comme illustré par la Fig.~\ref{fig:double_peak_delta_theta}. On constate que dans le cas d'un bon alignement, c'est-à-dire $\Delta \theta$ faible, la forme du second pic est très étroite. La forme de ce dernier peut donc servir de contrainte observationnelle. Cependant, le rapport de flux entre les deux pics ne dépend pas uniquement de $\Delta \theta$ mais aussi du signal de beaming (+ lentille gravitationnelle) appliqué à la courbe de lumière intrinsèque. En effet, la Fig.~\ref{fig:double_peak} montre la courbe de lumière observée avec deux pics dont le ratio de flux et le timing sont similaires au sursaut du 28 juillet 2018. On montre aussi l'astrométrie obtenue ainsi que trois images calculées avec \textsc{GYOTO} à trois temps d'observation particuliers ($t_{obs}=$ [$19$; $35$;$48,7$] min de gauche à droite). Il est aisé de constater que si la courbe intrinsèque est décalée vers la gauche (c'est-à-dire un $t_0$ plus petit) mais avec la même orbite observée, autrement dit le même beaming\footnote{On note qu'il faut adapter la valeur de $\varphi_0$.}, le ratio entre les deux pics sera plus faible puisque l'émission intrinsèque au moment du second pic sera plus faible. La largeur du second pic peut donc indiquer si $\Delta \theta$ est faible ($\leq 15 \degree$) mais pas le ratio de flux.

Il est important de garder en tête que les scénarios et exemples détaillés ci-dessus sont des cas particuliers avec des valeurs de paramètres sélectionnées à la main. Il y a de nombreuses dégénérescences qu'une analyse statistique de type MCMC révèlerait aisément (future étude). Les jeux de paramètres identifiés ici ne sont pas uniques. Cela signifie qu'il existe de nombreux minima locaux de $\chi^2$ dans l'espace des paramètres et qu'une attention particulière doit être portée concernant la première estimation (\textit{first guess} en anglais) d'un algorithme MCMC. De ce fait, une grille de l'espace des paramètres par laquelle le $\chi^2$ d'un jeu de données peut être évalué comme première approche pour déterminer les minima locaux semble être nécessaire en amont d'une étude MCMC (future étude).

\section{Limitations du modèle}\label{sec:limitations}
Bien que le modèle de plasmoïde~+~jet montre des résultats très encourageants pour expliquer les sursauts de Sagittarius~A* observés par GRAVITY, il est basé sur de nombreuses hypothèses, parfois très fortes. Certaines valeurs de paramètres sont même en contradiction avec les simulations. Dans cette section, on récapitule les différentes limitations du modèle actuel en présentant de possibles améliorations lorsque c'est nécessaire.

Premièrement, l'évolution des paramètres du plasma (densité, température sans dimension et champ magnétique) a été choisie pour être soit constante, soit linéaire, ce qui est très simplifié par rapport à un scénario réaliste. Cependant, on considère que ces évolutions sont très probablement fortement dépendantes des conditions initiales du flot d'accrétion, et sont donc faiblement contraintes. Ces hypothèses permettent, en contrepartie, de suivre de manière réaliste la distribution des électrons qui émettent le rayonnement synchrotron observé. On note cependant que la valeur de l'indice kappa utilisé est trop élevée pour les valeurs de magnétisation attendues autour de Sgr~A*. En effet, la Fig.~\ref{fig:kappa_value}, montrant la valeur de cet indice en fonction de la valeur de la magnétisation $\sigma$ et du paramètre $\beta$ du plasma (Eq.~\eqref{eq:kappa_index}), suggère une valeur de kappa autour de 2,8 pour une grande partie de l'espace des paramètres, qui plus est avec une magnétisation élevée comme attendue autour de Sgr~A*.

Un autre point important est le facteur de Lorentz maximum considéré. Premièrement, cette valeur peut être estimée à partir de la magnétisation du plasma en amont de la reconnexion (Eq.~3 de \cite{Ripperda2022}), qui est peu contrainte. De plus, il s'agit ici du facteur de Lorentz maximum des électrons au niveau du site d'accélération. Or, on s'intéresse aux électrons une fois qu'ils sont dans le point chaud. Les électrons ont donc refroidi par rayonnement synchrotron\footnote{Ce dernier n'étant pas forcément dirigé vers l'observateur à cause du beaming.} durant le temps nécessaire pour atteindre le point chaud. Ainsi, le facteur de Lorentz maximal des électrons au moment d'entrer dans le point chaud est plus faible que celui du site de reconnexion (qui est peu contraint). Cependant, lors de la fusion entre plasmoïdes, une nouvelle reconnexion magnétique a lieu (avec une magnétisation plus faible), pouvant accélérer à nouveau les électrons, mais avec une énergie maximale beaucoup plus faible. Au vu de ces éléments, ce paramètre est plus ou moins libre et, dans tous les cas, dégénéré avec l'indice kappa et la température sans dimension si l'on ne s'intéresse qu'aux sursauts en IR. En effet, les sursauts en rayons~X sont générés par des électrons de très hautes énergies, typiquement de l'ordre de grandeur de $\gamma \sim 10^6$. L'étude des sursauts en rayons~X offre donc une contrainte sur l'énergie maximale des électrons. Enfin, on note aussi que dans le cas de production de rayons~X par ces électrons très énergétiques, il peut y avoir de la production de paires électron-positron qui n'est pas prise en compte dans notre modélisation.

On modélise le point chaud par une sphère homogène par simplicité, à partir du plasmoïde circulaire observé dans les simulations 2D GRMHD \cite{Nathanail2020,Ripperda2020, Porth2021} et PIC \cite{Rowan2017, Ball2018, Werner2018}. L'aspect 3D de ce plasmoïde est en réalité cylindrique (cordes de flux) dans les simulations GRMHD \cite{Bransgrove2021, Nathanail2022, Ripperda2022} et PIC \cite{Nattila2021, Zhang2021, El_Mellah2023, Crinquand2022}. Ainsi, une géométrie réaliste de la source du sursaut est probablement plus complexe que dans notre modèle, mais n'est cependant pas pertinente, car nous suivons la position du centroïde du flux. D'autre part, la taille de la sphère est considérée comme constante au cours du temps. L’effet de cisaillement dans le flot d’accrétion est négligé. Or, la reconnexion produit de nombreux plasmoïdes de petites tailles qui fusionnent entre eux pour former des structures macroscopiques. Le temps caractéristique de fusion des plasmoïdes est de l'ordre de quelques temps d'Alfvén $t_A \sim r_g/c$ dans notre cas~\cite{Nalewajko2015}. Ce temps est très inférieur à $t_\mathrm{growth}$, qui correspond à la durée de reconnexion. De plus, ces structures macroscopiques ont une durée de vie limitée et finissent par se mélanger avec le reste du flot d'accrétion-éjection au bout d'un certain temps, de l'ordre de~$\sim 500 \, r_g/c$ \cite{Ripperda2022}.

La dynamique de notre point chaud est une éjection conique avec une vitesse radiale constante et une vitesse azimutale définie par les conditions initiales et la conservation du moment cinétique Newtonien. Bien que la forme de la trajectoire corresponde aux simulations, la vitesse radiale du modèle est un ordre de grandeur trop faible. En effet, la vitesse d'éjection des plasmoïdes attendue par les simulations est de l'ordre de $0.1$c, or avec une vitesse radiale aussi grande et la conservation du moment cinétique, la vitesse azimutale diminue très vite ($\dot{\varphi} \propto `\frac{1}{r(t)^2}$), et l'astrométrie résultante ne présente alors plus aucun mouvement orbital. Pour maintenir un mouvement orbital, c'est-à-dire une vitesse azimutale suffisamment importante, la vitesse radiale est donc beaucoup plus faible ($\sim 0.01$c). L'hypothèse de conservation du moment cinétique est donc en contradiction avec les simulations. Afin de résoudre ce problème, une solution serait de dériver une loi d'évolution des deux composantes de la vitesse (radiale et azimutale) à partir des simulations numériques. \cite{El_Mellah2023} a mesuré la vitesse radiale et azimutale des cordes de flux générées par la reconnexion au cours du temps (Fig. 8 de \cite{El_Mellah2023}). On constate ainsi que la vitesse azimutale décroit légèrement alors que la vitesse radiale augmente fortement initialement avant d'atteindre une asymptote. Une telle évolution peut être approchée par des formules analytiques, améliorant ainsi la dynamique du point chaud vis-à-vis des simulations numériques.

Le champ magnétique considéré dans le plasmoïde est isotrope. On ne prend donc pas en compte l'impact de la géométrie du champ magnétique sur les observables. La configuration du champ magnétique de l'état quiescent est a priori ordonnée verticalement dans le cas d'un jet fortement magnétisé. Le champ magnétique dans le plasmoïde, qui nous intéresse particulièrement ici, est moins contraint. La configuration du champ magnétique est fortement liée à la polarisation (voir Chap.~\ref{chap:Polarization}). Les propriétés observationnelles des sursauts de 2018 excluent une configuration purement toroïdale~\cite{Gravity2018}. Du point de vue des simulations \cite{Ripperda2022,El_Mellah2022}, la configuration du champ magnétique dans les tubes de flux est hélicoïdale (majoritairement poloïdale avec une composante toroïdale non nulle) du fait de la géométrie de la reconnexion. L'étude de cette configuration est donc la plus pertinente pour une future étude. On note cependant que dans les simulations 3D GRMHD de \cite{Ripperda2022}, un autre résultat de la reconnexion magnétique est l'éjection du disque avec la formation de larges tubes de champ magnétique vertical. \cite{Wielgus2022} ont développé un modèle de point chaud avec un champ magnétique vertical pour ajuster un sursaut radio de Sgr~A* observé avec ALMA. Tout comme il est possible que les deux domaines de fréquences (IR et radio resp.) observent les deux structures distinctes mentionnées au-dessus (cordes de flux et tube vertical resp.), il est aussi possible que la source de ces deux domaines de fréquences soit la même. Ainsi, un modèle de point chaud avec un champ magnétique vertical pour les sursauts en IR mériterait d'être étudié.

\begin{figure}
    \centering
    \resizebox{\hsize}{!}{\includegraphics{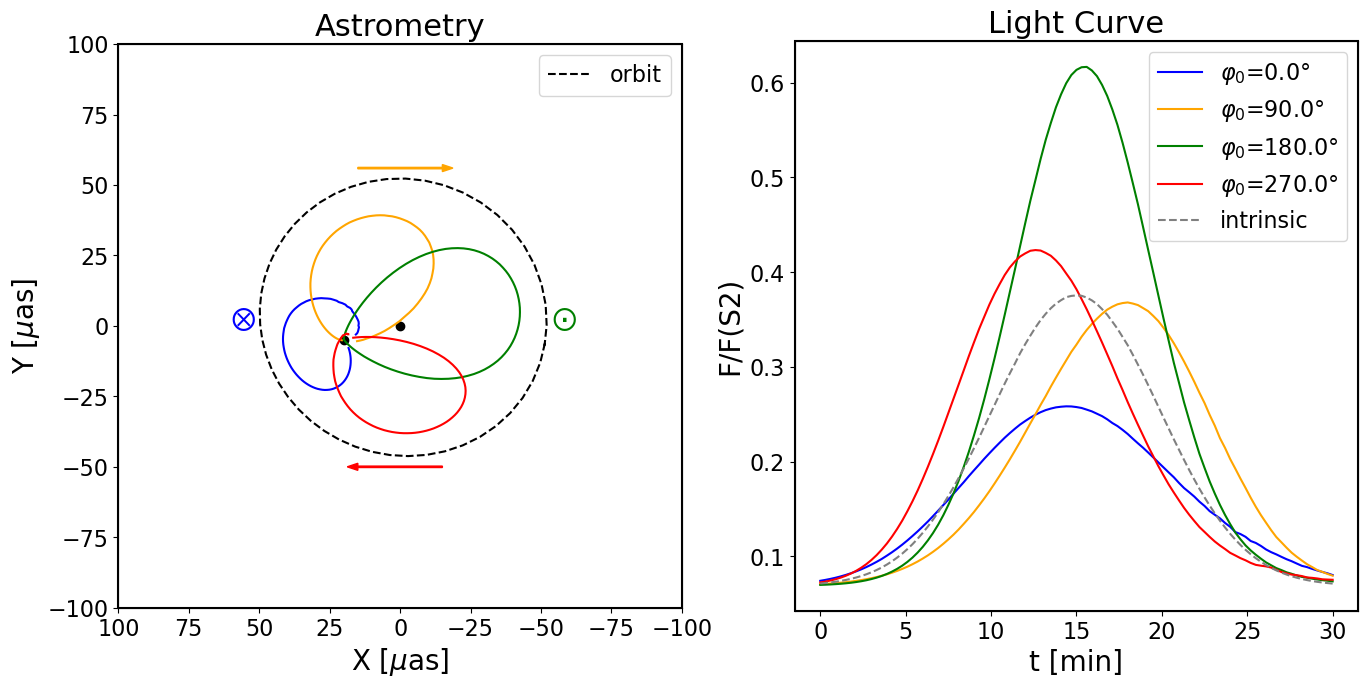}}
    \caption{Même chose qu'à la Fig.~\ref{fig:multi_phi0} avec un centroïde quiescent fortement décalé (en $X=20\, \mu$as, $Y=-5\, \mu$as) imitant un scénario où une partie du disque d'accrétion (responsable du flux quiescent) est éjecté par la reconnexion magnétique comme dans \cite{Ripperda2022}.}
    \label{fig:quiescent_bis}
\end{figure}

Comme dit précédemment, dans les simulations 3D GRMHD de \cite{Ripperda2022}, l'accrétion est stoppée durant un évènement de reconnexion accompagné par une éjection d'une partie du disque. Dans le cas où le flux quiescent provient du disque d'accrétion et non du jet comme on l'a considéré (voir Chap.~\ref{chap:modele_hotspot+jet}), le centroïde de l'état quiescent observé sera plus éloigné de la position du trou noir que dans notre modèle. Cela aura pour résultat de changer la forme des astrométries possibles comme illustré par la Fig.~\ref{fig:quiescent_bis}. On fait le choix d'un état quiescent axisymétrique à partir d'un modèle raisonnable, mais néanmoins améliorable afin de limiter le nombre de paramètres libres du modèle.

On note que le spin utilisé dans les calculs de tracé de rayon est fixé à zéro alors qu'un spin non nul est nécessaire pour générer un jet par le mécanisme de Blandford-Znajek (voir Chap.~\ref{chap:modele_hotspot+jet}). Ce choix vient du fait que la géométrie du tore utilisé pour l'état quiescent dépend fortement du spin. Par souci de cohérence, on garde donc une valeur nulle. Bien que ce paramètre soit très peu contraint, son impact sur le tracé de rayon est négligeable, avec un écart en flux entre un spin nul et un spin de 0,99 inférieur à $10 \%$, dans la majorité des cas, sauf dans le cas d'une lentille gravitationnelle. En effet, lors d'un effet de lentille gravitationnelle, les photons passent très près de l'anneau de photons qui est très sensible à la métrique. L'écart en flux entre $a=0$ et $a=0,99$ peut atteindre $\sim 30 \%$.

\part{Polarisation et Perspectives}
\chapter{Polarisation en Relativité Générale et dans GYOTO}\label{chap:Polarization}
\markboth{Polarisation en Relativité Générale et dans GYOTO}{Polarisation en Relativité Générale et dans GYOTO}
{
\hypersetup{linkcolor=black}
\minitoc 
}

\section{Polarisation de la lumière en général}
Pour commencer cette troisième partie, on s'intéresse à la polarisation, d'un point de vue théorique et numérique, dans l'optique d'étudier la polarisation des sursauts observés, mais aussi de toutes autres observables dont la source de rayonnement est proche d'un objet compact, nécessitant un traitement de la relativité générale.

\subsection{Définition}
\begin{figure}
    \centering
    \resizebox{\hsize}{!}{\includegraphics{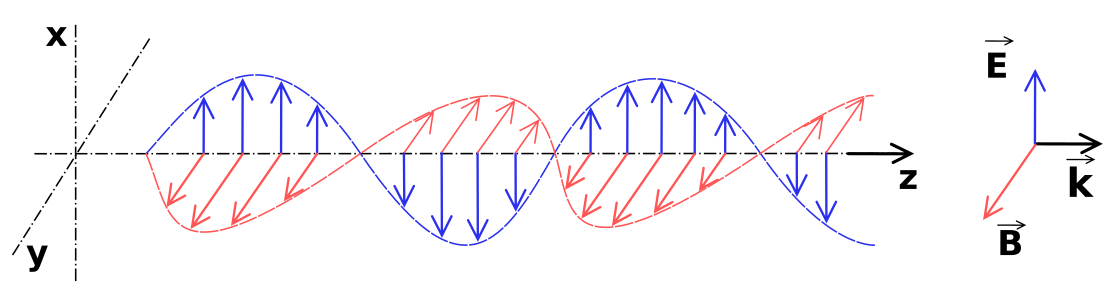}}
    \caption{Onde électromagnétique sinusoïdale se propageant le long de l'axe z, avec en bleu l'évolution du vecteur champ électrique et en rouge celle du vecteur champ magnétique de l'onde. Crédit : \href{https://commons.wikimedia.org/w/index.php?curid=2107870}{SuperManu}.}
    \label{fig:polar_linear}
\end{figure}

La lumière, qui est une onde électromagnétique, est une onde dite \textit{transverse} dans le vide, c'est-à-dire que les oscillations des champs électrique $\mathbf{E}$ et magnétique $\mathbf{B}$ qui la constituent sont perpendiculaires à la direction de propagation $\mathbf{K}$ ($\mathbf{E} \perp \mathbf{B} \perp \mathbf{K}$), comme illustré dans la Fig.~\ref{fig:polar_linear}.

\begin{definition}
    \textit{On définit le vecteur polarisation $\mathbf{F}$ d'une onde électromagnétique comme le vecteur orthogonal à la direction de propagation $\mathbf{K}$ et au vecteur champ magnétique $\mathbf{B}$, soit $\mathbf{F} = \mathbf{K} \times \mathbf{B}$. Dans le vide, ce vecteur est colinéaire au vecteur champ électrique.}
\end{definition}

Les sources de rayonnement de type corps noir, comme la majeure partie du rayonnement venant du Soleil, produisent de la lumière non polarisée, c'est-à-dire que le vecteur polarisation n'a pas de direction privilégiée. En effet, la lumière reçue en provenance de ce genre de sources est constituée de paquets d'ondes dont l'orientation du vecteur polarisation de ses constituants est aléatoire avec une distribution isotrope. A contrario, le rayonnement synchrotron, comme celui émis au voisinage de Sgr~A*, est fortement polarisé (voir section~\ref{sec:coef_synchrotron_polar}).

\begin{figure}
    \centering
    \resizebox{\hsize}{!}{\includegraphics{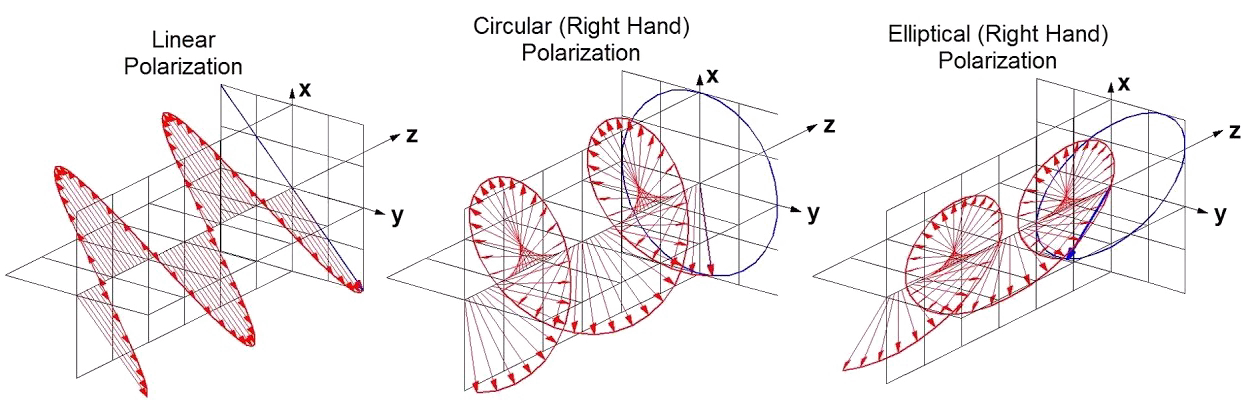}}
    \caption{Illustration des trois états de polarisation. De gauche à droite : linéaire, circulaire et elliptique. Dans cette représentation, la convention utilisée (sens horaire) est inversée par rapport à la convention usuelle définie par l'\href{https://www.iau.org/news/announcements/detail/ann16004/}{IAU}. Crédit : \href{https://www.youtube.com/watch?v=Q0qrU4nprB0}{meyavuz}.}
    \label{fig:polar_states}
\end{figure}

Selon la source de rayonnement, et le milieu traversé par l'onde avant d'atteindre l'observateur, la lumière reçue est soit non polarisée, soit polarisée selon un état particulier (polarisation linéaire, circulaire ou elliptique). Pour simplifier, on prend le cas d'une onde monochromatique unique sinusoïdale dans le vide. Lorsque, du point de vue d'un observateur le vecteur champ électrique oscille dans un plan contenant la direction de propagation - le plan x-z dans la Fig.~\ref{fig:polar_linear} - le vecteur polarisation oscille le long d'un axe (voir panneau de gauche de la Fig.~\ref{fig:polar_states}). On parle alors de polarisation linéaire. Si le champ électrique tourne dans un plan perpendiculaire à la direction de propagation, comme illustré au milieu de la Fig.~\ref{fig:polar_states} alors la polarisation est dite circulaire. Les deux états précédents décrivent la polarisation dans des cas spécifiques, le cas le plus général étant la polarisation elliptique (voir panneau de droite de la Fig.~\ref{fig:polar_states}).

\subsection{Base de polarisation}
\begin{figure}
    \centering
    \resizebox{0.6\hsize}{!}{\includegraphics{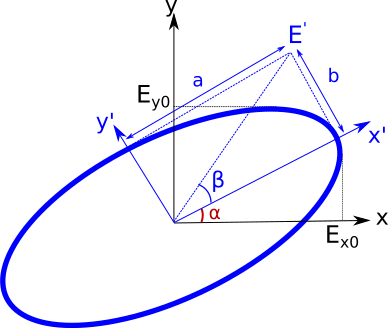}}
    \caption{Géométrie pour la polarisation elliptique. Les axes $x$ et $y$ constituent une base orthogonale dans le plan de polarisation. $\alpha$ est l'angle d'orientation définissant l'axe principal de l'ellipse et $\beta$ définit l'ellipticité, sa tangente étant égale au rapport de longueur des axes de l'ellipse. Crédit : Frédéric Vincent.}
    \label{fig:ellipse_polar}
\end{figure}

Afin de déterminer l'orientation du vecteur polarisation, il est nécessaire de définir une base. La polarisation se mesurant dans un plan, on considère un repère cartésien quelconque $(x,y)$ de l'écran de l'observateur. On prend le cas d'une polarisation elliptique d'une onde monochromatique. Dans le plan associé à l'ellipse $(x^\prime,y^\prime)$ et en utilisant une paramétrisation \href{https://en.wikipedia.org/wiki/Ellipse}{standard}, le champ électrique s'écrit
\begin{equation}
    \begin{aligned}
        \label{eq:ellipseframe}
        \mathbf{E} &= E_{x^\prime} \,\mathbf{e_{x^\prime}} + E_{y^\prime} \,\mathbf{e_{y^\prime}}, \\
        E_{x^\prime} &= E^\prime \cos \beta  \cos\left(\omega t\right), \\
        E_{y^\prime} &= E^\prime \sin \beta \sin\left(\omega t\right) \\
    \end{aligned}
\end{equation}
avec $\tan \beta$ le rapport de longueur des axes de l'ellipse, $E^{\prime 2}$ la somme des axes de l'ellipse au carré et $\omega$ la pulsation de l'onde monochromatique. L'orientation de l'ellipse dans le plan $(x,y)$ est, a priori, arbitraire et décrite par un angle $\alpha$ qui est l'inclinaison entre les plans $(x,y)$ et $(x^\prime,y^\prime)$, comme dans la Fig.~\ref{fig:ellipse_polar}. Dans la base $(x,y)$, le vecteur champ électrique s'écrit
\begin{equation}
    \begin{aligned}
        \label{eq:xyframe}
        \mathbf{E} &= E_x \,\mathbf{e_x} + E_y \,\mathbf{e_y}, \\ 
        E_x &= E_{0x} \cos\left(\omega t -\delta_x\right), \\
        E_y &= E_{0y} \cos\left(\omega t -\delta_y \right) \\
    \end{aligned}
\end{equation}
où $\delta_x$ et $\delta_y$ sont les phases. On peut ainsi définir la différence de phase entre les axes x et y par $\delta=\delta_x-\delta_y$. 

De plus, on peut exprimer les composantes de $\mathbf{E}$ dans la base $(x,y)$ à partir des composantes dans la base $(x^\prime,y^\prime)$ et l'angle $\alpha$ tel que
\begin{equation}
    \begin{aligned}
    \label{eq:xyframe_xyprime}
        E_{x} &= E_{x^\prime} \cos \alpha - E_{y^\prime} \sin \alpha, \\
        E_{y} &= E_{x^\prime} \sin \alpha + E_{y^\prime} \cos \alpha, \\
    \end{aligned}
\end{equation}
et l'on peut aussi réécrire \eqref{eq:ellipseframe}+\eqref{eq:xyframe_xyprime} de la manière suivante
\begin{equation}
    \label{eq:xyframe_dev}
    \begin{aligned}
        E_x &= E^\prime \left(\cos\alpha \cos\beta \, \cos \omega t - \sin\alpha\sin\beta\,\sin\omega t\right), \\
        E_y &= E^\prime \left(\sin\alpha \cos\beta \, \cos \omega t + \cos\alpha\sin\beta\,\sin\omega t\right). \\
    \end{aligned}
\end{equation}

En égalisant les équations \eqref{eq:xyframe_dev} et \eqref{eq:xyframe}, on obtient, par identification
\begin{equation}
    \begin{aligned}
        \label{eq:relations}
        E_{0x} \cos \delta_x &= E^\prime \cos\alpha \cos\beta, \\ 
        E_{0x} \sin \delta_x &= -E^\prime \sin\alpha \sin\beta, \\
        E_{0y} \cos \delta_y &= E^\prime \sin\alpha \cos\beta, \\
        E_{0y} \sin \delta_y &= E^\prime \cos\alpha \sin\beta, \\
        E_{0x}^2 &= E^{\prime 2} \left( \cos^2\alpha \cos^2\beta + \sin^2\alpha\sin^2\beta\right), \\
        E_{0y}^2 &= E^{\prime 2} \left( \sin^2\alpha \cos^2\beta + \cos^2\alpha\sin^2\beta\right). \\
    \end{aligned}
\end{equation}

On en déduit donc l'expression complexe des composantes du vecteur polarisation dans la base $(x,y)$ de l'observateur
\begin{equation}
    \begin{aligned}
        \bar{E}_x &= E_{0x} \exp [i(\omega t-\delta_x)], \\
        \bar{E}_y &= E_{0y} \exp [i (\omega t - \delta_y)]. \\
    \end{aligned}
\end{equation}

En prenant l'Ouest et le Nord\footnote{Pour avoir un trièdre direct avec le vecteur $\mathbf{K}$.} comme base (usuelle en astronomie), on peut ainsi définir l'angle de position du vecteur polarisation à l'instant $t$ noté EVPA (\textit{Electric Vector Position Angle}), calculé à partir du Nord et positif vers l'Est (voir Fig.~\ref{fig:convention}).

\subsection{Paramètres de Stokes}
Une seconde manière de représenter le vecteur polarisation est à travers les paramètres de Stokes\footnote{\href{https://en.wikipedia.org/wiki/Stokes\_parameters}{https://en.wikipedia.org/wiki/Stokes\_parameters}}, introduit en 1852. Cet ensemble de quatre paramètres pouvant être regroupés sous la forme d'un vecteur, appelé vecteur de Stokes, permet de décrire entièrement l'ellipse de la Fig.~\ref{fig:ellipse_polar}. Il ne s'agit rien de plus que d'une reparamétrisation de la polarisation. En effet, les précédents angles $\alpha$ et $\beta$ ne sont pas observables, contrairement aux paramètres de Stokes qui eux s'expriment comme des intensités spécifiques selon une direction particulière et sont donc observables. Ces paramètres sont définis à partir des composantes complexes du vecteur champ électrique $\mathbf{E}$ et sont liés aux paramètres $E^\prime$, $\alpha$ et $\beta$ tel que
\begin{equation}
    \begin{aligned}
        \label{eq:Stokes}
        I &\equiv \bar{E}_x \bar{E}_x^* + \bar{E}_y \bar{E}_y^* = E_{0x}^2 + E_{0y}^2 = E^{\prime 2}, \\
        Q &\equiv \bar{E}_x \bar{E}_x^* - \bar{E}_y \bar{E}_y^* = E_{0x}^2 - E_{0y}^2 = I \cos 2\alpha\,\cos 2\beta \\
        U &\equiv \bar{E}_x \bar{E}_y^* + \bar{E}_y \bar{E}_x^* = 2 E_{0x} \,E_{0y}\,\cos \delta = I \sin 2\alpha\,\cos 2\beta, \\
        V &\equiv  -i \left(\bar{E}_x \bar{E}_y^* - \bar{E}_y \bar{E}_x^*\right) = 2 E_{0x} \,E_{0y}\,\sin \delta = I \sin 2 \beta\\
    \end{aligned}
\end{equation}
avec la même notation qu'à la section précédente. Le signe négatif dans la définition de $V$ permet de mesurer l'angle de polarisation dans le sens trigonométrique en accord avec la convention de l'\href{https://www.iau.org/news/announcements/detail/ann16004/}{IAU} avec la dépendance en $+i\omega t$ de notre définition du champ électrique.

\begin{figure}
    \centering
    \resizebox{0.7\hsize}{!}{\includegraphics{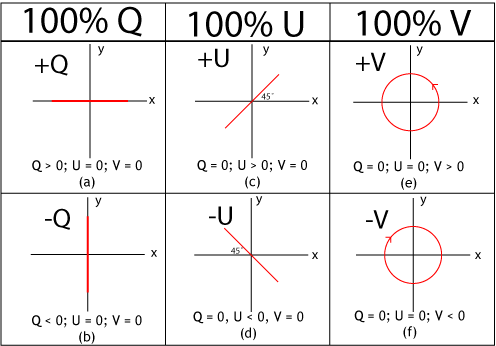}}
    \caption{Orientation du vecteur polarisation en fonction des valeurs des paramètres de Stokes. Crédit : Dan Moulton - \href{https://commons.wikimedia.org/w/index.php?curid=3319458}{https://commons.wikimedia.org/w/index.php?curid=3319458}}
    \label{fig:Stokes}
\end{figure}

Le premier paramètre $I$ correspond à l'intensité spécifique totale polarisée et non polarisée. Il s'agit de la quantité déjà discutée dans les chapitres précédents. On peut aisément déduire l'orientation du vecteur polarisation à partir des Eq.~\eqref{eq:Stokes} pour chaque paramètre de Stokes. Les paramètres $Q$ et $U$ décrivent la polarisation linéaire, avec $Q$ correspondant à un vecteur orienté horizontalement ou verticalement dans la base $(x,y)$ ($Q>0$ et $Q<0$ resp. avec $U=0$) et $U$ à un vecteur incliné à $\pm 45 \degree$ ($U>0$ et $U<0$ resp. avec $Q=0$), comme illustré dans la Fig.~\ref{fig:Stokes}. Le paramètre $V$, quant à lui, décrit la polarisation circulaire avec une polarisation dans le sens trigonométrique pour $V>0$ et dans le sens horaire pour $V<0$.

Comme il s'agit de grandeurs homogènes à une intensité spécifique, pouvant néanmoins être positives ou négatives, cette méthode de calcul de la polarisation est parfaitement adaptée à du tracé de rayon dans le sens inverse du temps (partant de l'observateur vers la source). L'orientation du vecteur polarisation, c'est-à-dire les paramètres de l'ellipse $\alpha$ et $\beta$ de la Fig.~\ref{fig:ellipse_polar}, peut être déterminée a posteriori à partir de l'ensemble des paramètres de Stokes $I$, $Q$, $U$ et $V$. Pour cela, on déduit des Eq.~\eqref{eq:Stokes} les relations suivantes
\begin{equation}
    \begin{aligned}
        \label{eq:polarangles}
        \tan 2 \alpha &= \frac{U}{Q}, \\
        \sin 2 \beta &= \frac{V}{I}. \\
    \end{aligned}
\end{equation}

De plus, à partir des Eq.~\eqref{eq:Stokes} et~\eqref{eq:polarangles}, on remarque que les paramètres $I$ et $V$ sont indépendants du choix de la base $(x,y)$ car ils ne dépendent pas de $\alpha$ : ils sont invariants par rotation contrairement à $Q$ et $U$.

\subsection{Degré de polarisation}
Dans le cas simple décrit précédemment d'une onde monochromatique unique entièrement polarisée, l'intensité totale peut être déterminée à partir des trois autres paramètres de Stokes tels que
\begin{equation}
    \label{eq:full_polar_Stokes_relation}
    I^2 = Q^2 + U^2 + V^2.
\end{equation}

Cependant, on n'observe jamais une onde unique monochromatique, mais un paquet d'ondes dont les phases sont indépendantes, se traduisant par une évolution du champ électrique plus complexe que dans la Fig.~\ref{fig:polar_states}. En effet, le champ électrique résultant est la somme des champs électriques des ondes individuelles. La polarisation en fonction du temps ne suit donc plus une ellipse bien définie. Il est donc utile de mesurer les paramètres de Stokes en faisant une moyenne temporelle (sur une période plus grande que la période d'oscillation)
\begin{equation}
    \begin{aligned}
        I &= \langle \bar{E}_x \bar{E}_x^* \rangle + \langle \bar{E}_y \bar{E}_y^*\rangle, \\
        Q &= \langle \bar{E}_x \bar{E}_x^*\rangle - \langle \bar{E}_y \bar{E}_y^* \rangle, \\
        U &= \langle \bar{E}_x \bar{E}_y^*\rangle + \langle \bar{E}_y \bar{E}_x^*\rangle, \\
        V &=  -i \left(\langle \bar{E}_x \bar{E}_y^* \rangle - \langle \bar{E}_y \bar{E}_x^*\rangle\right). \\
    \end{aligned}
\end{equation}

Si le vecteur polarisation est purement aléatoire, c'est-à-dire sans  aucune direction privilégiée, les paramètres de Stokes $Q$, $U$ et $V$ moyennés seront nuls, correspondant à une lumière non polarisée. A contrario, dans le cas d'une lumière polarisée, soit par son émission, soit par le milieu traversé (polariseur ou lame quart d'onde), la moyenne temporelle des paramètres de Stokes sera non nulle. Dans ce cas, l'Eq.~\eqref{eq:full_polar_Stokes_relation} devient
\begin{equation}
    \label{eq:partial_polar_Stokes_relation}
    I^2 \geq Q^2 + U^2 + V^2.
\end{equation}
On parle dans ce cas de lumière partiellement polarisée. On peut définir le degré de polarisation $d_P$ comme le rapport entre l'intensité spécifique de la lumière polarisée et l'intensité spécifique totale telles que
\begin{equation}
    d_P = \frac{\sqrt{Q^2 + U^2 + V^2}}{I}.
\end{equation}
La polarisation pouvant être découpée en deux composantes, la polarisation linéaire et circulaire, il est intéressant de définir leur degré de polarisation ($d_{lp}$ et $d_{cp}$ resp.) de la manière suivante
\begin{equation}
    \label{eq:linear_polar_fraction}
    d_{lp}=\frac{\sqrt{Q^2 + U^2}}{I}
\end{equation}
et 
\begin{equation}
    \label{eq:circular_polar_fraction}
    d_{cp}=\frac{|V|}{I}.
\end{equation}

\section{Polarisation en espace-temps courbe}
\subsection{Tenseur électromagnétique}
Cette partie est fortement inspirée des notes d'optique en relativité générale d'Eric Gourgoulhon.

On considère un espace-temps quadridimensionnel $\mathcal{M}$ défini par la métrique $\mathbf{g}$. Dans cet espace-temps, le champ électromagnétique peut être entièrement décrit par la 2-forme $\mathbf{F}$, c'est-à-dire un tenseur défini sur $\mathcal{M}$ antisymétrique ($F_{\alpha \beta} = -F_{\beta \alpha}$), indépendant du référentiel appelé \textit{tenseur électromagnétique} ou encore \textit{tenseur Faraday}. Ce dernier permet d'exprimer la 4-force de Lorentz appliquée à une particule chargée selon
\begin{equation}
    m u^\mu \nabla_\mu u_\alpha = q F_{\alpha \mu} u^\mu
\end{equation}
avec $m$ la masse de la particule, $q$ sa charge électrique, $\mathbf{u}$ sa 4-vitesse et $\mathbf{\nabla}$ la dérivation covariante. Les \textit{équations de Maxwell} relativistes décrivant le champ électromagnétique s'écrivent
\begin{equation}
    \begin{aligned}
        \label{eq:Maxwell_tensoriel}
        \mathbf{d F} &= 0 \\
        \boldsymbol{\nabla} \cdot \mathbf{F}^\# &= \mu_0 \mathbf{j}\footnote{Ici on utilise la convention $c=1$}
    \end{aligned}
\end{equation}
avec $\mathbf{j}$ le 4-vecteur courant électrique, $\mu_0$ la perméabilité diélectrique du vide et $\mathbf{F}^\#$ un tenseur contravariant déduit de $\mathbf{F}$ et de la métrique $\mathbf{g}$ telle que
\begin{equation}
    F^{\# \alpha \beta} = g^{\alpha \mu} g^{\beta \nu} F_{\mu \nu}.
\end{equation}
Écrit sous format indiciel et avec un système de coordonnées ($x^\alpha$), les Eqs.~\eqref{eq:Maxwell_tensoriel} deviennent
\begin{equation}
    \begin{aligned}
        \label{eq:Maxwell_indiciel}
        \frac{\partial F_{\beta \gamma}}{\partial x^\alpha} + \frac{\partial F_{\gamma \alpha}}{\partial x^\beta} +\frac{\partial F_{\alpha \beta}}{\partial x^\gamma} &= 0, \\
        \frac{1}{\sqrt{-g}} \frac{\partial}{\partial x^\mu} \left( \sqrt{-g} F^{\alpha \mu} \right) &= \mu_0\, j^\alpha
    \end{aligned}
\end{equation}
où $g$ est le déterminant de la métrique dans le système de coordonnées ($x^\alpha$).

De manière similaire à la physique classique où le champ électromagnétique peut être relié au potentiel du même nom, on peut définir un 4-potentiel $\mathbf{A}$ tel que
\begin{equation}
    \mathbf{F} = \mathbf{d A},
\end{equation}
soit, avec le système de coordonnées ($x^\alpha$)
\begin{equation}
    F_{\alpha \beta} = \frac{\partial A_\beta}{\partial x^\alpha} - \frac{\partial A_\alpha}{\partial x^\beta},
\end{equation}
qui est solution de la première équation de Maxwell~\eqref{eq:Maxwell_tensoriel}. Cependant, $\mathbf{A}$ n'est pas unique et nécessite un choix particulier appelé \textit{choix de jauge}. La plus simple, nommée \textit{jauge de Lorentz}\footnote{Nommée d'après le physicien Dannois Ludvig Valentin Lorenz (1829-1891), à ne pas confondre avec le physicien Néerlandais Hendrik A. Lorentz (1853-1928).}, étant définie par
\begin{equation}
    \begin{aligned}
        \boldsymbol{\nabla} \cdot \mathbf{A} &= 0 \\
        \Leftrightarrow \nabla_\mu A^\mu &= 0.
    \end{aligned}
\end{equation}

Dans le cas d'une onde plane monochromatique, le potentiel associé à l'onde est de la forme
\begin{equation}
    \mathbf{\hat{A}} = \mathbf{\hat{a}} e^{i \Phi},
\end{equation}
avec $\mathbf{\hat{a}}$ l'amplitude complexe\footnote{$\mathbf{\hat{a}}$ doit être un nombre complexe pour permettre tous les états de polarisation, notamment la polarisation circulaire.} dont la variation est supposée plus lente que celle de $\Phi$ (approximation de l'optique géométrique). Le tenseur Faraday devient
\begin{equation}
    \label{eq:Faraday_tensor}
    \begin{aligned}
        \hat{F}_{\alpha \beta} =& \nabla_\alpha \hat{A}_\beta - \nabla_\beta \hat{A}_\alpha \\
        =& e^{i\Phi} \nabla_\alpha \hat{a}_\beta + \hat{a}_\beta\,i \,e^{i\Phi}\,\nabla_\alpha \Phi - \left( e^{i\Phi} \nabla_\beta \hat{a}_\alpha + \hat{a}_\alpha\,i \,e^{i\Phi}\,\nabla_\beta \Phi\right) \\
        \approx& i\left(\hat{a}_\beta\,k_\alpha - \hat{a}_\alpha\,k_\beta \right)\,e^{i \Phi},
    \end{aligned}
\end{equation}
où l'on introduit le vecteur d'onde
\begin{equation}
    \mathbf{k} \equiv \boldsymbol{\nabla}{\Phi}.
\end{equation}
Les termes $\nabla_\alpha \hat{a}_\beta$ sont négligés en utilisant l'approximation de l'optique géométrique.

On peut ainsi définir le \textit{vecteur polarisation} unitaire
\begin{equation}
    \mathbf{\hat{f}} \equiv \frac{\mathbf{\hat{a}}}{a}
\end{equation}
où $a$ est un scalaire correspondant au module du vecteur complexe $\mathbf{\hat{a}}$. Ce vecteur est par définition perpendiculaire à $\mathbf{k}$ et transporté parallèlement le long de $\mathbf{k}$
\begin{equation}
    \mathbf{\hat{f}} \cdot \mathbf{k} = 0, \quad \boldsymbol{\nabla}_\mathbf{k} \mathbf{\hat{f}} =0.
\end{equation}

Le tenseur électromagnétique peut donc être réécrit de la manière suivante
\begin{equation}
    \hat{F}_{\alpha \beta} = i a \left(\hat{f}_\beta\,k_\alpha - \hat{f}_\alpha\,k_\beta \right)\,e^{i \Phi}.
\end{equation}
On en déduit que l'on peut ajouter n'importe quel multiple de vecteur tangent au photon $\mathbf{k}$ au vecteur polarisation $\mathbf{\hat{f}}$ tel que 
\begin{equation}
    \mathbf{\hat{f}} \longmapsto \mathbf{\hat{f}} + q \mathbf{k},
\end{equation}
où q est un champ scalaire arbitraire, sans changer le tenseur électromagnétique.

 
\subsection{Champs électrique et magnétique}
Contrairement au tenseur électromagnétique $\mathbf{\hat{F}}$ qui est indépendant du référentiel, les champs électrique $\mathbf{E}$ et magnétique $\mathbf{B}$ sont définis par rapport à un observateur $\mathcal{O}$ et sa 4-vitesse $\mathbf{u}$. La 1-forme (coordonnée covariante) du champ électrique $\mathbf{E}$ est reliée au tenseur électromagnétique par
\begin{equation}
    \label{eq:E_1form}
    E_\alpha = \Re (\hat{F}_{\alpha \mu} u^\mu).
\end{equation}
Le vecteur (coordonnée contravariante) champ magnétique est lui défini par
\begin{equation}
    \label{eq:B_vector}
    B^\alpha = \Re \left(-\frac{1}{2} \epsilon^{\alpha \mu \nu}_{\quad \  \rho} \hat{F}_{\mu \nu} u^\rho \right)
\end{equation}
où $\epsilon^{\alpha \mu \nu}_{\quad \  \rho}$ correspond à un tenseur de Levi-Civita dont les trois premiers indices ont été montés par le tenseur métrique $g^{\alpha \beta}$ tel que
\begin{equation}
    \epsilon^{\alpha \mu \nu}_{\quad \  \rho} = g^{\alpha \beta}\, g^{\mu \sigma}\, g^{\nu \xi}\, \epsilon_{\beta \sigma \xi \rho}
\end{equation}
où $\epsilon_{\beta \sigma \xi \rho}$ est le tenseur antisymétrique $4 \times 4$ de type (0,4)\footnote{Zéro coordonnée contravariante et quatre coordonnées covariantes.} de Levi-Civita, associé au tenseur métrique $\mathbf{g}$ dont le déterminant est $g$ dans une base donnée, et défini par
\begin{equation}
    \label{eq:Levi_Civita_tensor}
    \epsilon_{\alpha \beta \gamma \delta} = \pm \sqrt{-g} \left\{
    \begin{array}{ll}
        1 & \mbox{si } (\alpha, \beta, \gamma, \delta) \mbox{ est une permuation paire de (0,1,2,3)\footnotemark}\\
        -1 & \mbox{si } (\alpha, \beta, \gamma, \delta) \mbox{ est une permuation impaire de (0,1,2,3)\footnotemark}\\
        0 & \mbox{si au moins deux indices sont égaux}
    \end{array}
\right.
\end{equation}
Le signe dans l'Eq.~\eqref{eq:Levi_Civita_tensor} (avant la racine de $-g$) est déterminé par le type de base, positif si le repère est direct, négatif s'il est indirect.
\addtocounter{footnote}{-1}
\footnotetext{Exemple (0,1,2,3), (1,2,3,0), (2,3,0,1), ...}
\stepcounter{footnote}
\footnotetext{Exemple (3,2,1,0), (0,3,2,1), (1,0,3,2), ...}

Pour déterminer les coordonnées contravariantes du champ électrique mesuré par l'observateur $\mathcal{O}$, il suffit de monter l'indice avec la métrique
\begin{equation}
    E^\alpha = g^{\alpha \beta} E_\beta.
\end{equation}
Par construction, les champs électrique $\mathbf{E}$ et magnétique $\mathbf{B}$ sont orthogonaux à la 4-vitesse de l'observateur $\mathbf{u}$ tel que 
\begin{equation}
    \label{eq:orthogonality}
    E_\mu u^\mu = 0 \quad \mbox{et} \quad B_\mu u^\mu = 0.
\end{equation}
Comme $\mathbf{u}$ est un vecteur de genre temps, cela implique que les vecteurs $\mathbf{E}$ et $\mathbf{B}$ sont de genre espace.

Au vu des équations \eqref{eq:E_1form} à \eqref{eq:orthogonality}, le tenseur électromagnétique peut être entièrement défini par $\mathbf{u}$, $\mathbf{E}$ et $\mathbf{B}$ tel que
\begin{equation}
    F_{\alpha \beta} = \Re (\hat{F}_{\alpha \beta}) = u_\alpha E_\beta - E_\alpha u_\beta + \epsilon_{\mu \nu \alpha \beta}\, u^\mu \, B^\nu.
\end{equation}

Attention, les champs électrique et magnétique décrits ici sont ceux de l'onde électromagnétique reçue par un observateur $\mathcal{O}$, à ne pas confondre avec le champ magnétique d'un milieu quelconque.

\subsection{Transport parallèle de la base de l'observateur}
La définition du vecteur polarisation ci-dessus correspond à la définition en relativité générale à partir du tenseur électromagnétique. Or, on s'intéresse ici au vecteur polarisation $\mathbf{F_0}$ correspondant à l'orientation du vecteur champ électrique dans une base (que l'on va définir) de l'observateur calculée à partir du champ magnétique associé à l'onde incidente dans le référentiel de l'observateur ($\mathbf{F_0} = \mathbf{K_0} \times \mathbf{B_0}$). Ces deux définitions sont bien évidements reliées, cependant, la relation exacte entre ces deux quantités est non triviale et va au-delà du champ d'application de cette thèse. En effet, on fait le choix, de par la nature \textit{backward} du tracé de rayon de \textsc{Gyoto}, de définir la base de polarisation de l'observateur que l'on va transporter parallèlement jusqu'à la source d'émission où l'on déterminera la valeur des paramètres de Stokes. Ainsi, on ne traite à aucun moment le tenseur électromagnétique.

\begin{figure}
    \centering
    \resizebox{\hsize}{!}{\includegraphics{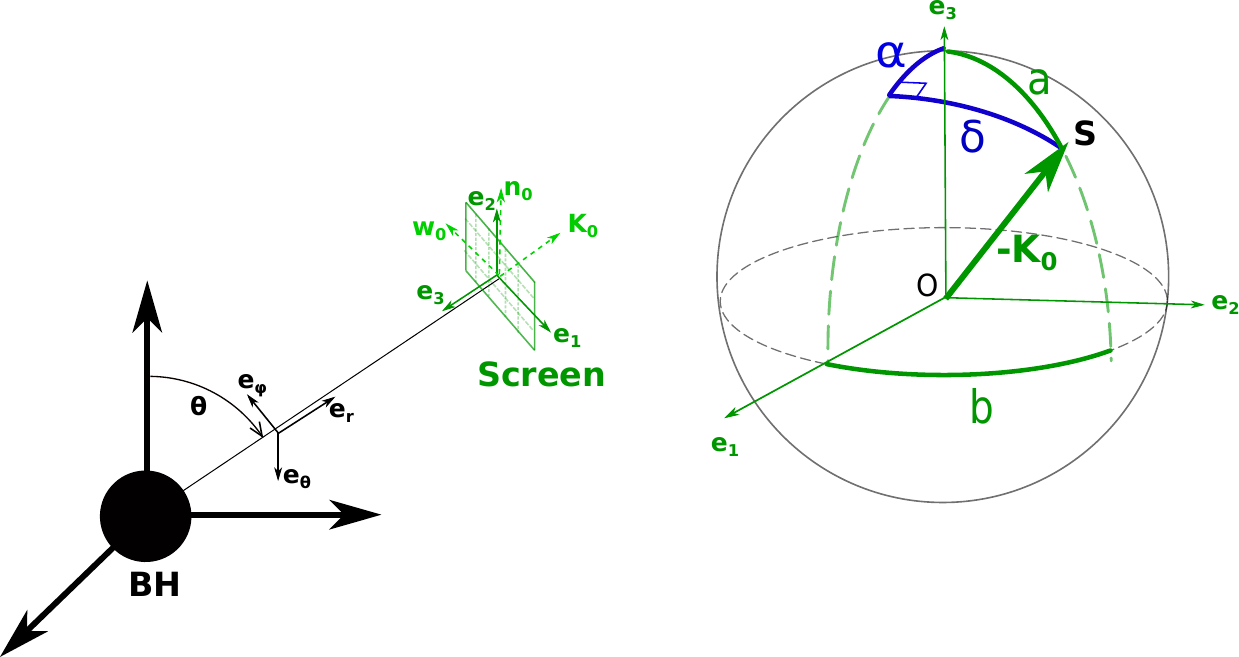}}
    \caption{Condition initiale du problème de tracé de rayons polarisés au niveau de l'écran de l'observateur distant avec à gauche la définition des différents vecteurs dans un espace-temps de Kerr avec un système de coordonnées sphériques. On note que cela est valable quels que soient le système de coordonnées et la métrique. L'écran est considéré comme ponctuel aux coordonnées $(r, \theta, \varphi)$. Le vecteur $\mathbf{e_3}$ est normal à l'écran et selon $-\mathbf{e_r}$. Pour chaque pixel de l'écran, les vecteurs définissant la base de polarisation $(\mathbf{K_0}, \mathbf{w_0}, \mathbf{n_0})$ dépendent de l'ascension droite $\alpha$ et de la déclinaison $\delta$ du pixel par rapport au pixel central. Crédit : Frédéric Vincent.}
    \label{fig:obs_frame}
\end{figure}

Les conditions initiales de l'intégration des géodésiques de genre lumière sont définies par les coordonnées de l'écran, considéré comme ponctuel, dans le système de coordonnées quelconque centré sur le trou noir\footnote{Cela est valable pour n'importe quels métrique et système de coordonnées.} (voir Chap.~\ref{chap:GYOTO}). Pour la suite, on considère un système de coordonnées sphérique avec la base orthonormale $(\mathbf{e_t}, \mathbf{e_r}, \mathbf{e_\theta}, \mathbf{e_\varphi})$. On définit le vecteur $\mathbf{e_3}$ comme le vecteur normal au plan de l'écran dirigé vers le trou noir ($\mathbf{e_3} = - \mathbf{e_r}$) et les vecteurs $\mathbf{e_1}$ et $\mathbf{e_2}$ forment ainsi le plan de l'écran avec $\mathbf{e_2}$ aligné sur l'axe de spin du trou noir ($\mathbf{e_2}=-\mathbf{e_\theta}$). Cependant, la direction de réception du photon $\mathbf{K_0}$ diffère de pixel à pixel (encodant les différentes directions de réception des photons sur le ciel). Pour un écran avec $N \times N$ pixels, le pixel d'indice $(i,j)$, avec $i$ et $j$ des entiers allant de $1$ à $N$, peut être associé à une ascension droite $\alpha$ et une déclinaison $\delta$\footnote{On choisit de décrire l'écran en coordonnées équatoriales de manière similaires aux observations. D'autres choix de coordonnées sont possibles.} (par rapport à la valeur du pixel central servant de référence), comme illustré dans la Fig.~\ref{fig:obs_frame}, et s'écrivent
\begin{equation}
    \begin{aligned}
        \alpha &= \frac{f}{N} \left( i-\frac{N+1}{2} \right), \\
        \delta &= \frac{f}{N} \left( j-\frac{N+1}{2} \right),
    \end{aligned}
\end{equation}
où $f$ est le champ de vue de l'écran. On peut en déduire les angles sphériques $a$ et $b$ correspondant aux coordonnées angulaires du photon incident sur le ciel local de l'observateur (voir Fig.~\ref{fig:obs_frame}) qui s'expriment selon
\begin{equation}
    \begin{aligned}
        \cos a &= \cos \alpha \, \cos \delta, \\
        \tan b &= \frac{\tan \alpha}{\sin \delta},
    \end{aligned}
\end{equation}
permettant de définir $\mathbf{K_0}$ dans le repère ($\mathbf{e_1}$, $\mathbf{e_2}$, $\mathbf{e_3}$)
\begin{equation}
    \label{eq:K_0}
    \mathbf{K_0} = -\sin a \cos b\, \mathbf{e_1} - \sin a \sin b\, \mathbf{e_2} - \cos a\, \mathbf{e_3},
\end{equation}
ainsi que les vecteurs Ouest $\mathbf{w_0}$ et Nord $\mathbf{n_0}$ définis vers le haut et la droite de l'écran resp. tel que
\begin{equation}
    \mathbf{w_0} = \left[ \sin^2 b (1- \cos a) - \cos a \right] \, \mathbf{e_1} + \sin b \cos b (1-\cos a)\, \mathbf{e_2} + \cos b \sin a \, \mathbf{e_3},
\end{equation}
\begin{equation}
    \label{eq:n_0}
    \mathbf{n_0} = - \sin b \cos b (1-\cos a)\, \mathbf{e_1} + \left[ \cos^2 b (1- \cos a) + \cos a \right] \, \mathbf{e_2} - \sin b \sin a \, \mathbf{e_3}.
\end{equation}

Ces trois vecteurs $(\mathbf{K_0}, \mathbf{w_0}, \mathbf{n_0})$ sont des vecteurs unitaires dans l'espace-temps plat de repos de l'observateur. On note que $\mathbf{K_0}$ correspond à la projection orthogonale à la 4-vitesse de l'observateur $\mathbf{u_0}$ du vecteur tangent au photon $\mathbf{k_0}$ qui constitue la condition initiale du tracé de rayon pour le pixel considéré.

On obtient donc pour le pixel central, où $a=b=0$, une direction purement radiale pour la direction du photon $\mathbf{K_0}^\mathrm{cen}=-\mathbf{e_3}$, ainsi que $\mathbf{w_0}^\mathrm{cen}=-\mathbf{e_1}$ et $\mathbf{n_0}^\mathrm{cen}=\mathbf{e_2}$.

\begin{figure}
    \centering
    \resizebox{0.6\hsize}{!}{\includegraphics{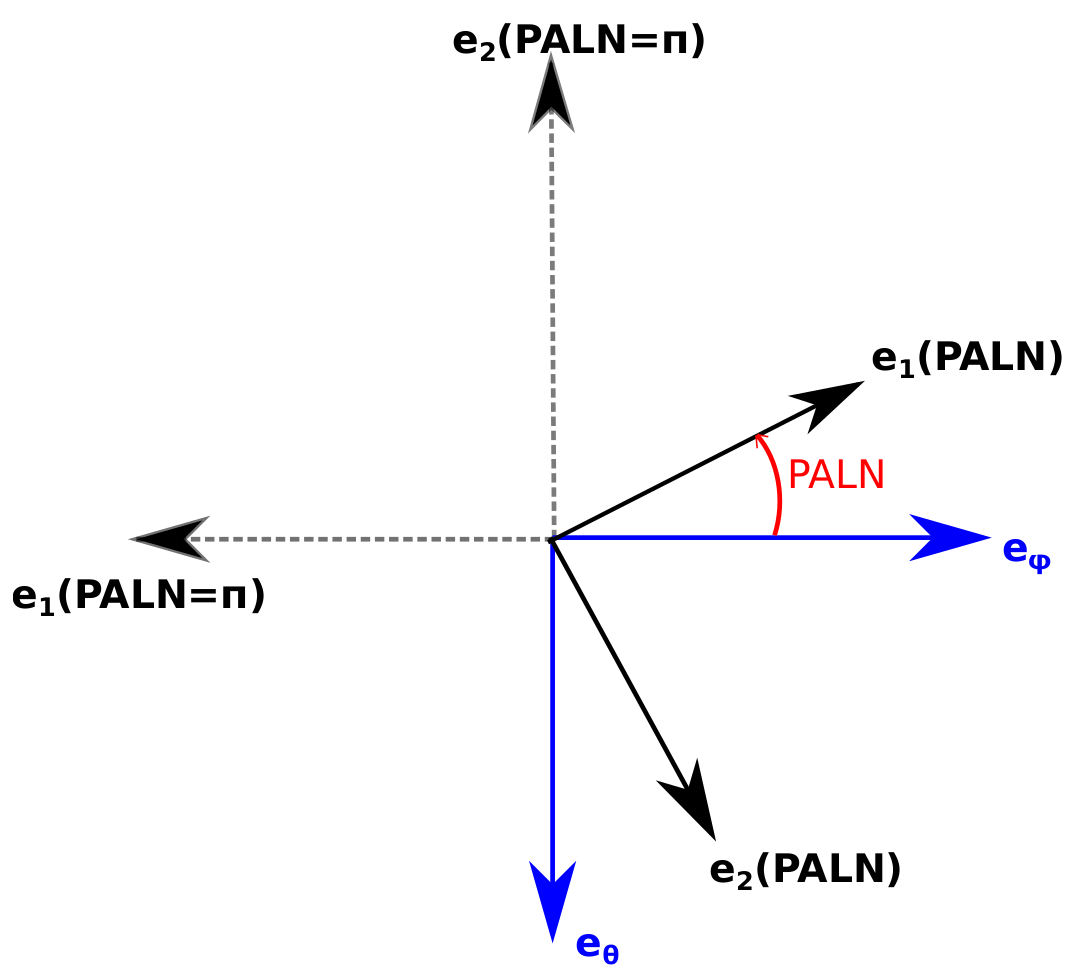}}
    \caption{Schéma du plan de l'écran illustrant la définition des vecteurs $\mathbf{e_1}$ et $\mathbf{e_2}$ en fonction des vecteurs de base $\mathbf{e_\varphi}$ et $\mathbf{e_\theta}$ ainsi que du PALN. Crédit : Frédéric Vincent.}
    \label{fig:paln}
\end{figure}

Pour simplifier, nous avons considéré précédemment que $\mathbf{e_2}$ est aligné sur le spin du trou noir, c'est-à-dire $\mathbf{e_2}=-\mathbf{e_\theta}$ qui correspond à un PALN de $180\degree$, comme illustré dans la Fig.~\ref{fig:paln}. Pour un PALN arbitraire, les vecteurs $\mathbf{e_1}$ et $\mathbf{e_2}$ deviennent
\begin{equation}
    \begin{aligned}
        \mathbf{e_1} &= \cos (\mathrm{PALN}) \, \mathbf{e_\varphi} - \sin (\mathrm{PALN}) \, \mathbf{e_\theta}, \\
        \mathbf{e_2} &= \sin (\mathrm{PALN}) \, \mathbf{e_\varphi} + \cos (\mathrm{PALN}) \, \mathbf{e_\theta}, \\
        \mathbf{e_3} &= -\mathbf{e_r}.
    \end{aligned}
\end{equation}

\begin{figure}
    \centering
    \resizebox{0.6\hsize}{!}{\includegraphics{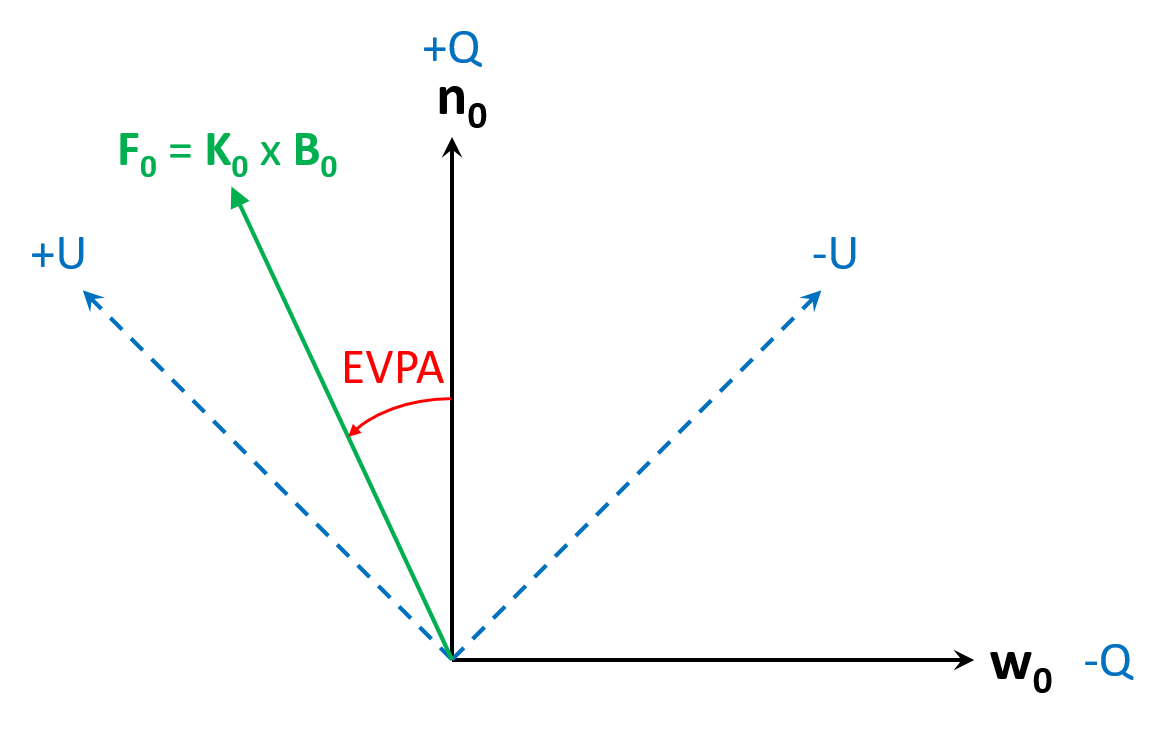}}
    \caption{Schéma explicitant la convention des paramètres de Stokes dans l'écran. L'orientation du vecteur polarisation dans la base de l'écran définit l'EVPA.}
    \label{fig:convention}
\end{figure}

On a donc défini la base de polarisation de l'observateur ($\mathbf{w_0}$,$\mathbf{n_0}$) dans laquelle sont déterminés les paramètres de Stokes observés avec la convention illustrée dans la Fig.~\ref{fig:convention}. Afin de calculer ces derniers en fonction du rayonnement de la source, il faut transporter parallèlement ces deux vecteurs de base le long de la géodésique de genre lumière du vecteur tangent au photon $\mathbf{k}$, de l'écran jusqu'à l'émetteur. On nomme $\mathbf{w}$ et $\mathbf{n}$ les vecteurs de la base de polarisation de l'écran transportés parallèlement le long de $\mathbf{k}$, l'indice $0$ marquant ainsi les conditions initiales au niveau de l'écran. L'intégration de la géodésique et le transport parallèle de $\mathbf{w}$ et $\mathbf{n}$ se fait en résolvant les équations suivantes
\begin{equation}
    \begin{aligned}
        \boldsymbol{\nabla_\mathbf{k}} \mathbf{k} &= \mathbf{0}, \\
        \boldsymbol{\nabla_\mathbf{k}} \mathbf{w} &= \mathbf{0}, \\
        \boldsymbol{\nabla_\mathbf{k}} \mathbf{n} &= \mathbf{0}. \\
    \end{aligned}
\end{equation}

\subsection{Projection dans le référentiel de l'émetteur}
Une fois transportée parallèlement de l'observateur jusqu'à l'émetteur, la base de polarisation de l'écran doit être projetée dans le référentiel de l'émetteur afin de déterminer les composantes du vecteur polarisation dans cette base. Cependant, deux vecteurs initialement orthogonaux, comme les vecteurs définissant la base de polarisation de l'écran $\mathbf{w}$ et $\mathbf{n}$, une fois projetés perpendiculairement à un troisième vecteur, ici, la 4-vitesse de l'émetteur $\mathbf{u}$, ne sont a priori plus orthogonaux, c'est-à-dire
\begin{equation}
        \perp_\mathbf{u} \mathbf{n}\, \cdot \perp_\mathbf{u} \mathbf{w} \neq 0
\end{equation}
où $\perp_\mathbf{u} \mathbf{n}$ est la projection orthogonale à $\mathbf{u}$ de $\mathbf{n}$ et s'exprime comme $\perp_\mathbf{u} \mathbf{n} = \mathbf{n} + (\mathbf{n} \cdot \mathbf{u}) \mathbf{u}$ (idem pour $\mathbf{w}$). On note que le signe entre les deux termes de droite est positif du fait que $\mathbf{u}$ est un vecteur de genre temps ($\mathbf{u} \cdot \mathbf{u} = -1$). On a donc
\begin{equation}
    \Leftrightarrow (\mathbf{n} + (\mathbf{n} \cdot \mathbf{u}) \mathbf{u}) \cdot (\mathbf{w} + (\mathbf{w} \cdot \mathbf{u}) \mathbf{u}) \neq 0.
\end{equation}
Or, pour déterminer correctement le vecteur polarisation dans la base ($\mathbf{w}$,$\mathbf{n}$), les vecteurs formant cette dernières doivent être orthogonaux dans la base de l'émetteur. Pour cela, on définit le vecteur $\mathbf{n}^\prime$ tel que
\begin{equation} \label{eq:decompo_nord}
    \mathbf{n}^\prime = \mathbf{n} + \alpha \mathbf{k}
\end{equation}
où $\alpha$ est un scalaire. Les vecteurs $\mathbf{n}^\prime$ et $\mathbf{k}$ sont orthogonaux du fait que $\mathbf{n} \cdot \mathbf{k} = 0$ et $\mathbf{k} \cdot \mathbf{k} = 0$. On calcule maintenant la projection orthogonale à la 4-vitesse de l'émetteur $\mathbf{u}$ de $\mathbf{n}^\prime$ 
\begin{equation}
    \begin{aligned}
        \perp_\mathbf{u} \mathbf{n}^\prime &= \perp_\mathbf{u} \mathbf{n} + \alpha \perp_\mathbf{u} \mathbf{k} \\
        &= \mathbf{n} + (\mathbf{n} \cdot \mathbf{u}) \mathbf{u} + \alpha \left( \mathbf{k} + (\mathbf{k} \cdot \mathbf{u}) \mathbf{u} \right).
    \end{aligned}
\end{equation}
En faisant le produit scalaire de ce dernier avec le vecteur $\mathbf{k}$, on obtient
\begin{equation}
    \begin{aligned}
        \perp_\mathbf{u} \mathbf{n}^\prime \, \cdot \mathbf{k} &= \mathbf{n} \cdot \mathbf{k} + (\mathbf{n} \cdot \mathbf{u}) (\mathbf{u} \cdot \mathbf{k}) + \alpha \left(\mathbf{k} \cdot \mathbf{k} + (\mathbf{k} \cdot \mathbf{u}) (\mathbf{u} \cdot \mathbf{k}) \right) \\
        0 &= (\mathbf{n} \cdot \mathbf{u}) (\mathbf{u} \cdot \mathbf{k}) + \alpha (\mathbf{k} \cdot \mathbf{u}) (\mathbf{u} \cdot \mathbf{k})
    \end{aligned}
\end{equation}
puisque $\mathbf{n} \cdot \mathbf{k} = \mathbf{k} \cdot \mathbf{k} = 0$. On a donc
\begin{equation}
    \begin{aligned}
        -\mathbf{n} \cdot \mathbf{u} &= \alpha (\mathbf{k} \cdot \mathbf{u}) \\
        \rightarrow \alpha &= - \frac{\mathbf{n} \cdot \mathbf{u}}{\mathbf{k} \cdot \mathbf{u}}.
    \end{aligned}
\end{equation}

On en déduit donc les expressions de $\mathbf{n}^\prime$ et $\mathbf{w}^\prime$ (de manière similaire), les vecteurs de la base de polarisation projetés orthogonalement à $\mathbf{u}$
\begin{equation}
    \label{eq:nord_west_prime}
    \begin{aligned}
        \mathbf{n}^\prime \, =& \, \mathbf{n} - \frac{\mathbf{n} \cdot \mathbf{u}}{\mathbf{k} \cdot \mathbf{u}}\, \mathbf{k} \\
        \mathbf{w}^\prime \, =& \, \mathbf{w} - \frac{\mathbf{w} \cdot \mathbf{u}}{\mathbf{k} \cdot \mathbf{u}}\, \mathbf{k} \\
    \end{aligned}
\end{equation}

\begin{figure}
    \centering
    \resizebox{\hsize}{!}{\includegraphics{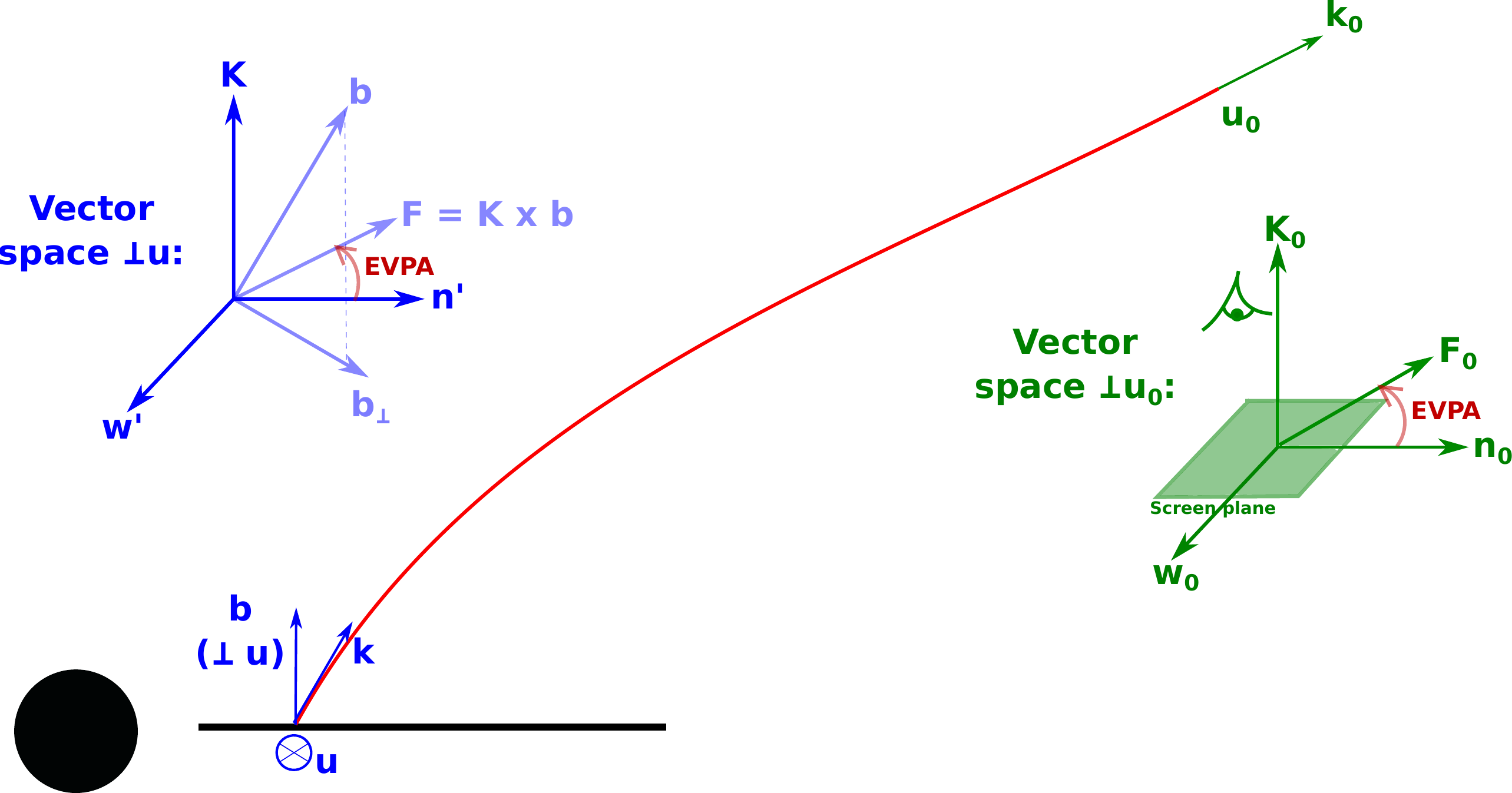}}
    \caption{Schéma récapitulant le problème du tracé de rayon en sens inverse du temps avec les référentiels de l'observateur (c-à-d perpendiculaire à la 4-vitesse $\mathbf{u_0}$ de ce dernier) en vert et de l'émetteur (perpendiculaire à la 4-vitesse $\mathbf{u}$ de la source) en bleu. Pour simplifier le schéma, le champ magnétique dans le référentiel du trou noir est considéré vertical et perpendiculaire à $\mathbf{u}$. Les vecteurs $\mathbf{n_0}$ et $\mathbf{w_0}$, correspondant au Nord et à l'Ouest resp., forment la base de polarisation de l'observateur. Ces derniers sont transportés parallèlement le long de la géodésique jusqu'à l'émetteur où ils sont projetés perpendiculairement à $\mathbf{u}$ grâce aux Eqs.~\eqref{eq:nord_west_prime}. Le vecteur polarisation $\mathbf{F}$ est défini (dans le cas du synchrotron) orthogonal à $\mathbf{K}$ et $\mathbf{b_\perp}$, respectivement la direction du photon dans le référentiel de l'émetteur et la projection orthogonale à $\mathbf{K}$ du champ magnétique $\mathbf{b}$. L'EVPA pour \textit{Electric Vector Position Angle} est l'angle du vecteur polarisation par rapport au Nord compté positivement vers l'Est. Crédit : Frédéric Vincent.}
    \label{fig:bases_polar}
\end{figure}

On note que ce qui nous intéresse, c'est l'angle entre le vecteur polarisation et le Nord (positif vers l'Est), soit $\mathbf{n} \cdot \mathbf{F}$. On a donc, d'après l'Eq.~\eqref{eq:decompo_nord},
\begin{equation}
    \mathbf{n} \cdot \mathbf{F} \, = \, \mathbf{n}^\prime \cdot \mathbf{F} + \alpha \mathbf{k} \cdot \mathbf{F}.
\end{equation}
En décomposant $\mathbf{k}$ en sa composante selon $\mathbf{u}$, et sa composante orthogonale à $\mathbf{u}$, que l'on nomme $\mathbf{K}$, tel que $\mathbf{k} = \mathbf{K} + \beta \mathbf{u}$, on peut déterminer $\mathbf{k} \cdot \mathbf{F}$ comme suit :
\begin{equation}
    \mathbf{k} \cdot \mathbf{F} = \mathbf{K} \cdot \mathbf{F} + \beta \mathbf{u} \cdot \mathbf{F}.
\end{equation}
Or, par définition $\mathbf{K} \cdot \mathbf{F} = 0$ et $\mathbf{u} \cdot \mathbf{F} = 0$ car $\mathbf{F}$ appartient à l'espace de repos de l'émetteur, orthogonal à $\mathbf{u}$. Ainsi, $\mathbf{k} \cdot \mathbf{F} = 0$, donc on a $\mathbf{n} \cdot \mathbf{F} \, = \, \mathbf{n}^\prime \cdot \mathbf{F}$ (idem pour $\mathbf{w}$). Il est aisé de vérifier que $\mathbf{n}^\prime$ et $\mathbf{w}^\prime$ sont bien orthogonaux entre eux et avec $\mathbf{K}$ normalisée. On a donc défini l'ensemble des vecteurs, dans le référentiel de l'émetteur, formant une base orthonormale directe ($\mathbf{K}$,$\mathbf{w}^\prime$,$\mathbf{n}^\prime$) comme illustré dans la partie en haut à gauche de la Fig.~\ref{fig:bases_polar}, dont on a besoin pour déterminer le vecteur de polarisation observé par un observateur lointain.

\section{Transfert radiatif polarisé en relativité générale}
\subsection{Équations}
Dans le cas non polarisé, où l'on s'intéresse uniquement à l'intensité spécifique totale invariante $\mathcal{I}$ définie par l'Eq.~\eqref{eq:conserv_Inu}, l'équation du transfert radiatif relativiste~\cite{Mihalas&Mihalas1984} est
\begin{equation}
    \label{eq:transfert_rad_relat}
    \frac{d\mathcal{I}}{d\lambda} = \mathcal{E} - \mathcal{A} \mathcal{I}
\end{equation}
avec $\mathcal{E}$ et $\mathcal{A}$ l'émission invariante et l'absorption invariante respectivement, et $\lambda$ le paramètre affine de la géodésique (à ne pas confondre avec la longueur d'onde !). La généralisation de cette équation dans le cas polarisé sur les quatre paramètres de Stokes, décrit par~\cite{Melrose1991}, s'écrit
\begin{equation}
    \label{eq:polarizedradtransf-frameinv}
 \frac{\dd}{\dd \lambda}\left( \begin{array}{c}
     I/\nu^3    \\
     Q/\nu^3    \\
     U/\nu^3    \\
     V/\nu^3    \end{array} \right) 
   =  
   \left( \begin{array}{c}
     j_I/\nu^2    \\
     j_Q/\nu^2    \\
     j_U/\nu^2    \\
     j_V/\nu^2    \end{array} \right)
   -
   \left( \begin{array}{cccc}
     \nu\,\alpha_I         &   \nu\,\alpha_Q  &  \nu\,\alpha_U  &  \nu\,\alpha_V   \\
     \nu\,\alpha_Q         &   \nu\,\alpha_I  &\nu\,r_V & -\nu\,r_U  \\
     \nu\,\alpha_U         &  -\nu\,r_V & \nu\,\alpha_I  & \nu\,r_Q  \\
     \nu\,\alpha_V          &   \nu\,r_U & -\nu\,r_Q &  \nu\,\alpha_I   \end{array} \right)
  \left( \begin{array}{c}
     I/\nu^3    \\
     Q/\nu^3    \\
     U/\nu^3    \\
     V/\nu^3    \end{array} \right)
\end{equation}
où les coefficients $j_X$, $\alpha_X$ et $r_X$ sont les coefficients d'émission, d'absorption et de rotation Faraday respectivement associés aux différents paramètres de Stokes avec les termes ($\times$ ou $/$) en $\nu$ permettant de traiter les quantités invariantes par changement de référentiel (voir Chap.~\ref{chap:GYOTO}). Dans un référentiel de l'émetteur arbitraire $(\mathbf{u},\mathbf{e_1},\mathbf{e_2},\mathbf{K})$, où les vecteurs $\mathbf{e_1}$,$\mathbf{e_2}$ et $\mathbf{K}$ sont orthogonaux à la 4-vitesse $\mathbf{u}$ de l'émetteur, on a
\begin{equation}
     \frac{\dd}{\dd s^{\mathrm{em}}}\left( \begin{array}{c}
     I^{\mathrm{em}}    \\
     Q^{\mathrm{em}}    \\
     U^{\mathrm{em}}    \\
     V^{\mathrm{em}}    \end{array} \right) 
   =  
   \left( \begin{array}{c}
     j_I^{\mathrm{em}}  \\
     j_Q^{\mathrm{em}}  \\
     j_U^{\mathrm{em}}  \\
     j_V^{\mathrm{em}}  \end{array} \right)
   -
   \left( \begin{array}{cccc}
     \alpha_I^{\mathrm{em}}        &   \alpha_Q^{\mathrm{em}}  &  \alpha_U^{\mathrm{em}}  &  \alpha_V^{\mathrm{em}}   \\
      \alpha_Q^{\mathrm{em}}       &   \alpha_I^{\mathrm{em}}  &r_V^{\mathrm{em}} & -r_U^{\mathrm{em}}  \\
      \alpha_U^{\mathrm{em}}       &  -r_V^{\mathrm{em}} & \alpha_I^{\mathrm{em}} & r_Q^{\mathrm{em}}  \\
      \alpha_V^{\mathrm{em}}       &  r_U^{\mathrm{em}} & -r_Q^{\mathrm{em}} &  \alpha_I^{\mathrm{em}}   \end{array} \right)
  \left( \begin{array}{c}
     I^{\mathrm{em}}    \\
     Q^{\mathrm{em}}    \\
     U^{\mathrm{em}}    \\
     V^{\mathrm{em}}    \end{array} \right).
   \label{eq:polarizedradtransf}
\end{equation}
Les vecteurs $\mathbf{e_1}$ et $\mathbf{e_2}$ sont pour le moment arbitraires et forment avec $\mathbf{u}$ et $\mathbf{K}$ une base orthonormée. La valeur des coefficients d'émission, d'absorption et de rotation Faraday dépendent de la base dans laquelle ils sont exprimés (voir section~\ref{sec:coef_synchrotron_polar}). On peut voir l'Eq.~\eqref{eq:polarizedradtransf} comme la version matricielle de l'Eq.~\eqref{eq:transfert_rad} avec un vecteur contenant l'émission dans les quatre paramètres de Stokes et une matrice 4x4 correspondant à l'absorption. Cette dernière contient à la fois les termes d'absorption et de rotation Faraday qui agissent comme de l'absorption du point de vue d'un unique paramètre de Stokes (polarisé). En effet, les coefficients de rotation, comme leur nom l'indique, traduisent une rotation du vecteur polarisation résultant en une conversion entre les paramètres de Stokes différents. Ainsi, ces derniers échangent de l'intensité spécifique d'un paramètre de Stokes à l'autre (de Q$\rightarrow$U par exemple).

\subsection{Coefficients synchrotron polarisés}\label{sec:coef_synchrotron_polar}
\subsubsection{Base naturelle d'expression des coefficients}
On s'intéresse à l'expression des coefficients polarisés $j_X$, $\alpha_X$ et $r_X$ dans le cas du rayonnement synchrotron. Le vecteur polarisation, correspondant, pour rappel, à l'orientation du vecteur champ électrique du photon, est orthogonal au champ magnétique ambiant et au vecteur d'onde. Il est alors judicieux de définir la projection orthogonale à $\mathbf{K}$ du champ magnétique ambiant dans le référentiel de l'émetteur\footnote{Il faut donc au préalable projeter le champ magnétique $\mathbf{B}$ dans le référentiel du trou noir dans le référentiel de l'émetteur $\mathbf{b} = \perp_\mathbf{u} \, B$. On note que, dans certains cas où la définition du champ magnétique se fait dans le référentiel de l'émetteur, cette étape n'est pas nécessaire.} $\mathbf{b}$ qui s'écrit
\begin{equation}
    \label{eq:b_perp}
    \boldsymbol{b_\perp} = \frac{\mathbf{b} - \left(\mathbf{b} \cdot \mathbf{K}\right) \mathbf{K}}{\vert \vert \mathbf{b} - \left(\mathbf{b} \cdot \mathbf{K}\right) \mathbf{K} \vert \vert}.
\end{equation}
Le vecteur $\boldsymbol{b_\perp}$ est donc un vecteur de genre espace normé (pratique pour le calcul d'angle) orthogonal à $\mathbf{K}$. On note le signe $-$ au numérateur et dénominateur qui traduit le fait que $\mathbf{K}$ est un vecteur de genre espace contrairement aux projections orthogonales à $\mathbf{u}$ où le signe est $+$ du fait de la nature de genre temps du vecteur $\mathbf{u}$.

\begin{figure}
    \centering
    \resizebox{0.4\hsize}{!}{\includegraphics{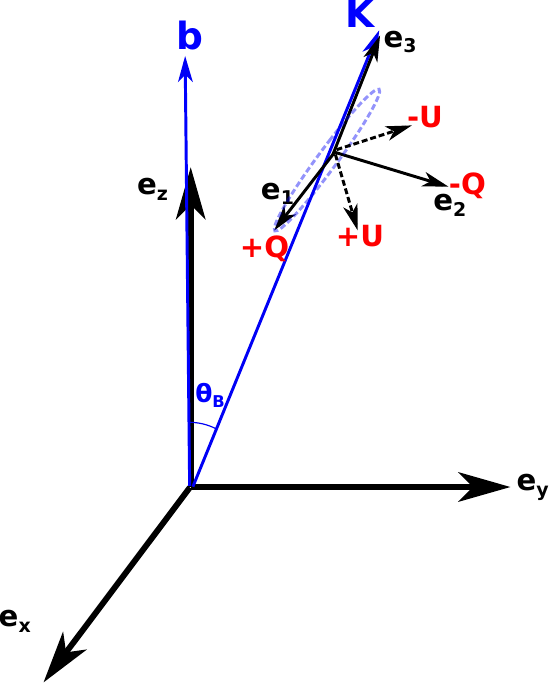}}
    \caption{Géométrie de la polarisation du rayonnement synchrotron avec les paramètres de Stokes associés. La convention utilisée pour l'orientation illustrée ici résulte en un coefficient d'émission positif pour Stokes $Q$. Certains auteurs, comme \cite{Pandya2016, Pandya2021}, utilisent une convention différente ($\mathbf{K} \in (z-x)$), résultant en un coefficient d'émission négatif pour Stokes Q. Crédit : Frédéric Vincent.}
    \label{fig:polar_frame}
\end{figure}

Du fait de cette géométrie décrite précédemment, et avec un choix judicieux de base pour l'expression des coefficients synchrotron, ces derniers peuvent être fortement simplifiés. En effet, en définissant le vecteur de base $\mathbf{e_1}$ à partir de $\boldsymbol{b_\perp}$ et $\mathbf{K}$ tels que
\begin{equation}\label{eq:e1_emmiter}
    \mathbf{e_1} = \boldsymbol{b_\perp} \times \mathbf{K}
\end{equation}
et le vecteur $\mathbf{e_2}$ tel que
\begin{equation}\label{eq:e2_emmiter}
    \mathbf{e_2} = \boldsymbol{b_\perp},
\end{equation}
le vecteur polarisation $\mathbf{F}$ est selon $\mathbf{e_1}$ correspondant ainsi à $j_Q > 0$ (horizontal) et $j_U = 0$ (voir Fig.~\ref{fig:polar_frame}). Attention, on note que l'on peut trouver dans la littérature $j_Q < 0$ dû au choix de la base\footnote{La différence de choix de base n'a aucune influence sur le signe de $I$ et $V$.}. Exprimés dans la base $(\mathbf{F},\boldsymbol{b_\perp},\mathbf{K})$, les coefficients d'émission, d'absorption et de rotation liés au paramètre de Stokes $U$ sont nuls ($j_U = \alpha_U = r_U = 0$), et l'expression des autres coefficients est détaillée plus bas.

On note que les coefficients $I$ et $Q$ sont symétriques par changement de signe du champ magnétique alors que $V$ est antisymétrique. En effet, $Q$ (tout comme $U$) marquant la polarisation linéaire, un changement de signe de $\mathbf{b}$ se traduit par $\mathbf{e_1} \longrightarrow - \mathbf{e_1}$ ce qui ne change pas l'orientation de la polarisation linéaire alors que pour la polarisation circulaire (Stokes $V$), cela se traduit par une inversion du sens de rotation. 

\subsubsection{Formules d'approximations pour des distributions standards} \label{sec:Pandya2021}
Comme dans le cas non polarisé, les intégrales sur la distribution des électrons rendent le calcul de ces coefficients coûteux en temps de calcul. Ainsi, dans le cas de distributions bien définies, comme les distributions thermiques, loi de puissance et kappa, les coefficients d'émission, d'absorption et de rotation Faraday peuvent être ajustés par des formules analytiques. \cite{Pandya2021} ont déterminé des formules analytiques pour l'ensemble de ces coefficients pour chacune des trois distributions citées précédemment avec la même méthodologie que \cite{Pandya2016} basée sur \cite{Leung2011}. On résume dans l'annexe~\ref{ap:coefs_synchrotron} les formules analytiques de chaque coefficient pour les trois distributions mentionnées en prenant notre convention d'orientation de base (se traduisant par $j_Q>0$ au lieu de $j_Q<0$) et nos notations.

\subsection{Lien entre les bases de polarisation} \label{sec:implementation_gyoto}
\begin{figure}
    \centering
    \resizebox{0.4\hsize}{!}{\includegraphics{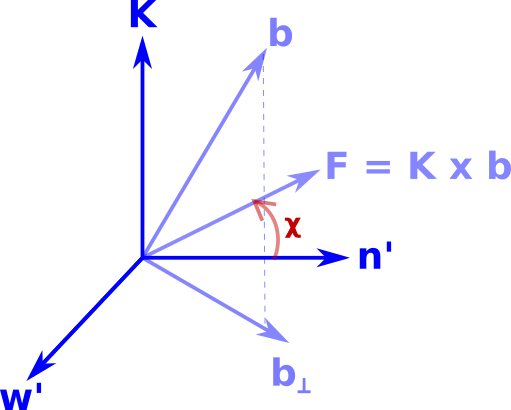}}
    \caption{Représentation des différentes bases de polarisation dans le référentiel de l'émetteur. La base $(\mathbf{K},\mathbf{w^\prime},\mathbf{n^\prime})$ est celle de l'observateur transporté parallèlement et $(\mathbf{F},\boldsymbol{b_\perp},\mathbf{K})$ est la base naturelle synchrotron. Le passage d'une base à l'autre se fait grâce à la matrice de rotation $\mathbf{R}\left(\chi\right)$ de l'Eq.~\eqref{eq:rot_chi}. Crédit : Frédéric Vincent.}
    \label{fig:chi_angle}
\end{figure}

Comme on l'a vu dans la section~\ref{sec:coef_synchrotron_polar}, la base naturelle dans laquelle les coefficients synchrotron s'expriment de manière simple ($j_U=\alpha_U=r_U=0$) est la base $(\mathbf{F},\boldsymbol{b_\perp},\mathbf{K})$, que l'on va désormais appeler $\mathcal{R}$. La base de polarisation de l'observateur transportée parallèlement et projetée dans le référentiel de l'émetteur est la base $(\mathbf{w^\prime},\mathbf{n^\prime},\mathbf{K})$, que l'on nomme $\mathcal{R}^\prime$. Ces deux bases n'ont a priori aucune raison d'être identiques, on constate cependant qu'elles ont le vecteur $\mathbf{K}$ en commun. Ces deux bases sont représentées dans la Fig.~\ref{fig:chi_angle}, où l'on constate aisément que le passage de $\mathcal{R}$ à $\mathcal{R}^\prime$ correspond à une rotation avec un angle $\chi$. On exprime donc la matrice de rotation $\mathbf{R}\left(\chi\right)$ telle que
\begin{equation}
    \label{eq:rot_chi}
    \mathbf{R}\left(\chi\right) = \left( \begin{array}{cccc}
     1        &   0           &  0          &  0   \\
     0        &   \cos 2\chi  & -\sin 2\chi & 0    \\
     0        &   \sin 2\chi  & \cos 2 \chi & 0    \\
     0        &   0           & 0           &  1   \end{array} \right).
\end{equation}

Dans le cas du rayonnement synchrotron, on connait le vecteur champ magnétique $\mathbf{b}$\footnote{Condition nécessaire pour calculer la polarisation synchrotron.} et donc sa projection orthogonale au vecteur $\mathbf{K}$, à savoir $\boldsymbol{b_\perp}$ (Eq.~\eqref{eq:b_perp}). On peut exprimer ce dernier dans la base $(\mathbf{w^\prime},\mathbf{n^\prime})$ de la manière suivante
\begin{equation}
    \boldsymbol{b_\perp} = \cos \chi \, \mathbf{w^\prime} + \sin \chi \, \mathbf{n^\prime}.
\end{equation}
On peut ainsi déterminer l'angle $\chi$ en calculant les produits scalaires des vecteurs $\mathbf{w^\prime}$ et $\mathbf{n^\prime}$ avec $\boldsymbol{b_\perp}$ tel que
\begin{equation}
    \tan \chi = \frac{\boldsymbol{b_\perp} \cdot \mathbf{n^\prime}}{\boldsymbol{b_\perp} \cdot \mathbf{w^\prime}}.
\end{equation}
et l'angle $\theta_B$ entre la direction du photon et le champ magnétique (utilisé dans les formules de la section~\ref{sec:coef_synchrotron_polar})
\begin{equation}
    \cos \theta_B = \mathbf{K} \cdot \mathbf{b},
\end{equation}
où on rappelle que $\mathbf{K}$ et $\mathbf{b}$ sont unitaires.

Enfin, l'équation finale du transfert radiatif dans le référentiel de l'émetteur est
\begin{equation}
    \label{eq:polarradtransf}
 \frac{\dd}{\dd s}\left( \begin{array}{c}
     I    \\
     Q    \\
     U    \\
     V    \end{array} \right) 
   =  \mathbf{R}\left(-\chi\right)
   \left( \begin{array}{c}
     j_I    \\
     j_Q    \\
     j_U    \\
     j_V    \end{array} \right)
   - \mathbf{R}\left(-\chi\right)
   \left( \begin{array}{cccc}
     \alpha_I    &   \alpha_Q  &  \alpha_U  &  \alpha_V   \\
     \alpha_Q    &   \alpha_I  & r_V        & -r_U        \\
     \alpha_U    &  -r_V       & \alpha_I   & r_Q         \\
     \alpha_V    &   r_U       & -r_Q       &  \alpha_I   \end{array} \right)
  \mathbf{R}\left(\chi\right)
  \left( \begin{array}{c}
     I    \\
     Q    \\
     U    \\
     V    \end{array} \right)
\end{equation}
où le vecteur de Stokes $(I,Q,U,V)^T$ est exprimé dans le référentiel $\mathcal{R}^\prime$ alors que le vecteur émission et la matrice d'absorption sont exprimés dans le référentiel $\mathcal{R}$, les matrices de rotation permettant de faire le passage de l'une à l'autre. Pour résoudre cette équation numériquement, on utilise la même méthode que dans le cas polarisé, mais avec des vecteurs (et matrices) au lieu de scalaires. Écrite sous format vectoriel, l'Eq.~\eqref{eq:polarradtransf} devient
\begin{equation}
    \label{eq:trans_polar_vect}
    \frac{\dd \mathbf{I}}{\dd s} = - \mathbf{K \mathbf{I}} + \mathbf{J}\footnotemark
\end{equation}
\footnotetext{Attention, les caractères en gras marquent des vecteurs/matrices carrées de longueur 4 pour les quatre paramètres de Stokes et non les quatre dimensions de l'espace-temps comme précédemment.}
\stepcounter{footnote}
où $\mathbf{I} = (I,Q,U,V)^T$ est le vecteur de paramètres de Stokes et
\begin{equation}
\mathbf{K} = 
\mathbf{R}\left(-\chi\right) \left( \begin{array}{cccc}
     \alpha_I        &   \alpha_Q  &  \alpha_U  &  \alpha_V   \\
      \alpha_Q          &   \alpha_I  &r_V & -r_U  \\
     \alpha_U          &  -r_V & \alpha_I & r_Q  \\
    \alpha_V          &  r_U & -r_Q &  \alpha_I   \end{array} \right) \mathbf{R}\left(\chi\right);
    \qquad
    \mathbf{J} = 
       \mathbf{R}\left(-\chi\right)
   \left(
   \begin{array}{c}
     j_I    \\
     j_Q    \\
    j_U       \\
   j_V       \end{array} \right).
\end{equation}
On peut réécrire $\mathbf{K}$ comme
\begin{equation}
\mathbf{K} = \left( \begin{array}{cccc}
     \alpha_I        &   \alpha_Q^\prime  &  \alpha_U^\prime  &  \alpha_V   \\
      \alpha_Q^\prime          &   \alpha_I  &r_V & -r_U^\prime  \\
     \alpha_U^\prime          &  -r_V & \alpha_I & r_Q^\prime  \\
    \alpha_V          &  r_U^\prime & -r_Q^\prime &  \alpha_I   \end{array} \right)
\end{equation}
après l'application des matrices de rotation avec
\begin{equation}
    \begin{aligned}
        \alpha_Q^\prime &= \alpha_Q \cos 2 \chi - \alpha_U \sin 2 \chi, \\
        \alpha_U^\prime &= \alpha_U \cos 2 \chi + \alpha_Q \sin 2 \chi, \\
        r_Q^\prime &= r_Q \cos 2 \chi - r_U \sin 2 \chi, \\
        r_U^\prime &= r_U \cos 2 \chi + r_Q \sin 2 \chi. \\
    \end{aligned}
\end{equation}

La solution générale de l'Eq.~\eqref{eq:trans_polar_vect} est
\begin{equation}
    \label{eq:sol_trans_polar}
    \mathbf{I}(s) = \int_{s_0}^{s} \exp \left(-\mathbf{K}(s-s^\prime)\right) \mathbf{J}(s^\prime) \dd s^\prime
\end{equation}
\begin{equation}
    \Leftrightarrow \mathbf{I} = \sum \exp \left(-\mathbf{K}(s-s^\prime)\right) \mathbf{J}(s^\prime) \dd s^\prime \quad \mathrm{ (numériquement)}
\end{equation}
On définit donc la matrice $\mathbf{O}$, correspondant à la matrice de transmission de la cellule, telle que
\begin{equation}
    \mathbf{O}(s,s^\prime) = \mathbf{O}(\delta s) = \exp  \left( -\mathbf{K}(s-s^\prime) \right).
\end{equation}
L'incrément de paramètres de Stokes de la cellule $i$ (voir Fig.~\ref{fig:schema_intreg_transfert_rad}) est donc
\begin{equation}
    \delta \mathbf{I} (s) = \mathbf{O}(\delta s) \mathbf{J} (s) \delta s,
\end{equation}
de manière très similaire au cas non polarisé. Cependant, l'exponentielle d'une matrice est non triviale. \cite{Landi1985} ont néanmoins déterminé une expression pour cette matrice qui s'écrit
\begin{equation}
    \begin{aligned}
        \mathbf{O}(\delta s) &= \mathrm{exp}\left(-\alpha_I \delta s\right) \left\{ \left[ \mathrm{cosh} \left(\Lambda_1 \delta s\right) + \cos \left( \Lambda_2 \delta s\right)\right] \mathbf{M_1}/2 - \sin\left( \Lambda_2 \delta s\right)\mathbf{M_2} \right.\\
        &\left.- \mathrm{sinh}\left( \Lambda_1 \delta s\right)\mathbf{M_3} +  \left[ \mathrm{cosh} \left(\Lambda_1 \delta s\right) - \cos \left( \Lambda_2 \delta s\right)\right] \mathbf{M_4}/2 \right\} \\ 
    \end{aligned}
\end{equation}
avec
\begin{equation}
    \begin{aligned}
        \mathbf{M_1} &= \mathbf{1}, \\
\mathbf{M_2} &= \frac{1}{\Theta}\left( \begin{array}{cccc}
     0       &   \Lambda_2\alpha_Q^\prime - \sigma \Lambda_1 r_Q^\prime  & \Lambda_2\alpha_U^\prime - \sigma \Lambda_1 r_U^\prime     &  \Lambda_2\alpha_V - \sigma \Lambda_1 r_V     \\
      \Lambda_2\alpha_Q^\prime - \sigma \Lambda_1 r_Q^\prime       &   0  & \sigma \Lambda_1 \alpha_V + \Lambda_2 r_V & -\sigma \Lambda_1 \alpha_U^\prime - \Lambda_2 r_U^\prime  \\
     \Lambda_2\alpha_U^\prime - \sigma \Lambda_1 r_U^\prime          &  -\sigma \Lambda_1 \alpha_V - \Lambda_2 r_V & 0 & \sigma \Lambda_1 \alpha_Q^\prime + \Lambda_2 r_Q^\prime \\
    \Lambda_2\alpha_V - \sigma \Lambda_1 r_V          &  \sigma \Lambda_1 \alpha_U^\prime + \Lambda_2 r_U^\prime &  -\sigma \Lambda_1 \alpha_Q^\prime - \Lambda_2 r_Q^\prime &  0  \end{array} \right),\\
\mathbf{M_3} &= \frac{1}{\Theta}\left( \begin{array}{cccc}
     0       &   \Lambda_1\alpha_Q^\prime + \sigma \Lambda_2 r_Q^\prime  & \Lambda_1\alpha_U^\prime + \sigma \Lambda_2 r_U^\prime   &  \Lambda_1\alpha_V + \sigma \Lambda_2 r_V     \\
      \Lambda_1\alpha_Q^\prime + \sigma \Lambda_2 r_Q^\prime      &   0  & - \sigma \Lambda_2 \alpha_V + \Lambda_1 r_V & \sigma \Lambda_2 \alpha_U^\prime - \Lambda_1 r_U^\prime  \\
     \Lambda_1\alpha_U^\prime + \sigma \Lambda_2 r_U^\prime        &  \sigma \Lambda_2 \alpha_V - \Lambda_1 r_V & 0 & -\sigma \Lambda_2 \alpha_Q^\prime + \Lambda_1 r_Q^\prime \\
    \Lambda_1\alpha_V + \sigma \Lambda_2 r_V        &  -\sigma \Lambda_2 \alpha_U^\prime + \Lambda_1 r_U^\prime &  \sigma \Lambda_2 \alpha_Q^\prime - \Lambda_1 r_Q^\prime &  0  \end{array} \right),\\
\mathbf{M_4} &= \frac{2}{\Theta} \\
&\times \left( \begin{array}{cccc}
     (\alpha^2 + r^2)/2       &   \alpha_V r_U^\prime - \alpha_U^\prime r_V  & \alpha_Q^\prime r_V - \alpha_V r_Q^\prime  & \alpha_U^\prime r_Q^\prime - \alpha_Q^\prime r_U^\prime   \\
      \alpha_U^\prime r_V - \alpha_V r_U^\prime      &   \alpha_Q^{\prime 2} + r_Q^{\prime 2} -  (\alpha^2 + r^2)/2    & \alpha_Q^\prime \alpha_U^\prime + r_Q^\prime r_U^\prime & \alpha_V \alpha_Q^\prime + r_V r_Q^\prime   \\
     \alpha_V r_Q^\prime - \alpha_Q^\prime r_V      &  \alpha_Q^\prime \alpha_U^\prime + r_Q^\prime r_U^\prime & \alpha_U^{\prime 2} + r_U^{\prime 2} -  (\alpha^2 + r^2)/2  & \alpha_U^\prime \alpha_V + r_U^\prime r_V \\
    \alpha_Q^\prime r_U^\prime - \alpha_U^\prime r_Q^\prime        &  \alpha_V \alpha_Q^\prime + r_V r_Q^\prime &  \alpha_U^\prime\alpha_V + r_U^\prime r_V &  \alpha_V^2 + r_V^2 -  (\alpha^2 + r^2)/2  \end{array} \right)\\
    \end{aligned}
\end{equation}
où
\begin{equation}
    \begin{aligned}
        \alpha^2 &= \alpha_Q^{\prime 2} + \alpha_U^{\prime 2} + \alpha_V^2, \\
        r^2 &= r_Q^{\prime 2} + r_U^{\prime 2} + r_V^2, \\
        \Lambda_1 &= \sqrt{\sqrt{\frac{1}{4}(\alpha^2 - r^2)^2 + (\alpha_Q^\prime r_Q^\prime + \alpha_U^\prime r_U^\prime + \alpha_V r_V)^2} + \frac{1}{2}(\alpha^2 - r^2)^2}, \\
        \Lambda_2 &= \sqrt{\sqrt{\frac{1}{4}(\alpha^2 - r^2)^2 + (\alpha_Q^\prime r_Q^\prime + \alpha_U^\prime r_U^\prime + \alpha_V r_V)^2} - \frac{1}{2}(\alpha^2 - r^2)^2}, \\
        \Theta &= \Lambda_1^2 + \Lambda_2^2, \\
\sigma &= \mathrm{sign}\left(\alpha_Q^\prime r_Q^\prime + \alpha_U^\prime r_U^\prime + \alpha_V r_V \right). \\
    \end{aligned}
\end{equation}

Enfin, tout comme dans le cas non polarisé, la matrice de transmission $\mathbf{T}$ est mise à jour à chaque pas d'intégration
\begin{equation}
    \mathbf{T} = \prod_{\mathrm{integration\: step}} \mathbf{O}(\delta s).
\end{equation}
Le terme $\mathbf{T} (0,0)$, correspondant à la transmission non polarisée, est utilisé comme condition limite de l'intégration du transfert radiatif et par extension de la géodésique (voir Chap.~\ref{chap:GYOTO}).

\section{Comparaison \textsc{Gyoto} - \textsc{Ipole}}
Afin de valider l'implémentation de la polarisation dans \textsc{Gyoto}, il est nécessaire de tester les résultats obtenus avec ce dernier dans des cas très simples et de comparer à d'autres codes similaires. La comparaison est effectuée avec le code \textsc{ipole}\footnote{\href{https://github.com/moscibrodzka/ipole}{https://github.com/moscibrodzka/ipole}} utilisé notamment par la collaboration EHT pour générer les images synthétiques à partir des simulations GRMHD \cite{EHT2022iii}. D'autres comparaisons sont prévues dans le futur avec le code \textsc{GRTRANS}~\cite{Dexter2016}.
\subsection{Tests simples}
Pour commencer, on effectue des tests simples, sans transfert radiatif, pour tester le transport parallèle et la projection dans le référentiel de l'émetteur, via la valeur de l'EVPA, pour lesquels le résultat attendu est évident. Pour cela, on considère une sphère de plasma optiquement mince ayant une orbite circulaire dans le plan équatorial d'un trou noir de Schwarzschild à $10$ $r_g$. On définit deux types de configurations magnétiques $\mathbf{b} = (b^t,b^r,b^\theta,b^\varphi)$ possibles :
\begin{itemize}
    \item[$\bullet$] Verticale avec 
\end{itemize}
\begin{equation}
    \label{eq:B_vertical}
    b^\alpha = \left\{
    \begin{array}{l}
        b^t = 0, \\
        b^r = \cos(\theta)/\sqrt{g_{rr}}, \\
        b^\theta = -\sin(\theta)/\sqrt{g_{\theta \theta}}, \\
        b^\varphi = 0.
    \end{array} \right.
\end{equation}

\begin{figure}
    \centering
    \resizebox{0.6\hsize}{!}{\includegraphics{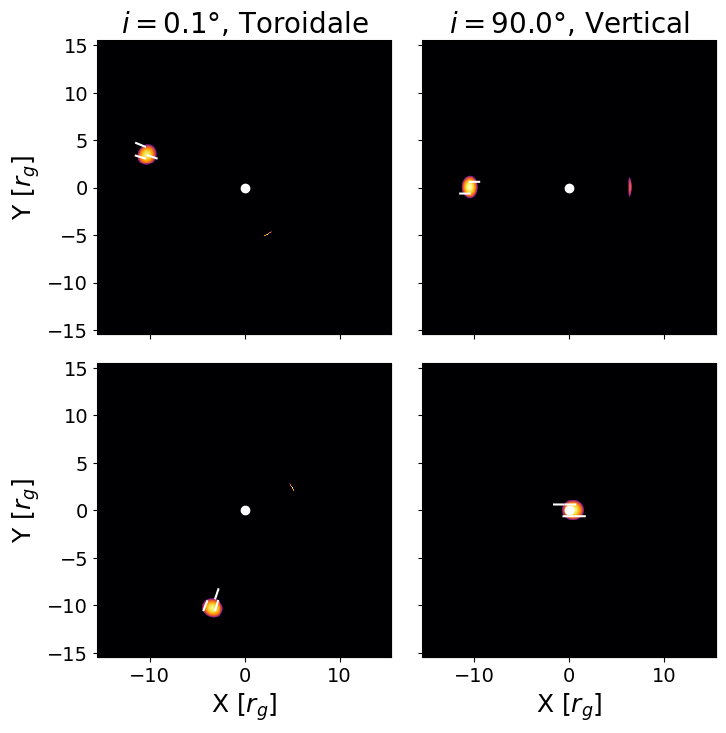}}
    \caption{Images et orientation du vecteur polarisation (traits blancs) d'une sphère de plasma orbitant un trou noir de Schwarzschild à $r=10$ $r_g$. La colonne de gauche montre deux images et la polarisation avec une inclinaison de $i=0,1\degree$ et une configuration magnétique toroïdale à des phases différentes. Dans la colonne de droite, l'inclinaison est de $i=90 \degree$ et le champ magnétique vertical. Le point blanc dans la colonne de gauche marque le centre des coordonnées afin de mieux visualiser l'aspect radial de la polarisation.}
    \label{fig:Test_polar_simple}
\end{figure}

\begin{itemize}
    \item[$\bullet$] Toroïdale avec 
\end{itemize}
\begin{equation}
    \label{eq:B_toroidal}
    b^\alpha = \left\{
    \begin{array}{l}
        b^t = \sqrt{\frac{g_{\varphi \varphi}}{g_{tt}} \frac{\Omega^2}{g_{tt} +g_{\varphi \varphi}\Omega^2}}, \\
        b^r = 0, \\
        b^\theta = 0, \\
        b^\varphi = \sqrt{\frac{g_{tt}}{g_{\varphi \varphi}} \frac{1}{g_{tt} + g_{\varphi \varphi}\Omega^2}}.
    \end{array} \right.
\end{equation}
avec $\Omega^2 = u^\varphi/u^t$. On note que les définitions de ces configurations de champ magnétique (en particulier celle toroïdale) sont valables pour un trou noir sans spin. On ne s'intéresse pas aux images secondaires et d'ordre plus élevé, car ayant fait au moins une demi orbite autour du trou noir, l'angle d'incidence du photon par rapport au champ magnétique est beaucoup moins triviale à estimer de façon intuitive, contrairement à celui de l'image primaire. À faible inclinaison ($i=0.1\degree$)\footnote{On rappelle que l'on ne peut pas considérer l'observateur le long de l'axe de spin/vertical du trou noir, c'est-à-dire $i=0 \degree$ pour la définition des autres angles.} et avec une configuration Toroïdale, la polarisation observée doit être approxivement radiale puisque $\mathbf{F} \propto \mathbf{K} \times \mathbf{b}$ (on a choisi cette configuration pour que $\mathbf{K}$ et $\mathbf{b}$ soient $\approx$ orthogonaux entre eux). De même, observé avec une inclinaison de $i=90\degree$, et avec une configuration magnétique verticale, la polarisation observée doit être horizontale. On utilise donc l'objet \textit{Star} de \textsc{Gyoto} en imposant de manière arbitraire $j_I=j_Q$ et $j_U=j_V=0$ (donc un rayonnement entièrement polarisé linéairement). Les coefficients d'absorption et de rotation sont considérés comme nuls pour s'affranchir des effets d'absorption et de dépolarisation via $r_V$ notamment. Le coefficient d'émission $j_I$ est calculé à partir de la fonction de Planck $B_\nu$ (choix arbitraire) pour une température de 6000 K\footnote{Cela n'a aucune incidence sur l'orientation du vecteur polarisation.}. On constate que l'on obtient bien la polarisation attendue, en regardant les images de la Fig.~\ref{fig:Test_polar_simple}. Dans celle-ci, la première colonne montre deux images de l'orientation du vecteur polarisation dans le cas ($i=0.1\degree$, Toroïdale) pour deux positions de la \textit{Star}. La seconde colonne, quant à elle, montre le cas ($i=90\degree$, Verticale), où la \textit{Star} est située à gauche du trou noir. Elle se déplace dans notre direction dans l'image du haut et entre l'observateur et le trou noir dans l'image du bas.

\subsection{Disque géométriquement épais}
Afin d'aller plus loin dans les tests et de vérifier la validité du code, on va maintenant chercher à comparer les résultats de \textsc{Gyoto} avec un autre code de tracé de rayon, \textsc{ipole}. De plus, on va utiliser un objet astrophysique plus complexe, tant géométriquement qu'optiquement comparé aux tests précédents. En effet, on va utiliser un disque géométriquement épais émettant du rayonnement synchrotron thermique observé en radio à 1,3 mm. On résume ci-après toutes les propriétés de la source du rayonnement. Celles-ci correspondent aux propriétés codées dans \textsc{ipole} pour assurer une vraie comparaison. On note qu'elles sont significativement différentes des propriétés du \textit{ThickDisk} historique de \textsc{Gyoto}. On note également que tous les tests ont été effectués avec un spin nul. Les paramètres sont résumés dans la table~\ref{tab:params_compare_ipole}.

\begin{table}
    \centering
    \begin{tabular}{ l c r }
        \hline
        \hline
        Paramètre & Symbole & Valeur\\
        \hline
        \textbf{Trou Noir} & &\\
        masse [$M_\odot$] & $M$ & $4,154 \times 10^6$ \\
        distance [kpc] & $d$ & $8,178$\\
        spin & $a$ & $0$\\
        \hline
        \textbf{Disque épais} & & \\
        rayon interne [$r_g$] & $r_{in}$ & $6$ \\
        densité de référence [cm$^{-3}$] & $N_0$ & $6 \times 10^6$\\
        température de référence sans dimension & $\Theta_0$ & $200$\\
        champ magnétique de référence [G] & $B_0$ & $100$\\
        épaisseur verticale & $\epsilon$ & $0,3$\\
        \hline
        \hline
    \end{tabular}
    \caption{Paramètres utilisés pour la comparaison entre \textsc{Gyoto} et \textsc{ipole} à l'aide d'un disque épais.}
    \label{tab:params_compare_ipole}
\end{table}

Tout d'abord, on définit le profil de la 4-vitesse du fluide $\mathbf{u} = (u^t,u^r,u^\theta, u^\varphi)$ autour du trou noir, suivant la prescription choisie dans \textsc{ipole}, telle que si $r > r_{ISCO}$
\begin{equation}
    u^\alpha = \left\{
    \begin{aligned}
        u^t &= \sqrt{\frac{-1}{g_{tt} + \Omega g_{\varphi \varphi}}} \\
        u^r &= 0 \\
        u^\theta &= 0 \\
        u^\varphi &= \Omega \sqrt{\frac{-1}{g_{tt} + \Omega g_{\varphi \varphi}}}
    \end{aligned}  \right.   
\end{equation}
et si $r < r_{ISCO}$
\begin{equation}
    u^\alpha = \left\{
    \begin{aligned}
        u^t &= -\sqrt{g^{tt}} \\
        u^r &= 0 \\
        u^\theta &= 0 \\
        u^\varphi &= 0
    \end{aligned}  \right. 
\end{equation}
avec $\Omega=r^{-1.5}$ la vitesse Képlérienne (puisque $a=0$).

Le disque, qui est considéré comme axisymétrique, est décrit par la densité $N_e$, la température sans dimension $\Theta_e$ et le champ magnétique de norme $B$ qui dépendent du rayon $r$ et de l'angle polaire $\theta$ tels que
\begin{equation}
    \begin{aligned}
        N_e &= N_0 \left( \frac{r}{r_g} \right)^{-3/2} \exp \left( -\frac{\cos^2 \theta}{2 \epsilon^2} \right), \\
        \Theta_e &= \Theta_0 \left( \frac{r}{r_g} \right)^{-0.84}, \\
        B &= B_0 \left( \frac{r}{r_g} \right)^{-1},
    \end{aligned}
\end{equation}
avec $N_0$ la densité, $\Theta_0$ la température sans dimension, $B_0$ la norme du champ magnétique de référence et $\epsilon = 0.3$ l'épaisseur verticale du disque. De plus, on considère que le disque s'étend de $r_\mathrm{in}=6\, r_g$ à l'infini en prenant une densité nulle pour $r<r_\mathrm{in}$\footnote{La condition de base dans \textsc{ipole} est différente, mais pour la comparaison, on applique cette condition pour les deux codes.}.

À partir de ces quantités et en supposant une distribution thermique des électrons, on peut en déduire la valeur de tous les coefficients synchrotron. Cependant, \textsc{ipole} n'utilise pas les formules d'approximation de \cite{Pandya2021} qui sont codées dans \textsc{Gyoto}, comme indiqué dans la section~\ref{sec:coef_synchrotron_polar}. Les formules pour les coefficients d'émission utilisées dans \textsc{ipole}, implémentées dans \textsc{Gyoto} pour les tests de transfert radiatif polarisé présentés ici, issues de~\cite{Dexter2016}, et résumées ci-après
\begin{equation}
    \label{eq:coef_synchro_ipole_a}
    \begin{aligned}
        j_I &= \frac{A}{\Theta_e^2} \times \left(2,5651(1+1,92 X^{-1/3} + 0,9977 X^{-2/3})\times \exp (-1,8899 X^{1/3})\right), \\
        j_Q &= \frac{A}{\Theta_e^2} \times \left(2,5651(1+0,93193 X^{-1/3} + 0,499873 X^{-2/3})\times \exp (-1,8899 X^{1/3})\right), \\
        j_U &= 0, \\
        j_V &= \frac{2A}{3 \Theta_e^3 \tan \theta_B} \\
        & \times \left((1,81348/X+3,42319 X^{-2/3}+0,0292545 X^{-1/2}+2,03773 X^{-1/3}) \exp(-1,8899 X^{1/3})\right), \\
    \end{aligned}
\end{equation}
avec $X=\nu/ \nu_s$, $\nu_s = 3 \Theta_e^2 \nu_B \sin \theta /2+1$, $A=N_e e^2 \nu/(2 c \sqrt{3})$, et $\theta_B$ l'angle entre le champ magnétique $\mathbf{b}$ et le vecteur direction du photon $\mathbf{K}$. Les coefficients d'absorption sont calculés à partir des coefficients d'émission et la loi de Kirchhoff. Les coefficients de rotation Faraday s'écrivent
\begin{equation}
    r_Q = \frac{N_e e^2 \nu_B^2 \sin^2 \theta_B}{m_e c \nu^3} \, f_m \times \left( \frac{K_1(\Theta_e^{-1})}{K_2(\Theta_e^{-1})}+6\Theta_e \right)
\end{equation}
avec
\begin{equation}
\begin{aligned}
    f_m &= f_0 + \frac{1}{2} \left(0,011 \exp (-1,69 X^{1/2})-0,003135 X^{4/3} \right) \left(1+ \tanh (10 \ln(0,6648 X^{-1/2}))\right), \\
    f_0 &= 2,011 \exp (-19,78 X^{-0,5175}) - \cos (39,89 X^{-1/2}) \exp (-70,16 X^{-0,6})-0,011 \exp (-1,69 X^{-1/2})
\end{aligned}
\end{equation}
et
\begin{equation}
    \begin{aligned}
        r_V = \frac{2 \pi \nu}{c} \frac{W \Omega_0}{(2 \pi \nu)^3} \cos \theta_B
        &\times \left\{
        \begin{array}{ll}
            \frac{\left( - \ln (\frac{1}{2 \Theta_e}) - 0,5772\right) -J_e}{2 \Theta_e^2}, & \mbox{ pour $\Theta_e > 3$}\\
            \frac{K_0(\Theta_e^{-1})-J_e}{K_2(\Theta_e^{-1})}, & \mbox{ pour $0,2 < \Theta_e \leq 3$}\\
            1, & \mbox{ pour $\Theta_e < 0,2$}
        \end{array}
        \right.
    \end{aligned}
\end{equation}
avec 
\begin{equation}
    \label{eq:coef_synchro_ipole_z}
    \begin{aligned}
        W &= \frac{4 \pi N_e e^2}{m_e}, \\
        \Omega_0 &= 2 \pi \nu_B, \\
        J_e &= 0,43793091 \ln (1+0,00185777 X_e^{1,50316886}), \\
        X_e &= \Theta_e \left( \sqrt{2} \sin \theta_B \frac{10^3 \cdot \Omega_0}{2 \pi \nu} \right)^{1/2}.
    \end{aligned}
\end{equation}

\begin{figure}
    \centering
    \resizebox{0.8\hsize}{!}{\includegraphics{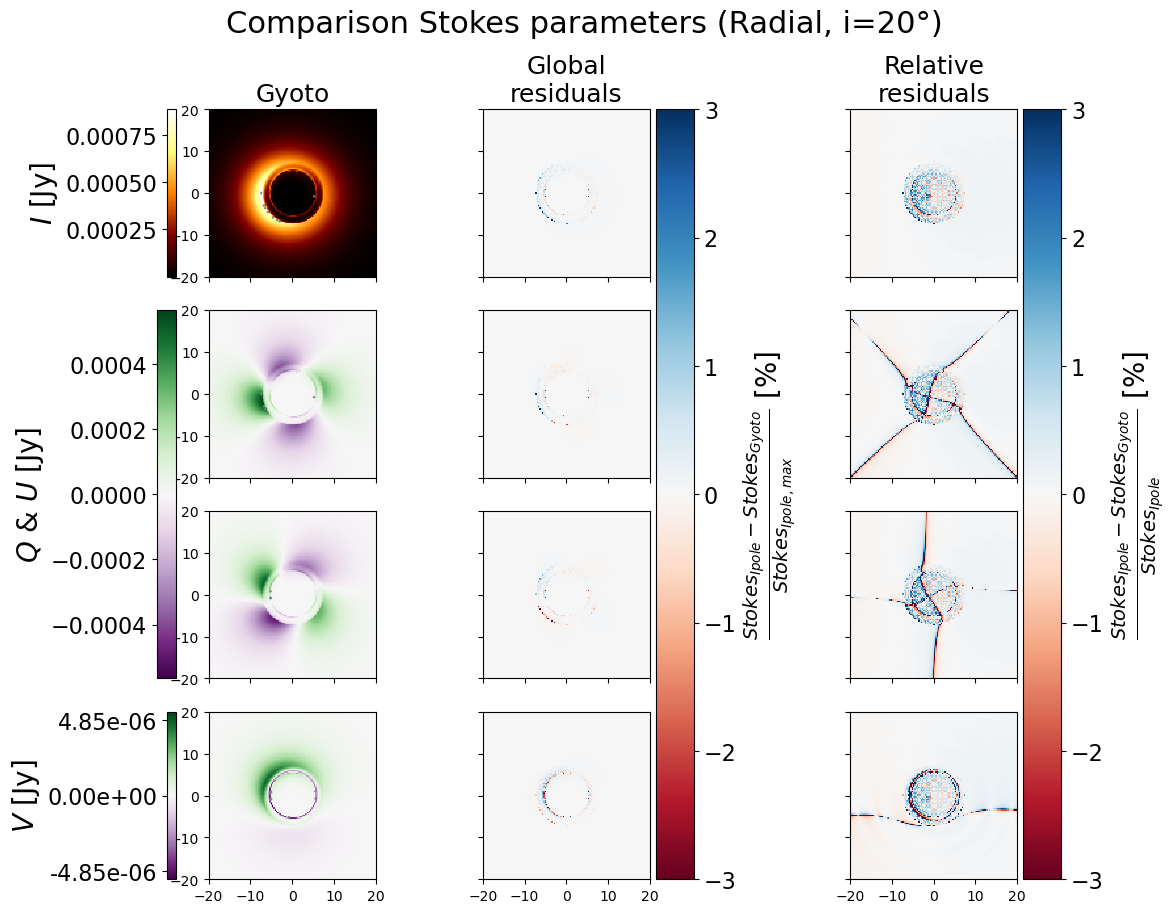}}
    \caption{Comparaison des images polarisées produites avec \textsc{ipole} et \textsc{Gyoto}. La première colonne montre l'image d'un disque épais, tel que définit plus haut, observé avec une inclinaison de $20 \degree$ et un champ magnétique radiale, dans les quatre paramètres de Stokes ($I$, $Q$, $U$ et $V$) en Jansky. L'échelle de couleur des paramètres $Q$ et $U$ est commune (puisqu'ils définissent la polarisation linéaire). La colonne centrale montre l'erreur globale pour chaque pixel entre \textsc{ipole} et \textsc{Gyoto}, normalisé par la valeur maximale (d'\textsc{ipole}). Enfin, la dernière colonne montre l'erreur relative de chaque pixel pour chaque paramètre de Stokes. Le niveau de $3\%$ d'erreur globale est marqué par les contours gris dans la première colonne.}
    \label{fig:compareIpole_Radial}
\end{figure}

\begin{figure}
    \centering
    \resizebox{0.8\hsize}{!}{\includegraphics{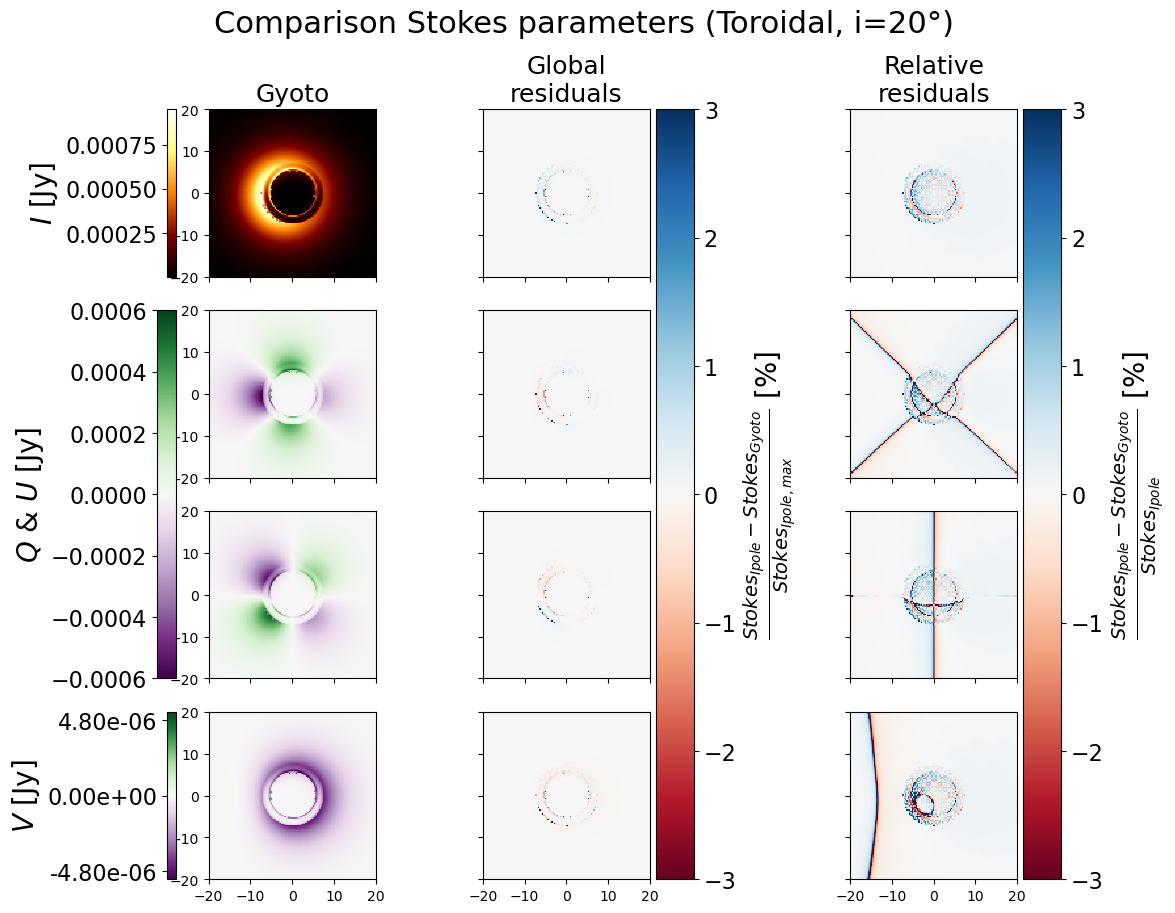}}
    \caption{Même chose qu'à la Fig.~\ref{fig:compareIpole_Radial}, avec une configuration magnétique toroïdale.}
    \label{fig:compareIpole_Toroidal}

    \vspace{1cm}
    
    \resizebox{0.8\hsize}{!}{\includegraphics{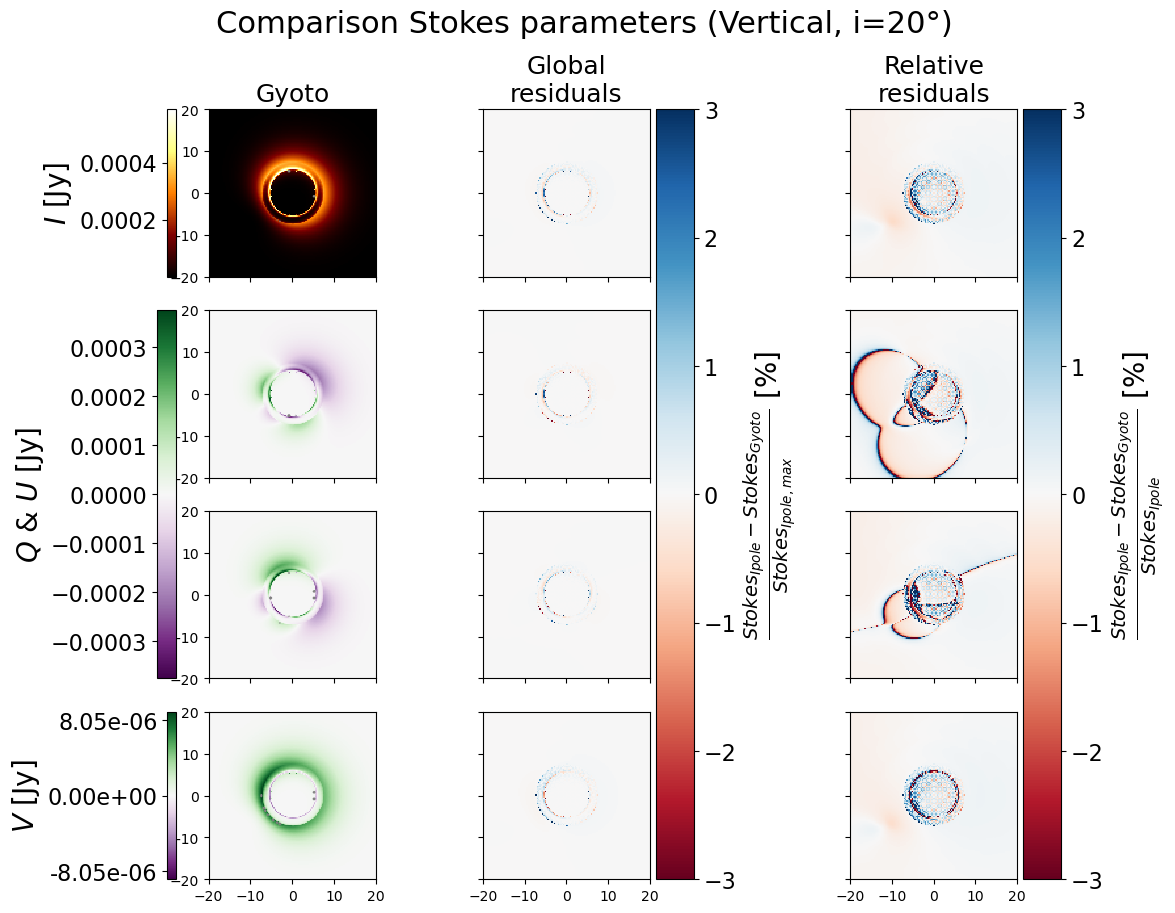}}
    \caption{Même chose qu'à la Fig.~\ref{fig:compareIpole_Radial}, avec une configuration magnétique verticale.}
    \label{fig:compareIpole_Vertical}
\end{figure}

Pour ces tests, on considère trois configurations pour le champ magnétique : toroïdal, radial et vertical, observés avec une faible inclinaison de $20 \degree$. Comme pour les coefficients synchrotron, on adopte la définition de \textsc{ipole} pour le vecteur champ magnétique disponible \href{https://github.com/moscibrodzka/ipole/blob/ipole-v2.0/model_analytic.c}{ici}. On peut ainsi comparer les résultats de \textsc{Gyoto} et \textsc{ipole} en prenant la même configuration. Les Figs.~\ref{fig:compareIpole_Radial}~à~\ref{fig:compareIpole_Vertical} montrent les images pour les quatre paramètres de Stokes dans la première colonne (la première ligne étant l'intensité totale $I$, la seconde Stokes $Q$, la troisième Stokes $U$ et la dernière la polarisation circulaire Stokes $V$). La seconde colonne montre la carte d'erreur globale normalisée définie comme
\begin{equation}
    Err_{X,\mathrm{global}} [i,j] = \left\vert \frac{\mathrm{Stokes}_X^\mathrm{ipole}[i,j] - \mathrm{Stokes}_X^\mathrm{Gyoto}[i,j]}{\mathrm{max}(\mathrm{Stokes}_X^\mathrm{ipole})} \right\vert
\end{equation}
pour le pixel $[i,j]$, et la dernière colonne montre la carte d'erreur relative. 

On constate que dans l'ensemble des configurations (toroïdale, radiale, et verticale) on obtient un très bon accord entre les deux codes avec une erreur normalisée médiane inférieure à $0,1 \%$ (la statistique des erreurs est résumé dans la Table~\ref{tab:erreurs}). La différence des flux intégrés pour chaque paramètre de Stokes est aussi de l'ordre de $0,1 \%$ sauf dans le cas radial pour Stokes $U$ où la différence atteint $\sim 1\%$ en raison de quelques pixels avec une forte erreur relative mais dont la contribution est négligeable. On constate dans les Figs.~\ref{fig:compareIpole_Radial}~à~\ref{fig:compareIpole_Vertical} l'excellent accord entre \textsc{Gyoto} et \textsc{ipole} (colonne du milieu) ainsi que des lignes particulières pour lesquelles l'erreur relative (colonne de droite) se démarquent. Cependant, ces erreurs relatives élevées se produisent majoritairement là où le flux est proche de zéro. Ainsi, dans les cartes d'erreur normalisées, ces erreurs deviennent négligeables. Lorsque l'on calcule la polarisation de l'ensemble de l'image, en faisant la somme des pixels, la contribution des zones précédentes est donc d'autant plus négligeable.

Une seconde zone pour laquelle l'erreur relative est plus importante que la médiane (mais toujours de l'ordre de $\sim 1\%$) est l'anneau de photons. En effet, les photons issus de ce dernier sont passés proche du trou noir, là où le potentiel gravitationnel est le plus fort et avec une trajectoire complexe, sensible au schéma et au pas d'intégration de la géodésique. De plus amples tests sont prévus pour diagnostiquer les calculs de polarisation pour l'anneau de photons. Enfin, la zone correspondant à l'ombre du trou noir est aussi une source d'erreur relative de l'ordre de $\leq 10\%$ pour quelques pixels uniquement, cependant, comme dans le cas des lignes, le flux est proche de zéro, la contribution de cette zone au flux intégré est donc négligeable. Ces tests ont aussi été menés avec une forte inclinaison ($i=80 \degree$) et résumés dans l'annexe~\ref{ap:Test_Polar}.

\begin{table}
    \centering
    \begin{tabular}{lcccc}
        \hline
        \hline
        Paramètre de Stokes & $I$ & $Q$ & $U$ & $V$\\
        \hline
        Radial, $i=20\degree$ & & & & \\
        erreur relative moyenne [$\%$] & $0,061$ & $0,004$ & $0,097$ & $0,092$ \\
        erreur relative \textit{standard deviation} [$\%$] & $0,75$ & $4,87$ & $5,02$ & $8,44$ \\
        erreur normalisée moyenne [$\%$] & $0,012$ & $7,51.10^{-4}$ & $-0,0011$ & $0,0068$ \\
        erreur normalisée \textit{standard deviation} [$\%$] & 0,181 & 0,122 & 0,160 & 0,381 \\
        \hline
        Toroïdal, $i=20\degree$ & & & & \\
        erreur relative moyenne [$\%$] & 0,057 & $6,82.10^{-3}$ & 0,311 & 0,081 \\
        erreur relative \textit{standard deviation} [$\%$] & 0,578 & 4,08 & 34,0 & 5,76 \\
        erreur normalisée moyenne [$\%$] & 0,011 & $1,09.10^{-3}$ & $5,69.10^{-5}$ & $-9,98.10^{-3}$ \\
        erreur normalisée \textit{standard deviation} [$\%$] & 0,234 & 0,187 & 0,177 & 0,233 \\
        \hline
        Vertical, $i=20\degree$ & & & & \\
        erreur relative moyenne [$\%$] & $3,88.10^{-3}$ & -0,774 & $4,29.10^{-3}$ & -0,027 \\
        erreur relative \textit{standard deviation} [$\%$] & 0,729 & 90.4 & 6,46 & 9,27 \\
        erreur normalisée moyenne [$\%$] & $4,08.10^{-3}$ & $-4,98.10^{-3}$ & $2,56.10^{-3}$ & $8,27.10^{-3}$ \\
        erreur normalisée \textit{standard deviation} [$\%$] & 0,283 & 0,301 & 0,172 & 0,295 \\
        \hline
        Radial, $i=80\degree$ & & & & \\
        erreur relative moyenne [$\%$] & 0,048 & 0,081 & 0,055 & 0,215 \\
        erreur relative \textit{standard deviation} [$\%$] & 0,257 & 4,39 & 1,96 & 19,8 \\
        erreur normalisée moyenne [$\%$] & $3,86.10^{-3}$ & $1,97.10^{-3}$ & $-5,14.10^{-4}$ & $-2,47.10^{-3}$ \\
        erreur normalisée \textit{standard deviation} [$\%$] & 0,069 & 0,059 & 0,062 & 0,150 \\
        \hline
        Toroïdal, $i=80\degree$ & & & & \\
        erreur relative moyenne [$\%$] & 0,038 & 0,042 & 0,083 & 0,050 \\
        erreur relative \textit{standard deviation} [$\%$] & 0,188 & 0,651 & 0,052 & 2,040 \\
        erreur normalisée moyenne [$\%$] & $3,66.10^{-3}$ & $3,30.10^{-3}$ & $5,11.10^{-4}$ & $1,54.10^{-3}$ \\
        erreur normalisée \textit{standard deviation} [$\%$] & 0,050 & 0,051 & 0,107 & 0,052 \\
        \hline
        Vertical, $i=80\degree$ & & & & \\
        erreur relative moyenne [$\%$] & 0,043 & 0,052 & -0,117 & 0,124 \\
        erreur relative \textit{standard deviation} [$\%$] & 0,213 & 1,98 & 22,6 & 10,9 \\
        erreur normalisée moyenne [$\%$] & $3,86.10^{-3}$ & $-4,52.10^{-3}$ & $1,74.10^{-3}$ & $1,72.10^{-3}$ \\
        erreur normalisée \textit{standard deviation} [$\%$] & 0,053 & 0,049 & 0,103 & 0,073 \\        
        \hline
    \end{tabular}
    \caption{Statistiques d'erreurs entre les images obtenues avec \textsc{Gyoto} et \textsc{ipole}. L'erreur des paramètres intégrés sur tout le champ de vue est inférieure à $1\%$ pour l'ensemble des configurations.}
    \label{tab:erreurs}
\end{table}

\begin{figure}
    \centering
    \resizebox{\hsize}{!}{\includegraphics{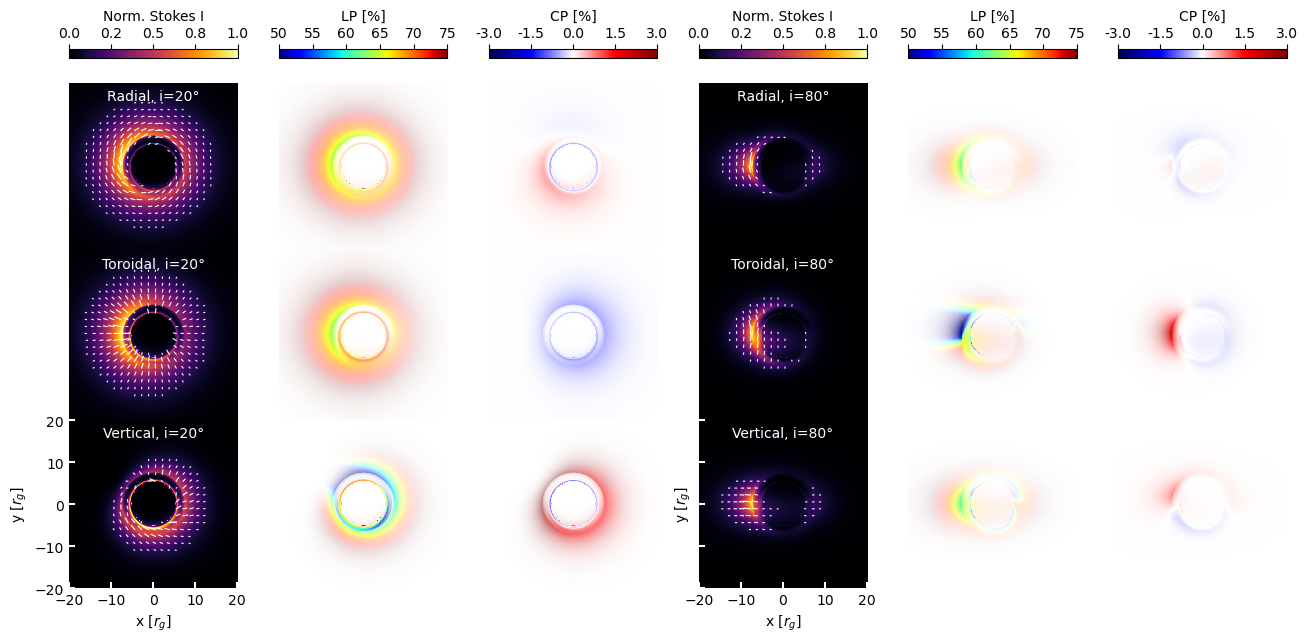}}
    \caption{Images d'un disque épais autour d'un trou noir de Schwarzschild, identique à la définition de \textsc{ipole}, calculé avec \textsc{Gyoto} polarisé. Les colonnes 1 et 4 montrent l'intensité totale normalisée pour trois configurations du champ magnétique : radial (1ère ligne), toroïdal (2ème ligne) et vertical (3ème ligne), à des inclinaisons de $20\degree$ et $80\degree$ respectivement. L'orientation du vecteur polarisation est représentée par les traits blancs. Les colonnes 2 et 5 montrent la fraction de polarisation linéaire $LP = (Q^2 + U^2)^{1/2}/I$ et les colonnes 3 et 6 la fraction de polarisation circulaire algébrique $CP=V/I$. Les échelles et codes couleur correspondent à la Fig.~2 de \cite{Vos2022}.}
    \label{fig:compareIpole_incli}
\end{figure}

Une fois le code validé pour une faible inclinaison, on cherche aussi à faire une comparaison pour une forte inclinaison. La Fig.~\ref{fig:compareIpole_incli} montre l'image d'intensité totale avec l'orientation des vecteurs polarisation (traits blancs) pour les trois configurations magnétiques mentionnées plus haut avec une inclinaison de $20 \degree$ pour la première colonne et une inclinaison de $80\degree$ pour la quatrième colonne. Les colonnes 2 et 5 montrent la fraction de polarisation linéaire aux inclinaisons respectives, et les colonnes 3 et 6 montre la polarisation circulaire. Les valeurs des axes et les codes couleur sont choisis pour correspondre à ceux de la Fig.~2 de~\cite{Vos2022}. Cette dernière et notre Fig.~\ref{fig:compareIpole_incli} montrent un très bon accord entre \textsc{Gyoto} et \textsc{ipole} permettant de valider cette première série de tests pour la version polarisée de \textsc{Gyoto}.

Il est aussi intéressant de reproduire les mêmes tests avec le calcul des coefficients synchrotron issus de~\cite{Pandya2021}, présentés dans la section~\ref{sec:Pandya2021}. On obtient dans ce cas une erreur relative médiane inférieure ou égale à $10\%$. Bien que significativement supérieures, ces erreurs au niveau des flux de chaque pixel des images s'expliquent par l'erreur des formules analytiques de~\cite{Pandya2021} et de \textsc{ipole} par rapport aux valeurs exactes. Les deux méthodes utilisent des formules d'ajustement différentes en fonction de la température (on s'intéresse ici uniquement à une distribution thermique), avec potentiellement des domaines de validité différents (le domaine de validité des formules utilisées dans \textsc{ipole} n'est pas connu) et une erreur maximale de $30 \%$ pour les formules de~\cite{Pandya2021}. Il est alors normal d'atteindre une erreur de l'ordre de $1-10 \%$ sur le flux après intégration de l'équation de transfert radiatif.

\subsection{Boucles Q-U d'un point chaud}
\begin{figure}
    \centering
    \resizebox{0.4\hsize}{!}{\includegraphics{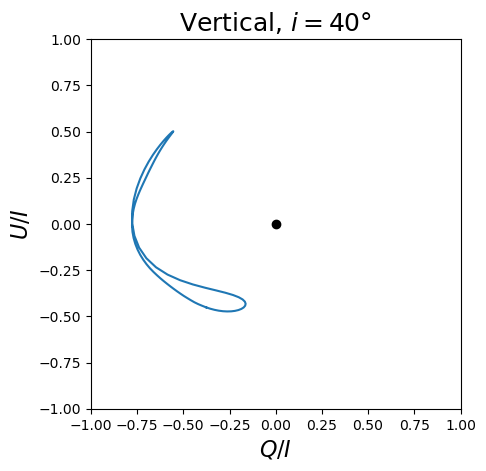}}
    \caption{Évolution de l'état de polarisation linéaire dans le plan $Q-U$ (normalisé par l'intensité totale $I$) d'un point chaud ayant un profil de densité et température Gaussien, un champ magnétique vertical, orbitant à $r=11$ $r_g$ d'un trou noir de Schwarzschild ayant une vitesse Képlérienne, et observé avec une inclinaison de $40 \degree$.}
    \label{fig:QU_loop_Vertical_40deg}
\end{figure}

Après validation de la version polarisée de \textsc{Gyoto}, il est intéressant d'étudier la polarisation d'un point chaud dans un cas simple. En effet, l'intensité totale, ainsi que les autres paramètres de Stokes, dépendent de la localisation de la source dans le plan du ciel et sont donc variables avec le temps. Une représentation particulière (couramment utilisée) de la variabilité temporelle de la polarisation est de tracer l'état de polarisation dans le plan $Q-U$ avec le paramètre Stokes $Q$ sur l'axe horizontal et le paramètre Stokes $U$ sur l'axe vertical. Il est important de ne pas confondre l'orientation du vecteur polarisation dans le plan du ciel et la représentation dans le plan $Q-U$ ! Cette dernière est particulièrement adaptée lorsque le rayonnement observé est variable\footnote{La variabilité pouvant être uniquement due aux effets de RG.}. La variabilité temporelle de la polarisation se traduit par un déplacement dans le plan $Q-U$ formant des motifs caractéristiques plus ou moins complexes en fonction de la géométrie du système (configuration magnétique, inclinaison, mouvement orbital, etc). On parle alors de \textit{boucle de polarisation} dans le plan $Q-U$, dont la première publication pour les sursauts de Sgr~A* a été faite par~\cite{Marrone2006}. On note que la forme de ces boucles n'est pas forcément circulaire. Cependant, dans certaines configurations, la courbe dans le plan $Q-U$ ne forme pas une boucle, mais des motifs plus complexes, comme ceux illustrés dans la Fig.~\ref{fig:QU_loop_Vertical_40deg}. On utilise tout de même le terme de boucle de polarisation.

On définit notre point chaud de manière similaire au Chap.~\ref{chap:modele_hotspot+jet} en utilisant l'objet \textit{Blob} de \textsc{Gyoto}, où l'on considère une orbite circulaire (Képlérienne) dans le plan équatorial d'une sphère de plasma uniforme dont la densité et la température évoluent en fonction du temps suivant les Eqs.~\eqref{eq:ne(t)}~et~\eqref{eq:Te(t)}. Pour simplifier, on considère ici une densité et une température constantes en prenant un très grand temps caractéristique pour la modulation Gaussienne $t_\sigma \gg P_\mathrm{orb}$ se traduisant par une émission intrinsèque constante. Les trois configurations du champ magnétique considérées, à savoir radial, toroïdal et vertical, sont définies directement en coordonnées Boyer-Lindquist tel que
\begin{equation}
    b^\alpha = \left\{
    \begin{array}{l}
        b^t = 0, \\
        b^r = 1/\sqrt{g_{rr}}, \\
        b^\theta = 0, \\
        b^\varphi = 0
    \end{array} \right.
\end{equation}
dans le cas radial et via les Eqs.~\eqref{eq:B_vertical} et~\eqref{eq:B_toroidal} pour les cas vertical et toroïdal respectivement. En ce qui concerne l'émission, on va considérer un rayonnement synchrotron issu d'une population thermique d'électrons dans un premier temps, puis on va comparer les boucles obtenues avec un rayonnement synchrotron issu d'une distribution kappa. De plus, on reprend les paramètres du tableau~\ref{tab:quiescent table params} pour le trou noir. Les paramètres du point chaud sont résumés dans le tableau~\ref{tab:hotspot_polar}.

\begin{table}[]
    \centering
    \begin{tabular}{ l c r }
        \hline
        \hline
        Paramètre & Symbole & Valeur\\
        \hline
        \textbf{Écran} & &\\
        nombre de pixels & N$_\mathrm{pix}$ & $1024$\\
        Champ de vue [$\mu$as] & fov & $200$\\
        Longueur d'onde d'observation [m] & $\lambda_\mathrm{obs}$ & $1,3 \times 10^{-3}$\\
        PALN [$\degree$] & $\Omega$ & $180$\\
        \textbf{Orbite} & & \\
        rayon orbital [$r_g$] & $r$ & $9$\\
        angle initial azimutal [$\degree$] & $\varphi_0$ & $0$\\
        \textbf{Physique} & & \\
        rayon du point chaud [$r_g$] & R & $1$\\
        Gaussian sigma [min] & $t_\sigma$ & $10000$\\
        magnétisation & $\sigma$ & $0,01$\\
        densité [cm$^{-3}$] & $n_e$ & $10^7$\\
        température [K] & $T_e$ & $10^{11}$\\
        indice distribution-$\kappa$ & $\kappa$ & $3,5$\\
        \hline
        \hline
    \end{tabular}
    \caption{Paramètres du point chaud pour générer les boucles $Q-U$ des Figs.~\ref{fig:QU_loops_20deg}. On note qu'avec une partie non thermique, le flux dans le cas d'une distribution kappa est supérieur au cas thermique.}
    \label{tab:hotspot_polar}
\end{table}

\begin{figure}
    \centering
    \resizebox{0.9\hsize}{!}{\includegraphics{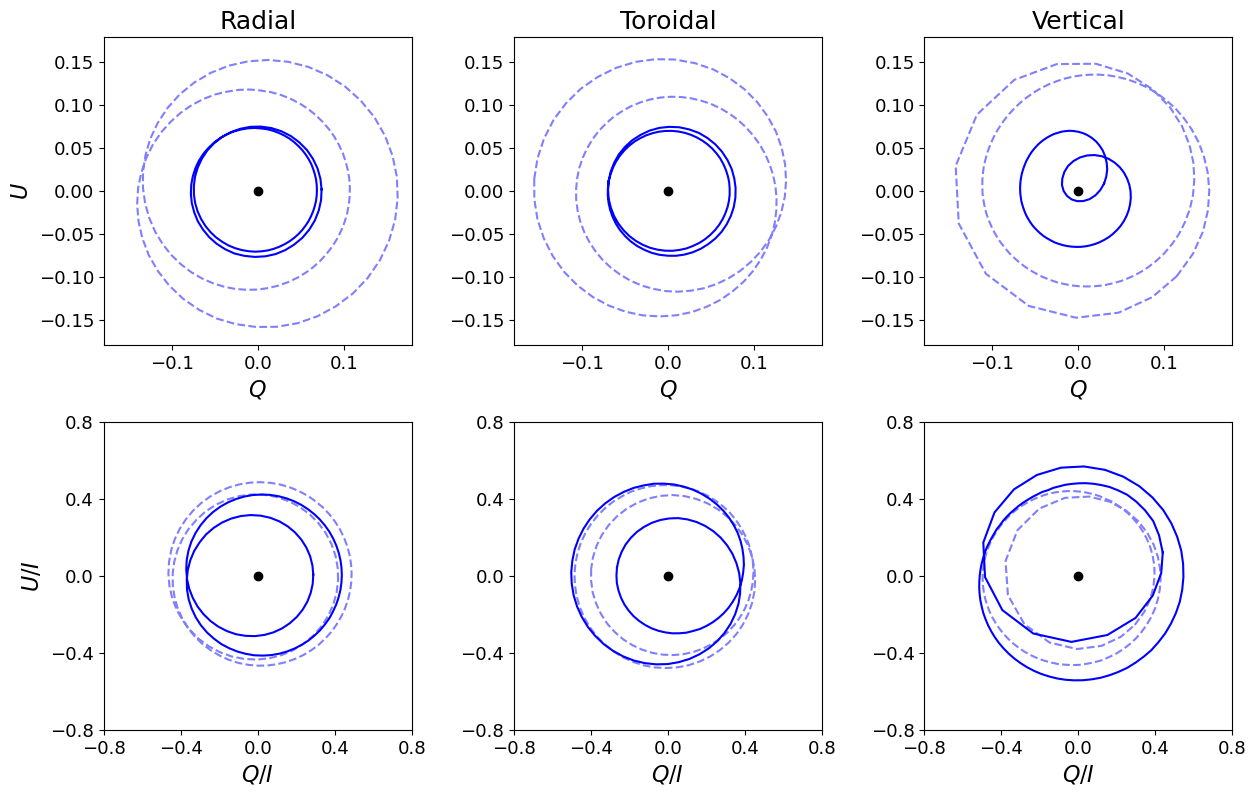}}
    \caption{Évolution de l'état de polarisation linéaire dans le plan $Q-U$ en haut (en Jy), et dans le plan $Q-U$ normalisé par l'intensité totale $I$ (en bas) d'un point chaud orbitant à $r=11$ $r_g$ d'un trou noir de Schwarzschild avec une vitesse Képlérienne, pour les trois configurations de champ magnétique (radial à gauche, toroïdale au milieu et vertical à droite), pour une inclinaison de $i=20 \degree$. La distribution considérée pour les calculs des coefficients synchrotron est thermique en traits pleins et kappa en tirets.}
    \label{fig:QU_loops_20deg}
\end{figure}

La Fig.~\ref{fig:QU_loops_20deg} montre les boucles de polarisation dans le plan $Q-U$ en haut, et dans le plan $Q-U$ normalisé par l'intensité totale $I$ en bas. Les colonnes représentent chacune des trois configurations magnétiques (radial pour la première colonne, toroïdal pour la seconde et vertical pour la troisième), à faible inclinaison ($i=20 \degree$). La distribution des électrons considérée est thermique en trait plein et suit une distribution kappa en tirets. On remarque que les courbes sont très différentes entre les deux lignes, c'est-à-dire entre le plan $Q-U$ et le plan $Q-U$ normalisé, montrant l'importance du choix du plan dans la représentation des boucles de polarisation. On constate aisément que la forme générale des boucles de polarisation, à savoir la circularité, le rapport de surface entre les deux boucles et la position du point d'intersection entre ces dernières, dépend de la configuration magnétique en plus de l'inclinaison. Ainsi, la mesure de la polarisation, tracée dans l'espace $Q-U$, permet de contraindre ces deux paramètres. Cependant, bien que la forme générale soit un indice fort de la configuration magnétique et de l'inclinaison, la position exacte du point d'intersection et le rapport de surface des deux boucles sont très dépendants du modèle considéré et de la distribution des électrons. En effet, on présente ici les boucles de polarisation de notre modèle de point chaud analytique décrit dans le Chap.~\ref{chap:modele_hotspot+jet}, que l'on peut comparer à d'autres modèles existants, comme celui de \cite{Vos2022} qui considère une sphère avec un profil gaussien et non une sphère uniforme. Les résultats sont similaires à faible inclinaison (Fig.~\ref{fig:QU_loops_20deg}) dans les cas toroïdal et radial, mais présentent néanmoins des différences dans les détails dues aux choix de la modélisation.

On constate aussi, en comparant les boucles de polarisation pour les deux distributions d'électrons considérées (thermique en trait plein et kappa en tirets), que le choix de la distribution des électrons influence significativement le rapport de surface des boucles, en particulier dans le cas vertical, à $20 \degree$ d'inclinaison. 

Pour conclure, la polarisation est et sera encore plus importante à l'avenir pour contraindre la physique des flots d'accrétion autour des trous noirs et particulièrement pour les sursauts de Sgr~A*.

\chapter{Ouvertures et conclusion}\label{chap:ouvertures}
\markboth{Ouvertures et conclusion}{Ouvertures et conclusion}
{
\hypersetup{linkcolor=black}
\minitoc 
}

\section{Perspectives d'évolutions du modèle de point chaud issu de la reconnexion magnétique}
\subsection{Contraindre les paramètres}
Le modèle de point chaud basé sur la reconnexion magnétique présenté au Chap.~\ref{chap:Plasmoid Flare model} montre des résultats très encourageants. Comme ce modèle est basé sur une simulation de l'évolution de la distribution des électrons, il est peu judicieux de vouloir ajuster les données "à l'aveugle", avec un algorithme explorant l'espace des paramètres de manière aléatoire, de type MCMC. En effet, pour chaque variation d'un des paramètres physiques, une nouvelle simulation de l'évolution de la distribution va être lancée avec les nouveaux paramètres. Or, le pas réalisé par l'algorithme entre deux points de l'espace des paramètres peut être très faible, ne justifiant pas une nouvelle simulation. De plus, ce genre d'algorithme est très sensible aux valeurs initiales et risque de converger vers un minimum local au lieu de la "vraie" solution. Enfin, ce processus doit être réitéré pour chaque jeu de données.

Ainsi, il est plus judicieux de faire une grille sur l'espace des paramètres avec un pas déterminé et connu. Pour chaque point de cette grille, on peut donc évaluer le $\chi^2$ de nos données et déterminer le minimum global ainsi que des minima locaux à partir desquels on peut lancer un algorithme MCMC. Comme on l'a vu au Chap.~\ref{chap:Plasmoid Flare model}, il y a de nombreuses dégénérescences entre les paramètres : la grille ne pouvant être infiniment fine, il est possible qu'un minimum, a priori local, donne une meilleure solution après un algorithme MCMC que le minimum global identifié par la grille. Le gros avantage de cette méthodologie est que pour chaque jeu de données, on a juste à évaluer le $\chi^2$ pour chaque point de la grille permettant de réduire significativement l'espace des paramètres à explorer.

Cependant, le nombre de paramètres de ce modèle est très élevé : onze, sans compter l'inclinaison, le PALN, la position réelle et la masse de Sgr~A*. Le nombre de dimensions de l'espace des paramètres est donc trop élevé, même pour une grille, au vu du temps de calcul nécessaire pour générer une astrométrie et une courbe de lumière synthétiques. Il faut donc contraindre au maximum les paramètres qui peuvent l'être à partir d'autres observations ou des simulations. La masse de Sgr~A* est maintenant bien contrainte à partir des observations du mouvement des étoiles S autour de Sgr~A*~\cite{Gravity2020a}. Plusieurs observations, comprenant les sursauts de 2018 observés par GRAVITY~\cite{Gravity2018} et les résultats de~\cite{EHT2022a}, tendent vers une faible inclinaison $i =  160 \degree \pm 10 \degree$ sans donner de contraintes solides sur le PALN ($\Omega \sim 115\degree - 160 \degree$). Un des sursauts observés en 2018 présentait deux pics que l'on peut interpréter avec notre modèle comme un effet du beaming et de lentille gravitationnelle permettant de contraindre l'alignement entre l'inclinaison et l'angle polaire d'éjection du point chaud $\theta$. Il est possible de contraindre, dans une certaine mesure, la valeur du champ magnétique à travers le temps de refroidissement synchrotron. \cite{von_Fellenberg2023} ont déterminé le profil moyen des sursauts de Sgr~A* pouvant être décrit par deux exponentielles (une croissante et une décroissante). Le temps caractéristique de la seconde peut donc être utilisé pour contraindre le champ magnétique. Concernant les autres paramètres, on peut s'intéresser aux résultats des simulations numériques GRMHD et (GR)PIC. En effet, les simulations PIC montrent que l'indice de la loi de puissance de la distribution des électrons à haute énergie, relié à notre paramètre $\kappa$, dépend de la magnétisation du plasma en amont de la reconnexion à travers l'Eq.~\eqref{eq:kappa_index} comme illustré dans la Fig.~\ref{fig:kappa_value}. On constate que la valeur de $\kappa$ peut être raisonnablement fixée à 2,8 pour des valeurs vraisemblables de magnétisation autour de Sgr~A*. Cela pose cependant un problème par rapport aux sursauts en rayons~X (voir section~\ref{sec:multi_lambda}). D'une manière générale, les simulations GRMHD et GRPIC obtiennent une vitesse radiale d'éjection de l'ordre de $v_r = 0,1$~c, incompatible avec nos hypothèses actuelles (voir Chap.~\ref{chap:Plasmoid Flare model}). \cite{El_Mellah2023} a déterminé le profil de vitesse (radial et azimutal) des tubes de flux générés par la reconnexion et éjectés dans la magnétosphère. Cependant, ces profils sont dépendants du modèle et du spin, impliquant la nécessité d'effectuer de plus amples recherches avant de contraindre la vitesse d'éjection. Enfin, le temps de croissance $t_\mathrm{growth}$ peut raisonnablement être fixé à $100\, r_g/c$ et la densité à celle de l'état quiescent.

On peut donc ainsi réduire le nombre de paramètres entièrement libres à cinq : le rayon orbital initial, le temps d'observation initial, l'angle azimutal et la vitesse azimutale initiaux et le PALN, auxquels on peut rajouter l'inclinaison. Ces paramètres ne sont pas complètement libres, mais ont un intervalle de valeurs plus large que les autres. L'espace des paramètres est bien plus raisonnable, ce qui permet la création d'une grille en un temps de calcul réaliste (encore à déterminer) comparé à précédemment.

Cependant, le problème du lancement d'une nouvelle simulation d'évolution de la distribution des électrons pour chaque pas de l'algorithme MCMC est toujours présent. Pour cela, le recours à l'interpolation entre les points d'une (autre) grille uniquement liée à \texttt{EMBLEM}, c'est-à-dire l'évolution de la distribution et des coefficients synchrotron, pourrait économiser beaucoup de temps de calcul.

\subsection{Polarisation}
Comme on l'a vu au Chap.~\ref{chap:Polarization}, la polarisation est très sensible à la configuration du champ magnétique et à l'inclinaison, mais aussi au modèle utilisé (sphère de densité uniforme ou avec un profil gaussien, entre autres) à travers les propriétés des boucles $Q-U$. La reconnexion magnétique correspond à une réorganisation de la topologie des lignes de champs magnétiques. Le produit de la reconnexion, les tubes de flux, ont donc des configurations magnétiques particulières, soit hélicoïdales avec l'axe principal dans le plan $(\mathbf{e_\theta}, \mathbf{e_\varphi})$~\cite{Ripperda2022, El_Mellah2023}, soit verticales~\cite{Ripperda2022}. L'inclusion de la configuration du champ magnétique dans le modèle de point chaud est donc un axe d'amélioration naturel.

Au moment de la rédaction de ces lignes, \texttt{EMBLEM} suppose un champ magnétique isotrope pour le calcul d'évolution de la distribution des électrons et le calcul des coefficients synchrotron (qui sont deux calculs indépendants). Dans le cas de l'évolution de la distribution des électrons, l'hypothèse d'un champ magnétique isotrope permet de ne considérer qu'une seule valeur de champ magnétique pour le refroidissement synchrotron qui, normalement, se calcule à partir de la composante du champ magnétique perpendiculaire à la vitesse (voir Chap.~\ref{chap:modele_hotspot+jet}). Cette condition reste valable dans le cas d'une distribution isotrope des électrons (hypothèse qui est déjà faite). Ainsi, considérer une configuration particulière du champ magnétique (non isotrope) n'a pas d'influence sur le calcul d'évolution de la distribution des électrons. Cependant, dans le cas du calcul des coefficients synchrotron, l'angle entre le champ magnétique et le vecteur direction de propagation du photon $\theta_B$ ne peut plus être moyenné puisque ni le champ magnétique, ni le vecteur direction de propagation du photon ne sont isotropes (seul le photon atteignant l'observateur nous intéresse). Il faut donc calculer les coefficients synchrotron pour un certain nombre ($\sim 100$) de valeurs de $\theta_B$ que l'on va ensuite interpoler pour les valeurs particulières dans \textsc{Gyoto} (la dépendance en $\theta_B$ étant non triviale car elle intervient dans une borne d'une intégrale). On se retrouve donc à faire une interpolation tridimensionnelle sur la fréquence, le temps et l'angle $\theta_B$. De plus, cela augmente significativement la taille des fichiers générés par \texttt{EMBLEM} qui sont aussi plus nombreux puisqu'en plus des coefficients d'émission et d'absorption non polarisés, on rajoute les coefficients polarisés (Stokes $Q$, $U$ et $V$) ainsi que les coefficients de rotation Faraday. On passe donc de deux tableaux 2D à 11 tableaux 3D qu'il faut interpoler.

Le calcul de l'ensemble des coefficients polarisés pour une distribution arbitraire ainsi que l'adaptation du code \texttt{EMBLEM} pour rajouter la dépendance à $\theta_B$ est un travail en cours au moment de cette rédaction.

\subsection{Multi-longueurs d'onde}\label{sec:multi_lambda}
En plus de la polarisation, l'autre caractéristique fondamentale des sursauts de Sgr~A* pouvant permettre de différencier les modèles et de mieux comprendre la physique des flots d'accrétion, est l'aspect multi-longueurs d'onde. Comme on l'a vu au Chap.~\ref{chap:Sgr~A* flares}, les sursauts de Sgr~A* sont aussi observés en rayons~X et en radio avec des propriétés particulières comme la présence ou l'absence de contrepartie en rayons~X des sursauts en IR, les délais entre les différentes longueurs d'onde (voir Figs.~\ref{fig:Boyce2022} et \cite{Fazio2018}), les indices spectraux dans les différentes bandes, etc. La question du lien entre les sursauts IR et radios fait toujours débat. Même en partant du principe que les sursauts aux différentes longueurs d'onde sont liés, il n'est pas évident que la source exacte du rayonnement à chacune de ces longueurs d'onde soit identique.

Dans l'hypothèse où les sursauts de Sgr~A* sont générés par un événement de reconnexion magnétique (dans le disque ou dans la magnétosphère), on peut envisager plusieurs scénarios pour expliquer les propriétés multi-longueurs d'onde observées.
\begin{itemize}
    \item[$\bullet$] Scénario à \textbf{une seule zone} (point chaud) : 
\end{itemize}

\begin{figure}
    \centering
    \resizebox{\hsize}{!}{\includegraphics{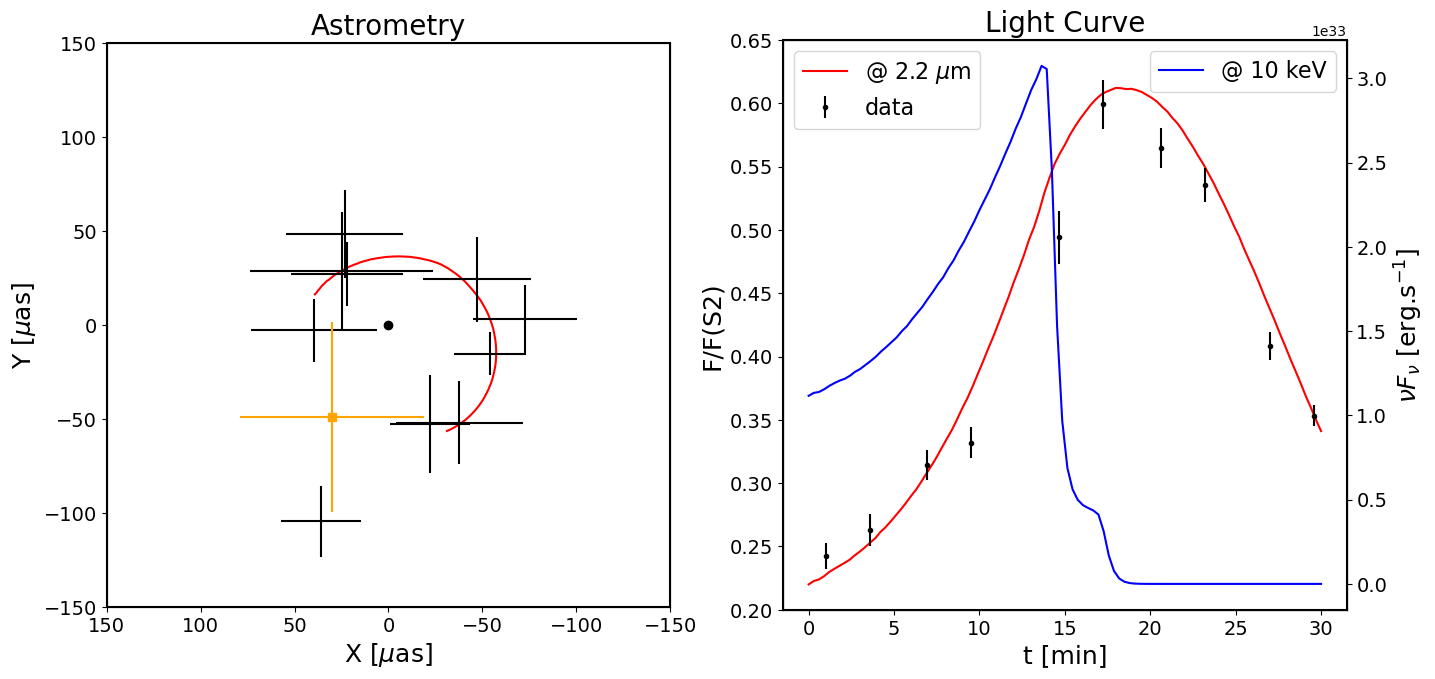}}
    \caption{Données et modèle du sursaut du 22 juillet 2018 observé en NIR. Le panneau de gauche montre l'astrométrie du sursaut tandis que le panneau de droite montre les courbes de lumière observées. Les courbes en rouge correspondent au modèle de point chaud issu de la reconnexion à $2,2\, \mu$m, avec les paramètres listés dans les tableaux~\ref{tab:emblem} et \ref{tab:plasmoid_orbital_params}. La courbe bleue dans le panneau de droite correspond à la courbe de lumière du même modèle à $10$ keV (rayons~X). Les données en noir sont uniquement à $2,2\, \mu$m. Le point noir dans le panneau de gauche représente la position de Sgr~A* dans \textsc{GYOTO} et la croix orange représente la position de Sgr~A* mesurée à travers l'orbite de S2.}
    \label{fig:Test_X_ray}
\end{figure}

Dans ce modèle, l'ensemble du rayonnement, à toutes les fréquences, est issu d'une seule zone, un tube de flux avec une population non thermique d'électrons (loi de puissance ou loi kappa), refroidissant par rayonnement synchrotron, comme dans notre modèle du Chap.~\ref{chap:Plasmoid Flare model}. La Fig.~\ref{fig:Test_X_ray} montre l'astrométrie et la courbe de lumière observées par GRAVITY en noir, ainsi que le modèle de point chaud issu de la reconnexion avec les paramètres des tableaux~\ref{tab:emblem} et \ref{tab:plasmoid_orbital_params}, comme dans le Chap.~\ref{chap:Plasmoid Flare model}, auxquels on a rajouté la courbe de lumière observée du modèle à $10$~keV (rayons~X) en bleu. On constate que le pic de la courbe de lumière en rayons~X précède celui en IR de $\sim 6$ min, ce qui est comparable aux observations (voir Fig.~\ref{fig:Boyce2022}). Néanmoins, le flux en rayons~X de notre modèle chute beaucoup trop rapidement comparé aux données de sursauts observés en NIR et rayons~X, qui suggèrent des taux de croissance et de décroissance similaires. Cela est normal au vu de notre modèle étant donné que l'on stoppe brutalement l'injection d'électrons dans le point chaud après $t_\mathrm{growth}$. Les électrons émettant du rayonnement~X sont très énergétiques, $\gamma \sim 10^{5-6}$, et ont donc un temps de refroidissement synchrotron très court ($\sim 10$ sec) comparé au temps d'observation. Cela résulte en une décroissance rapide du flux observé. Pour pallier à ce problème, au lieu de brutalement stopper l'injection, on peut faire diminuer le $\gamma_\mathrm{max}$ de l'injection pour $t>t_\mathrm{growth}$, rendant plus lente la décroissance à haute énergie. En effet, il est peu probable que le moteur de reconnexion soit constant et stoppe brutalement, il est plus vraisemblable qu'il perde de la puissance avant de s'arrêter (réflexion en cours et à vérifier avec les simulations). Dans le domaine radio, le flux intrinsèque augmente continuellement puisque les électrons à plus haute énergie, en se refroidissant, se concentrent aux basses énergies dont l'émission est dans le domaine radio. À cela s'ajoute la modulation du beaming comme on l'a vu au Chap.~\ref{chap:modele_hotspot+jet}. On peut aisément supposer que la décroissance du flux radio vient de la perte de cohérence de la zone d'émission (mixage avec le reste du flot d'accrétion) qui n'est pas modélisée ici.

\begin{itemize}
    \item[$\bullet$] Scénario à \textbf{deux zones} (nappe de courant + tube de flux) : 
\end{itemize}

Un second scénario envisageable, toujours dans le contexte de la reconnexion magnétique, est un modèle à deux zones. La première zone est la nappe de courant où a lieu la reconnexion et l'accélération des électrons à très haute énergie. Cette population d'électrons est responsable de l'émission en rayons~X des sursauts. Avant d'arriver dans la seconde zone qui est un point chaud, ces électrons ont refroidi à des énergies plus modérées et vont ainsi émettre en IR et radio. Lors de la fusion des tubes de flux avec le point chaud, une reconnexion secondaire peut avoir lieu, mais avec un flux magnétique réduit, ce qui résulte en une accélération de particules à des énergies plus faibles que la première reconnexion. Les électrons présents dans le point chaud peuvent avoir une distribution étendue comme celle considérée au Chap.~\ref{chap:Plasmoid Flare model}, mais avec une énergie maximale plus faible. On a donc une distinction très claire entre la source des sursauts en rayons~X et celle des sursauts en IR/radio.

\begin{itemize}
    \item[$\bullet$] Scénario à \textbf{deux zones} (deux tubes de flux distincts) : 
\end{itemize}

\begin{figure}
    \centering
    \resizebox{0.5\hsize}{!}{\includegraphics{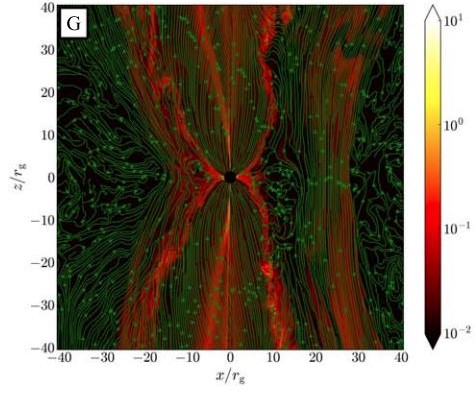}}
    \caption{Instantané de la température sans dimension du flot d’accrétion-éjection dans le plan méridional d’une simulation 3D GRMHD, identifié comme un état post-sursaut, mais pas encore quiescent. Crédit : \cite{Ripperda2022}.}
    \label{fig:Ripperda2022bis}
\end{figure}

Dans le scénario précédent, les sursauts en IR et radio sont liés à la même zone d'émission modélisée par un point chaud. Ainsi, la configuration du champ magnétique pour les deux domaines de longueurs d'onde, que l'on peut déterminer à partir de la polarisation (voir Chap.~\ref{chap:Polarization}), doit être identique. \cite{Wielgus2022} ont montré que les sursauts radio peuvent être modélisés par un point chaud avec une configuration magnétique verticale. Dans le cas des sursauts IR, \cite{Gravity2018} ayant exclu la configuration toroïdale, les configurations purement verticale, purement radiale et poloidale sont donc encore possibles. Si les configurations magnétiques déduites de la polarisation en radio et IR sont différentes, alors cela signifie que leurs zones d'émission sont différentes. Les simulations 3D GRMHD de \cite{Ripperda2022} montrent qu'en plus de former des plasmoïdes/tubes de flux, la reconnexion éjecte une partie du disque en créant une "barrière" magnétique verticale (en $X\sim 20\, r_g$ dans la Fig.~\ref{fig:Ripperda2022bis}) orbitant autour du trou noir de manière similaire au point chaud modélisé par \cite{Wielgus2022}. L'éjection du disque au moment de la reconnexion résulte en une diminution du flux en radio, qui est ensuite suivie par une augmentation due à cette structure avec un champ magnétique verticale alimenté progressivement en matière. Les sursauts en IR et rayons~X sont, quant à eux, produits comme dans le modèle à une zone par le tube de flux issu directement de la reconnexion. Ce scénario a l'avantage d'expliquer à la fois la diminution du flux radio au moment des sursauts IR et~X, suivi d'un sursaut radio après un certain délai (on voit dans la Fig.~\ref{fig:Ripperda2022bis} que le plasma dans les tubes de flux a eu le temps de se refroidir comparé à la première ligne de la Fig.~\ref{fig:Ripperda2022}), et d'une différence (hypothétique, mais possible) de configuration de champ magnétique entre IR et radio que l'on peut déduire de la polarisation.

\begin{itemize}
    \item[$\bullet$] Scénario à \textbf{trois zones} : 
\end{itemize}

Enfin, on peut aussi envisager que chaque longueur d'onde a sa propre zone d'émission. En reprenant les zones définies dans les scénarios précédents, on a donc les sursauts en rayons~X produits par la nappe de courant, ceux en IR par le tube de flux issu directement de la reconnexion (notre modèle de point chaud amélioré) et ceux en radio par le tube de flux magnétique vertical modélisable par un point chaud \cite{Wielgus2022}. Cependant, ce scénario est complexe à modéliser correctement, et il offre un trop grand nombre de libertés, ce qui risque de se traduire par un sur-ajustement des données.\\

Les scénarios présentés ici sont des ébauches issues des différentes contraintes observationnelles et nécessitent chacun une étude approfondie.

\section{Autres Projets}
Durant la thèse, d'autres projets ont été réalisés ou sont en cours. Les deux principaux sont présentés brièvement ici.

\subsection{Interface entre simulations et \textsc{Gyoto}}
Comme on l'a vu précédemment, \textsc{Gyoto} est un code de tracé de rayons permettant de générer des images de sources évoluant dans un espace-temps courbe (ou plat, mais l'intérêt est plus réduit). La plupart des sources de rayonnement dans \textsc{Gyoto} sont des modèles de natures différentes : étoiles, disques d'accrétion, jets, etc. Cependant, il est aussi intéressant de produire des images d'un flot d'accrétion généré non pas par un modèle analytique, mais par les simulations (GRMHD par exemple). En effet, les modèles de disque sont le plus souvent axisymétriques auxquels on peut rajouter une sur-densité (par exemple). Un modèle a l'avantage de ne pas nécessiter des milliers d'heures de calcul en amont, mais il est toutefois intéressant de générer les observables à partir des résultats des simulations. En effet, dû aux effets relativistes explicités dans les Chap.~\ref{chap:GYOTO} et~\ref{chap:modele_hotspot+jet}, les résultats des simulations GRMHD, à savoir les cartes (2D ou 3D) de densité ou de température par exemple au cours du temps, ne peuvent pas être directement traduits en flux ou astrométrie observés. Ainsi, pour étudier la variabilité d'un disque d'accrétion simulé, il faut utiliser un code de tracé de rayons comme \textsc{Gyoto}.

Le lien entre les résultats de simulations GRMHD et \textsc{Gyoto} n'est pas trivial. En effet, les grandeurs des simulations sont des quantités sans dimension qu'il faut correctement normaliser avant de les mettre à la bonne échelle. De plus, les quantités sont déterminées sur une grille, celle de la simulation. Or, la probabilité que la trajectoire des photons passe par ces points de grille est nulle, ce qui nécessite une interpolation des quantités sur le nombre de dimensions de la simulation. Cette interface entre des simulations avec \href{https://amrvac.org/}{\textsc{AMRVAC}} et \textsc{Gyoto} existait déjà dans le cadre de simulations hydrodynamiques en RG. Le travail effectué a été d'étendre l'interface pour des simulations GRMHD, prenant en compte le champ magnétique, en 2.5D, c'est-à-dire les quantités hydrodynamiques en 2D ($r$, $\varphi$) et le champ magnétique en 3D. Les résultats de cette collaboration faite avec l'équipe \textit{Astrophysique des hautes énergies} du laboratoire AstroParticules et Cosmologie (Raphaël Mignon-Risse et Peggy Varnière) sont rapportés dans \cite{Mignon_Risse2021} (dont je suis co-auteur).

\subsection{Étude statistique des sursauts}
Le nombre de paramètres du modèle de point chaud issu de la reconnexion est élevé, et les temps de calcul sont relativement longs, ce qui rend difficile un ajustement. Il faut donc contraindre au maximum les paramètres. Cependant, chaque sursaut a un effet de beaming différent qui complexifie l'analyse. En normalisant l'ensemble des sursauts et en les superposant, \cite{von_Fellenberg2023} ont déduit des propriétés statistiques des sursauts en IR et rayons~X (voir Fig.~\ref{fig:von_Fellenberg2023}). Ils ont par la suite créé un modèle de point chaud orbitant dans le plan équatorial du trou noir à vitesse Képlérienne avec une émission variable (deux exponentielles), qu'ils ont utilisé pour générer un grand nombre de courbes de lumière. Ces courbes de lumières ont été normalisées et superposées de manière identique aux données des sursauts. Les propriétés statistiques ainsi obtenues sur les courbes de lumière synthétiques ont été comparées aux propriétés des données permettant de contraindre les paramètres de l'émission intrinsèque (les temps caractéristiques). 

Comme les courbes de lumière sont normalisées, le processus de rayonnement n'a pas d'importance en première approximation. Le spectre utilisé est un spectre plat. Or, on a vu au Chap.~\ref{chap:modele_hotspot+jet} l'influence du spectre d'émission de la source sur les observables. En prolongement de \cite{von_Fellenberg2023}, on utilise le modèle de point chaud issu de la reconnexion, et notamment la combinaison \texttt{EMBLEM}-\textsc{Gyoto} pour générer les courbes de lumière synthétiques. Afin d'économiser du temps de calcul, les deux codes sont utilisés de manière distincte. En effet, on utilise \texttt{EMBLEM} pour générer des courbes de lumière intrinsèques en variant ses paramètres et on utilise \textsc{Gyoto} pour déterminer la modification par effets relativistes de manière indépendante. Les courbes de lumière ainsi obtenues sont la convolution de la courbe de lumière intrinsèque avec les effets relativistes.

\begin{figure}
    \centering
    \resizebox{0.6\hsize}{!}{\includegraphics{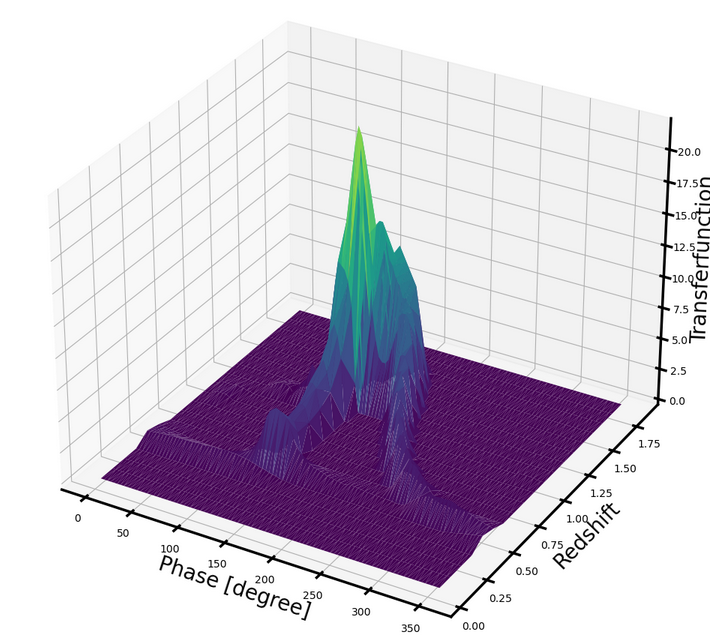}}
    \caption{Fonction de transfert en fonction du décalage spectral et de la phase orbitale, issue de cartes de décalage spectral générées par \textsc{Gyoto} d'un point chaud en orbite à $r=6$ $r_g$ avec une vitesse azimutale $\dot{\varphi}=0,066$ rad.$t_g^{-1}$ et une inclinaison de $i=90 \degree$. Crédit : Sebastiano von Fellenberg.}
    \label{fig:transfert_function}
\end{figure}

L'intérêt principal de cette étude est d'avoir une source avec une dépendance spectrale soutenue par un processus physique que l'on considère réaliste. Ainsi, pour déterminer le flux observé au temps $t$, on calcule d'abord une carte de décalage spectral de la source à la phase $\phi$ de son orbite (toujours dans le plan équatorial, mais avec une vitesse arbitraire) avec \textsc{Gyoto}. Cela nous permet de déterminer la fonction de transfert, c'est-à-dire la puissance en fonction de la fréquence, qui dans le cas de l'orbite d'un point chaud se résume à deux pics plus ou moins larges correspondants au décalage spectral de l'image primaire et secondaire (voir Fig.~\ref{fig:transfert_function}). La fonction de transfert permet de déterminer la fréquence d'émission de chacune des images que l'on va utiliser pour calculer l'intensité observée à partir de la densité spectrale d'énergie.

Les paramètres influençant la fonction de transfert (qui contient tout les effets relativistes) sont l'inclinaison $i$, le rayon de l'orbite $r$ et la vitesse orbitale $v_\varphi$. En ce qui concerne l'émission intrinsèque, les paramètres libres sont le champ magnétique (pour déterminer le temps caractéristique de décroissance), l'indice de la loi d'injection $\alpha$ (voir Eq.~\eqref{eq:injterm}), et la température sans dimension $\Theta_e$ (la densité max et l'indice kappa sont dégénérés avec la température). Ce projet est en cours de réalisation avec Sebastiano von Fellenberg et Michi Bauböck.

\section{Conclusion}
Sagittarius~A*, le trou noir supermassif au centre de la Voie Lactée, est le trou noir avec la plus grande taille angulaire dans le ciel ($\sim 50\, \mu$as), ce qui en fait un laboratoire idéal pour étudier les flots d'accrétion autour des trous noirs et les effets de la Relativité Générale. Il est aussi une cible idéale pour tester le théorème de calvitie des trous noirs (\textit{no hair theorem}). Depuis sa découverte en 1974 par \cite{Balick1974} comme source radio extrêmement compacte, associée à un trou noir supermassif, Sgr~A* est régulièrement observé à différentes longueurs d'onde, du domaine radio aux rayons~X en passant par les IR. Les observations des quelques secondes d'arc autour de Sgr~A* en NIR ont révélé une population d'étoiles, les étoiles-S, orbitant autour du trou noir avec des périodes de l'ordre de quelques dizaines d'années. Le suivi de ces étoiles a permis de déterminer la masse de Sgr~A* à $M_{BH}=(4,297 \pm 0.016) \times 10^6 M_\odot$ pour une distance de $R_0 = (8.277 \pm 9)$~pc~\citep{GRAVITY2022a}, mais aussi de confirmer des effets prédits par la Relativité Générale comme la précession de Schwarzschild et le décalage vers le rouge gravitationnel~\cite{Gravity2020b}.\\

En 2001, \cite{Baganoff2001} ont détecté une émission brillante en rayons~X appelée sursaut, provenant de Sgr~A*. Deux ans plus tard, \cite{Genzel2003} rapportent la détection d'un sursaut en NIR. Ces détections ont suscité l'intérêt de la communauté qui a alors entamé un vaste effort pour tenter d'expliquer l'origine de ces sursauts. Les efforts ainsi réalisés, tant du point de vue purement théorique qu'au niveau des simulations numériques, ont mené à différents scénarios capables d'expliquer ces sursauts. Grâce à l'avènement de l'optique adaptative et de l'interférométrie optique, de nouvelles observations, notamment avec l'instrument GRAVITY/VLTI ayant une précision astrométrique de $\sim$ 30-50 $\mu$as, ont vu le jour, ajoutant de nouvelles contraintes pour la modélisation des sursauts de Sgr~A*. En effet, en 2018, la Collaboration GRAVITY a observé trois sursauts brillants dont la source présente un mouvement orbital \cite{Gravity2018, Wielgus2022}. Depuis lors, certains modèles se démarquent, dont un modèle dit de point chaud pouvant avoir diverses origines physiques. L'une d'entre elles commence à se distinguer des autres et fait l'objet d'un effort important de recherche à travers le monde : la reconnexion magnétique.\\

L'objectif de cette thèse était d'étudier et de développer des modèles pour expliquer les propriétés des sursauts en NIR de Sgr~A* observés par GRAVITY. Étant donné que les sursauts se produisent près de l'horizon des événements du trou noir, à environ 10 fois le rayon gravitationnel ($r_g$), la courbure de l'espace-temps a un impact significatif sur la trajectoire et la fréquence des photons émis par la source des sursauts. Afin de tenir compte de tous les effets de la relativité générale et restreinte, nous avons utilisé le code de tracé de rayons \textsc{Gyoto}, qui permet de générer des images et des spectres d'une source modélisée dans un espace-temps courbe.

Le premier modèle étudié dans le chapitre~\ref{chap:modele_hotspot+jet} est une amélioration du modèle du point chaud proposé par \cite{Gravity2018}. Dans ce modèle, le point chaud est défini comme une sphère homogène de plasma constituée d'électrons suivant une distribution kappa (un cœur thermique et une loi de puissance à haute énergie), émettant du rayonnement synchrotron. Le point chaud décrit une orbite circulaire autour du trou noir avec une vitesse Képlérienne. Afin de simuler une sur-densité qui aurait été chauffée puis refroidie en se mélangeant avec le reste du flot d'accrétion, nous appliquons une modulation gaussienne sur la densité et la température, ce qui se traduit par une émission gaussienne dans le référentiel de l'émetteur.

Cependant, la variabilité observée est une combinaison de cette variabilité intrinsèque avec les effets relativistes. L'introduction d'une variabilité intrinsèque de la source de rayonnement ajoute des effets connus, tels que la différence de temps d'émission entre l'image primaire et l'image secondaire du fait de la différence de distance parcourue. Lorsque le temps caractéristique de la variabilité intrinsèque est de l'ordre de la période orbitale et correspond à la période du beaming, l'effet du temps de propagation a un impact important sur l'astrométrie. En effet, la différence de temps d'émission combinée au beaming peut, dans certaines configurations, équilibrer les contributions des images primaire et secondaire dans le calcul de l'astrométrie, alors que la plupart du temps, l'image primaire domine nettement.

Ce modèle de point chaud, bien que simple, mais avec un processus de rayonnement plus réaliste que le modèle de \cite{Gravity2018} et une variabilité intrinsèque, permet de comprendre l'influence des différents effets relativistes sur les observables, à savoir l'astrométrie et la courbe de lumière. De plus, ce modèle présente des résultats encourageants pour l'ajustement des données GRAVITY. Cependant, tout comme le modèle de \cite{Gravity2018}, il ne parvient pas à expliquer la vitesse super-Képlérienne observée en astrométrie, puisque nous avons supposé une vitesse Képlérienne dans ce modèle.\\

En plus d'un point chaud pour modéliser la source des sursauts, il est intéressant de modéliser l'état quiescent de Sgr~A*, cependant, les contraintes actuelles ne permettent pas de converger vers un unique modèle quiescent. De plus, la nature exacte de la source du rayonnement quiescent (disque ou jet ou les deux) n'a pas d'impact sur les observables GRAVITY, vu que la source n'est pas résolue et que le plasma est optiquement mince. Ce qui nous importe ici est l'impact de l'inclusion de l'état quiescent (que l'on modélise à $2,2\, \mu$m par un jet), son centroïde et son flux, sur l'astrométrie observée. L'influence sur la courbe de lumière est assez triviale, vu que l'on rajoute un flux constant.

L'astrométrie mesurée correspond au mouvement du centroïde du flux, c'est-à-dire au barycentre de l'ensemble du rayonnement émis pondéré par le flux de chaque source. Ainsi, lorsque le flux du point chaud (image primaire et/ou secondaire) est plus faible ou comparable au flux de l'état quiescent, le centroïde de l'image composée du jet et du point chaud est décalé (plus proche du jet) par rapport à celui de l'image contenant uniquement le point chaud. En début et en fin de sursaut, lorsque la condition précédente est vérifiée, l'astrométrie se rapproche du centroïde du jet (fixe). Cela a pour effet de décaler le centre de l'orbite apparente (affectée par le centroïde du jet) par rapport à la position du trou noir, qui correspond approximativement (à faible inclinaison) au centre de l'orbite apparente du point chaud, sans prendre en compte l'état quiescent. Ce décalage est marginalement observé par GRAVITY pour chacun des trois sursauts, mais avec des orientations différentes dans le ciel pour chaque sursaut. L'inclusion d'un modèle quiescent dans le calcul de l'astrométrie (et de la courbe de lumière) permet d'expliquer cette dernière observation.\\

Le modèle de point chaud entièrement analytique précédent (avec la modulation gaussienne) est prometteur et montre des résultats globalement satisfaisants, sauf en ce qui concerne la vitesse orbitale qui est présumée Képlérienne, alors que les observations de GRAVITY tendent vers une vitesse super-Képlérienne. De plus, le processus physique à l'origine de la surdensité et du chauffage n'est pas traité. Ainsi, un second modèle de point chaud a été étudié et développé, basé sur la reconnexion magnétique.

En utilisant à la fois des contraintes observationnelles et des simulations numériques de type GR(R)MHD et (GR)PIC, nous avons développé un modèle de point chaud semi-analytique correspondant au produit final de la reconnexion (et non la reconnexion elle-même !) : un plasmoïde / tube de flux macroscopique. En effet, la reconnexion magnétique, qui correspond à la dissipation de flux magnétique via une réorganisation de la topologie des lignes de champ, produit des chaînes de plasmoïdes microscopiques qui fusionnent les uns avec les autres pour former un îlot magnétique, le point chaud, accumulant les particules accélérées au site de reconnexion. Les particules accélérées peuvent atteindre de très hautes énergies, avec une distribution typiquement non thermique. Ensuite, ces particules se refroidissent par rayonnement synchrotron. La puissance émise, la fréquence d'émission ainsi que le taux de refroidissement dépendent de l'énergie des particules. Pour cette raison, nous faisons évoluer la distribution des particules (ici, des électrons) en considérant l'injection d'électrons accélérés par la reconnexion et leur refroidissement par rayonnement synchrotron, via le code d'évolution cinétique \texttt{EMBLEM}. Ce dernier nous permet a posteriori de déterminer les coefficients synchrotron en fonction du temps, pour calculer le flux émis par le point chaud. Le mouvement considéré pour le point chaud est une éjection conique, c'est-à-dire avec un angle polaire $\theta$ constant et une vitesse radiale $v_r$ positive. Cette dynamique est issue des résultats de simulations GRPIC de reconnexion magnétique dans la magnétosphère du trou noir \cite{El_Mellah2022}. Elle est également soutenue par des simulations GRMHD \cite{Ripperda2020, Ripperda2022} où la reconnexion a lieu dans le plan équatorial, mais où les produits de la reconnexion magnétique, les plasmoïdes, sont éjectés dans la magnétosphère.

Un autre résultat important de \cite{El_Mellah2022} est que la reconnexion a lieu au point-Y (dans la magnétosphère) de la dernière ligne de champ magnétique fermée appelée séparatrice. La vitesse orbitale de la séparatrice est la vitesse Képlérienne du flot d'accrétion au niveau de son point d'ancrage dans le disque, qui se situe à un rayon plus faible que le point-Y. Ainsi, la vitesse de la séparatrice au niveau du point-Y et, par extension, des plasmoïdes, est super-Képlérienne et dépend, entre autres, du spin du trou noir.

De plus, le mouvement hors du plan équatorial offre la possibilité d'observer l'effet de lentille gravitationnelle lorsque le point chaud est derrière le trou noir par rapport à l'observateur. Cette configuration est uniquement possible lorsque l'inclinaison est proche de 90° pour une orbite dans le plan équatorial, ce qui n'est pas le cas pour Sgr~A* où l'inclinaison est de $\sim 20\degree$. Cet effet de lentille peut se traduire par un second pic dans la courbe de lumière, similaire à celui observé par GRAVITY le 28 juillet 2018. Ainsi, ce modèle, construit à partir des résultats de simulations numériques, est capable d'expliquer les propriétés astrométriques et les courbes de lumière des sursauts détectés par GRAVITY en 2018.\\

GRAVITY a également mesuré la polarisation des sursauts de 2018, qui dépend fortement de la configuration magnétique. Cependant, dans sa version actuelle, le modèle de point chaud basé sur la reconnexion magnétique utilisant \texttt{EMBLEM} suppose un champ magnétique isotrope pour le calcul des coefficients synchrotron. Par conséquent, ce modèle n'est pas en mesure de modéliser la polarisation des sursauts.

En outre, l'ajout de la polarisation dans \textsc{Gyoto}, comme présenté au Chap.\ref{chap:Polarization}, est une récente évolution. En effet, le problème de la polarisation en espace-temps courbe est significativement plus complexe qu'en espace-temps plat. Lors du tracé de rayon en sens inverse du temps réalisé par \textsc{Gyoto}, il est nécessaire de définir la base de polarisation de l'observateur. Cette base doit ensuite être transportée (parallèlement) jusqu'à l'émetteur le long de la trajectoire du photon, où elle sera projetée dans le référentiel de l'émetteur. Cette opération n'est pas aussi évidente qu'il n'y paraît (voir Chap.\ref{chap:Polarization}). Une fois dans le référentiel de l'émetteur, et en connaissant la valeur des coefficients dans la base naturelle du rayonnement synchrotron (qui est différente de la base de l'observateur transportée parallèlement), on peut calculer la valeur des paramètres de Stokes qui définissent la polarisation dans la base de l'observateur en résolvant l'équation du transfert radiatif polarisé.

Une fois implémenté, le code \textsc{Gyoto} a été testé en comparant ses résultats à ceux obtenus à partir d'un autre code de tracé de rayon polarisé : \textsc{ipole}. Les résultats montrent un excellent accord entre les deux codes dans des conditions identiques.\\

Ainsi, bien que le modèle de point chaud issu de la reconnexion magnétique soit déjà très prometteur, son évolution naturelle et évidente est l'ajout de la polarisation en prenant en compte une ou plusieurs configurations magnétiques particulières. En effet, la polarisation est une caractéristique importante des sursauts de Sgr~A*, et son inclusion dans le modèle permettrait de mieux correspondre aux observations.

De plus, les sursauts de Sgr~A* sont également observés à d'autres longueurs d'onde, tels que les rayons~X et les ondes radio, présentant des propriétés particulières (voir Chap.\ref{chap:Sgr~A* flares}). Par conséquent, la capacité du modèle à expliquer à la fois les données astrométriques en NIR, la polarisation en NIR et en radio, ainsi que les propriétés temporelles des courbes de lumière à toutes les longueurs d'onde, est vitale pour confirmer le scénario de la reconnexion magnétique comme origine des sursauts de Sgr~A*.

\newpage
\thispagestyle{plain}
\mbox{}
\newpage

@ARTICLE{Genzel2010,
       author = {{Genzel}, Reinhard and {Eisenhauer}, Frank and {Gillessen}, Stefan},
        title = "{The Galactic Center massive black hole and nuclear star cluster}",
      journal = {Reviews of Modern Physics},
     keywords = {98.35.Jk, Galactic center bar circumnuclear matter and bulge, Astrophysics - Astrophysics of Galaxies},
         year = 2010,
        month = oct,
       volume = {82},
       number = {4},
        pages = {3121-3195},
          doi = {10.1103/RevModPhys.82.3121},
archivePrefix = {arXiv},
       eprint = {1006.0064},
 primaryClass = {astro-ph.GA},
       adsurl = {https://ui.adsabs.harvard.edu/abs/2010RvMP...82.3121G},
      adsnote = {Provided by the SAO/NASA Astrophysics Data System}
}

@ARTICLE{Pawsey1955,
       author = {{Pawsey}, J.~L.},
        title = "{A Catalogue of Reliably Known Discrete Sources of Cosmic Radio Waves.}",
      journal = {\apj},
         year = 1955,
        month = jan,
       volume = {121},
        pages = {1},
          doi = {10.1086/145957},
       adsurl = {https://ui.adsabs.harvard.edu/abs/1955ApJ...121....1P},
      adsnote = {Provided by the SAO/NASA Astrophysics Data System}
}

@article{Murchikova2019,
author = {Murchikova, Elena and Phinney, E. and Pancoast, Anna and Blandford, Roger},
year = {2019},
month = {06},
pages = {83-86},
title = {A cool accretion disk around the Galactic Centre black hole},
volume = {570},
journal = {Nature},
doi = {10.1038/s41586-019-1242-z}
}

@INPROCEEDINGS{Kassim1999,
       author = {{Kassim}, N.~E. and {Larosa}, T.~N. and {Lazio}, T.~J.~W. and {Hyman}, S.~D.},
        title = "{Wide Field Radio Imaging of the Galactic Center}",
    booktitle = {The Central Parsecs of the Galaxy},
         year = 1999,
       editor = {{Falcke}, Heino and {Cotera}, Angela and {Duschl}, Wolfgang J. and {Melia}, Fulvio and {Rieke}, Marcia J.},
       series = {Astronomical Society of the Pacific Conference Series},
       volume = {186},
        month = jun,
        pages = {403},
       adsurl = {https://ui.adsabs.harvard.edu/abs/1999ASPC..186..403K},
      adsnote = {Provided by the SAO/NASA Astrophysics Data System}
}

@ARTICLE{Yusef-Zadeh1986,
       author = {{Yusef-Zadeh}, F. and {Morris}, Mark and {Slee}, O.~B. and {Nelson}, G.~J.},
        title = "{Nonthermal Radio Emission from the Galactic Center Arc}",
      journal = {\apj},
     keywords = {Galactic Nuclei, Milky Way Galaxy, Nonthermal Radiation, Radio Sources (Astronomy), Nebulae, Optical Thickness, Polarization Characteristics, Astrophysics, GALAXIES: NUCLEI, NEBULAE: GENERAL, RADIATION MECHANISMS, RADIO SOURCES: GENERAL},
         year = 1986,
        month = nov,
       volume = {310},
        pages = {689},
          doi = {10.1086/164719},
       adsurl = {https://ui.adsabs.harvard.edu/abs/1986ApJ...310..689Y},
      adsnote = {Provided by the SAO/NASA Astrophysics Data System}
}

@ARTICLE{Roberts1993,
       author = {{Roberts}, D.~A. and {Goss}, W.~M.},
        title = "{Multiconfiguration VLA H92 alpha Observations of Sagittarius A West at 1 Arcsecond Resolution}",
      journal = {\apjs},
     keywords = {Galactic Nuclei, Gas Dynamics, H Ii Regions, Milky Way Galaxy, Velocity Distribution, H Alpha Line, Sagittarius Constellation, Temperature Distribution, Very Large Array (Vla), Astrophysics, GALAXY: CENTER, ISM: INDIVIDUAL NAME: SAGITTARIUS A, ISM: MAGNETIC FIELDS, RADIO LINES: GENERAL},
         year = 1993,
        month = may,
       volume = {86},
        pages = {133},
          doi = {10.1086/191773},
       adsurl = {https://ui.adsabs.harvard.edu/abs/1993ApJS...86..133R},
      adsnote = {Provided by the SAO/NASA Astrophysics Data System}
}

@ARTICLE{Christopher2005,
       author = {{Christopher}, M.~H. and {Scoville}, N.~Z. and {Stolovy}, S.~R. and {Yun}, Min S.},
        title = "{HCN and HCO$^{+}$ Observations of the Galactic Circumnuclear Disk}",
      journal = {\apj},
     keywords = {Galaxy: Center, ISM: Kinematics and Dynamics, ISM: Molecules, Radio Continuum: ISM, Radio Lines: ISM, Stars: Formation, Astrophysics},
         year = 2005,
        month = mar,
       volume = {622},
       number = {1},
        pages = {346-365},
          doi = {10.1086/427911},
archivePrefix = {arXiv},
       eprint = {astro-ph/0502532},
 primaryClass = {astro-ph},
       adsurl = {https://ui.adsabs.harvard.edu/abs/2005ApJ...622..346C},
      adsnote = {Provided by the SAO/NASA Astrophysics Data System}
}

@INPROCEEDINGS{Schodel2007,
       author = {{Schodel}, Rainer and {Eckart}, Andreas},
        title = "{The structure of the nuclear stellar cluster of the Milky Way}",
    booktitle = {Black Holes from Stars to Galaxies -- Across the Range of Masses},
         year = 2007,
       editor = {{Karas}, Vladimir and {Matt}, Giorgio},
       volume = {238},
        month = apr,
        pages = {187-190},
          doi = {10.1017/S1743921307004942},
       adsurl = {https://ui.adsabs.harvard.edu/abs/2007IAUS..238..187S},
      adsnote = {Provided by the SAO/NASA Astrophysics Data System}
}

@ARTICLE{Baganoff2001,
       author = {{Baganoff}, F.~K. and {Bautz}, M.~W. and {Brandt}, W.~N. and {Chartas}, G. and {Feigelson}, E.~D. and {Garmire}, G.~P. and {Maeda}, Y. and {Morris}, M. and {Ricker}, G.~R. and {Townsley}, L.~K. and {Walter}, F.},
        title = "{Rapid X-ray flaring from the direction of the supermassive black hole at the Galactic Centre}",
      journal = {\nat},
     keywords = {Astrophysics},
         year = 2001,
        month = sep,
       volume = {413},
       number = {6851},
        pages = {45-48},
          doi = {10.1038/35092510},
archivePrefix = {arXiv},
       eprint = {astro-ph/0109367},
 primaryClass = {astro-ph},
       adsurl = {https://ui.adsabs.harvard.edu/abs/2001Natur.413...45B},
      adsnote = {Provided by the SAO/NASA Astrophysics Data System}
}

@ARTICLE{Baganoff2003,
       author = {{Baganoff}, F.~K. and {Maeda}, Y. and {Morris}, M. and {Bautz}, M.~W. and {Brandt}, W.~N. and {Cui}, W. and {Doty}, J.~P. and {Feigelson}, E.~D. and {Garmire}, G.~P. and {Pravdo}, S.~H. and {Ricker}, G.~R. and {Townsley}, L.~K.},
        title = "{Chandra X-Ray Spectroscopic Imaging of Sagittarius A* and the Central Parsec of the Galaxy}",
      journal = {\apj},
     keywords = {Accretion, Accretion Disks, Black Hole Physics, Galaxies: Active, Galaxy: Center, X-Rays: ISM, X-Rays: Stars, Astrophysics},
         year = 2003,
        month = jul,
       volume = {591},
       number = {2},
        pages = {891-915},
          doi = {10.1086/375145},
archivePrefix = {arXiv},
       eprint = {astro-ph/0102151},
 primaryClass = {astro-ph},
       adsurl = {https://ui.adsabs.harvard.edu/abs/2003ApJ...591..891B},
      adsnote = {Provided by the SAO/NASA Astrophysics Data System}
}

@ARTICLE{Muno2004,
       author = {{Muno}, M.~P. and {Baganoff}, F.~K. and {Bautz}, M.~W. and {Feigelson}, E.~D. and {Garmire}, G.~P. and {Morris}, M.~R. and {Park}, S. and {Ricker}, G.~R. and {Townsley}, L.~K.},
        title = "{Diffuse X-Ray Emission in a Deep Chandra Image of the Galactic Center}",
      journal = {\apj},
     keywords = {Galaxy: Center, X-Rays: ISM, Astrophysics},
         year = 2004,
        month = sep,
       volume = {613},
       number = {1},
        pages = {326-342},
          doi = {10.1086/422865},
archivePrefix = {arXiv},
       eprint = {astro-ph/0402087},
 primaryClass = {astro-ph},
       adsurl = {https://ui.adsabs.harvard.edu/abs/2004ApJ...613..326M},
      adsnote = {Provided by the SAO/NASA Astrophysics Data System}
}

@ARTICLE{Lo1983,
       author = {{Lo}, K.~Y. and {Claussen}, M.~J.},
        title = "{High-resolution observations of ionized gas in central 3 parsecs of the Galaxy: possible evidence for infall}",
      journal = {\nat},
     keywords = {Galactic Nuclei, Gravitational Effects, Interstellar Gas, Ionized Gases, Milky Way Galaxy, Astronomical Maps, Astronomical Models, Gas Streams, High Resolution, Nebulae, Astrophysics},
         year = 1983,
        month = dec,
       volume = {306},
       number = {5944},
        pages = {647-651},
          doi = {10.1038/306647a0},
       adsurl = {https://ui.adsabs.harvard.edu/abs/1983Natur.306..647L},
      adsnote = {Provided by the SAO/NASA Astrophysics Data System}
}

@ARTICLE{Maeda2002,
       author = {{Maeda}, Y. and {Baganoff}, F.~K. and {Feigelson}, E.~D. and {Morris}, M. and {Bautz}, M.~W. and {Brandt}, W.~N. and {Burrows}, D.~N. and {Doty}, J.~P. and {Garmire}, G.~P. and {Pravdo}, S.~H. and {Ricker}, G.~R. and {Townsley}, L.~K.},
        title = "{A Chandra Study of Sagittarius A East: A Supernova Remnant Regulating the Activity of Our Galactic Center?}",
      journal = {\apj},
     keywords = {Galaxy: Center, ISM: Individual: Name: Sagittarius A East, ISM: Supernova Remnants, X-Rays: ISM, Astrophysics},
         year = 2002,
        month = may,
       volume = {570},
       number = {2},
        pages = {671-687},
          doi = {10.1086/339773},
archivePrefix = {arXiv},
       eprint = {astro-ph/0102183},
 primaryClass = {astro-ph},
       adsurl = {https://ui.adsabs.harvard.edu/abs/2002ApJ...570..671M},
      adsnote = {Provided by the SAO/NASA Astrophysics Data System}
}

@ARTICLE{Gillessen2009,
       author = {{Gillessen}, S. and {Eisenhauer}, F. and {Trippe}, S. and {Alexander}, T. and {Genzel}, R. and {Martins}, F. and {Ott}, T.},
        title = "{Monitoring Stellar Orbits Around the Massive Black Hole in the Galactic Center}",
      journal = {\apj},
     keywords = {black hole physics, astrometry, Galaxy: center, infrared: stars, Astrophysics},
         year = 2009,
        month = feb,
       volume = {692},
       number = {2},
        pages = {1075-1109},
          doi = {10.1088/0004-637X/692/2/1075},
archivePrefix = {arXiv},
       eprint = {0810.4674},
 primaryClass = {astro-ph},
       adsurl = {https://ui.adsabs.harvard.edu/abs/2009ApJ...692.1075G},
      adsnote = {Provided by the SAO/NASA Astrophysics Data System}
}

@ARTICLE{Blum2003,
       author = {{Blum}, R.~D. and {Ramirez}, Solange V. and {Sellgren}, K. and {Olsen}, K.},
        title = "{Really Cool Stars and the Star Formation History at the Galactic Center}",
      journal = {\apj},
     keywords = {Galaxy: Center, Stars: AGB and Post-AGB, Stars: Late-Type, Stars: Supergiants, Astrophysics},
         year = 2003,
        month = nov,
       volume = {597},
       number = {1},
        pages = {323-346},
          doi = {10.1086/378380},
archivePrefix = {arXiv},
       eprint = {astro-ph/0307291},
 primaryClass = {astro-ph},
       adsurl = {https://ui.adsabs.harvard.edu/abs/2003ApJ...597..323B},
      adsnote = {Provided by the SAO/NASA Astrophysics Data System}
}

@ARTICLE{Trippe2008,
       author = {{Trippe}, S. and {Gillessen}, S. and {Gerhard}, O.~E. and {Bartko}, H. and {Fritz}, T.~K. and {Maness}, H.~L. and {Eisenhauer}, F. and {Martins}, F. and {Ott}, T. and {Dodds-Eden}, K. and {Genzel}, R.},
        title = "{Kinematics of the old stellar population at the Galactic centre}",
      journal = {\aap},
     keywords = {Galaxy: center, Galaxy: kinematics and dynamics, stars: kinematics, infrared: stars, Astrophysics},
         year = 2008,
        month = dec,
       volume = {492},
       number = {2},
        pages = {419-439},
          doi = {10.1051/0004-6361:200810191},
archivePrefix = {arXiv},
       eprint = {0810.1040},
 primaryClass = {astro-ph},
       adsurl = {https://ui.adsabs.harvard.edu/abs/2008A&A...492..419T},
      adsnote = {Provided by the SAO/NASA Astrophysics Data System}
}

@ARTICLE{Schodel2009,
       author = {{Schodel}, R. and {Merritt}, D. and {Eckart}, A.},
        title = "{The nuclear star cluster of the Milky Way: proper motions and mass}",
      journal = {\aap},
     keywords = {instrumentation: adaptive optics, techniques: high angular resolution, stars: kinematics, Galaxy: center, Galaxy: structure, Astrophysics - Astrophysics of Galaxies},
         year = 2009,
        month = jul,
       volume = {502},
       number = {1},
        pages = {91-111},
          doi = {10.1051/0004-6361/200810922},
archivePrefix = {arXiv},
       eprint = {0902.3892},
 primaryClass = {astro-ph.GA},
       adsurl = {https://ui.adsabs.harvard.edu/abs/2009A&A...502...91S},
      adsnote = {Provided by the SAO/NASA Astrophysics Data System}
}

@ARTICLE{Genzel2000,
       author = {{Genzel}, R. and {Pichon}, C. and {Eckart}, A. and {Gerhard}, O.~E. and {Ott}, T.},
        title = "{Stellar dynamics in the Galactic Centre: proper motions and anisotropy}",
      journal = {\mnras},
     keywords = {CELESTIAL MECHANICS, STELLAR DYNAMICS, STARS: KINEMATICS, GALAXY: CENTRE, GALAXY: KINEMATICS AND DYNAMICS, Astrophysics},
         year = 2000,
        month = sep,
       volume = {317},
       number = {2},
        pages = {348-374},
          doi = {10.1046/j.1365-8711.2000.03582.x},
archivePrefix = {arXiv},
       eprint = {astro-ph/0001428},
 primaryClass = {astro-ph},
       adsurl = {https://ui.adsabs.harvard.edu/abs/2000MNRAS.317..348G},
      adsnote = {Provided by the SAO/NASA Astrophysics Data System}
}

@ARTICLE{Paumard2006,
       author = {{Paumard}, T. and {Genzel}, R. and {Martins}, F. and {Nayakshin}, S. and {Beloborodov}, A.~M. and {Levin}, Y. and {Trippe}, S. and {Eisenhauer}, F. and {Ott}, T. and {Gillessen}, S. and {Abuter}, R. and {Cuadra}, J. and {Alexander}, T. and {Sternberg}, A.},
        title = "{The Two Young Star Disks in the Central Parsec of the Galaxy: Properties, Dynamics, and Formation}",
      journal = {\apj},
     keywords = {Galaxy: Center, Stars: Early-Type, Stars: Formation, Stars: Luminosity Function, Mass Function, Stellar Dynamics, Astrophysics},
         year = 2006,
        month = jun,
       volume = {643},
       number = {2},
        pages = {1011-1035},
          doi = {10.1086/503273},
archivePrefix = {arXiv},
       eprint = {astro-ph/0601268},
 primaryClass = {astro-ph},
       adsurl = {https://ui.adsabs.harvard.edu/abs/2006ApJ...643.1011P},
      adsnote = {Provided by the SAO/NASA Astrophysics Data System}
}

@ARTICLE{Lu2009,
       author = {{Lu}, J.~R. and {Ghez}, A.~M. and {Hornstein}, S.~D. and {Morris}, M.~R. and {Becklin}, E.~E. and {Matthews}, K.},
        title = "{A Disk of Young Stars at the Galactic Center as Determined by Individual Stellar Orbits}",
      journal = {\apj},
     keywords = {black hole physics, Galaxy: center, infrared: stars, techniques: high angular resolution, Astrophysics},
         year = 2009,
        month = jan,
       volume = {690},
       number = {2},
        pages = {1463-1487},
          doi = {10.1088/0004-637X/690/2/1463},
archivePrefix = {arXiv},
       eprint = {0808.3818},
 primaryClass = {astro-ph},
       adsurl = {https://ui.adsabs.harvard.edu/abs/2009ApJ...690.1463L},
      adsnote = {Provided by the SAO/NASA Astrophysics Data System}
}

@ARTICLE{Kocsis2011,
       author = {{Kocsis}, Bence and {Tremaine}, Scott},
        title = "{Resonant relaxation and the warp of the stellar disc in the Galactic Centre}",
      journal = {\mnras},
     keywords = {celestial mechanics, Galaxy: centre, Galaxy: nucleus, Astrophysics - Galaxy Astrophysics},
         year = 2011,
        month = mar,
       volume = {412},
       number = {1},
        pages = {187-207},
          doi = {10.1111/j.1365-2966.2010.17897.x},
archivePrefix = {arXiv},
       eprint = {1006.0001},
 primaryClass = {astro-ph.GA},
       adsurl = {https://ui.adsabs.harvard.edu/abs/2011MNRAS.412..187K},
      adsnote = {Provided by the SAO/NASA Astrophysics Data System}
}

@ARTICLE{Nayakshin2006,
       author = {{Nayakshin}, Sergei and {Dehnen}, Walter and {Cuadra}, Jorge and {Genzel}, Reinhard},
        title = "{Weighing the young stellar discs around Sgr A*}",
      journal = {\mnras},
     keywords = {accretion, accretion discs, stars: formation, Galaxy: centre, galaxies: active, Astrophysics},
         year = 2006,
        month = mar,
       volume = {366},
       number = {4},
        pages = {1410-1414},
          doi = {10.1111/j.1365-2966.2005.09906.x},
archivePrefix = {arXiv},
       eprint = {astro-ph/0511830},
 primaryClass = {astro-ph},
       adsurl = {https://ui.adsabs.harvard.edu/abs/2006MNRAS.366.1410N},
      adsnote = {Provided by the SAO/NASA Astrophysics Data System}
}

@ARTICLE{Ghez2005,
       author = {{Ghez}, A.~M. and {Salim}, S. and {Hornstein}, S.~D. and {Tanner}, A. and {Lu}, J.~R. and {Morris}, M. and {Becklin}, E.~E. and {Duchene}, G.},
        title = "{Stellar Orbits around the Galactic Center Black Hole}",
      journal = {\apj},
     keywords = {Black Hole Physics, Galaxy: Center, Galaxy: Kinematics and Dynamics, Infrared: Stars, Techniques: High Anular Resolution, Astrophysics},
         year = 2005,
        month = feb,
       volume = {620},
       number = {2},
        pages = {744-757},
          doi = {10.1086/427175},
archivePrefix = {arXiv},
       eprint = {astro-ph/0306130},
 primaryClass = {astro-ph},
       adsurl = {https://ui.adsabs.harvard.edu/abs/2005ApJ...620..744G},
      adsnote = {Provided by the SAO/NASA Astrophysics Data System}
}

@ARTICLE{Eisenhauer2005,
       author = {{Eisenhauer}, F. and {Genzel}, R. and {Alexander}, T. and {Abuter}, R. and {Paumard}, T. and {Ott}, T. and {Gilbert}, A. and {Gillessen}, S. and {Horrobin}, M. and {Trippe}, S. and {Bonnet}, H. and {Dumas}, C. and {Hubin}, N. and {Kaufer}, A. and {Kissler-Patig}, M. and {Monnet}, G. and {Strobele}, S. and {Szeifert}, T. and {Eckart}, A. and {Schodel}, R. and {Zucker}, S.},
        title = "{SINFONI in the Galactic Center: Young Stars and Infrared Flares in the Central Light-Month}",
      journal = {\apj},
     keywords = {Black Hole Physics, Galaxy: Center, Galaxy: Structure, Infrared: Stars, Techniques: Spectroscopic, Astrophysics},
         year = 2005,
        month = jul,
       volume = {628},
       number = {1},
        pages = {246-259},
          doi = {10.1086/430667},
archivePrefix = {arXiv},
       eprint = {astro-ph/0502129},
 primaryClass = {astro-ph},
       adsurl = {https://ui.adsabs.harvard.edu/abs/2005ApJ...628..246E},
      adsnote = {Provided by the SAO/NASA Astrophysics Data System}
}

@ARTICLE{Schodel2003,
       author = {{Schodel}, R. and {Genzel}, R. and {Ott}, T. and {Eckart}, A.},
        title = "{The Galactic Center stellar cluster: The central arcsecond}",
      journal = {Astronomische Nachrichten Supplement},
     keywords = {Astrophysics},
         year = 2003,
        month = sep,
       volume = {324},
       number = {1},
        pages = {535-541},
          doi = {10.1002/asna.200385048},
archivePrefix = {arXiv},
       eprint = {astro-ph/0304197},
 primaryClass = {astro-ph},
       adsurl = {https://ui.adsabs.harvard.edu/abs/2003ANS...324..535S},
      adsnote = {Provided by the SAO/NASA Astrophysics Data System}
}

@ARTICLE{GRAVITY2022a,
       author = {{GRAVITY Collaboration} and {Abuter}, R. and {Aimar}, N. and {Amorim}, A. and {Ball}, J. and {Baubock}, M. and {Berger}, J.~P. and {Bonnet}, H. and {Bourdarot}, G. and {Brandner}, W. and {Cardoso}, V. and {Clenet}, Y. and {Dallilar}, Y. and {Davies}, R. and {de Zeeuw}, P.~T. and {Dexter}, J. and {Drescher}, A. and {Eisenhauer}, F. and {Forster Schreiber}, N.~M. and {Foschi}, A. and {Garcia}, P. and {Gao}, F. and {Gendron}, E. and {Genzel}, R. and {Gillessen}, S. and {Habibi}, M. and {Haubois}, X. and {Heissel}, G. and {Henning}, T. and {Hippler}, S. and {Horrobin}, M. and {Jochum}, L. and {Jocou}, L. and {Kaufer}, A. and {Kervella}, P. and {Lacour}, S. and {Lapeyrere}, V. and {Le Bouquin}, J. -B. and {Lena}, P. and {Lutz}, D. and {Ott}, T. and {Paumard}, T. and {Perraut}, K. and {Perrin}, G. and {Pfuhl}, O. and {Rabien}, S. and {Shangguan}, J. and {Shimizu}, T. and {Scheithauer}, S. and {Stadler}, J. and {Stephens}, A.~W. and {Straub}, O. and {Straubmeier}, C. and {Sturm}, E. and {Tacconi}, L.~J. and {Tristram}, K.~R.~W. and {Vincent}, F. and {von Fellenberg}, S. and {Widmann}, F. and {Wieprecht}, E. and {Wiezorrek}, E. and {Woillez}, J. and {Yazici}, S. and {Young}, A.},
        title = "{Mass distribution in the Galactic Center based on interferometric astrometry of multiple stellar orbits}",
      journal = {\aap},
     keywords = {black hole physics, instrumentation: interferometers, Galaxy: center, Astrophysics - Astrophysics of Galaxies, Astrophysics - Instrumentation and Methods for Astrophysics, General Relativity and Quantum Cosmology},
         year = 2022,
        month = jan,
       volume = {657},
          eid = {L12},
        pages = {L12},
          doi = {10.1051/0004-6361/202142465},
archivePrefix = {arXiv},
       eprint = {2112.07478},
 primaryClass = {astro-ph.GA},
       adsurl = {https://ui.adsabs.harvard.edu/abs/2022A&A...657L..12G},
      adsnote = {Provided by the SAO/NASA Astrophysics Data System}
}

@ARTICLE{Gravity2020a,
       author = {{GRAVITY Collaboration} and {Abuter}, R. and {Amorim}, A. and {Baubock}, M. and {Berger}, J.~P. and {Bonnet}, H. and {Brandner}, W. and {Cardoso}, V. and {Clenet}, Y. and {de Zeeuw}, P.~T. and {Dexter}, J. and {Eckart}, A. and {Eisenhauer}, F. and {Forster Schreiber}, N.~M. and {Garcia}, P. and {Gao}, F. and {Gendron}, E. and {Genzel}, R. and {Gillessen}, S. and {Habibi}, M. and {Haubois}, X. and {Henning}, T. and {Hippler}, S. and {Horrobin}, M. and {Jimenez-Rosales}, A. and {Jochum}, L. and {Jocou}, L. and {Kaufer}, A. and {Kervella}, P. and {Lacour}, S. and {Lapeyrere}, V. and {Le Bouquin}, J. -B. and {Lena}, P. and {Nowak}, M. and {Ott}, T. and {Paumard}, T. and {Perraut}, K. and {Perrin}, G. and {Pfuhl}, O. and {Rodriguez-Coira}, G. and {Shangguan}, J. and {Scheithauer}, S. and {Stadler}, J. and {Straub}, O. and {Straubmeier}, C. and {Sturm}, E. and {Tacconi}, L.~J. and {Vincent}, F. and {von Fellenberg}, S. and {Waisberg}, I. and {Widmann}, F. and {Wieprecht}, E. and {Wiezorrek}, E. and {Woillez}, J. and {Yazici}, S. and {Zins}, G.},
        title = "{Detection of the Schwarzschild precession in the orbit of the star S2 near the Galactic centre massive black hole}",
      journal = {\aap},
     keywords = {black hole physics, Galaxy: nucleus, gravitation, relativistic processes, Astrophysics - Astrophysics of Galaxies, Astrophysics - Instrumentation and Methods for Astrophysics, General Relativity and Quantum Cosmology},
         year = 2020,
        month = apr,
       volume = {636},
          eid = {L5},
        pages = {L5},
          doi = {10.1051/0004-6361/202037813},
archivePrefix = {arXiv},
       eprint = {2004.07187},
 primaryClass = {astro-ph.GA},
       adsurl = {https://ui.adsabs.harvard.edu/abs/2020A&A...636L...5G},
      adsnote = {Provided by the SAO/NASA Astrophysics Data System}
}

@ARTICLE{Gravity2020b,
       author = {{GRAVITY Collaboration} and {Abuter}, R. and {Amorim}, A. and {Baubock}, M. and {Berger}, J.~B. and {Bonnet}, H. and {Brandner}, W. and {Cardoso}, V. and {Clenet}, Y. and {de Zeeuw}, P.~T. and {Dallilar}, Y. and {Dexter}, J. and {Eckart}, A. and {Eisenhauer}, F. and {Forster Schreiber}, N.~M. and {Garcia}, P. and {Gao}, F. and {Gendron}, E. and {Genzel}, R. and {Gillessen}, S. and {Habibi}, M. and {Haubois}, X. and {Henning}, T. and {Hippler}, S. and {Horrobin}, M. and {Jimenez-Rosales}, A. and {Jochum}, L. and {Jocou}, L. and {Kaufer}, A. and {Kervella}, P. and {Lacour}, S. and {Lapeyrere}, V. and {Le Bouquin}, J. -B. and {Lena}, P. and {Nowak}, M. and {Ott}, T. and {Paumard}, T. and {Perraut}, K. and {Perrin}, G. and {Pfuhl}, O. and {Ponti}, G. and {Rodriguez Coira}, G. and {Shangguan}, J. and {Scheithauer}, S. and {Stadler}, J. and {Straub}, O. and {Straubmeier}, C. and {Sturm}, E. and {Tacconi}, L.~J. and {Vincent}, F. and {von Fellenberg}, S.~D. and {Waisberg}, I. and {Widmann}, F. and {Wieprecht}, E. and {Wiezorrek}, E. and {Woillez}, J. and {Yazici}, S. and {Zins}, G.},
        title = "{The flux distribution of Sgr A*}",
      journal = {\aap},
     keywords = {Galaxy: center, black hole physics, accretion, accretion disks, Astrophysics - Astrophysics of Galaxies, Astrophysics - Instrumentation and Methods for Astrophysics, General Relativity and Quantum Cosmology},
         year = 2020,
        month = jun,
       volume = {638},
          eid = {A2},
        pages = {A2},
          doi = {10.1051/0004-6361/202037717},
archivePrefix = {arXiv},
       eprint = {2004.07185},
 primaryClass = {astro-ph.GA},
       adsurl = {https://ui.adsabs.harvard.edu/abs/2020A&A...638A...2G},
      adsnote = {Provided by the SAO/NASA Astrophysics Data System}
}

@ARTICLE{Gravity2020c,
       author = {{GRAVITY Collaboration} and {Baubock}, M. and {Dexter}, J. and {Abuter}, R. and {Amorim}, A. and {Berger}, J.~P. and {Bonnet}, H. and {Brandner}, W. and {Clenet}, Y. and {Coude Du Foresto}, V. and {de Zeeuw}, P.~T. and {Duvert}, G. and {Eckart}, A. and {Eisenhauer}, F. and {Forster Schreiber}, N.~M. and {Gao}, F. and {Garcia}, P. and {Gendron}, E. and {Genzel}, R. and {Gerhard}, O. and {Gillessen}, S. and {Habibi}, M. and {Haubois}, X. and {Henning}, T. and {Hippler}, S. and {Horrobin}, M. and {Jimenez-Rosales}, A. and {Jocou}, L. and {Kervella}, P. and {Lacour}, S. and {Lapeyrere}, V. and {Le Bouquin}, J. -B. and {Lena}, P. and {Ott}, T. and {Paumard}, T. and {Perraut}, K. and {Perrin}, G. and {Pfuhl}, O. and {Rabien}, S. and {Rodriguez Coira}, G. and {Rousset}, G. and {Scheithauer}, S. and {Stadler}, J. and {Sternberg}, A. and {Straub}, O. and {Straubmeier}, C. and {Sturm}, E. and {Tacconi}, L.~J. and {Vincent}, F. and {von Fellenberg}, S. and {Waisberg}, I. and {Widmann}, F. and {Wieprecht}, E. and {Wiezorrek}, E. and {Woillez}, J. and {Yazici}, S.},
        title = "{Modeling the orbital motion of Sgr A*'s near-infrared flares}",
      journal = {\aap},
     keywords = {black hole physics, Galaxy: center, accretion, accretion disks, Astrophysics - High Energy Astrophysical Phenomena},
         year = 2020,
        month = mar,
       volume = {635},
          eid = {A143},
        pages = {A143},
          doi = {10.1051/0004-6361/201937233},
archivePrefix = {arXiv},
       eprint = {2002.08374},
 primaryClass = {astro-ph.HE},
       adsurl = {https://ui.adsabs.harvard.edu/abs/2020A&A...635A.143G},
      adsnote = {Provided by the SAO/NASA Astrophysics Data System}
}

@ARTICLE{Balick1974,
       author = {{Balick}, B. and {Brown}, R.~L.},
        title = "{Intense sub-arcsecond structure in the galactic center.}",
      journal = {\apj},
     keywords = {Fine Structure, Galactic Nuclei, Galactic Radio Waves, Microwave Emission, Radio Sources (Astronomy), Brightness Temperature, Galactic Structure, Graphs (Charts), Ionized Gases, Microwave Interferometers, Milky Way Galaxy, Radiant Flux Density, Astrophysics},
         year = 1974,
        month = dec,
       volume = {194},
        pages = {265-270},
          doi = {10.1086/153242},
       adsurl = {https://ui.adsabs.harvard.edu/abs/1974ApJ...194..265B},
      adsnote = {Provided by the SAO/NASA Astrophysics Data System}
}

@ARTICLE{Lynden1971,
       author = {{Lynden-Bell}, D. and {Rees}, M.~J.},
        title = "{On quasars, dust and the galactic centre}",
      journal = {\mnras},
         year = 1971,
        month = jan,
       volume = {152},
        pages = {461},
          doi = {10.1093/mnras/152.4.461},
       adsurl = {https://ui.adsabs.harvard.edu/abs/1971MNRAS.152..461L},
      adsnote = {Provided by the SAO/NASA Astrophysics Data System}
}

@ARTICLE{Wu2015,
       author = {{Wu}, Xue-Bing and {Wang}, Feige and {Fan}, Xiaohui and {Yi}, Weimin and {Zuo}, Wenwen and {Bian}, Fuyan and {Jiang}, Linhua and {McGreer}, Ian D. and {Wang}, Ran and {Yang}, Jinyi and {Yang}, Qian and {Thompson}, David and {Beletsky}, Yuri},
        title = "{An ultraluminous quasar with a twelve-billion-solar-mass black hole at redshift 6.30}",
      journal = {\nat},
     keywords = {Astrophysics - Astrophysics of Galaxies},
         year = 2015,
        month = feb,
       volume = {518},
       number = {7540},
        pages = {512-515},
          doi = {10.1038/nature14241},
archivePrefix = {arXiv},
       eprint = {1502.07418},
 primaryClass = {astro-ph.GA},
       adsurl = {https://ui.adsabs.harvard.edu/abs/2015Natur.518..512W},
      adsnote = {Provided by the SAO/NASA Astrophysics Data System}
}

@ARTICLE{Wollman1977,
       author = {{Wollman}, E.~R. and {Geballe}, T.~R. and {Lacy}, J.~H. and {Townes}, C.~H. and {Rank}, D.~M.},
        title = "{Ne II 12.8 micron emission from the galactic center. II.}",
      journal = {\apjl},
     keywords = {Emission Spectra, Galactic Nuclei, Infrared Spectra, Milky Way Galaxy, Neon, Astronomical Spectroscopy, Far Infrared Radiation, Galactic Structure, Red Shift, Stellar Spectra, Astrophysics},
         year = 1977,
        month = dec,
       volume = {218},
        pages = {L103-L107},
          doi = {10.1086/182585},
       adsurl = {https://ui.adsabs.harvard.edu/abs/1977ApJ...218L.103W},
      adsnote = {Provided by the SAO/NASA Astrophysics Data System}
}

@ARTICLE{Shen2005,
       author = {{Shen}, Zhi-Qiang and {Lo}, K.~Y. and {Liang}, M. -C. and {Ho}, Paul T.~P. and {Zhao}, J. -H.},
        title = "{A size of \raisebox{-0.5ex}\textasciitilde1AU for the radio source Sgr A* at the centre of the Milky Way}",
      journal = {\nat},
     keywords = {Astrophysics},
         year = 2005,
        month = nov,
       volume = {438},
       number = {7064},
        pages = {62-64},
          doi = {10.1038/nature04205},
archivePrefix = {arXiv},
       eprint = {astro-ph/0512515},
 primaryClass = {astro-ph},
       adsurl = {https://ui.adsabs.harvard.edu/abs/2005Natur.438...62S},
      adsnote = {Provided by the SAO/NASA Astrophysics Data System}
}

@ARTICLE{Witzel2018,
       author = {{Witzel}, G. and {Martinez}, G. and {Hora}, J. and {Willner}, S.~P. and {Morris}, M.~R. and {Gammie}, C. and {Becklin}, E.~E. and {Ashby}, M.~L.~N. and {Baganoff}, F. and {Carey}, S. and {Do}, T. and {Fazio}, G.~G. and {Ghez}, A. and {Glaccum}, W.~J. and {Haggard}, D. and {Herrero-Illana}, R. and {Ingalls}, J. and {Narayan}, R. and {Smith}, H.~A.},
        title = "{Variability Timescale and Spectral Index of Sgr A* in the Near Infrared: Approximate Bayesian Computation Analysis of the Variability of the Closest Supermassive Black Hole}",
      journal = {\apj},
     keywords = {accretion, accretion disks, black hole physics, Galaxy: center, methods: statistical, radiation mechanisms: non-thermal, Astrophysics - High Energy Astrophysical Phenomena},
         year = 2018,
        month = aug,
       volume = {863},
       number = {1},
          eid = {15},
        pages = {15},
          doi = {10.3847/1538-4357/aace62},
archivePrefix = {arXiv},
       eprint = {1806.00479},
 primaryClass = {astro-ph.HE},
       adsurl = {https://ui.adsabs.harvard.edu/abs/2018ApJ...863...15W},
      adsnote = {Provided by the SAO/NASA Astrophysics Data System}
}

@ARTICLE{Do2019,
       author = {{Do}, Tuan and {Witzel}, Gunther and {Gautam}, Abhimat K. and {Chen}, Zhuo and {Ghez}, Andrea M. and {Morris}, Mark R. and {Becklin}, Eric E. and {Ciurlo}, Anna and {Hosek}, Matthew, Jr. and {Martinez}, Gregory D. and {Matthews}, Keith and {Sakai}, Shoko and {Schodel}, Rainer},
        title = "{Unprecedented Near-infrared Brightness and Variability of Sgr A*}",
      journal = {\apjl},
     keywords = {Supermassive black holes, Low-luminosity active galactic nuclei, Near infrared astronomy, Galactic center, 1663, 2033, 1093, 565, Astrophysics - Astrophysics of Galaxies, Astrophysics - High Energy Astrophysical Phenomena},
         year = 2019,
        month = sep,
       volume = {882},
       number = {2},
          eid = {L27},
        pages = {L27},
          doi = {10.3847/2041-8213/ab38c3},
archivePrefix = {arXiv},
       eprint = {1908.01777},
 primaryClass = {astro-ph.GA},
       adsurl = {https://ui.adsabs.harvard.edu/abs/2019ApJ...882L..27D},
      adsnote = {Provided by the SAO/NASA Astrophysics Data System}
}

@ARTICLE{Carter1968,
       author = {{Carter}, Brandon},
        title = "{Global Structure of the Kerr Family of Gravitational Fields}",
      journal = {Physical Review},
         year = 1968,
        month = oct,
       volume = {174},
       number = {5},
        pages = {1559-1571},
          doi = {10.1103/PhysRev.174.1559},
       adsurl = {https://ui.adsabs.harvard.edu/abs/1968PhRv..174.1559C},
      adsnote = {Provided by the SAO/NASA Astrophysics Data System}
}

@ARTICLE{Levin&Perez-Giz2008,
       author = {{Levin}, Janna and {Perez-Giz}, Gabe},
        title = "{A periodic table for black hole orbits}",
      journal = {\prd},
     keywords = {97.60.Lf, 04.70.-s, 95.30.Sf, Black holes, Physics of black holes, Relativity and gravitation, General Relativity and Quantum Cosmology, Astrophysics},
         year = 2008,
        month = may,
       volume = {77},
       number = {10},
          eid = {103005},
        pages = {103005},
          doi = {10.1103/PhysRevD.77.103005},
archivePrefix = {arXiv},
       eprint = {0802.0459},
 primaryClass = {gr-qc},
       adsurl = {https://ui.adsabs.harvard.edu/abs/2008PhRvD..77j3005L},
      adsnote = {Provided by the SAO/NASA Astrophysics Data System}
}

@BOOK{Mihalas&Mihalas1984,
       author = {{Mihalas}, Dimitri and {Weibel Mihalas}, Barbara},
        title = "{Foundations of radiation hydrodynamics}",
         year = 1984,
       adsurl = {https://ui.adsabs.harvard.edu/abs/1984frh..book.....M},
      adsnote = {Provided by the SAO/NASA Astrophysics Data System}
}

@ARTICLE{Do2009,
       author = {{Do}, T. and {Ghez}, A.~M. and {Morris}, M.~R. and {Yelda}, S. and {Meyer}, L. and {Lu}, J.~R. and {Hornstein}, S.~D. and {Matthews}, K.},
        title = "{A Near-Infrared Variability Study of the Galactic Black Hole: A Red Noise Source with NO Detected Periodicity}",
      journal = {\apj},
     keywords = {black hole physics, Galaxy: center, techniques: high angular resolution, Astrophysics},
         year = 2009,
        month = feb,
       volume = {691},
       number = {2},
        pages = {1021-1034},
          doi = {10.1088/0004-637X/691/2/1021},
archivePrefix = {arXiv},
       eprint = {0810.0446},
 primaryClass = {astro-ph},
       adsurl = {https://ui.adsabs.harvard.edu/abs/2009ApJ...691.1021D},
      adsnote = {Provided by the SAO/NASA Astrophysics Data System}
}

@ARTICLE{Gravity2018,
       author = {{Gravity Collaboration} and {Abuter}, R. and {Amorim}, A. and {Baubock}, M. and {Berger}, J.~P. and {Bonnet}, H. and {Brandner}, W. and {Clenet}, Y. and {Coude Du Foresto}, V. and {de Zeeuw}, P.~T. and {Deen}, C. and {Dexter}, J. and {Duvert}, G. and {Eckart}, A. and {Eisenhauer}, F. and {Forster Schreiber}, N.~M. and {Garcia}, P. and {Gao}, F. and {Gendron}, E. and {Genzel}, R. and {Gillessen}, S. and {Guajardo}, P. and {Habibi}, M. and {Haubois}, X. and {Henning}, Th. and {Hippler}, S. and {Horrobin}, M. and {Huber}, A. and {Jimenez-Rosales}, A. and {Jocou}, L. and {Kervella}, P. and {Lacour}, S. and {Lapeyrere}, V. and {Lazareff}, B. and {Le Bouquin}, J. -B. and {Lena}, P. and {Lippa}, M. and {Ott}, T. and {Panduro}, J. and {Paumard}, T. and {Perraut}, K. and {Perrin}, G. and {Pfuhl}, O. and {Plewa}, P.~M. and {Rabien}, S. and {Rodriguez-Coira}, G. and {Rousset}, G. and {Sternberg}, A. and {Straub}, O. and {Straubmeier}, C. and {Sturm}, E. and {Tacconi}, L.~J. and {Vincent}, F. and {von Fellenberg}, S. and {Waisberg}, I. and {Widmann}, F. and {Wieprecht}, E. and {Wiezorrek}, E. and {Woillez}, J. and {Yazici}, S.},
        title = "{Detection of orbital motions near the last stable circular orbit of the massive black hole SgrA*}",
      journal = {\aap},
     keywords = {Galaxy: center, black hole physics, gravitation, relativistic processes, Astrophysics - Astrophysics of Galaxies},
         year = 2018,
        month = oct,
       volume = {618},
          eid = {L10},
        pages = {L10},
          doi = {10.1051/0004-6361/201834294},
archivePrefix = {arXiv},
       eprint = {1810.12641},
 primaryClass = {astro-ph.GA},
       adsurl = {https://ui.adsabs.harvard.edu/abs/2018A&A...618L..10G},
      adsnote = {Provided by the SAO/NASA Astrophysics Data System}
}

@ARTICLE{Genzel2003,
       author = {{Genzel}, R. and {Schodel}, R. and {Ott}, T. and {Eckart}, A. and {Alexander}, T. and {Lacombe}, F. and {Rouan}, D. and {Aschenbach}, B.},
        title = "{Near-infrared flares from accreting gas around the supermassive black hole at the Galactic Centre}",
      journal = {\nat},
     keywords = {Astrophysics},
         year = 2003,
        month = oct,
       volume = {425},
       number = {6961},
        pages = {934-937},
          doi = {10.1038/nature02065},
archivePrefix = {arXiv},
       eprint = {astro-ph/0310821},
 primaryClass = {astro-ph},
       adsurl = {https://ui.adsabs.harvard.edu/abs/2003Natur.425..934G},
      adsnote = {Provided by the SAO/NASA Astrophysics Data System}
}

@ARTICLE{Nayakshin2004,
       author = {{Nayakshin}, S. and {Cuadra}, J. and {Sunyaev}, R.},
        title = "{X-ray flares from Sgr A$^{*}$: Star-disk interactions?}",
      journal = {\aap},
     keywords = {Galaxy: center, X-rays: galaxies, accretion, accretion disks, black hole physics, Astrophysics},
         year = 2004,
        month = jan,
       volume = {413},
        pages = {173-188},
          doi = {10.1051/0004-6361:20031537},
archivePrefix = {arXiv},
       eprint = {astro-ph/0304126},
 primaryClass = {astro-ph},
       adsurl = {https://ui.adsabs.harvard.edu/abs/2004A&A...413..173N},
      adsnote = {Provided by the SAO/NASA Astrophysics Data System}
}

@ARTICLE{Gezari2014,
       author = {{Gezari}, Suvi},
        title = "{The tidal disruption of stars by supermassive black holes}",
      journal = {Physics Today},
         year = 2014,
        month = may,
       volume = {67},
       number = {5},
        pages = {37-42},
          doi = {10.1063/PT.3.2382},
       adsurl = {https://ui.adsabs.harvard.edu/abs/2014PhT....67e..37G},
      adsnote = {Provided by the SAO/NASA Astrophysics Data System}
}

@ARTICLE{Rees1988,
       author = {{Rees}, Martin J.},
        title = "{Tidal disruption of stars by black holes of {}10$^{6}$-{}10$^{8}$ solar masses in nearby galaxies}",
      journal = {\nat},
     keywords = {Active Galactic Nuclei, Black Holes (Astronomy), Stellar Mass, Quasars, Red Shift, Star Distribution, Stellar Flares, Tides, Astrophysics},
         year = 1988,
        month = jun,
       volume = {333},
       number = {6173},
        pages = {523-528},
          doi = {10.1038/333523a0},
       adsurl = {https://ui.adsabs.harvard.edu/abs/1988Natur.333..523R},
      adsnote = {Provided by the SAO/NASA Astrophysics Data System}
}

@ARTICLE{Lovelace2014,
       author = {{Lovelace}, R.~V.~E. and {Romanova}, M.~M.},
        title = "{Rossby wave instability in astrophysical discs}",
      journal = {Fluid Dynamics Research},
     keywords = {Astrophysics - Solar and Stellar Astrophysics},
         year = 2014,
        month = aug,
       volume = {46},
       number = {4},
          eid = {041401},
        pages = {041401},
          doi = {10.1088/0169-5983/46/4/041401},
archivePrefix = {arXiv},
       eprint = {1312.4572},
 primaryClass = {astro-ph.SR},
       adsurl = {https://ui.adsabs.harvard.edu/abs/2014FlDyR..46d1401L},
      adsnote = {Provided by the SAO/NASA Astrophysics Data System}
}

@ARTICLE{Hamaus2009,
       author = {{Hamaus}, N. and {Paumard}, T. and {Muller}, T. and {Gillessen}, S. and {Eisenhauer}, F. and {Trippe}, S. and {Genzel}, R.},
        title = "{Prospects for Testing the Nature of Sgr A*'s Near-Infrared Flares on the Basis of Current Very Large Telescope{\textemdash}and Future Very Large Telescope Interferometer{\textemdash}Observations}",
      journal = {\apj},
     keywords = {astrometry, black hole physics, Galaxy: center, gravitational lensing, techniques: interferometric, Astrophysics},
         year = 2009,
        month = feb,
       volume = {692},
       number = {1},
        pages = {902-916},
          doi = {10.1088/0004-637X/692/1/902},
archivePrefix = {arXiv},
       eprint = {0810.4947},
 primaryClass = {astro-ph},
       adsurl = {https://ui.adsabs.harvard.edu/abs/2009ApJ...692..902H},
      adsnote = {Provided by the SAO/NASA Astrophysics Data System}
}

@ARTICLE{Aimar2023,
       author = {{Aimar}, N. and {Dmytriiev}, A. and {Vincent}, F.~H. and {El Mellah}, I. and {Paumard}, T. and {Perrin}, G. and {Zech}, A.},
        title = "{Magnetic reconnection plasmoid model for Sagittarius A* flares}",
      journal = {arXiv e-prints},
     keywords = {Astrophysics - High Energy Astrophysical Phenomena},
         year = 2023,
        month = jan,
          eid = {arXiv:2301.11874},
        pages = {arXiv:2301.11874},
          doi = {10.48550/arXiv.2301.11874},
archivePrefix = {arXiv},
       eprint = {2301.11874},
 primaryClass = {astro-ph.HE},
       adsurl = {https://ui.adsabs.harvard.edu/abs/2023arXiv230111874A},
      adsnote = {Provided by the SAO/NASA Astrophysics Data System}
}

@BOOK{RL86,
       author = {{Rybicki}, George B. and {Lightman}, Alan P.},
        title = "{Radiative Processes in Astrophysics}",
         year = 1986,
       adsurl = {https://ui.adsabs.harvard.edu/abs/1986rpa..book.....R},
      adsnote = {Provided by the SAO/NASA Astrophysics Data System}
}

@ARTICLE{Ball2021,
       author = {{Ball}, David and {Ozel}, Feryal and {Christian}, Pierre and {Chan}, Chi-Kwan and {Psaltis}, Dimitrios},
        title = "{A Plasmoid model for the Sgr A* Flares Observed With Gravity and CHANDRA}",
      journal = {\apj},
     keywords = {Black Hole physics, High energy astrophysics, 159, 739, Astrophysics - High Energy Astrophysical Phenomena},
         year = 2021,
        month = aug,
       volume = {917},
       number = {1},
          eid = {8},
        pages = {8},
          doi = {10.3847/1538-4357/abf8ae},
archivePrefix = {arXiv},
       eprint = {2005.14251},
 primaryClass = {astro-ph.HE},
       adsurl = {https://ui.adsabs.harvard.edu/abs/2021ApJ...917....8B},
      adsnote = {Provided by the SAO/NASA Astrophysics Data System}
}

@ARTICLE{Vincent2014,
       author = {{Vincent}, F.~H. and {Paumard}, T. and {Perrin}, G. and {Varniere}, P. and {Casse}, F. and {Eisenhauer}, F. and {Gillessen}, S. and {Armitage}, P.~J.},
        title = "{Distinguishing an ejected blob from alternative flare models at the Galactic Centre with GRAVITY}",
      journal = {\mnras},
     keywords = {black hole physics, instrumentation: interferometers, astrometry, Galaxy: centre, Astrophysics - High Energy Astrophysical Phenomena},
         year = 2014,
        month = jul,
       volume = {441},
       number = {4},
        pages = {3477-3487},
          doi = {10.1093/mnras/stu812},
archivePrefix = {arXiv},
       eprint = {1404.6149},
 primaryClass = {astro-ph.HE},
       adsurl = {https://ui.adsabs.harvard.edu/abs/2014MNRAS.441.3477V},
      adsnote = {Provided by the SAO/NASA Astrophysics Data System}
}

@ARTICLE{Johnson1997,
       author = {{Johnson}, Peter D.},
        title = "{2. Synchrotron Radiation}",
      journal = {Atomic},
         year = 1997,
        month = jan,
       volume = {29},
        pages = {23-43},
          doi = {10.1016/S0076-695X(08)60611-0},
       adsurl = {https://ui.adsabs.harvard.edu/abs/1997ExMPS..29...23J},
      adsnote = {Provided by the SAO/NASA Astrophysics Data System}
}

@ARTICLE{Bower2019,
       author = {{Bower}, Geoffrey C. and {Dexter}, Jason and {Asada}, Keiichi and {Brinkerink}, Christiaan D. and {Falcke}, Heino and {Ho}, Paul and {Inoue}, Makoto and {Markoff}, Sera and {Marrone}, Daniel P. and {Matsushita}, Satoki and {Moscibrodzka}, Monika and {Nakamura}, Masanori and {Peck}, Alison and {Rao}, Ramprasad},
        title = "{ALMA Observations of the Terahertz Spectrum of Sagittarius A*}",
      journal = {\apjl},
     keywords = {accretion, accretion disks, black hole physics, galaxies: nuclei, Galaxy: center, Astrophysics - High Energy Astrophysical Phenomena},
         year = 2019,
        month = aug,
       volume = {881},
       number = {1},
          eid = {L2},
        pages = {L2},
          doi = {10.3847/2041-8213/ab3397},
archivePrefix = {arXiv},
       eprint = {1907.08319},
 primaryClass = {astro-ph.HE},
       adsurl = {https://ui.adsabs.harvard.edu/abs/2019ApJ...881L...2B},
      adsnote = {Provided by the SAO/NASA Astrophysics Data System}
}

@ARTICLE{Penrose1969,
       author = {{Penrose}, Roger},
        title = "{Gravitational Collapse: the Role of General Relativity}",
      journal = {Nuovo Cimento Rivista Serie},
         year = 1969,
        month = jan,
       volume = {1},
        pages = {252},
       adsurl = {https://ui.adsabs.harvard.edu/abs/1969NCimR...1..252P},
      adsnote = {Provided by the SAO/NASA Astrophysics Data System}
}

@ARTICLE{Blandford-Znajek1977,
       author = {{Blandford}, R.~D. and {Znajek}, R.~L.},
        title = "{Electromagnetic extraction of energy from Kerr black holes.}",
      journal = {\mnras},
     keywords = {Black Holes (Astronomy), Electromagnetic Fields, Energy Sources, Rotating Matter, Active Galactic Nuclei, Astrophysics, Electron-Positron Pairs, Pair Production, Astrophysics},
         year = 1977,
        month = may,
       volume = {179},
        pages = {433-456},
          doi = {10.1093/mnras/179.3.433},
       adsurl = {https://ui.adsabs.harvard.edu/abs/1977MNRAS.179..433B},
      adsnote = {Provided by the SAO/NASA Astrophysics Data System}
}

@ARTICLE{Narayan1995,
       author = {{Narayan}, Ramesh and {Yi}, Insu and {Mahadevan}, Rohan},
        title = "{Explaining the spectrum of Sagittarius A$^{*}$ with a model of an accreting black hole}",
      journal = {\nat},
         year = 1995,
        month = apr,
       volume = {374},
       number = {6523},
        pages = {623-625},
          doi = {10.1038/374623a0},
       adsurl = {https://ui.adsabs.harvard.edu/abs/1995Natur.374..623N},
      adsnote = {Provided by the SAO/NASA Astrophysics Data System}
}

@ARTICLE{Cheng2011,
       author = {{Cheng}, K. -S. and {Chernyshov}, D.~O. and {Dogiel}, V.~A. and {Ko}, C. -M. and {Ip}, W. -H.},
        title = "{Origin of the Fermi Bubble}",
      journal = {\apjl},
     keywords = {black hole physics, galaxies: jets, Galaxy: halo, radiation mechanisms: non-thermal, Astrophysics - High Energy Astrophysical Phenomena},
         year = 2011,
        month = apr,
       volume = {731},
       number = {1},
          eid = {L17},
        pages = {L17},
          doi = {10.1088/2041-8205/731/1/L17},
archivePrefix = {arXiv},
       eprint = {1103.1002},
 primaryClass = {astro-ph.HE},
       adsurl = {https://ui.adsabs.harvard.edu/abs/2011ApJ...731L..17C},
      adsnote = {Provided by the SAO/NASA Astrophysics Data System}
}

@ARTICLE{Zhu2019,
       author = {{Zhu}, Zhenlin and {Li}, Zhiyuan and {Morris}, Mark R. and {Zhang}, Shuo and {Liu}, Siming},
        title = "{A Deep Chandra View of a Candidate Parsec-scale Jet from the Galactic Center Supermassive Black Hole}",
      journal = {\apj},
     keywords = {black hole physics, Galaxy: center, ISM: jets and outflows, radiation mechanisms: non-thermal, X-rays: individual: G359.944‑0.052, Astrophysics - High Energy Astrophysical Phenomena},
         year = 2019,
        month = apr,
       volume = {875},
       number = {1},
          eid = {44},
        pages = {44},
          doi = {10.3847/1538-4357/ab0e05},
archivePrefix = {arXiv},
       eprint = {1811.00906},
 primaryClass = {astro-ph.HE},
       adsurl = {https://ui.adsabs.harvard.edu/abs/2019ApJ...875...44Z},
      adsnote = {Provided by the SAO/NASA Astrophysics Data System}
}

@ARTICLE{Yusef-Zadeh2020,
       author = {{Yusef-Zadeh}, F. and {Royster}, M. and {Wardle}, M. and {Cotton}, W. and {Kunneriath}, D. and {Heywood}, I. and {Michail}, J.},
        title = "{Evidence for a jet and outflow from Sgr A*: a continuum and spectral line study}",
      journal = {\mnras},
     keywords = {accretion, accretion discs, black hole physics, Galaxy: centre, galaxies: jets, Astrophysics - High Energy Astrophysical Phenomena, Astrophysics - Astrophysics of Galaxies},
         year = 2020,
        month = dec,
       volume = {499},
       number = {3},
        pages = {3909-3931},
          doi = {10.1093/mnras/staa2399},
archivePrefix = {arXiv},
       eprint = {2008.04317},
 primaryClass = {astro-ph.HE},
       adsurl = {https://ui.adsabs.harvard.edu/abs/2020MNRAS.499.3909Y},
      adsnote = {Provided by the SAO/NASA Astrophysics Data System}
}

@ARTICLE{Vincent2019,
       author = {{Vincent}, F.~H. and {Abramowicz}, M.~A. and {Zdziarski}, A.~A. and {Wielgus}, M. and {Paumard}, T. and {Perrin}, G. and {Straub}, O.},
        title = "{Multi-wavelength torus-jet model for Sagittarius A*}",
      journal = {\aap},
     keywords = {Galaxy: center, accretion, accretion disks, black hole physics, relativistic processes, Astrophysics - High Energy Astrophysical Phenomena, General Relativity and Quantum Cosmology},
         year = 2019,
        month = apr,
       volume = {624},
          eid = {A52},
        pages = {A52},
          doi = {10.1051/0004-6361/201834946},
archivePrefix = {arXiv},
       eprint = {1902.01175},
 primaryClass = {astro-ph.HE},
       adsurl = {https://ui.adsabs.harvard.edu/abs/2019A&A...624A..52V},
      adsnote = {Provided by the SAO/NASA Astrophysics Data System}
}

@ARTICLE{Ressler2023,
       author = {{Ressler}, Sean M. and {White}, Christopher J. and {Quataert}, Eliot},
        title = "{Wind-Fed GRMHD Simulations of Sagittarius A*: Tilt and Alignment of Jets and Accretion Discs, Electron Thermodynamics, and Multi-Scale Modeling of the Rotation Measure}",
      journal = {arXiv e-prints},
     keywords = {Astrophysics - High Energy Astrophysical Phenomena},
         year = 2023,
        month = mar,
          eid = {arXiv:2303.15503},
        pages = {arXiv:2303.15503},
archivePrefix = {arXiv},
       eprint = {2303.15503},
 primaryClass = {astro-ph.HE},
       adsurl = {https://ui.adsabs.harvard.edu/abs/2023arXiv230315503R},
      adsnote = {Provided by the SAO/NASA Astrophysics Data System}
}

@ARTICLE{Nättilä2021,
       author = {{Nattila}, Joonas and {Beloborodov}, Andrei M.},
        title = "{Radiative Turbulent Flares in Magnetically Dominated Plasmas}",
      journal = {\apj},
     keywords = {Plasma astrophysics, High energy astrophysics, Astrophysical magnetism, Computational astronomy, Compact objects, Nonthermal radiation sources, 1261, 739, 102, 293, 288, 1119, Astrophysics - High Energy Astrophysical Phenomena, Physics - Plasma Physics},
         year = 2021,
        month = nov,
       volume = {921},
       number = {1},
          eid = {87},
        pages = {87},
          doi = {10.3847/1538-4357/ac1c76},
archivePrefix = {arXiv},
       eprint = {2012.03043},
 primaryClass = {astro-ph.HE},
       adsurl = {https://ui.adsabs.harvard.edu/abs/2021ApJ...921...87N},
      adsnote = {Provided by the SAO/NASA Astrophysics Data System}
}

@ARTICLE{Moscibrodzka2013,
       author = {{Moscibrodzka}, Monika and {Falcke}, Heino},
        title = "{Coupled jet-disk model for Sagittarius A*: explaining the flat-spectrum radio core with GRMHD simulations of jets}",
      journal = {\aap},
     keywords = {accretion, accretion disks, black hole physics, magnetohydrodynamics (MHD), radiative transfer, Galaxy: center, Astrophysics - High Energy Astrophysical Phenomena},
         year = 2013,
        month = nov,
       volume = {559},
          eid = {L3},
        pages = {L3},
          doi = {10.1051/0004-6361/201322692},
archivePrefix = {arXiv},
       eprint = {1310.4951},
 primaryClass = {astro-ph.HE},
       adsurl = {https://ui.adsabs.harvard.edu/abs/2013A&A...559L...3M},
      adsnote = {Provided by the SAO/NASA Astrophysics Data System}
}

@ARTICLE{Ressler2017,
       author = {{Ressler}, S.~M. and {Tchekhovskoy}, A. and {Quataert}, E. and {Gammie}, C.~F.},
        title = "{The disc-jet symbiosis emerges: modelling the emission of Sagittarius A* with electron thermodynamics}",
      journal = {\mnras},
     keywords = {accretion, accretion discs, black hole physics, MHD, relativistic processes, Galaxy: centre, Astrophysics - High Energy Astrophysical Phenomena},
         year = 2017,
        month = may,
       volume = {467},
       number = {3},
        pages = {3604-3619},
          doi = {10.1093/mnras/stx364},
archivePrefix = {arXiv},
       eprint = {1611.09365},
 primaryClass = {astro-ph.HE},
       adsurl = {https://ui.adsabs.harvard.edu/abs/2017MNRAS.467.3604R},
      adsnote = {Provided by the SAO/NASA Astrophysics Data System}
}

@ARTICLE{Davelaar2018,
       author = {{Davelaar}, J. and {Moscibrodzka}, M. and {Bronzwaer}, T. and {Falcke}, H.},
        title = "{General relativistic magnetohydrodynamical {\ensuremath{\kappa}}-jet models for Sagittarius A*}",
      journal = {\aap},
     keywords = {black hole physics, accretion, accretion disks, acceleration of particles, radiation mechanisms: non-thermal, radiative transfer, Astrophysics - High Energy Astrophysical Phenomena},
         year = 2018,
        month = apr,
       volume = {612},
          eid = {A34},
        pages = {A34},
          doi = {10.1051/0004-6361/201732025},
archivePrefix = {arXiv},
       eprint = {1712.02266},
 primaryClass = {astro-ph.HE},
       adsurl = {https://ui.adsabs.harvard.edu/abs/2018A&A...612A..34D},
      adsnote = {Provided by the SAO/NASA Astrophysics Data System}
}

@ARTICLE{EHT2022a,
       author = {{Event Horizon Telescope Collaboration} and {Akiyama}, Kazunori and {Alberdi}, Antxon and {Alef}, Walter and {Algaba}, Juan Carlos and {Anantua}, Richard and {Asada}, Keiichi and {Azulay}, Rebecca and {Bach}, Uwe and {Baczko}, Anne-Kathrin and {Ball}, David and {Balokovic}, Mislav and {Barrett}, John and {Baubock}, Michi and {Benson}, Bradford A. and {Bintley}, Dan and {Blackburn}, Lindy and {Blundell}, Raymond and {Bouman}, Katherine L. and {Bower}, Geoffrey C. and {Boyce}, Hope and {Bremer}, Michael and {Brinkerink}, Christiaan D. and {Brissenden}, Roger and {Britzen}, Silke and {Broderick}, Avery E. and {Broguiere}, Dominique and {Bronzwaer}, Thomas and {Bustamante}, Sandra and {Byun}, Do-Young and {Carlstrom}, John E. and {Ceccobello}, Chiara and {Chael}, Andrew and {Chan}, Chi-kwan and {Chatterjee}, Koushik and {Chatterjee}, Shami and {Chen}, Ming-Tang and {Chen}, Yongjun and {Cheng}, Xiaopeng and {Cho}, Ilje and {Christian}, Pierre and {Conroy}, Nicholas S. and {Conway}, John E. and {Cordes}, James M. and {Crawford}, Thomas M. and {Crew}, Geoffrey B. and {Cruz-Osorio}, Alejandro and {Cui}, Yuzhu and {Davelaar}, Jordy and {Laurentis}, Mariafelicia De and {Deane}, Roger and {Dempsey}, Jessica and {Desvignes}, Gregory and {Dexter}, Jason and {Dhruv}, Vedant and {Doeleman}, Sheperd S. and {Dougal}, Sean and {Dzib}, Sergio A. and {Eatough}, Ralph P. and {Emami}, Razieh and {Falcke}, Heino and {Farah}, Joseph and {Fish}, Vincent L. and {Fomalont}, Ed and {Ford}, H. Alyson and {Fraga-Encinas}, Raquel and {Freeman}, William T. and {Friberg}, Per and {Fromm}, Christian M. and {Fuentes}, Antonio and {Galison}, Peter and {Gammie}, Charles F. and {Garcia}, Roberto and {Gentaz}, Olivier and {Georgiev}, Boris and {Goddi}, Ciriaco and {Gold}, Roman and {Gomez-Ruiz}, Arturo I. and {Gomez}, Jose L. and {Gu}, Minfeng and {Gurwell}, Mark and {Hada}, Kazuhiro and {Haggard}, Daryl and {Haworth}, Kari and {Hecht}, Michael H. and {Hesper}, Ronald and {Heumann}, Dirk and {Ho}, Luis C. and {Ho}, Paul and {Honma}, Mareki and {Huang}, Chih-Wei L. and {Huang}, Lei and {Hughes}, David H. and {Ikeda}, Shiro and {Impellizzeri}, C.~M. Violette and {Inoue}, Makoto and {Issaoun}, Sara and {James}, David J. and {Jannuzi}, Buell T. and {Janssen}, Michael and {Jeter}, Britton and {Jiang}, Wu and {Jimenez-Rosales}, Alejandra and {Johnson}, Michael D. and {Jorstad}, Svetlana and {Joshi}, Abhishek V. and {Jung}, Taehyun and {Karami}, Mansour and {Karuppusamy}, Ramesh and {Kawashima}, Tomohisa and {Keating}, Garrett K. and {Kettenis}, Mark and {Kim}, Dong-Jin and {Kim}, Jae-Young and {Kim}, Jongsoo and {Kim}, Junhan and {Kino}, Motoki and {Koay}, Jun Yi and {Kocherlakota}, Prashant and {Kofuji}, Yutaro and {Koch}, Patrick M. and {Koyama}, Shoko and {Kramer}, Carsten and {Kramer}, Michael and {Krichbaum}, Thomas P. and {Kuo}, Cheng-Yu and {Bella}, Noemi La and {Lauer}, Tod R. and {Lee}, Daeyoung and {Lee}, Sang-Sung and {Leung}, Po Kin and {Levis}, Aviad and {Li}, Zhiyuan and {Lico}, Rocco and {Lindahl}, Greg and {Lindqvist}, Michael and {Lisakov}, Mikhail and {Liu}, Jun and {Liu}, Kuo and {Liuzzo}, Elisabetta and {Lo}, Wen-Ping and {Lobanov}, Andrei P. and {Loinard}, Laurent and {Lonsdale}, Colin J. and {Lu}, Ru-Sen and {Mao}, Jirong and {Marchili}, Nicola and {Markoff}, Sera and {Marrone}, Daniel P. and {Marscher}, Alan P. and {Marti-Vidal}, Ivan and {Matsushita}, Satoki and {Matthews}, Lynn D. and {Medeiros}, Lia and {Menten}, Karl M. and {Michalik}, Daniel and {Mizuno}, Izumi and {Mizuno}, Yosuke and {Moran}, James M. and {Moriyama}, Kotaro and {Moscibrodzka}, Monika and {Muller}, Cornelia and {Mus}, Alejandro and {Musoke}, Gibwa and {Myserlis}, Ioannis and {Nadolski}, Andrew and {Nagai}, Hiroshi and {Nagar}, Neil M. and {Nakamura}, Masanori and {Narayan}, Ramesh and {Narayanan}, Gopal and {Natarajan}, Iniyan and {Nathanail}, Antonios and {Fuentes}, Santiago Navarro and {Neilsen}, Joey and {Neri}, Roberto and {Ni}, Chunchong and {Noutsos}, Aristeidis and {Nowak}, Michael A. and {Oh}, Junghwan and {Okino}, Hiroki and {Olivares}, Hector and {Ortiz-Leon}, Gisela N. and {Oyama}, Tomoaki and {Ozel}, Feryal and {Palumbo}, Daniel C.~M. and {Paraschos}, Georgios Filippos and {Park}, Jongho and {Parsons}, Harriet and {Patel}, Nimesh and {Pen}, Ue-Li and {Pesce}, Dominic W. and {Pietu}, Vincent and {Plambeck}, Richard and {PopStefanija}, Aleksandar and {Porth}, Oliver and {Potzl}, Felix M. and {Prather}, Ben and {Preciado-Lopez}, Jorge A. and {Psaltis}, Dimitrios and {Pu}, Hung-Yi and {Ramakrishnan}, Venkatessh and {Rao}, Ramprasad and {Rawlings}, Mark G. and {Raymond}, Alexander W. and {Rezzolla}, Luciano and {Ricarte}, Angelo and {Ripperda}, Bart and {Roelofs}, Freek and {Rogers}, Alan and {Ros}, Eduardo and {Romero-Canizales}, Cristina and {Roshanineshat}, Arash and {Rottmann}, Helge and {Roy}, Alan L. and {Ruiz}, Ignacio and {Ruszczyk}, Chet and {Rygl}, Kazi L.~J. and {Sanchez}, Salvador and {Sanchez-Arguelles}, David and {Sanchez-Portal}, Miguel and {Sasada}, Mahito and {Satapathy}, Kaushik and {Savolainen}, Tuomas and {Schloerb}, F. Peter and {Schonfeld}, Jonathan and {Schuster}, Karl-Friedrich and {Shao}, Lijing and {Shen}, Zhiqiang and {Small}, Des and {Sohn}, Bong Won and {SooHoo}, Jason and {Souccar}, Kamal and {Sun}, He and {Tazaki}, Fumie and {Tetarenko}, Alexandra J. and {Tiede}, Paul and {Tilanus}, Remo P.~J. and {Titus}, Michael and {Torne}, Pablo and {Traianou}, Efthalia and {Trent}, Tyler and {Trippe}, Sascha and {Turk}, Matthew and {van Bemmel}, Ilse and {van Langevelde}, Huib Jan and {van Rossum}, Daniel R. and {Vos}, Jesse and {Wagner}, Jan and {Ward-Thompson}, Derek and {Wardle}, John and {Weintroub}, Jonathan and {Wex}, Norbert and {Wharton}, Robert and {Wielgus}, Maciek and {Wiik}, Kaj and {Witzel}, Gunther and {Wondrak}, Michael F. and {Wong}, George N. and {Wu}, Qingwen and {Yamaguchi}, Paul and {Yoon}, Doosoo and {Young}, Andre and {Young}, Ken and {Younsi}, Ziri and {Yuan}, Feng and {Yuan}, Ye-Fei and {Zensus}, J. Anton and {Zhang}, Shuo and {Zhao}, Guang-Yao and {Zhao}, Shan-Shan and {Agurto}, Claudio and {Allardi}, Alexander and {Amestica}, Rodrigo and {Araneda}, Juan Pablo and {Arriagada}, Oriel and {Berghuis}, Jennie L. and {Bertarini}, Alessandra and {Berthold}, Ryan and {Blanchard}, Jay and {Brown}, Ken and {Cardenas}, Mauricio and {Cantzler}, Michael and {Caro}, Patricio and {Castillo-Dominguez}, Edgar and {Chan}, Tin Lok and {Chang}, Chih-Cheng and {Chang}, Dominic O. and {Chang}, Shu-Hao and {Chang}, Song-Chu and {Chen}, Chung-Chen and {Chilson}, Ryan and {Chuter}, Tim C. and {Ciechanowicz}, Miroslaw and {Colin-Beltran}, Edgar and {Coulson}, Iain M. and {Crowley}, Joseph and {Degenaar}, Nathalie and {Dornbusch}, Sven and {Duran}, Carlos A. and {Everett}, Wendeline B. and {Faber}, Aaron and {Forster}, Karl and {Fuchs}, Miriam M. and {Gale}, David M. and {Geertsema}, Gertie and {Gonzalez}, Edouard and {Graham}, Dave and {Gueth}, Frederic and {Halverson}, Nils W. and {Han}, Chih-Chiang and {Han}, Kuo-Chang and {Hasegawa}, Yutaka and {Hernandez-Rebollar}, Jose Luis and {Herrera}, Cristian and {Herrero-Illana}, Ruben and {Heyminck}, Stefan and {Hirota}, Akihiko and {Hoge}, James and {Hostler Schimpf}, Shelbi R. and {Howie}, Ryan E. and {Huang}, Yau-De and {Jiang}, Homin and {Jinchi}, Hao and {John}, David and {Kimura}, Kimihiro and {Klein}, Thomas and {Kubo}, Derek and {Kuroda}, John and {Kwon}, Caleb and {Lacasse}, Richard and {Laing}, Robert and {Leitch}, Erik M. and {Li}, Chao-Te and {Liu}, Ching-Tang and {Liu}, Kuan-Yu and {Lin}, Lupin C. -C. and {Lu}, Li-Ming and {Mac-Auliffe}, Felipe and {Martin-Cocher}, Pierre and {Matulonis}, Callie and {Maute}, John K. and {Messias}, Hugo and {Meyer-Zhao}, Zheng and {Montana}, Alfredo and {Montenegro-Montes}, Francisco and {Montgomerie}, William and {Moreno Nolasco}, Marcos Emir and {Muders}, Dirk and {Nishioka}, Hiroaki and {Norton}, Timothy J. and {Nystrom}, George and {Ogawa}, Hideo and {Olivares}, Rodrigo and {Oshiro}, Peter and {Perez-Beaupuits}, Juan Pablo and {Parra}, Rodrigo and {Phillips}, Neil M. and {Poirier}, Michael and {Pradel}, Nicolas and {Qiu}, Richard and {Raffin}, Philippe A. and {Rahlin}, Alexandra S. and {Ramirez}, Jorge and {Ressler}, Sean and {Reynolds}, Mark and {Rodriguez-Montoya}, Ivan and {Saez-Madain}, Alejandro F. and {Santana}, Jorge and {Shaw}, Paul and {Shirkey}, Leslie E. and {Silva}, Kevin M. and {Snow}, William and {Sousa}, Don and {Sridharan}, T.~K. and {Stahm}, William and {Stark}, Anthony A. and {Test}, John and {Torstensson}, Karl and {Venegas}, Paulina and {Walther}, Craig and {Wei}, Ta-Shun and {White}, Chris and {Wieching}, Gundolf and {Wijnands}, Rudy and {Wouterloot}, Jan G.~A. and {Yu}, Chen-Yu and {Yu (于威)}, Wei and {Zeballos}, Milagros},
        title = "{First Sagittarius A* Event Horizon Telescope Results. I. The Shadow of the Supermassive Black Hole in the Center of the Milky Way}",
      journal = {\apjl},
     keywords = {Black holes, Kerr black holes, Rotating black holes, Heterodyne interferometry, Galactic center, 162, 886, 1406, 726, 565},
         year = 2022,
        month = may,
       volume = {930},
       number = {2},
          eid = {L12},
        pages = {L12},
          doi = {10.3847/2041-8213/ac667410.3847/2041-8213/ac6675},
       adsurl = {https://ui.adsabs.harvard.edu/abs/2022ApJ...930L..12E},
      adsnote = {Provided by the SAO/NASA Astrophysics Data System}
}

@ARTICLE{EHT2022b,
       author = {{Event Horizon Telescope Collaboration} and {Akiyama}, Kazunori and {Alberdi}, Antxon and {Alef}, Walter and {Carlos Algaba}, Juan and {Anantua}, Richard and {Asada}, Keiichi and {Azulay}, Rebecca and {Bach}, Uwe and {Baczko}, Anne-Kathrin and {Ball}, David and {Balokovic}, Mislav and {Barrett}, John and {Baubock}, Michi and {Benson}, Bradford A. and {Bintley}, Dan and {Blackburn}, Lindy and {Blundell}, Raymond and {Bouman}, Katherine L. and {Bower}, Geoffrey C. and {Boyce}, Hope and {Bremer}, Michael and {Brinkerink}, Christiaan D. and {Brissenden}, Roger and {Britzen}, Silke and {Broderick}, Avery E. and {Broguiere}, Dominique and {Bronzwaer}, Thomas and {Bustamante}, Sandra and {Byun}, Do-Young and {Carlstrom}, John E. and {Ceccobello}, Chiara and {Chael}, Andrew and {Chan}, Chi-kwan and {Chatterjee}, Koushik and {Chatterjee}, Shami and {Chen}, Ming-Tang and {Chen}, Yongjun and {Cheng}, Xiaopeng and {Cho}, Ilje and {Christian}, Pierre and {Conroy}, Nicholas S. and {Conway}, John E. and {Cordes}, James M. and {Crawford}, Thomas M. and {Crew}, Geoffrey B. and {Cruz-Osorio}, Alejandro and {Cui}, Yuzhu and {Davelaar}, Jordy and {De Laurentis}, Mariafelicia and {Deane}, Roger and {Dempsey}, Jessica and {Desvignes}, Gregory and {Dexter}, Jason and {Dhruv}, Vedant and {Doeleman}, Sheperd S. and {Dougal}, Sean and {Dzib}, Sergio A. and {Eatough}, Ralph P. and {Emami}, Razieh and {Falcke}, Heino and {Farah}, Joseph and {Fish}, Vincent L. and {Fomalont}, Ed and {Ford}, H. Alyson and {Fraga-Encinas}, Raquel and {Freeman}, William T. and {Friberg}, Per and {Fromm}, Christian M. and {Fuentes}, Antonio and {Galison}, Peter and {Gammie}, Charles F. and {Garcia}, Roberto and {Gentaz}, Olivier and {Georgiev}, Boris and {Goddi}, Ciriaco and {Gold}, Roman and {Gomez-Ruiz}, Arturo I. and {Gomez}, Jose L. and {Gu}, Minfeng and {Gurwell}, Mark and {Hada}, Kazuhiro and {Haggard}, Daryl and {Haworth}, Kari and {Hecht}, Michael H. and {Hesper}, Ronald and {Heumann}, Dirk and {Ho}, Luis C. and {Ho}, Paul and {Honma}, Mareki and {Huang}, Chih-Wei L. and {Huang}, Lei and {Hughes}, David H. and {Ikeda}, Shiro and {Violette Impellizzeri}, C.~M. and {Inoue}, Makoto and {Issaoun}, Sara and {James}, David J. and {Jannuzi}, Buell T. and {Janssen}, Michael and {Jeter}, Britton and {Jiang}, Wu and {Jimenez-Rosales}, Alejandra and {Johnson}, Michael D. and {Jorstad}, Svetlana and {Joshi}, Abhishek V. and {Jung}, Taehyun and {Karami}, Mansour and {Karuppusamy}, Ramesh and {Kawashima}, Tomohisa and {Keating}, Garrett K. and {Kettenis}, Mark and {Kim}, Dong-Jin and {Kim}, Jae-Young and {Kim}, Jongsoo and {Kim}, Junhan and {Kino}, Motoki and {Koay}, Jun Yi and {Kocherlakota}, Prashant and {Kofuji}, Yutaro and {Koch}, Patrick M. and {Koyama}, Shoko and {Kramer}, Carsten and {Kramer}, Michael and {Krichbaum}, Thomas P. and {Kuo}, Cheng-Yu and {Bella}, Noemi La and {Lauer}, Tod R. and {Lee}, Daeyoung and {Lee}, Sang-Sung and {Leung}, Po Kin and {Levis}, Aviad and {Li}, Zhiyuan and {Lico}, Rocco and {Lindahl}, Greg and {Lindqvist}, Michael and {Lisakov}, Mikhail and {Liu}, Jun and {Liu}, Kuo and {Liuzzo}, Elisabetta and {Lo}, Wen-Ping and {Lobanov}, Andrei P. and {Loinard}, Laurent and {Lonsdale}, Colin J. and {Lu}, Ru-Sen and {Mao}, Jirong and {Marchili}, Nicola and {Markoff}, Sera and {Marrone}, Daniel P. and {Marscher}, Alan P. and {Marti-Vidal}, Ivan and {Matsushita}, Satoki and {Matthews}, Lynn D. and {Medeiros}, Lia and {Menten}, Karl M. and {Michalik}, Daniel and {Mizuno}, Izumi and {Mizuno}, Yosuke and {Moran}, James M. and {Moriyama}, Kotaro and {Moscibrodzka}, Monika and {Muller}, Cornelia and {Mus}, Alejandro and {Musoke}, Gibwa and {Myserlis}, Ioannis and {Nadolski}, Andrew and {Nagai}, Hiroshi and {Nagar}, Neil M. and {Nakamura}, Masanori and {Narayan}, Ramesh and {Narayanan}, Gopal and {Natarajan}, Iniyan and {Nathanail}, Antonios and {Navarro Fuentes}, Santiago and {Neilsen}, Joey and {Neri}, Roberto and {Ni}, Chunchong and {Noutsos}, Aristeidis and {Nowak}, Michael A. and {Oh}, Junghwan and {Okino}, Hiroki and {Olivares}, Hector and {Ortiz-Leon}, Gisela N. and {Oyama}, Tomoaki and {Ozel}, Feryal and {Palumbo}, Daniel C.~M. and {Filippos Paraschos}, Georgios and {Park}, Jongho and {Parsons}, Harriet and {Patel}, Nimesh and {Pen}, Ue-Li and {Pesce}, Dominic W. and {Pietu}, Vincent and {Plambeck}, Richard and {PopStefanija}, Aleksandar and {Porth}, Oliver and {Potzl}, Felix M. and {Prather}, Ben and {Preciado-Lopez}, Jorge A. and {Psaltis}, Dimitrios and {Pu}, Hung-Yi and {Ramakrishnan}, Venkatessh and {Rao}, Ramprasad and {Rawlings}, Mark G. and {Raymond}, Alexander W. and {Rezzolla}, Luciano and {Ricarte}, Angelo and {Ripperda}, Bart and {Roelofs}, Freek and {Rogers}, Alan and {Ros}, Eduardo and {Romero-Canizales}, Cristina and {Roshanineshat}, Arash and {Rottmann}, Helge and {Roy}, Alan L. and {Ruiz}, Ignacio and {Ruszczyk}, Chet and {Rygl}, Kazi L.~J. and {Sanchez}, Salvador and {Sanchez-Arguelles}, David and {Sanchez-Portal}, Miguel and {Sasada}, Mahito and {Satapathy}, Kaushik and {Savolainen}, Tuomas and {Schloerb}, F. Peter and {Schonfeld}, Jonathan and {Schuster}, Karl-Friedrich and {Shao}, Lijing and {Shen}, Zhiqiang and {Small}, Des and {Sohn}, Bong Won and {SooHoo}, Jason and {Souccar}, Kamal and {Sun}, He and {Tazaki}, Fumie and {Tetarenko}, Alexandra J. and {Tiede}, Paul and {Tilanus}, Remo P.~J. and {Titus}, Michael and {Torne}, Pablo and {Traianou}, Efthalia and {Trent}, Tyler and {Trippe}, Sascha and {Turk}, Matthew and {van Bemmel}, Ilse and {van Langevelde}, Huib Jan and {van Rossum}, Daniel R. and {Vos}, Jesse and {Wagner}, Jan and {Ward-Thompson}, Derek and {Wardle}, John and {Weintroub}, Jonathan and {Wex}, Norbert and {Wharton}, Robert and {Wielgus}, Maciek and {Wiik}, Kaj and {Witzel}, Gunther and {Wondrak}, Michael F. and {Wong}, George N. and {Wu}, Qingwen and {Yamaguchi}, Paul and {Yoon}, Doosoo and {Young}, Andre and {Young}, Ken and {Younsi}, Ziri and {Yuan}, Feng and {Yuan}, Ye-Fei and {Zensus}, J. Anton and {Zhang}, Shuo and {Zhao}, Guang-Yao and {Zhao}, Shan-Shan and {Chan}, Tin Lok and {Qiu}, Richard and {Ressler}, Sean and {White}, Chris},
        title = "{First Sagittarius A* Event Horizon Telescope Results. V. Testing Astrophysical Models of the Galactic Center Black Hole}",
      journal = {\apjl},
     keywords = {Black hole physics, Galactic center, 159, 565},
         year = 2022,
        month = may,
       volume = {930},
       number = {2},
          eid = {L16},
        pages = {L16},
          doi = {10.3847/2041-8213/ac6672},
       adsurl = {https://ui.adsabs.harvard.edu/abs/2022ApJ...930L..16E},
      adsnote = {Provided by the SAO/NASA Astrophysics Data System}
}

@ARTICLE{Ripperda2022,
       author = {{Ripperda}, B. and {Liska}, M. and {Chatterjee}, K. and {Musoke}, G. and {Philippov}, A.~A. and {Markoff}, S.~B. and {Tchekhovskoy}, A. and {Younsi}, Z.},
        title = "{Black Hole Flares: Ejection of Accreted Magnetic Flux through 3D Plasmoid-mediated Reconnection}",
      journal = {\apjl},
     keywords = {641, 1261, 1964, 739, 162, Astrophysics - High Energy Astrophysical Phenomena, General Relativity and Quantum Cosmology, Physics - Plasma Physics},
         year = 2022,
        month = jan,
       volume = {924},
       number = {2},
          eid = {L32},
        pages = {L32},
          doi = {10.3847/2041-8213/ac46a1},
archivePrefix = {arXiv},
       eprint = {2109.15115},
 primaryClass = {astro-ph.HE},
       adsurl = {https://ui.adsabs.harvard.edu/abs/2022ApJ...924L..32R},
      adsnote = {Provided by the SAO/NASA Astrophysics Data System}
}

@ARTICLE{Pandya2016,
       author = {{Pandya}, Alex and {Zhang}, Zhaowei and {Chandra}, Mani and {Gammie}, Charles F.},
        title = "{Polarized Synchrotron Emissivities and Absorptivities for Relativistic Thermal, Power-law, and Kappa Distribution Functions}",
      journal = {\apj},
     keywords = {plasmas, polarization, radiation mechanisms: general, radiative transfer, relativistic processes, Astrophysics - High Energy Astrophysical Phenomena},
         year = 2016,
        month = may,
       volume = {822},
       number = {1},
          eid = {34},
        pages = {34},
          doi = {10.3847/0004-637X/822/1/34},
archivePrefix = {arXiv},
       eprint = {1602.08749},
 primaryClass = {astro-ph.HE},
       adsurl = {https://ui.adsabs.harvard.edu/abs/2016ApJ...822...34P},
      adsnote = {Provided by the SAO/NASA Astrophysics Data System}
}

@ARTICLE{Tagger2006,
       author = {{Tagger}, Michel and {Melia}, Fulvio},
        title = "{A Possible Rossby Wave Instability Origin for the Flares in Sagittarius A*}",
      journal = {\apjl},
     keywords = {Accretion, Accretion Disks, Black Hole Physics, Galaxy: Center, Instabilities, Magnetohydrodynamics: MHD, Plasmas, Astrophysics},
         year = 2006,
        month = jan,
       volume = {636},
       number = {1},
        pages = {L33-L36},
          doi = {10.1086/499806},
archivePrefix = {arXiv},
       eprint = {astro-ph/0511520},
 primaryClass = {astro-ph},
       adsurl = {https://ui.adsabs.harvard.edu/abs/2006ApJ...636L..33T},
      adsnote = {Provided by the SAO/NASA Astrophysics Data System}
}

@INPROCEEDINGS{Haniff2003,
       author = {{Haniff}, C.},
        title = "{An introduction to interferometry}",
    booktitle = {EAS Publications Series},
         year = 2003,
       editor = {{Perrin}, G. and {Malbet}, F.},
       series = {EAS Publications Series},
       volume = {6},
        month = jan,
        pages = {3},
          doi = {10.1051/eas:2003001},
       adsurl = {https://ui.adsabs.harvard.edu/abs/2003EAS.....6....3H},
      adsnote = {Provided by the SAO/NASA Astrophysics Data System}
}

@ARTICLE{EHT2019,
       author = {{Event Horizon Telescope Collaboration} and {Akiyama}, Kazunori and {Alberdi}, Antxon and {Alef}, Walter and {Asada}, Keiichi and {Azulay}, Rebecca and {Baczko}, Anne-Kathrin and {Ball}, David and {Balokovic}, Mislav and {Barrett}, John and {Bintley}, Dan and {Blackburn}, Lindy and {Boland}, Wilfred and {Bouman}, Katherine L. and {Bower}, Geoffrey C. and {Bremer}, Michael and {Brinkerink}, Christiaan D. and {Brissenden}, Roger and {Britzen}, Silke and {Broderick}, Avery E. and {Broguiere}, Dominique and {Bronzwaer}, Thomas and {Byun}, Do-Young and {Carlstrom}, John E. and {Chael}, Andrew and {Chan}, Chi-kwan and {Chatterjee}, Shami and {Chatterjee}, Koushik and {Chen}, Ming-Tang and {Chen}, Yongjun and {Cho}, Ilje and {Christian}, Pierre and {Conway}, John E. and {Cordes}, James M. and {Crew}, Geoffrey B. and {Cui}, Yuzhu and {Davelaar}, Jordy and {De Laurentis}, Mariafelicia and {Deane}, Roger and {Dempsey}, Jessica and {Desvignes}, Gregory and {Dexter}, Jason and {Doeleman}, Sheperd S. and {Eatough}, Ralph P. and {Falcke}, Heino and {Fish}, Vincent L. and {Fomalont}, Ed and {Fraga-Encinas}, Raquel and {Freeman}, William T. and {Friberg}, Per and {Fromm}, Christian M. and {Gomez}, Jose L. and {Galison}, Peter and {Gammie}, Charles F. and {Garcia}, Roberto and {Gentaz}, Olivier and {Georgiev}, Boris and {Goddi}, Ciriaco and {Gold}, Roman and {Gu}, Minfeng and {Gurwell}, Mark and {Hada}, Kazuhiro and {Hecht}, Michael H. and {Hesper}, Ronald and {Ho}, Luis C. and {Ho}, Paul and {Honma}, Mareki and {Huang}, Chih-Wei L. and {Huang}, Lei and {Hughes}, David H. and {Ikeda}, Shiro and {Inoue}, Makoto and {Issaoun}, Sara and {James}, David J. and {Jannuzi}, Buell T. and {Janssen}, Michael and {Jeter}, Britton and {Jiang}, Wu and {Johnson}, Michael D. and {Jorstad}, Svetlana and {Jung}, Taehyun and {Karami}, Mansour and {Karuppusamy}, Ramesh and {Kawashima}, Tomohisa and {Keating}, Garrett K. and {Kettenis}, Mark and {Kim}, Jae-Young and {Kim}, Junhan and {Kim}, Jongsoo and {Kino}, Motoki and {Koay}, Jun Yi and {Koch}, Patrick M. and {Koyama}, Shoko and {Kramer}, Michael and {Kramer}, Carsten and {Krichbaum}, Thomas P. and {Kuo}, Cheng-Yu and {Lauer}, Tod R. and {Lee}, Sang-Sung and {Li}, Yan-Rong and {Li}, Zhiyuan and {Lindqvist}, Michael and {Liu}, Kuo and {Liuzzo}, Elisabetta and {Lo}, Wen-Ping and {Lobanov}, Andrei P. and {Loinard}, Laurent and {Lonsdale}, Colin and {Lu}, Ru-Sen and {MacDonald}, Nicholas R. and {Mao}, Jirong and {Markoff}, Sera and {Marrone}, Daniel P. and {Marscher}, Alan P. and {Marti-Vidal}, Ivan and {Matsushita}, Satoki and {Matthews}, Lynn D. and {Medeiros}, Lia and {Menten}, Karl M. and {Mizuno}, Yosuke and {Mizuno}, Izumi and {Moran}, James M. and {Moriyama}, Kotaro and {Moscibrodzka}, Monika and {Muller}, Cornelia and {Nagai}, Hiroshi and {Nagar}, Neil M. and {Nakamura}, Masanori and {Narayan}, Ramesh and {Narayanan}, Gopal and {Natarajan}, Iniyan and {Neri}, Roberto and {Ni}, Chunchong and {Noutsos}, Aristeidis and {Okino}, Hiroki and {Olivares}, Hector and {Ortiz-Leon}, Gisela N. and {Oyama}, Tomoaki and {Ozel}, Feryal and {Palumbo}, Daniel C.~M. and {Patel}, Nimesh and {Pen}, Ue-Li and {Pesce}, Dominic W. and {Pietu}, Vincent and {Plambeck}, Richard and {PopStefanija}, Aleksandar and {Porth}, Oliver and {Prather}, Ben and {Preciado-Lopez}, Jorge A. and {Psaltis}, Dimitrios and {Pu}, Hung-Yi and {Ramakrishnan}, Venkatessh and {Rao}, Ramprasad and {Rawlings}, Mark G. and {Raymond}, Alexander W. and {Rezzolla}, Luciano and {Ripperda}, Bart and {Roelofs}, Freek and {Rogers}, Alan and {Ros}, Eduardo and {Rose}, Mel and {Roshanineshat}, Arash and {Rottmann}, Helge and {Roy}, Alan L. and {Ruszczyk}, Chet and {Ryan}, Benjamin R. and {Rygl}, Kazi L.~J. and {Sanchez}, Salvador and {Sanchez-Arguelles}, David and {Sasada}, Mahito and {Savolainen}, Tuomas and {Schloerb}, F. Peter and {Schuster}, Karl-Friedrich and {Shao}, Lijing and {Shen}, Zhiqiang and {Small}, Des and {Sohn}, Bong Won and {SooHoo}, Jason and {Tazaki}, Fumie and {Tiede}, Paul and {Tilanus}, Remo P.~J. and {Titus}, Michael and {Toma}, Kenji and {Torne}, Pablo and {Trent}, Tyler and {Trippe}, Sascha and {Tsuda}, Shuichiro and {van Bemmel}, Ilse and {van Langevelde}, Huib Jan and {van Rossum}, Daniel R. and {Wagner}, Jan and {Wardle}, John and {Weintroub}, Jonathan and {Wex}, Norbert and {Wharton}, Robert and {Wielgus}, Maciek and {Wong}, George N. and {Wu}, Qingwen and {Young}, Ken and {Young}, Andre and {Younsi}, Ziri and {Yuan}, Feng and {Yuan}, Ye-Fei and {Zensus}, J. Anton and {Zhao}, Guangyao and {Zhao}, Shan-Shan and {Zhu}, Ziyan and {Algaba}, Juan-Carlos and {Allardi}, Alexander and {Amestica}, Rodrigo and {Anczarski}, Jadyn and {Bach}, Uwe and {Baganoff}, Frederick K. and {Beaudoin}, Christopher and {Benson}, Bradford A. and {Berthold}, Ryan and {Blanchard}, Jay M. and {Blundell}, Ray and {Bustamente}, Sandra and {Cappallo}, Roger and {Castillo-Dominguez}, Edgar and {Chang}, Chih-Cheng and {Chang}, Shu-Hao and {Chang}, Song-Chu and {Chen}, Chung-Chen and {Chilson}, Ryan and {Chuter}, Tim C. and {Cordova Rosado}, Rodrigo and {Coulson}, Iain M. and {Crawford}, Thomas M. and {Crowley}, Joseph and {David}, John and {Derome}, Mark and {Dexter}, Matthew and {Dornbusch}, Sven and {Dudevoir}, Kevin A. and {Dzib}, Sergio A. and {Eckart}, Andreas and {Eckert}, Chris and {Erickson}, Neal R. and {Everett}, Wendeline B. and {Faber}, Aaron and {Farah}, Joseph R. and {Fath}, Vernon and {Folkers}, Thomas W. and {Forbes}, David C. and {Freund}, Robert and {Gomez-Ruiz}, Arturo I. and {Gale}, David M. and {Gao}, Feng and {Geertsema}, Gertie and {Graham}, David A. and {Greer}, Christopher H. and {Grosslein}, Ronald and {Gueth}, Frederic and {Haggard}, Daryl and {Halverson}, Nils W. and {Han}, Chih-Chiang and {Han}, Kuo-Chang and {Hao}, Jinchi and {Hasegawa}, Yutaka and {Henning}, Jason W. and {Hernandez-Gomez}, Antonio and {Herrero-Illana}, Ruben and {Heyminck}, Stefan and {Hirota}, Akihiko and {Hoge}, James and {Huang}, Yau-De and {Impellizzeri}, C.~M. Violette and {Jiang}, Homin and {Kamble}, Atish and {Keisler}, Ryan and {Kimura}, Kimihiro and {Kono}, Yusuke and {Kubo}, Derek and {Kuroda}, John and {Lacasse}, Richard and {Laing}, Robert A. and {Leitch}, Erik M. and {Li}, Chao-Te and {Lin}, Lupin C. -C. and {Liu}, Ching-Tang and {Liu}, Kuan-Yu and {Lu}, Li-Ming and {Marson}, Ralph G. and {Martin-Cocher}, Pierre L. and {Massingill}, Kyle D. and {Matulonis}, Callie and {McColl}, Martin P. and {McWhirter}, Stephen R. and {Messias}, Hugo and {Meyer-Zhao}, Zheng and {Michalik}, Daniel and {Montana}, Alfredo and {Montgomerie}, William and {Mora-Klein}, Matias and {Muders}, Dirk and {Nadolski}, Andrew and {Navarro}, Santiago and {Neilsen}, Joseph and {Nguyen}, Chi H. and {Nishioka}, Hiroaki and {Norton}, Timothy and {Nowak}, Michael A. and {Nystrom}, George and {Ogawa}, Hideo and {Oshiro}, Peter and {Oyama}, Tomoaki and {Parsons}, Harriet and {Paine}, Scott N. and {Penalver}, Juan and {Phillips}, Neil M. and {Poirier}, Michael and {Pradel}, Nicolas and {Primiani}, Rurik A. and {Raffin}, Philippe A. and {Rahlin}, Alexandra S. and {Reiland}, George and {Risacher}, Christopher and {Ruiz}, Ignacio and {Saez-Madain}, Alejandro F. and {Sassella}, Remi and {Schellart}, Pim and {Shaw}, Paul and {Silva}, Kevin M. and {Shiokawa}, Hotaka and {Smith}, David R. and {Snow}, William and {Souccar}, Kamal and {Sousa}, Don and {Sridharan}, T.~K. and {Srinivasan}, Ranjani and {Stahm}, William and {Stark}, Anthony A. and {Story}, Kyle and {Timmer}, Sjoerd T. and {Vertatschitsch}, Laura and {Walther}, Craig and {Wei}, Ta-Shun and {Whitehorn}, Nathan and {Whitney}, Alan R. and {Woody}, David P. and {Wouterloot}, Jan G.~A. and {Wright}, Melvin and {Yamaguchi}, Paul and {Yu}, Chen-Yu and {Zeballos}, Milagros and {Zhang}, Shuo and {Ziurys}, Lucy},
        title = "{First M87 Event Horizon Telescope Results. I. The Shadow of the Supermassive Black Hole}",
      journal = {\apjl},
     keywords = {accretion, accretion disks, black hole physics, galaxies: active, galaxies: individual: M87, galaxies: jets, gravitation, Astrophysics - Astrophysics of Galaxies, Astrophysics - High Energy Astrophysical Phenomena, General Relativity and Quantum Cosmology},
         year = 2019,
        month = apr,
       volume = {875},
       number = {1},
          eid = {L1},
        pages = {L1},
          doi = {10.3847/2041-8213/ab0ec7},
archivePrefix = {arXiv},
       eprint = {1906.11238},
 primaryClass = {astro-ph.GA},
       adsurl = {https://ui.adsabs.harvard.edu/abs/2019ApJ...875L...1E},
      adsnote = {Provided by the SAO/NASA Astrophysics Data System}
}

@ARTICLE{Waisberg2019,
       author = {{Waisberg}, Idel and {Dexter}, Jason and {Olivier-Petrucci}, Pierre and {Dubus}, Guillaume and {Perraut}, Karine},
        title = "{Collimated radiation in SS 433. Constraints from spatially resolved optical jets and Cloudy modeling of the optical bullets}",
      journal = {\aap},
     keywords = {techniques: interferometric, line: formation, binaries: close, stars: jets, stars: individual: SS 433, Astrophysics - High Energy Astrophysical Phenomena, Astrophysics - Astrophysics of Galaxies, Astrophysics - Solar and Stellar Astrophysics},
         year = 2019,
        month = apr,
       volume = {624},
          eid = {A127},
        pages = {A127},
          doi = {10.1051/0004-6361/201834747},
archivePrefix = {arXiv},
       eprint = {1811.12564},
 primaryClass = {astro-ph.HE},
       adsurl = {https://ui.adsabs.harvard.edu/abs/2019A&A...624A.127W},
      adsnote = {Provided by the SAO/NASA Astrophysics Data System}
}

@ARTICLE{Zhu2016,
       author = {{Zhu}, Chunming and {Liu}, Rui and {Alexander}, David and {McAteer}, R.~T. James},
        title = "{Observation of the Evolution of a Current Sheet in a Solar Flare}",
      journal = {\apjl},
     keywords = {Sun: corona, Sun: flares, Astrophysics - Solar and Stellar Astrophysics},
         year = 2016,
        month = apr,
       volume = {821},
       number = {2},
          eid = {L29},
        pages = {L29},
          doi = {10.3847/2041-8205/821/2/L29},
archivePrefix = {arXiv},
       eprint = {1603.07062},
 primaryClass = {astro-ph.SR},
       adsurl = {https://ui.adsabs.harvard.edu/abs/2016ApJ...821L..29Z},
      adsnote = {Provided by the SAO/NASA Astrophysics Data System}
}

@ARTICLE{Mandrini1996,
       author = {{Mandrini}, C.~H. and {Demoulin}, P. and {Van Driel-Gesztelyi}, L. and {Schmieder}, B. and {Cauzzi}, G. and {Hofmann}, A.},
        title = "{3D Magnetic Reconnection at an X-Ray Bright Point}",
      journal = {\solphys},
     keywords = {Flare, Magnetic Reconnection, Coronal Loop, Photospheric Magnetic Field, Free Magnetic Energy},
         year = 1996,
        month = sep,
       volume = {168},
       number = {1},
        pages = {115-133},
          doi = {10.1007/BF00145829},
       adsurl = {https://ui.adsabs.harvard.edu/abs/1996SoPh..168..115M},
      adsnote = {Provided by the SAO/NASA Astrophysics Data System}
}

@article{Jel2023,
author = {Jel, P. and Bárta, M. and Holesovickach, V},
year = {2023},
month = {05},
pages = {},
title = {MHD SIMULATIONS IN PLASMA PHYSICS}
}

@ARTICLE{Boyce2022,
       author = {{Boyce}, H. and {Haggard}, D. and {Witzel}, G. and {von Fellenberg}, S. and {Willner}, S.~P. and {Becklin}, E.~E. and {Do}, T. and {Eckart}, A. and {Fazio}, G.~G. and {Gurwell}, M.~A. and {Hora}, J.~L. and {Markoff}, S. and {Morris}, M.~R. and {Neilsen}, J. and {Nowak}, M. and {Smith}, H.~A. and {Zhang}, S.},
        title = "{Multiwavelength Variability of Sagittarius A* in 2019 July}",
      journal = {\apj},
     keywords = {Galactic center, Black hole physics, Accretion, Non-thermal radiation sources, Supermassive black holes, 565, 159, 14, 1119, 1663, Astrophysics - High Energy Astrophysical Phenomena},
         year = 2022,
        month = may,
       volume = {931},
       number = {1},
          eid = {7},
        pages = {7},
          doi = {10.3847/1538-4357/ac6104},
archivePrefix = {arXiv},
       eprint = {2203.13311},
 primaryClass = {astro-ph.HE},
       adsurl = {https://ui.adsabs.harvard.edu/abs/2022ApJ...931....7B},
      adsnote = {Provided by the SAO/NASA Astrophysics Data System}
}

@ARTICLE{El_Mellah2022,
       author = {{El Mellah}, I. and {Cerutti}, B. and {Crinquand}, B. and {Parfrey}, K.},
        title = "{Spinning black holes magnetically connected to a Keplerian disk. Magnetosphere, reconnection sheet, particle acceleration, and coronal heating}",
      journal = {\aap},
     keywords = {acceleration of particles, magnetic reconnection, black hole physics, radiation mechanisms: non-thermal, methods: numerical, relativistic processes, Astrophysics - High Energy Astrophysical Phenomena},
         year = 2022,
        month = jul,
       volume = {663},
          eid = {A169},
        pages = {A169},
          doi = {10.1051/0004-6361/202142847},
archivePrefix = {arXiv},
       eprint = {2112.03933},
 primaryClass = {astro-ph.HE},
       adsurl = {https://ui.adsabs.harvard.edu/abs/2022A&A...663A.169E},
      adsnote = {Provided by the SAO/NASA Astrophysics Data System}
}

@ARTICLE{Dexter2020,
       author = {{Dexter}, J. and {Tchekhovskoy}, A. and {Jimenez-Rosales}, A. and {Ressler}, S.~M. and {Baubock}, M. and {Dallilar}, Y. and {de Zeeuw}, P.~T. and {Eisenhauer}, F. and {von Fellenberg}, S. and {Gao}, F. and {Genzel}, R. and {Gillessen}, S. and {Habibi}, M. and {Ott}, T. and {Stadler}, J. and {Straub}, O. and {Widmann}, F.},
        title = "{Sgr A* near-infrared flares from reconnection events in a magnetically arrested disc}",
      journal = {\mnras},
     keywords = {accretion, accretion discs, black hole physics, MHD, polarization, radiative transfer, Galaxy: centre, Astrophysics - High Energy Astrophysical Phenomena, Astrophysics - Astrophysics of Galaxies},
         year = 2020,
        month = oct,
       volume = {497},
       number = {4},
        pages = {4999-5007},
          doi = {10.1093/mnras/staa2288},
archivePrefix = {arXiv},
       eprint = {2006.03657},
 primaryClass = {astro-ph.HE},
       adsurl = {https://ui.adsabs.harvard.edu/abs/2020MNRAS.497.4999D},
      adsnote = {Provided by the SAO/NASA Astrophysics Data System}
}

@article{dmytriiev2021,
    author = {Dmytriiev, A and Sol, H and Zech, A},
    title = "{Connecting steady emission and very high energy flaring states in blazars: the case of Mrk 421}",
    journal = {Monthly Notices of the Royal Astronomical Society},
    volume = {505},
    number = {2},
    pages = {2712-2730},
    year = {2021},
    month = {05},
    abstract = "{Various attempts have been made in the literature at describing the origin and the physical mechanisms behind flaring events in blazars with radiative emission models, but detailed properties of multiwavelength (MWL) light curves still remain difficult to reproduce. We have developed a versatile radiative code, based on a time-dependent treatment of particle acceleration, escape, and radiative cooling, allowing us to test different scenarios to connect the continuous low-state emission self-consistently with that during flaring states. We consider flares as weak perturbations of the quiescent state and apply this description to the 2010 February MWL flare of Mrk 421, the brightest very high energy (VHE) flare ever detected from this archetypal blazar, focusing on interpretations with a minimum number of free parameters. A general criterion is obtained, which disfavours a one-zone model connecting low and high state under our assumptions. A two-zone model combining physically connected acceleration and emission regions yields a satisfactory interpretation of the available time-dependent MWL light curves and spectra of Mrk 421, although certain details remain difficult to reproduce. The two-zone scenario finally proposed for the complex quiescent and flaring VHE emitting region involves both Fermi-I and Fermi-II acceleration mechanisms, respectively, at the origin of the quiescent and flaring emission.}",
    issn = {0035-8711},
    doi = {10.1093/mnras/stab1445},
    url = {https://doi.org/10.1093/mnras/stab1445},
    eprint = {https://academic.oup.com/mnras/article-pdf/505/2/2712/38611659/stab1445.pdf},
}

@ARTICLE{Ripperda2020,
       author = {{Ripperda}, Bart and {Bacchini}, Fabio and {Philippov}, Alexander A.},
        title = "{Magnetic Reconnection and Hot Spot Formation in Black Hole Accretion Disks}",
      journal = {\apj},
     keywords = {Black Hole physics, Accretion, Magnetohydrodynamics, General relativity, Plasma astrophysics, 159, 14, 1964, 641, 1261, Astrophysics - High Energy Astrophysical Phenomena, General Relativity and Quantum Cosmology, Physics - Plasma Physics},
         year = 2020,
        month = sep,
       volume = {900},
       number = {2},
          eid = {100},
        pages = {100},
          doi = {10.3847/1538-4357/ababab},
archivePrefix = {arXiv},
       eprint = {2003.04330},
 primaryClass = {astro-ph.HE},
       adsurl = {https://ui.adsabs.harvard.edu/abs/2020ApJ...900..100R},
      adsnote = {Provided by the SAO/NASA Astrophysics Data System}
}

@ARTICLE{Ball_model,
       author = {{Ball}, David and {Ozel}, Feryal and {Christian}, Pierre and {Chan}, Chi-Kwan and {Psaltis}, Dimitrios},
        title = "{A Plasmoid model for the Sgr A* Flares Observed With Gravity and CHANDRA}",
      journal = {\apj},
     keywords = {Black Hole physics, High energy astrophysics, 159, 739, Astrophysics - High Energy Astrophysical Phenomena},
         year = 2021,
        month = aug,
       volume = {917},
       number = {1},
          eid = {8},
        pages = {8},
          doi = {10.3847/1538-4357/abf8ae},
archivePrefix = {arXiv},
       eprint = {2005.14251},
 primaryClass = {astro-ph.HE},
       adsurl = {https://ui.adsabs.harvard.edu/abs/2021ApJ...917....8B},
      adsnote = {Provided by the SAO/NASA Astrophysics Data System}
}

@ARTICLE{Vranic2016,
       author = {{Vranic}, M. and {Martins}, J.~L. and {Fonseca}, R.~A. and {Silva}, L.~O.},
        title = "{Classical radiation reaction in particle-in-cell simulations}",
      journal = {Computer Physics Communications},
     keywords = {Radiation reaction, Particle-in-cell, Laser-matter interactions, Relativistic electron motion, Physics - Plasma Physics},
         year = 2016,
        month = jul,
       volume = {204},
        pages = {141-151},
          doi = {10.1016/j.cpc.2016.04.002},
archivePrefix = {arXiv},
       eprint = {1502.02432},
 primaryClass = {physics.plasm-ph},
       adsurl = {https://ui.adsabs.harvard.edu/abs/2016CoPhC.204..141V},
      adsnote = {Provided by the SAO/NASA Astrophysics Data System}
}

@ARTICLE{Ripperda2019,
       author = {{Ripperda}, B. and {Bacchini}, F. and {Porth}, O. and {Most}, E.~R. and {Olivares}, H. and {Nathanail}, A. and {Rezzolla}, L. and {Teunissen}, J. and {Keppens}, R.},
        title = "{General-relativistic Resistive Magnetohydrodynamics with Robust Primitive-variable Recovery for Accretion Disk Simulations}",
      journal = {\apjs},
     keywords = {Black hole physics, accretion, magnetohydrodynamics, general relativity, computational methods, plasma astrophysics, 159, 14, 1964, 641, 1965, 1261, Physics - Computational Physics, Astrophysics - High Energy Astrophysical Phenomena, Astrophysics - Instrumentation and Methods for Astrophysics, General Relativity and Quantum Cosmology, Physics - Plasma Physics},
         year = 2019,
        month = sep,
       volume = {244},
       number = {1},
          eid = {10},
        pages = {10},
          doi = {10.3847/1538-4365/ab3922},
archivePrefix = {arXiv},
       eprint = {1907.07197},
 primaryClass = {physics.comp-ph},
       adsurl = {https://ui.adsabs.harvard.edu/abs/2019ApJS..244...10R},
      adsnote = {Provided by the SAO/NASA Astrophysics Data System}
}

@ARTICLE{Crinquand2022,
       author = {{Crinquand}, Benjamin and {Cerutti}, Benoit and {Dubus}, Guillaume and {Parfrey}, Kyle and {Philippov}, Alexander},
        title = "{Synthetic Images of Magnetospheric Reconnection-Powered Radiation around Supermassive Black Holes}",
      journal = {\prl},
     keywords = {Astrophysics - High Energy Astrophysical Phenomena},
         year = 2022,
        month = nov,
       volume = {129},
       number = {20},
          eid = {205101},
        pages = {205101},
          doi = {10.1103/PhysRevLett.129.205101},
archivePrefix = {arXiv},
       eprint = {2202.04472},
 primaryClass = {astro-ph.HE},
       adsurl = {https://ui.adsabs.harvard.edu/abs/2022PhRvL.129t5101C},
      adsnote = {Provided by the SAO/NASA Astrophysics Data System}
}

@ARTICLE{Uzdensky2005,
       author = {{Uzdensky}, Dmitri A.},
        title = "{Force-Free Magnetosphere of an Accretion Disk-Black Hole System. II. Kerr Geometry}",
      journal = {\apj},
     keywords = {Accretion, Accretion Disks, Black Hole Physics, Galaxies: Active, Magnetic Fields, Magnetohydrodynamics: MHD, Astrophysics},
         year = 2005,
        month = feb,
       volume = {620},
       number = {2},
        pages = {889-904},
          doi = {10.1086/427180},
archivePrefix = {arXiv},
       eprint = {astro-ph/0410715},
 primaryClass = {astro-ph},
       adsurl = {https://ui.adsabs.harvard.edu/abs/2005ApJ...620..889U},
      adsnote = {Provided by the SAO/NASA Astrophysics Data System}
}

@ARTICLE{de_Gouveia_dal_Pino2005,
       author = {{de Gouveia dal Pino}, E.~M. and {Lazarian}, A.},
        title = "{Production of the large scale superluminal ejections of the microquasar GRS 1915+105 by violent magnetic reconnection}",
      journal = {\aap},
     keywords = {acceleration of particles, accretion, accretion disks, black hole physics, magnetic fields},
         year = 2005,
        month = oct,
       volume = {441},
       number = {3},
        pages = {845-853},
          doi = {10.1051/0004-6361:20042590},
       adsurl = {https://ui.adsabs.harvard.edu/abs/2005A&A...441..845D},
      adsnote = {Provided by the SAO/NASA Astrophysics Data System}
}

@ARTICLE{Parfrey2015,
       author = {{Parfrey}, K. and {Giannios}, D. and {Beloborodov}, A.~M.},
        title = "{Black hole jets without large-scale net magnetic flux.}",
      journal = {\mnras},
     keywords = {black hole physics, MHD, gamma-ray burst: general, galaxies: active, galaxies: jets, X-rays: binaries, Astrophysics - High Energy Astrophysical Phenomena},
         year = 2015,
        month = jan,
       volume = {446},
        pages = {L61-L65},
          doi = {10.1093/mnrasl/slu162},
archivePrefix = {arXiv},
       eprint = {1410.0374},
 primaryClass = {astro-ph.HE},
       adsurl = {https://ui.adsabs.harvard.edu/abs/2015MNRAS.446L..61P},
      adsnote = {Provided by the SAO/NASA Astrophysics Data System}
}

@ARTICLE{Crinquand2021,
       author = {{Crinquand}, B. and {Cerutti}, B. and {Dubus}, G. and {Parfrey}, K. and {Philippov}, A.},
        title = "{Synthetic gamma-ray light curves of Kerr black hole magnetospheric activity from particle-in-cell simulations}",
      journal = {\aap},
     keywords = {black hole physics, magnetic fields, acceleration of particles, plasmas, radiation mechanisms: non-thermal, methods: numerical, Astrophysics - High Energy Astrophysical Phenomena},
         year = 2021,
        month = jun,
       volume = {650},
          eid = {A163},
        pages = {A163},
          doi = {10.1051/0004-6361/202040158},
archivePrefix = {arXiv},
       eprint = {2012.09733},
 primaryClass = {astro-ph.HE},
       adsurl = {https://ui.adsabs.harvard.edu/abs/2021A&A...650A.163C},
      adsnote = {Provided by the SAO/NASA Astrophysics Data System}
}

@ARTICLE{Nalewajko2015,
       author = {{Nalewajko}, Krzysztof and {Uzdensky}, Dmitri A. and {Cerutti}, Benoit and {Werner}, Gregory R. and {Begelman}, Mitchell C.},
        title = "{On the Distribution of Particle Acceleration Sites in Plasmoid-dominated Relativistic Magnetic Reconnection}",
      journal = {\apj},
     keywords = {acceleration of particles, magnetic reconnection, Astrophysics - High Energy Astrophysical Phenomena},
         year = 2015,
        month = dec,
       volume = {815},
       number = {2},
          eid = {101},
        pages = {101},
          doi = {10.1088/0004-637X/815/2/101},
archivePrefix = {arXiv},
       eprint = {1508.02392},
 primaryClass = {astro-ph.HE},
       adsurl = {https://ui.adsabs.harvard.edu/abs/2015ApJ...815..101N},
      adsnote = {Provided by the SAO/NASA Astrophysics Data System}
}

@ARTICLE{Wielgus2022,
       author = {{Wielgus}, M. and {Moscibrodzka}, M. and {Vos}, J. and {Gelles}, Z. and {Marti-Vidal}, I. and {Farah}, J. and {Marchili}, N. and {Goddi}, C. and {Messias}, H.},
        title = "{Orbital motion near Sagittarius A$^{*}$ . Constraints from polarimetric ALMA observations}",
      journal = {\aap},
     keywords = {Galaxy: nucleus, Galaxy: center, black hole physics, gravitational lensing: strong, polarization, magnetic reconnection, Astrophysics - High Energy Astrophysical Phenomena},
         year = 2022,
        month = sep,
       volume = {665},
          eid = {L6},
        pages = {L6},
          doi = {10.1051/0004-6361/202244493},
archivePrefix = {arXiv},
       eprint = {2209.09926},
 primaryClass = {astro-ph.HE},
       adsurl = {https://ui.adsabs.harvard.edu/abs/2022A&A...665L...6W},
      adsnote = {Provided by the SAO/NASA Astrophysics Data System}
}

@ARTICLE{El_Mellah2023,
       author = {{El Mellah}, I. and {Cerutti}, B. and {Crinquand}, B.},
        title = "{Reconnection-driven flares in 3D black hole magnetospheres -- A scenario for hot spots around Sagittarius A*}",
      journal = {arXiv e-prints},
     keywords = {Astrophysics - High Energy Astrophysical Phenomena},
         year = 2023,
        month = may,
          eid = {arXiv:2305.01689},
        pages = {arXiv:2305.01689},
          doi = {10.48550/arXiv.2305.01689},
archivePrefix = {arXiv},
       eprint = {2305.01689},
 primaryClass = {astro-ph.HE},
       adsurl = {https://ui.adsabs.harvard.edu/abs/2023arXiv230501689E},
      adsnote = {Provided by the SAO/NASA Astrophysics Data System}
}

@ARTICLE{Cemeljic2022,
       author = {{{\v{C}}emeljic}, Miljenko and {Yang}, Hai and {Yuan}, Feng and {Shang}, Hsien},
        title = "{Formation of Episodic Jets and Associated Flares from Black Hole Accretion Systems}",
      journal = {\apj},
     keywords = {High energy astrophysics, 739, Astrophysics - High Energy Astrophysical Phenomena},
         year = 2022,
        month = jul,
       volume = {933},
       number = {1},
          eid = {55},
        pages = {55},
          doi = {10.3847/1538-4357/ac70cc},
archivePrefix = {arXiv},
       eprint = {2205.10531},
 primaryClass = {astro-ph.HE},
       adsurl = {https://ui.adsabs.harvard.edu/abs/2022ApJ...933...55C},
      adsnote = {Provided by the SAO/NASA Astrophysics Data System}
}

@ARTICLE{Nathanail2020,
       author = {{Nathanail}, Antonios and {Fromm}, Christian M. and {Porth}, Oliver and {Olivares}, Hector and {Younsi}, Ziri and {Mizuno}, Yosuke and {Rezzolla}, Luciano},
        title = "{Plasmoid formation in global GRMHD simulations and AGN flares}",
      journal = {\mnras},
     keywords = {accretion, accretion discs, black hole physics, magnetic reconnection, Astrophysics - High Energy Astrophysical Phenomena, General Relativity and Quantum Cosmology},
         year = 2020,
        month = jun,
       volume = {495},
       number = {2},
        pages = {1549-1565},
          doi = {10.1093/mnras/staa1165},
archivePrefix = {arXiv},
       eprint = {2002.01777},
 primaryClass = {astro-ph.HE},
       adsurl = {https://ui.adsabs.harvard.edu/abs/2020MNRAS.495.1549N},
      adsnote = {Provided by the SAO/NASA Astrophysics Data System}
}

@ARTICLE{Porth2021,
       author = {{Porth}, O. and {Mizuno}, Y. and {Younsi}, Z. and {Fromm}, C.~M.},
        title = "{Flares in the Galactic Centre - I. Orbiting flux tubes in magnetically arrested black hole accretion discs}",
      journal = {\mnras},
     keywords = {accretion, accretion discs, black hole physics, magnetic field, MHD, methods: numerical, Astrophysics - High Energy Astrophysical Phenomena},
         year = 2021,
        month = apr,
       volume = {502},
       number = {2},
        pages = {2023-2032},
          doi = {10.1093/mnras/stab163},
archivePrefix = {arXiv},
       eprint = {2006.03658},
 primaryClass = {astro-ph.HE},
       adsurl = {https://ui.adsabs.harvard.edu/abs/2021MNRAS.502.2023P},
      adsnote = {Provided by the SAO/NASA Astrophysics Data System}
}

@ARTICLE{Rowan2017,
       author = {{Rowan}, Michael E. and {Sironi}, Lorenzo and {Narayan}, Ramesh},
        title = "{Electron and Proton Heating in Transrelativistic Magnetic Reconnection}",
      journal = {\apj},
     keywords = {acceleration of particles, accretion, accretion disks, galaxies: jets, magnetic reconnection, radiation mechanisms: non-thermal, X-rays: binaries, Astrophysics - High Energy Astrophysical Phenomena, Physics - Plasma Physics},
         year = 2017,
        month = nov,
       volume = {850},
       number = {1},
          eid = {29},
        pages = {29},
          doi = {10.3847/1538-4357/aa9380},
archivePrefix = {arXiv},
       eprint = {1708.04627},
 primaryClass = {astro-ph.HE},
       adsurl = {https://ui.adsabs.harvard.edu/abs/2017ApJ...850...29R},
      adsnote = {Provided by the SAO/NASA Astrophysics Data System}
}

@ARTICLE{Ball2018,
       author = {{Ball}, David and {Sironi}, Lorenzo and {Ozel}, Feryal},
        title = "{Electron and Proton Acceleration in Trans-relativistic Magnetic Reconnection: Dependence on Plasma Beta and Magnetization}",
      journal = {\apj},
     keywords = {accretion, accretion disks, galaxies: jets, magnetic reconnection, radiation mechanisms: nonthermal, X-rays: binaries, Astrophysics - High Energy Astrophysical Phenomena},
         year = 2018,
        month = jul,
       volume = {862},
       number = {1},
          eid = {80},
        pages = {80},
          doi = {10.3847/1538-4357/aac820},
archivePrefix = {arXiv},
       eprint = {1803.05556},
 primaryClass = {astro-ph.HE},
       adsurl = {https://ui.adsabs.harvard.edu/abs/2018ApJ...862...80B},
      adsnote = {Provided by the SAO/NASA Astrophysics Data System}
}

@ARTICLE{Bransgrove2021,
       author = {{Bransgrove}, Ashley and {Ripperda}, Bart and {Philippov}, Alexander},
        title = "{Magnetic Hair and Reconnection in Black Hole Magnetospheres}",
      journal = {\prl},
     keywords = {Astrophysics - High Energy Astrophysical Phenomena, General Relativity and Quantum Cosmology, Physics - Plasma Physics},
         year = 2021,
        month = jul,
       volume = {127},
       number = {5},
          eid = {055101},
        pages = {055101},
          doi = {10.1103/PhysRevLett.127.055101},
archivePrefix = {arXiv},
       eprint = {2109.14620},
 primaryClass = {astro-ph.HE},
       adsurl = {https://ui.adsabs.harvard.edu/abs/2021PhRvL.127e5101B},
      adsnote = {Provided by the SAO/NASA Astrophysics Data System}
}

@ARTICLE{Nathanail2022,
       author = {{Nathanail}, Antonios and {Mpisketzis}, Vasilis and {Porth}, Oliver and {Fromm}, Christian M. and {Rezzolla}, Luciano},
        title = "{Magnetic reconnection and plasmoid formation in three-dimensional accretion flows around black holes}",
      journal = {\mnras},
     keywords = {black hole physics, magnetic reconnection, accretion, accretion discs, magnetohydrodynamics, Astrophysics - High Energy Astrophysical Phenomena, General Relativity and Quantum Cosmology, Physics - Plasma Physics},
         year = 2022,
        month = jul,
       volume = {513},
       number = {3},
        pages = {4267-4277},
          doi = {10.1093/mnras/stac1118},
archivePrefix = {arXiv},
       eprint = {2111.03689},
 primaryClass = {astro-ph.HE},
       adsurl = {https://ui.adsabs.harvard.edu/abs/2022MNRAS.513.4267N},
      adsnote = {Provided by the SAO/NASA Astrophysics Data System}
}

@ARTICLE{Nattila2021,
       author = {{Nattila}, Joonas and {Beloborodov}, Andrei M.},
        title = "{Radiative Turbulent Flares in Magnetically Dominated Plasmas}",
      journal = {\apj},
     keywords = {Plasma astrophysics, High energy astrophysics, Astrophysical magnetism, Computational astronomy, Compact objects, Nonthermal radiation sources, 1261, 739, 102, 293, 288, 1119, Astrophysics - High Energy Astrophysical Phenomena, Physics - Plasma Physics},
         year = 2021,
        month = nov,
       volume = {921},
       number = {1},
          eid = {87},
        pages = {87},
          doi = {10.3847/1538-4357/ac1c76},
archivePrefix = {arXiv},
       eprint = {2012.03043},
 primaryClass = {astro-ph.HE},
       adsurl = {https://ui.adsabs.harvard.edu/abs/2021ApJ...921...87N},
      adsnote = {Provided by the SAO/NASA Astrophysics Data System}
}

@ARTICLE{Zhang2021,
       author = {{Zhang}, Hao and {Sironi}, Lorenzo and {Giannios}, Dimitrios},
        title = "{Fast Particle Acceleration in Three-dimensional Relativistic Reconnection}",
      journal = {\apj},
     keywords = {739, Astrophysics - High Energy Astrophysical Phenomena},
         year = 2021,
        month = dec,
       volume = {922},
       number = {2},
          eid = {261},
        pages = {261},
          doi = {10.3847/1538-4357/ac2e08},
archivePrefix = {arXiv},
       eprint = {2105.00009},
 primaryClass = {astro-ph.HE},
       adsurl = {https://ui.adsabs.harvard.edu/abs/2021ApJ...922..261Z},
      adsnote = {Provided by the SAO/NASA Astrophysics Data System}
}

@ARTICLE{Werner2018,
       author = {{Werner}, G.~R. and {Uzdensky}, D.~A. and {Begelman}, M.~C. and {Cerutti}, B. and {Nalewajko}, K.},
        title = "{Non-thermal particle acceleration in collisionless relativistic electron-proton reconnection}",
      journal = {\mnras},
     keywords = {acceleration of particles, accretion, accretion discs, magnetic reconnection, relativistic processes, BL Lacertae objects: general, X-rays: binaries, Astrophysics - High Energy Astrophysical Phenomena},
         year = 2018,
        month = feb,
       volume = {473},
       number = {4},
        pages = {4840-4861},
          doi = {10.1093/mnras/stx2530},
archivePrefix = {arXiv},
       eprint = {1612.04493},
 primaryClass = {astro-ph.HE},
       adsurl = {https://ui.adsabs.harvard.edu/abs/2018MNRAS.473.4840W},
      adsnote = {Provided by the SAO/NASA Astrophysics Data System}
}

@ARTICLE{Dodds-Eden2009,
       author = {{Dodds-Eden}, K. and {Porquet}, D. and {Trap}, G. and {Quataert}, E. and {Haubois}, X. and {Gillessen}, S. and {Grosso}, N. and {Pantin}, E. and {Falcke}, H. and {Rouan}, D. and {Genzel}, R. and {Hasinger}, G. and {Goldwurm}, A. and {Yusef-Zadeh}, F. and {Clenet}, Y. and {Trippe}, S. and {Lagage}, P. -O. and {Bartko}, H. and {Eisenhauer}, F. and {Ott}, T. and {Paumard}, T. and {Perrin}, G. and {Yuan}, F. and {Fritz}, T.~K. and {Mascetti}, L.},
        title = "{Evidence for X-Ray Synchrotron Emission from Simultaneous Mid-Infrared to X-Ray Observations of a Strong Sgr A* Flare}",
      journal = {\apj},
     keywords = {accretion, accretion disks, black hole physics, Galaxy: center, infrared: general, radiation mechanisms: general, X-rays: general, Astrophysics - Galaxy Astrophysics, Astrophysics - High Energy Astrophysical Phenomena},
         year = 2009,
        month = jun,
       volume = {698},
       number = {1},
        pages = {676-692},
          doi = {10.1088/0004-637X/698/1/676},
archivePrefix = {arXiv},
       eprint = {0903.3416},
 primaryClass = {astro-ph.GA},
       adsurl = {https://ui.adsabs.harvard.edu/abs/2009ApJ...698..676D},
      adsnote = {Provided by the SAO/NASA Astrophysics Data System}
}

@ARTICLE{Eckart2006,
       author = {{Eckart}, A. and {Baganoff}, F.~K. and {Schodel}, R. and {Morris}, M. and {Genzel}, R. and {Bower}, G.~C. and {Marrone}, D. and {Moran}, J.~M. and {Viehmann}, T. and {Bautz}, M.~W. and {Brandt}, W.~N. and {Garmire}, G.~P. and {Ott}, T. and {Trippe}, S. and {Ricker}, G.~R. and {Straubmeier}, C. and {Roberts}, D.~A. and {Yusef-Zadeh}, F. and {Zhao}, J.~H. and {Rao}, R.},
        title = "{The flare activity of Sagittarius A*. New coordinated mm to X-ray observations}",
      journal = aap},
     keywords = {Astrophysics},
         year = 2006,
        month = may,
       volume = {450},
       number = {2},
        pages = {535-555},
          doi = {10.1051/0004-6361:20054418},
archivePrefix = {arXiv},
       eprint = {astro-ph/0512440},
 primaryClass = {astro-ph},
       adsurl = {https://ui.adsabs.harvard.edu/abs/2006A&A...450..535E},
      adsnote = {Provided by the SAO/NASA Astrophysics Data System}
}

@ARTICLE{Yusef-Zadeh2006,
       author = {{Yusef-Zadeh}, F. and {Bushouse}, H. and {Dowell}, C.~D. and {Wardle}, M. and {Roberts}, D. and {Heinke}, C. and {Bower}, G.~C. and {Vila-Vilaro}, B. and {Shapiro}, S. and {Goldwurm}, A. and {Belanger}, G.},
        title = "{A Multiwavelength Study of Sgr A*: The Role of Near-IR Flares in Production of X-Ray, Soft {\ensuremath{\gamma}}-Ray, and Submillimeter Emission}",
      journal = {\apj},
     keywords = {Accretion, Accretion Disks, Black Hole Physics, Galaxies: Nuclei, Galaxy: Center, Astrophysics},
         year = 2006,
        month = jun,
       volume = {644},
       number = {1},
        pages = {198-213},
          doi = {10.1086/503287},
archivePrefix = {arXiv},
       eprint = {astro-ph/0510787},
 primaryClass = {astro-ph},
       adsurl = {https://ui.adsabs.harvard.edu/abs/2006ApJ...644..198Y},
      adsnote = {Provided by the SAO/NASA Astrophysics Data System}
}

@ARTICLE{von_Fellenberg2023,
       author = {{von Fellenberg}, S.~D. and {Witzel}, G. and {Baubock}, M. and {Chung}, H. -H. and {Aimar}, N. and {Bordoni}, M. and {Drescher}, A. and {Eisenhauer}, F. and {Genzel}, R. and {Gillessen}, S. and {Marchili}, N. and {Paumard}, T. and {Perrin}, G. and {Ott}, T. and {Ribeiro}, D.~C. and {Ros}, E. and {Vincent}, F. and {Widmann}, F. and {Willner}, S.~P. and {Anton Zensus}, J.},
        title = "{General relativistic effects and the near-infrared and X-ray variability of Sgr A* I}",
      journal = {\aap},
     keywords = {acceleration of particles, accretion, accretion disks, black hole physics, Galaxy: center, Astrophysics - High Energy Astrophysical Phenomena},
         year = 2023,
        month = jan,
       volume = {669},
          eid = {L17},
        pages = {L17},
          doi = {10.1051/0004-6361/202245575},
archivePrefix = {arXiv},
       eprint = {2301.02558},
 primaryClass = {astro-ph.HE},
       adsurl = {https://ui.adsabs.harvard.edu/abs/2023A&A...669L..17V},
      adsnote = {Provided by the SAO/NASA Astrophysics Data System}
}

@ARTICLE{Eckart2006a,
       author = {{Eckart}, A. and {Baganoff}, F.~K. and {Schodel}, R. and {Morris}, M. and {Genzel}, R. and {Bower}, G.~C. and {Marrone}, D. and {Moran}, J.~M. and {Viehmann}, T. and {Bautz}, M.~W. and {Brandt}, W.~N. and {Garmire}, G.~P. and {Ott}, T. and {Trippe}, S. and {Ricker}, G.~R. and {Straubmeier}, C. and {Roberts}, D.~A. and {Yusef-Zadeh}, F. and {Zhao}, J.~H. and {Rao}, R.},
        title = "{The flare activity of Sagittarius A*. New coordinated mm to X-ray observations}",
      journal = {\aap},
     keywords = {Astrophysics},
         year = 2006,
        month = may,
       volume = {450},
       number = {2},
        pages = {535-555},
          doi = {10.1051/0004-6361:20054418},
archivePrefix = {arXiv},
       eprint = {astro-ph/0512440},
 primaryClass = {astro-ph},
       adsurl = {https://ui.adsabs.harvard.edu/abs/2006A&A...450..535E},
      adsnote = {Provided by the SAO/NASA Astrophysics Data System}
}

@ARTICLE{Michail2021,
       author = {{Michail}, Joseph M. and {Wardle}, Mark and {Yusef-Zadeh}, Farhad and {Kunneriath}, Devaky},
        title = "{Multiwavelength Observations of Sgr A*. I. 2019 July 18}",
      journal = {\apj},
     keywords = {565, 16, 1663, Astrophysics - High Energy Astrophysical Phenomena, Astrophysics - Astrophysics of Galaxies},
         year = 2021,
        month = dec,
       volume = {923},
       number = {1},
          eid = {54},
        pages = {54},
          doi = {10.3847/1538-4357/ac2d2c},
archivePrefix = {arXiv},
       eprint = {2107.09681},
 primaryClass = {astro-ph.HE},
       adsurl = {https://ui.adsabs.harvard.edu/abs/2021ApJ...923...54M},
      adsnote = {Provided by the SAO/NASA Astrophysics Data System}
}

@ARTICLE{Boyce2022,
       author = {{Boyce}, H. and {Haggard}, D. and {Witzel}, G. and {von Fellenberg}, S. and {Willner}, S.~P. and {Becklin}, E.~E. and {Do}, T. and {Eckart}, A. and {Fazio}, G.~G. and {Gurwell}, M.~A. and {Hora}, J.~L. and {Markoff}, S. and {Morris}, M.~R. and {Neilsen}, J. and {Nowak}, M. and {Smith}, H.~A. and {Zhang}, S.},
        title = "{Multiwavelength Variability of Sagittarius A* in 2019 July}",
      journal = {\apj},
     keywords = {Galactic center, Black hole physics, Accretion, Non-thermal radiation sources, Supermassive black holes, 565, 159, 14, 1119, 1663, Astrophysics - High Energy Astrophysical Phenomena},
         year = 2022,
        month = may,
       volume = {931},
       number = {1},
          eid = {7},
        pages = {7},
          doi = {10.3847/1538-4357/ac6104},
archivePrefix = {arXiv},
       eprint = {2203.13311},
 primaryClass = {astro-ph.HE},
       adsurl = {https://ui.adsabs.harvard.edu/abs/2022ApJ...931....7B},
      adsnote = {Provided by the SAO/NASA Astrophysics Data System}
}

@ARTICLE{Fazio2018,
       author = {{Fazio}, G.~G. and {Hora}, J.~L. and {Witzel}, G. and {Willner}, S.~P. and {Ashby}, M.~L.~N. and {Baganoff}, F. and {Becklin}, E. and {Carey}, S. and {Haggard}, D. and {Gammie}, C. and {Ghez}, A. and {Gurwell}, M.~A. and {Ingalls}, J. and {Marrone}, D. and {Morris}, M.~R. and {Smith}, H.~A.},
        title = "{Multiwavelength Light Curves of Two Remarkable Sagittarius A* Flares}",
      journal = {\apj},
     keywords = {accretion, accretion disks, black hole physics, Galaxy: center, infrared: general, submillimeter: general, X-rays: individual: Sgr Aa, Astrophysics - High Energy Astrophysical Phenomena},
         year = 2018,
        month = sep,
       volume = {864},
       number = {1},
          eid = {58},
        pages = {58},
          doi = {10.3847/1538-4357/aad4a2},
archivePrefix = {arXiv},
       eprint = {1807.07599},
 primaryClass = {astro-ph.HE},
       adsurl = {https://ui.adsabs.harvard.edu/abs/2018ApJ...864...58F},
      adsnote = {Provided by the SAO/NASA Astrophysics Data System}
}

@ARTICLE{Gravity2021,
       author = {{GRAVITY Collaboration} and {Abuter}, R. and {Amorim}, A. and {Baubock}, M. and {Baganoff}, F. and {Berger}, J.~P. and {Boyce}, H. and {Bonnet}, H. and {Brandner}, W. and {Clenet}, Y. and {Davies}, R. and {de Zeeuw}, P.~T. and {Dexter}, J. and {Dallilar}, Y. and {Drescher}, A. and {Eckart}, A. and {Eisenhauer}, F. and {Fazio}, G.~G. and {Forster Schreiber}, N.~M. and {Foster}, K. and {Gammie}, C. and {Garcia}, P. and {Gao}, F. and {Gendron}, E. and {Genzel}, R. and {Ghisellini}, G. and {Gillessen}, S. and {Gurwell}, M.~A. and {Habibi}, M. and {Haggard}, D. and {Hailey}, C. and {Harrison}, F.~A. and {Haubois}, X. and {Heissel}, G. and {Henning}, T. and {Hippler}, S. and {Hora}, J.~L. and {Horrobin}, M. and {Jimenez-Rosales}, A. and {Jochum}, L. and {Jocou}, L. and {Kaufer}, A. and {Kervella}, P. and {Lacour}, S. and {Lapeyrere}, V. and {Le Bouquin}, J. -B. and {Lena}, P. and {Lowrance}, P.~J. and {Lutz}, D. and {Markoff}, S. and {Mori}, K. and {Morris}, M.~R. and {Neilsen}, J. and {Nowak}, M. and {Ott}, T. and {Paumard}, T. and {Perraut}, K. and {Perrin}, G. and {Ponti}, G. and {Pfuhl}, O. and {Rabien}, S. and {Rodriguez-Coira}, G. and {Shangguan}, J. and {Shimizu}, T. and {Scheithauer}, S. and {Smith}, H.~A. and {Stadler}, J. and {Stern}, D.~K. and {Straub}, O. and {Straubmeier}, C. and {Sturm}, E. and {Tacconi}, L.~J. and {Vincent}, F. and {von Fellenberg}, S.~D. and {Waisberg}, I. and {Widmann}, F. and {Wieprecht}, E. and {Wiezorrek}, E. and {Willner}, S.~P. and {Witzel}, G. and {Woillez}, J. and {Yazici}, S. and {Young}, A. and {Zhang}, S. and {Zins}, G.},
        title = "{Constraining particle acceleration in Sgr A$^{{\ensuremath{\star}}}$ with simultaneous GRAVITY, Spitzer, NuSTAR, and Chandra observations}",
      journal = {\aap},
     keywords = {Galaxy: center, accretion, accretion disks, black hole physics, Astrophysics - High Energy Astrophysical Phenomena},
         year = 2021,
        month = oct,
       volume = {654},
          eid = {A22},
        pages = {A22},
          doi = {10.1051/0004-6361/202140981},
archivePrefix = {arXiv},
       eprint = {2107.01096},
 primaryClass = {astro-ph.HE},
       adsurl = {https://ui.adsabs.harvard.edu/abs/2021A&A...654A..22G},
      adsnote = {Provided by the SAO/NASA Astrophysics Data System}
}

@BOOK{Melrose1991,
       author = {{Melrose}, D.~B. and {McPhedran}, R.~C.},
        title = "{Electromagnetic Processes in Dispersive Media}",
         year = 1991,
       adsurl = {https://ui.adsabs.harvard.edu/abs/1991epdm.book.....M},
      adsnote = {Provided by the SAO/NASA Astrophysics Data System}
}

@ARTICLE{Landi1985,
       author = {{Landi Degl'Innocenti}, E. and {Landi Degl'Innocenti}, M.},
        title = "{On the solution of the radiative transfer equations for polarized radiation}",
      journal = {\solphys},
     keywords = {Polarized Radiation, Radiative Transfer, Stellar Atmospheres, Chromosphere, Computational Astrophysics, Matrices (Mathematics), Stokes Law Of Radiation, Astrophysics},
         year = 1985,
        month = jun,
       volume = {97},
        pages = {239-250},
          doi = {10.1007/BF00165988},
       adsurl = {https://ui.adsabs.harvard.edu/abs/1985SoPh...97..239L},
      adsnote = {Provided by the SAO/NASA Astrophysics Data System}
}

@ARTICLE{Leung2011,
       author = {{Leung}, Po Kin and {Gammie}, Charles F. and {Noble}, Scott C.},
        title = "{Numerical Calculation of Magnetobremsstrahlung Emission and Absorption Coefficients}",
      journal = {\apj},
     keywords = {methods: numerical, radiation mechanisms: general},
         year = 2011,
        month = aug,
       volume = {737},
       number = {1},
          eid = {21},
        pages = {21},
          doi = {10.1088/0004-637X/737/1/21},
       adsurl = {https://ui.adsabs.harvard.edu/abs/2011ApJ...737...21L},
      adsnote = {Provided by the SAO/NASA Astrophysics Data System}
}

@ARTICLE{Pandya2021,
       author = {{Marszewski}, Andrew and {Prather}, Ben S. and {Joshi}, Abhishek V. and {Pandya}, Alex and {Gammie}, Charles F.},
        title = "{Updated Transfer Coefficients for Magnetized Plasmas}",
      journal = {\apj},
     keywords = {Radiative transfer, Polarimetry, Plasma astrophysics, Radiative processes, Relativistic disks, 1335, 1278, 1261, 2055, 1388, Astrophysics - High Energy Astrophysical Phenomena},
         year = 2021,
        month = nov,
       volume = {921},
       number = {1},
          eid = {17},
        pages = {17},
          doi = {10.3847/1538-4357/ac1b28},
archivePrefix = {arXiv},
       eprint = {2108.10359},
 primaryClass = {astro-ph.HE},
       adsurl = {https://ui.adsabs.harvard.edu/abs/2021ApJ...921...17M},
      adsnote = {Provided by the SAO/NASA Astrophysics Data System}
}

@ARTICLE{Vos2022,
       author = {{Vos}, J. and {Moscibrodzka}, M.~A. and {Wielgus}, M.},
        title = "{Polarimetric signatures of hot spots in black hole accretion flows}",
      journal = {\aap},
     keywords = {black hole physics, relativistic processes, radiative transfer, methods: numerical, Astrophysics - High Energy Astrophysical Phenomena, General Relativity and Quantum Cosmology},
         year = 2022,
        month = dec,
       volume = {668},
          eid = {A185},
        pages = {A185},
          doi = {10.1051/0004-6361/202244840},
archivePrefix = {arXiv},
       eprint = {2209.09931},
 primaryClass = {astro-ph.HE},
       adsurl = {https://ui.adsabs.harvard.edu/abs/2022A&A...668A.185V},
      adsnote = {Provided by the SAO/NASA Astrophysics Data System}
}

@ARTICLE{Gelles2021,
       author = {{Gelles}, Zachary and {Himwich}, Elizabeth and {Johnson}, Michael D. and {Palumbo}, Daniel C.~M.},
        title = "{Polarized image of equatorial emission in the Kerr geometry}",
      journal = {\prd},
     keywords = {General Relativity and Quantum Cosmology, Astrophysics - High Energy Astrophysical Phenomena, High Energy Physics - Theory},
         year = 2021,
        month = aug,
       volume = {104},
       number = {4},
          eid = {044060},
        pages = {044060},
          doi = {10.1103/PhysRevD.104.044060},
archivePrefix = {arXiv},
       eprint = {2105.09440},
 primaryClass = {gr-qc},
       adsurl = {https://ui.adsabs.harvard.edu/abs/2021PhRvD.104d4060G},
      adsnote = {Provided by the SAO/NASA Astrophysics Data System}
}

@ARTICLE{Hofmann1993,
       author = {{Hofmann}, R. and {Eckart}, A. and {Genzel}, R. and {Drapatz}, S.},
        title = "{High Resolution K-Band Images of the Galactic Centre}",
      journal = {\apss},
     keywords = {Angular Resolution, Galactic Nuclei, Infrared Imagery, Radio Sources (Astronomy), Cameras, High Resolution, Ultrahigh Frequencies, Astrophysics},
         year = 1993,
        month = jul,
       volume = {205},
       number = {1},
        pages = {1-4},
          doi = {10.1007/BF00657949},
       adsurl = {https://ui.adsabs.harvard.edu/abs/1993Ap&SS.205....1H},
      adsnote = {Provided by the SAO/NASA Astrophysics Data System}
}

@BOOK{Frank2002,
       author = {{Frank}, Juhan and {King}, Andrew and {Raine}, Derek J.},
        title = "{Accretion Power in Astrophysics: Third Edition}",
         year = 2002,
       adsurl = {https://ui.adsabs.harvard.edu/abs/2002apa..book.....F},
      adsnote = {Provided by the SAO/NASA Astrophysics Data System}
}

@BOOK{Merritt2013,
       author = {{Merritt}, David},
        title = "{Dynamics and Evolution of Galactic Nuclei}",
         year = 2013,
       adsurl = {https://ui.adsabs.harvard.edu/abs/2013degn.book.....M},
      adsnote = {Provided by the SAO/NASA Astrophysics Data System}
}

@ARTICLE{Shakura1973,
       author = {{Shakura}, N.~I. and {Sunyaev}, R.~A.},
        title = "{Black holes in binary systems. Observational appearance.}",
      journal = {\aap},
         year = 1973,
        month = jan,
       volume = {24},
        pages = {337-355},
       adsurl = {https://ui.adsabs.harvard.edu/abs/1973A&A....24..337S},
      adsnote = {Provided by the SAO/NASA Astrophysics Data System}
}

@ARTICLE{Yuan2014,
       author = {{Yuan}, Feng and {Narayan}, Ramesh},
        title = "{Hot Accretion Flows Around Black Holes}",
      journal = {\araa},
     keywords = {Astrophysics - High Energy Astrophysical Phenomena},
         year = 2014,
        month = aug,
       volume = {52},
        pages = {529-588},
          doi = {10.1146/annurev-astro-082812-141003},
archivePrefix = {arXiv},
       eprint = {1401.0586},
 primaryClass = {astro-ph.HE},
       adsurl = {https://ui.adsabs.harvard.edu/abs/2014ARA&A..52..529Y},
      adsnote = {Provided by the SAO/NASA Astrophysics Data System}
}

@ARTICLE{Yuan2003,
       author = {{Yuan}, Feng and {Quataert}, Eliot and {Narayan}, Ramesh},
        title = "{Nonthermal Electrons in Radiatively Inefficient Accretion Flow Models of Sagittarius A*}",
      journal = {\apj},
     keywords = {Accretion, Accretion Disks, Black Hole Physics, Galaxies: Active, Galaxy: Center, Radiation Mechanisms: Nonthermal, Radiation Mechanisms: Thermal, Astrophysics},
         year = 2003,
        month = nov,
       volume = {598},
       number = {1},
        pages = {301-312},
          doi = {10.1086/378716},
archivePrefix = {arXiv},
       eprint = {astro-ph/0304125},
 primaryClass = {astro-ph},
       adsurl = {https://ui.adsabs.harvard.edu/abs/2003ApJ...598..301Y},
      adsnote = {Provided by the SAO/NASA Astrophysics Data System}
}

@INPROCEEDINGS{Marrone2006,
       author = {{Marrone}, Daniel P. and {Moran}, James M. and {Zhao}, Jun-Hui and {Rao}, Ramprasad},
        title = "{The Submillimeter Polarization of Sgr A*}",
     keywords = {Astrophysics},
    booktitle = {Journal of Physics Conference Series},
         year = 2006,
       series = {Journal of Physics Conference Series},
       volume = {54},
        month = dec,
        pages = {354-362},
          doi = {10.1088/1742-6596/54/1/056},
archivePrefix = {arXiv},
       eprint = {astro-ph/0607432},
 primaryClass = {astro-ph},
       adsurl = {https://ui.adsabs.harvard.edu/abs/2006JPhCS..54..354M},
      adsnote = {Provided by the SAO/NASA Astrophysics Data System}
}

@ARTICLE{Brinkerink2015,
       author = {{Brinkerink}, Christiaan D. and {Falcke}, Heino and {Law}, Casey J. and {Barkats}, Denis and {Bower}, Geoffrey C. and {Brunthaler}, Andreas and {Gammie}, Charles and {Impellizzeri}, C.~M. Violette and {Markoff}, Sera and {Menten}, Karl M. and {Moscibrodzka}, Monika and {Peck}, Alison and {Rushton}, Anthony P. and {Schaaf}, Reinhold and {Wright}, Melvyn},
        title = "{ALMA and VLA measurements of frequency-dependent time lags in Sagittarius A*: evidence for a relativistic outflow}",
      journal = {\aap},
     keywords = {accretion, accretion disks, black hole physics, radiation mechanisms: non-thermal, Galaxy: center, galaxies: jets, Astrophysics - Astrophysics of Galaxies, Astrophysics - High Energy Astrophysical Phenomena},
         year = 2015,
        month = apr,
       volume = {576},
          eid = {A41},
        pages = {A41},
          doi = {10.1051/0004-6361/201424783},
archivePrefix = {arXiv},
       eprint = {1502.03423},
 primaryClass = {astro-ph.GA},
       adsurl = {https://ui.adsabs.harvard.edu/abs/2015A&A...576A..41B},
      adsnote = {Provided by the SAO/NASA Astrophysics Data System}
}

@ARTICLE{Bower2015,
       author = {{Bower}, Geoffrey C. and {Markoff}, Sera and {Dexter}, Jason and {Gurwell}, Mark A. and {Moran}, James M. and {Brunthaler}, Andreas and {Falcke}, Heino and {Fragile}, P. Chris and {Maitra}, Dipankar and {Marrone}, Dan and {Peck}, Alison and {Rushton}, Anthony and {Wright}, Melvyn C.~H.},
        title = "{Radio and Millimeter Monitoring of Sgr A*: Spectrum, Variability, and Constraints on the G2 Encounter}",
      journal = {\apj},
     keywords = {accretion, accretion disks, black hole physics, galaxies: active, galaxies: jets, Galaxy: center, Astrophysics - High Energy Astrophysical Phenomena},
         year = 2015,
        month = mar,
       volume = {802},
       number = {1},
          eid = {69},
        pages = {69},
          doi = {10.1088/0004-637X/802/1/69},
archivePrefix = {arXiv},
       eprint = {1502.06534},
 primaryClass = {astro-ph.HE},
       adsurl = {https://ui.adsabs.harvard.edu/abs/2015ApJ...802...69B},
      adsnote = {Provided by the SAO/NASA Astrophysics Data System}
}

@ARTICLE{Liu2016,
       author = {{Liu}, Hauyu Baobab and {Wright}, Melvyn C.~H. and {Zhao}, Jun-Hui and {Mills}, Elisabeth A.~C. and {Requena-Torres}, Miguel A. and {Matsushita}, Satoki and {Martin}, Sergio and {Ott}, Jurgen and {Morris}, Mark R. and {Longmore}, Steven N. and {Brinkerink}, Christiaan D. and {Falcke}, Heino},
        title = "{The 492 GHz emission of Sgr A* constrained by ALMA}",
      journal = {\aap},
     keywords = {techniques: polarimetric, black hole physics, polarization, radiation mechanisms: non-thermal, Galaxy: nucleus, Astrophysics - High Energy Astrophysical Phenomena, Astrophysics - Astrophysics of Galaxies},
         year = 2016,
        month = sep,
       volume = {593},
          eid = {A44},
        pages = {A44},
          doi = {10.1051/0004-6361/201628176},
archivePrefix = {arXiv},
       eprint = {1604.00599},
 primaryClass = {astro-ph.HE},
       adsurl = {https://ui.adsabs.harvard.edu/abs/2016A&A...593A..44L},
      adsnote = {Provided by the SAO/NASA Astrophysics Data System}
}

@ARTICLE{von_Fellenberg2018,
       author = {{von Fellenberg}, Sebastiano D. and {Gillessen}, Stefan and {Gracia-Carpio}, Javier and {Fritz}, Tobias K. and {Dexter}, Jason and {Baubock}, Michi and {Ponti}, Gabriele and {Gao}, Feng and {Habibi}, Maryam and {Plewa}, Philipp M. and {Pfuhl}, Oliver and {Jimenez-Rosales}, Alejandra and {Waisberg}, Idel and {Widmann}, Felix and {Ott}, Thomas and {Eisenhauer}, Frank and {Genzel}, Reinhard},
        title = "{A Detection of Sgr A* in the Far Infrared}",
      journal = {\apj},
     keywords = {accretion, accretion disks, black hole physics, Galaxy: center, Astrophysics - Astrophysics of Galaxies},
         year = 2018,
        month = aug,
       volume = {862},
       number = {2},
          eid = {129},
        pages = {129},
          doi = {10.3847/1538-4357/aacd4b},
archivePrefix = {arXiv},
       eprint = {1806.07395},
 primaryClass = {astro-ph.GA},
       adsurl = {https://ui.adsabs.harvard.edu/abs/2018ApJ...862..129V},
      adsnote = {Provided by the SAO/NASA Astrophysics Data System}
}

@ARTICLE{Dexter2009,
       author = {{Dexter}, Jason and {Agol}, Eric},
        title = "{A Fast New Public Code for Computing Photon Orbits in a Kerr Spacetime}",
      journal = {\apj},
     keywords = {accretion, accretion disks, black hole physics, radiative transfer, relativity, Astrophysics - High Energy Astrophysical Phenomena},
         year = 2009,
        month = may,
       volume = {696},
       number = {2},
        pages = {1616-1629},
          doi = {10.1088/0004-637X/696/2/1616},
archivePrefix = {arXiv},
       eprint = {0903.0620},
 primaryClass = {astro-ph.HE},
       adsurl = {https://ui.adsabs.harvard.edu/abs/2009ApJ...696.1616D},
      adsnote = {Provided by the SAO/NASA Astrophysics Data System}
}

@ARTICLE{Dexter2016,
       author = {{Dexter}, Jason},
        title = "{A public code for general relativistic, polarised radiative transfer around spinning black holes}",
      journal = {\mnras},
     keywords = {accretion, accretion discs, black hole physics, radiative transfer, relativistic processes, Galaxy: centre, galaxies: jets, Astrophysics - High Energy Astrophysical Phenomena},
         year = 2016,
        month = oct,
       volume = {462},
       number = {1},
        pages = {115-136},
          doi = {10.1093/mnras/stw1526},
archivePrefix = {arXiv},
       eprint = {1602.03184},
 primaryClass = {astro-ph.HE},
       adsurl = {https://ui.adsabs.harvard.edu/abs/2016MNRAS.462..115D},
      adsnote = {Provided by the SAO/NASA Astrophysics Data System}
}

@ARTICLE{Eckart2004,
       author = {{Eckart}, A. and {Baganoff}, F.~K. and {Morris}, M. and {Bautz}, M.~W. and {Brandt}, W.~N. and {Garmire}, G.~P. and {Genzel}, R. and {Ott}, T. and {Ricker}, G.~R. and {Straubmeier}, C. and {Viehmann}, T. and {Schodel}, R. and {Bower}, G.~C. and {Goldston}, J.~E.},
        title = "{First simultaneous NIR/X-ray detection of a flare from Sgr A*}",
      journal = {\aap},
     keywords = {black hole physics, X-rays: general, infrared: general, accretion, accretion disks, Galaxy: center, Galaxy: nucleus, Astrophysics},
         year = 2004,
        month = nov,
       volume = {427},
        pages = {1-11},
          doi = {10.1051/0004-6361:20040495},
archivePrefix = {arXiv},
       eprint = {astro-ph/0403577},
 primaryClass = {astro-ph},
       adsurl = {https://ui.adsabs.harvard.edu/abs/2004A&A...427....1E},
      adsnote = {Provided by the SAO/NASA Astrophysics Data System}
}

@ARTICLE{Falcke1998,
       author = {{Falcke}, Heino and {Goss}, W.~M. and {Matsuo}, Hiroshi and {Teuben}, Peter and {Zhao}, Jun-Hui and {Zylka}, Robert},
        title = "{The Simultaneous Spectrum of Sagittarius A* from 20 Centimeters to 1 Millimeter and the Nature of the Millimeter Excess}",
      journal = {\apj},
     keywords = {Black Hole Physics, Galaxies: Nuclei, Galaxy: Center, ISM: Individual: Name: Sagittarius A, Radio Continuum: ISM, Astrophysics},
         year = 1998,
        month = may,
       volume = {499},
       number = {2},
        pages = {731-734},
          doi = {10.1086/305687},
archivePrefix = {arXiv},
       eprint = {astro-ph/9801085},
 primaryClass = {astro-ph},
       adsurl = {https://ui.adsabs.harvard.edu/abs/1998ApJ...499..731F},
      adsnote = {Provided by the SAO/NASA Astrophysics Data System}
}

@ARTICLE{Ghisellini1991,
       author = {{Ghisellini}, Gabriele and {Svensson}, Roland},
        title = "{The synchrotron and cyclo-synchrotron absorption cross-section}",
      journal = {\mnras},
     keywords = {Absorption Cross Sections, Cyclotron Radiation, Synchrotron Radiation, Charged Particles, Einstein Equations, Photons, Physics (General)},
         year = 1991,
        month = oct,
       volume = {252},
        pages = {313-318},
          doi = {10.1093/mnras/252.3.313},
       adsurl = {https://ui.adsabs.harvard.edu/abs/1991MNRAS.252..313G},
      adsnote = {Provided by the SAO/NASA Astrophysics Data System}
}

@ARTICLE{Hoshino2022,
       author = {{Hoshino}, Masahiro},
        title = "{Efficiency of nonthermal particle acceleration in magnetic reconnection}",
      journal = {Physics of Plasmas},
     keywords = {Astrophysics - High Energy Astrophysical Phenomena, Astrophysics - Solar and Stellar Astrophysics, Physics - Plasma Physics, Physics - Space Physics},
         year = 2022,
        month = apr,
       volume = {29},
       number = {4},
          eid = {042902},
        pages = {042902},
          doi = {10.1063/5.0086316},
archivePrefix = {arXiv},
       eprint = {2203.15169},
 primaryClass = {astro-ph.HE},
       adsurl = {https://ui.adsabs.harvard.edu/abs/2022PhPl...29d2902H},
      adsnote = {Provided by the SAO/NASA Astrophysics Data System}
}

@ARTICLE{Johnson2022,
       author = {{Johnson}, Grant and {Kilian}, Patrick and {Guo}, Fan and {Li}, Xiaocan},
        title = "{Particle Acceleration in Magnetic Reconnection with Ad Hoc Pitch-angle Scattering}",
      journal = {\apj},
     keywords = {Plasma astrophysics, Solar flares, Solar corona, Solar magnetic reconnection, Space plasmas, Relativistic jets, Pulsar wind nebulae, 1261, 1496, 1483, 1504, 1544, 1390, 2215, Astrophysics - Solar and Stellar Astrophysics, Astrophysics - High Energy Astrophysical Phenomena, Physics - Plasma Physics, Physics - Space Physics},
         year = 2022,
        month = jul,
       volume = {933},
       number = {1},
          eid = {73},
        pages = {73},
          doi = {10.3847/1538-4357/ac7143},
archivePrefix = {arXiv},
       eprint = {2205.08600},
 primaryClass = {astro-ph.SR},
       adsurl = {https://ui.adsabs.harvard.edu/abs/2022ApJ...933...73J},
      adsnote = {Provided by the SAO/NASA Astrophysics Data System}
}

@ARTICLE{Klion2023,
       author = {{Klion}, Hannah and {Jambunathan}, Revathi and {Rowan}, Michael E. and {Yang}, Eloise and {Willcox}, Donald and {Vay}, Jean-Luc and {Lehe}, Remi and {Myers}, Andrew and {Huebl}, Axel and {Zhang}, Weiqun},
        title = "{Particle-in-Cell Simulations of Relativistic Magnetic Reconnection with Advanced Maxwell Solver Algorithms}",
      journal = {arXiv e-prints},
     keywords = {Astrophysics - High Energy Astrophysical Phenomena, Computer Science - Computational Engineering, Finance, and Science},
         year = 2023,
        month = apr,
          eid = {arXiv:2304.10566},
        pages = {arXiv:2304.10566},
          doi = {10.48550/arXiv.2304.10566},
archivePrefix = {arXiv},
       eprint = {2304.10566},
 primaryClass = {astro-ph.HE},
       adsurl = {https://ui.adsabs.harvard.edu/abs/2023arXiv230410566K},
      adsnote = {Provided by the SAO/NASA Astrophysics Data System}
}

@ARTICLE{Meyer2006,
       author = {{Meyer}, L. and {Eckart}, A. and {Schodel}, R. and {Duschl}, W.~J. and {Mu{\v{z}}ic}, K. and {Dov{\v{c}}iak}, M. and {Karas}, V.},
        title = "{Near-infrared polarimetry setting constraints on the orbiting spot model for Sgr A* flares}",
      journal = {\aap},
     keywords = {black hole physics, accretion, accretion disks, Galaxy: center, Astrophysics},
         year = 2006,
        month = dec,
       volume = {460},
       number = {1},
        pages = {15-21},
          doi = {10.1051/0004-6361:20065925},
archivePrefix = {arXiv},
       eprint = {astro-ph/0610104},
 primaryClass = {astro-ph},
       adsurl = {https://ui.adsabs.harvard.edu/abs/2006A&A...460...15M},
      adsnote = {Provided by the SAO/NASA Astrophysics Data System}
}

@ARTICLE{Schoeffler2023,
       author = {{Schoeffler}, K.~M. and {Grismayer}, T. and {Uzdensky}, D. and {Silva}, L.~O.},
        title = "{High-energy synchrotron flares powered by strongly radiative relativistic magnetic reconnection: 2D and 3D PIC simulations}",
      journal = {\mnras},
     keywords = {magnetic reconnection, radiation: dynamics, relativistic processes, stars: magnetars, (transients:) gamma-ray bursts, Astrophysics - High Energy Astrophysical Phenomena, Physics - Plasma Physics},
         year = 2023,
        month = aug,
       volume = {523},
       number = {3},
        pages = {3812-3839},
          doi = {10.1093/mnras/stad1588},
archivePrefix = {arXiv},
       eprint = {2303.16643},
 primaryClass = {astro-ph.HE},
       adsurl = {https://ui.adsabs.harvard.edu/abs/2023MNRAS.523.3812S},
      adsnote = {Provided by the SAO/NASA Astrophysics Data System}
}

@ARTICLE{Selvi2022,
       author = {{Selvi}, Sebastiaan and {Porth}, Oliver and {Ripperda}, Bart and {Bacchini}, Fabio and {Sironi}, Lorenzo and {Keppens}, Rony},
        title = "{Effective resistivity in relativistic collisionless plasmoid-mediated reconnection}",
      journal = {arXiv e-prints},
     keywords = {Astrophysics - High Energy Astrophysical Phenomena, Physics - Plasma Physics},
         year = 2022,
        month = nov,
          eid = {arXiv:2211.04553},
        pages = {arXiv:2211.04553},
          doi = {10.48550/arXiv.2211.04553},
archivePrefix = {arXiv},
       eprint = {2211.04553},
 primaryClass = {astro-ph.HE},
       adsurl = {https://ui.adsabs.harvard.edu/abs/2022arXiv221104553S},
      adsnote = {Provided by the SAO/NASA Astrophysics Data System}
}

@MISC{Sun2018,
       author = {{Sun}, Mouyuan and {Grier}, C.~J. and {Peterson}, B.~M.},
        title = "{PyCCF: Python Cross Correlation Function for reverberation mapping studies}",
     keywords = {Software},
 howpublished = {Astrophysics Source Code Library, record ascl:1805.032},
         year = 2018,
        month = may,
          eid = {ascl:1805.032},
        pages = {ascl:1805.032},
archivePrefix = {ascl},
       eprint = {1805.032},
       adsurl = {https://ui.adsabs.harvard.edu/abs/2018ascl.soft05032S},
      adsnote = {Provided by the SAO/NASA Astrophysics Data System}
}

@ARTICLE{Trippe2007,
       author = {{Trippe}, S. and {Paumard}, T. and {Ott}, T. and {Gillessen}, S. and {Eisenhauer}, F. and {Martins}, F. and {Genzel}, R.},
        title = "{A polarized infrared flare from Sagittarius A* and the signatures of orbiting plasma hotspots}",
      journal = {\mnras},
     keywords = {accretion, accretion discs, black hole physics, Galaxy: centre, Astrophysics},
         year = 2007,
        month = mar,
       volume = {375},
       number = {3},
        pages = {764-772},
          doi = {10.1111/j.1365-2966.2006.11338.x},
archivePrefix = {arXiv},
       eprint = {astro-ph/0611737},
 primaryClass = {astro-ph},
       adsurl = {https://ui.adsabs.harvard.edu/abs/2007MNRAS.375..764T},
      adsnote = {Provided by the SAO/NASA Astrophysics Data System}
}

@ARTICLE{Vincent2011,
       author = {{Vincent}, F.~H. and {Paumard}, T. and {Perrin}, G. and {Mugnier}, L. and {Eisenhauer}, F. and {Gillessen}, S.},
        title = "{Performance of astrometric detection of a hotspot orbiting on the innermost stable circular orbit of the Galactic Centre black hole}",
      journal = {\mnras},
     keywords = {black hole physics, instrumentation: interferometers, astrometry, Galaxy: centre, Astrophysics - Astrophysics of Galaxies},
         year = 2011,
        month = apr,
       volume = {412},
       number = {4},
        pages = {2653-2664},
          doi = {10.1111/j.1365-2966.2010.18084.x},
archivePrefix = {arXiv},
       eprint = {1011.5439},
 primaryClass = {astro-ph.GA},
       adsurl = {https://ui.adsabs.harvard.edu/abs/2011MNRAS.412.2653V},
      adsnote = {Provided by the SAO/NASA Astrophysics Data System}
}

@ARTICLE{torus+jet,
       author = {{Vincent}, F.~H. and {Abramowicz}, M.~A. and {Zdziarski}, A.~A. and {Wielgus}, M. and {Paumard}, T. and {Perrin}, G. and {Straub}, O.},
        title = "{Multi-wavelength torus-jet model for Sagittarius A*}",
      journal = {\aap},
     keywords = {Galaxy: center, accretion, accretion disks, black hole physics, relativistic processes, Astrophysics - High Energy Astrophysical Phenomena, General Relativity and Quantum Cosmology},
         year = 2019,
        month = apr,
       volume = {624},
          eid = {A52},
        pages = {A52},
          doi = {10.1051/0004-6361/201834946},
archivePrefix = {arXiv},
       eprint = {1902.01175},
 primaryClass = {astro-ph.HE},
       adsurl = {https://ui.adsabs.harvard.edu/abs/2019A&A...624A..52V},
      adsnote = {Provided by the SAO/NASA Astrophysics Data System}
}

@ARTICLE{Chernoglazov2023,
       author = {{Chernoglazov}, Alexander and {Hakobyan}, Hayk and {Philippov}, Alexander A.},
        title = "{High-Energy Radiation and Ion Acceleration in Three-dimensional Relativistic Magnetic Reconnection with Strong Synchrotron Cooling}",
      journal = {arXiv e-prints},
     keywords = {Astrophysics - High Energy Astrophysical Phenomena, Physics - Plasma Physics},
         year = 2023,
        month = may,
          eid = {arXiv:2305.02348},
        pages = {arXiv:2305.02348},
          doi = {10.48550/arXiv.2305.02348},
archivePrefix = {arXiv},
       eprint = {2305.02348},
 primaryClass = {astro-ph.HE},
       adsurl = {https://ui.adsabs.harvard.edu/abs/2023arXiv230502348C},
      adsnote = {Provided by the SAO/NASA Astrophysics Data System}
}

@ARTICLE{EHT2022iii,
       author = {{Event Horizon Telescope Collaboration} and {Akiyama}, Kazunori and {Alberdi}, Antxon and {Alef}, Walter and {Algaba}, Juan Carlos and {Anantua}, Richard and {Asada}, Keiichi and {Azulay}, Rebecca and {Bach}, Uwe and {Baczko}, Anne-Kathrin and {Ball}, David and {Balokovic}, Mislav and {Barrett}, John and {Baubock}, Michi and {Benson}, Bradford A. and {Bintley}, Dan and {Blackburn}, Lindy and {Blundell}, Raymond and {Bouman}, Katherine L. and {Bower}, Geoffrey C. and {Boyce}, Hope and {Bremer}, Michael and {Brinkerink}, Christiaan D. and {Brissenden}, Roger and {Britzen}, Silke and {Broderick}, Avery E. and {Broguiere}, Dominique and {Bronzwaer}, Thomas and {Bustamante}, Sandra and {Byun}, Do-Young and {Carlstrom}, John E. and {Ceccobello}, Chiara and {Chael}, Andrew and {Chan}, Chi-kwan and {Chatterjee}, Koushik and {Chatterjee}, Shami and {Chen}, Ming-Tang and {Chen}, Yongjun and {Cheng}, Xiaopeng and {Cho}, Ilje and {Christian}, Pierre and {Conroy}, Nicholas S. and {Conway}, John E. and {Cordes}, James M. and {Crawford}, Thomas M. and {Crew}, Geoffrey B. and {Cruz-Osorio}, Alejandro and {Cui}, Yuzhu and {Davelaar}, Jordy and {De Laurentis}, Mariafelicia and {Deane}, Roger and {Dempsey}, Jessica and {Desvignes}, Gregory and {Dexter}, Jason and {Dhruv}, Vedant and {Doeleman}, Sheperd S. and {Dougal}, Sean and {Dzib}, Sergio A. and {Eatough}, Ralph P. and {Emami}, Razieh and {Falcke}, Heino and {Farah}, Joseph and {Fish}, Vincent L. and {Fomalont}, Ed and {Ford}, H. Alyson and {Fraga-Encinas}, Raquel and {Freeman}, William T. and {Friberg}, Per and {Fromm}, Christian M. and {Fuentes}, Antonio and {Galison}, Peter and {Gammie}, Charles F. and {Garcia}, Roberto and {Gentaz}, Olivier and {Georgiev}, Boris and {Goddi}, Ciriaco and {Gold}, Roman and {Gomez-Ruiz}, Arturo I. and {Gomez}, Jose L. and {Gu}, Minfeng and {Gurwell}, Mark and {Hada}, Kazuhiro and {Haggard}, Daryl and {Haworth}, Kari and {Hecht}, Michael H. and {Hesper}, Ronald and {Heumann}, Dirk and {Ho}, Luis C. and {Ho}, Paul and {Honma}, Mareki and {Huang}, Chih-Wei L. and {Huang}, Lei and {Hughes}, David H. and {Ikeda}, Shiro and {Impellizzeri}, C.~M. Violette and {Inoue}, Makoto and {Issaoun}, Sara and {James}, David J. and {Jannuzi}, Buell T. and {Janssen}, Michael and {Jeter}, Britton and {Jiang}, Wu and {Jimenez-Rosales}, Alejandra and {Johnson}, Michael D. and {Jorstad}, Svetlana and {Joshi}, Abhishek V. and {Jung}, Taehyun and {Karami}, Mansour and {Karuppusamy}, Ramesh and {Kawashima}, Tomohisa and {Keating}, Garrett K. and {Kettenis}, Mark and {Kim}, Dong-Jin and {Kim}, Jae-Young and {Kim}, Jongsoo and {Kim}, Junhan and {Kino}, Motoki and {Koay}, Jun Yi and {Kocherlakota}, Prashant and {Kofuji}, Yutaro and {Koch}, Patrick M. and {Koyama}, Shoko and {Kramer}, Carsten and {Kramer}, Michael and {Krichbaum}, Thomas P. and {Kuo}, Cheng-Yu and {La Bella}, Noemi and {Lauer}, Tod R. and {Lee}, Daeyoung and {Lee}, Sang-Sung and {Leung}, Po Kin and {Levis}, Aviad and {Li}, Zhiyuan and {Lico}, Rocco and {Lindahl}, Greg and {Lindqvist}, Michael and {Lisakov}, Mikhail and {Liu}, Jun and {Liu}, Kuo and {Liuzzo}, Elisabetta and {Lo}, Wen-Ping and {Lobanov}, Andrei P. and {Loinard}, Laurent and {Lonsdale}, Colin J. and {Lu}, Ru-Sen and {Mao}, Jirong and {Marchili}, Nicola and {Markoff}, Sera and {Marrone}, Daniel P. and {Marscher}, Alan P. and {Marti-Vidal}, Ivan and {Matsushita}, Satoki and {Matthews}, Lynn D. and {Medeiros}, Lia and {Menten}, Karl M. and {Michalik}, Daniel and {Mizuno}, Izumi and {Mizuno}, Yosuke and {Moran}, James M. and {Moriyama}, Kotaro and {Moscibrodzka}, Monika and {Muller}, Cornelia and {Mus}, Alejandro and {Musoke}, Gibwa and {Myserlis}, Ioannis and {Nadolski}, Andrew and {Nagai}, Hiroshi and {Nagar}, Neil M. and {Nakamura}, Masanori and {Narayan}, Ramesh and {Narayanan}, Gopal and {Natarajan}, Iniyan and {Nathanail}, Antonios and {Fuentes}, Santiago Navarro and {Neilsen}, Joey and {Neri}, Roberto and {Ni}, Chunchong and {Noutsos}, Aristeidis and {Nowak}, Michael A. and {Oh}, Junghwan and {Okino}, Hiroki and {Olivares}, Hector and {Ortiz-Leon}, Gisela N. and {Oyama}, Tomoaki and {Ozel}, Feryal and {Palumbo}, Daniel C.~M. and {Paraschos}, Georgios Filippos and {Park}, Jongho and {Parsons}, Harriet and {Patel}, Nimesh and {Pen}, Ue-Li and {Pesce}, Dominic W. and {Pietu}, Vincent and {Plambeck}, Richard and {PopStefanija}, Aleksandar and {Porth}, Oliver and {Potzl}, Felix M. and {Prather}, Ben and {Preciado-Lopez}, Jorge A. and {Psaltis}, Dimitrios and {Pu}, Hung-Yi and {Ramakrishnan}, Venkatessh and {Rao}, Ramprasad and {Rawlings}, Mark G. and {Raymond}, Alexander W. and {Rezzolla}, Luciano and {Ricarte}, Angelo and {Ripperda}, Bart and {Roelofs}, Freek and {Rogers}, Alan and {Ros}, Eduardo and {Romero-Canizales}, Cristina and {Roshanineshat}, Arash and {Rottmann}, Helge and {Roy}, Alan L. and {Ruiz}, Ignacio and {Ruszczyk}, Chet and {Rygl}, Kazi L.~J. and {Sanchez}, Salvador and {Sanchez-Arguelles}, David and {Sanchez-Portal}, Miguel and {Sasada}, Mahito and {Satapathy}, Kaushik and {Savolainen}, Tuomas and {Schloerb}, F. Peter and {Schonfeld}, Jonathan and {Schuster}, Karl-Friedrich and {Shao}, Lijing and {Shen}, Zhiqiang and {Small}, Des and {Sohn}, Bong Won and {SooHoo}, Jason and {Souccar}, Kamal and {Sun}, He and {Tazaki}, Fumie and {Tetarenko}, Alexandra J. and {Tiede}, Paul and {Tilanus}, Remo P.~J. and {Titus}, Michael and {Torne}, Pablo and {Traianou}, Efthalia and {Trent}, Tyler and {Trippe}, Sascha and {Turk}, Matthew and {van Bemmel}, Ilse and {van Langevelde}, Huib Jan and {van Rossum}, Daniel R. and {Vos}, Jesse and {Wagner}, Jan and {Ward-Thompson}, Derek and {Wardle}, John and {Weintroub}, Jonathan and {Wex}, Norbert and {Wharton}, Robert and {Wielgus}, Maciek and {Wiik}, Kaj and {Witzel}, Gunther and {Wondrak}, Michael F. and {Wong}, George N. and {Wu}, Qingwen and {Yamaguchi}, Paul and {Yoon}, Doosoo and {Young}, Andre and {Young}, Ken and {Younsi}, Ziri and {Yuan}, Feng and {Yuan}, Ye-Fei and {Zensus}, J. Anton and {Zhang}, Shuo and {Zhao}, Guang-Yao and {Zhao}, Shan-Shan},
        title = "{First Sagittarius A* Event Horizon Telescope Results. III. Imaging of the Galactic Center Supermassive Black Hole}",
      journal = {\apjl},
     keywords = {Aperture synthesis, Black holes, Galactic center, Radio astronomy, Very long baseline interferometry, High angular resolution, 53, 162, 565, 1338, 1769, 2167},
         year = 2022,
        month = may,
       volume = {930},
       number = {2},
          eid = {L14},
        pages = {L14},
          doi = {10.3847/2041-8213/ac6429},
       adsurl = {https://ui.adsabs.harvard.edu/abs/2022ApJ...930L..14E},
      adsnote = {Provided by the SAO/NASA Astrophysics Data System}
}

@ARTICLE{Chiarberge&Guisellini1999,
       author = {{Chiaberge}, Marco and {Ghisellini}, Gabriele},
        title = "{Rapid variability in the synchrotron self-Compton model for blazars}",
      journal = {\mnras},
     keywords = {Astrophysics},
         year = 1999,
        month = jul,
       volume = {306},
       number = {3},
        pages = {551-560},
          doi = {10.1046/j.1365-8711.1999.02538.x},
archivePrefix = {arXiv},
       eprint = {astro-ph/9810263},
 primaryClass = {astro-ph},
       adsurl = {https://ui.adsabs.harvard.edu/abs/1999MNRAS.306..551C},
      adsnote = {Provided by the SAO/NASA Astrophysics Data System}
}

@ARTICLE{Ressler2018,
       author = {{Ressler}, S.~M. and {Quataert}, E. and {Stone}, J.~M.},
        title = "{Hydrodynamic simulations of the inner accretion flow of Sagittarius A* fuelled by stellar winds}",
      journal = {\mnras},
     keywords = {accretion, accretion discs, black hole physics, hydrodynamics, stars: Wolf-Rayet, Galaxy: centre, X-rays: ISM, Astrophysics - High Energy Astrophysical Phenomena},
         year = 2018,
        month = aug,
       volume = {478},
       number = {3},
        pages = {3544-3563},
          doi = {10.1093/mnras/sty1146},
archivePrefix = {arXiv},
       eprint = {1805.00474},
 primaryClass = {astro-ph.HE},
       adsurl = {https://ui.adsabs.harvard.edu/abs/2018MNRAS.478.3544R},
      adsnote = {Provided by the SAO/NASA Astrophysics Data System}
}

@ARTICLE{Heywood2022,
       author = {{Heywood}, I. and {Rammala}, I. and {Camilo}, F. and {Cotton}, W.~D. and {Yusef-Zadeh}, F. and {Abbott}, T.~D. and {Adam}, R.~M. and {Adams}, G. and {Aldera}, M.~A. and {Asad}, K.~M.~B. and {Bauermeister}, E.~F. and {Bennett}, T.~G.~H. and {Bester}, H.~L. and {Bode}, W.~A. and {Botha}, D.~H. and {Botha}, A.~G. and {Brederode}, L.~R.~S. and {Buchner}, S. and {Burger}, J.~P. and {Cheetham}, T. and {de Villiers}, D.~I.~L. and {Dikgale-Mahlakoana}, M.~A. and {du Toit}, L.~J. and {Esterhuyse}, S.~W.~P. and {Fanaroff}, B.~L. and {February}, S. and {Fourie}, D.~J. and {Frank}, B.~S. and {Gamatham}, R.~R.~G. and {Geyer}, M. and {Goedhart}, S. and {Gouws}, M. and {Gumede}, S.~C. and {Hlakola}, M.~J. and {Hokwana}, A. and {Hoosen}, S.~W. and {Horrell}, J.~M.~G. and {Hugo}, B. and {Isaacson}, A.~I. and {Jozsa}, G.~I.~G. and {Jonas}, J.~L. and {Joubert}, A.~F. and {Julie}, R.~P.~M. and {Kapp}, F.~B. and {Kenyon}, J.~S. and {Kotze}, P.~P.~A. and {Kriek}, N. and {Kriel}, H. and {Krishnan}, V.~K. and {Lehmensiek}, R. and {Liebenberg}, D. and {Lord}, R.~T. and {Lunsky}, B.~M. and {Madisa}, K. and {Magnus}, L.~G. and {Mahgoub}, O. and {Makhaba}, A. and {Makhathini}, S. and {Malan}, J.~A. and {Manley}, J.~R. and {Marais}, S.~J. and {Martens}, A. and {Mauch}, T. and {Merry}, B.~C. and {Millenaar}, R.~P. and {Mnyandu}, N. and {Mokone}, O.~J. and {Monama}, T.~E. and {Mphego}, M.~C. and {New}, W.~S. and {Ngcebetsha}, B. and {Ngoasheng}, K.~J. and {Ockards}, M.~T. and {Oozeer}, N. and {Otto}, A.~J. and {Passmoor}, S.~S. and {Patel}, A.~A. and {Peens-Hough}, A. and {Perkins}, S.~J. and {Ramaila}, A.~J.~T. and {Ramanujam}, N.~M.~R. and {Ramudzuli}, Z.~R. and {Ratcliffe}, S.~M. and {Robyntjies}, A. and {Salie}, S. and {Sambu}, N. and {Schollar}, C.~T.~G. and {Schwardt}, L.~C. and {Schwartz}, R.~L. and {Serylak}, M. and {Siebrits}, R. and {Sirothia}, S.~K. and {Slabber}, M. and {Smirnov}, O.~M. and {Sofeya}, L. and {Taljaard}, B. and {Tasse}, C. and {Tiplady}, A.~J. and {Toruvanda}, O. and {Twum}, S.~N. and {van Balla}, T.~J. and {van der Byl}, A. and {van der Merwe}, C. and {Van Tonder}, V. and {Van Wyk}, R. and {Venter}, A.~J. and {Venter}, M. and {Wallace}, B.~H. and {Welz}, M.~G. and {Williams}, L.~P. and {Xaia}, B.},
        title = "{The 1.28 GHz MeerKAT Galactic Center Mosaic}",
      journal = {\apj},
     keywords = {565, 571, 1346, Astrophysics - Astrophysics of Galaxies},
         year = 2022,
        month = feb,
       volume = {925},
       number = {2},
          eid = {165},
        pages = {165},
          doi = {10.3847/1538-4357/ac449a},
archivePrefix = {arXiv},
       eprint = {2201.10541},
 primaryClass = {astro-ph.GA},
       adsurl = {https://ui.adsabs.harvard.edu/abs/2022ApJ...925..165H},
      adsnote = {Provided by the SAO/NASA Astrophysics Data System}
}

@ARTICLE{Haniff2007a,
       author = {{Haniff}, Chris},
        title = "{An introduction to the theory of interferometry}",
      journal = {\nar},
     keywords = {07.60.Ly, 42.25.-p, 42.25.Hz, 42.30.-d, 42.30.Kq, 95.55.-n, 95.55.Br, 95.75.Kk, Interferometers, Wave optics, Interference, Imaging and optical processing, Fourier optics, Astronomical and space-research instrumentation, Astrometric and interferometric instruments, Interferometry},
         year = 2007,
        month = oct,
       volume = {51},
       number = {8-9},
        pages = {565-575},
          doi = {10.1016/j.newar.2007.06.002},
       adsurl = {https://ui.adsabs.harvard.edu/abs/2007NewAR..51..565H},
      adsnote = {Provided by the SAO/NASA Astrophysics Data System}
}

@ARTICLE{Haniff2007b,
       author = {{Haniff}, Chris},
        title = "{Ground-based optical interferometry: A practical primer}",
      journal = {\nar},
     keywords = {07.60.Ly, 42.15.Eq, 42.25.Hz, 42.68.Bz, 95.55.-n, 95.55.Br, 95.75.Kk, Interferometers, Optical system design, Interference, Atmospheric turbulence effects, Astronomical and space-research instrumentation, Astrometric and interferometric instruments, Interferometry},
         year = 2007,
        month = oct,
       volume = {51},
       number = {8-9},
        pages = {583-596},
          doi = {10.1016/j.newar.2007.06.004},
       adsurl = {https://ui.adsabs.harvard.edu/abs/2007NewAR..51..583H},
      adsnote = {Provided by the SAO/NASA Astrophysics Data System}
}

@INPROCEEDINGS{Mignon_Risse2021,
       author = {{Mignon-Risse}, R. and {Aimar}, N. and {Varniere}, P. and {Casse}, F. and {Vincent}, F.},
        title = "{A Possible Instability Origin for the Flares in Sagittarius A*: Linking Simulations and Observations}",
     keywords = {Sagittarius A*, flare, submillimeter},
    booktitle = {SF2A-2021: Proceedings of the Annual meeting of the French Society of Astronomy and Astrophysics. Eds.: A. Siebert},
         year = 2021,
       editor = {{Siebert}, A. and {Baillie}, K. and {Lagadec}, E. and {Lagarde}, N. and {Malzac}, J. and {Marquette}, J. -B. and {N'Diaye}, M. and {Richard}, J. and {Venot}, O.},
        month = dec,
        pages = {113-116},
       adsurl = {https://ui.adsabs.harvard.edu/abs/2021sf2a.conf..113M},
      adsnote = {Provided by the SAO/NASA Astrophysics Data System}
}

@ARTICLE{Jiang2023,
       author = {{Jiang}, Hong-Xuan and {Mizuno}, Yosuke and {Fromm}, Christian M. and {Nathanail}, Antonios},
        title = "{Two-temperature GRMHD simulations of black hole accretion flows with multiple magnetic loops}",
      journal = {\mnras},
     keywords = {accretion disc, black hole physics, magnetic reconnection, methods: numerical, Astrophysics - High Energy Astrophysical Phenomena},
         year = 2023,
        month = jun,
       volume = {522},
       number = {2},
        pages = {2307-2324},
          doi = {10.1093/mnras/stad1106},
archivePrefix = {arXiv},
       eprint = {2304.06230},
 primaryClass = {astro-ph.HE},
       adsurl = {https://ui.adsabs.harvard.edu/abs/2023MNRAS.522.2307J},
      adsnote = {Provided by the SAO/NASA Astrophysics Data System}
}

@ARTICLE{Nathanail2022a,
       author = {{Nathanail}, Antonios and {Mpisketzis}, Vasilis and {Porth}, Oliver and {Fromm}, Christian M. and {Rezzolla}, Luciano},
        title = "{Magnetic reconnection and plasmoid formation in three-dimensional accretion flows around black holes}",
      journal = {\mnras},
     keywords = {black hole physics, magnetic reconnection, accretion, accretion discs, magnetohydrodynamics, Astrophysics - High Energy Astrophysical Phenomena, General Relativity and Quantum Cosmology, Physics - Plasma Physics},
         year = 2022,
        month = jul,
       volume = {513},
       number = {3},
        pages = {4267-4277},
          doi = {10.1093/mnras/stac1118},
archivePrefix = {arXiv},
       eprint = {2111.03689},
 primaryClass = {astro-ph.HE},
       adsurl = {https://ui.adsabs.harvard.edu/abs/2022MNRAS.513.4267N},
      adsnote = {Provided by the SAO/NASA Astrophysics Data System}
}

@ARTICLE{Nathanail2022b,
       author = {{Nathanail}, Antonios and {Dhang}, Prasun and {Fromm}, Christian M.},
        title = "{Magnetic field structure in the vicinity of a supermassive black hole in low-luminosity galaxies: the case of Sgr A*}",
      journal = {\mnras},
     keywords = {Black hole physics, Magnetohydrodynamics, Sgr A*, Astrophysics - High Energy Astrophysical Phenomena},
         year = 2022,
        month = jul,
       volume = {513},
       number = {4},
        pages = {5204-5210},
          doi = {10.1093/mnras/stac1276},
archivePrefix = {arXiv},
       eprint = {2205.12287},
 primaryClass = {astro-ph.HE},
       adsurl = {https://ui.adsabs.harvard.edu/abs/2022MNRAS.513.5204N},
      adsnote = {Provided by the SAO/NASA Astrophysics Data System}
}

@ARTICLE{Yoon2020,
       author = {{Yoon}, D. and {Chatterjee}, K. and {Markoff}, S.~B. and {van Eijnatten}, D. and {Younsi}, Z. and {Liska}, M. and {Tchekhovskoy}, A.},
        title = "{Spectral and imaging properties of Sgr A* from high-resolution 3D GRMHD simulations with radiative cooling}",
      journal = {\mnras},
     keywords = {accretion, accretion discs, black hole physics, MHD, methods: numerical, stars: jets, galaxies: individual: (SgrA*), Astrophysics - High Energy Astrophysical Phenomena},
         year = 2020,
        month = dec,
       volume = {499},
       number = {3},
        pages = {3178-3192},
          doi = {10.1093/mnras/staa3031},
archivePrefix = {arXiv},
       eprint = {2009.14227},
 primaryClass = {astro-ph.HE},
       adsurl = {https://ui.adsabs.harvard.edu/abs/2020MNRAS.499.3178Y},
      adsnote = {Provided by the SAO/NASA Astrophysics Data System}
}

@ARTICLE{Broderick2006,
       author = {{Broderick}, Avery E. and {Loeb}, Abraham},
        title = "{Imaging optically-thin hotspots near the black hole horizon of Sgr A* at radio and near-infrared wavelengths}",
      journal = {\mnras},
     keywords = {black hole physics, polarization, techniques: interferometric, Galaxy: centre, infrared: general, submillimetre, Astrophysics},
         year = 2006,
        month = apr,
       volume = {367},
       number = {3},
        pages = {905-916},
          doi = {10.1111/j.1365-2966.2006.10152.x},
archivePrefix = {arXiv},
       eprint = {astro-ph/0509237},
 primaryClass = {astro-ph},
       adsurl = {https://ui.adsabs.harvard.edu/abs/2006MNRAS.367..905B},
      adsnote = {Provided by the SAO/NASA Astrophysics Data System}
}

@article{Li2013,
author = {Li, Min and Tang, Hai-Bin and Ren, Jun-Xue and York, Thomas},
year = {2013},
month = {10},
pages = {3502-},
title = {Modeling of plasma processes in the slowly diverging magnetic fields at the exit of an applied-field magnetoplasmadynamic thruster},
volume = {20},
journal = {Physics of Plasmas},
doi = {10.1063/1.4824619}
}

@INPROCEEDINGS{Marrone2006,
       author = {{Marrone}, Daniel P. and {Moran}, James M. and {Zhao}, Jun-Hui and {Rao}, Ramprasad},
        title = "{The Submillimeter Polarization of Sgr A*}",
     keywords = {Astrophysics},
    booktitle = {Journal of Physics Conference Series},
         year = 2006,
       series = {Journal of Physics Conference Series},
       volume = {54},
        month = dec,
        pages = {354-362},
          doi = {10.1088/1742-6596/54/1/056},
archivePrefix = {arXiv},
       eprint = {astro-ph/0607432},
 primaryClass = {astro-ph},
       adsurl = {https://ui.adsabs.harvard.edu/abs/2006JPhCS..54..354M},
      adsnote = {Provided by the SAO/NASA Astrophysics Data System}
}

@ARTICLE{EHT2019,
       author = {{Event Horizon Telescope Collaboration} and {Akiyama}, Kazunori and {Alberdi}, Antxon and {Alef}, Walter and {Asada}, Keiichi and {Azulay}, Rebecca and {Baczko}, Anne-Kathrin and {Ball}, David and {Balokovic}, Mislav and {Barrett}, John and {Bintley}, Dan and {Blackburn}, Lindy and {Boland}, Wilfred and {Bouman}, Katherine L. and {Bower}, Geoffrey C. and {Bremer}, Michael and {Brinkerink}, Christiaan D. and {Brissenden}, Roger and {Britzen}, Silke and {Broderick}, Avery E. and {Broguiere}, Dominique and {Bronzwaer}, Thomas and {Byun}, Do-Young and {Carlstrom}, John E. and {Chael}, Andrew and {Chan}, Chi-kwan and {Chatterjee}, Shami and {Chatterjee}, Koushik and {Chen}, Ming-Tang and {Chen}, Yongjun and {Cho}, Ilje and {Christian}, Pierre and {Conway}, John E. and {Cordes}, James M. and {Crew}, Geoffrey B. and {Cui}, Yuzhu and {Davelaar}, Jordy and {De Laurentis}, Mariafelicia and {Deane}, Roger and {Dempsey}, Jessica and {Desvignes}, Gregory and {Dexter}, Jason and {Doeleman}, Sheperd S. and {Eatough}, Ralph P. and {Falcke}, Heino and {Fish}, Vincent L. and {Fomalont}, Ed and {Fraga-Encinas}, Raquel and {Freeman}, William T. and {Friberg}, Per and {Fromm}, Christian M. and {Gomez}, Jose L. and {Galison}, Peter and {Gammie}, Charles F. and {Garcia}, Roberto and {Gentaz}, Olivier and {Georgiev}, Boris and {Goddi}, Ciriaco and {Gold}, Roman and {Gu}, Minfeng and {Gurwell}, Mark and {Hada}, Kazuhiro and {Hecht}, Michael H. and {Hesper}, Ronald and {Ho}, Luis C. and {Ho}, Paul and {Honma}, Mareki and {Huang}, Chih-Wei L. and {Huang}, Lei and {Hughes}, David H. and {Ikeda}, Shiro and {Inoue}, Makoto and {Issaoun}, Sara and {James}, David J. and {Jannuzi}, Buell T. and {Janssen}, Michael and {Jeter}, Britton and {Jiang}, Wu and {Johnson}, Michael D. and {Jorstad}, Svetlana and {Jung}, Taehyun and {Karami}, Mansour and {Karuppusamy}, Ramesh and {Kawashima}, Tomohisa and {Keating}, Garrett K. and {Kettenis}, Mark and {Kim}, Jae-Young and {Kim}, Junhan and {Kim}, Jongsoo and {Kino}, Motoki and {Koay}, Jun Yi and {Koch}, Patrick M. and {Koyama}, Shoko and {Kramer}, Michael and {Kramer}, Carsten and {Krichbaum}, Thomas P. and {Kuo}, Cheng-Yu and {Lauer}, Tod R. and {Lee}, Sang-Sung and {Li}, Yan-Rong and {Li}, Zhiyuan and {Lindqvist}, Michael and {Liu}, Kuo and {Liuzzo}, Elisabetta and {Lo}, Wen-Ping and {Lobanov}, Andrei P. and {Loinard}, Laurent and {Lonsdale}, Colin and {Lu}, Ru-Sen and {MacDonald}, Nicholas R. and {Mao}, Jirong and {Markoff}, Sera and {Marrone}, Daniel P. and {Marscher}, Alan P. and {Marti-Vidal}, Ivan and {Matsushita}, Satoki and {Matthews}, Lynn D. and {Medeiros}, Lia and {Menten}, Karl M. and {Mizuno}, Yosuke and {Mizuno}, Izumi and {Moran}, James M. and {Moriyama}, Kotaro and {Moscibrodzka}, Monika and {Muller}, Cornelia and {Nagai}, Hiroshi and {Nagar}, Neil M. and {Nakamura}, Masanori and {Narayan}, Ramesh and {Narayanan}, Gopal and {Natarajan}, Iniyan and {Neri}, Roberto and {Ni}, Chunchong and {Noutsos}, Aristeidis and {Okino}, Hiroki and {Olivares}, Hector and {Ortiz-Leon}, Gisela N. and {Oyama}, Tomoaki and {Ozel}, Feryal and {Palumbo}, Daniel C.~M. and {Patel}, Nimesh and {Pen}, Ue-Li and {Pesce}, Dominic W. and {Pietu}, Vincent and {Plambeck}, Richard and {PopStefanija}, Aleksandar and {Porth}, Oliver and {Prather}, Ben and {Preciado-Lopez}, Jorge A. and {Psaltis}, Dimitrios and {Pu}, Hung-Yi and {Ramakrishnan}, Venkatessh and {Rao}, Ramprasad and {Rawlings}, Mark G. and {Raymond}, Alexander W. and {Rezzolla}, Luciano and {Ripperda}, Bart and {Roelofs}, Freek and {Rogers}, Alan and {Ros}, Eduardo and {Rose}, Mel and {Roshanineshat}, Arash and {Rottmann}, Helge and {Roy}, Alan L. and {Ruszczyk}, Chet and {Ryan}, Benjamin R. and {Rygl}, Kazi L.~J. and {Sanchez}, Salvador and {Sanchez-Arguelles}, David and {Sasada}, Mahito and {Savolainen}, Tuomas and {Schloerb}, F. Peter and {Schuster}, Karl-Friedrich and {Shao}, Lijing and {Shen}, Zhiqiang and {Small}, Des and {Sohn}, Bong Won and {SooHoo}, Jason and {Tazaki}, Fumie and {Tiede}, Paul and {Tilanus}, Remo P.~J. and {Titus}, Michael and {Toma}, Kenji and {Torne}, Pablo and {Trent}, Tyler and {Trippe}, Sascha and {Tsuda}, Shuichiro and {van Bemmel}, Ilse and {van Langevelde}, Huib Jan and {van Rossum}, Daniel R. and {Wagner}, Jan and {Wardle}, John and {Weintroub}, Jonathan and {Wex}, Norbert and {Wharton}, Robert and {Wielgus}, Maciek and {Wong}, George N. and {Wu}, Qingwen and {Young}, Ken and {Young}, Andre and {Younsi}, Ziri and {Yuan}, Feng and {Yuan}, Ye-Fei and {Zensus}, J. Anton and {Zhao}, Guangyao and {Zhao}, Shan-Shan and {Zhu}, Ziyan and {Algaba}, Juan-Carlos and {Allardi}, Alexander and {Amestica}, Rodrigo and {Anczarski}, Jadyn and {Bach}, Uwe and {Baganoff}, Frederick K. and {Beaudoin}, Christopher and {Benson}, Bradford A. and {Berthold}, Ryan and {Blanchard}, Jay M. and {Blundell}, Ray and {Bustamente}, Sandra and {Cappallo}, Roger and {Castillo-Dominguez}, Edgar and {Chang}, Chih-Cheng and {Chang}, Shu-Hao and {Chang}, Song-Chu and {Chen}, Chung-Chen and {Chilson}, Ryan and {Chuter}, Tim C. and {Cordova Rosado}, Rodrigo and {Coulson}, Iain M. and {Crawford}, Thomas M. and {Crowley}, Joseph and {David}, John and {Derome}, Mark and {Dexter}, Matthew and {Dornbusch}, Sven and {Dudevoir}, Kevin A. and {Dzib}, Sergio A. and {Eckart}, Andreas and {Eckert}, Chris and {Erickson}, Neal R. and {Everett}, Wendeline B. and {Faber}, Aaron and {Farah}, Joseph R. and {Fath}, Vernon and {Folkers}, Thomas W. and {Forbes}, David C. and {Freund}, Robert and {Gomez-Ruiz}, Arturo I. and {Gale}, David M. and {Gao}, Feng and {Geertsema}, Gertie and {Graham}, David A. and {Greer}, Christopher H. and {Grosslein}, Ronald and {Gueth}, Frederic and {Haggard}, Daryl and {Halverson}, Nils W. and {Han}, Chih-Chiang and {Han}, Kuo-Chang and {Hao}, Jinchi and {Hasegawa}, Yutaka and {Henning}, Jason W. and {Hernandez-Gomez}, Antonio and {Herrero-Illana}, Ruben and {Heyminck}, Stefan and {Hirota}, Akihiko and {Hoge}, James and {Huang}, Yau-De and {Impellizzeri}, C.~M. Violette and {Jiang}, Homin and {Kamble}, Atish and {Keisler}, Ryan and {Kimura}, Kimihiro and {Kono}, Yusuke and {Kubo}, Derek and {Kuroda}, John and {Lacasse}, Richard and {Laing}, Robert A. and {Leitch}, Erik M. and {Li}, Chao-Te and {Lin}, Lupin C. -C. and {Liu}, Ching-Tang and {Liu}, Kuan-Yu and {Lu}, Li-Ming and {Marson}, Ralph G. and {Martin-Cocher}, Pierre L. and {Massingill}, Kyle D. and {Matulonis}, Callie and {McColl}, Martin P. and {McWhirter}, Stephen R. and {Messias}, Hugo and {Meyer-Zhao}, Zheng and {Michalik}, Daniel and {Montana}, Alfredo and {Montgomerie}, William and {Mora-Klein}, Matias and {Muders}, Dirk and {Nadolski}, Andrew and {Navarro}, Santiago and {Neilsen}, Joseph and {Nguyen}, Chi H. and {Nishioka}, Hiroaki and {Norton}, Timothy and {Nowak}, Michael A. and {Nystrom}, George and {Ogawa}, Hideo and {Oshiro}, Peter and {Oyama}, Tomoaki and {Parsons}, Harriet and {Paine}, Scott N. and {Penalver}, Juan and {Phillips}, Neil M. and {Poirier}, Michael and {Pradel}, Nicolas and {Primiani}, Rurik A. and {Raffin}, Philippe A. and {Rahlin}, Alexandra S. and {Reiland}, George and {Risacher}, Christopher and {Ruiz}, Ignacio and {Saez-Madain}, Alejandro F. and {Sassella}, Remi and {Schellart}, Pim and {Shaw}, Paul and {Silva}, Kevin M. and {Shiokawa}, Hotaka and {Smith}, David R. and {Snow}, William and {Souccar}, Kamal and {Sousa}, Don and {Sridharan}, T.~K. and {Srinivasan}, Ranjani and {Stahm}, William and {Stark}, Anthony A. and {Story}, Kyle and {Timmer}, Sjoerd T. and {Vertatschitsch}, Laura and {Walther}, Craig and {Wei}, Ta-Shun and {Whitehorn}, Nathan and {Whitney}, Alan R. and {Woody}, David P. and {Wouterloot}, Jan G.~A. and {Wright}, Melvin and {Yamaguchi}, Paul and {Yu}, Chen-Yu and {Zeballos}, Milagros and {Zhang}, Shuo and {Ziurys}, Lucy},
        title = "{First M87 Event Horizon Telescope Results. I. The Shadow of the Supermassive Black Hole}",
      journal = {\apjl},
     keywords = {accretion, accretion disks, black hole physics, galaxies: active, galaxies: individual: M87, galaxies: jets, gravitation, Astrophysics - Astrophysics of Galaxies, Astrophysics - High Energy Astrophysical Phenomena, General Relativity and Quantum Cosmology},
         year = 2019,
        month = apr,
       volume = {875},
       number = {1},
          eid = {L1},
        pages = {L1},
          doi = {10.3847/2041-8213/ab0ec7},
archivePrefix = {arXiv},
       eprint = {1906.11238},
 primaryClass = {astro-ph.GA},
       adsurl = {https://ui.adsabs.harvard.edu/abs/2019ApJ...875L...1E},
      adsnote = {Provided by the SAO/NASA Astrophysics Data System}
}

@ARTICLE{Scoville2003,
       author = {{Scoville}, N.~Z. and {Stolovy}, S.~R. and {Rieke}, M. and {Christopher}, M. and {Yusef-Zadeh}, F.},
        title = "{Hubble Space Telescope Pa{\ensuremath{\alpha}} and 1.9 Micron Imaging of Sagittarius A West}",
      journal = {\apj},
     keywords = {ISM: Dust, Extinction, Galaxy: Center, ISM: H II Regions, Infrared: ISM, Astrophysics},
         year = 2003,
        month = sep,
       volume = {594},
       number = {1},
        pages = {294-311},
          doi = {10.1086/376790},
archivePrefix = {arXiv},
       eprint = {astro-ph/0305350},
 primaryClass = {astro-ph},
       adsurl = {https://ui.adsabs.harvard.edu/abs/2003ApJ...594..294S},
      adsnote = {Provided by the SAO/NASA Astrophysics Data System}
}

@ARTICLE{Fritz2011,
       author = {{Fritz}, T.~K. and {Gillessen}, S. and {Dodds-Eden}, K. and {Lutz}, D. and {Genzel}, R. and {Raab}, W. and {Ott}, T. and {Pfuhl}, O. and {Eisenhauer}, F. and {Yusef-Zadeh}, F.},
        title = "{Line Derived Infrared Extinction toward the Galactic Center}",
      journal = {\apj},
     keywords = {dust, extinction, Galaxy: center, Galaxy: fundamental parameters, Astrophysics - Astrophysics of Galaxies},
         year = 2011,
        month = aug,
       volume = {737},
       number = {2},
          eid = {73},
        pages = {73},
          doi = {10.1088/0004-637X/737/2/73},
archivePrefix = {arXiv},
       eprint = {1105.2822},
 primaryClass = {astro-ph.GA},
       adsurl = {https://ui.adsabs.harvard.edu/abs/2011ApJ...737...73F},
      adsnote = {Provided by the SAO/NASA Astrophysics Data System}
}

@ARTICLE{Muno2003,
       author = {{Muno}, M.~P. and {Baganoff}, F.~K. and {Bautz}, M.~W. and {Brandt}, W.~N. and {Broos}, P.~S. and {Feigelson}, E.~D. and {Garmire}, G.~P. and {Morris}, M.~R. and {Ricker}, G.~R. and {Townsley}, L.~K.},
        title = "{A Deep Chandra Catalog of X-Ray Point Sources toward the Galactic Center}",
      journal = {\apj},
     keywords = {Catalogs, Galaxy: Center, X-Rays: General, Astrophysics},
         year = 2003,
        month = may,
       volume = {589},
       number = {1},
        pages = {225-241},
          doi = {10.1086/374639},
archivePrefix = {arXiv},
       eprint = {astro-ph/0301371},
 primaryClass = {astro-ph},
       adsurl = {https://ui.adsabs.harvard.edu/abs/2003ApJ...589..225M},
      adsnote = {Provided by the SAO/NASA Astrophysics Data System}
}

@ARTICLE{Paumard2004,
       author = {{Paumard}, T. and {Maillard}, J. -P. and {Morris}, M.},
        title = "{Kinematic and structural analysis of the <ASTROBJ>Minispiral</ASTROBJ> in the Galactic Center from BEAR spectro-imagery}",
      journal = {\aap},
     keywords = {Galaxy: center, ISM: individual objects: Sgr A West, ISM: kinematics and dynamics, infrared: ISM, Instrumentation: interferometers, line: profiles, Astrophysics},
         year = 2004,
        month = oct,
       volume = {426},
        pages = {81-96},
          doi = {10.1051/0004-6361:20034209},
archivePrefix = {arXiv},
       eprint = {astro-ph/0405197},
 primaryClass = {astro-ph},
       adsurl = {https://ui.adsabs.harvard.edu/abs/2004A&A...426...81P},
      adsnote = {Provided by the SAO/NASA Astrophysics Data System}
}

@ARTICLE{PTA2023,
       author = {{Antoniadis}, J. and {Babak}, S. and {Bak Nielsen}, A. -S. and {Bassa}, C.~G. and {Berthereau}, A. and {Bonetti}, M. and {Bortolas}, E. and {Brook}, P.~R. and {Burgay}, M. and {Caballero}, R.~N. and {Chalumeau}, A. and {Champion}, D.~J. and {Chanlaridis}, S. and {Chen}, S. and {Cognard}, I. and {Desvignes}, G. and {Falxa}, M. and {Ferdman}, R.~D. and {Franchini}, A. and {Gair}, J.~R. and {Goncharov}, B. and {Graikou}, E. and {Griessmeier}, J. -M. and {Guillemot}, L. and {Guo}, Y.~J. and {Hu}, H. and {Iraci}, F. and {Izquierdo-Villalba}, D. and {Jang}, J. and {Jawor}, J. and {Janssen}, G.~H. and {Jessner}, A. and {Karuppusamy}, R. and {Keane}, E.~F. and {Keith}, M.~J. and {Kramer}, M. and {Krishnakumar}, M.~A. and {Lackeos}, K. and {Lee}, K.~J. and {Liu}, K. and {Liu}, Y. and {Lyne}, A.~G. and {McKee}, J.~W. and {Main}, R.~A. and {Mickaliger}, M.~B. and {Nitu}, I.~C. and {Parthasarathy}, A. and {Perera}, B.~B.~P. and {Perrodin}, D. and {Petiteau}, A. and {Porayko}, N.~K. and {Possenti}, A. and {Samajdar}, H. Quelquejay Leclere A. and {Sanidas}, S.~A. and {Sesana}, A. and {Shaifullah}, G. and {Speri}, L. and {Spiewak}, R. and {Stappers}, B.~W. and {Susarla}, S.~C. and {Theureau}, G. and {Tiburzi}, C. and {van der Wateren}, E. and {Vecchio}, A. and {Venkatraman Krishnan}, V. and {Verbiest}, J.~P.~W. and {Wang}, J. and {Wang}, L. and {Wu}, Z.},
        title = "{The second data release from the European Pulsar Timing Array I. The dataset and timing analysis}",
      journal = {arXiv e-prints},
     keywords = {Astrophysics - High Energy Astrophysical Phenomena, Astrophysics - Astrophysics of Galaxies, Astrophysics - Instrumentation and Methods for Astrophysics, General Relativity and Quantum Cosmology},
         year = 2023,
        month = jun,
          eid = {arXiv:2306.16224},
        pages = {arXiv:2306.16224},
          doi = {10.48550/arXiv.2306.16224},
archivePrefix = {arXiv},
       eprint = {2306.16224},
 primaryClass = {astro-ph.HE},
       adsurl = {https://ui.adsabs.harvard.edu/abs/2023arXiv230616224A},
      adsnote = {Provided by the SAO/NASA Astrophysics Data System}
}

@ARTICLE{Alfven1976,
       author = {{Alfven}, H.},
        title = "{On frozen-in field lines and field-line reconnection}",
      journal = {\jgr},
     keywords = {Earth Magnetosphere, Geomagnetism, Lines Of Force, Magnetic Fields, Atmospheric Models, Magnetic Coils, Nomenclatures, Solar Wind, Geophysics},
         year = 1976,
        month = aug,
       volume = {81},
       number = {22},
        pages = {4019},
          doi = {10.1029/JA081i022p04019},
       adsurl = {https://ui.adsabs.harvard.edu/abs/1976JGR....81.4019A},
      adsnote = {Provided by the SAO/NASA Astrophysics Data System}
}

@INPROCEEDINGS{Sweet1958,
       author = {{Sweet}, P.~A.},
        title = "{The Neutral Point Theory of Solar Flares}",
    booktitle = {Electromagnetic Phenomena in Cosmical Physics},
         year = 1958,
       editor = {{Lehnert}, B.},
       volume = {6},
        month = jan,
        pages = {123},
       adsurl = {https://ui.adsabs.harvard.edu/abs/1958IAUS....6..123S},
      adsnote = {Provided by the SAO/NASA Astrophysics Data System}
}

@ARTICLE{Parker1957,
       author = {{Parker}, E.~N.},
        title = "{Sweet's Mechanism for Merging Magnetic Fields in Conducting Fluids}",
      journal = {\jgr},
     keywords = {MAGNETIC FIELDS, THEORY},
         year = 1957,
        month = dec,
       volume = {62},
       number = {4},
        pages = {509-520},
          doi = {10.1029/JZ062i004p00509},
       adsurl = {https://ui.adsabs.harvard.edu/abs/1957JGR....62..509P},
      adsnote = {Provided by the SAO/NASA Astrophysics Data System}
}

@ARTICLE{Zweibel2009,
       author = {{Zweibel}, Ellen G. and {Yamada}, Masaaki},
        title = "{Magnetic Reconnection in Astrophysical and Laboratory Plasmas}",
      journal = {\araa},
         year = 2009,
        month = sep,
       volume = {47},
       number = {1},
        pages = {291-332},
          doi = {10.1146/annurev-astro-082708-101726},
       adsurl = {https://ui.adsabs.harvard.edu/abs/2009ARA&A..47..291Z},
      adsnote = {Provided by the SAO/NASA Astrophysics Data System}
}

\begin{singlespace}
\setlength\labelalphawidth{0em}
\small\printbibliography[heading=bibintoc,title=Bibliographie]
\end{singlespace}

\pagestyle{fancy}
    \fancyhf{} 
    \fancyhead[LE]{\selectfont\nouppercase{\leftmark}}
    \fancyhead[RO]{\selectfont\nouppercase{\rightmark}}
    \fancyfoot[C]{\thepage}

\newpage
\thispagestyle{plain}
\mbox{}
\newpage 
\begin{appendices}
\addcontentsline{toc}{part}{Annexes}
\chapter{Article A\&A : Magnetic reconnection plasmoid model for Sagittarius A* flares} \label{ap:Paper}

Article accepté pour publication dans la revue \textit{Astronomy} \& \textit{Astrophysics} le 27 Janvier 2023 \cite{Aimar2023}.
\includepdf[pages={1-}, scale=0.9,pagecommand={\thispagestyle{fancy}}]{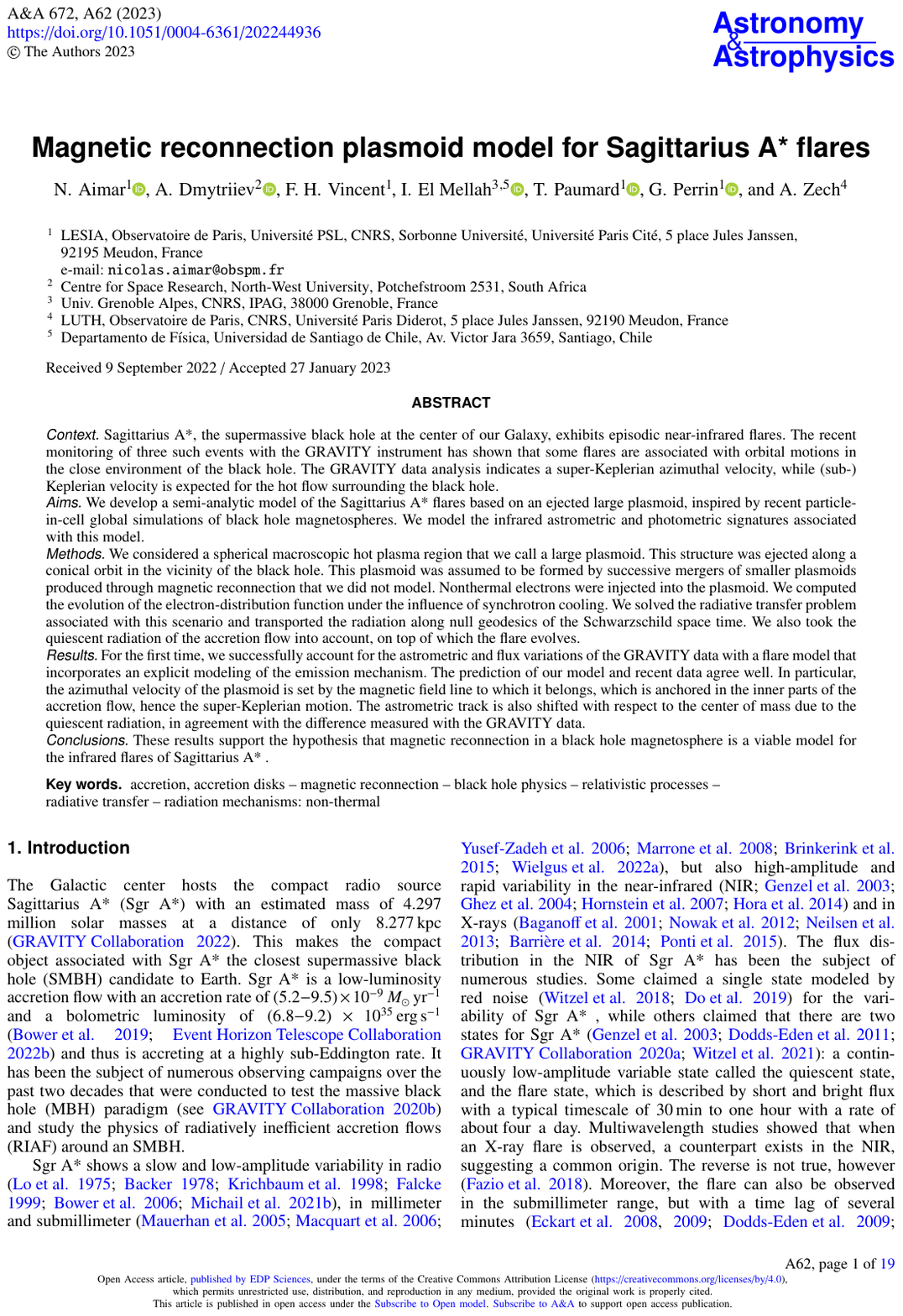}
\chapter{Conversion CGS-SI} \label{ap:Units}
\section{Expressions de formules entre CGS et SI}
Dans certains cas, en particulier liée à l'électromagnétique, l'expression des formules peut changer en fonction du système d'unité utilisé. Cela vient de la définition de la perméabilité magnétique $\mu_0$ et permittivité diélectrique $\epsilon_0$ du vide qui diffèrent dans ces deux types de système d'unité. En effet, pour simplifier les équations et éviter d'introduire de nouvelles grandeurs, la perméabilité et permittivité ont définies telles que $\mu_0 = \epsilon_0 = 1$ dans le système CGS (centimètre-gramme-seconde). Il existe différent type de système CGS avec des définitions différentes, on exprime ici la définition la plus commune pour étudier le genre de système de cette thèse.

De ce fait, l'expression des formules liées à ces grandeurs dépendent du système. Voici quelques exemples, notamment les formules utilisées dans les Chaps.~\ref{chap:modele_hotspot+jet} et \ref{chap:Plasmoid Flare model} :
\begin{table}[ht]
    \centering
    \begin{tabular}{l r c c}
         & & CGS & SI \\
        Rayon classique de l'électron & $r_0$ =  & $e^2/m_e c^2$ & $\frac{e^2/m_e c^2}{4 \pi \epsilon_0}$ \\
        Densité d'énergie magnétique & $U_B$ = & $B^2/8\pi$ & $ B^2/2\mu_0$ \\
        Équation de Maxwell-Gauss & $\vec{\nabla} \cdot \vec{E}$ = & $4 \pi \rho$ & $\frac{\rho}{\epsilon_0}$ \\
        Équation de Maxwell-Thomson & $\vec{\nabla} \cdot \vec{B}$ = & $0$ & $0$ \\
        Équation de Maxwell-Faraday & $\vec{\nabla} \times \vec{E}$ = & $-\frac{1}{c} \frac{\partial \vec{B}}{\partial t}$ & $-\frac{\partial \vec{B}}{\partial t}$ \\
        Équation de Maxwell-Ampère & $\vec{\nabla} \times \vec{B}$ = & $\frac{1}{c} \left(4 \pi \vec{J} + \frac{\partial \vec{E}}{\partial t} \right)$ & $\mu_0 \left(\vec{J} + \epsilon_0\frac{\partial \vec{E}}{\partial t} \right)$ \\
    \end{tabular}
\end{table}
\newpage
\thispagestyle{plain}
\mbox{}
\newpage
\chapter{Formules d'approximations pour les coefficients synchrotron} \label{ap:coefs_synchrotron}
\section{Coefficients non polarisés}
On donne ici les formules d'approximation des coefficients synchrotron ajusté par \cite{Pandya2016} pour des distributions thermiques, en loi de puissance et kappa. Pour chacune des trois distributions, le coefficient d'émission synchrotron peut s'écrire \cite{Pandya2016}
\begin{equation} \label{eq:ap_emis}
    j_\nu = \frac{N_e e^2 \nu_{crit}}{c} \mathbf{J_S} \left( \frac{\nu}{\nu_{crit}}, \alpha \right),
\end{equation}
et le coefficient d'absorption
\begin{equation}\label{eq:ap_absor}
    \alpha_\nu = \frac{N_e e^2}{\nu m_e c} \mathbf{A_S} \left( \frac{\nu}{\nu_{crit}}, \alpha \right),
\end{equation}
où $J_S$ et $A_S$ sont respectivement les coefficients d'émission et d'absorption sans dimension qui dépendent de la distribution considérée et $\alpha$ l'angle entre la direction d'émission et le champ magnétique. Ainsi, toute la complexité de la double intégration et de la dépendance à la distribution
est contenue dans les coefficients sans dimension qui sont ici des formules d’ajustement~\cite{Pandya2016}.

Pour une \textbf{distribution thermique}, le coefficient d'émission sans dimension s'écrit
\begin{equation} \label{eq:emission_thermique_Pandya_thermique}
    J_S^{th} = e^{-X^{1/3}} \frac{\sqrt{2} \pi}{27} \sin{\alpha} (X^{1/2} + 2^{11/12} X^{1/6})^2,
\end{equation}
et le coefficient d'absorption s'obtient à partir de la loi de \textit{Kirchhoff}
\begin{equation}
    A_S^{th} = \frac{J_S}{B_\nu} = J_S \frac{m_e c^2 \nu_{crit}}{2 h \nu^2} (e^{h \nu /(kT)}-1)
\end{equation}
où $B_\nu$ est la fonction de Planck et $X=\nu/\nu_{crit}$. Ces formules sont valables pour $3 < \Theta_e < 40$ avec une erreur relative maximale de $35\%$.\\

Pour une \textbf{distribution en loi de puissance}, ces coefficients s'écrivent
\begin{equation} \label{eq:emission_thermique_Pandya_PL}
    J_S^{PL} =\frac{3^{p/2}(p-1) \sin{\alpha}}{2 (p+1)(\gamma_{min}^{1-p}-\gamma_{max}^{1-p})} \times \Gamma \left( \frac{3p-1}{12} \right) \Gamma \left( \frac{3p+19}{12} \right) \left( \frac{\nu}{\nu_{crit} \sin{\alpha}} \right)^{-(p-1)/2},
\end{equation}
\begin{equation}
    A_S^{PL} = \frac{3^{(p+1)/2}(p-1)}{4 (\gamma_{min}^{1-p}-\gamma_{max}^{1-p})} \times \Gamma \left( \frac{3p+12}{12} \right) \Gamma \left( \frac{3p+22}{12} \right) \left( \frac{\nu}{\nu_{crit} \sin{\alpha}} \right)^{-(p+2)/2}
\end{equation}
pour $\gamma_{min}^2 < \nu/\nu_{crit} < \gamma_{max}^2$ et $1,5 < p < 6,5$ avec une erreur relative maximale de $35\%$.\\

Enfin, pour une \textbf{distribution Kappa}, on distingue deux régimes de fréquences en fonction de la valeur de $X_\kappa = \nu/\nu_\kappa$ avec $\nu_\kappa = \nu_{crit} (w \kappa)^2$. Dans le régime des basses fréquences
\begin{equation}
    J_{S,lo} = X_\kappa^{1/3} \sin{\alpha} \frac{4 \pi \Gamma(\kappa-4/3)}{3^{7/3} \Gamma (\kappa-2)}
\end{equation}
et dans le régime des hautes fréquences
\begin{equation}
    J_{S,hi} = X_\kappa^{-(\kappa-2)/2} \sin{\alpha}\ 3^{(\kappa-1)/2} \frac{(\kappa-2)(\kappa-1)}{4} \Gamma \left( \frac{\kappa}{4} - \frac{1}{3} \right) \Gamma \left( \frac{\kappa}{4} + \frac{4}{3} \right).
\end{equation}

Le coefficient d'émission sans dimension final s'écrit
\begin{equation}
    J_S^\kappa = (J_{S,lo}^{-x}+J_{S,hi}^{-x})^{-1/x}
\end{equation}
avec $x=3\kappa^{-3/2}$. De manière similaire pour le coefficient d'absorption, il s'écrit dans les basses fréquences
\begin{multline}
    A_{S,lo} = X_\kappa^{-2/3} 3^{1/6} \frac{10}{41} \frac{2 \pi}{(w \kappa)^{10/3-\kappa}} \frac{(\kappa-2)(\kappa-1)\kappa}{3\kappa-1}\\
    \times \Gamma\left( \frac{5}{3} \right) \ _2F_1 \left( \kappa-\frac{1}{3}, \kappa+1, \kappa+\frac{2}{3}, -\kappa w \right)
\end{multline}
où $_2F_1$ est une fonction hypergéométrique. Pour les hautes fréquences, il s'écrit
\begin{equation}
    A_{S,hi} = X_\kappa^{-(1+\kappa}/2 \frac{\pi^{3/2}}{3} \frac{(\kappa-2)(\kappa-1) \kappa}{(w \kappa)^3}\\
    \times \left( \frac{2\Gamma(2+\kappa/2)}{2+\kappa} -1 \right) \left( \left(\frac{3}{\kappa}\right)^{19/4}+\frac{3}{5} \right).
\end{equation}

Le coefficient d'absorption sans dimension final s'écrit
\begin{equation}
    A_S^\kappa = (A_{S,lo}^{-x}+A_{S,hi}^{-x})^{-1/x}
\end{equation}
avec $x=\left( -\frac{7}{4} + \frac{8}{5}\kappa \right)^{-43/50}$. Ces approximations sont valables pour $3 < w < 40$ et $2,5 < \kappa < 7,5$ avec une erreur relative maximale de $40\%$.

\section{Coefficients polarisés}
On garde les mêmes expressions pour les coefficients d'émission et d'absorption que dans le cas non polarisé (Eqs.~\eqref{eq:ap_emis} et \eqref{eq:ap_absor}) en nommant $\theta_B$ l'angle entre la direction d'émission et le champ magnétique. $S$ fait ici référence à l'un des quatre paramètres de Stokes. Les formules ci-dessous proviennent de~\cite{Pandya2021}.

\begin{itemize}
    \item[] \textbf{Distribution thermique :}
\end{itemize}
Cette distribution est paramétrée par la température sans dimension $\Theta_e$ et la norme du champ magnétique $B$ pour le calcul de la fréquence cyclotron (Eq.~\eqref{eq:cyclotron_freq}). Les coefficients d'émission sans dimension sont
\begin{equation}
    J_S = \exp(-X^{1/3}) \times \left\{
    \begin{array}{ll}
        \frac{\sqrt{2} \pi}{27} \sin \theta_B (X^{1/2}+2^{11/12}X^{1/6})^2, & \mbox{ (Stokes $I$)}\\
        \frac{\sqrt{2} \pi}{27} \sin \theta_B (X^{1/2}+\frac{7 \Theta_e^{24/25}+35}{10 \Theta_e^{24/25}+75}2^{11/12}X^{1/6})^2, & \mbox{ (Stokes $Q$)}\\
        0, & \mbox{ (Stokes $U$)}\\
        \frac{1}{\Theta_e} \cos{\theta_B} (\frac{\pi}{3} + \frac{\pi}{3}X^{1/3}+\frac{2}{300}X^{1/2}+\frac{2 \pi}{19}X^{2/3}). & \mbox{ (Stokes $V$)}
    \end{array}
    \right.
\end{equation}
avec $X=\nu/\nu_s$ et $\nu_s=(2/9)\nu_c \sin{\theta_B} \Theta_e^2$.
Les coefficients d'absorption sont déterminés à partir de la loi de Kirchhoff
\begin{equation}
    j_s - \alpha B_\nu = 0,
\end{equation}
avec $B_\nu$ la fonction de Planck d'un corps noir. Les coefficients de rotation Faraday s'écrivent
\begin{equation}
    r_Q = \frac{N_e e^2 \nu_c^2 \sin^2 \theta_B}{m_e c \nu^3} f_m(X) \left[ \frac{K_1(\Theta_e^{-1})}{K_2(\Theta_e^{-1})} +6\Theta_e \right],
\end{equation}
où
\begin{equation}
    f_m(X) = f_0(X) + \left[ 0,011 \exp(-1,69X^{-1/2})-0,003135X^{4/3} \right] \left( \frac{1}{2} \left[1+\tanh(10 \ln(0,6648X^{-1/2}))\right] \right),
\end{equation}
avec
\begin{equation}
    f_0(X) = 2,011 \exp(-19,78X^{-0,5175})- \cos(39,89X^{-1/2})\exp(-70,16X^{-0,6})-0,011\exp(-1,69X^{-1/2}),
\end{equation}
et
\begin{equation}
    r_V = \frac{2 N_e e^2 \nu_c}{m_e c \nu^2} \frac{K_0(\Theta_e^{-1})- \Delta J_5(X)}{K_2(\Theta_e^{-1}} \cos \theta_B,
\end{equation}
avec
\begin{equation}
    \Delta J_5(X) = 0,4379 \ln(1+1,3414X^{-0,7515})
\end{equation}
et $K_n$ la fonction de Bessel modifié du second type et d'ordre $n$. Ces expressions sont valables pour $\nu/\nu_c \gg 1$.

\begin{itemize}
    \item[] \textbf{Distribution en loi de puissance :}
\end{itemize}
Comme son nom l'indique, cette distribution peut être paramétrée par un indice de loi de puissance $p$, un facteur de Lorentz minimal $\gamma_\mathrm{min}$ et maximal $\gamma_\mathrm{max}$ et le champ magnétique comme précédemment. Les coefficients d'émission sans dimension s'écrivent
\begin{equation}
\begin{aligned}
    J_S &= \frac{3^{p/2}(p-1)\sin \theta_B}{2(p+1)(\gamma_\mathrm{min}^{1-p}-\gamma_\mathrm{max}^{1-p})} \\
    &\times \Gamma\left(\frac{3p-1}{12} \right) \Gamma\left(\frac{3p+19}{12} \right) \left(\frac{\nu}{\nu_c \sin \theta_B} \right)^{-(p-1)/2}\\
    &\times \left\{
    \begin{array}{ll}
        1, & \mbox{ (Stokes $I$)}\\
        \frac{p+1}{p+7/3}, & \mbox{ (Stokes $Q$)}\\
        0, & \mbox{ (Stokes $U$)}\\
        \frac{171}{250} \frac{p^{49/100}}{\tan \theta_B} \left( \frac{\nu}{3\nu_c \sin \theta_B} \right)^{-1/2}, & \mbox{ (Stokes $V$)}
    \end{array}
    \right.
\end{aligned}
\end{equation}
les coefficients d'absorption sans dimension
\begin{equation}
\begin{aligned}
    A_S &= \frac{3^{(p+1)/2}(p-1)}{4(\gamma_\mathrm{min}^{1-p}-\gamma_\mathrm{max}^{1-p})} \\
    &\times \Gamma\left(\frac{3p+21}{12} \right) \Gamma\left(\frac{3p+22}{12} \right) \left(\frac{\nu}{\nu_c \sin \theta_B} \right)^{-(p+21)/2} \\
    &\times \left\{
    \begin{array}{ll}
        1, & \mbox{ (Stokes $I$)}\\
        (\frac{17}{500}p-\frac{43}{1250})^{43/500}, & \mbox{ (Stokes $Q$)}\\
        0, & \mbox{ (Stokes $U$)}\\
        (\frac{71}{100}p+\frac{22}{625})^{197/500} (\frac{31}{10}(\sin \theta_B)^{-48/25}-\frac{31}{10})^{64/125} \left(\frac{\nu}{\nu_c \sin \theta_B} \right)^{-1/2} \mathrm{sgn}(\cos \theta_B). & \mbox{ (Stokes $V$)}
    \end{array}
    \right.
\end{aligned}
\end{equation}
et enfin les coefficients de rotation Faraday
\begin{equation}
    r_Q = r_\perp \left( \frac{\nu_c \sin \theta_B}{\nu} \right)^3 \, \frac{\gamma_\mathrm{min}^{2-p}}{(p/2)-1} \left[ 1- \left( \frac{2 \nu_c \gamma_\mathrm{min}^2 \sin \theta_B}{3 \nu} \right)^{p/2-1} \right],
\end{equation}
\begin{equation}
    r_V = 2 r_\perp \frac{p+2}{p+1} \left( \frac{\nu_c \sin \theta_B}{\nu} \right)^2 \gamma_\mathrm{min}^{-(p+1)} \ln(\gamma_\mathrm{min}) \cot \theta_B,
\end{equation}
où
\begin{equation}
    r_\perp = \frac{N_e e^2}{m_e c \nu_c \sin \theta_B} (p-1) \left[ \gamma_\mathrm{min}^{1-p}-\gamma_\mathrm{max}^{1-p} \right]^{-1}.
\end{equation}
Ces formules analytiques d'ajustement sont valides pour $\gamma_\mathrm{min} \leq 10^2$ et $\nu/\nu_c \gg 1$. 

\begin{itemize}
    \item[] \textbf{Distribution kappa :}
\end{itemize}
Comme dit dans le Chap.~\ref{chap:modele_hotspot+jet}, cette distribution est une combinaison entre le deux premières avec une jonction fluide. Elle se caractérise par la température sans dimension ici appelée $w$ et l'indice $\kappa$, lié à l'indice de la loi de puissance $p$ par $\kappa = p+1$, et le champ magnétique $B$. Pour cette distribution, on distingue deux cas limite, \textit{fréquence faible} et \textit{fréquence élevée} déterminés par $X_\kappa = \nu/\nu_\kappa$ où $\nu_\kappa = \nu_c (w \kappa)^2 \sin \theta_B$. Les coefficients d'émission sans dimension dans le cas de fréquences faibles sont
\begin{equation}
\begin{aligned}
    J_{S,lo} &= X_\kappa^{1/3} \sin \theta_B \frac{4 \pi \Gamma(\kappa-4/3)}{3^{7/3}\Gamma(\kappa-2)} \\
    &\times \left\{
    \begin{array}{ll}
        1, & \mbox{ (Stokes $I$)}\\
        \frac{1}{2}, & \mbox{ (Stokes $Q$)}\\
        0, & \mbox{ (Stokes $U$)}\\
        \left( \frac{3}{4} \right)^2 \left[ (\sin \theta_B)^{-12/5} -1 \right]^{12/25} \frac{\kappa^{-66/125}}{w} X_\kappa^{-7/20}, & \mbox{ (Stokes $V$)}
    \end{array}
    \right.
\end{aligned}
\end{equation}
et dans le cas des hautes fréquences sont
\begin{equation}
\begin{aligned}
    J_{S,hi} &= X_\kappa^{-(\kappa-2)/2} \sin \theta_B 3^{(\kappa-1)/2} \frac{(\kappa-2)(\kappa-1)}{4} \Gamma\left( \frac{\kappa}{4} - \frac{1}{3} \right) \Gamma\left( \frac{\kappa}{4} + \frac{4}{3} \right)\\
    &\times \left\{
    \begin{array}{ll}
        1, & \mbox{ (Stokes $I$)}\\
        \left[ (\frac{4}{5})^2 + \frac{\kappa}{50} \right], & \mbox{ (Stokes $Q$)}\\
        0, & \mbox{ (Stokes $U$)}\\
        \left( \frac{7}{8} \right)^2 \left[ (\sin \theta_B)^{-5/2} -1 \right]^{11/25} \frac{\kappa^{-11/25}}{w} X_\kappa^{-1/2}. & \mbox{ (Stokes $V$)}
    \end{array}
    \right.
\end{aligned}
\end{equation}
Les coefficients d'émission sans dimension finaux sont calculés de la manière suivante
\begin{equation}
    J_S = \left\{
    \begin{array}{ll}
        \left( J_{S,lo}^{-x} + J_{S,hi}^{-x} \right)^{-1/x}, & \mbox{ (Stokes $I$)}\\
        \left( J_{S,lo}^{-x} + J_{S,hi}^{-x} \right)^{-1/x}, & \mbox{ (Stokes $Q$)}\\
        \left( J_{S,lo}^{-x} + J_{S,hi}^{-x} \right)^{-1/x} \mathrm{sgn}( \cos \theta_B), & \mbox{ (Stokes $V$)}
    \end{array}
    \right.
\end{equation}
avec
\begin{equation}
    x = \left\{
    \begin{array}{ll}
        3 \kappa^{-3/2}, & \mbox{ (Stokes $I$)}\\
        \frac{37}{10} \kappa^{-8/5}, & \mbox{ (Stokes $Q$)}\\
        3 \kappa^{-3/2}. & \mbox{ (Stokes $V$)}
    \end{array}
    \right.
\end{equation}

Pour les coefficients d'absorption sans dimension, on a
\begin{equation}
\begin{aligned}
    A_{S,lo} &= X_\kappa^{-2/3} 3^{1/6} \frac{2 \pi}{(w\kappa)^{10/3-\kappa}} \frac{(\kappa-2)(\kappa-1)\kappa}{3\kappa-1} \\
    &\times \Gamma \left(\frac{5}{3} \right) _2F_1 \left(\kappa-\frac{1}{3}, \kappa+1, \kappa+\frac{2}{3}, -w\kappa \right) \\
    &\times \left\{
    \begin{array}{ll}
        1, & \mbox{ (Stokes $I$)}\\
        \frac{25}{48}, & \mbox{ (Stokes $Q$)}\\
        0, & \mbox{ (Stokes $U$)}\\
        \frac{77}{100 w} \left[ (\sin \theta_B)^{-114/50} -1 \right]^{223/500} X_\kappa^{-7/20} \kappa^{-7/10}, & \mbox{ (Stokes $V$)}
    \end{array}
    \right.
\end{aligned}
\end{equation}
et
\begin{equation}
\begin{aligned}
    A_{S,hi} &= X_\kappa^{-(1+\kappa)/2} \frac{\pi^{3/2}}{3} \frac{(\kappa-2)(\kappa-1)}{(w \kappa)^3} \\
    &\times \left( \frac{\Gamma (2+\kappa/2)}{2+ \kappa} - 1 \right)\\
    &\times \left\{
    \begin{array}{ll}
        \left( \frac{3}{\kappa} \right)^{19/4} + \frac{3}{5}, & \mbox{ (Stokes $I$)}\\
        \left( 21^2 \kappa^{-(12/5)^2} + \frac{11}{20} \right), & \mbox{ (Stokes $Q$)}\\
        0, & \mbox{ (Stokes $U$)}\\
        \frac{143}{10} w^{-116/125} \left[ (\sin \theta_B)^{-41/20} -1 \right]^{1/2} \left[ 13^2 \kappa^{-8} +\frac{13}{2500}\kappa - \frac{263}{2500 + \frac{47}{200 \kappa}} \right] X_\kappa^{-1/2}. & \mbox{ (Stokes $V$)}
    \end{array}
    \right.
\end{aligned}
\end{equation}
Les coefficients d'émission sans dimension finaux sont calculés de la manière suivante
\begin{equation}
    A_S = \left\{
    \begin{array}{ll}
        \left( A_{S,lo}^{-x} + A_{S,hi}^{-x} \right)^{-1/x}, & \mbox{ (Stokes $I$)}\\
        \left( A_{S,lo}^{-x} + A_{S,hi}^{-x} \right)^{-1/x}, & \mbox{ (Stokes $Q$)}\\
        \left( A_{S,lo}^{-x} + A_{S,hi}^{-x} \right)^{-1/x} \mathrm{sgn}( \cos \theta_B), & \mbox{ (Stokes $V$)}
    \end{array}
    \right.
\end{equation}
avec
\begin{equation}
    x = \left\{
    \begin{array}{ll}
        \left( -\frac{7}{4+\frac{8}{5} \kappa} \right)^{-43/50}, & \mbox{ (Stokes $I$)}\\
        \frac{7}{5} \kappa^{-23/20}, & \mbox{ (Stokes $Q$)}\\
        \frac{61}{50} \kappa^{-142/125} + \frac{7}{1000}. & \mbox{ (Stokes $V$)}
    \end{array}
    \right.
\end{equation}

Enfin, les coefficients de rotation Faraday ont été ajustés pour quatre valeurs spécifiques de $\kappa$
\begin{equation}
\begin{aligned}
    r_Q &= \frac{N_e e^2 \nu_c^2 \sin^2 \theta_B}{m_e c \nu^3} f(X_\kappa) \\
    &\times \left\{
    \begin{array}{ll}
        17 w - 3 \sqrt{w} + 7 \sqrt{w} \exp(-5w), & \mbox{ ($\kappa=3,5$)}\\
        \frac{46}{3} w - \frac{5}{3} \sqrt{w} + 7 \sqrt{w} \exp(-5w), & \mbox{ ($\kappa=4,0$)}\\
        14 w - \frac{13}{8} \sqrt{w} + \frac{9}{2} \sqrt{w} \exp(-5w), & \mbox{ ($\kappa=4,5$)}\\
        \frac{25}{2} w - \sqrt{w} + 5 \sqrt{w} \exp(-5w), & \mbox{ ($\kappa=5,0$)}
    \end{array}
    \right.
\end{aligned}
\end{equation}
avec
\begin{equation}
    f(X) = \left\{
    \begin{array}{ll}
        1 - \exp(-\frac{X^{0,84}}{30}) - \sin(\frac{X}{10}) \exp(-\frac{3 X^{0,471}}{2}), & \mbox{ ($\kappa=3,5$)}\\
        1 - \exp(-\frac{X^{0,84}}{18}) - \sin(\frac{X}{6}) \exp(-\frac{7 X^{0,5}}{4}), & \mbox{ ($\kappa=4,0$)}\\
        1 - \exp(-\frac{X^{0,84}}{12}) - \sin(\frac{X}{4}) \exp(-2X^{0,525}), & \mbox{ ($\kappa=4,5$)}\\
        1 - \exp(-\frac{X^{0,84}}{8}) - \sin(\frac{3X}{8}) \exp(-\frac{9 X^{0,541}}{4}), & \mbox{ ($\kappa=5,0$)}
    \end{array}
    \right.
\end{equation}
et
\begin{equation}
\begin{aligned}
    r_V &= \frac{2N_e e^2 \nu_c \cos \theta_B}{m_e c \nu^2} \frac{K_0(\Theta_e^{-1}}{K_2(\Theta_e^{-1}} g(X_\kappa) \\
    &\times \left\{
    \begin{array}{ll}
        \frac{w^2 + 2w +1}{(25/8)w^2+4w+1}, & \mbox{ ($\kappa=3,5$)}\\
        \frac{w^2 + 54w +50}{(30/11)w^2+134w+50}, & \mbox{ ($\kappa=4,0$)}\\
        \frac{w^2 + 43w +38}{(7/3)w^2+(185/2)w+38}, & \mbox{ ($\kappa=4,5$)}\\
        \frac{w + (13/14)}{2w+(13/14)}, & \mbox{ ($\kappa=5,0$)}
    \end{array}
    \right.
\end{aligned}
\end{equation}
avec
\begin{equation}
    g(X) = \left\{
    \begin{array}{ll}
        1 - 0,17 \ln (1+0,447 X^{-1/2}), & \mbox{ ($\kappa=3,5$)}\\
        1 - 0,17 \ln (1+0,391 X^{-1/2}), & \mbox{ ($\kappa=4,0$)}\\
        1 - 0,17 \ln (1+0,348 X^{-1/2}), & \mbox{ ($\kappa=4,5$)}\\
        1 - 0,17 \ln (1+0,313 X^{-1/2}), & \mbox{ ($\kappa=5,0$)}
    \end{array}
    \right.
\end{equation}
Ces ajustements sont valables pour $X_\kappa \gg 10^-1$ et $\nu / \nu_c \gg 1$.
\newpage
\thispagestyle{plain}
\mbox{}
\newpage
\chapter{Complément tests de comparaison \textsc{Gyoto}-\textsc{ipole}.} \label{ap:Test_Polar}

\begin{figure}
    \centering
    \resizebox{0.8\hsize}{!}{\includegraphics{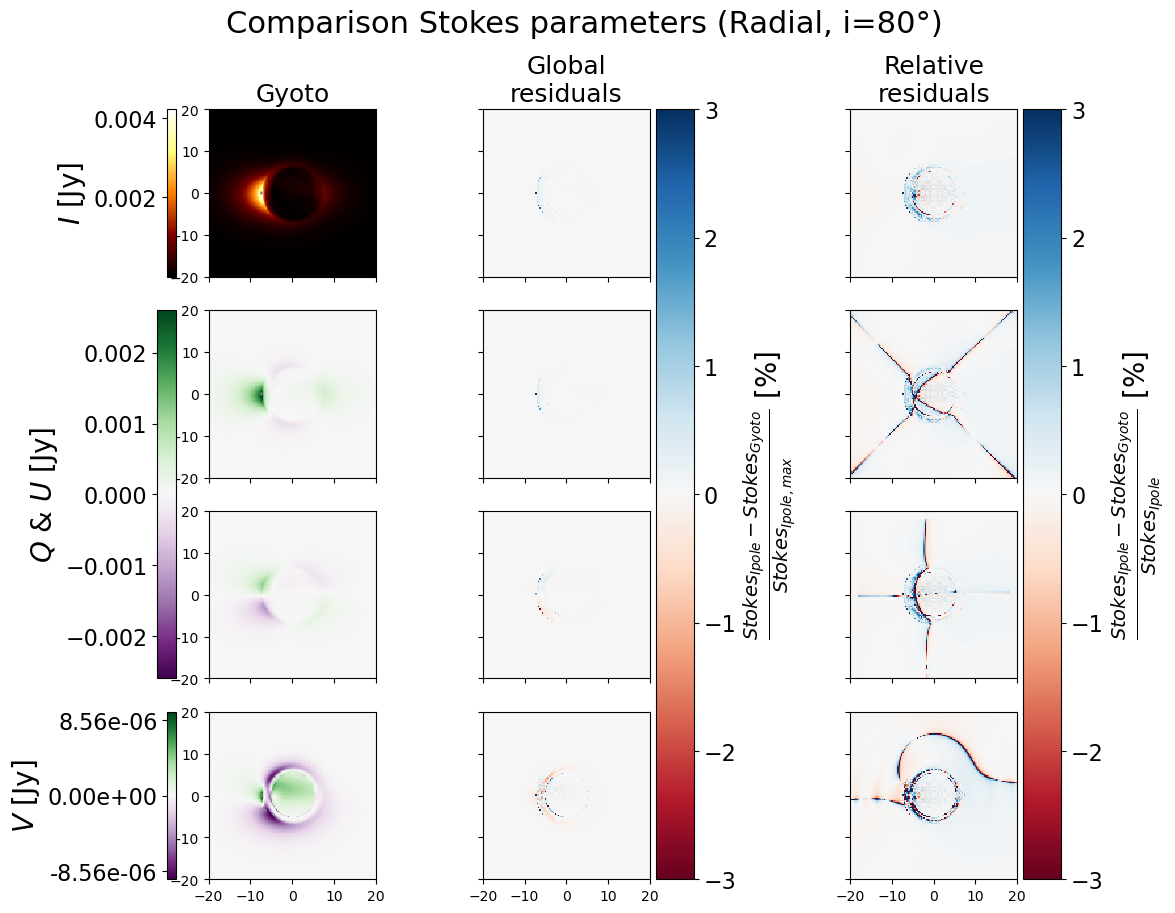}}
    \caption{Comparaison des images polarisées produite avec \textsc{ipole} et \textsc{Gyoto}. La première colonne montre l'image d'un disque épais, tel que définit plus haut, observé avec une inclinaison de $80 \degree$ et un champ magnétique radiale, dans les quatre paramètres de Stokes ($I$, $Q$, $U$ et $V$) en Jansky. L'échelle de couleur des paramètres $Q$ et $U$ est commune (puisqu'ils définissent la polarisation linéaire). La colonne centrale montre l'erreur globale pour chaque pixel entre \textsc{ipole} et \textsc{Gyoto} normalisé par la valeur maximale (d'\textsc{ipole}). Enfin, la dernière colonne montre l'erreur relative de chaque pixel pour chaque paramètre de Stokes. Le niveau de $3\%$ d'erreur globale est marqué par les contours gris dans la première colonne.}
    \label{fig:compareIpole_Radial_80deg}
\end{figure}

\begin{figure}
    \centering
    \resizebox{0.8\hsize}{!}{\includegraphics{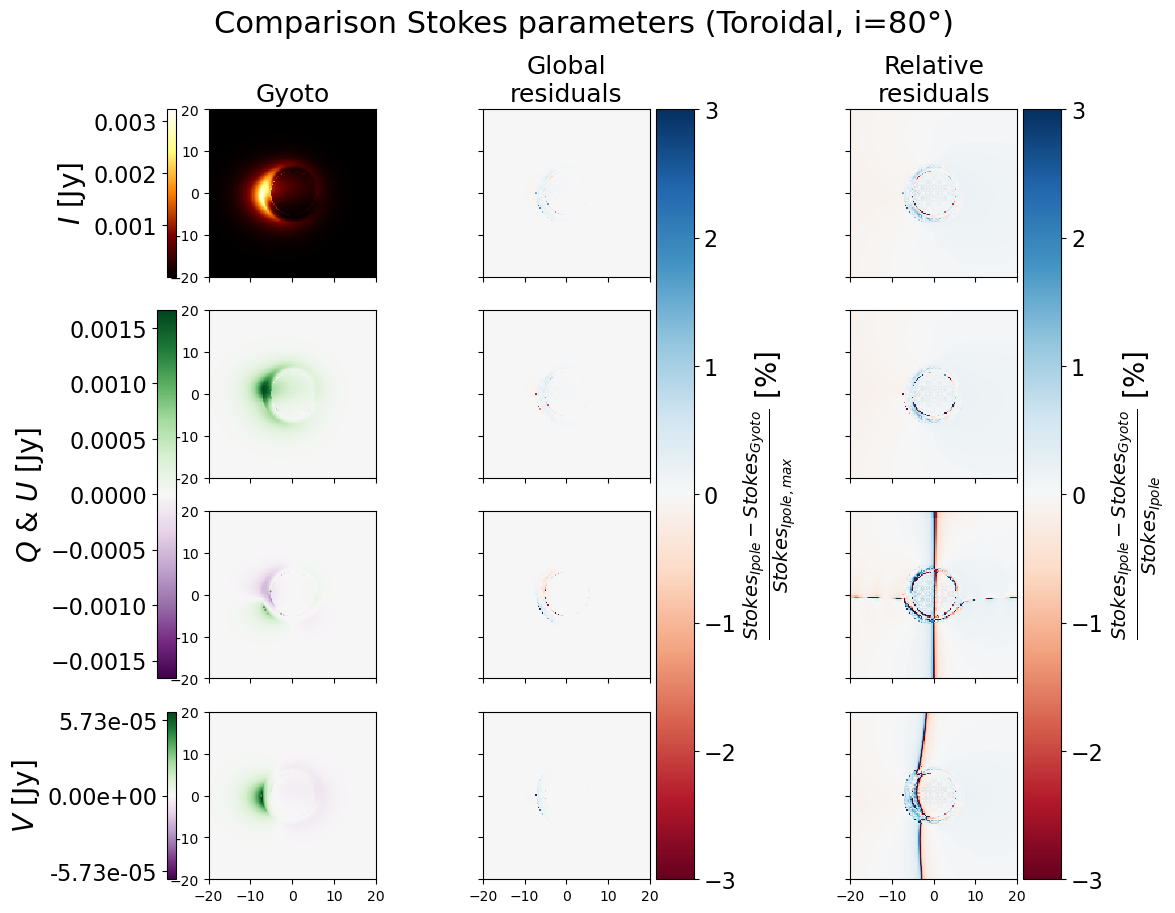}}
    \caption{Même chose qu'à la Fig.~\ref{fig:compareIpole_Radial_80deg} avec une configuration magnétique toroïdale.}
    \label{fig:compareIpole_Toroidal_80deg}

    \vspace{1cm}
    
    \resizebox{0.8\hsize}{!}{\includegraphics{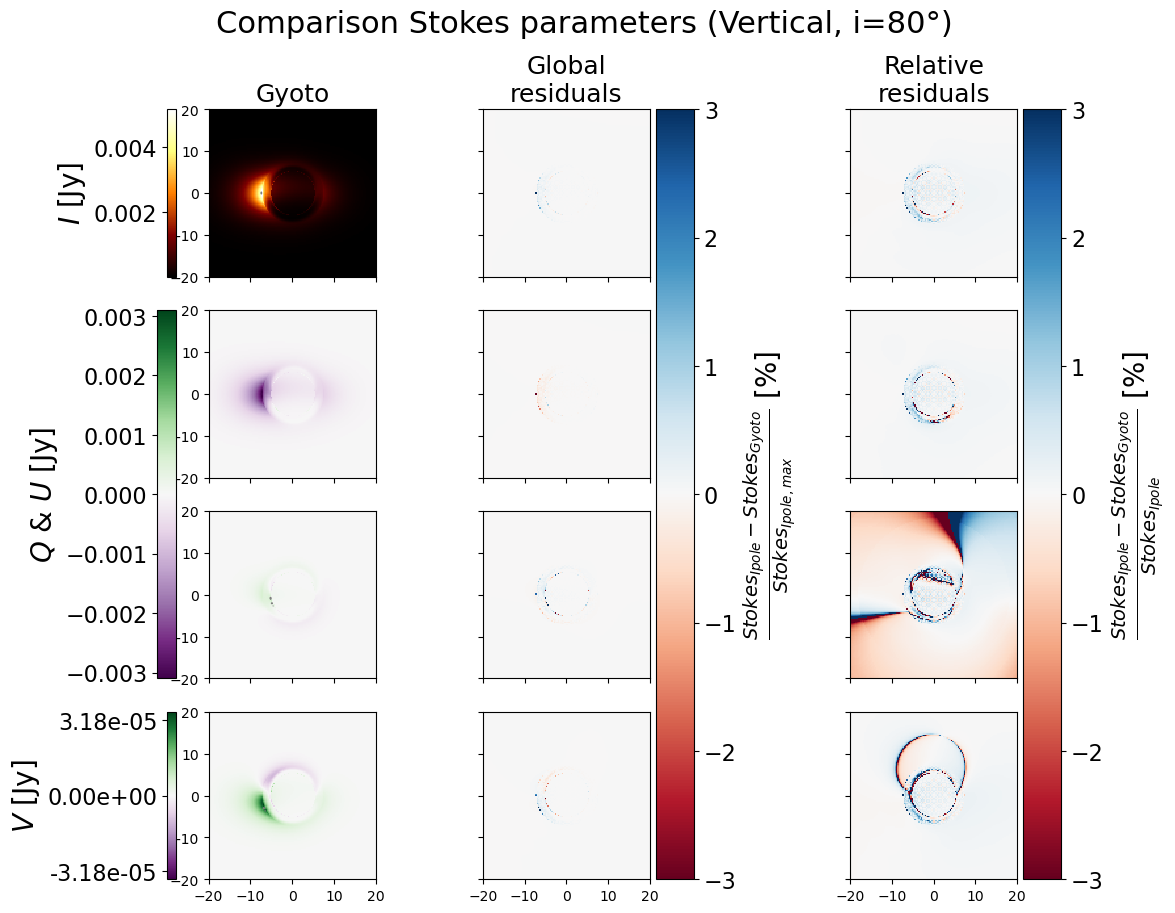}}
    \caption{Même chose qu'à la Fig.~\ref{fig:compareIpole_Radial_80deg} avec une configuration magnétique verticale.}
    \label{fig:compareIpole_Vertical_80deg}
\end{figure}
\end{appendices}

\newpage
\thispagestyle{plain}
\mbox{}
\newpage
\newpage
\thispagestyle{plain}
\mbox{}
\newpage

\end{document}